\documentclass[letterpaper,12pt,titlepage,oneside,final]{book}
 
\newcommand{\href}[1]{#1} 

\usepackage{ifthen}
\newboolean{PrintVersion}
\setboolean{PrintVersion}{false}

\usepackage{amsmath,amssymb,amstext} 
\usepackage{graphicx} 

\usepackage[pdftex,pagebackref=false,hypertexnames=false]{hyperref} 
\usepackage[capitalize]{cleveref} 
\crefname{appendix}{Appendix}{Appendices}

\hypersetup{
    plainpages=false,       
    unicode=false,          
    pdftoolbar=true,        
    pdfmenubar=true,        
    pdffitwindow=false,     
    pdfstartview={FitH},    
    pdftitle={Rigorous Security Proofs for Practical Quantum Key Distribution},    
pdfauthor={Devashish Tupkary},    
    pdfsubject={Physics},  
    pdfkeywords={qkd} {quantum information} {cryptography}, 
    pdfnewwindow=true,      
    colorlinks=true,        
    linkcolor=blue,         
    citecolor=green,        
    filecolor=magenta,      
    urlcolor=cyan           
}
\ifthenelse{\boolean{PrintVersion}}{   
\hypersetup{	
    citecolor=black,%
    filecolor=black,%
    linkcolor=black,%
    urlcolor=black}
}{} 

\usepackage[automake,toc,abbreviations]{glossaries-extra} 

\let\origdoublepage\cleardoublepage
\newcommand{\clearemptydoublepage}{%
  \clearpage{\pagestyle{empty}\origdoublepage}}
\let\cleardoublepage\clearemptydoublepage

\makeglossaries

\usepackage{amsmath}
\usepackage{mathtools}
\usepackage{graphicx}
\usepackage{amsfonts}
\usepackage{color}
\usepackage{microtype} 
\usepackage{subcaption} 
\usepackage{array}
\usepackage{tabularx}
\usepackage{booktabs} 

\usepackage{xspace}
\usepackage{enumitem}  
\usepackage{verbatim} 

\usepackage{xparse} 
\usepackage{xspace}
\usepackage{braket}
\usepackage{physics}
\usepackage{bm}

\usepackage{float}
\usepackage{mdframed}
\usepackage{enumitem}  
\makeindex
\usepackage{amsthm}
\usepackage{thmtools}
\usepackage{thm-restate}

\usepackage{amsfonts}
\usepackage{color}
\usepackage{microtype} 

\crefname{enumi}{}{} 
\usepackage{epigraph}
\newtheorem{theorem}{Theorem}[section]
\newtheorem{lemma}{Lemma}[section]
\newtheorem{corollary}{Corollary}[section]

\newtheorem{remark}{Remark}[section]

\theoremstyle{definition} 
\newtheorem{definition}{Definition}[section]

\newtheorem{prot}{Protocol}

\newcommand{\Pos}{\operatorname{Pos}}
\newcommand{\term}[1]{\textbf{#1}} 

\newcommand{\pf}{\operatorname{\mathtt{Pur}}} 

\newcommand{\id}{\mathbb{I}} 
\DeclareMathOperator{\idmap}{\mathord{\rm id}} 

\newcommand{\CPTP}{\operatorname{CPTP}}

\newcommand{\dop}[1]{\operatorname{S}_{#1}} 

\newcommand{\numstates}{\bar{d}_A}
\newcommand{\nummeas}{\bar{d}_B}
\newcommand{\unitary}{\operatorname{U}}
\newcommand{\isometry}{\operatorname{V}}
\newcommand{\linear}{\operatorname{L}}
\newcommand{\mapset}{\operatorname{T}}

\newcommand{\vac}{\ket{\mathrm{vac}}}

\newcommand{\Svirt}{\widetilde{S}}

\newcommand{\Fobs}{\mathbf{F}^{\mathrm{obs}}}
\newcommand{\Fbar}{\mathbf{\bar{F}}}

\newcommand{\Renyi}{R\'{e}nyi} 
\NewDocumentCommand{\Halpha}{o}{%
  \ensuremath{\widetilde{H}^{\uparrow\IfValueT{#1}{\,,#1}}_\alpha}%
}
  \NewDocumentCommand{\HalphaDown}{o}{%
\ensuremath{\widetilde{H}^{\downarrow\IfValueT{#1}{\,,#1}}_\alpha}%
  }

\NewDocumentCommand{\Hmin}{o}{%
  \ensuremath{H_{\mathrm{min}}^{\IfValueTF{#1}{#1}{\varepsilon}}}%
}

\NewDocumentCommand{\Hmax}{o}{%
  \ensuremath{H_{\mathrm{max}}^{\IfValueTF{#1}{#1}{\varepsilon}}}%
}

\newcommand{\permset}[1]{S_{#1}^\mathrm{perm}}

\newcommand{\Xbasis}{\mathsf{X}}
\newcommand{\Zbasis}{\mathsf{Z}}
\newcommand{\Kbasis}{\mathsf{K}}

\newcommand{\testgenflag}{z}
\newcommand{\test}{\mathtt{test}}
\newcommand{\gen}{\mathtt{gen}}
\newcommand{\beamsplit}{t}
\newcommand{\tracedist}[1]{\frac{1}{2} \norm{#1}_1}

\newcommand{\Sclick}{S_\mathrm{click}}
\newcommand{\nFSS}{N_\mathrm{FSS}}

\newcommand{\EVcost }{\ceil{\log(\frac{1}{\epsEV})}}

\newcommand{\bstat}{b_\mathrm{stat}}
\newcommand{\freq}{\operatorname{freq}}

\newcommand{\eps}{\varepsilon} 
\newcommand{\epssecure}{\eps^\mathrm{secure}}
\newcommand{\epscorr}{\eps^\mathrm{correct}}
\newcommand{\epssecret}{\eps^\mathrm{secret}}
\newcommand{\epsEV}{\eps_\mathrm{EV}}
\newcommand{\epsPA}{\eps_\mathrm{PA}}
\newcommand{\epsAT}{\eps_\mathrm{AT}} 
\newcommand{\epstilde}{\widetilde{\varepsilon}}

\newcommand{\epsATb}{\varepsilon_\text{AT-b}}

\newcommand{\epsATa}{\varepsilon_\text{AT-a}}

\newcommand{\epsATc}{\varepsilon_\text{AT-c}}
\newcommand{\epsdecoy}{\varepsilon_\text{AT-d}}
\newcommand{\epsATsingle}{\varepsilon_\text{AT-s}}

\newcommand{\Esecdef}{\bm{E}}
\newcommand{\Eve}{\bm{\hat{E}}}
 \newcommand{\keyspace}{\mathcal{K}}
\newcommand{\OlAlB}{\Omega_{l_A,l_B}}
\newcommand{\idealmap}{\mathcal{R}_\mathrm{ideal}}
\newcommand{\QKDprotocol}{\mathcal{P}_\mathrm{QKD}}
\newcommand{\epsbar}{\bar{\varepsilon}}

\newcommand{\CP}{\widehat{C}} 
\newcommand{\flagkey}{F_\mathrm{K}}
\newcommand{\flagEV}{F_\mathrm{EV}}
\newcommand{\flagEC}{F_\mathrm{EC}}

\newcommand{\CEC}{C_\mathrm{EC}}
\newcommand{\CEV}{C_\mathrm{EV}}

\newcommand{\HEV}{H_\mathrm{EV}}
\newcommand{\HPA}{H_\mathrm{PA}}

\newcommand{\announcementfunction}{f_\mathrm{ann}}
\newcommand{\keymapfunction}{f_\mathrm{kmap}}

\newcommand{\Ashield}{\hat{A}}
\newcommand{\Ameas}{\bar{A}}
\newcommand{\leak}{\lambda_\mathrm{EC}}
\newcommand{\leakfixed}{\lambda_\mathrm{EC}^\mathrm{fixed}} 
\newcommand{\lkey}{\ell} 
\newcommand{\QKDmapfullwithoutBobMeas}[1]{{\bar{\mathcal{G}}}^{\mathrm{full}}_{#1}}

\newcommand{\protMap}[1]{\mathcal{E}^{\left(#1\right)}_\mathrm{QKD}} 
\newcommand{\protMapId}[1]{\mathcal{E}^{(#1),\mathrm{ideal}}_\mathrm{QKD}} 

\newcommand{\protMapbeforeSR}[1]{\widetilde{\mathcal{E}}^{\left(#1\right)}_\mathrm{QKD}} 
\newcommand{\protMapIdbeforeSR}[1]{\widetilde{\mathcal{E}}^{(#1),\mathrm{ideal}}_\mathrm{QKD}} 

\newcommand{\protMapSquash}[1]{\bar{\mathcal{E}}^{\left(#1\right)}_\mathrm{QKD}} 

\newcommand{\lfixed}{\ell_{\mathrm{fixed}}} 
\newcommand{\PAstring}{\mbf{S}} 
\newcommand{\singleRoundBot}{\bot}

\newcommand{\cP}{\hat{c}}

\newcommand{\cobs}{\cP_1^n}
\newcommand{\allpublic}{\CP_1^n \CEC \CEV \HPA \HEV}

\newcommand{\CPhat}{\widetilde{C}}

\newcommand{\hashfamily}[2]{\mathcal{F}_{\mathrm{hash}}\left(#1,#2\right)}
\newcommand{\idealhashfamily}[2]{\mathcal{F}^\mathrm{ideal}_{\mathrm{hash}}\left(#1,#2\right)}

\newcommand{\block}{M_\mathrm{block}}

\newcommand{\accept}{\texttt{accept}\xspace}
\newcommand{\abort}{\texttt{abort}\xspace}
\newcommand{\authabort}{\texttt{auth-abort}\xspace}

\newcommand{\lprePA}{l_\mathrm{pre\text{-}PA}}

\newcommand{\acceptanceset}{\mathcal{S}_\mathrm{acc}}
\newcommand{\feasibleset}{\mathcal{S}_\mathrm{feas}}
\newcommand{\variableset}{\mathcal{V}}
\newcommand{\OmegaEV}{\Omega_\mathrm{EV}}
\newcommand{\OmegaAT}{\Omega_\mathrm{AT}}

\newcommand{\mbf}[1]{\mathbf{#1}} 
\newcommand{\floor}[1]{\left\lfloor #1 \right\rfloor} 
\newcommand{\ceil}[1]{\left\lceil #1 \right\rceil} 

\renewcommand{\complement}{\mathrm{C}} 

\newcommand{\QKDpostprocessingmap}{\mathcal{M}_\mathrm{CPP}} 
\newcommand{\QKDGmapfull}[1]{{\mathcal{G}}^{\mathrm{full}}_{#1}} 

\newcommand{\QKDGmap}[1]{\mathcal{G}_{#1}} 

\newcommand{\kappalowerfixed}[1]{\kappa^\mathrm{fixed,L}_{#1}}
\newcommand{\kappaupperfixed}[1]{\kappa^\mathrm{fixed,U}_{#1}}
\newcommand{\kappalowervariable}[1]{\kappa^\mathrm{var,L}_{#1}}
\newcommand{\kappauppervariable}[1]{\kappa^\mathrm{var,U}_{#1}}

\newcommand{\cost}{g_{n,x}} 

\newcommand{\C}[1]{{\mathbb{C}^{#1}}}
\newcommand{\Sym}[1]{\mathrm{Sym}^n\left(#1\right)}

\newcommand{\dimSym}[1]{\dim\left(\Sym{\C{#1}} \right)} 
\newcommand{\perm}{\pi} 
\newcommand{\permutation}[2]{{P^{#1,#2}_\perm}} 

\newcommand{\promise}{\sigma_A}

\newcommand{\state}{\rho_{A_1^nB_1^n}}

\newcommand{\blockdim}[2]{{d_{#1}^{#2}}} 

\newcommand{\measChannel}[1]{\mathcal{M}^{\mathrm{meas}}_{#1}}

\newcommand{\attack}[1]{\mathcal{A}_{#1}} 
\newcommand{\attackSquash}[1]{{\attack{#1}^{\mathrm{Sq}}}}

\newcommand{\cnotobs}{\vec{p}}
\newcommand{\EURset}{\mathcal{V}}
\newcommand{\event}[1]{\Omega{(#1)} }
\newcommand{\decoyparams}{\nXmu{\allk},\nKmu{\allk},\eXmu{\allk},\eZ}

\newcommand{\EURbeta}{\beta}

\newcommand{\pzA}{p^{(A)}_{(\Zbasis)}}
\newcommand{\pxA}{p^{(A)}_{(\Xbasis)}}
\newcommand{\pzB}{p^{(B)}_{(\Zbasis)}}
\newcommand{\pxB}{p^{(B)}_{(\Xbasis)}}
\newcommand{\AlicePOVM}[2]{\Gamma^{(A)}_{(#1,#2)}}
\newcommand{\BobPOVM}[2]{\Gamma^{(B)}_{(#1,#2)}}
\newcommand{\AliceBobPOVM}[2]{\Gamma_{(#1),(#2)}}
\newcommand{\AliceBobPOVMFilter}[2]{\tilde{F}_{(#1),(#2)}}
\newcommand{\AliceBobPOVMsecond}[2]{G^{\mathrm{con}}_{(#1),(#2)}}
\newcommand{\con}{\mathrm{con}}

\newcommand{\AliceBobPOVMsecondsandwich}[2]{G^{\text{con,F}}_{(#1),(#2)}}
\newcommand{\AliceBobPOVMprimeFilter}[2]{{F}_{(#1),(#2)}}

\newcommand{\nX}{n_{\Xbasis}}
\newcommand{\nXrv}{\bm{\nX}}

\newcommand{\nXmu}[1]{n_{\Xbasis,\mu_{#1}}}
\newcommand{\nXmurv}[1]{\bm{\nXmu{#1}}}
\newcommand{\nXph}[1]{n_{\Xbasis,{#1}}}
\newcommand{\nXphrv}[1]{\bm{\nXph{#1}}}

\newcommand{\nXneqmu}[1]{n_{\Xbasis_{\neq},\mu_{#1}}}
\newcommand{\nXneqmurv}[1]{\bm{\nXneqmu{#1}}}

\newcommand{\nXperf}{\tilde{n}_{\Xbasis}}
\newcommand{\nXperfrv}{\bm{\nXperf}}
\newcommand{\eX}{e^{\mathrm{obs}}_{\Xbasis}  }
\newcommand{\eXrv}{\bm{\eX}}

\newcommand{\eXmu}[1]{e^{\text{obs}}_{\Xbasis,\mu_{#1}}}
\newcommand{\eXmurv}[1]{\bm{\eXmu{#1}}}
\newcommand{\eXph}[1]{e^{\text{obs}}_{\Xbasis,#1}}
\newcommand{\eXphrv}[1]{\bm{\eXph{#1}}}

\newcommand{\eXperf}{\tilde{e}^{\text{obs}}_{\Xbasis \Xbasis}  }
\newcommand{\eXperfrv}{\bm{\eXperf}}
\newcommand{\X}{\bm{X}}

\newcommand{\Y}{\bm{Y}}

\newcommand{\fserf}{f_\mathrm{serf}}

\newcommand{\eZ}{e^{\text{obs}}_{\Zbasis} }

\newcommand{\eph}{e^{\mathrm{key}}_{\Xbasis} } 
\newcommand{\ephrv}{\bm{\eph}}
\newcommand{\ephph}[1]{e^\mathrm{key}_{\Xbasis,{#1}}}
\newcommand{\ephphrv}[1]{\bm{\ephph{#1}}}

\newcommand{\ephperf}{\tilde{e}^{\text{key}}_{\Zbasis \Xbasis} } 
\newcommand{\ephperfrv}{\bm{\ephperf}}

\newcommand{\ephwierd}{\tilde{e}^\text{key}_{\Xbasis \Xbasis}}
\newcommand{\ephwierdrv}{\bm{\ephwierd}}

\newcommand{\params}{\nX,\nK,\eX,\eZ}

\newcommand{\nK}{n_{\Kbasis}}
\newcommand{\nKrv}{\bm{\nK}}
\newcommand{\nKperf}{\tilde{n}_{\Kbasis}}
\newcommand{\nKperfrv}{\bm{\nKperf}}

\newcommand{\nKmu}[1]{n_{\Kbasis,\mu_{#1}}}
\newcommand{\nKmurv}[1]{\bm{\nKmu{#1}}}
\newcommand{\nKph}[1]{n_{\Kbasis,{#1}}}
\newcommand{\nKphrv}[1]{\bm{\nKph{#1}}}

\newcommand{\Bound}{\mathcal{B}}
\newcommand{\Boundbasiczero}{\Bound_{0,0}}
\newcommand{\Boundbasicdelta}{\Bound_{\deltaone,\deltatwo}}

\newcommand{\deltaone}{\delta_1}
\newcommand{\deltatwo}{\delta_2}

\newcommand{\serfbound}{\zeta}
\newcommand{\qualityfactor}{c_q}
\newcommand{\proj}{\mathrm{\Pi}}

\newcommand{\num}{\bm{N}}
\newcommand{\dc}{\text{dc}} 

\newcommand{\binfunction}[3]{F_\mathrm{bin}(#1,#2,#3)}

\newcommand{\cone}{c_1}
\newcommand{\ctwo}{c_2}

\newcommand{\nint}{N_\mathrm{int}}
\newcommand{\allk}{\vec{k}}

\newcommand{\constraints}{\mathcal{S}_\text{constraints}}
\newcommand{\Bounddecoymin}[1]{\Bound^\text{decoy}_{\text{min}-#1}}
\newcommand{\Bounddecoymax}[1]{\Bound^\text{decoy}_{\text{max}-#1}}

\newcommand{\nO}{n_O}
\newcommand{\nOrv}{\bm{\nO}}
\newcommand{\nOmu}[1]{n_{O,\mu_{#1}}}
\newcommand{\nOmurv}[1]{\bm{\nOmu{#1}}}

\newcommand{\nOph}[1]{n_{O,{#1}}}
\newcommand{\nOphrv}[1]{\bm{\nOph{#1}}}

\newcommand{\etadet}{\eta_{\text{det} }}
\newcommand{\dcprob}{d_{\text{det} }}
\newcommand{\etachar}{\Delta_{\eta}}
\newcommand{\dcchar}{\Delta_{\dc}}

\newcommand{\dMin}{d_{\text{min}}} 
\newcommand{\dMax}{d_{\text{max}}} 
\newcommand{\dMultAvg}{d_{\text{mult}}} 

\newcommand{\etaMax}{\eta_{\text{max}}} 
\newcommand{\etaMin}{\eta_{\text{min}}} 
\newcommand{\etaRenorm}{r_{\eta}} 
\newcommand{\BobPOVMswap}[2]{\Gamma^{(B),(\text{swap})}_{(#1,#2)}}

\newcommand{\mode}{\mathbf{d}}
\newcommand{\transpose}{{\mathsf T}}

\newcommand{\QKDGmapfullbeforeSR}[1]{\widetilde{{\mathcal{G}}}^{\mathrm{full}}_{#1}}

\newcommand{\sigmaconstraint}[1]{\sigma_{A_{#1}}^{(#1)}}

\newcommand{\attackset}[1]{\bm{\mathcal{A}}^\mathrm{set}_{#1}} 
\newcommand{\Qset}[1]{\bm{\mathcal{Q}}^\mathrm{set}_{#1}} 
\newcommand{\Qattack}[1]
{\mathcal{Q}_{#1}}

\newcommand{\sourcesymbol}{\sigma} 

\newcommand{\fhat}{\hat{f}}

\newcommand{\fhatfull}{\fhat_\mathrm{full}}

\newcommand{\fhatfullQKD}{\fhat^\mathrm{QKD}_\mathrm{full}}

\newcommand{\kappafuncgeneric}[4]{\kappa\left( #1 ,\; #2, \; #3, \; #4 \right)}

\newcommand{\kappaQKDfunc}[3]{\kappa^\mathrm{QKD}\left( #1 , #2,  #3 \right)}

\newcommand{\ndecoy}{{N_{\mathrm{ph}}}} 

\newcommand{\preservedSubspace}{\Pi_{m\leq \nFSS}}
\newcommand{\nonpreservedSubspace}{\Pi_{m> \nFSS}}
\newcommand{\pibar}{{\overline{\Pi}}} 
\newcommand{\pibarFlag}{\pibar^{\mathrm{Sq}}} 
\newcommand{\Gammacc}{\Gamma_{\mathrm{o}}} 
\newcommand{\lambdamin}{{\lambda_{\min}}}

\newcommand{\realistic}{practical}

\newcommand{\coreQKDprotocol}{\widetilde{\mathcal{P}}_\mathrm{QKD}}
\newcommand{\APPprotocol}{\mathcal{P}_\mathrm{APP}}
\newcommand{\delayedAPPprotocol}{\mathcal{P}^\mathrm{del}_\mathrm{APP}}
\newcommand{\delayedQKDprotocol}{\mathcal{P}^\mathrm{del}_\mathrm{QKD}}
\newcommand{\worldreal}{\mathcal{W}^\mathrm{real}_\mathrm{auth}}
\newcommand{\worlddelreal}{\mathcal{W}^\mathrm{real}_\mathrm{del-auth}}
\newcommand{\worldhonest}{\mathcal{W}^\mathrm{hon}_\mathrm{auth}}
\newcommand{\worldvirtual}{\mathcal{W}^\mathrm{virt}}
\newcommand{\projkeymap}{\Pi^{l^\prime_A,l^\prime_B}_{K_A K_B} }

\newcommand{\finalacceptabortevent}[2]{\Omega^\mathrm{fin-dec}_{(#1,#2)}}

\newcommand{\CAs}{\bm{{C_{A \rightarrow E}}} }
\newcommand{\CBs}{\bm{{C_{B \rightarrow E}}} }
\newcommand{\CAr}{\bm{{C_{E \rightarrow A}}} }
\newcommand{\CBr}{\bm{{C_{E \rightarrow B}}} }

\newcommand{\Cauth}{\bm{C_\mathrm{auth}}}
\newcommand{\CAsauth}{\bm{{C^\mathrm{auth}_{A \rightarrow E}}} }
\newcommand{\CBsauth}{ \bm{{C^\mathrm{auth}_{B \rightarrow E}}} }
\newcommand{\CArauth}{\bm{{C^\mathrm{auth}_{E \rightarrow A}}} }
\newcommand{\CBrauth}{\bm{ {C^\mathrm{auth}_{E \rightarrow B}}} }

\newcommand{\transcript}{\mathcal{T}}

\newcommand{\authmap}
{\mathcal{E}^\mathrm{repl}_\mathrm{auth}}

\newcommand{\authcommmap}{\mathcal{E}_\mathrm{comm}}
\newcommand{\authupdatemap}{\mathcal{E}_\mathrm{update}}

\newcommand{\delauthcommmap}{\mathcal{E}^\mathrm{del}_\mathrm{comm}}
\newcommand{\delauthupdatemap}{\mathcal{E}^\mathrm{del}_\mathrm{update}}
\newcommand{\EfinalQKD}{\bm{E_\mathrm{fin}}}
 \newcommand{\CfinalQKD}{\bm{C_\mathrm{fin}}}  
 \newcommand{\Efinal}{\bm{E^\prime_{\mathrm{fin}}}} 
\newcommand{\Cfinal}{\bm{C^\prime_{\mathrm{fin}}}} 
\newcommand{\Ccorr}{\widetilde{C}}

\newcommand{\OlAlBprime}{\Omega_{l^\prime_A,l^\prime_B}}
\newcommand{\Onice}{\Omega_{\mathrm{auth\text{-}hon}}}
\newcommand{\Onicedel}{\Omega_{\mathrm{dauth\text{-}hon}}}

 \newcommand{\poissonian}[2]{p^\mathrm{poiss}_{#1}(#2)}

\usepackage{tikz}
\usepackage{tikz-cd}

\usetikzlibrary{shapes.geometric,positioning,calc,arrows,arrows.meta}

\tikzstyle{arrow}       = [thick, ->, >=stealth]
\tikzstyle{dashedarrow} = [thick, ->, >=stealth, dashed]
\tikzstyle{connect}     = [thick, <->, >=stealth]

\tikzstyle{processLarge} =
  [rectangle, rounded corners, minimum width=2cm, minimum height=1cm,
   text centered, draw=black, align=center]

\tikzstyle{virtualprocess} =
  [rectangle, rounded corners, minimum width=3cm, minimum height=1cm,
   text centered, draw=black, dashed, align=center]

\tikzstyle{emptyprocess} =
  [rectangle, minimum width=3cm, minimum height=1cm, text centered, align=center]

\tikzstyle{processSmall} =
  [rectangle, line width=0.3mm, minimum width=1cm, minimum height=1cm,
   text centered, text width=1cm, draw=black]

\tikzstyle{point} = [circle, inner sep=0pt, minimum size=3pt, fill=black]

\tikzstyle{channel} =
  [rectangle, line width=0.7mm, minimum width=5cm, minimum height=3cm,
   text centered, draw=black, align=center]

\tikzset{
  pics/detectorLarge/.style args={#1,#2,#3}{
    code={
      \draw[line width=0.7mm] (0,-2) -- (0,2);
      \draw[line width=0.7mm] (0,2) -- ++(1,0);
      \draw[line width=0.7mm] (0,-2) -- ++(1,0);
      \draw[line width=0.7mm] (1,2) arc (90:-90:2);

      \node[#3, align=center] (#1) at (1.3,0) {\scalebox{0.9}{\huge #2}};
    }
  }
}

\tikzset{
  pics/detectorSmall/.style args={#1,#2,#3}{
    code={
      \draw[line width=0.5mm] (0,-1) -- (0,1);
      \draw[line width=0.5mm] (0,1) -- ++(0.1,0);
      \draw[line width=0.5mm] (0,-1) -- ++(0.1,0);
      \draw[line width=0.5mm] (0.1,1) arc (90:-90:1);

      \coordinate (#1) at (0,0);            
      \coordinate (#1-out) at (1.1,0);      

      \node[#3, align=center] (#1) at (0.5,0) {#2};
    }
  }
}

\tikzset{
  pics/beamsplitter/.style args={#1,#2,#3}{
    code={
      \coordinate (A) at (0.5,0.5);
      \coordinate (B) at (-0.5,-0.5);
      \coordinate (#1) at (0,0);
      \draw[line width=0.3mm] (A) -- (B);
      \node[align=center,font=\fontsize{9}{9}\selectfont] at #3 {#2};
    }
  }
}

\tikzset{
  pics/pulse/.style args={#1}{
    code={
      \draw[#1] plot[smooth,tension=1]
        coordinates {(0,0) (0.3,0.2) (0.5,1) (0.7,0.2) (1,0)};
    }
  }
}

\newlength\colsep   
\newlength\offsetX  
\newlength\offsetY  

\begin{document}

\pagestyle{empty}
\pagenumbering{roman}

\begin{titlepage}
        \begin{center}
        \vspace*{1.0cm}

        \Huge
        {\bf Rigorous Security Proofs for Practical Quantum Key Distribution  }   

        \vspace*{1.0cm}

        \normalsize
        by \\

        \vspace*{1.0cm}

        \Large
        Devashish Tupkary \\

        \vspace*{3.0cm}

        \normalsize
        A thesis \\
        presented to the University of Waterloo \\ 
        in fulfillment of the \\
        thesis requirement for the degree of \\
        Doctor of Philosophy \\
        in \\
        Physics (Quantum Information) \\

        \vspace*{2.0cm}

        Waterloo, Ontario, Canada, 2026 \\

        \vspace*{1.0cm}

        \copyright\ Devashish Tupkary 2026 \\
        \end{center}
\end{titlepage}

\pagestyle{plain}
\setcounter{page}{2}

\cleardoublepage 
\phantomsection    
 
\addcontentsline{toc}{chapter}{Examining Committee Membership}
\begin{center}\textbf{Examining Committee Membership}\end{center}
  \noindent
The following served on the Examining Committee for this thesis. The decision of the Examining Committee is by majority vote.
  \bigskip
  
  \noindent
\begin{tabbing}
Internal-External Member: \=  \kill 
External Examiner: \>  Marco Tomamichel \\ 
\> Professor, Dept. of Electrical and Computer Engineering,\\
\>National University of Singapore \\
\end{tabbing} 
  \bigskip
  
  \noindent
\begin{tabbing}
Internal-External Member: \=  \kill 
Supervisor(s): \> Norbert L\"{u}tkenhaus \\
\> Professor, Dept. of Physics and Astronomy, \\ \> University of Waterloo 
\end{tabbing}
  \bigskip
  
  \noindent
  \begin{tabbing}
Internal-External Member: \=  \kill 
Internal Member: \> Kevin Resch \\
\> Professor, Dept. of Physics and Astronomy, \\
\> University of Waterloo \\
\end{tabbing}
  \bigskip
  
  \noindent
\begin{tabbing}
Internal-External Member: \=  \kill 
Internal-External Member: \> Graeme Smith \\
\> Associate Professor, Dept. of Applied Mathematics,\\
\> University of Waterloo \\
\end{tabbing}
  \bigskip
  
  \noindent
\begin{tabbing}
Internal-External Member: \=  \kill 
Other Member(s): \> Michele Mosca \\
\> Professor, Deparment of Combinatorics \& Optimization, \\
\>University of Waterloo \\
\end{tabbing}

\cleardoublepage
\phantomsection    

 \addcontentsline{toc}{chapter}{Author's Declaration}
 \begin{center}\textbf{Author's Declaration}\end{center}

 \noindent
 \noindent  
 This thesis consists of material all of which I authored or co-authored: see Statement of Contributions included in the thesis. This is a true copy of the thesis, including any required final revisions, as accepted by my examiners.
  \bigskip
  
  \noindent
I understand that my thesis may be made electronically available to the public.

\cleardoublepage
\phantomsection    

\addcontentsline{toc}{chapter}{Statement of Contributions}
\begin{center}\textbf{Statement of Contributions}\end{center}
This thesis is based on the following works, of which I am first author:
\begin{itemize}
\item Ref.~\cite{tupkary_security_2024} (\cref{chap:variable}) : Devashish Tupkary, Ernest Y.-Z. Tan, and Norbert L\"utkenhaus. \textit{Security proof for
variable-length quantum key distribution.} \href{https://journals.aps.org/prresearch/abstract/10.1103/PhysRevResearch.6.023002}{Phys. Rev. Research 6, 023002}, April
2024.
\item Ref.~\cite{tupkary_phase_2024} (\cref{chap:EUR}) : Devashish Tupkary, Shlok Nahar, Pulkit Sinha, and Norbert L\"utkenhaus. \textit{Phase error
rate estimation in QKD with imperfect detectors.} \href{https://quantum-journal.org/papers/q-2025-12-11-1937/}{Quantum 9, 1937}, December 2025.

\item Ref.~\cite{inprep_BDR3} (\cref{chap:MEAT}) : Devashish Tupkary, Shlok Nahar, Amir Arqand, Ernest Y.-Z Tan, and Norbert L\"{u}tkenhaus. \textit{A rigorous and complete security proof of decoy-state BB84 quantum key distribution.} \href{https://arxiv.org/abs/2601.18035}{arXiv}, January 2026. 
\item Ref.~\cite{inprep_authentication} (\cref{chap:classicalauthentication}) : Devashish Tupkary, Shlok Nahar, and Ernest Y.-Z. Tan. \textit{Authentication in Security Proofs for Quantum Key Distribution.} \href{https://arxiv.org/abs/2601.17960}{arXiv}, January 2026. 
\end{itemize}

\noindent This thesis is also based on the following works, of which I am second author:
\begin{itemize}
    \item Ref.~\cite{nahar_postselection_2024} (\cref{chap:postselection}) : Shlok Nahar, Devashish Tupkary, Yuming Zhao, Norbert L\"utkenhaus, and Ernest Y.-Z.
Tan. \textit{Postselection Technique for Optical Quantum Key Distribution with Improved
de Finetti Reductions}. \href{https://journals.aps.org/prxquantum/abstract/10.1103/PRXQuantum.5.040315}{PRX Quantum 5, 040315}, October 2024. Shlok Nahar is primarily responsible for the  de Finetti reductions, and incorporating optics in the postselection technique. I, along with Ernest Y.-Z. Tan, am responsible for the rigorous application of the postselection technique to QKD security analysis. 
\item Ref.~\cite{Kamin2025} (\cref{chap:variable,chap:postselection}) : Lars Kamin, Devashish Tupkary, and Norbert L\"utkenhaus. \textit{Improved finite-size effects
in QKD protocols with applications to decoy-state QKD}. \href{https://arxiv.org/abs/2502.05382}{arXiv}, February 2025. Lars Kamin is responsible for the finite-size analysis of decoy-state protocols. I am responsible for adapting those methods to variable-length protocols. 
\end{itemize}

\noindent This thesis also adapts small portions of Ref.~\cite{tupkary2025qkdsecurityproofsdecoystate}, of which I am first author. 

\noindent This thesis uses code written as a part of the  Open QKD Security software package \cite{burniston_software_2024}. It also uses  additional code written by Lars Kamin, John Burniston, and Shlok Nahar for specific publications. In particular, the code used for plots in \cref{chap:MEAT} was entirely written by Lars Kamin and John Burniston as a part of the work Ref.~\cite{kamin_renyi_2025}, in which I was not involved. The code used for the decoy-state plots in \cref{chap:variable,chap:postselection} was mostly written by Lars Kamin as a part of Ref.~\cite{Kamin2025}. The code used for plots in \cref{chap:EUR} was written collaboratively by the author and Shlok Nahar.
\cleardoublepage
\phantomsection    

\addcontentsline{toc}{chapter}{Abstract}
\begin{center}\textbf{Abstract}\end{center}
This thesis is concerned with the rigorous security analysis of practical Quantum Key Distribution (QKD) protocols, using a variety of modern proof techniques. Throughout, the emphasis is on mathematical rigor across a wide range of security proof frameworks.

We begin by presenting a security proof for variable-length QKD protocols against IID collective attacks, which represents the first such result for generic QKD protocols. We then show that this analysis can be lifted to hold against coherent attacks by an adversary, using the postselection technique. In doing so, we extend the application of the postselection technique to practical QKD protocols, and resolve a long-standing flaw in the method, thereby placing its application to QKD on a rigorous mathematical footing. 

We next study security proofs based on entropic uncertainty relations. These proofs proceed by bounding the so-called ``phase error rate", using the observed statistics available in the actual protocol. All known methods of bounding the phase error rate require strong assumptions on hardware: namely, that all detectors have exactly equal probability of detection. This renders these security analysis inapplicable to practical QKD scenarios.  We show that such phase error rates can be bounded even when detectors are imperfect and only approximately characterized. This resolves a long-standing well-known open problem of nearly two decades, and renders this proof technique applicable to realistic scenarios. 

We then study security proofs using the recently obtained marginal-constrained entropy accumulation theorem, and obtain a highly rigorous and general result for the security analysis for practical QKD protocols. Most importantly, the proof is constructed in a transparent and self-contained manner, and is designed to be a key ingredient in certification efforts for QKD. Moreover, it can be easily modified to apply to other protocols of interest, and to device imperfections and side-channels.

We also revisit the assumptions on authentication traditionally made in QKD security analyses, which assume that all classical messages are delivered faithfully and on time, without any aborts. We show that these assumptions are generally unrealistic, and that adopting realistic authentication assumptions necessitates a modification of both the standard QKD security definition and the corresponding security analysis. However, under mild and easily satisfied protocol design conditions, security under realistic authentication can be reduced to the usual idealized setting. As a result, existing QKD security proofs can be lifted to the realistic authentication setting with only a minor protocol modification.

A distinctive feature of this thesis is its unified presentation of multiple major QKD security proof frameworks using consistent protocol descriptions and notation. This first-of-its-kind treatment enables direct comparison and contrast between different approaches, a perspective that is often obscured when these techniques are developed in isolation. Consequently, this work is intended not only as a collection of new technical results, but also as a pedagogical reference for understanding rigorous security analysis in quantum key distribution.

\cleardoublepage
\phantomsection    

\addcontentsline{toc}{chapter}{Acknowledgements}
\begin{center}\textbf{Acknowledgements}\end{center}

It is difficult to find the right words to describe the contributions of my supervisor, Norbert L\"utkenhaus, to this thesis. Throughout my time at IQC, he has been unfailingly supportive of my work, always generous with his time despite an exceptionally busy schedule, and an exemplary mentor in every sense. I invariably left his office in better spirits than when I entered. Perhaps the highest praise I can offer is that my years in graduate school have been among the most fulfilling and rewarding years of my life, and this is due in large part to his guidance and mentorship. If I am ever in the position to advise graduate students of my own, I will strive to emulate his approach as closely as I can.

This thesis owes an enormous debt to Ernest Y. Z. Tan. Ernest is not merely competent, but \emph{scarily} so - the kind of competence that made one wonder how anyone gets there in the first place. A proof that passed Ernest’s scrutiny had a genuinely negligible probability of being flawed. Almost everything I know about information theory, along with a substantial portion of the technical skills utilized in this thesis, I owe to him. Our discussions were always immensely valuable, and he had an uncanny ability to identify subtle issues long before they became apparent to others - a feature I exploited often in my work.

Another person who deserves special mention is Shlok Nahar, who has the distinction\footnote{It is perhaps debatable whether this truly counts as a distinction.} of being my single biggest collaborator during my PhD. Shlok contributed to nearly all of the work I did throughout graduate school; more importantly, he and I contributed in  complementary ways, and at times with  wildly different intuitions and perspectives. We got into disagreements on a weekly basis, and their resolution typically left both of us slightly wiser.
I am deeply grateful to have had him as a collaborator.

Overall, this thesis benefited greatly from being part of the OQCT group, which has a remarkably strong, vibrant, and intellectually generous community. I am especially grateful to Zhiyao Wang, Aodh\'an Corrigan, Jerome Wiesseman, John Burniston, Lars Kamin, Amir Arqand, and Florian Kanitschar, with whom I had numerous discussions that shaped my thinking in both direct and subtle ways.
In particular, Lars Kamin and John Burniston wrote a substantial body of open source code that is utilized in this thesis. I am also very grateful to Guillermo Curr\'as-Lorenzo, Margarida Pereira, and Victor Zapatero for valuable discussions and fruitful collaborations.

I would like to thank my PhD advisory committee members: Professor Thomas Jennewein, Professor Michele Mosca, Professor Graeme Smith and my supervisor for giving helpful feedback and time towards my research progress. I would also like to thank the members of my PhD examination committee for their time and effort in reading this thesis and for serving on the examining committee.

I would also like to thank Michele Roche and Chin Lee for being exceptionally efficient organizers and for handling a a large variety of bureaucratic tasks.  Once they told me they would take care of something, I felt genuinely free to delete that task from my brain.

Finally, I would like to thank my family, both back home and the larger one I found here in Waterloo. This family is far too big to name individually, but I would certainly not have had even half as much fun without the countless walks, chai sessions, dinners, trips, and movie nights we shared. I feel incredibly fortunate to have spent these years surrounded by such an amazing group of people.
\cleardoublepage
\phantomsection    

\addcontentsline{toc}{chapter}{Dedication}
\begin{center}\textbf{Dedication}\end{center}

\emph{To Aai, Baba, and Tai.}
\cleardoublepage
\phantomsection    

\renewcommand\contentsname{Table of Contents}
\tableofcontents
\cleardoublepage
\phantomsection    

\addcontentsline{toc}{chapter}{List of Figures}
\listoffigures
\cleardoublepage
\phantomsection		

\addcontentsline{toc}{chapter}{List of Tables}
\listoftables
\cleardoublepage
\phantomsection		



\pagenumbering{arabic}

\chapter{Introduction}
\label{chap:introduction}
\epigraph{Where I try to explain to a layperson why their taxes should fund this thesis; where I lay bare some of the existential questions that have arisen along the
way to this thesis; and where I explain the best way to read this thesis, and justify its length.
}{}
Human beings have an inherent need to communicate. From spoken language to writing, and from the telegraph to the internet, a large fraction of technological progress has been driven by the desire to transmit information reliably across distance and time. Writing allows ideas to persist across generations; telephones allow voices to travel across continents; and modern digital communication platforms enable instantaneous global interaction. Beyond communication itself lies a deeper and equally fundamental requirement: the need for \emph{private} communication. Humans must be able to share information selectively, with the assurance that only intended recipients can access it. Throughout history, private communication has been essential for diplomacy, trade, personal relationships, and military strategy. 

In a world with just simple face-to-face interactions, this private communication would be as simple as checking that nobody is hiding in the bushes. As communication technologies evolve beyond face-to-face interactions, so too does the need to protect messages from interception.

The art and science of enabling secure communication in the presence of adversaries is known as \emph{cryptography}.\footnote{The word ``secure'' is often used to mean a variety of things in cryptography, depending on whether one is concerned with confidentiality, integrity, authenticity, or something else.} Evidence of cryptographic techniques appears across many ancient civilizations. One of the earliest known examples dates back to around 1900~BC, found in an inscription in the tomb of an Egyptian nobleman, where unusual hieroglyphic substitutions were used to conceal meaning \cite{redhat_BriefHistory}. In ancient India, the \emph{Arthashastra}, written somewhere around 300BC  attributed to Kautilya, describes espionage practices and explicitly mentions the use of ``secret writing'' for covert communication. In the Roman era, Julius Caesar famously employed what is now known as the Caesar cipher to transmit messages to his generals.

The Caesar cipher is a simple substitution cipher, in which each letter of the plain text (the message to be communicated) is shifted by a fixed number of positions in the alphabet, to yield the cipher text. While historically significant, such schemes rely on the secrecy of the encryption method itself. Once the method is known, the cipher can be broken easily, for example through frequency analysis of letters in the underlying language. In the case of the Caesar cipher, one can simply try shifting the letters in reverse, and checking which shifts lead to meaningful plain texts. This limitation illustrates an important principle, later formalized as Kerckhoffs’ principle: security should not depend on keeping the system itself secret.

A major conceptual advance came during the Renaissance with the work of Blaise de Vigenère, who introduced ciphers that made explicit use of a secret \emph{key}. A secret key is any piece of information  which is known to the communicating parties, but unknown to the eavesdropper. In the Vigenère cipher, this key is combined with the plain text letters using modular arithmetic. The Vigenère cipher employs a short key that is repeated periodically, a feature that was later refined by the Vernam cipher (also known as the one-time pad), in which the key was required to be as long as the message itself. In both these cases, even if the encryption procedure is known, this does not necessarily suffice to break the cipher, since correct decryption additionally requires knowledge of the secret key.

The twentieth century, and in particular the Second World War, marked a turning point in cryptography. Mechanical and electromechanical cipher devices such as the German Enigma machine and the Japanese Purple cipher were deployed at unprecedented scale. Their cryptanalysis by Allied efforts demonstrated both the power and the limitations of classical cryptographic techniques, and helped establish cryptography as a scientific discipline rather than a collection of ad hoc methods.

This transition was completed by the work of Claude Shannon. In his seminal paper \cite{shannon_communicationtheory_1949}, Shannon placed proved \emph{perfect secrecy} - that is, where the cipher text tells you nothing at all about the underlying plain text - is achievable if and only if the encryption key is at least as long as the message and is used only once. The resulting scheme, known as the one-time pad, is provably secure against any adversary, regardless of computational power (as long as the key is perfectly secure). 

Thus, secure communication can be reduced to the problem of \emph{key distribution}. If two parties can somehow share a secret key securely, then they can use the one-time pad to protect their messages. Conversely, if key distribution is insecure, no encryption scheme can compensate for it, since the eavesdropper is assumed to know everything about the encryption process except the key. This observation lies at the heart of modern cryptography and directly motivates the subject of this thesis.

\section{Traditional Classical Cryptography}

Modern cryptography operates in a fundamentally digital setting. Information is represented as bits, processed by classical computers, and transmitted over classical communication networks such as the internet. Cryptographic algorithms are implemented as software or hardware procedures that manipulate bit strings, and adversaries are modeled as entities with access to the certain communication infrastructure and computational resources, subject to specified limitations.

In this setting, the central challenge is enabling two distant parties to communicate securely over a public channel that may be fully monitored by an adversary. Early symmetric-key cryptographic systems assumed that communicating parties already shared a secret key, established presumably by sending a trusted courier, or during a prior meeting. While effective in small or controlled environments, this assumption does not scale to large networks, where secure pre-distribution of keys becomes logistically infeasible.

A major breakthrough occurred in the 1970s with the invention of public-key cryptography. Protocols such as Diffie–Hellman key exchange \cite{diffie_newdirections_1976} and RSA \cite{rsa_1978} allow two parties to establish a shared secret over an open, insecure but authenticated classical channel. By \emph{authenticated}, we mean that an adversary may eavesdrop on all communicated messages but cannot impersonate either party or modify transmitted messages\footnote{While this is a standard assumption in cryptography, in practice, an adversary can always delay, block, or interrupt messages, and any concrete authentication scheme allows a nonzero probability of successful forgery. These subtleties are discussed briefly in the context of QKD in \cref{chap:classicalauthentication}.}. These schemes rely on mathematical problems believed to be computationally hard\footnote{That is, have runtime that is exponential in the size of the input.}, such as the discrete logarithm problem or integer factorization. Crucially, this hardness is a \emph{belief} rather than a proven fact. Moreover, this belief is known to be invalid in the presence of large-scale quantum computers, as we discuss below. Historically, such beliefs about computational hardness have repeatedly turned out to be  optimistic. For example, when RSA was introduced in 1977, key sizes on the order of 512 bits were widely considered secure for the foreseeable future. However, advances in algorithms, hardware, and large-scale collaborative computation led to the successful factorization of a 512-bit RSA modulus in 1999---less than a quarter century after the scheme’s proposal---rendering such key sizes  insecure \cite{rsa512}. Similar episodes have occurred for other cryptographic assumptions, which shows that confidence in computational hardness is inherently provisional and subject to revision as techniques evolve.

In practice, modern secure communication systems use hybrid encryption architectures. Public-key cryptography is employed only to establish a short symmetric key, after which efficient symmetric-key algorithms such as AES \cite{aes} are used to encrypt bulk data. This approach combines the scalability of public-key cryptography with the efficiency of symmetric encryption.

Crucially, the security of all such classical cryptographic systems, especially public-key cryptography is \emph{computational}. Their guarantees rely on assumptions about the limitations of adversaries’ computational power, and their abilities to solve certain mathematical problems. If these assumptions fail, the security of the system collapses. 

\section{Quantum Computers}

The conditional nature of computational security became particularly apparent with the emergence of quantum computing. In 1994, Peter Shor discovered a quantum algorithm capable of efficiently factoring large integers and computing discrete logarithms \cite{Shor_1997}. Shor’s algorithm implies that a sufficiently large quantum computer would completely break widely deployed public-key cryptosystems, including RSA and elliptic-curve cryptography, which underpins the overwhelming majority of public-key deployments on today’s Internet.

This discovery was both surprising and consequential. It demonstrated that cryptographic security based solely on classical computational hardness is vulnerable to advances in computing paradigms. In response, significant effort has been devoted to the development of post-quantum cryptography (PQC): classical cryptographic schemes designed to remain secure against quantum adversaries. These schemes are based on problems believed to be hard for both classical and quantum computers. The ongoing NIST post-quantum cryptography standardization process \cite{nistNISTreleases} reflects the global importance of this transition.

Despite these advances, post-quantum cryptography remains computational in nature. Its security continues to rely on unproven assumptions about algorithmic hardness.

\section{Quantum Key Distribution}
In parallel with the development of post-quantum cryptography, a fundamentally different approach to secure communication emerged. In 1984, Bennett and Brassard proposed a new protocol, which now bears their names \cite{Bennett_2014} (and for which they were awarded the Turing Award); it uses the principles of quantum mechanics to establish\footnote{Note that the protocol, and QKD in general, \emph{establishes} secret keys rather than distributing pre-existing ones. A more accurate name would therefore be quantum key establishment. Alas, it is now too late.
} secret keys. This work gave birth to the field of quantum key distribution. The central idea is to encode information into quantum states of light in such a way, that any attempt at stealing this information results in a disturbance of those states. This  can then be detected by the honest communicating parties, which can abort the protocol when necessary.

Quantum key distribution (QKD) enables two parties to establish a shared secret key with information-theoretic security, even in the presence of an adversary with unlimited computational power. Security follows from the properties of information implied by physical principles, such as the impossibility of perfectly copying unknown quantum states and the unavoidable disturbance caused by measurement (and by assumptions on the functioning of the hardware).

Over the past four decades, QKD has evolved from a theoretical proposal into a mature field with experimental demonstrations over optical fiber, free-space links, and satellite-based platforms. Commercial QKD systems exist today, and efforts are underway to standardize and certify QKD technologies. \cref{chap:MEAT} in this thesis is a result of one such effort.

\section{``PQC vs QKD"}

There is an ongoing and at times contentious exchange regarding the respective roles of post-quantum cryptography (PQC) and quantum key distribution (QKD). This discussion raises fundamental questions about the long-term foundations of cryptographic security\footnote{and has raised many existential questions for this author.} and is therefore worth briefly addressing here. A comprehensive treatment is beyond the scope of this thesis, and the topic has been examined in considerably greater depth elsewhere; we refer the reader, for example, to Refs.~\cite{renner_debateqkdrebuttalnsa_2023,stebila_caseforqkd_2010,bundPositionPaper,Bernstein2017,nsa_qkd,etsi_eg203310_2016,ioannou_newspin_2011}. Nevertheless, since this thesis is concerned with the security analysis of QKD protocols, it would be inappropriate to omit this discussion entirely.

Proponents of PQC emphasize that many real-world cryptographic failures are not caused by breakthroughs in the underlying mathematical assumptions, but rather by more mundane engineering issues, such as software vulnerabilities, side-channel leakage, poor operational security, and social engineering. In this context, it is argued that QKD introduces a new class of implementation challenges, including novel side-channel attacks and system-level vulnerabilities, for which there is comparatively less operational experience and confidence. They further note that QKD comes with practical constraints: it requires a dedicated quantum channel, and it addresses only the problem of key distribution. As a result, it must be combined with additional mechanisms for authentication and encryption, often relying on classical cryptographic components.
From this perspective, QKD
may be intellectually appealing, but it requires specialized hardware and (in practice)
a pre-shared seed key to authenticate classical communication\footnote{Although authentication is required for classical cryptographic schemes as well. Even when using computationally secure authentication, QKD still provides strong ``everlasting security" \cite{unruh_everlasting_2013} guarantees under the assumption that the authentication remains secure during the execution of the protocol \cite{ioannou_newspin_2011,mosca_qkdinclassical_2013}.}. Moreover, present-day QKD
deployments face distance limitations and can be vulnerable to implementation and
side-channel attacks. These considerations are used to argue that QKD is too expensive and operationally cumbersome to be broadly useful. 
Finally, discussions around timelines for cryptographically relevant quantum computers remain highly uncertain, with expert assessments often extending well beyond a decade~\cite{mosca_quantumthreat_2024}. However, this consideration is not central to the role of QKD, which is not intended solely as a countermeasure to specific quantum algorithms such as Shor's algorithm, but rather as a means of providing cryptographic diversity and mitigating future risks, whether classical or quantum.

In contrast, proponents of QKD emphasize that it promises information-theoretically secure key establishment—something that purely classical approaches such as PQC cannot provide.\footnote{This information-theoretic security applies to the mathematical model of the protocol with modelling assumptions, and does not automatically extend to imperfect physical implementations. However, note that this is the case for the implementation of \emph{any} cryptographic protocol, whether classical or quantum. } They further argue that, for long-term confidentiality, future breakthroughs in cryptanalysis cannot be ruled out, and that it is therefore valuable to have a key-establishment mechanism whose security does not rest on computational assumptions. This perspective is often formalized through the so-called Mosca inequality, which highlights the risk that the time until large-scale quantum attacks become feasible may be shorter than the combined data shelf-life and cryptographic migration time \cite{etsi_eg203310_2016,mosca_quantumthreat_2024}. Proponents additionally argue that although current QKD systems remain costly, these costs are expected to decrease as the technology matures and deployments scale. Furthermore, while present-day QKD systems are limited in transmission distance, the eventual development of quantum repeaters \cite{Azuma_quantumrepeaters_2023} has the potential to overcome this limitation. Until such technology becomes available, long-distance QKD can be realized through trusted-node networks.

The author of this thesis adopts a deliberately agnostic position with respect to this ongoing debate. As will be seen throughout this work, QKD enables the establishment of information-theoretic security—at least at the level of a precisely specified protocol description under explicit modeling assumptions on the relevant hardware implementing the protocol. Such security guarantees are fundamentally unattainable using purely classical cryptographic methods, and this constitutes a distinctive scientific and engineering achievement that, in itself, justifies continued study of the field.

This perspective is further motivated by the historical difficulty of reliably forecasting the longevity of classical cryptographic assumptions. Experience has repeatedly shown that predictions about when particular cryptosystems will become insecure are highly uncertain and often  optimistic.  In this sense, information-theoretic security offers a qualitatively different form of assurance—one that is not contingent on assumptions about future computational capabilities—and provides a compelling rationale for investigating QKD alongside classical and post-quantum approaches.

At the same time, QKD systems typically entail higher costs, require dedicated hardware, and are susceptible to their own class of side-channel attacks that target imperfections in practical implementations. Consequently, substantial theoretical and experimental effort is required to obtain satisfactory security proofs for mathematical models of QKD systems that are both rigorous and realistic, in the sense that they accurately capture the relevant features of the underlying physical implementations. In fact, the primary  goal of this thesis is to help advance the theoretical state of the art in this direction.

Whether the additional security guarantees offered by QKD justify its practical cost, operational complexity, and deployment constraints is ultimately not a scientific question, but one to be decided by economic, regulatory, and strategic considerations. In this thesis, we do not attempt to adjudicate this trade-off. Rather, the objective is to ensure that the security claims made by QKD protocols are stated as precisely, transparently, and rigorously as possible, and that the assumptions underlying these claims are clearly articulated and directly connected to realistic deployment scenarios. 
Arguments concerning whether the eventual market for QKD technologies will be large or small lie outside the scope of this thesis. In this sense, the value of the results presented here does not depend on market outcomes, but rather on their ability to sharpen our understanding of QKD and to provide a rigorous basis upon which future engineering, certification, or deployment decisions may be made.

\section{Organization of this thesis}

This thesis is focused on the \emph{rigorous} security analysis of \emph{practical} QKD
protocols using a \emph{variety} of proof techniques. Each emphasized word in this sentence
is deliberate and reflects a central theme.

\paragraph*{Rigorous.} There exists a large body of academic literature on QKD security proofs; however, many
existing works contain gaps, implicit assumptions, or unresolved technical issues. This thesis builds on substantial prior work while aiming to address these
deficiencies by presenting a careful, explicit, and internally consistent analysis that
avoids such omissions and errors.  For instance, in \cref{chap:postselection} we fix a longstanding flaw in the postselection technique for QKD.
Moreover, there is currently no single paper that fully specifies a practical QKD
protocol in all of its operational detail and then provides a  self-contained
and rigorous security analysis of that protocol (see Refs.~\cite{tupkary2025qkdsecurityproofsdecoystate,mizutani2025protocolleveldescriptionselfcontainedsecurity} for recent attempts at filling this gap). Our own attempt to provide such an analysis is also very recent, and presented in \cref{chap:MEAT}.

\paragraph*{Practical.} It should be noted that a highly formal and fully rigorous security analysis of a QKD protocol does exist \cite{tomamichel_largely_2017}. However, that analysis applies to an idealized protocol that is not designed for practical implementation and abstracts away many operational and physical details relevant to real-world systems.

In contrast, this thesis focuses explicitly on protocols - such as the decoy-state BB84 protocol - that are intended to be implementable in practices, and also makes contributions to the security analysis in the presence of hardware imperfections (\cref{chap:EUR}). The analysis developed here, especially in \cref{chap:MEAT} is sufficiently general to accommodate a wide range of protocol variants and implementation choices.

\paragraph*{Variety.} Finally, modern QKD security theory offers a wide range of proof techniques, most of which
are often studied in isolation. A distinctive feature of this thesis is that it covers
three of the four major proof techniques used in contemporary QKD security analysis \cite{tupkary2025qkdsecurityproofsdecoystate}.\footnote{The phase error correction approach, which is not covered here, is closely related
to the entropic uncertainty relations approach studied in this thesis; see
Ref.~\cite{tsurumaru_leftover_2020}. Both approaches rely on obtaining suitable bounding the same quantity, the so called phase error rate.} As a result, this work is intended to be both
comparative and pedagogical, and to serve as a reference for readers interested in
rigorous security analyses of practical QKD protocols, across multiple proof techniques.

\noindent The thesis is organized as follows:

 In \cref{chap:background}, we introduce the notation and conventions from quantum information theory used throughout this thesis, review relevant entropic quantities and their basic properties, and provide a brief overview of the aspects of quantum optics needed to model practical QKD implementations.

In \cref{chap:qkdbackground}, we cover the basics of QKD. We present the generic protocol framework studied in this thesis, state the security definition, and review core tools used in security analyses of practical protocols. This includes, in particular, the use of the Leftover Hashing Lemma, and other tools such as source-replacement schemes, squashing maps, and source maps. 

In \cref{chap:variable} we start covering new results. We study variable-length security and the additional conceptual and technical issues that arise when the output key length is not a fixed value, but depends on the protocol observations. We present a generic security proof for variable-length QKD protocols and highlight the specific modifications relative to the standard fixed-length analysis. We also show that these modifications do not introduce a substantial performance penalty.

\cref{chap:postselection} studies the postselection technique and its application to practical QKD. The postselection technique \cite{christandl_postselection_2009,nahar_postselection_2024} is a generic tool that can be used to reduce the security analysis of protocols satisfying a certain ``permutation-invariant" property, to security analysis against IID collective attacks. We resolve several issues that must be addressed before this technique can be rigorously applied to practical optical, prepare-and-measure implementations. In particular, this includes the role of infinite-dimensional Hilbert spaces (and the resulting dimension dependence of the reduction) as well as a patch to a critical gap in the original argument applying this technique to QKD.  

\cref{chap:EUR} develops security proofs based on entropic uncertainty relations. A central object in this approach is the phase error rate and the challenge of bounding it using experimentally accessible data. All prior treatments relied on highly idealized detector models - for example, assuming perfectly identical detection probabilities - in order to obtain such bounds, and consequently fail to tolerate even infinitesimal deviations from this idealized setting. As a result, obtaining meaningful bounds on the phase error rate in the presence of realistic detector imperfections has remained a well-known open problem for nearly two decades. In this thesis, we provide a solution to this problem by introducing new results on sampling measurements from quantum states. These results render entropic uncertainty relation–based  (and phase error correction-based) security proofs  applicable to practical detector scenarios.

 \cref{chap:MEAT} presents security analyses based on the marginal-constrained entropy accumulation theorem. This chapter contains a rigorous and complete security proof for a well specified decoy-state BB84 protocol. Building on decades of prior work, we combine the relevant ingredients into a single coherent framework that yields tight key-rate bounds and provides a clear path toward incorporating realistic implementation imperfections. Beyond establishing the security of a specific protocol, this chapter develops a general and modular framework that can be readily adapted to a broad class of QKD protocols. The framework unifies all major ingredients required for the analysis of realistic QKD protocols, including the analysis of classical authentication and classical processing, source-replacement schemes, finite-size analysis, source maps, squashing maps, and decoy-state techniques. In doing so, this work consolidates a diverse range of techniques scattered across the QKD literature into a unified formalism, representing a general and rigorous treatment of QKD security, which can be reused for further analyses. 

 \cref{chap:classicalauthentication} discusses classical authentication and its role in QKD protocols. Standard QKD analyses often assume an ideal authenticated classical channel in which messages are always delivered correctly (albeit with arbitrary delay). In practice, practical authentication mechanisms instead guarantee that either a message is accepted as authentic or the receiver outputs a special failure symbol. We explain how this discrepancy necessitates a careful treatment of asymmetric abort events (where one party aborts while the other does not), and we show how (under appropriate conditions) the resulting security analysis in the realistic authentication setting can be reduced to the security analysis of the standard, idealized authentication setting. 

\cref{chap:conclusion} concludes the thesis with a summary of contributions, a discussion of open problems and future directions, and a comparative perspective on the proof techniques developed throughout the thesis.

\section{How to read this thesis}

This thesis is written using largely consistent notation throughout, and any changes to notation are explicitly highlighted when they are unavoidable, as is occasionally necessary given the breadth of topics covered. More importantly, it is written in a modular fashion. Each of \cref{chap:EUR,chap:MEAT,chap:variable,chap:classicalauthentication,chap:postselection} presents a distinct research contribution and may be read independently of the others. All necessary background material is consolidated in \cref{chap:background,chap:qkdbackground}. \cref{chap:background} provides general mathematical background material; readers who are already well versed in quantum information theory may choose to skip or skim this chapter. \cref{chap:qkdbackground} introduces QKD-specific background and may similarly be skimmed by readers already familiar with QKD. 

Readers interested in a particular proof technique may therefore focus on the corresponding chapter: \cref{chap:postselection} for the postselection technique, \cref{chap:EUR} for entropic uncertainty relations, and \cref{chap:MEAT} for the entropy accumulation approach. \cref{chap:classicalauthentication} addresses classical authentication and its role in QKD, and can be read independently of the security proof techniques developed elsewhere in the thesis.

Furthermore, our notation closely follows that of the recent review on QKD security proofs in Ref.~\cite{tupkary2025qkdsecurityproofsdecoystate}, which may make this thesis a useful technical companion to that work.

\chapter{Background} \label{chap:background}
\epigraph{
Where we set up basic background in quantum information theory; where we discuss
a large number of statements about several different entropies; and where we
introduce the quantum description of light and associated hardware required to understand QKD.
}{}

\section{Quantum Information Theory}

This section reviews essential concepts from quantum information theory \cite{watrous_theory_2018} and establishes the notation and conventions used throughout this thesis. We use capital letters such as $A$, $B$, and $C$ to denote \emph{registers}, which represent physical systems capable of storing information. For a sequence of registers, $A_j$ denotes the $j$th register, and we write $A_i^j$ to denote the collection of registers $A_i, \dots, A_j$.  The register $A$ is  further associated with a  complex Hilbert space $\mathcal{H}_A$, which provides the linear-algebraic structure used to represent states, measurements, and transformations of the system. Throughout this thesis, we will typically work with finite-dimensional registers; cases involving infinite-dimensional registers will be clear from the context.

For such spaces, we define the following sets of operators:

\begin{itemize}
  \item $\linear(A,B)$: the set of all linear operators acting on $\mathcal{H}_A$ to produce vectors in $\mathcal{H}_B$, that is,  
  \[
  \linear(A,B) = \{ X : \mathcal{H}_A \to \mathcal{H}_B \mid X \text{ is  linear} \}.
  \]
  This space is itself a finite-dimensional complex vector space of dimension $(\dim \mathcal{H}_A)(\dim \mathcal{H}_B)$. We use $\linear(A)$ to denote $\linear(A,A)$.

  \item $\Pos(A)$: the set of (hermitian) positive semidefinite operators on $\mathcal{H}_A$, i.e.,
  \[
  \Pos(A) = \{ X \in L(A) : X = X^{\dagger},\, X \ge 0 \}.
  \]

  \item $\dop{\leq}(A)$: the set of subnormalized states on $A$, i.e., positive semidefinite operators with trace at most one:
  \[
  \dop{\leq}(A) = \{ \rho_A \in \Pos(A) : \Tr[\rho_A] \le 1 \}.
  \]

  \item $\dop{=}(A)$: the set of normalized quantum states (density operators) on $A$, i.e.,
  \[
  \dop{=}(A) = \{ \rho_A \in \Pos(A) : \Tr[\rho_A] = 1 \}.
  \]

  \item $\unitary(A)$: the set of unitary operators on $\mathcal{H}_A$, that is,
  \[
  \unitary(A) = \{ U \in \linear(A) : U^{\dagger} U = U U^{\dagger} = \id_A \}.
  \]
  These represent reversible physical transformations acting on register $A$.
  \item $\isometry(A,B)$: the set of isometries from $A$ to $B$, defined as
  \[
  \isometry(A,B) = \{ V : \mathcal{H}_A \to \mathcal{H}_B \text{ linear} \;|\; V^{\dagger} V = \id_A \}.
  \]
  Isometries preserve inner products but need not be surjective; they model quantum evolutions that embed system $A$ into a larger system $B$.
\end{itemize}
If $A$ is a register with associated Hilbert space $\mathcal{H}_A$, we will sometimes use  $\Pos(\mathcal{H}_A)$ and  $\Pos(A)$ interchangeably. This notational choice simplifies expressions in later chapters.

\subsection{Quantum States}

\begin{definition}
A (possibly subnormalized) \term{quantum state} on a register $A$ is represented by a positive semidefinite operator $\rho_A \in \Pos(A)$ with $\Tr(\rho_A) \le 1$, where subscripts specify the registers on which the state exists. When $\Tr(\rho_A) = 1$, the state is said to be normalized.
\end{definition}

\begin{definition}
    A state is \term{pure} if it can be written as $\rho_A = \ket{\psi}\bra{\psi}$ for some unit vector $\ket{\psi} \in \mathcal{H}_A$. Otherwise, it is mixed.
\end{definition}

Composite systems are described by tensor products: for registers $A$ and $B$, the joint space is $\mathcal{H}_{AB} = \mathcal{H}_A \otimes \mathcal{H}_B$.
A joint state $\rho_{AB}$ on $AB$ has marginals given by the partial trace, e.g., $\rho_A = \Tr_B(\rho_{AB})$.
 For a multipartite state $\rho_{ABC\cdots}$, we write $\rho_A$ to denote its marginal on register $A$. Conversely, for a state $\rho_A$, we use $\rho_{AB}$ to denote some extension of $\rho_A$ to an additional register $B$.

\begin{definition}
A state $\rho \in \dop{\leq}(CQ)$ is said to be \term{classical on $C$} (with respect to a specified basis on $C$) if it is in the form\footnote{Throughout this thesis, we adopt the convention that whenever a
summation $\sum_c$ is written without an explicit specification of the
summation range, it is understood to be taken over all admissible values
of the variable $c$, with the corresponding domain being implicitly
determined by the surrounding context.} 
\begin{align}
\rho_{CQ} = \sum_c \lambda_c \ketbra{c} \otimes \sigma_c,
\label{eq:cq}
\end{align}
for some normalized states $\sigma_c \in \dop{=}(Q) $ and weights $\lambda_c \geq 0$, with $\ket{c}$ being the specified basis states on $C$. In most circumstances, we will not explicitly specify this ``classical basis'' of $C$, leaving it to be implicitly defined by context. Throughout this thesis, all variants of the register $C$ (including decorated versions such as $\hat{C}$, $\tilde{C}$, $\CEC$ etc.) are taken to be classical registers.

\end{definition}

In the protocols studied in this thesis, different operations are often performed depending on which events occur during the execution of the protocol. It is therefore necessary to condition on classical events in order to describe such conditional operations precisely. An event on a register $C$ is simply a subset of the possible values that the register $C$ can take. For any event $\Omega$, we use $\Omega^\complement$ to denote its complement. We introduce the relevant definitions below.

\begin{definition}\label{def:cond}
(Conditioning on classical events) For a state $\rho \in \dop{\leq}(CQ)$ classical on $C$, written in the form
$\rho_{CQ} = \sum_c \lambda_c \ketbra{c}{c} \otimes \sigma_c$ 
for some $\sigma_c \in \dop{=}(Q)$ and $\lambda_c \geq 0$,
and an event $\Omega$ defined on the register $C$, we will define a corresponding \term{partial state} and \term{conditional state} as, respectively,
\begin{align}
\rho_{\land\Omega} \coloneqq \lambda_c \sum_{c\in\Omega} \ketbra{c}{c} \otimes \sigma_c, \qquad\qquad \rho_{|\Omega} \coloneqq \frac{\Tr{\rho}}{\Tr{\rho_{\land\Omega}}} \rho_{\land\Omega} = \frac{
\sum_{c} \lambda_c
}{\sum_{c\in\Omega} \lambda_c}\rho_{\land\Omega} .
\end{align}
We refer to $\rho_{\wedge \Omega}$ as the state $\rho$ being partial\footnote{One could instead also think of it as being ``subnormalized conditioned" on the event $\Omega$.} on $\Omega$. The process of taking partial states is commutative and ``associative'', in the sense that for any events $\Omega,\Omega'$ we have $(\rho_{\land\Omega})_{\land\Omega'} = (\rho_{\land\Omega'})_{\land\Omega} = \rho_{\land(\Omega\land\Omega')}$. On the other hand, some disambiguating parentheses are needed when combined with taking conditional states (due to the normalization factors).
\end{definition}
Given these definitions, for a normalized state $\rho \in \dop{=}(CQ)$ that is classical on $C$, we can write it in the form
\begin{align}\label{eq:cqstateprobs}
\rho_{CQ} = \sum_c \rho(c) \ketbra{c}{c} \otimes \rho_{Q|c},
\end{align}
where $\rho(c)$ denotes the probability of register $C$ storing $c$ according to $\rho$, and $\rho_{Q|c}$ can indeed be interpreted as the corresponding conditional state on $Q$, i.e.~$\rho_{Q|c} = \tr_C[{\rho_{|\Omega_{C=c}}}]$ where $\Omega_{C=c}$ is the event $C$ stores $c$. 

\begin{definition}\label{def:purify}
For registers $Q,Q'$ with $\dim(Q) \leq \dim(Q')$, a \term{purifying function for $Q$ onto $Q'$} is a function $\pf: \dop{\leq}(Q) \to \dop{\leq}(QQ')$ such that for any state $\rho_Q$, the state $\pf(\rho_Q)$ is a purification of $\rho_Q$ onto the register $Q'$, i.e.~a (possibly subnormalized) rank-$1$ operator such that $\tr_{Q'}[{\pf(\rho_Q)}] = \rho_Q$.
\end{definition}
Note that a purifying function is \emph{not} a channel (i.e.~CPTP map), for instance, because it is necessarily nonlinear. Furthermore, all purifications of a given state $\rho_Q$ are isometrically related \cite[Proposition 2.29]{watrous_theory_2018}, 
in the sense that if $\ket{\psi}_{RQ}$ and $\ket{\phi}_{R'Q}$ are two purifications of $\rho_Q$, 
then there exists an isometry $V \in \isometry(R,R')$
such that 
\[
(\,\id_Q \otimes V\,)\ket{\psi}_{QR} = \ket{\phi}_{QR'}.
\]

\subsection{Measurements}
In quantum mechanics, measurements are described mathematically by positive operator-valued measures (POVMs). Each POVM is specified by a collection of positive semidefinite operators that together represent the possible outcomes of the measurement. 

\begin{definition}[POVM]
A \term{positive operator-valued measure (POVM)} on the register $A$ with outcome set $\mathcal{O}$ is a collection of operators
\[
\{\Gamma_k : k \in \mathcal{O}\} \subseteq \Pos(A)
\]
satisfying
\[
\sum_{k \in \mathcal{O}} \Gamma_k = \id_A.
\]
Each operator $\Gamma_k$ is referred to as a POVM element and corresponds to a distinct measurement outcome $k \in \mathcal{O}$. The probability that outcome $k$ occurs when the system in a state $\rho \in S_{=}(A)$ is measured using the above POVM,  is given by:
\[
p(k) = \Tr(\Gamma_k \rho).
\]
\end{definition}

In this general framework, projective measurements correspond to POVMs whose elements are orthogonal projections. In more general settings, such as when the measurement apparatus involves ancillary systems or coarse-graining of outcomes, POVMs provide the correct mathematical description of what is physically observed.

\subsection{Quantum Channels}
\begin{definition}
    A \term{quantum channel} is a linear map $
\Phi: L(A) \to L(B)
$
that is completely positive and trace preserving. A linear map on the operator space is completely positive if, for every auxiliary register $R$, the extended map $\Phi \otimes\idmap_R$ maps positive operators to positive operators.
A channel is trace-preserving if $\Tr[\Phi(X)] = \Tr[X]$ for all $X \in L(A)$. 
\end{definition}

We denote the set of CPTP maps from registers $A$ to $B$ via $\CPTP(A,B)$. By Stinespring’s theorem \cite[Proposition 2.20]{watrous_theory_2018}, every quantum channel can be represented as
\[
\Phi(\rho_A) = \Tr_E \big( U_{AE} (\rho_A \otimes \ket{0}\!\bra{0}E) U_{AE}^\dagger \big),
\]
for some environment system $E$ and unitary operator $U_{AE}$ acting jointly on $A$ and $E$.

The following lemma is a standard result stating that any extension of a state $\rho_A$ can be obtained by applying an appropriate quantum channel to the purifying subsystem of some purification of $\rho_A$.

\begin{lemma}
    \label{lemma:pur_to_ext}
    Let $\rho_A\in \dop{=}(A)$ be a density operator. Furthermore, let $\rho_{AB}\in \dop{=}(AB)$ and $\rho_{AA'}\in \dop{=}(AA')$ be any extension and purification of $\rho_A$, respectively. Then
    \begin{align}
\exists\mathcal{N}\in\CPTP(A',B)\;\operatorname{s.t.}\;\rho_{AB}=\mathcal{N}[\rho_{AA'}].
    \end{align}
    \begin{proof}
        Let $\rho_{ABR}$ be any purification of $\rho_{AB}$, then by the isometric equivalence of purifications, there exists an isometry $V:A'\to BR$ such that
        \begin{align}
            \rho_{ABR}=V\rho_{AA'}V^\dagger.
        \end{align}
        Therefore, $\rho_{AB}=\Tr_R\left[V\rho_{AA'}V^\dagger\right]$. The claim follows by setting $\mathcal{N}$ to be a channel which first applies the isometry $V$, and then traces out the $R$ register.
    \end{proof}
\end{lemma}

We will often work with channels that implement some measurement and stores the outcomes in a new register. These channels are defined below. 

\begin{definition}[Measurement channel] \label{def:measurementchannels}
Let $\{\Gamma_k\}^m_{k=1}\subset \Pos(A)$ be a POVM. Let $X$ be a classical register of dimension $m$. We define the \term{measurement channel} $\measChannel{\{\Gamma_k\}}\in\CPTP(A,X)$ corresponding to the POVM $\{\Gamma_k\}_{k=1}^m$ to be the quantum channel which acts on any state as
\begin{equation}
\measChannel{\{\Gamma_k\}} \left[ \rho \right] = \sum_{k=1}^m \Tr\left[\Gamma_k \rho \right]\ketbra{k}_X.
\end{equation}
\end{definition}

\subsection{Norms and Distances}
We now define several norms and distance measures that are required for the mathematical machinery used in this thesis. In practice, we only make essential use of the Schatten $1$-norm (\cref{eq:schatten}) and the associated trace distance or $1$-norm. The remaining notions are included primarily for completeness, as they are needed to define the purified distance that appears in the definition of the smooth min-entropy. The detailed properties of these additional quantities are not directly utilized in this thesis. We refer to the reader to Ref.~\cite{tomamichel_quantum_2016} for more details. 

\begin{definition}(Schatten $p$-norm)\label{def:schatten}
     For  every linear operator $X$, we define its \term{Schatten $p$-norm} as
    \begin{align}
        \label{eq:schatten}
        \norm{X}_p\coloneq\left(\Tr{\abs{X}^p}\right)^{1/p},
    \end{align}
    where $p\in[1,\infty)$ and $\abs{X}=\sqrt{X^\dagger X}$.
\end{definition}

\begin{definition}(Generalized Trace Distance \cite[Eq.(3.23)]{tomamichel_quantum_2016})
Let $\rho_A, \sigma_A \in \dop{\leq}(A)$ be two (possibly subnormalized) quantum states on a register $A$. 
The \term{generalized trace distance} between $\rho_A$ and $\sigma_A$ is defined as
\begin{equation}\label{eq:generalizedtracedistance}
T(\rho_A, \sigma_A) 
  \coloneqq \frac{1}{2} \bigl\| \rho_A - \sigma_A \bigr\|_1 
  + \frac{1}{2} \bigl| \Tr[\rho_A] - \Tr[\sigma_A] \bigr|,
\end{equation}
where $\| X \|_1 = \Tr[\sqrt{X^{\dagger} X}]$ denotes the trace norm or Schatten $1$-norm. 
\end{definition}
If $\rho_A, \sigma_A \in S_{=}(A)$ are normalized quantum states, then 
$\Tr[\rho_A] = \Tr[\sigma_A] = 1$, and the generalized trace distance reduces to the usual 
\term{trace distance}
\[
T(\rho_A, \sigma_A) = \frac{1}{2}\|\rho_A - \sigma_A\|_1.
\]
In this case, $T(\rho_A, \sigma_A)$ quantifies the statistical distinguishability between 
$\rho_A$ and $\sigma_A$: if one is given a single copy of a quantum state that is either $\rho_A$ or $\sigma_A$, 
each occurring with apriori probability $\frac{1}{2}$, 
then the optimal probability of correctly identifying the state by any quantum measurement is
\[
p_{\text{guess}}(\rho_A, \sigma_A) 
  = \frac{1}{2}\bigl(1 + T(\rho_A, \sigma_A)\bigr).
\]
Thus, the trace distance quantifies the maximum distinguishing advantage between two quantum states.

\begin{definition}(Generalized Fidelity and Purified Distance \cite[Definition 3.7,3.8]{tomamichel_quantum_2016})\label{def:purifieddistance}
Let $\rho_A, \sigma_A \in \dop{\leq}(A)$ be two subnormalized quantum states. 
The \term{generalized fidelity} between them is defined as
\begin{equation}
F(\rho_A, \sigma_A) 
  \coloneqq \left( \bigl\| \sqrt{\rho_A}\sqrt{\sigma_A} \bigr\|_1 
  + \sqrt{ \bigl(1 - \Tr[\rho_A]\bigr)\bigl(1 - \Tr[\sigma_A]\bigr) }\right)^2.
  \end{equation}
  The corresponding \term{purified distance} is given by
\begin{equation}
P(\rho_A, \sigma_A) \coloneqq \sqrt{\,1 - F(\rho_A, \sigma_A)\,}.
\end{equation}
For normalized states, the second term in $F(\rho_A, \sigma_A)$ vanishes, 
and these definitions coincide with the standard fidelity and purified distance.
\end{definition}
The generalized trace distance and generalized fidelity are related by the Fuchs-van de Graff inequalities, in exactly the same manner as the usual trace distance and fideltiy. 
\begin{lemma}(Fuchs–van de Graaf Inequalities \cite[Lemma 3.5]{tomamichel_quantum_2016}, \cite{fuchs_cryptographicdistinguishability_1999})
For all subnormalized states $\rho_A, \sigma_A \in S_{\leq}(A)$,
\begin{equation}
 T(\rho_A, \sigma_A) \leq P(\rho_A,\sigma_A) \leq \sqrt{2 T(\rho_A, \sigma_A)}.
\end{equation}
\end{lemma}

\section{Entropies and their properties}
\label{sec:entropies}

Entropy measures quantify uncertainty, randomness, and correlations in quantum systems,
and they play a central role in modern security proofs for QKD.
In this thesis, we will primarily work with three kinds of entropies \cite{tomamichel_quantum_2016}:
the von Neumann entropy, the family {\Renyi} entropies, and smooth min- and max-entropies.

Operationally, these entropies admit interpretations in terms of fundamental information-processing tasks. For our purposes, we rely on the fact that {\Renyi} entropies and smooth min-entropies quantify the amount of near-uniform randomness that can be extracted in the presence of quantum side information, which is precisely the quantity that enters QKD security proofs (see \cref{lemma:LHL}). Furthermore, under suitable conditions, both of these entropies can be related to the von Neumann entropy. The latter not only characterizes the asymptotic performance of QKD protocols, but is also the quantity that is often computed in our numerical key rate evaluations, for instance in \cref{chap:postselection,chap:variable}.

We will now state the definitions and properties of these entropies which are required later in the thesis. We stress that the definitions are included for completeness, our analysis relies only on structural properties such as relations between conditioned and unconditioned entropies, chain rules, data-processing inequalities, and similar identities. Throughout this thesis, all logarithms (i.e, $\log(x)$) are taken to base $2$.

\subsection{Von Neumann entropy}
\label{subsec:vonNeumann}

\begin{definition}[Von Neumann entropy]
Let $\rho_A \in \dop{=}(A)$ be a density operator. The \term{von Neumann entropy} of
$\rho_A$ is defined as
\begin{equation}
\label{eq:vonNeumannEntropy}
H(A)_\rho \;:=\; - \Tr\!\left[ \rho_A \log \rho_A \right].
\end{equation}
\end{definition}

For a bipartite state $\rho_{AB} \in \dop{=}(AB)$, we define the \term{conditional von
Neumann entropy} by
\begin{equation}
\label{eq:conditionalVonNeumannEntropy}
H(A|B)_\rho \;:=\; H(AB)_\rho - H(B)_\rho.
\end{equation}
The conditional entropy can be negative, but is non-negative if $A$ is classical.

\begin{lemma}[Averaging over conditioning on events]
\label{lemma:vnClassicalConditioningAnalogue}
Let $\rho_{ABCY}$ be a state that is classical on $Y,C$:
\begin{equation}
\rho_{ABCY}
=
\sum_{y \in \Lambda} p(y)\,\rho_{ABC\mid y}\otimes \ket{y}\!\bra{y}_Y,
\end{equation}
where $p(y)$ is a probability distribution over $\Lambda$, and $Y$ can be generated from $C$ (more precisely: $Y \leftrightarrow C \leftrightarrow AB$ forms a Markov chain), where $\{ \ket{y} \}_{y\in\Lambda}$ is an orthonormal basis for $Y$ and each
$\rho_{ABC\mid y}\in\dop{=}(ABC)$. Then the von Neumann conditional entropy satisfies
\begin{equation}
\label{eq:vnConditioningAverage}
H(A | BC) = H(A|BCY)_\rho
=
\sum_{y \in \Lambda} p(y)\, H(A|BC)_{\rho_{\mid y}}.
\end{equation}
\end{lemma}
\begin{proof} The proof follows from  the definition of the entropies and simple algebraic manipulations \end{proof}.

\subsection{Renyi Entropies}

	\begin{definition}[{\Renyi} entropy] \label{def:renyientropy}
		For $\rho \in \dop=(AB)$, and $\alpha \in (0,1) \cup (1,\infty)$, the \term{sandwiched {\Renyi} entropy} of $A$ given $B$ for a state $\rho_{AB}$ is given by
		\begin{equation}
			\Halpha (A|B)_\rho \coloneqq \max_{\sigma_B \in S_\circ(B)} \Halpha(A|B)_{\rho | \sigma}	,
				\end{equation}
		where
		\begin{equation}
			\begin{aligned}
				&\Halpha(A|B)_{\rho| \sigma} \coloneqq \\
				& \begin{cases*}
				\frac{1}{1-\alpha} \log \Tr \left[   \left( \sigma_B^{\frac{1-\alpha}{2\alpha}} \rho_{AB}  \sigma_B^{\frac{1-\alpha}{2\alpha}} \right)^\alpha \right] & if $\rho \in A \otimes \operatorname{supp}(\sigma)$ ,\\
					-\infty     & otherwise.
				\end{cases*}
			\end{aligned}
		\end{equation}
	\end{definition}
	Note that the sandwiched {\Renyi} Entropy may be notated differently in different works: for instance, it is denoted as $H_\alpha(A|B)$ in Ref.~\cite{dupuis_privacy_2023} (Definition 1), and $\widetilde{H}^{\uparrow}_{A|B}$ in Ref.~\cite{tomamichel_quantum_2016} (Definition 5.2), and $H^{\uparrow}_\alpha(A|B)$ in Ref.~\cite{dupuis_entropy_2020} (Definition B.1).  We will now state several results regarding the {\Renyi} entropy that we utilize in this thesis.   We start with results on data processing inequalities. 

\begin{lemma}\label{lemma:DPI}
	(Data-processing \cite[Theorem~1]{frank2013monotonicity}) Let $\rho \in \dop{=}(Q Q')$, take any $\alpha\in [\frac{1}{2},1)\cup(1,\infty]$. Then for any channel $\mathcal{E} \in \CPTP(Q',Q'')$, 
	\begin{align}
		\Halpha(Q|Q'')_{\mathcal{E}[\rho]} \geq \Halpha(Q|Q')_{\rho}.
	\end{align}
	If $\mathcal{E}$ is an isometry, then we have equality in the above bound.
\end{lemma}

The following lemma relates the entropy on a state conditioned on an event, to the entropy of the state without any conditioning. 
\begin{lemma}\label{lemma:conditioning}
    (\cite[Lemma~B.5]{dupuis_entropy_2020})
    Let $\rho_{ABC} \in \dop{=}(ABC)$ be classical on $C$, such that $\rho_{ABC} = \sum_c p_c \ketbra{c}{c} \otimes \rho_{AB|c}$ for some probability distribution $\{p_c\}$ and normalized conditional states $\rho_{AB_{|c}}$. Then, for each $c$ and any $\alpha \in (1, \infty)$, we have:
\begin{align}\label{eq:conditioning}
\Halpha(A|B)_{\rho_{|c}}\geq \Halpha(A|B)_\rho-\frac{\alpha}{\alpha-1}\log\left(\frac{1}{p_c}\right)
\end{align}
\end{lemma}

The following lemma allows us to split off classical registers by subtracting the number of bits of those registers. 
\begin{lemma}\label{lemma:EC_cost}
    (\cite[Proposition~2.9]{LWD16})
    Let $\rho_{ABC} \in \dop{=}(ABC)$ be classical on $C$. Then, for any $\alpha\in(0,\infty)$, we have:
    \begin{align}
        \label{eq:EC_cost}
        \Halpha(A|BC)_\rho\geq \Halpha(A|B)_\rho-\log 
        |C|
    \end{align}
\end{lemma}
The following lemma states that {\Renyi} entropy is additive across tensor products.

\begin{lemma}(Additivity of {\Renyi} Entropy, \cite[Corollary 5.2]{tomamichel_quantum_2016} ) \label{lemma:additivity}
		For any two states $\rho_{AB} \in \dop{=}(AB), \sigma_{CD} \in \dop{=}(CD)$, and $\alpha \geq \frac{1}{2}$, we have
		\begin{equation}
			\Halpha(AC | BD)_{\rho \otimes \sigma}  = \Halpha(A|B)_\rho + \Halpha(C|D)_\sigma.
			\end{equation}
	
	\end{lemma}

The next lemma relates the {\Renyi} entropy to the von Neumann entropy. It is used along with the previous lemma in \cref{chap:variable} to reduce the $n$-round {\Renyi} entropy to $n$ times the single-round von Neumann entropy.

	\begin{lemma}(\cite[Lemma B.9]{dupuis_entropy_2020})\label{lemma:contrenyi}
		For any $\rho_{AB}\in \dop{=} (AB)$, and $1 < \alpha < 1+1/ \log(1+2d_A)$, we have
		\begin{equation}
			\Halpha(A|B)_\rho >  H(A|B)_\rho - (\alpha-1)\log^2(1+2d_A).
		\end{equation}
	\end{lemma}

The following lemma is analogous to \cref{lemma:vnClassicalConditioningAnalogue}, where we average over conditional entropies of states conditioned on events. 
	
	\begin{lemma}  \label{lemma:renyiweightedaverage}
		Let $\rho_{ABCY} = \sum_{y \in \Lambda} p(y) \rho_{ABC| y} \otimes \ketbra{y} \in \dop{=}(ABCY)$ be classical in $Y,C$, where $p(y)$ is a probability distribution over $\Lambda$, and $Y$ can be generated from $C$ (more precisely: $Y \leftrightarrow C \leftrightarrow AB$ forms a Markov chain).  Let $\Lambda^\prime \subseteq \Lambda$. Then, 
		\begin{equation}
			  \sum_{y \in \Lambda^\prime}  p(y)	2^{-\frac{(\alpha-1)}{\alpha} \Halpha( A| BC)_{\rho_{ |y} } } \leq 2^{-\frac{(\alpha-1)}{\alpha} \Halpha( A| BC)_\rho } .
		\end{equation}
	\end{lemma} 
	\begin{proof} We have
		\begin{equation}
			\sum_{y \in \Lambda^\prime}  p(y)	2^{-\frac{(\alpha-1)}{\alpha} \Halpha( A| BC)_{\rho_{ |y} } } \leq \sum_{y \in \Lambda}  p(y)	2^{-\frac{(\alpha-1)}{\alpha} \Halpha( A| BC)_{\rho_{ |y} } },
		\end{equation}
		since we only add positive terms to the expression to go from the LHS to the RHS. Now, on the RHS, $p(y)$ is a normalized probability distribution function over $\Lambda$. Therefore, we can directly use \cite[Proposition 5.1]{tomamichel_quantum_2016}, and we obtain
			\begin{equation}
			\sum_{y \in \Lambda}  p(y)	2^{-\frac{(\alpha-1)}{\alpha} \Halpha( A| BC)_{\rho_{ |y} } } =	2^{-\frac{(\alpha-1)}{\alpha} \Halpha( A| BCY)_{\rho}  }.
		\end{equation}
		Since $Y$ can be generated from $C$,  the fact that $\Halpha( A| BCY) = \Halpha(A|BC)$ follows by applying the data-processing inequality for {\Renyi} entropy  (\cref{lemma:DPI}) in both directions $YC \rightarrow C$ and $C \rightarrow YC$. Therefore, the claim follows.
	\end{proof}

\subsection{Smoothed min Entropies}

\begin{definition}[Conditional Min-Entropy]
Let $\rho_{AB} \in \dop{\leq}(AB)$ be a (possible subnormalized) bipartite quantum state.  
The \term{conditional min-entropy} of $A$ given $B$ is defined as
\begin{equation}
H_\mathrm{\min}(A|B)_\rho 
  \coloneqq - \inf_{\sigma_B \in \dop{\leq}(B)}
    \inf \{ \lambda \in \mathbb{R} : \rho \le 2^{\lambda} \sigma \}.
\end{equation}
Equivalently, $H_{\min}(A|B)_\rho$ is the largest real number $\lambda$ such that
\[
\rho_{AB} \le 2^{-\lambda}\,\id_A \otimes \sigma_B
\]
for some $\sigma_B \in \dop{\leq}(B)$.
\end{definition}

Intuitively, the conditional min-entropy quantifies how unpredictable $A$ is to an observer 
holding the quantum system $B$.  
In the special case where $A$ is classical, it characterizes the optimal probability of 
correctly guessing the value of $A$ given access to $B$:
\[
p_{\text{opt-guess}}(A|B)_\rho = 2^{-H_{\min}(A|B)_\rho}.
\]
Hence, a larger min-entropy corresponds to a smaller guessing probability and thus to 
greater secrecy of $A$ conditioned on $B$.

\begin{definition}[Smooth Entropies]
For $\varepsilon \ge 0$, the \term{$\varepsilon$-smoothed conditional min-entropy} of $A$ given $B$ 
is defined as
\begin{equation}
\Hmin(A|B)_\rho 
  \coloneqq \sup_{\tilde{\rho}_{AB} \in \mathcal{B}^{\varepsilon}(\rho_{AB})}
      H_{\min}(A|B)_{\tilde{\rho}},
\end{equation}
where $\mathcal{B}^{\varepsilon}(\rho_{AB})$ denotes the set of all subnormalized states 
$\tilde{\rho}_{AB} \in \dop{\leq}(AB)$ that are $\varepsilon$-close to $\rho_{AB}$ in purified distance:
\begin{equation}
\mathcal{B}^{\varepsilon}(\rho_{AB})
  = \{\, \tilde{\rho}_{AB} \in \dop{\leq}(AB) : 
        P(\tilde{\rho}_{AB}, \rho_{AB}) \le \varepsilon \,\}.
\end{equation}
Similarly, the \term{$\varepsilon$-smoothed conditional max-entropy} of $A$ given $B$ 
is defined as
\begin{equation}
\Hmax(A|B)_\rho 
  \coloneqq \inf_{\tilde{\rho}_{AB} \in \mathcal{B}^{\varepsilon}(\rho_{AB})}
      H_{\max}(A|B)_{\tilde{\rho}},
\end{equation}
\end{definition}

The smoothing parameter $\varepsilon$ allows for small perturbations of the state 
$\rho_{AB}$ and captures a notion of approximate secrecy:  
$\Hmin(A|B)_\rho$ quantifies the number of nearly uniform and independent bits 
that can be extracted from $A$ given access to $B$, up to an error $\varepsilon$.  We will work with smooth entropies in \cref{chap:EUR} where we will utilize the following statement known as entropic uncertainty relations:

\begin{theorem}[Entropic uncertainty relation with quantum side information  \cite{tomamichel_uncertainty_2011}]
\label{thm:EUR}
Let $\rho_{ABE}\in\dop{=}(ABE)$\footnote{We can also extend this statement to subnormalized states, but we do not require it in this thesis.} be a tripartite state. Consider two POVMs on $A$, given by $\{M_x\}_{x\in\mathcal{X}}$, and $\{N_z\}_{z\in\mathcal{Z}}$, and refer to them as $\Xbasis$ and $\Zbasis$ respectively. Let $X$ and $Z$ be the classical registers that store the classical outcomes obtained by measuring register $A$ with these these POVMs respectively (while leaving the other registers unchanged),
so that the corresponding post-measurement classical--quantum states are
\begin{equation}
\label{eq:EUR_postX}
\rho_{XBE}=\sum_{x\in\mathcal{X}} \ketbra{x}_X \otimes
\Tr_A \left[(M_x \otimes \id_{BE})\rho_{ABE}\right],
\end{equation}
and
\begin{equation}
\label{eq:EUR_postZ}
\rho_{ZBE}=\sum_{z\in\mathcal{Z}} \ketbra{z}_Z \otimes
\Tr_A\left[(N_z\otimes \id_{BE})\rho_{ABE}) \right].
\end{equation}
Then, for any $\varepsilon\in(0,1)$,
\begin{equation}
\label{eq:EUR_statement}
\Hmin(Z|E)_{\rho}
+
\Hmax(X|B)_{\rho}
\;\ge\;
\log\!\left(\frac{1}{c_q}\right),
\end{equation}
where the \term{measurement incompatibility constant} $c_q$ is defined by
\begin{equation}
\label{eq:EUR_cq_def}
c_q
\;:=\;
\max_{x\in\mathcal{X},\,z\in\mathcal{Z}}
\left\lVert \sqrt{M_x}\,\sqrt{N_z}\right\rVert_\infty^{2}.
\end{equation}
\end{theorem}

The quantity $c_q$ quantifies the overlap (or compatibility) of the two POVMs.
If the two POVMs are nearly compatible (large overlap), then $c_q$ is close
to $1$ and the bound in \cref{eq:EUR_statement} becomes weak. Conversely, if the
measurements are highly incompatible (small overlap), then $c_q$ is small and the right-hand
side $\log(1/c_q)$ is large, forcing a strong tradeoff: high predictability of the $\mathsf{X}$
outcome given $E$ implies low predictability of the $\mathsf{Z}$ outcome given $B$, and vice versa. We will see how this is utilized in the QKD security analysis in \cref{subsec:EUR} and \cref{chap:EUR}. We will now state some lemmas that we need to work with these entropies.

The following lemma can be used to remove classical registers from the smooth min-entropy, as long as one subtracts the number of bits of those registers, analogous to \cref{lemma:EC_cost}.

\begin{lemma}(\cite[Lemma 6.8]{tomamichel_quantum_2016})
Let $\rho_{AEC} \in \dop{\leq}(AEC)$ be a (possibly subnormalized) state classical on $C$.  Then, for every $\varepsilon \in (0,1)$,
\begin{equation}
  \Hmin(A|E C)_\rho 
  \geq \Hmin(A|E)_\rho \;-\; \log |C|.
\end{equation}
\end{lemma}

The following lemma is used to relate the entropy conditioned on an event to the entropy without conditioning on the event. Notice that it concerns states that are \emph{partial} on an event (not \emph{conditioned}), i.e, the states that include the events are not normalized.

\begin{lemma}[{\cite[Lemma 10]{tomamichel_largely_2017}}] \label{lemma:gettingridofsubnormalizedconditioning}
Let $\rho_{AEC} \in \dop{\leq}(AEC)$ be a (possibly subnormalized) state classical on $C$, and let $\Omega$ be any event on $C$.  Then, for every $\varepsilon \in (0,\sqrt{\Pr(\Omega)_\rho})$, we have:
\begin{equation}
  \Hmin(A|E C)_{\rho \wedge \Omega} 
  \geq \Hmin(A|EC)_\rho
\end{equation}
\end{lemma}

\begin{lemma}[{Data processing inequality for the smooth min-entropy \cite[Theorem 6.2]{tomamichel_quantum_2016})}]
\label{lemma:DPIsmoothmin}
Let $\rho_{AQ} \in \dop{\leq}(AQ)$ be a (possibly subnormalized) quantum state, and let 
$\mathcal{E} \in \CPTP(Q,Q')$.
Then, for every $\varepsilon\in[ 0,\sqrt{\Tr[\rho]) } )$,
\begin{equation}
\begin{aligned}
 & \Hmin[\varepsilon](A \mid Q)_{(\rho_{AQ})}
  \;\le\;
  \Hmin[\varepsilon](A \mid Q')_{(\id_{A}\otimes\mathcal{E})[\rho_{AQ}]}, \\
 & \Hmax[\varepsilon](A \mid Q)_{(\rho_{AQ})}
  \;\le\;
  \Hmax[\varepsilon](A \mid Q')_{(\id_{A}\otimes\mathcal{E})[\rho_{AQ}]}.
  \end{aligned}
\end{equation}
\end{lemma}

\section{Quantum Optics} \label{sec:backgroundquantumoptics}
QKD protocols are implemented using quantum optics, with information encoded in
photons of light. As a result, it is necessary to understand the basic principles
of quantum optics in order to construct accurate mathematical models of the
quantum operations performed by the honest parties. In particular, we must
understand the mathematical descriptions of states prepared by Alice and the measurements performed by Bob.
For an introductory treatment of quantum optics, see Ref.~\cite{gerry_introductoryquantumoptics_2023}.

\subsection{Optical modes}

In classical electrodynamics, an optical mode refers to an element of a chosen
orthonormal basis of solutions to Maxwell’s equations for the electromagnetic
field (or equivalently, for the vector potential) in vacuum, subject to
appropriate boundary conditions. A general classical field configuration can be
expressed as a linear combination of such modes.

Upon quantization, the electromagnetic field is promoted to an operator-valued
field. In the canonical quantization procedure, the complex amplitudes
associated with each classical mode are replaced by annihilation and creation
operators. Each optical mode is then mathematically equivalent to an
independent quantum harmonic oscillator.

It is important to emphasize that an optical mode is simply a choice of basis
used to describe the quantum optical system. In principle, one may choose any
complete orthonormal mode basis (for example, plane-wave modes, cavity modes,
or wavepacket modes). Each choice comes equipped with mode functions that
determine how physical quantities of interest such as the electric and magnetic
field operators vary in space and time. For the purposes of this thesis, we do
not need to work explicitly with the spatial structure of the mode functions,
and we therefore focus only on the associated mode operators.

Let $\hat{a}$ and $\hat{a}^\dagger$ denote the annihilation and creation
operators of a given optical mode, and let $\vac$ denote the vacuum state,
corresponding to the absence of photons in that mode. These operators satisfy
the canonical commutation relation
\begin{equation}
    [\hat{a}, \hat{a}^\dagger] = 1 .
\end{equation}

In this thesis, we consider state-preparation procedures that produce either
(i) ideal single-photon qubit states\footnote{A single photon state over two modes is exactly identical to qubit.} or (ii) phase-randomized weak coherent
states. Ideal single-photon states are useful as a conceptual model and as a
benchmark for security analyses, while weak coherent states provide an accurate
description of practical optical sources (lasers) used in quantum key distribution.
To describe both cases, we first introduce Fock  states.

\subsubsection{Fock states}

Fock states, also known as photon-number states, are eigenstates of the number
operator $\hat{a}^\dagger \hat{a}$. We denote these states by $\ket{n}$,
where $N \in \mathbb{N}$ denotes the eigenvalue of the state. That is, they satisfy
\begin{equation}
    \hat{a}^\dagger \hat{a} \ket{N} = N \ket{N}.
\end{equation}
The vacuum state is given by $\vac$, where we reserve the use of $\ket{0}$ to describe one of the basis states of a qubit, and use $\vac$ to denote the zero photon vacuum state instead. Higher-number states can
be generated by repeated application of the creation operator,
\begin{equation}
    \ket{N} = \frac{(\hat{a}^\dagger)^N}{\sqrt{N!}} \vac.
\end{equation}
The set $\{ \ket{N} : N \in \mathbb{N} \}$ forms a complete orthonormal basis for
the Hilbert space of a single optical mode.

The action of the annihilation and creation operators on Fock states is given by
\begin{align}
    \hat{a} \ket{N} &= \sqrt{N}\,\ket{N-1}, \\
    \hat{a}^\dagger \ket{N} &= \sqrt{N+1}\,\ket{N+1},
\end{align}
for all $n \in \mathbb{N}$, with the convention that $\hat{a}\vac = 0$.

\subsubsection{Single-photon states}

A single-photon state corresponds to the Fock state $\ket{1}$. In practice,
photons possess additional degrees of freedom, such as polarization, temporal
mode, or spatial mode. When restricting attention to a fixed optical mode, the
single-photon subspace is one-dimensional. However, when additional degrees of
freedom are considered—most notably polarization—the single-photon subspace can
be used to encode a qubit.

For example, fixing all degrees of freedom except polarization, one may define
two orthogonal single-photon states $\ket{1}_H = \hat{a}^\dagger_H \vac $ and $\ket{1}_V = \hat{a}^\dag_V \vac$ corresponding to
horizontal and vertical polarization, respectively. An arbitrary single-photon
qubit state can then be written as a superposition
\[
    \alpha \ket{1}_H + \beta \ket{1}_V,
\]
where $\alpha, \beta \in \mathbb{C}$ and $|\alpha|^2 + |\beta|^2 = 1$.

\subsubsection{Coherent states}

Coherent states are defined as eigenstates of the annihilation operator. For a
complex amplitude $\alpha \in \mathbb{C}$, the coherent state $\ket{\alpha}$
satisfies
\begin{equation}
    \hat{a} \ket{\alpha} = \alpha \ket{\alpha}.
\end{equation}
Coherent states can be expressed in the Fock basis as
\begin{equation} \label{eq:coherentstate}
    \ket{\alpha}
    \coloneq e^{-|\alpha|^2/2}
      \sum_{N=0}^{\infty}
      \frac{\alpha^N}{\sqrt{N!}} \ket{N}.
\end{equation}
The mean photon number of the coherent state is given by
\begin{equation}
    \mu := \langle \alpha | \hat{a}^\dagger \hat{a} | \alpha \rangle = |\alpha|^2.
\end{equation}
One can write $\alpha$ as $ \alpha = \sqrt{\mu}\,e^{i\theta}$ where $\theta \in [0,2\pi]$ corresponds to the phase and $\mu \in \mathbb{R}$ corresponds to the intensity of the coherent state. 

In many practical settings the optical phase is not fixed but is instead
uniformly randomized. We refer to the resulting state as a \emph{
phase-randomized} coherent state. It is
defined as the mixture
\begin{equation}
\label{eq:phase_randomized_def}
    \rho^{\mathrm{PR}}_{\mu}
    := \frac{1}{2\pi}\int_{0}^{2\pi} d\theta \;
    \ket{\sqrt{\mu}e^{i\theta}}\!\bra{\sqrt{\mu}e^{i\theta}} .
\end{equation}
Using the Fock-basis expansion of coherent states (\cref{eq:coherentstate}),
one finds that phase randomization removes the off-diagonal coherences in the
photon-number basis, yielding a state that is diagonal in the Fock basis:
\begin{equation}
\label{eq:phase_randomized_poisson}
    \rho^{\mathrm{PR}}_{\mu}
    = \sum_{N=0}^{\infty} \poissonian{\mu}{N}\,\ketbra{N},
\end{equation}
where $\poissonian{\mu}{N}$ is the Poisson distribution with mean $\mu$,
\begin{equation}
\label{eq:poisson_dist}
 \poissonian{\mu}{N} = e^{-\mu}\frac{\mu^N}{N!}, \qquad N \in \mathbb{N}.
\end{equation}
In particular, $\rho^{\mathrm{PR}}_{\mu}$ may be interpreted as a classical
mixture of $N$-photon Fock states, where the photon number $N$ is distributed
according to $\poissonian{\mu}{N}$.

\subsection{Detectors}
\label{sec:detectors}
In QKD, Bob’s measurement apparatus is  implemented using standard linear-optical
elements such as beam splitters (BS) and polarizing beam splitters (PBS) followed
by (single-photon) threshold detectors. In this thesis we model these components
at a level sufficient to describe the POVMs induced on the relevant optical
modes, as is standard for QKD security analysis. 
\subsubsection{Beam splitters}
\label{sec:beamsplitter}

\begin{figure}[t]
\centering
\begin{tikzpicture}[x=1cm,y=1cm]

    \pic at (0,0) {beamsplitter={BS,{BS},(0.25,-0.55)}};

    \draw (-2,0) -- (-0,0);
    \node[left] at (-2,0) {$\hat a_{\mathrm{in}}$};

    \draw (0,-2) -- (0,-0);
    \node[below] at (0,-2) {$\hat b_{\mathrm{in}}$};

    \draw (0,0) -- (2,0);
    \node[right] at (2,0) {$\hat a_{\mathrm{out}}$};

    \draw (0,0) -- (0,2);
    \node[above] at (0,2) {$\hat b_{\mathrm{out}}$};

\end{tikzpicture}
\caption{Schematic of a beam splitter with two input ports (left and bottom) and two output ports (right and top).}
\label{fig:beamsplitter_ports}
\end{figure}
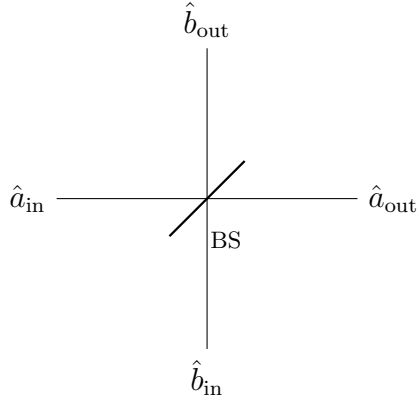

A (lossless) beam splitter is a two-mode linear optical element that mixes two
input modes into two output modes via a unitary transformation (see \cref{fig:beamsplitter_ports}). Let $\hat{a}$ and
$\hat{b}$ denote annihilation operators for two input spatial modes. A beam
splitter with amplitude transmissivity $t$ and reflectivity $r$ (satisfying
$|t|^2+|r|^2=1$) induces the following mode transformation
\begin{equation}
\label{eq:BS_transform}
\begin{pmatrix}
\hat{a}_{\mathrm{out}} \\
\hat{b}_{\mathrm{out}}
\end{pmatrix}
=
\begin{pmatrix}
t & r \\
-r^* & t^*
\end{pmatrix}
\begin{pmatrix}
\hat{a}_{\mathrm{in}} \\
\hat{b}_{\mathrm{in}}
\end{pmatrix}.
\end{equation}
A common convention for a symmetric $50{:}50$ beam splitter is
$t = 1/\sqrt{2}$ and $r = i/\sqrt{2}$, for which
\[
\hat{a}_{\mathrm{out}}=\frac{\hat{a}_{\mathrm{in}}+i\hat{b}_{\mathrm{in}}}{\sqrt{2}},
\qquad
\hat{b}_{\mathrm{out}}=\frac{i\hat{a}_{\mathrm{in}}+\hat{b}_{\mathrm{in}}}{\sqrt{2}}.
\]
Beam splitters are used throughout polarization-based BB84 receivers, to select a measurement basis passively or to route light between
different interferometric paths.

\subsubsection{Polarizing beam splitters}
\label{sec:PBS}

\begin{figure}[t]
\centering
\begin{tikzpicture}[x=1cm,y=1cm]

    \pic at (0,0) {beamsplitter={PBS,{PBS},(0.25,-0.55)}};

    \draw (-2,0) -- (-0,0);
    \node[left] at (-2,0) {$ x_H \hat{a}_{H} + x_V \hat{a}_{V}$};

    \draw (0,0) -- (2,0);
    \node[right] at (2,0) {$\hat a_{H}$};

    \draw (0,0) -- (0,2);
    \node[above] at (0,2) {$\hat a_{V}$};

\end{tikzpicture}
  \caption{Idealized polarizing beam splitter (PBS) that routes the horizontal and vertical polarization components of the input mode  into separate spatial output modes.}
    \label{fig:pbs}
\end{figure}
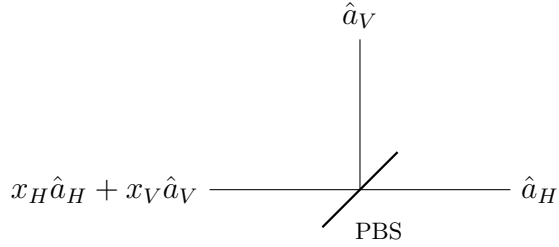

A polarizing beam splitter (PBS) separates two orthogonal polarizations (e.g.,
horizontal $H$ and vertical $V$) into different spatial output modes (see \cref{fig:pbs}). At an
idealized level, a PBS acts as a polarization-dependent router: it transmits one
polarization and reflects the orthogonal polarization. In a polarization-encoded
BB84 receiver, a PBS is typically used to map the two orthogonal polarization
outcomes to two separate detectors.

\subsubsection{Threshold detectors}
\label{subsubsec:threshold}

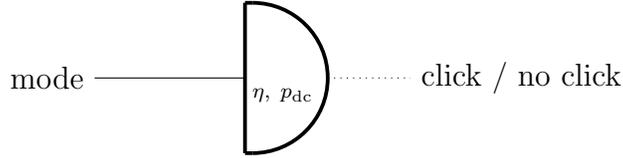
\begin{figure}[t]
    \centering
    \begin{tikzpicture}[x=1cm,y=1cm]

    \draw (-2,0) -- (0,0);
    \node[left] at (-2,0) {mode};

    \pic at (0,0)
      {detectorSmall={det,{\scriptsize $\eta,\;p_{\mathrm{dc}}$},below}};

    \draw[dotted] (det-out) -- (2.2,0);
    \node[right] at (2.2,0) {click / no click};

\end{tikzpicture}
    \caption{Schematic of a threshold (on/off) detector. In the imperfect model, $\eta$ denotes the overall detection efficiency and $p_{\mathrm{dc}}$ the dark-count probability per detection window.}
    \label{fig:threshold_detector}
\end{figure}

A threshold detector is a binary (on/off) photodetector: it does not measure the
photon number, but only reports whether at least one photon was detected (see \cref{fig:threshold_detector}). In the
idealized perfect case (unit efficiency, no dark counts), the measurement has
outcomes \emph{no click} and \emph{click}, described by the POVM
\begin{equation}
\label{eq:ideal_threshold_povm}
\Gamma_{\mathrm{noclick}} \coloneq \ketbra{\mathrm{vac}}, \qquad
\Gamma_{\mathrm{click}} \coloneq \id - \ketbra{\mathrm{vac}} = \sum_{n=1}^{\infty} \ketbra{n},
\end{equation}
where the operators act on the Hilbert space of the measured optical mode, that is, the mode going into the detector.

Realistic threshold detectors are imperfect. Two dominant non-idealities are:
\begin{itemize}
    \item \term{Finite detection efficiency:} the probability that an incoming
    single photon produces a detection event is typically less than $1$. This
    effect is modeled by a parameter $\eta \in [0,1]$, referred to as the
    (overall) detection efficiency.
    \item \term{Dark counts:} the detector may produce a click even in the
    absence of incident photons. Experimentally, dark counts are often reported
    as a rate (e.g., counts per second). For theoretical analysis, however, it is
    convenient to work with the probability of a dark count occurring within a
    given detection time window. We denote this probability by
    $p_{\mathrm{dc}} \in [0,1]$.
\end{itemize}

These effects are commonly modeled as follows. Loss is treated as an effective
beam splitter of transmissivity $\eta$ placed in front of an otherwise ideal
threshold detector, so that each photon is independently transmitted to the
detector with probability $\eta$. Dark counts are modeled as classical
post-processing: independently of the incoming optical state, the detector
produces a click with probability $p_{\mathrm{dc}}$ during the detection time
window.

Taken together, this model implies that, when the detector is illuminated by an
$n$-photon Fock state $\ket{n}$, the probability of obtaining no click is
\begin{equation}
\label{eq:noclick_prob}
\Pr(\mathrm{noclick}\mid n)
= (1-p_{\mathrm{dc}})\,(1-\eta)^n,
\end{equation}
i.e., no dark count occurs and all $n$ photons fail to be detected. Equivalently,
the corresponding POVM elements describing the imperfect threshold detector are
\begin{equation}
\label{eq:threshold_povm_imperfect}
\Gamma^\mathrm{noisy}_{\mathrm{noclick}}
= (1-p_{\mathrm{dc}})\sum_{n=0}^{\infty}(1-\eta)^n \ketbra{n},
\qquad
\Gamma^\mathrm{noisy}_{\mathrm{click}} = \id - \Gamma^\mathrm{noisy}_{\mathrm{noclick}}.
\end{equation}
This simple model is widely used in QKD security analyses, as it accurately
captures the statistics of threshold detection while avoiding the introduction
of additional internal detector degrees of freedom.

Having covered the basics of quantum optics, we will specify the relevant details of state preparation and measurement in QKD protocols in \cref{subsec:protocolvariations}. 

\section{Miscellaneous}
In this section we collect some remaining miscellaneous concepts. We start with the definition of universal$_2$ hashing, which is a critical ingredient of QKD protocols. 

\begin{definition}\label{def:2universal}($\delta$-almost-universal$_2$ and universal$_2$ hash families) 
Given a domain $D$ and a finite codomain $D'$, a \term{$\delta$-almost-universal$_2$} ($\delta$-AU$_2$) hash family consists of a set $\mathcal{F}$ of functions $f:D \to D'$, together with a probability distribution $P_\mathrm{HASH}$ over $\mathcal{F}$. The defining property is that, if $h$ is sampled according to $P_\mathrm{HASH}$, then for all distinct $x,y \in D$, the probability of a collision is bounded as, 
\begin{align}\label{eq:AUdefn}
\Pr[h(x) = h(y)] \leq \delta,
\end{align}
where the probability is taken over the choice of $h$. A special case is \term{universal$_2$ hashing}, which corresponds to
\begin{align}
\delta = \frac{1}{|D'|}.
\end{align}
A further special case is \term{ideal universal$_2$ hashing}, which corresponds to $\delta = \frac{1}{|D'|}$ and equality holding in \cref{eq:AUdefn}, i.e.~for all distinct $x,y \in D$,
\begin{align}
\Pr[h(x) = h(y)] = \frac{1}{|D'|}.
\end{align}
\end{definition}

It is straightforward to show that the family of maps from
$m$ bits to $n$ bits defined by matrix multiplication, after choosing an $n \times m$ binary matrix uniformly
at random, is an ideal universal$_2$ family of hash functions. However, this
construction is not typically used in QKD applications, as it requires a seed of
length $mn$ bits to specify the matrix. Instead, structured families of universal hash functions with much smaller seed
requirements are employed. In particular, Toeplitz hashing is known to be ideal
universal$_2$ \cite{mansour_computational_1993}, while requiring only $(m+n-1)$ bits
of seed randomness.%
\footnote{A Toeplitz matrix is a matrix whose entries are constant along each
diagonal. That is, an $n \times m$ binary Toeplitz matrix $T$ satisfies
$T_{i,j} = t_{i-j}$ for some binary sequence $\{t_k\}$, so the entire matrix is
fully specified by its first row and first column, requiring $m+n-1$ bits in
total.}

We will also work with frequency distributions and probability distributions, which are defined below. 

\begin{definition}\label{def:freq}
(Frequency distributions) For a string $z_1^n\in\mathcal{Z}^n$ on some alphabet $\mathcal{Z}$, $\freq_{z_1^n}$ denotes the following probability distribution on $\mathcal{Z}$:
\begin{align}
\freq_{z_1^n}(z) \coloneq \frac{\text{number of occurrences of $z$ in $z_1^n$}}{n} .
\end{align}
\end{definition}
We denote the set of probability distributions over an alphabet $\mathcal{Z}$, via $\mathbb{P}(\mathcal{Z})$. Note that we have, by definition, $\freq_{z_1^n} \in \mathbb{P}(\mathcal{Z})$.

\subsection{Numerics} \label{subsec:numerics}

We will often be tasked with solving optimization problems of the form
\begin{equation}
\begin{aligned}
\inf_\rho \; & f(\rho) \\
\text{subject to: } \; & \text{linear constraints on } \rho, \\
& \rho > 0
\end{aligned}
\end{equation}
where $f(\rho)$ is a convex function. Loosely speaking, this can be interpreted as optimizing over the worst-case attack by an adversary that is compatible with the observed statistics. The objective function $f(\rho)$ is typically some single-round entropic quantity of interest, $\rho>0$ ensures that we get a valid quantum state, and compatibility with observations is enforced via linear constraints on $\rho$.

The main challenge in performing these computations is that we require a value that is guaranteed to be a \emph{lower bound} on the true infimum. Consequently, some common optimization approaches, such as gradient descent, are not immediately suitable, as they do not provide guarantees of global optimality. Fortunately, several works have developed numerical methods that aim to produce values below the true QKD infimum \cite{kamin_renyi_2025,winick_reliable_2018,hu2022robust,navarro_finite_2025,he2024qics,lorente2025quantum,chung2025generalized,kossmann2024optimising,wang2019characterising} for a variety of relevant optimization problems.

In this thesis, we will \emph{not} go into the details or nuances of performing these computations, and instead rely on established work that addresses these tasks. Nevertheless, we note that this topic can be subtle and technically involved. Practical constraints such as numerical precision and solver tolerances often complicate attempts to obtain reliable lower bounds. Moreover, various modifications (with appropriate justification) are frequently applied to reduce the complexity of the problem before it is passed to a numerical routine. We do not discuss these details in this thesis, and instead refer to the specific papers that develop these techniques for more details. In particular, we primarily rely on Ref.~\cite{winick_reliable_2018} for the computations in \cref{chap:variable,chap:postselection}, and its extension to {\Renyi} entropies \cite{kamin_renyi_2025} in \cref{chap:MEAT}.

\chapter{Basics of Quantum Key Distribution} 
\epigraph{
Where we explain what QKD is and why it works; where we introduce the central
tools used in practical QKD security analysis; and where we outline the range of
proof techniques available for QKD security analysis.
}{}

\label{chap:qkdbackground}
\begin{figure}
    \centering
    \includegraphics[width=1\linewidth]{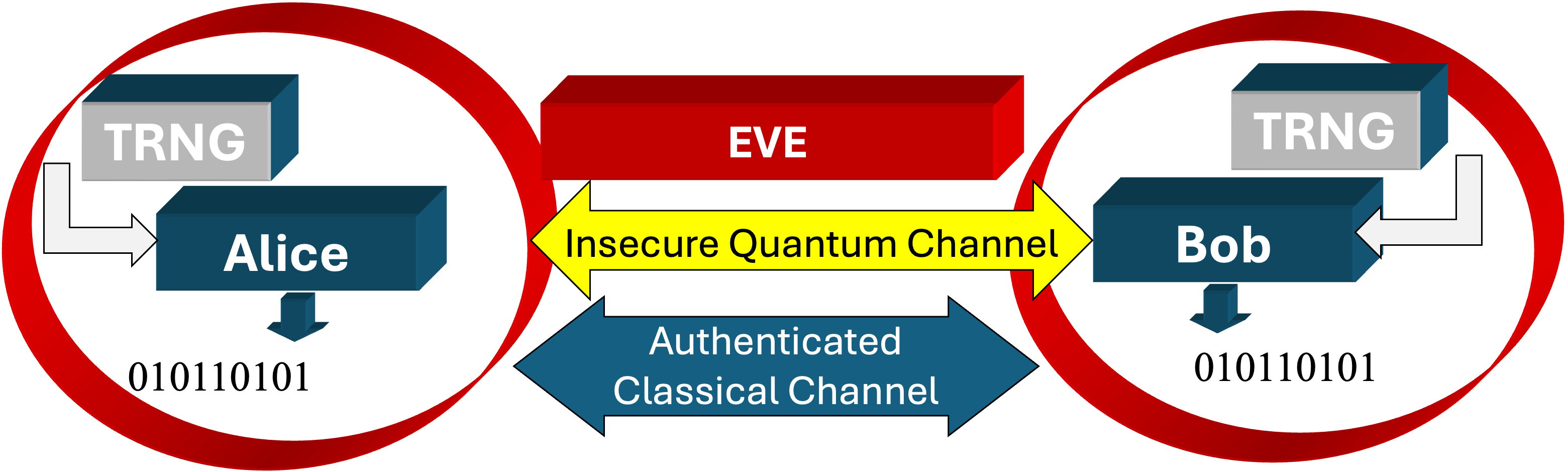}
    \caption{Schematic of a quantum key distribution (QKD) protocol. The task is to
establish a shared secret key between two distant parties, Alice and Bob. The
protocol relies on trusted quantum devices operated within secure perimeters,
access to local true random number generators (TRNGs), an authenticated
classical channel, and an insecure quantum channel that may be fully controlled
by an adversary. Figure from \cite{NL_lecturenotes}.}
    \label{fig:QKDschematic}
\end{figure}

In this chapter, we cover the basic concepts and tools of QKD security analysis
that are used throughout this thesis. We restrict our attention to
prepare-and-measure QKD protocols, as these are the primary focus of this thesis.
Nevertheless, many of the tools presented here also apply to
entanglement-based protocols, and in some cases can be extended to
measurement-device-independent (MDI) QKD \cite{lo_Measurementdeviceindependent_2012}.

Quantum key distribution (QKD) enables two distant parties, Alice and Bob, to
establish a shared secret key by exchanging quantum states and authenticated
classical messages (see \cref{fig:QKDschematic}). Both Alice and Bob are assumed to have access to local randomness if needed\footnote{Atleast one party is required to utilize local randomness to create the initial raw data in QKD.}, which they utilize in making random choices. This randomness is assumed to be perfect, and uncorrelated to the adversary. Alice uses this randomness to prepare quantum states and send them to Bob, who
performs measurements (potentially using his local randomness) on the received systems. The quantum channel connecting Alice and Bob is assumed to be completely insecure: the adversary Eve is allowed
to perform arbitrary operations on the quantum systems leaving Alice’s laboratory before they enter
 Bob’s laboratory. The intuitive reason why QKD is possible
is that Alice sends quantum states that are not all mutually orthogonal.
Consequently, no party (including Eve) can perfectly distinguish all possible
signal states sent by Alice. 

However, secrecy alone is not sufficient: Alice and Bob must also end up with
\emph{identical} keys. At first glance, this appears contradictory, since Bob
must be able to correctly infer Alice’s data, while we have just argued that no
measurement can perfectly distinguish all of Alice’s states. This tension is
resolved as follows. After Alice has sent the quantum states and Bob has
performed his measurements, Alice announces (over a classical channel) which
\emph{subset} each signal state belongs to. These subsets are typically chosen so that
the states within each subset are perfectly distinguishable. In some rounds,
Bob’s measurement is compatible with the announced subset, allowing him to infer
Alice’s data with high accuracy; in other rounds, it is not, and those rounds are discarded.
This procedure necessarily requires classical communication between Alice and
Bob.

Classical communication is also required for several other steps of the
protocol. Alice and Bob must inspect parts of their data to decide whether to
accept or abort the protocol, and to determine the length of the final key they
aim to generate. Moreover, in any practical implementation, noise and
imperfections lead to discrepancies between Alice’s and Bob’s data, making
it necessary to perform classical post-processing steps such as error
correction, error verification, and privacy amplification. These procedures are
discussed in detail in \cref{sec:protdescriptionqkdbackground}.

Throughout this thesis, Alice and Bob are assumed to have access to an
authenticated classical channel. Except in \cref{chap:classicalauthentication},
this assumption is taken to mean that all classical messages sent by one honest
party are eventually received correctly by the other party, without modification.
The adversary is allowed to read all classical messages, but is assumed not to
tamper with them. This is a standard assumption in QKD security proof literature. In
\cref{chap:classicalauthentication}, we examine this assumption in detail, and show that while it is unrealistic, the security analysis undertaken with this unrealistic assumption can then be subsequently lifted to hold under weaker and more realistic assumptions (see \cref{remark:authenticationsecdef} in the next section).

This chapter is organized as follows. In \cref{sec:securitydefinition}, we
introduce the security definition of QKD, and define the two requirements of \emph{secrecy} and \emph{correctness}. In
\cref{sec:protdescriptionqkdbackground}, we present a generic description for
prepare-and-measure QKD protocols, which we analyze in subsequent chapters. In \cref{sec:statesandmeasurements}, we  describe the states prepared, measurements performed, announcements made, and sifting procedures for the protocols studied in this thesis. In \cref{sec:correctnessviaEV} we show that correctness is satisfied for the protocol we study, and thus for the rest of this thesis, we focus on proving secrecy. In \cref{sec:LHL}, we state
the Leftover Hashing Lemma and explain its central role in proving the secrecy of QKD
protocols. In
\cref{subsec:boundingentropy}, we explain the various kinds of entropic bounds that are needed to use the Leftover Hashing Lemma, and briefly discuss how such bounds can be obtained. In \cref{sec:qkdtoolbox}, we present some tools that are widespread in QKD security analysis, such as source-replacement schemes \cite{bennett_quantum_1992,curty_entanglement_2004}, source maps \cite{gottesman_security_2004}, squashing maps \cite{tsurumaru_security_2008,beaudry_squashing_2008,tsurumaru_squash_2010,gittsovich_squashing_2014,zhang_security_2021}. Finally, in \cref{sec:qkdprooftechniques} we briefly discuss the various QKD proof techniques, which are expanded upon in later chapters.

\section{Security Definition} \label{sec:securitydefinition}
Let us focus on the output state of a generic QKD protocol, defined on the registers $K_A K_B \Esecdef$.  
Alice and Bob possess classical registers $K_A$ and $K_B$, which encode keys of arbitrary length.  
This is formalized by modelling $K_A$ as a direct sum
\[
    K_A = \bigoplus_{l_A} K_A^{l_A},
\]
where $K_A^{l_A}$ is a classical register holding a key of length $l_A$.  
An analogous decomposition is used for $K_B$.  
We treat any party aborting as outputting a key of length~$0$, represented by a special symbol~$\bot$.  
The register $\Esecdef$ denotes all of Eve's information at the end of the protocol, and may include a copy of all classical communication.

For the entirety of this thesis, except in \cref{chap:classicalauthentication}, we consider a setting in which the QKD protocol always terminates with Alice and Bob outputting keys of the \emph{same} length. In the communication model we adopt (where all messages are transmitted faithfully, and delivered some  time after they are sent), such a  requirement can be enforced by appropriate protocol design, for instance by having Alice and Bob exchange their respective output key length values, ensuring that the lengths match.

\begin{remark} \label{remark:authenticationsecdef}
    Realistic channels do not provide such strong guarantees, since Eve can always block, reorder, or delay classical messages.  
Nevertheless, in \cref{chap:classicalauthentication} we show that any security analysis carried out under the assumption of honest authentication (i.e., all messages are delivered faithfully at some point after being sent, with no adversarial interference) can be \emph{lifted} to the more realistic setting where Eve may attack the classical authentication mechanism.  This lifting procedure is entirely generic, and does not depend on the details of the QKD protocol.  For simplicity, we therefore adopt the standard honest authentication assumption throughout the initial parts of this thesis, and later show in \cref{chap:classicalauthentication} how this assumption can be removed. 
\end{remark}

In this case, the output state can be written as
\begin{equation} \label{cref:realoutputstatesymmetric}
\rho^\text{real}_{K_A K_B \Esecdef} \coloneq \bigoplus_{l} \Pr(\Omega_{\mathrm{len}=l}) \rho^\mathrm{real}_{K_A^{l} K_B^{l}  \Esecdef | \Omega_{\mathrm{len}=l}}
\end{equation}
where $\Omega_{\mathrm{len}=l}$ denotes the event that Alice and Bob produce a key of length $l$. The ideal output state $\rho^\mathrm{ideal}_{K_A K_B  \Esecdef}$ is defined to be the one obtained by acting a map $\idealmap \in \CPTP(K_A K_B, K_A K_B)$ acting on the real output state $\rho^\mathrm{real}_{K_A K_B  \Esecdef}$ as
\begin{equation}
       \rho^\mathrm{ideal}_{K_A K_B  \Esecdef} \coloneq \idealmap\left[\rho^\mathrm{real}_{K_A K_B  \Esecdef} \right]. 
\end{equation}
The map $\idealmap$ looks at the length of the keys stored in registers $K_A, K_B$ to compute $l$, and  replaces the $K_A,K_B$ registers with the ideal state\footnote{If $l=0$, then $\tau^{l,l}_{K_A K_B} = \ketbra{\bot \bot}_{K_A K_B}$.} 
  \begin{equation}\label{eq:taukAkBsymmetric}
\begin{aligned}
\tau^{l, l}_{K_A K_B} = \displaystyle \frac{1}{2^{l}} \sum_{k \in \{0,1\}^{l}} \ketbra{kk}_{K_A K_B}
\end{aligned}
\end{equation}
Thus intuitively, any key obtained from the ideal state is safe to use, since it is ``secret'' against any side-information registers.  We can now state the security definition of QKD.

\begin{definition}(QKD Security with symmetric aborts \cite{ben-or_universal_2004,portmann_security_2022}) \label{def:qkdsecuritysymmetric}
Let $\QKDprotocol$ be a QKD protocol, and let $\rho^\text{real}_{K_A K_B  \Esecdef}$ be the output state of the QKD protocol, and let $\mathcal{W}\left( \QKDprotocol \right)$ denote the set of possible output states of the QKD protocol. Suppose that the output key lengths for all output states in $\mathcal{W}\left( \QKDprotocol \right)$ are always equal.  Let $\rho^\text{ideal}_{K_A K_B  \Esecdef}$ be the ideal output state, obtained by acting the map $\idealmap$ on the actual output state. That is
\begin{equation} \label{eq:secdefrealidealsymmetric}
    \begin{aligned}
        \rho^\text{real}_{K_A K_B  \Esecdef} &\coloneq \bigoplus_{l} \Pr(\Omega_{\mathrm{len}=l}) \rho^\text{real}_{K_A^{l} K_B^{l} \Esecdef| \Omega_{\mathrm{len}=l}}, \\
        \rho^\text{ideal}_{K_A K_B  \Esecdef}  &\coloneq \idealmap \left[ \rho^\text{real}_{K_A K_B  \Esecdef}   \right].
    \end{aligned}
\end{equation}
Then, the (variable-length) QKD protocol is $\epssecure$-secure if, for all output state $\rho^\text{real}_{K_A K_B \Esecdef} \in \mathcal{W}(\QKDprotocol)$\footnote{Throughout this thesis, we will often work with equivalent descriptions of the QKD protocol and of the security definition. These descriptions are equivalent due to the fact that they all characterize the same correct set of possible output states.}, the following inequality is satisfied\footnote{Note that in Ref.~\cite{ferradini2025definingsecurityquantumkey}, the trace norm appearing in the security definition is not divided by $2$. In contrast, the typical security definition \cite{ben-or_universal_2004,portmann_security_2022} includes the explicit factor of $1/2$.  The definition used in Ref.~\cite{ferradini2025definingsecurityquantumkey} is deliberate and well motivated within that work; we stress that the difference amounts only to an overall factor of $2$ in the  security parameter.}:
\begin{equation} \label{eq:tracedistsecdefsymmetric}
    \tracedist{ \rho^\text{real}_{K_A K_B \Esecdef} -  \rho^\text{ideal}_{K_A K_B  \Esecdef}} \leq \epssecure.
    \end{equation}
If one specializes to a case where the QKD protocol outputs a key of some \emph{fixed} length $l_\mathrm{fixed}$ conditioned on a particular acceptance event $\Omega_\mathrm{acc}$, and aborts otherwise (i.e.~produces a key of zero bits upon the event $\Omega^\complement_\mathrm{acc}$),  then we refer to it as a fixed-length protocol. In this case, the states in \cref{eq:secdefrealidealsymmetric} simplify further to the special form 
\begin{equation}
    \begin{aligned}
        \rho^\text{real}_{K_A K_B \Esecdef} &\coloneq  \Pr(\Omega_\mathrm{acc}) \rho^\text{real}_{K_A K_B  \Esecdef| \Omega_\mathrm{acc}} + \Pr(\Omega^\complement_\mathrm{acc}) \rho^\text{real}_{K_A K_B \Esecdef| \Omega^\complement_\mathrm{acc}},  \\
        \rho^\text{ideal}_{K_A K_B  \Esecdef}  &\coloneq \idealmap \left[ \rho^\text{real}_{K_A K_B \Esecdef}   \right],
    \end{aligned}
\end{equation}
and thus the above $\epssecure$-security condition (for a fixed-length QKD protocol) simplifies to the condition that all output states $\rho^\text{real}_{K_A K_B  \Esecdef} \in \mathcal{W}(\QKDprotocol)$ satisfy
\begin{equation} \label{eq:tracedistsecdeffixed}
    \tracedist{ \rho^\text{real}_{K_A K_B  \Esecdef} -  \rho^\text{ideal}_{K_A K_B  \Esecdef}}  = \Pr(\Omega_\mathrm{acc})  \tracedist{ \rho^\text{real}_{K_A K_B  \Esecdef | \Omega_\mathrm{acc}} -  \rho^\text{ideal}_{K_A K_B \Esecdef| \Omega_\mathrm{acc}}} \leq \epssecure,
\end{equation} 
since the distance between the real and ideal states is zero when the protocol aborts ($\Omega^\complement_\mathrm{acc}$ occurs).
\end{definition}
Thus, a QKD protocol is said to be secure if it always outputs a state that is
close to an ideal state. In fact, the above definition can be shown to be
universally composable \cite{portmann_security_2022,ben-or_universal_2004,broadbent_2023},
meaning that the generated key can be safely used as a subroutine in arbitrary
cryptographic applications without compromising security, even when composed
with other protocols and executed concurrently. While composability is a subtle concept which we do not discuss here, one can already see its essential
content directly from the trace-distance criterion. In particular, it is easy to show that replacing
the ideal QKD output state with the real output state at any point within a
larger cryptographic protocol can only change the probability of an event (in the larger cryptographic protocol) by
at most $\epssecure$. Thus, for the purposes of the security analysis of the larger protocol, one can assume that one has access to the ideal QKD state.

\subsection{Performance}

Notice that the security definition itself says nothing how often a key is produced. In particular, the probability that the protocol accepts or aborts depends on the attack strategy employed by Eve, and no restrictions are imposed on this behaviour. From a purely security-theoretic perspective, a protocol that aborts all the time is therefore perfectly secure.

\epigraph{``A plane that doesn’t fly usually doesn’t crash.''}{Renato Renner \cite{arrazolablog}.}

For this reason, security alone is not a sufficient criterion for judging the usefulness of a QKD protocol. One must additionally consider its performance. 

This requires specifying an honest behaviour of the protocol, which specifies the realistic behaviour of the protocol when no adversary is present. A common performance criterion is \emph{robustness}, which guarentees that the protocol does not abort too often for the specified honest behaviour. In this case, one shows that, under a specified honest model, the protocol accepts with high probability at least $1-\varepsilon_{\mathrm{robust}}$. 

An alternative, and often more informative, figure of merit is the \emph{expected key rate}. This is the average performance of the protocol over multiple runs, under the assumed honest behaviour. The expected key rate captures both the probability of acceptance and the amount of key generated upon acceptance.\footnote{Note that the expected key rate comes with its own subtleties. Consider any QKD protocol that is $\varepsilon$-secure, with some expected key length. Now, modify this protocol so that Alice first tosses a coin with probability $\varepsilon_\mathrm{coin}$ of landing tails. If the coin lands heads, she implements the original QKD protocol; if it lands tails, she simply tells Bob, and they insert an arbitrarily long string of zeros into the key register. This modified protocol can be easily shown to be $(\varepsilon+\varepsilon_\mathrm{coin})$-secure. However, the expected key rate can now be made arbitrarily large. Clearly, one would not want to use such a protocol, because in the special case where the coin lands tails, Alice and Bob would generate an extremely long but essentially meaningless key, and in fact would never need to run a new QKD protocol again.}

\subsection{Secrecy and Correctness} \label{subsec:correctnessandsecrecy}
The security requirement for QKD is typically broken down into two simpler requirements of secrecy and correctness, which we define below. We state the definitions for variable-length protocols, since fixed-length protocols can simply be treated as a special case of variable-length protocols where there are only two possible output lengths, $\lfixed$  and $0$.

\begin{definition}[Correctness and Secrecy] \label{def:qkdcorrectnessandsecrecysymmetric}
Let $\QKDprotocol$ be a QKD protocol, and let $\rho^\text{real}_{K_A K_B  \Esecdef}$ be the output state of the QKD protocol, and let $\mathcal{W}\left( \QKDprotocol \right)$ denote the set of possible output states of the QKD protocol, i.e, consider the same setting as in \cref{def:qkdsecuritysymmetric}. The QKD protocol is $\epssecret$-secret if, for all output state $\rho^\text{real}_{K_A K_B \Esecdef} \in \mathcal{W}(\QKDprotocol)$, the following inequality is satisfied:
\begin{equation} \label{eq:tracedistsecrecy}
    \tracedist{ \rho^\text{real}_{K_A  \Esecdef} -  \rho^\text{ideal}_{K_A   \Esecdef}} \leq \epssecret.
    \end{equation}

The QKD protocol is $\epscorr$-correct, if, for all output state $\rho^\text{real}_{K_A K_B \Esecdef} \in \mathcal{W}(\QKDprotocol)$, the following inequality is satisfied:
\begin{equation} \label{eq:tracedistcorrect}
   \Pr(K_A \neq K_B)_{ \rho^\text{real}_{K_A  K_B \Esecdef} } \leq \epscorr.
    \end{equation}
\end{definition}
The secrecy requirement ensures that Alice's key is secret and unknown to Eve. The correctness requirement ensures that Alice and Bob have the same key. Both requirements can be combined to obtain the required security property for QKD protocols via the following lemma, proved in Appendix.~\ref{Appendix:misc}.

\begin{restatable}{lemma}{correctnessandsecrecy}
(Correctness and Secrecy imply Security) \label{lemma:securityfromcorrandsecrecy}
 A QKD protocol that is $\epssecret$-secret, and $\epscorr$-correct, is $(\epssecret+\epscorr)$-secure.
\end{restatable}

\section{Protocol Description} \label{sec:protdescriptionqkdbackground}
In this section, we describe a generic QKD protocol that will serve as the basis for our security analysis throughout this thesis. A more general and formal version of this protocol will be introduced in \cref{chap:MEAT}, where it will be analyzed in detail. We deliberately begin with a simpler formulation here for two reasons. First, the analyses in \cref{chap:EUR,chap:variable,chap:postselection} is unable to handle the general version of this protocol which can be tackled using the full generality of the MEAT \cite{arqand_marginal_2025}, and thus introducing all of its components at this stage would only obscure the key ideas. Second, the level of formality and precision in \cref{chap:MEAT} is substantially higher than in the other chapters, since \cref{chap:MEAT} is intended to provide a complete and rigorous security analysis. Thus, we omit some technical details that are included in that complete analysis which are not needed in \cref{chap:EUR,chap:variable,chap:postselection}. Importantly, the simpler protocol presented here, and its security analysis in \cref{chap:EUR,chap:variable,chap:postselection} is fully rigorous and mathematically sound; it merely omits some additional formal machinery. We nevertheless maintain notation that is  consistent with the general formulation in \cref{chap:MEAT}.

We assume that the protocol begins with Alice and Bob having access to  local randomness stored in classical registers, which they use to implement the steps of the QKD protocol. The adversary is uncorrelated to this randomness, which is only accessible to the honest parties, and is never leaked to the adversary (even after the termination of the QKD protocol).\footnote{While this randomness is sometimes used to generate states or classical messages that are publicly released at some point (when explicitly specified in the protocol), what we mean by this requirement is that the ``raw'' values of the randomness are never made accessible to the adversary.} In this thesis, we do not explicitly describe the generation and use of these local random numbers in the QKD protocol execution; they are implicitly utilized in choosing the signal states sent, basis choices, and choosing seeds for hashing. We note that the assumption of perfect random numbers can be relaxed by utilizing random number generators that are $\eps_\text{rng}$-close  to perfect, in the composable security framework.

\begin{prot}[QKD Protocol] \label{prot:qkdprotocol} 
\leavevmode \\
    \begin{enumerate}
    
 \item  For rounds $j$ from $1$ to $n$, Alice and Bob perform the following operations:

 	\begin{enumerate}
    \item  \textbf{State Preparation: } \label{protstepbackground:stateprep} At time $t_j^\mathrm{A}$, Alice prepares a state $\sigma_k \in \dop{=}(A'_j)$ out of $\numstates$ possible states,  with probability $p_k$, and sends them to Bob through an insecure quantum channel. She stores the label $k$ for her choice of the signal state in a classical register \(X_j\). This step requires the use of local randomness. Alice sends these states in sequence, and we therefore have $t^\mathrm{A}_j < t^\mathrm{A}_{j+1}$. 
    \item \textbf{Measurement: } \label{protstepbackground:meas} At time $t_j^\mathrm{B}$, Bob performs a measurement using  a POVM  \(\{\Gamma_k^{(B_j)}\}_{k=1 \dots \nummeas}\), obtains one of $\nummeas$ possible outcomes, and stores his results in a classical register \(Y_j\)
    . Depending on the exact detection setup used, this step may require the use of local randomness. Bob performs these measurements in sequence, and we therefore have $t^\mathrm{B}_j < t^\mathrm{B}_{j+1}$.

\begin{remark} \label{remark:protocolpermutation} \textbf{Optional permutation step:} Alice (or Bob) picks a random permutation and announces it. Both parties then apply this permutation to their classical data. This step is required only for protocols analyzed using the postselection technique in \cref{chap:postselection} to satisfy certain permutation invariance requirements ( \cref{lemma:satisfyingpermutationinvariance}), and therefore appears exclusively in that chapter; it also entails an additional public announcement. However, since the postselection technique ultimately reduces the security analysis to that of IID collective attacks, it suffices to analyze security against IID collective attacks alone. In that setting, this permutation step plays no role, as argued in \cref{lemma:permutationdoesnotmatter}, and can be ignored\footnote{However, it still needs to be implemented in the protocol, in order to justify the usage of the postselection technique.}. Consequently, we do need to explicitly introduce notation or registers for this step.

\end{remark}
    
    \item \textbf{Public announcement: } At some time after all states have been sent and measured, Alice and Bob engage in interactive public announcements, using authenticated classical channel. These public announcements result in a classical register $\CP_j$.

    \item \textbf{Sifting and key map: } Alice maps her local data stored in $ X_j$, along with the public annoucements $\CP_j$, to her private register $S_j$. She then applies a deterministic rule, based solely on $\CP_j$, to discard certain rounds from $S_j$. This produces the (potentially shortened) pre-amplification string, which is stored in the register $\PAstring$.\footnote{Thus, the register $\PAstring$ takes values from the set of all possible strings of length less than or equal to $n$, and composed of symbols from the alphabet of the register $S_j$ that are \textit{not} discarded. In most common scenarios, $S_j$ takes values in $\{0,1,\singleRoundBot\}$ (based on $\CP_j$,$X_j$), and the last outcome denotes that the round is going to be discarded. Thus, $\PAstring$ stores a binary string of length up to $n$ bits.} In this thesis, we consider protocols where  Alice’s remapping and discarding are defined so that whenever a value of $S_j$ will be discarded (based on $\CP_j$), $S_j$ is set to a fixed placeholder symbol $\singleRoundBot$. Since Alice determines the mapping to $S_j$, she can directly encode discard rounds in this way.\footnote{In principle, this means that Alice could implement the discarding step by inspecting the sequence $S_1^n$ alone, without separately referring to the announcements $\CP_1^n$.}

    \end{enumerate}
  After all the announcements and sifting and key map operations are completed, the state of the protocol is given by 
\begin{equation}
    \rho_{ \PAstring S_1^n X_1^n Y_1^n  \CP_1^n   \Eve  }.
    \end{equation} 
 where $\Eve$ denotes Eve's quantum side information at the end of her attack on the $n$ rounds. Alice and Bob move on to the next step in the protocol only after all public announcements are completed (which also implies that all signal preparation and measurements are completed). Eve is assumed to have access to public announcements $\CP_1^n$.

\item \label{step:varlengthdecision} \textbf{Variable length decision:} Alice and Bob use predetermined functions of the  public announcements $\CP_1^n$ to determine parameters for error correction and key generation. Often these functions only depend on the frequency distribution $\Fobs \coloneq \freq(\cobs)$ of the public announcements, where we use $\cobs$ to denote the value stored in the classical register $\CP_1^n$.
\begin{enumerate}
    \item Alice computes the value $\leak(\Fobs) \in \mathbb{N}$ which determines the number of possible transcripts possible in the error-correction protocol.
    \item Alice computes the length of the output key to be produced, given by $\lkey(\Fobs) \in \mathbb{N}$, using a predetermined function $\lkey(\cdot)$. The functions $\lkey(\Fobs)$ and $\leak(\Fobs)$ are closely related through the security proof.
    \item These values may be computed by both parties, or computed by one party and then sent to the other. Since Eve knows $\Fobs$, revealing these values via public announcements to Eve does not give her any new information.
\end{enumerate}

The state in the protocol, conditioned on observing a specific value in the $\CP_1^n$ registers, is given by
\begin{equation}
    \rho_{ \PAstring S_1^n X_1^n Y_1^n  \CP_1^n  \Eve | \Omega(\cobs) },
    \end{equation}
    where $\Omega(\cobs)$ denotes the event that  the value $\cobs$ is observed in the $\CP_1^n$ registers. Note that Bob can also discard the same rounds as Alice to obtain a string $\bm{Y}$ of the same length as  $\PAstring. $.

	\item \textbf{Error correction:} Alice and Bob perform an error-correction protocol that results in Bob outputting a guess for Alice's pre-amplification string. This error-correction protocol is such that the number of possible transcripts is the protocol is upper bounded by $2^{\leak(\Fobs)}$.\footnote{For the simplest protocols, where Alice sends an error syndrome of exactly $\leak(\Fobs)$ bits to Bob, the value $\leak(\Fobs)$ coincides with the number of bits leaked. For protocols where Alice may send up to some maximum number of bits, one instead sets the maximum number of bits to $\leak(\Fobs)-1$, so that the total number of possible transcripts is at most $ 2 + 2^2 + \dots + 2^{\leak(\Fobs)-1} = 2^{\leak(\Fobs)} - 2 \leq 2^{\leak(\Fobs)}$.} The protocol results in Bob outputting a guess $\PAstring_B$ for Alice's pre-amplification string. Note that since the location of the discarded rounds are known to both Bob and Alice, this protocol is designed such that $\PAstring_B$ and $\PAstring$ have exactly the same length (typically by having Bob start with the string $\bm{Y}$).

The state in the protocol, conditioned on observing a specific value in the $\CP_1^n$ registers, is given by
\begin{equation}
    \rho_{ \PAstring \PAstring_B \CP_1^n  \CEC \flagkey \flagEC  \Eve | \Omega(\cobs) }.
    \end{equation}
   where $\Omega(\cobs)$ denotes the event that $\CP_1^n$ registers take the value $\cobs$. Note that here we omit the registers $Y_1^n$, since they are no longer needed for the rest of the protocol. This is tantamount to either deleting them, or ensuring that this register is not leaked to the adversary at any point (even after the completion of the protocol). 
    
         \item  \textbf{Error verification:}  Alice and Bob perform error-verification by computing universal$_2$ hashes on their respective pre-amplification strings $(\PAstring, \PAstring_B)$. The hash families map the input strings to $\ceil{\log(1/\epsEV)}$ output bits. They do so by sharing the hash seed in register $\HEV$, and having one party share the computed hash value with the other party in the register $\CEV$. We use $\OmegaEV$ to refer to the event where the  hash values match. Alice and Bob abort the protocol if $\OmegaEV$ does not occur. 

         The state in the protocol, conditioned on $\Omega(\cobs)$ and error-verification passing, is given by
\begin{equation}
    \rho_{ \PAstring \PAstring_B \CP_1^n  \CEC \CEV  \HEV \Eve | \Omega(\cobs) \wedge \OmegaEV }.
    \end{equation}
 
where $\wedge$ denotes the logical `and' operator and $\OmegaEV$ denotes the event that error-verification passes, i.e hash values matched.

 \item \textbf{Privacy amplification:}  Alice and Bob perform \emph{ideal}  universal$_2$\footnote{Recall that ideal universal$_2$ (see \cref{def:2universal}) corresponds to the case where the collision probability of the hash is \emph{equal} to $1/|D'|$, where $D'$ is the codomain of the hash family. Note that this requirement can be relaxed to universal$_2$ by sacrificing one additional bit of output key, see \cref{remark:LHLequalityissue} for a discussion.} hashing on their their pre-amplification strings to generate the output key. That is, they first compute the length $\lprePA$ of strings stored in $\PAstring,\PAstring_B$, and then choose a hash function from an ideal universal$_2$ family mapping $\lprePA$ bits to $\lkey(\cobs)$ bits.  The hash choice is publicly  communicated in the register $\HPA$, and the chosen hash function is applied to $\PAstring, \PAstring_B$ to obtain the final output registers $K_A,K_B$.

 The state in the protocol, conditioned on $\Omega(\cobs) \wedge \OmegaEV$, is given by 
\begin{equation}
    \rho_{K_A K_B  \CP_1^n  \CEC \CEV \HEV \HPA \Eve  | \Omega(\cobs) \wedge \OmegaEV }.
    \end{equation}

     \end{enumerate}
 \end{prot}

\subsection{Protocol Variations} \label{subsec:protocolvariations}
We now comment on several protocol variations that will be studied throughout this thesis. 

\paragraph{On-the-fly announcements:} First, observe that the protocol described above assumes that all public announcements occur \emph{after} all quantum signals have been transmitted and measured. If these public announcements before the all quantum signals have been sent and measured, then it allows Eve to potentially adapt her attack based on these announcements. Such announcements are referred to as ``on-the-fly" announcements (see \cite[Section 5.1.2]{tupkary2025qkdsecurityproofsdecoystate} for a discussion). This restriction, forbidding on-the-fly announcements, greatly simplifies the security analysis and is, in fact, essential for the postselection-based proof in \cref{chap:postselection}. Although such a restriction is not inherently required by the EUR or phase-error approaches, our analyses in \cref{chap:EUR} likewise assume it for technical convenience. In contrast, the MEAT framework places no such limitation, and therefore \cref{chap:MEAT} allows on-the-fly announcements. 

\paragraph{Fixed-length protocols:} Second, the protocol above is formulated as a variable-length protocol. A simpler special case, which is easier to analyze, is the fixed-length version. This is obtained by imposing the following restrictions on the \nameref{prot:qkdprotocol} from \cref{sec:protdescriptionqkdbackground}:

\begin{itemize}
\item The variable-length decision step in the protocol is restricted to a simple \term{acceptance test} form, in which Alice computes the observed frequency distribution $\Fobs = \freq(\cobs)$ on the classical string $\cobs$, and checks whether this frequency distribution lies inside some predetermined set ($\acceptanceset$) of frequency vectors often called an \term{acceptance set}. If it does, we refer to this as the acceptance test ``passing'', and the protocol produces an output key of length $\lfixed$ (assuming error-verification also passes). Else, the protocol aborts.

\item We require that there exists a \emph{fixed} constant value $\leakfixed \in \mathbb{N}$, such that the number of possible transcripts in the error correction step (conditioned on the acceptance test passing) is at most $2^{\leakfixed}$.

\item At the privacy amplification step, Alice and Bob instead check only whether the acceptance test passed, and the error-verification passed. If both these events hold, Alice and Bob accept the protocol overall, and hash to final keys of length $\lfixed$; otherwise they abort the protocol. Note that this means the event $\Omega_\mathrm{acc}$ of the protocol accepting overall is exactly the joint event 
\begin{align}
\Omega_\mathrm{acc} = \OmegaAT \land \OmegaEV,
\end{align}
and thus we may instead denote it as $\OmegaAT \land \OmegaEV$ in contexts where that is more convenient.
\end{itemize}

Having explained the high-level structure of the protocol that we analyze in later chapters above, we now consider some additional details that needed to obtain a concrete realization of the QKD protocol. This includes

\begin{itemize}
\item the precise set of signal states sent and the measurements performed, which we specify in the next section;
\item the exact form of the public announcements, and how they are computed from the underlying data generated by state preparation and measurement; 
\item the parameters governing state preparation (e.g., probabilities, intensities, and related settings), which are only specified on a per-chapter basis; 
\item the specific error-correction and privacy-amplification choices such as the exact error-correction protocol, hash families, etc. Beyond the aspects already stated above, further implementation details of these classical subroutines do not affect the security analysis, and we therefore do not specify them.  
\end{itemize}

The following section can also be skipped on the first reading: readers can return to this section when they encounter security analysis that required those details.

\section{BB84 signal states, measurements, announcements and sifting} \label{sec:statesandmeasurements}

We consider two families of protocols: qubit-based BB84 \cite{Bennett_2014} and decoy-state BB84 \cite{Hwang_qkdiwthhighloss_2003,Lo_decoystate_2005,wang_beating_2005}.  
In both cases, Alice prepares quantum states that encode classical information
in one of two mutually unbiased bases, conventionally denoted by $\Xbasis$ and
$\Zbasis$. If Bob uses an active measurement setup, he chooses to make a measurement in either $\Xbasis$ or $\Zbasis$. If he uses a passive detection setup, a beam splitter is used instead of an active basis choice. We consider polarization as the encoding degree of freedom.

\subsection{States Sent by Alice} \label{subsec:alicestatessent}

We now specify the signal states $\{\sigma_k\}_k$ and preparation probabilities
$\{p_k\}_k$ appearing in Step~\ref{protstepbackground:stateprep} of the
\nameref{prot:qkdprotocol}.    It is
convenient to decompose the state preparation label $k$ as
\[
k = (a, \testgenflag),
\]
for the qubit BB84 setting or, in the decoy-state setting,
\[
k = (a, \mu, \testgenflag),
\]
where $a$ denotes the polarization choice, $\mu$ the intensity setting, and
$\testgenflag \in \{\test,\gen\}$ indicates whether the
round is designated as a ``test" round or `` generation" round. In test rounds, Alice and Bob announce their exact state sent, and measurement outcome obtained, and thus no key is generated from these rounds.  In the generation rounds, Alice and Bob only announce their basis choice, and Bob announces whether he obtained he obtained a no-click outcome (no detectors clicked), or a click outcome (some detectors clicked). 

The choice of $k$ proceeds as follows.  First, Alice selects 
$\testgenflag$, determining whether the round is a test (with probability $\gamma$) or key-generation (with probability $(1-\gamma)$)
round.  Conditioned on this choice, she samples the remaining components of $k$
according to a probability distribution that may depend on $\testgenflag$. This is denoted via $p_{a|\testgenflag}$ or $p_{a,\mu | \testgenflag}$.

\paragraph{Qubit BB84.}
For qubit BB84, the polarization alphabet is
\[
a \in \{0,1,+,-\},
\]
corresponding to the computational ($\Zbasis$) and Hadamard ($\Xbasis$) bases.
The signal states are given by
\[
\sigma_{(a,\testgenflag)} = \ketbra{a}{a},
\]
where $\ket{0} ,\ket{1}$ corresponds to $\Zbasis$ basis and $\ket{+}  = \frac{\ket{0} + \ket{1}}{\sqrt{2}},\ket{-}= \frac{\ket{0} - \ket{1}}{\sqrt{2}}$ corresponds to $\Xbasis$ basis.
The prepared quantum state depends only on the polarization choice $a$ and is
independent of $\testgenflag$; the latter is recorded purely as classical side
information in Alice’s register.

\paragraph{Decoy-state BB84.}
For decoy-state BB84, the polarization alphabet is
\[
a \in \{H,V,A,D\},
\]
partitioned into the rectilinear basis $\{H,V\}$ and the diagonal basis
$\{A,D\}$.  Alice additionally chooses an intensity setting $\mu$ from a finite
set of possible intensities. We assume that Alice prepares fully phase-randomized coherent states with
perfect polarization encoding.  The corresponding signal states are
\begin{equation} \label{eq:AliceSignalStatesDescriptionCompact}
    \sigma_{(a,\mu,\testgenflag)}
    = \sum_{N=0}^{\infty} e^{-\mu}\frac{\mu^N}{N!}\ketbra{N}_a ,
\end{equation}
where $\ketbra{N}_a$ denotes an $N$-photon Fock state in which all photons have
polarization $a$.  As in the qubit case, the prepared quantum state depends only
on $(a,\mu)$ and is independent of whether the round is a test or
key-generation round. 

In both protocols, Alice records the full label $k$ in her classical register
$X_j$ for round $j$.  Consequently, the alphabet of $X_j$ coincides with the set
of all possible signal-state labels $k$. Note that Alice preparing a single qubit can be equivalently thought of as Alice producing a single photon, of the appropriate polarization.

\subsection{Measurements Performed by Bob} \label{subsec:bobmeasurements}

We now specify Bob’s measurement in Step~\ref{protstepbackground:meas} of the
\nameref{prot:qkdprotocol}. As the same measurement is performed in
every round, we suppress the round index $j$ throughout. Bob stores the exact measurement outcome in his classical register $Y_j$.

\paragraph{Qubit Measurements.}
In the qubit BB84 protocol with loss, Bob’s measurement acts on a
three-dimensional Hilbert space $
\widetilde{\mathcal{H}}_B \cong \mathbb{C}^3$, where the subspace spanned by $\ket{0},\ket{1}$ corresponds to the logical
qubit, and an additional orthogonal basis state $\vac$ represents a
no-detection (loss) event.

Upon choosing a basis $b_B \in \{\Zbasis,\Xbasis\}$, Bob performs a
three-outcome projective measurement. If $b_B=\Zbasis$, the POVM is given by 
\[
\left\{ \ketbra{0}{0},\ \ketbra{1}{1},\ \ketbra{\mathrm{vac}}{\mathrm{vac}} \right\},
\]
and if $b_B=\Xbasis$, the POVM is
\[
\left\{ \ketbra{+}{+},\ \ketbra{-}{-},\ \ketbra{\mathrm{vac}}{\mathrm{vac}}  \right\}.
\]

\paragraph{Active detection setup with threshold detectors}

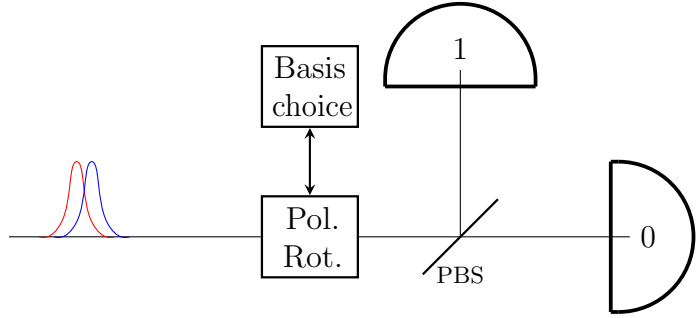
\begin{figure}
\centering
\begin{tikzpicture}[node distance=2cm]

\node[processSmall] (Basis) at (0,2) {Basis\\choice};
\node[processSmall] (PolRot) at (0,0) {Pol.\\Rot.};

\draw (-4,0) -- (PolRot.west);

\draw[connect] (Basis) -> (PolRot);

\pic at (2,0) {beamsplitter={pbs,PBS,(0,-0.5)}};
\pic at (4,0) {detectorSmall={det0,$0$,black}};
\pic[rotate=90] at (2,2) {detectorSmall={det1,$1$,black}};

\draw (PolRot) -- (pbs);
\draw (pbs) -- (det0);
\draw (pbs) -- (det1);

\pic at (-3.6,0) {pulse={red}};
\pic at (-3.4,0) {pulse={blue}};

\end{tikzpicture}
\caption{Schematic of an active detection setup using threshold detectors.}
\label{fig:bb84active}
\end{figure}
We model an active detection setup (see \cref{fig:bb84active}) as follows.  Bob receives an optical signal
described by annihilation operators $\hat{a}_H$ and $\hat{a}_V$ corresponding to
horizontal and vertical polarization modes.  Bob actively chooses a basis
$b_B \in \{\Zbasis,\Xbasis\}$ and applies a basis-dependent linear-optical
transformation that maps the input modes to a pair of output modes
$\hat{b}_0^{(b_B)}$ and $\hat{b}_1^{(b_B)}$, which are then measured using
threshold detectors.

For $b_B=\Zbasis$, we take the output modes to be
\begin{equation} \label{eq:ActiveModesZ}
    \hat{b}^{(\Zbasis)}_0 = \hat{a}_H,
    \qquad
    \hat{b}^{(\Zbasis)}_1 = \hat{a}_V .
\end{equation}
For the diagonal basis $b_B=\Xbasis$, Bob applies a $45^\circ$ polarization
rotation, yielding the output modes
\begin{equation} \label{eq:ActiveModesX}
    \hat{b}^{(\Xbasis)}_0 = \frac{1}{\sqrt{2}}(\hat{a}_H + \hat{a}_V),
    \qquad
    \hat{b}^{(\Xbasis)}_1 = \frac{1}{\sqrt{2}}(\hat{a}_H - \hat{a}_V).
\end{equation}

Each output mode is monitored by a threshold detector.  For any subset
$\Sclick \subseteq \{0,1\}$, the POVM element corresponding to exactly the
detectors in $\Sclick$ clicking is
\begin{equation} \label{eq:ActivePOVMGeneral}
    M_{\Sclick}^{(B,b_B)}
    =
    \sum_{\{N_j \ge 1\}_{j \in \Sclick}}
    \frac{1}{\prod_{j\in\Sclick} N_j!}
    \left(
        \prod_{j\in\Sclick} (\hat{b}^{(b_B)\dagger}_j)^{N_j}
    \right)
    \ketbra{\mathrm{vac}}
    \left(
        \prod_{j\in\Sclick} (\hat{b}^{(b_B)}_j)^{N_j}
    \right),
\end{equation}
and the no-click outcome is
\[
M_{\emptyset}^{(B,b_B)} = \ketbra{\mathrm{vac}}.
\]
Bob records the complete outcome (including the chosen basis) in his classical
register $Y_j$.

In this thesis, we plot key rates for the active basis-choice scenario of
the decoy-state BB84 protocol. To do so, we consider a scenario in which
Bob maps his double-click outcomes in each basis to single-click outcomes
at random. This classical post-processing step is required for the
appropriate use of squashing maps, as explained in
\cref{subsec:squashing}. As a result, in each basis Bob effectively has
three outcomes, rather than four. The corresponding POVM elements can be obtained by straightforward linear combinations of POVM elements defined above. 

For the passive BB84 setup, we do not explicitly plot key rates. Where
possible, however, we explain how our results can be applied to passive
implementations as well. In particular, all analyses in this thesis
except for the entropic uncertainty relation (EUR) analysis in
\cref{chap:EUR} can be straightforwardly extended to passive setups.

\paragraph{Passive setup with Threshold detectors.}

\begin{figure}
   \begin{tikzpicture}[node distance=2cm]

\pic at (0,0) {beamsplitter={bs5050,BS:$\beamsplit/(1-\beamsplit)$,(0,-0.6)}};

\pic at (0,2) {beamsplitter={pbsHV,PBS,(-0.5,0)}};
\pic at (2,2) {detectorSmall={detH,$H$,black}};
\pic[rotate=90] at (0,4) {detectorSmall={detV,$V$,black}};

\node[processSmall] (PolRot) at (2,0) {Pol.\\Rot.};
\pic at (6,0) {beamsplitter={pbsAD,PBS,(0,-0.5)}};
\pic[rotate=90] at (6,2) {detectorSmall={detA,$A$,black}};
\pic at (8,0) {detectorSmall={detD,$D$,black}};

\draw (-4,0) -- (bs5050);
\draw (bs5050) -- (pbsHV);
\draw (pbsHV) -- (detH);
\draw (pbsHV) -- (detV);

\draw (bs5050) -- (PolRot);
\draw (PolRot) -- (pbsAD);
\draw (pbsAD) -- (detA);
\draw (pbsAD) -- (detD);

\pic at (-3.6,0) {pulse={red}};
\pic at (-3.4,0) {pulse={blue}};

\end{tikzpicture}

\caption{Schematic of the passive detection setup using theshold detectors.} \label{fig:bb84passive}
\end{figure}
In the passive detection setup (see \cref{fig:bb84passive}),  Bob implements a passive basis choice using a beam splitter with splitting ratio $\beamsplit \in [0,1]$. One output arm is sent directly to a polarizing beam splitter followed by two detectors. The other arm is sent through a polarization rotator, and is then measured using a polarizing beam splitter followed by two detectors. The first arm can be viewed as implementing a $\Zbasis$-basis measurement, while the second can be viewed as implementing a $45^\circ$-rotated $\Xbasis$-basis measurement. The output modes can be written in terms of the input modes via
\begin{equation} \label{eq:BobOutputModesCompact}
    \begin{aligned}
        \hat{b}_H &= \sqrt{\beamsplit}\, \hat{a}_H, \\
        \hat{b}_V &= \sqrt{\beamsplit}\, \hat{a}_V, \\
        \hat{b}_D &= \sqrt{\tfrac{1-\beamsplit}{2}}\, (\hat{a}_H + \hat{a}_V), \\
        \hat{b}_A &= \sqrt{\tfrac{1-\beamsplit}{2}}\, (\hat{a}_H - \hat{a}_V).
    \end{aligned}
\end{equation}
There are $16$ possible click patterns in total. Among these, there are
four possible single-click events. These single-clicks can be used to
assign a basis $b_B$ to Bob’s outcome, noting that this basis assignment
is not an active choice by Bob, but is instead inferred from the
measurement data. In addition, there are two possible double-click patterns, one associated
with each basis. All remaining outcomes correspond to cross-click events,
in which more than one detector clicks across different bases.

\begin{remark} \label{remark:bobpovmform}
The exact operator-level form of these POVM elements is not always directly
relevant for the QKD security analysis.  The main property we will rely on is
that they are jointly block-diagonal with respect to the total photon number.
This structure is used in the construction of squashing maps (see \cref{subsec:squashing}).  Such maps allow us to restrict
attention to effective POVMs supported only on a finite photon-number
subspace—sometimes even the single-photon (qubit) subspace—without affecting
security.
\end{remark}

\subsection{Public Announcements and Sifting}

Our protocols involve the following procedure for public announcements and
sifting. For the active BB84 case, we assume that double-clicks are being randomly mapped to single-clicks.

\begin{itemize}
    \item \textbf{Test rounds.}
    On all test rounds, Alice and Bob both publicly announce their exact state
    preparation and measurement outcomes.  They set $S_j = \bot$, and these
    rounds are discarded and not used for key generation.

    \item \textbf{Key-generation rounds.}
    On all key generation rounds,  Bob publicly announces whether
    he obtained a detection event outcome or not.  

    \begin{itemize}
        \item If Bob does not obtain a detection event, he makes no further
        announcement.  Alice sets $S_j = \bot$, and the round is discarded. 

        \item If Bob obtains a detection event, he publicly announces his
        measurement basis. Alice announces her basis. If these bases match, Alice then assigns a bit
        value to $S_j$ based on her state preparation, and Bob assigns a
        corresponding bit value to his local data. If they do not match, Alice sets $S_j$ to $\bot$, and the round is discarded.
    \end{itemize}
\end{itemize}

For passive protocols, Alice and Bob announce all relevant information in
test rounds. In key generation rounds, they follow the above procedure when Bob obtains either a single-click outcome or a
no-click outcome. If Bob instead obtains a multi-click outcome, he
announces this fact and the corresponding round is discarded.

The remainder of the protocol proceeds exactly as in
\nameref{prot:qkdprotocol}.  We have now specified enough details of the protocol to provide a clear overall picture of its operation. Note that in \cref{chap:MEAT} we will specify the protocol in considerably more detail. We now turn to the security analysis of the protocol, beginning with correctness, which is addressed in the next section.

\subsection{Motivation for decoy-state protocols} \label{subsec:motivationfordecoy}
Here, we briefly motivate why decoy-state protocols are important for practical QKD implementations.
The qubit BB84 protocol is conceptually natural: it relies on sending qubit states that are not perfectly distinguishable. A single photon constitutes a qubit because its polarization degree of freedom spans a two-dimensional Hilbert space. However, the qubit BB84 protocol is highly impractical to implement directly, as it would require preparing exactly one photon on demand with a precise polarization. It is also very difficult to ensure that Bob measures a qubit state, since Eve is free to send any number of photons into Bob's lab. Instead, practical systems rely on phase-randomized weak coherent pulse (WCP) sources for technological feasibility, which emit states with a Poissonian photon number distribution (\cref{eq:AliceSignalStatesDescriptionCompact}). Moreover, Bob uses threshold detectors, which do not resolve photon number. 

In such implementations, any pulse for which Alice’s source emits a multiphoton state is fundamentally insecure. This is because an adversary can perform a photon-number-splitting (PNS) attack
\cite{Lutkenhaus_estimates_1999,Bennett_experimentalquantumcryptography_1992,Brassard_limitationsonpractical_2000,Lutkenhaus_security_2000}.
In this attack, Eve deterministically splits off one photon from a multiphoton pulse and stores it in a quantum memory, while forwarding the remaining photons to Bob. Since Bob uses threshold detectors and cannot resolve photon number, he cannot detect that the pulse has been modified, nor that the total photon number has been reduced. After Alice and Bob publicly announce their basis choices, Eve measures her stored photon in the correct basis, thereby learning the encoded bit without changing Alice's and Bob's observations.

Moreover, by exploiting channel loss, Eve can selectively suppress single-photon signals ( which are immune to the PNS attack) while preferentially transmitting compromised multiphoton pulses, all while reproducing the expected detection statistics at Bob’s side. As a result, the Poissonian photon number statistics of WCP sources severely degrade the achievable key rate unless explicitly addressed. Decoy-state protocols \cite{Lo_decoystate_2005,Hwang_qkdiwthhighloss_2003,wang_beating_2005} are employed to mitigate this vulnerability.

To build intuition, consider a hypothetical source that emits a \emph{definite} photon number and also announces it to Alice. Alice and Bob could then group their observed data based on the photon number and directly extract the single-photon contribution, while pessimistically discarding all multiphoton rounds. This prevents Eve from concealing a PNS attack by mixing single- and multiphoton behavior.

In practice, the photon number is not announced. The decoy-state method provides a statistical substitute by varying the intensity of the WCP source, thereby producing different photon-number distributions. Crucially, Eve cannot distinguish an $m$-photon pulse originating from different intensity settings, as her interaction can depend only on the photon number and not on Alice’s choice of intensity, which is unknown at the time of the attack. Thus, the observations on rounds with different intensities can be used to infer something about Eve's attack on rounds with specific photon number (typically zero-photon or one-photon rounds). This is the essence of the decoy-state idea. 

Overall, the task remains fundamentally the same: to infer properties of Eve’s attack, and hence the amount of information she may have gained, using the observed statistics. Decoy-state protocols simply provide a richer set of states, measurements, and observed data with which to perform this inference. As we will see in \cref{sec:LHL,subsec:boundingentropy}, these observations are ultimately used to bound an appropriate entropic quantity that quantifies secrecy.

\section{Correctness via error-verification} \label{sec:correctnessviaEV}
It is straightforward to prove that correctness is satisfied in our protocol description, which we will do now.  
In fact, this argument relies solely on the properties of the error-verification step in our \nameref{prot:qkdprotocol}, and is largely independent of the rest of the protocol.

\begin{lemma}[Correctness is satisfied] \label{lemma:correctnessissatisfied}
The \nameref{prot:qkdprotocol} is $\epsEV$-correct.
\end{lemma}
\begin{proof}
 This proof is a simple consequence of the error-verification step from \nameref{prot:qkdprotocol}, and the proof of correctness can be found in many works. For any output state of the QKD protocol, we have
    \begin{equation}
        \begin{aligned}
            \Pr(K_A \neq K_B) &= \Pr(K_A \neq K_B \wedge \OmegaEV) + \Pr(K_A \neq K_B \wedge \OmegaEV^\complement) \\
            &=\Pr(K_A \neq K_B \wedge \OmegaEV) \\
            &\leq \Pr(\PAstring \neq \PAstring_B \wedge \OmegaEV) \\
            &= \Pr(\OmegaEV | \PAstring \neq \PAstring_B)\Pr(\PAstring \neq \PAstring_B) \\
            &\leq \Pr(\OmegaEV | \PAstring \neq \PAstring_B ) \\
            &\leq 2^{-\left(\ceil{\log(1/\epsEV)}  \right)} \\
            &\leq \epsEV,
        \end{aligned}
    \end{equation}
    where the first line follows from the properties of probability, and the second line follows from the fact that $\OmegaEV^\complement \implies K_A = K_B$, since the protocol aborts. The third line follows from that fact that $K_A \neq K_B \implies \PAstring\neq \PAstring_B$; in other words, the output keys being unequal necessarily implies that the pre-amplification strings are unequal. The fourth and fifth lines follow from the properties of probability.  The sixth line follows from the fact that error-verification involves checking universal$_2$ hashes, and the final line follows from simple algebra.
\end{proof}

Thus, to complete the security analysis of \nameref{prot:qkdprotocol}, it remains only to establish the secrecy requirement.  This is the technically challenging part of the proof.  
We therefore turn next to the Leftover Hashing Lemma, which serves as a key tool in this analysis.

\section{Leftover Hashing Lemma} \label{sec:LHL}
The Leftover Hashing Lemma is a central tool in QKD security analysis.   It provides a method for converting classical data that is not perfectly secret into data that is \emph{close} to perfectly secret, by applying a hash function chosen from a universal$_2$ hash family.  
The quality of this extraction is determined by an appropriate entropic measure of the underlying state, typically expressed in terms of either Rényi entropies or smooth min-entropies.  Note that there are many variants of the Leftover Hashing Lemma, and when we refer to the LHL, we mean the broader family of such results.

\begin{lemma}(Leftover-hashing Lemma (LHL). \cite[Theorem~8]{dupuis_privacy_2023} and \cite[Proposition 8]{tomamichel_largely_2017}.)
\label{lemma:LHL}
Let $\rho_{AE} \in \dop{=}(AE)$ be a classical-quantum state, and $(\mathcal{H}_{\mathcal{A}\rightarrow\mathcal{Z}},p_h)$ be a  family of \emph{ideal} universal$_2$ hash functions with $\mathcal{K}=\{0,1\}^l$. Consider the state $\rho_{KEH}$ that is obtained when $h$ is a function drawn from that family with probability $p_h$, and applied to the register $A$ to obtain the register $K$, and the choice of the function is stored in $H$. Then, we have
\begin{align}
    \label{eq:privacy_amp_renyi}
  \frac{1}{2}\left\|\rho_{KEH}-\frac{1}{|\mathcal{K}|}\id_K\otimes\rho_{EH} \right\|_1 \leq 2^{\frac{1-\alpha}{\alpha}\big(\Halpha(A|E)_\rho-l + 2\big)},
\end{align}
for $\alpha\in(1,2)$\footnote{For the entirety of this thesis, unless otherwise specified, the {\Renyi} parameter  $\alpha\in(1,2)$, since that is the regime relevant for QKD.}
If $\rho_{AE} \in \dop{\leq}(AE)$ is a subnormalized state, and $\epsbar \in [0,\sqrt{\Tr[\rho_{AE}]})$,  we also have, 
\begin{align}
    \label{eq:privacy_min}
  \frac{1}{2}\left\|\rho_{KEH}-\frac{1} {|\mathcal{K}|}\id_K\otimes\rho_{EH} \right\|_1 \leq  2^{-\frac{1}{2}(\Hmin{\epsbar}(A|E)_\rho - l + 2)} + 2\epsbar,
\end{align}
\end{lemma}

In the \nameref{prot:qkdprotocol}, let us consider the state just before privacy amplification, denoted
\[
    \rho_{\PAstring \PAstring_B \CP_1^n \CEC \CEV \HEV \Eve},
\]
where, for the moment, we omit the conditioning on the events.  The Leftover Hashing Lemma asserts that once we obtain a lower bound on the relevant entropy of this state, the extractable key length~$l$ can be chosen to be approximately equal to this entropy. 

Of course, in a cryptographic protocol, Eve is not kind enough to reveal the attack she has performed, and we would not trust her even if she did. Therefore the state for which the entropy must be evaluated is not known to Alice and Bob. Instead, one performs some kind of  worst-case analysis over all states compatible with the observed data, which we describe in the next section. This analysis addresses how the associated entropies can be bounded, and which types of bounds are fundamentally impossible to obtain. Indeed, the bulk of QKD security analysis consists of appropriately bounding the relevant entropic quantity which is then utilized in the LHL.

Finally, note that Alice and Bob must communicate classically to ensure that they apply the same hash function to $\PAstring$ and $\PAstring_B$. This choice of hash function is assumed to be revealed to the adversary, both in the protocol and in the statement of the Leftover Hashing Lemma, but only after Alice announces it and not beforehand.

\begin{remark}\label{remark:LHLequalityissue}
When considering hash families that output $l$-bits, one must be careful to distinguish between \emph{universal\(_2\)} hashing (defined by requiring the collision probability to be $\leq 2^{-l}$) and \emph{ideal universal\(_2\)} hashing (where the collision probability is exactly $2^{-l}$), a subtle difference that is often missed in the literature. In particular, the {\Renyi} leftover hashing lemma stated in Ref.~\cite{dupuis_privacy_2023}, as well as the smooth min-entropy version in Ref.~\cite{tomamichel_largely_2017}, require \emph{ideal} universal$_2$ hashing.  
This ideal requirement can in fact be relaxed: one may use ordinary universal$_2$ hashing instead, obtaining essentially the same bound at the cost of losing only one or two bits of key, as shown in Ref.~\cite{kamin_phd_2026}. For this thesis, we restrict our analysis and statements to protocols that utilize ideal universal$_2$ hashing in the privacy amplification step. Note that Toeplitz hashing is ideal universal$_2$ \cite{mansour_computational_1993,krawzyck_LFSR-based_1994}.
\end{remark}

\subsection{Bounding the entropy in Leftover Hashing Lemma} \label{subsec:boundingentropy}
We now consider the various kinds of entropic bounds one may attempt to obtain for use in the LHL. In particular, we also discuss which of them are possible and which of them are fundamentally impossible.  
For simplicity, we restrict attention to fixed-length attacks, where a $\lfixed$-bit key is produced upon the event $\Omega_\mathrm{acc}$, and key of length $0$ is produced otherwise ($\Omega^\complement_\mathrm{acc}$).
A useful tool for building intuition is the \term{intercept--resend attack}, in which Eve stores everything leaving Alice's lab and forwards only arbitrary ``garbage'' signals to Bob. (Note that more involved versions of this attack exist, but this simpler version suffices for the point we wish to make here).   
In this case, once the public announcements $\CP_1^n$ are revealed, Eve typically knows Alice's raw key $S_1^n$ exactly, and therefore also knows the pre-amplification string $\PAstring$.

\begin{remark} \label{remark:variablelengthinputPA}
Recall that in our protocol description the universal$_2$ hashing is applied to the register~$\PAstring$.  
In many of the examples below, however, we will instead bound the entropy of the register~$S_1^n$, because it is typically more convenient to analyze that register. However, notice that these registers are equivalent from Eve's perspective: she has access to the public announcements $\CP_1^n$, she can freely transform $\PAstring$ to $S_1^n$ and vice versa.  Thus, it is natural to expect the relevant entropy quantities to also be exactly equal. We show this formally in \cref{chap:variable,chap:MEAT}, and use these registers interchangeably for the sake of pedagogy here. 
\end{remark}

\begin{itemize}
    \item We may attempt to obtain bounds of the form
\[
    \Hmin[\epsbar]\bigl(S_1^n \mid \CP_1^n \CEC \CEV \HEV \Eve\bigr)_{\rho} \geq \term{constant}
\]
However, a moment of thought shows that, due to the possibility of an intercept--resend attack (where Eve gets full information on the $S_1^n$),  any such bound must be essentially trivial, and the constant must be close to $0$.\footnote{It is typically not exactly zero, due to the smoothing.}

\item Instead, one may attempt to bound the entropy of the state conditioned on acceptance, that is:
\[
    \Hmin[\epsbar]\bigl(S_1^n \mid \CP_1^n \CEC \CEV \HEV \Eve \bigr)_{\rho |  \Omega_\mathrm{acc} } \geq \term{constant}.
\]
However, again,  the intercept--resend attack renders the above bound trivial: Eve gets a perfect copy of $S_1^n$, \textit{regardless of any event} that is conditioned on. Note that this problem can be avoided by having the smoothing parameter itself depend on $\Pr(\Omega_\mathrm{acc})$.\footnote{Typically via $\epsbar \approx \sqrt{ \cdot / \Pr(\Omega_\mathrm{acc})}$, so the smoothing parameter grows larger as acceptance probability grows smaller.} We will see an example of such an approach in our proof of variable-lengths security using the EUR approach in \cref{chap:EUR}.

\item One could try subnormalized conditioning, that is
\[
    \Hmin[\epsbar]\bigl(S_1^n \mid \CP_1^n \CEC \CEV \HEV \Eve \bigr)_{\rho \wedge \Omega_\mathrm{acc} } \geq \term{constant}.
\]
Non-trivial bounds of the above form are indeed possible, and are obtained when considering fixed-length protocols using the postselection technique (\cref{chap:postselection}), and the EUR technique (\cref{chap:EUR}). Furthermore, it turns out that such bounds actually suffice for proving the security of QKD protocols, precisely because the 
security requirement also includes a $\Pr(\Omega_\mathrm{acc})$ prefactor in the trace distance term (see \cref{def:qkdsecuritysymmetric}).
\item  Another possibility is to consider a bound of the form
\[
    \Hmin[\epsbar]\bigl(S_1^n \mid \CP_1^n \CEC \CEV \HEV \Eve \bigr)_{\rho | \Omega_\mathrm{acc} } \geq \term{constant} - \term{pre-factor} \times  \log( \frac{1}{\Pr( \Omega_\mathrm{acc})} ).
\]
Here the right-hand side explicitly depends on the probability of the conditioning event.  
Such bounds are possible, precisely because the logarithmic penalty term. Bounds of this type typically arise in proofs relying on the EAT \cite{metger_generalised_2022,dupuis_entropy_2020}.\footnote{Although our analysis using MEAT \cite{arqand_marginal_2025} in \cref{chap:MEAT} will follow a somewhat different path.}
\end{itemize}

The key point is that one must condition on appropriate events, and do so in the correct way, in order to obtain meaningful lower bounds on the entropy that can be used in a rigorous security proof. We will see concrete examples of this principle throughout this thesis. These are different ways of formalizing the intuition that, while the entropy is not be large for all possible states that can arise, the set of all states can be divided into two categories: those for which the entropy bound is high, and those for which the protocol aborts with high probability. In both cases, security is obtained.

\section{Practical QKD Toolbox}\label{sec:qkdtoolbox}
We will now cover some standard tools that are utilized in QKD security analysis, such as the source-replacement schemes \cite{bennett_quantum_1992,curty_entanglement_2004}, source maps \cite{gottesman_security_2004}, squashing maps \cite{tsurumaru_security_2008,beaudry_squashing_2008,tsurumaru_squash_2010,gittsovich_squashing_2014,zhang_security_2021}. Note that apart from the source-replacement scheme, the remaining tools  are only required when dealing with optical implementations of QKD, and are utilized in reducing infinite-dimensional state preparation and measurement operations to finite dimensions. Accordingly, the reader may safely defer a detailed study of these tools until they are encountered later in the thesis. They can also be used more generally to incorporate imperfections \cite{gottesman_security_2004,nahar2025imperfect,nahar_imperfect_2023,curras_securityquantumkeydistribution_2025}.

\subsection{Source-Replacement Schemes} \label{subsec:sourcereplacement}
The source-replacement scheme is a technique that can be used to describe
Alice's preparation of states in $\dop{=}(A')$ equivalently as her creating a pure,
entangled state across $\dop{=}(\Ameas \Ashield A')$, and then performing
measurements on $\Ameas$. The $A'$ system then behaves as required, whereas
the $\Ashield$ system is referred to as the \emph{shield} system, and is required to correctly describe the preparation of mixed states. This is a
system that is not measured by Alice, Bob, or Eve.  Note that this is a purely
theoretical device, and the actual protocol does not need to change in any
way. It is useful because it allows us to think of prepare-and-measure protocols as
entanglement-based protocols. The proof proceeds by explicit construction. The formal lemma describing the use of the source-replacement scheme is stated below.

\begin{lemma}[Source-Replacement and Shield Systems]
\label{lemma:backgroundsourcereplacementandshield}
Consider a procedure that prepares one out of $\numstates$ possible states $\sigma_k \in \dop{=}(A')$ with
probability $p_k$ and records the outcome choice $k$ in the register $X$. Let $\ket{\sigma_k}_{\Ashield A'}$ be a purification of
$\sigma_k$, and let $\{\ket{k}_{\Ameas}\}_i$ be an orthonormal basis on the register $\Ameas$ of dimension $\numstates$. 
Then, the same procedure can be equivalently described by the preparation of
the pure state
\begin{equation}
    \sigma_{\Ameas \Ashield A'}
    = \sum_{i,k = 1}^{\numstates} \sqrt{p_i p_k} 
      \ketbra{i}{k}_{\Ameas}
      \otimes \ketbra{\sigma_i}{\sigma_k}_{\Ashield A'} ,
\end{equation}
 followed by projectively measuring the $\Ameas$ system with
$\{ \ketbra{k}{k} \}_k$ and storing the outcome in the $X$ system, and tracing out the $\Ashield$ system. That is, the state obtained on the $XA'$ registers for both procedures is identical.
\end{lemma}
\begin{proof}
The proof follows from explicit computation. Measuring the $ \sigma_{\Ameas \Ashield A'}$ with the given POVM gives us 
\begin{equation}
    \sigma_{X \Ashield A'}
    = \sum_{k=1}^{\numstates}  p_k
      \ketbra{k}{k}_{X}
      \otimes \ketbra{\sigma_k}{\sigma_k}_{\Ashield A'}.\end{equation}
Tracing out $\Ashield$ gives us the state describing the original state preparation
\begin{equation}
    \sigma_{X A'}
    = \sum_{k=1}^{\numstates}  p_k
      \ketbra{k}{k}_{X}
      \otimes (\sigma_k)_{A'}.     
      \end{equation}
\end{proof}

Thus, we can equivalently describe the signal-preparation phase of the
\nameref{prot:qkdprotocol} as Alice preparing a global source-replaced state
$\sigma_{A_1^n (A')_1^n} = \otimes_{1}^n \sigma_{A A'}$ (where we identify $A = \Ameas \Ashield$). After
Eve's attack, this state becomes $\sigma_{A_1^n B_1^n \Eve}$, which is then
followed by measurements by Alice and Bob, along with subsequent protocol steps. This is
exactly the structure one would expect in a typical entanglement-based
protocol, in that both Alice and Bob perform measurements on a received state. Consequently, we may instead prove security for the protocol in
which Alice and Bob simply perform their measurements, and imagine that the
state $\sigma_{A_1^n B_1^n \Eve}$ is supplied directly by Eve.\footnote{
Note that when computing key rates using numerical methods, to reduce complexity, the source-replacement scheme is often implemented only for the actual states sent without any $\test$ or $\gen$ label, with these labels instead being assigned by Alice after state preparation according to some probability distribution. For example, if $k=(a,\testgenflag)$, then Alice can prepare states corresponding to $a$ first, and \emph{then} assign $\testgenflag$ according to $\Pr(\testgenflag | a)$. The two pictures are equivalent.}

Notice that in the construction above, we have 
\begin{equation} \label{eq:sigmaAentire}
    \sigma_{A_1^n} = \bigotimes_{j=1}^n \sigma_{A},
\end{equation}
where
\begin{equation} \label{eq:sigmaA}
    \sigma_A
    = \sum_{i,k} \sqrt{p_i p_k}\,
      \braket{\sigma_i}{\sigma_k}_{\Ashield A'}
      \ketbra{i}{k}_{\Ameas}
\end{equation}
is a fixed operator. Thus, Eve is not allowed to supply \emph{any} arbitrary
state of her choosing; she must supply a state whose marginal on $A_1^n$ is an IID state that matches \cref{eq:sigmaA,eq:sigmaAentire}.
 This reflects the fact that, in the
original protocol, the state preparation is a trusted operation performed by
Alice. We refer to this requirement as the \emph{fixed-marginal promise} or
the \emph{fixed-marginal constraint}. This will play an important role in our analysis of the postselection technique in \cref{chap:postselection}.

\subsubsection{Security Definition Revisited}

It is convenient to now view the \nameref{prot:qkdprotocol} in terms of the 
source-replaced state, together with a CPTP map 
$\protMap{\lkey}$ that performs all operations involved in the QKD protocol. That is, $\protMap{\lkey}$ measures the
$A_1^n B_1^n$ systems, performs all classical processing and public announcements in the registers
$\allpublic$, and finally produces the output key in the registers
$K_A K_B$. In this picture, the ideal QKD map is given by 
$\idealmap \circ \protMapId{\lkey}$; that is, it runs the original QKD 
protocol and then replaces the key registers with ideal ones, as described in 
\cref{sec:securitydefinition}. Note that the superscript $\lkey$ denotes the \emph{function} that determines 
the output key length from the public announcements. If one is interested in 
fixed-length protocols, we tolerate a slight abuse of notation and replace 
$\lkey$ with the constant value $\lfixed$.

\begin{definition}[Equivalent formulation of QKD protocol] \label{def:equivalentQKDprotocol}
At this stage, it is convenient to describe an instance of the 
\nameref{prot:qkdprotocol} by the pair 
$\{\protMap{\lkey}, \sigma_A\}$, where 
$\protMap{\lkey} \in \CPTP(A_1^n B_1^n , \allpublic)$ implements the QKD 
protocol described in \nameref{prot:qkdprotocol} (after replacing Alice's 
state-preparation operations with measurements). The ideal QKD protocol is then given by  the pair 
$\{\protMapId{\lkey}, \sigma_A\}$, where $\protMapId{\lkey} = \idealmap \circ \protMap{\lkey}$.
Furthermore, we use $\{\protMap{\lfixed}, \sigma_A\}$ to describe a fixed-length variant of \nameref{prot:qkdprotocol} (with the ideal protocol defined analogously). 
\end{definition}

Note that the fact that the above constitutes an equivalent definition comes from the fact that the set of possible output states for the above scenario is exactly identical to the original QKD protocol. This allows us to state the following equivalent definition of secrecy for a 
prepare-and-measure QKD protocol. We only require the secrecy definition 
here, although the full security definition can be reformulated in the same 
manner. These equivalent formulations will then be used in \cref{chap:variable,chap:postselection,chap:MEAT}.

\begin{definition}[$\epssecret$-secret PMQKD protocol with fixed marginal $\sigma_A$] \label{def:epsSecPromise}
Consider the \nameref{prot:qkdprotocol} represented as 
$\{\protMap{\lkey}, \sigma_A\}$.\footnote{One can replace $\protMap{\lkey}$ with $\protMap{\lfixed}$ if one is interested in fixed-length protocols.} The protocol is 
\textit{$\epssecret$-secret with fixed marginal $\sigma_A$} if
\begin{equation}
    \begin{aligned}
        \label{eq:epsSec}
        &\tracedist{
            \left(
                \left( \Tr_{K_B} \circ \protMap{\lkey}
                - \Tr_{K_B} \circ \protMapId{\lkey} \right)
                \otimes \idmap_{\Eve}
            \right)
           \left[ \rho_{A_1^n B_1^n \Eve} \right]
        }
        \leq \epssecret, \\
        &\forall\, \rho_{A_1^n B_1^n \Eve}\in \dop{=}(A_1^n B_1^n \Eve)
        \text{ such that }
     \rho_{A_1^n}
        = \left( \sigma_A \right)^{\otimes n}.
    \end{aligned}
\end{equation}
\end{definition}

If one further specializes to IID collective attacks, where Eve is required to perform the same operations on each round, as we will in \cref{chap:postselection,chap:variable}, then we obtain the following secrecy definition.

\begin{definition}  [$\epssecret$-secret PMQKD protocol with fixed marginal $\sigma_A$ against IID collective attacks] \label{def:epsSecPromiseIIDcollective}

Consider the \nameref{prot:qkdprotocol} represented as 
$\{\protMap{\lkey}, \sigma_A\}$.\footnote{One can replace $\protMap{\lkey}$ with $\protMap{\lfixed}$ if one is interested in fixed-length protocols.} The protocol is 
\textit{$\epssecret$-secret with fixed marginal $\sigma_A$} against IID collective attacks if
\begin{equation}
    \begin{aligned}
        \label{eq:epsSecIID}
        &\tracedist{
            \left(
                \left( \Tr_{K_B} \circ \protMap{\lkey}
                - \Tr_{K_B} \circ \protMapId{\lkey} \right)
                \otimes \idmap_{E_1^n}
            \right)
           \left[ \rho_{ABE}^{\otimes n} \right]
        }
        \leq \epssecret, \\
        &\forall\, \rho_{ABE} \in \dop{=}(ABE)
        \text{ such that }
        \Tr_{B E}\!\left[ \rho_{A B E} \right]
        =  \sigma_A .
    \end{aligned}
\end{equation}
\end{definition}

\subsection{Source Maps} \label{subsec:sourcemaps}

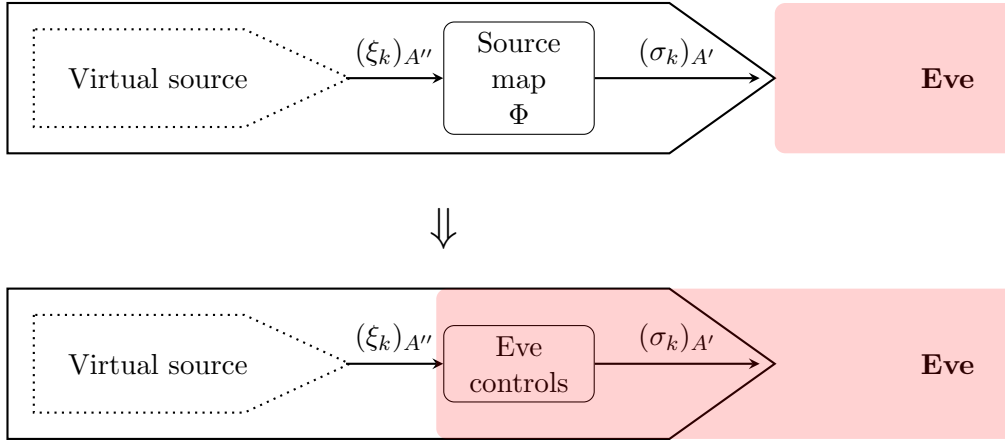
\begin{figure}[ht]
\centering
\begin{tikzpicture}[font=\small]


\tikzset{
  laser/.style={
    draw=black, thick, dotted, fill=none,
    trapezium, trapezium left angle=90, trapezium right angle=70,
    minimum height=1.0cm, minimum width=2.4cm,
    inner sep=2pt, outer sep=0pt,
    align=center
  },
  evebox/.style={
    processLarge, fill=none, draw=black
  }
}

\coordinate (Tleft)  at (-6,  2.2);
\coordinate (Tright) at ( 5.5, 2.2);
\coordinate (TbotL)  at (-6,  0.2);
\coordinate (TbotR)  at ( 5.5, 0.2);

\draw[draw=black, thick, fill=none]
  (Tleft) --
  (TbotL) --
  (2.8,0.2) --
  (4.2,1.2) --
  (2.8,2.2) --
  cycle;

\def\VTinset{0.35} 
\def\VTshrink{0.35} 
\def\VTshrinkX{5.3}  

\coordinate (VT_L)  at ({-6+\VTinset}, {1.2+1.0-\VTshrink});
\coordinate (VT_Lb) at ({-6+\VTinset}, {1.2-1.0+\VTshrink});
\coordinate (VT_Mb) at ({2.8-\VTinset-\VTshrinkX}, {1.2-1.0+\VTshrink});
\coordinate (VT_T)  at ({4.2-\VTinset-\VTshrinkX}, {1.2});
\coordinate (VT_Mt) at ({2.8-\VTinset-\VTshrinkX}, {1.2+1.0-\VTshrink});

\draw[draw=black, thick, dotted, fill=none]
  (VT_L) -- (VT_Lb) -- (VT_Mb) -- (VT_T) -- (VT_Mt) -- cycle;

\node[align=center] at (-4.0,1.2) {Virtual source};

\node[evebox] (SrcMapTop) at (0.8,1.2) {Source\\map\\$\Phi$};

\draw[arrow] (-1.5,1.2) -- (SrcMapTop.west)
  node[midway, above] {$(\xi_{k})_{A''}$};

\draw[arrow] (SrcMapTop.east) -- (4.0,1.2)
  node[midway, above] {$(\sigma_k)_{A'}$};

\node[align=center] at (6.5,1.2) {\textbf{Eve}};

\node at (-0.2,-0.8) {\scalebox{1.6}{$\Downarrow$}};
\def\TopEveX{4.2}
\def\TopYmax{2.2}
\def\TopYmin{0.2}

\path[fill=red, opacity=0.18, draw=none, rounded corners=4pt]
  (\TopEveX,\TopYmin) rectangle (7.4,\TopYmax);

\coordinate (Bleft)  at (-6, -1.6);
\coordinate (Bright) at ( 5.5,-1.6);
\coordinate (BbotL)  at (-6, -3.6);
\coordinate (BbotR)  at ( 5.5,-3.6);

\draw[draw=black, thick, fill=none]
  (Bleft) --
  (BbotL) --
  (2.8,-3.6) --
  (4.2,-2.6) --
  (2.8,-1.6) --
  cycle;

\def\VBinset{0.35}
\def\VBshrink{0.35}
\def\VBshrinkX{5.3}
\coordinate (VB_L)  at ({-6+\VBinset}, {-2.6+1.0-\VBshrink});
\coordinate (VB_Lb) at ({-6+\VBinset}, {-2.6-1.0+\VBshrink});
\coordinate (VB_Mb) at ({2.8-\VBinset-\VBshrinkX}, {-2.6-1.0+\VBshrink});
\coordinate (VB_T)  at ({4.2-\VBinset-\VBshrinkX}, {-2.6});
\coordinate (VB_Mt) at ({2.8-\VBinset-\VBshrinkX}, {-2.6+1.0-\VBshrink});

\draw[draw=black, thick, dotted, fill=none]
  (VB_L) -- (VB_Lb) -- (VB_Mb) -- (VB_T) -- (VB_Mt) -- cycle;

\node[align=center] at (-4.0,-2.6) {Virtual source};

\node[evebox] (EveCtrl) at (0.8,-2.6) {Eve\\controls};

\draw[arrow] (-1.5,-2.6) -- (EveCtrl.west)
  node[midway, above] {$(\xi_{k})_{A''}$};

\draw[arrow] (EveCtrl.east) -- (4.0,-2.6)
  node[midway, above] {$(\sigma_{k})_{A'}$};

\node[align=center] at (6.5,-2.6) {\textbf{Eve}};

\def\BotEveX{-0.3}
\def\BotYmax{-1.6}
\def\BotYmin{-3.6}

\path[fill=red, opacity=0.18, draw=none, rounded corners=4pt]
  (\BotEveX,\BotYmin) rectangle (7.4,\BotYmax);

\end{tikzpicture}
\caption{Schematic illustrating the use of a source map. A virtual source prepares $(\xi_{k})_{A''}$, which is mapped to the real emitted state $(\sigma_{k})_{A'}$ by a source map $\Psi$. Then, one can ``give" Eve control over the source map (meaning that she is allowed to perform any operation she wants in place of the source map). The security of the latter implies security of the former, as argued rigorously in \cref{lemma:sourcemapsecurityMEAT}.}
\label{fig:sourcemap_schematic}
\end{figure}

The main idea behind source maps is fairly intuitive, as described in \cref{fig:sourcemap_schematic}. The real source is first modelled as a virtual source followed by a quantum channel called the ``source map". The source map is then ``given" to Eve: a step that only gives her more power. In particular, Eve is allowed to replace the source map with any CPTP map of her choice. Consequently, any security statement proved under the assumption that Alice prepares $\{(\xi_k)_{A''}\}_k$ remains valid for the actual protocol in which Alice prepares $\{(\sigma_k)_{A'}\}_k$, since one of Eve's attacks in the virtual source setting involves her applying the source map $\Psi$ as the first step of her attack. The security analysis can then be restricted to the more convenient virtual source.  We rigorously apply source maps to QKD in \cref{chap:MEAT,chap:postselection}, and defer to those chapters for a detailed and rigorous treatment. Here, we instead focus on providing intuition. 

\begin{definition}[Source map] \label{def:sourcemap}
Let $\{(\sigma_k)_{A'}\}_k$ be the (possibly infinite-dimensional) family of states that Alice prepares in the actual protocol. Let $A''$ be a (typically finite-dimensional) register, and let $\{(\xi_k)_{A''}\}_k$ be a family of states on $A''$.  
A \term{source map} from the virtual source $\{(\xi_k)_{A''}\}_k$ to the actual source $\{(\sigma_k)_{A'}\}_k$ is a quantum channel
$
\Psi \in \CPTP(A'' , A')
$
such that
\begin{equation}
\label{eq:sourcemapconditionbackground}
\sigma_k = \Psi[\xi_k]
\qquad\text{for all } k .
\end{equation}
\end{definition}

We note that the registers $A'$ and $A''$ are treated interchangeably throughout this thesis. In particular, any statement involving the register $A'$ describing the real state preparation can also be applied to $A''$ (describing virtual state preparation).

\subsubsection{Tagging map} \label{subsec:taggingsourcemap}
An important class of source maps that is frequently used in the security analysis of decoy-state protocols is \emph{tagging} \cite{gottesman_security_2004}. Tagging applies when the real source emits states that are block-diagonal in a suitable decomposition, such as photon number. In this case, the real source can be replaced by a virtual source that operates as follows: with some probability, it emits a state supported on a finite photon-number subspace (corresponding to photon numbers below a fixed threshold), and with the remaining probability, it emits a classical \emph{tag} that is readable by Eve and reveals the exact state that was prepared. 

The purpose of the finite cut-off is to isolate a finite-dimensional subspace that can be handled explicitly in the security proof, while treating all higher photon-number components pessimistically by assuming that they are fully known to Eve. For example, in polarization-encoded decoy-state QKD, the source prepares fully phase-randomized coherent states, which are block-diagonal in the total photon number across all polarizations. Such a source can therefore be replaced by a virtual source that probabilistically emits a vacuum or low-photon-number state within a fixed cut-off, or otherwise emits a tagged signal that is assumed to be completely compromised.

\begin{lemma}[Tagged laser source]\label{lemma:tagging}
    Let $\sigma_{(a,\mu,\testgenflag)} = \sum_{N=0}^\infty e^{-\mu}\frac{\mu^N}{N!}\ketbra{N}_a$ be the real state prepared by Alice corresponding to setting choice $a,\mu$, where $\ket{N}_a$ denotes a $N$-photon Fock state in the polarization $a$. Let the virtual state prepared by Alice be
    \begin{align} \label{eq:taggedStates}
        \xi_{(a,\mu,\testgenflag)} = \sum_{N=0}^\ndecoy e^{-\mu}\frac{\mu^N}{N!} \ketbra{N}_a + \left(1-\sum_{N=0}^\ndecoy e^{-\mu}\frac{\mu^N}{N!}\right) \ketbra{a,\mu},
    \end{align}
    where $\{\ket{a,\mu}\}_{a,\mu}$ form an orthonormal basis for a space orthogonal to the span of $\{\ket{N}\}_{N=0}^\ndecoy$.
    Then there exists a source map $\Psi_{\mathrm{tag}}$ (see \cref{def:sourcemap}) such that $\Psi_{\mathrm{tag}}[\xi_{(a,\mu,\testgenflag})] = \sigma_{(a,\mu,\testgenflag)}$ for all setting choices $a,\mu,\testgenflag$.
\end{lemma}
\begin{proof}
    Define $\Psi_{\mathrm{tag}}$ to be the channel that projects onto $\ketbra{a,\mu}$ and prepares  $$\frac{1}{\left(1-\sum_{N=0}^\ndecoy e^{-\mu}\frac{\mu^N}{N!}\right)}\sum_{N=\ndecoy+1}^\infty e^{-\mu}\frac{\mu^N}{N!} \ketbra{N}_a$$ for all $a$, $\mu$; and that acts as the identity on the space spanned by $\{\ket{N}\}_{N=0}^\ndecoy$. It is then straightforward to verify that $\Psi_{\mathrm{tag}}[\xi_{a,\mu}] = \sigma_{a,\mu}$.
\end{proof}

\subsubsection{Security Definition Revisited (yet again)}
We present yet another equivalent way to view and reason with  \nameref{prot:qkdprotocol} and its security definition, that focuses on the state preparation aspect.  
Recall that a prepare-and-measure protocol involves Alice preparing states $\sigma_k \in \dop{=}(A')$ with probability $p_k$ in each round. That is, such protocols can be described as follows:
\begin{enumerate}
    \item  Alice first prepares the state $
\sigma_{X_1^n (A')_1^n}
=
\left(
\sum_k p_k \ketbra{k}_X \otimes (\sigma_k)_{A'}
\right)^{\otimes n}$.
\item This state is then subjected to Eve’s attack $\attack{} \in \CPTP((A')_1^n, B_1^n \Eve)$, resulting in the post attack state $\rho_{X_1^n B_1^n \Eve}$.
\item This is followed by Bob’s measurement and the remaining steps of the \nameref{prot:qkdprotocol}, which can be described by a map $
\protMapbeforeSR{\lkey} \in \CPTP(X_1^n B_1^n , \allpublic)$.
\end{enumerate}

Equivalently, if we decide to replace Alice's state preparation procedure according to the source-replacement scheme, then we obtain the following equivalent description:

\begin{enumerate}
\item Alice first prepares the state $\rho_{\Ameas_1^n \Ashield_1^n (A')_1^n}$ as described in \cref{lemma:shieldandsourcereplacement}. We identify $A$ with the systems $\Ameas,\Ashield$ that do not leave Alice's lab.

\item  This state is then subjected to Eve’s attack $\attack{} \in \CPTP((A')_1^n, B_1^n \Eve)$, resulting in the post attack state $\rho_{A_1^n B_1^n \Eve}$.
\item This is followed by both Alice's and Bob’s measurement and the remaining steps of the \nameref{prot:qkdprotocol}, which can be described by a map
\begin{equation}
\protMapbeforeSR{\lkey} \in \CPTP(X_1^n B_1^n , \allpublic).
\end{equation}
\end{enumerate}
 
Note that the only difference between $\protMap{\lkey}$  and $\protMapbeforeSR{\lkey}$ (defined here) is that the former implements the QKD protocol with Alice's state preparation being described with the source-replacement scheme, and therefore involves an additional measurement on Alice's $\Ameas$ system to create her local register $X$. In particular, we have 
\begin{equation}
\protMap{\lkey}
=
\protMapbeforeSR{\lkey} \circ\left( \measChannel{ \{\ketbra{k}\}} \right)^{\otimes n},
\end{equation}
where $\measChannel{ \{\ketbra{k}\}}  \in \CPTP(A,X)$ is a channel that measures the $A$ systems of the source-replaced state and stores the measurement outcome in the $X$ register. Putting all of this together, we obtain the following equivalent definition. 

\begin{definition}[Equivalent formulation of QKD protocol] \label{def:equivalentQKDprotocolbeforeSR}
At this stage, it is convenient to describe an instance of the 
\nameref{prot:qkdprotocol} by the pair 
$\{\protMapbeforeSR{\lkey}, \sigma_{XA'}\}$, where 
$\protMapbeforeSR{\lkey} \in \CPTP(X_1^n B_1^n , \allpublic)$ implements the QKD 
protocol described in \nameref{prot:qkdprotocol}. The \emph{ideal} QKD protocol is then given by  the pair 
$\{\protMapIdbeforeSR{\lkey}, \sigma_{XA'}\}$, where $\protMapIdbeforeSR{\lkey} = \idealmap \circ \protMapbeforeSR{\lkey}$. The definition here is related to \cref{def:equivalentQKDprotocol} via
\begin{equation}
    \sigma_{XA'} = \sum_k p_k \ketbra{k}_X \otimes (\sigma_k)_{A'}
\end{equation}
which represents Alice's state preparation, and 
\begin{equation}
    \protMap{\lkey} = \protMapbeforeSR{\lkey} \circ \measChannel{ \{\ketbra{k}\}} 
\end{equation}
where $\measChannel{ \{\ketbra{k}\}} $ is the channel that performs Alice's measurements on the source-replaced state. The ideal QKD protocol is then given by the pair $\{\protMapIdbeforeSR{\lkey},\sigma_{XA'} \}$, where $
\protMapIdbeforeSR{\lkey} = \idealmap \circ \protMapbeforeSR{\lkey}$.

\end{definition}

This allows us to state the following equivalent definition of secrecy for a 
prepare-and-measure QKD protocol. We only require the secrecy definition 
here, although the full security definition can be reformulated in the same 
manner.

\begin{definition}[$\epssecret$-secret PMQKD protocol ]
Consider the \nameref{prot:qkdprotocol} represented as 
$\{\protMapbeforeSR{\lkey}, \sigma_{XA'}\}$, where $\sigma_{XA'}$ represents Alice's signal preparation in each round.  The protocol is 
\term{$\epssecret$-secret } if
\begin{equation}
    \begin{aligned}
        &\tracedist{
            \left(
                \left( \Tr_{K_B} \circ \protMapbeforeSR{\lkey}
                - \Tr_{K_B} \circ \protMapIdbeforeSR{\lkey} \right)
                \otimes \attack{}
            \right)
           \left[ \sigma_{X_1^n (A')_1^n } \right]
        }
        \leq \epssecret, \\
        &\forall\, \attack{} \in \CPTP( (A')_1^n, B_1^n \Eve)
    \end{aligned}
\end{equation}
\label{def:epsSecPromisechannelversion}
\end{definition}

\subsection{Squashing Maps} \label{subsec:squashing}

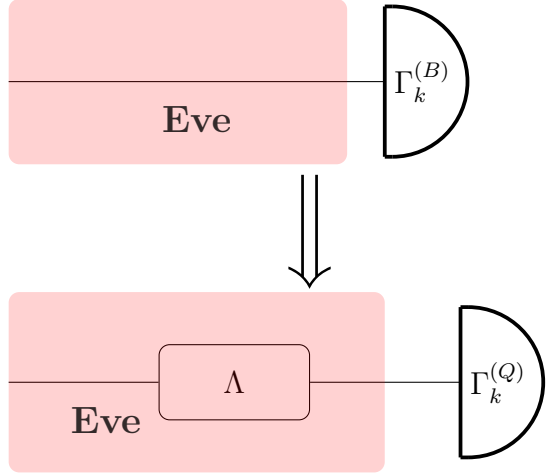
\begin{figure}[ht!]
    \centering
 \begin{tikzpicture}

\begin{scope}[shift={(1,0)}]
    \pic {detectorSmall={detinf, $\Gamma^{(B)}_k$, black}};
    \draw (-5,0) -- ([xshift=-0.0cm]detinf.west);

    \def\TopEveX{-5.0}
    \def\TopYmax{1.1}
    \def\TopYmin{-1.1}
    \path[fill=red, opacity=0.18, draw=none, rounded corners=4pt]
      (\TopEveX,\TopYmin) rectangle (-0.5,\TopYmax);

    \node[opacity=0.8, font=\large] at (-2.5,-0.5) {\textbf{Eve}};
\end{scope}

\begin{scope}[shift={(-1,-4)}]
    \node[processLarge] (SquashingMap) {$\Lambda$};

    \pic at ([xshift=2.0cm]SquashingMap.east)
        {detectorSmall={detfinideal, $\Gamma^{(Q)}_k$, black}};

    \draw (-3,0) -- (SquashingMap.west);
    \draw (SquashingMap.east) -- ([xshift=0.0cm]detfinideal.west);

    \def\BotEveX{-3.0}
    \def\BotYmax{1.2}
    \def\BotYmin{-1.2}
    \path[fill=red, opacity=0.18, draw=none, rounded corners=4pt]
      (\BotEveX,\BotYmin) rectangle (2,\BotYmax);

    \node[opacity=0.8, text width=3cm, align=center, font=\large]
        at ([xshift=-0.7cm,yshift=-0.5cm]SquashingMap.west) {\textbf{Eve}};
\end{scope}

\node at (0,-2) {\scalebox{2}{$\Big\Downarrow$}};

\end{tikzpicture}
    \caption{An infinite-dimensional POVM can be
    modelled as a squashing map $\Lambda$ followed by a finite-dimensional
    POVM. Giving the squashing map $\Lambda$ to Eve allows us to
    restrict our analysis to the finite-dimensional POVM.}
    \label{fig:squashing}
\end{figure}

Squashing maps are used to modify detection setups to more convenient ones for theoretical analysis \cite{beaudry_squashing_2008,tsurumaru_security_2008,tsurumaru_squash_2010,fung_universal_2011,gittsovich_squashing_2014,zhang_security_2021,upadhyaya_dimension_2021, nahar_postselection_2024} . As with source maps, we defer the
formal statements regarding the use of squashing maps to \cref{chap:MEAT,chap:postselection}, and describe them intuitively here (see \cref{fig:squashing}). The main idea behind squashing maps is that the measurement statistics of a
(possibly infinite-dimensional) system can be reproduced by a two-step
procedure. First, a squashing map $\Lambda$ is applied to the
incoming state, mapping it to a potentially lower-dimensional register. This is then
followed by a measurement on that register.

\begin{definition}[Squashing map] \label{def:squashing}
Let $\{\Gamma^{(B)}_k\}_{k=1}^{\nummeas}$ be a POVM acting on a (possibly infinite-dimensional) register $B$.
Let $Q$ be a (typically finite-dimensional) register, together with a POVM
$\{\Gamma^{(Q)}_k\}_{k=1}^{\nummeas}$ acting on $Q$. Then,  a 
\term{squashing map} from the former POVM to the latter POVM is a quantum channel $\Lambda \in \CPTP(B, Q)$, such that
\begin{equation}
\label{eq:squashingmapdef}
\Tr\!\left[\Gamma^{(B)}_k \rho\right]
=
\Tr\!\left[\Gamma^{(Q)}_k \, \Lambda[\rho]\right]
\qquad
\forall\, \rho \in \dop{=}(B), \ \forall\, k = 1,\dots,\nummeas .
\end{equation}
\end{definition}

Thus, instead of performing measurements with $\{\Gamma^{(B)}_k\}_k$, one can instead equivalently perform the squashing map and perform measurements using $\{\Gamma^{(Q)}_k\}_k$. The argument then proceeds by ``giving" the squashing map to Eve, and allowing her to perform any operation she wants (see \cref{fig:squashing}). We discuss a variety of squashing maps below:
\begin{itemize}
    \item \textbf{Simple Squasher: } 
The first squashing map proposed in
\cite{beaudry_squashing_2008,gittsovich_squashing_2014} was the
\emph{simple squasher}, which reduced an active-basis–choice,
polarization-encoded detection setup with perfect detectors to an
effective qubit measurement (with double-click events randomly assigned to
$0$ or $1$). We will use this squasher in \cref{chap:variable,chap:postselection,chap:MEAT} for key rate calculations. In particular, this allows us to reduce the active-choice threshold detection setup  to the one with qubit measurements (with an additional dimension for loss). Both these measurements are described in \cref{subsec:alicestatessent}. Its main drawback is that it requires detection efficiencies and dark count rates of all detectors to be identical.
\item \textbf{The flag-state squasher: } The most versatile squashing construction currently used is the \emph{flag-state squasher} \cite{zhang_security_2021}, which applies to POVMs that are block-diagonal. It does not require all detectors to be identical, and in fact can be used even with imperfectly characterized detectors \cite{nahar2025imperfect}. Importantly, it covers detection setups based on threshold detectors, which are block-diagonal in the photon-number basis, as described in \cref{sec:statesandmeasurements,sec:backgroundquantumoptics}.

We do not go into the technical details here, but briefly note that the flag-state squasher operates by splitting the Hilbert space into two subspaces. In the preserved subspaces (typically corresponding to low photon numbers), the incoming state is left unchanged and passed through. For higher photon numbers, the state is measured and the output consists of a classical flag encoding the measurement outcome.

The main nuance to note is that if Eve’s attack were completely unrestricted, then a protocol formulated directly in terms of the squashed POVM elements could not guarantee security. This is due to the existence of classical flags. Eve could simply measure the incoming state herself and then send Bob a flag encoding the outcome she obtained. Eve would then have full knowledge of Bob’s measurement results, and no secrecy could be established. Consequently, this naive approach cannot yield meaningful security statements. Instead, one must impose suitable restrictions to prevent all information from being leaked through the flag registers.\footnote{In \cref{fig:squashing}, this can be interpreted as not giving Eve complete control over the squashing map, but rather only restricted control.}

There are several complementary ways of formalizing  such restrictions \cite{zhang_security_2021,Kamin2025,wang2025phase}, all of which rely on the fact that high photon-number components have a non-negligible probability of producing multi-click events. While we do not perform explicit computations in this thesis, we indicate how various results can nevertheless be applied in conjunction with the flag-state squasher.

\item \textbf{The weight-preserving flag-state squasher: }
The weight-preserving flag-state squasher is a modification of the
flag-state squasher proposed in Ref.~\cite{nahar_postselection_2024}. Its
main motivation is to sidestep the difficulties that arise when combining
the postselection technique with the standard flag-state squasher, in
particular the challenge of enforcing the required flag-state squasher
restrictions on the flag space. We defer a detailed discussion of this issue to
\cref{chap:postselection}.
\end{itemize}

Having discussed the main tools of QKD security analysis, we give  a brief explanation of the various proof technique approaches in the next section.
\begin{remark}\label{remark:natureofreduction}
We emphasize that, when using these tools, the precise nature of the reduction statements is important. The most
useful reductions are those established directly at the level of protocol
security; that is, security of the original QKD protocol follows from the security
of a modified QKD protocol that is more convenient to analyze (because it replaces infinite-dimensional signal states with finite-dimensional ones, say). In this thesis, we will only carry out such reductions. This approach is
particularly appealing because it allows one to perform the reduction first,
and then apply \emph{any} proof technique of choice to complete the remaining QKD
security analysis. 

In contrast, many works only establish the relevant reduction at a later stage of
the analysis, for instance at the level of the single-round optimization in the final key rate expression. Such reductions often interact in subtle ways with the chosen
proof technique (see \cref{remark:infinitedimensions}), and are less general. 
\end{remark}

\section{QKD proof techniques} \label{sec:qkdprooftechniques}
In this section, we give a brief pedagogical overview of various proof techniques for QKD which are utilized at length throughout this thesis. This section is based on Ref.~\cite{tupkary2025qkdsecurityproofsdecoystate}. 

\subsection{Entropic Uncertainty Relations} \label{subsec:EUR}
Proofs that rely on the entropic uncertainty relations \cite{tomamichel_uncertainty_2011} utilize the following statement to obtain a bound on the smooth min-entropy of the pre-amplification string, which is then used in the Leftover Hashing Lemma (see \cref{sec:LHL}): that for any state $\rho_{A^mB^m \Eve }$, 
\begin{equation} \label{eq:eur}
    \Hmin[\bar{\epsilon}](S_1^m | \Eve)_{\mathcal{E}_{\Zbasis}(\rho)} + H_\mathrm{max}^{\bar{\epsilon}}( \Svirt_1^m | B_1^m)_{\mathcal{E}_{\Xbasis}(\rho)} \geq m c_q
\end{equation}
where $\mathcal{E}_{\Zbasis}(\rho)$ is a channel that measures the $A^m$ system in the $\Zbasis$ basis, and $\mathcal{E}_{\Xbasis}(\rho)$ is a channel that measures the $A^m$ system in the $\Xbasis$ basis, and $c_q$ is a parameter that depends on these measurements POVMs \footnote{The $\Zbasis$ and $\Xbasis$ basis here need not be taken literally. One can use the EUR statement for  \textit{any} two POVMs, one corresponding to $\Zbasis$ and $\Xbasis$. (To obtain a useful statement, one must ensure that $c_q>0$ for the two POVMs considered). For the BB84 protocol with ideal source, these correspond to the canonical $\Zbasis$ and $\Xbasis$ basis measurements.}. Here, Alice's measurement POVMs and her reduced state are obtained from the source-replacement scheme (see also \cref{subsec:sourcereplacement}).

In the actual protocol, the $\Zbasis$ measurements are performed, which determines (a part of) the pre-amplification string whose min-entropy we wish to bound.
The max entropy term is then suitably bounded by obtaining an estimate (upper bound) on the number of ``phase errors". This refers to the number of errors corresponding to (fictitious) measurements, where Alice and Bob measure these rounds in the complementary basis $\Xbasis$ instead of $\Zbasis$. Note that only one measurement ($\Zbasis$) is performed in the actual protocol. Thus, the central task in this approach is to properly define the phase error rate via fictitious measurements, and then bound it via suitable statistics arguments. We will study this technique in \cref{chap:EUR}

Note that one typically only applies the EUR analysis on a part of the  pre-amplification string $\PAstring$\footnote{That is, the part that comes from rounds where Alice prepares single-photons. Rounds where Alice sends a multi-photon pulses cannot lead to secrecy due to the photon number splitting attack \cite{Lutkenhaus_estimates_1999,Bennett_experimentalquantumcryptography_1992,Brassard_limitationsonpractical_2000,Lutkenhaus_security_2000}, which manifests here as the $c_q$ term becoming zero.}.  Thus, one must also relate the smooth min-entropy obtained via the EUR analysis to the smooth min-entropy of the actual pre-amplification string. This is done with the use of suitable chain rules.

\subsection{Postselection technique}
\label{subsec:postselection}

The postselection technique \cite{christandl_postselection_2009,nahar_postselection_2024} is a proof technique that reduces the analysis of coherent attacks to the analysis of IID collective attacks, which we will cover in \cref{chap:postselection}. It is composed primarily of three parts:
\begin{enumerate}
    \item First, given a permutation invariant protocol, the permutation invariance property is used to reduce the analysis from all possible states $\rho_{A^nB^n}$ that could be shared by Alice and Bob to permutationally invariant states $\bar{\rho}_{A^nB^n}$ (with Eve holding a purification in the $\Eve$ register).
    \item Then, one applies a de Finetti theorem for permutationally invariant states, of the form
    \begin{align} \label{eq:deFinettiTheorem}
        \bar{\rho}_{A^nB^n} \leq g_{n,x} \tau_{A^nB^n} = g_{n,x} \int d\sigma \sigma_{AB}^{\otimes n}.
    \end{align}
    In the above formula, $\tau_{A^nB^n}$ is a particular mixture of IID states, while the $g_{n,x}$ term is defined by the formula $g_{n,x} = \binom{n+x-1}{x-1}$ (where $x$ is a value that depends on the dimensions of Alice and Bob's subsystems), and affects the final security parameters and key length in the security analysis against coherent attacks. The above statement is used to reduce the security of the permutational invariant state $\bar{\rho}_{A^nB^n}$ to the security of a \textit{purification} of $\tau_{A^nB^n}$, with some costs to the security parameter and key length that depend on $g_{n,x}$. 
    \item Finally, a finite-size security proof is obtained against \emph{all} IID states in the mixture $\tau_{A^nB^n}$. Note that \textit{any} proof method can be used for the IID security proof: we will see one such method in \cref{chap:variable}. This is then used to prove the security of the purification of $\tau_{A^nB^n}$.
\end{enumerate}

Essentially, the proof can be viewed as ``lifting'' the security proof against IID collective attacks to one against coherent attacks, at the price of some penalties to the key length and security parameter. It is important to note that this lift is independent of the details of the proof technique used for the security analysis against IID collective attacks. 

\subsection{Entropy accumulation based proofs} \label{subsec:EAT}
There are a variety of accumulation theorems, such as the entropy accumulation theorem (EAT) \cite{dupuis_entropy_2020,dupuis_entropy_2019}, generalized entropy accumulation theorem (GEAT) \cite{metger_generalised_2022,metger_security_2023}  generalized {\Renyi } entropy accumulation theorem \cite{arqand_generalized_2024}, and marginal-constrained entropy accumulation theorems (MEAT) \cite{inprep_vanhimbeeck_tight_2024,fawzi_additivity_2025,arqand_marginal_2025}. We will utilize the MEAT in \cref{chap:MEAT}, since that variant is the most suitable for the analysis of prepare-and-measure QKD protocols.  In contrast, the EAT cannot be directly applied to prepare-and-measure protocols, and the GEAT and GREAT suffer from having to impose severe restrictions on the rate at which Alice can prepare and transmit the signal states.

Abstractly, entropy accumulation theorems obtain a lower bound on the \(n\)-round entropy (smooth-min and/or \Renyi) of a sequential process in terms of a single-round quantity of the form\footnote{The correction term here depends on the probability of the event $\Omega_\mathrm{acc}$. Technically, the sequential process outputs a string different from $\PAstring$, which goes through a discarding procedure, before resulting in the pre-amplification string  $\PAstring$. However, this discrepancy can be resolved as discussed earlier in \cref{remark:variablelengthinputPA}.}
\begin{equation}
    \Hmin[\epsbar](\PAstring | \CP_1^n \Eve )_{\rho|\OmegaAT} \geq n h_\mathrm{single-round} -\text{corrections},
\end{equation}
where \(\PAstring\) are the secret registers containing the pre-amplification string, and $\Eve$ denotes all of Eve's quantum side information registers. Here $n$ is the total number of rounds in the QKD protocol, and $h_\mathrm{single-round}$ is a value that can be computed in terms of a minimization problem only involving single rounds of the protocol. 

Thus, the security analysis within these proof techniques typically consists of justifying that the QKD protocol being implemented can be analyzed as a sequence of channels, and showing that these channels satisfy the conditions required to apply the relevant EAT theorem of interest. The task then reduces to the proper evaluation of the single-round quantity.

\subsection{Phase error correction}
This last proof technique is notable in that it does \emph{not} go through the approach of bounding entropic quantities and using the Leftover hashing Lemma. It will not be discussed in this thesis, except via its connection to the EUR method, that is described below.

The phase error correction based proof technique \cite{koashi_simple_2005,koashi_simple_2009}, sometimes referred to as a proof based on complementarity, is a proof technique based on the original Shor-Preskill \cite{shor_simple_2000} and the GLLP \cite{gottesman_proof_2003} proofs.
In this technique, the security statement is related to the probability of a particular (virtual) phase error correction protocol succeeding. This, in turn is related to an upper bound on the number of ``phase errors", which are define similarly as in the EUR approach. Thus, there is a immense degree of structural  similarities between the EUR-based security proof approaches and the phase error correction-based security proof approaches. We do not discuss this exhaustively here, and refer the reader Ref.~\cite[Section 6.2.2]{tupkary2025qkdsecurityproofsdecoystate} and Ref.~\cite{tsurumaru_leftover_2020} instead. However, we note that our results in \cref{chap:EUR} can be applied using phase error correction based proof technique as well.

\chapter{Variable-length Quantum Key Distribution} \label{chap:variable}

\epigraph{
Where we let QKD protocols decide their output key length; show that such
protocols are necessary for practical implementations; investigate the
various issues that arise when  we let the lengths of strings appearing in the protocol
vary; and make our lives easier by sticking to IID collective attacks.
}{}

Security proofs for QKD protocols are typically proven in the ``fixed-length'' scenario (see \cref{subsec:protocolvariations}), where Alice and Bob either produce a key of a fixed length, or abort the protocol  \cite{tomamichel_largely_2017,george_numerical_2021,bunandar_numerical_2020,renner_security_2005,rusca_finite-key_2018,lim_concise_2014,wiesemann_consolidated_2024}. Such protocols accept and produce a key of fixed length if and only if their observed statistics belong to some predetermined ``acceptance set".  Otherwise, the protocol aborts.  Such protocols have two main disadvantages.

First, in order to ensure that the protocol accepts with high probability for honest behaviour, the acceptance set $\acceptanceset$ (see \cref{subsec:protocolvariations}) needs to be chosen carefully. Typically, the acceptance set is chosen to be the set of statistics that are close to what is expected from honest behaviour \cite{george_numerical_2021,bunandar_numerical_2020,renner_security_2005}. This requires Alice and Bob to know the honest behaviour of the channel connecting Alice and Bob, \emph{before} a run of the QKD protocol. In many practical scenarios, such as ground-to-satellite QKD \cite{bourgoin_Comprehensive_2013,dequal_Feasibility_2021,liao_Satellitetoground_2017,trinh_Statistical_2022,sidhu_finite_2022}, it is difficult to know the behaviour of the channel in advance, since the behaviour depends on weather. In fact, this can be a problem even in fibre-based setups \cite{Wang_twinfield_2022,Clivati_coherent_2022,Dynes_stability_2012}.

Second, even if the honest behaviour is known, the \emph{size} of the acceptance set affects the length of the final key that can be produced. This reflects the fact that the key has to be secure for the \textit{worst-case} event that accepts. Larger acceptance sets have a higher probability of accepting on any given run of the QKD protocol, but lead to a shorter length of the final key, since they include worse accept events. We will see this shortly in \cref{sec:fixedlengthsecurity}. In particular, if users choose a large acceptance set, and then find that their observed statistics are much better than expected, they are \emph{not} allowed to produce a larger key.
Thus, there is a trade-off between fixed-length protocols that accept with high probability, and which produce a large key on accepting.

A variable-length QKD  protocol is one that allows users to adjust the length of the key generated based upon the observed statistics during the protocol \cite{ben-or_universal_2004,portmann_security_2022}.  This eliminates the trade-off described above.  It also does not require the expected behaviour of the channel to be known in advance, simplifying implementations.  In fact, variable-length protocols are almost invariably the ones implemented
in the laboratory. Nevertheless, it is only relatively recently that rigorous
security analyses for such protocols have been developed. The disconnect
between practical implementations and the theoretical literature is perhaps
best summarized by the following remark (produced with permission):

\epigraph{``Norbert, I am shocked. I've always thought that one can decide on the
  final key length as a function of the observed error rate in any given
  QKD run."}{Giles Brassard, of BB84 fame.}

In this chapter, we present a security proof for variable-length QKD protocols against IID collective attacks.  This proof can then be lifted to hold against coherent attacks using the postselection technique \cite{christandl_postselection_2009}, specifically \cite[Corollary 4.1]{nahar_postselection_2024}. This lift is discussed in \cref{chap:postselection}. 

We will begin by first looking at security proofs for fixed-length protocols in \cref{sec:fixedlengthsecurity}. In \cref{sec:variablelengthsecurity} we will modify this analysis, and obtain a security proof for variable-length protocols. In \cref{sec:constructingsetsestimators} we explain the application of these results to the \nameref{prot:qkdprotocol}. In \cref{sec:variableapplicationtoQubitbb84,sec:variableapplicationtoDecoybb84} we will apply our results to compute key rates for the qubit BB84 and decoy-state BB84 protocols, and compare the performance of fixed-length and variable-length protocols. We will see that the variable-length protocols offer significant advantages. Finally, in \cref{sec:variablelengthPA} we will present a subtle but important gap in the analysis of privacy amplification on variable-length input strings, and resolve it. This chapter is primarily based on Ref.~\cite{tupkary_security_2024}, along with certain elements from Ref.~\cite{Kamin2025}.

\subsubsection{Discussion on Variable-length proofs across proof techniques}
Variable-length security proofs have been obtained in several prior proof techniques. Ref.~\cite{hayashi_concise_2012} is an early work in the phase-error–correction framework that presents a security proof for qubit BB84, but it relies on several unsatisfactory assumptions (for instance, it assumes that Alice’s pre-amplification string is perfectly uniform). A proof without such assumptions within the phase error correction framework can be found in \cite[Chapter 3]{kawakami_security_nodate}.
In the EUR framework, a proof was given in \cite[Supplementary Note A]{curras-lorenzo_tight_2021}. We will present a  and slightly more general proof for the EUR method in \cref{chap:EUR}, essentially following the same steps as these prior works. For EAT-based methods, the MEAT framework \cite{arqand_marginal_2025} naturally accommodates variable-length protocols, and variable-length proofs can therefore be found in Refs.~\cite{kamin_renyi_2025,inprep_BDR3}. 

 \section{Proving Fixed-Length Security} \label{sec:fixedlengthsecurity}

Recall our description of the fixed-length variant of \nameref{prot:qkdprotocol}
from \cref{sec:protdescriptionqkdbackground,subsec:protocolvariations}. After
signal transmission and measurements, Alice and Bob compute the frequency
distribution $\Fobs = \operatorname{freq}(\cobs)$. If they find that
$\Fobs \in \acceptanceset$, the acceptance test passes, and we refer to this
event as $\OmegaAT$. Otherwise, they abort. Alice and Bob may also abort during
the error-verification steps, so the overall acceptance event is
$\OmegaAT \wedge \OmegaEV$. Furthermore, using source replacement, the
signal-preparation phase of \nameref{prot:qkdprotocol} can be viewed as Alice
performing measurements on the source-replaced state. The entire fixed-length
protocol is represented by $\{ \protMap{\lfixed}, \sigma_A \}$
(see \cref{def:equivalentQKDprotocol}), where we recall that $\sigma_{A}$ denotes the marginal on $A$, and $\protMap{\lfixed}$ is the QKD protocol map that implements measurements, public announcements, and classical postprocessing, and outputs the key registers $K_A,K_B$. 
Recalling the equivalent secrecy definition from \cref{def:epsSecPromiseIIDcollective}, we are required to show that
\begin{equation}
    \begin{aligned}
        &\tracedist{
            \left(
                \left( \Tr_{K_B} \circ \protMap{\lfixed}
                - \Tr_{K_B} \circ \protMapId{\lfixed} \right)
                \otimes \idmap_{E_1^n}
            \right)
            \left[ \rho_{ABE}^{\otimes n} \right]
        }
        \leq \epssecret, \\
        &\forall\, \rho_{ABE} \in \dop{=}(ABE)
        \text{ such that }
        \Tr_{BE}\!\left( \rho_{ABE} \right)
        = \sigma_A.
    \end{aligned}
\end{equation}

For our analysis, we require the following structural property of the QKD maps
$\protMap{\lkey}$.

\begin{lemma}[Evolution of states in $\{ \protMap{\lkey}, \sigma_A \}$]
    The evolution of states in the protocol $\{ \protMap{\lkey}, \sigma_A \}$
    can be written as
    \[
        \protMap{\lkey}
        = \QKDpostprocessingmap \circ \left( \QKDGmap{}^{\otimes n} \right),
    \]
    where $\QKDGmap{} \in \CPTP(AB, S Y \CP)$ performs the measurements and
    classical announcements in each round, and
    $\QKDpostprocessingmap \in \CPTP(S_1^n Y_1^n \CP_1^n, K_A K_B \allpublic)$
    performs the subsequent post-processing.
\end{lemma}
\begin{proof}
    Follows from tje \nameref{prot:qkdprotocol}, and the source-replacement scheme from \cref{subsec:sourcereplacement}.
\end{proof}

A fixed-length security proof \cite{renner_security_2005,george_numerical_2021} proceeds by dividing all possible input states
into two categories (see also \cite[Section~4]{tupkary2025qkdsecurityproofsdecoystate}: those that cause the protocol to abort with high
probability, and those that do not. States that abort with high probability
require no further analysis. For the remaining states, one proves that the
required entropic bounds hold.

In particular, we first construct the \emph{feasible set}
$\feasibleset \subseteq \dop{=}(AB)$ with the following properties:

\begin{itemize}
    \item If the initial state $\rho_{AB}$ (with Eve holding the purifying system $E$) is not in the feasible set, then the probability of obtaining observations lying in the acceptance set (i.e, $\Fobs \in \acceptanceset$) is at most $\epsAT$:
    \begin{equation}
        \rho \notin \feasibleset
        \;\implies\;
 \Pr(\OmegaAT) = \Pr(\Fobs \in \acceptanceset)_\rho \leq \epsAT.
    \end{equation}
For all such states, we will show that security holds because the protocol will abort with high probability.
    \item For all remaining states, we compute a lower bound on the relevant entropic quantity, which then determines the output length $\lfixed$ to be
    \begin{equation} \label{eq:lfixedvalue}
        \begin{aligned}
            \lfixed &\leq \max \Bigg\{ 0, \\
            &\floor{
                n \inf_{\nu \in \Sigma (\feasibleset) }
                \Halpha(S | \CP E)_{\nu}
                - \leakfixed
                - \EVcost 
                - \frac{\alpha}{\alpha - 1}
                 \log\left( \frac{1}{ \epsPA} \right) 
                  + 2 } \Bigg\},\\
    \Sigma(\feasibleset) &\coloneq \left\{\; \QKDGmap{}[\omega_{ABE}] \; \mid  \omega_A = \sigma_A, \omega_{AB} \in \feasibleset \right\},
        \end{aligned}
    \end{equation}
    where $E$ is purifying register. 
The set $\Sigma(\feasibleset)$ corresponds to the set of possible states that can be obtained after signal transmission, measurements, public announcements, and sifting, for a single round of the QKD protocol, when the the starting state $\omega_{AB}$ belongs to the feasible set. 
\end{itemize}

We will explicitly construct this feasible set for a QKD protocol in \cref{sec:variableapplicationtoQubitbb84}, using simple concentration inequalities. We can now write down the security statement for fixed-length protocols \cite{george_numerical_2021,renner_security_2005}.

\begin{theorem}[Fixed-length security statement of $\{ \protMap{\lfixed}, \sigma_A \}$ for IID collective attacks] \label{theorem:fixedlengthIID}
Consider the (fixed-length) \nameref{prot:qkdprotocol} given by $\{ \protMap{\lfixed}, \sigma_A \}$, where $\lfixed$ is given by \cref{eq:lfixedvalue}. Then, the protocol is $\max\{\epsPA,\epsAT\}$-secret against IID collective attacks, $(\max\{\epsPA,\epsAT\}+\epsEV)$-secure against IID collective attacks (see \cref{def:epsSecPromiseIIDcollective}).
\end{theorem}

\begin{proof}
We have already shown that the protocol $\epsEV$-correct in \cref{lemma:correctnessissatisfied}, and that correctness and secrecy imply security in \cref{lemma:securityfromcorrandsecrecy}. Thus, we only need to show that secrecy holds against IID collective attacks, i.e, \cref{def:epsSecPromiseIIDcollective} is satisfied. We proceed by a case analysis on the state $\rho_{AB}$, which is the state shared by Alice and Bob after Eve's attack. Eve can without loss of generality be assumed to hold some purification of this state, denoted by $\rho_{ABE}$.\footnote{If Eve wishes to hold an extension rather than a purification, we may
always consider a scenario in which she instead holds a purification of that
extension and simply ignores the additional purifying system. This can also be
formalized using suitable data-processing inequalities for the relevant entropies, since there always exists a quantum channel mapping the purifying
system to the extension system (\cref{lemma:pur_to_ext}).} Thus, the QKD protocol starts with the state $\rho^{\otimes n}_{ABE}$, which is run through the channels $\QKDGmap{}^{\otimes n}$ and $\QKDpostprocessingmap$, to produce the final output state.

Note that it suffices to consider only those states that satisfy the marginal constraint, namely $\rho_A = \sigma_A$.

\paragraph*{States not in the feasible set:} For all states $\rho_{AB} \notin \feasibleset$, the secrecy requirement is satisfied, since the protocol aborts with high probability. We have
\begin{equation}
    \begin{aligned}
  \Delta &= \tracedist{
            \left(
                \left( \Tr_{K_B} \circ \protMap{\lfixed}
                - \Tr_{K_B} \circ \protMapId{\lfixed} \right)
                \otimes \idmap_{E_1^n}
            \right)
            \left[ \rho_{ABE}^{\otimes n} \right]
        }
       \\
&= \Pr(\OmegaAT \wedge \OmegaEV) \tracedist{\rho^{(\lfixed)}_{K_A  \allpublic E_1^n| \OmegaAT \wedge \OmegaEV} - \rho^{(\lfixed),\mathrm{ideal}}_{K_A \allpublic E_1^n| \OmegaAT \wedge \OmegaEV}} \\
&\leq \Pr(\OmegaAT \wedge \OmegaEV)  \\
&\leq \Pr(\OmegaAT )  \\
&\leq \epsAT
\end{aligned}
        \end{equation}
where the first two lines follow simply by rewriting the expression, the third line follows from the fact that the trace distance is upper bounded by $1$, the fourth line follows from the properties of probability, and the final line follows from the fact that $\rho_{AB} \notin \feasibleset$.

\paragraph{States in the feasible set:} For all states $\rho_{AB} \in \feasibleset$, we use the Leftover Hashing Lemma, and various manipulations of the entropic quantities, to show that the value of $\lfixed$ specified in \cref{eq:lfixedvalue} ensures secrecy. This can be seen by the following chain of inequalities:

\begin{align}
\label{eq:iidLHLfixed:a}
\Delta
&= \tracedist{
  \left(
    \left(
      \Tr_{K_B} \circ \protMap{\lfixed}
      - \Tr_{K_B} \circ \protMapId{\lfixed}
    \right)
    \otimes \idmap_{E_1^n}
  \right)
  \left[ \rho_{ABE}^{\otimes n} \right]
} \\[0.5em]
\label{eq:iidLHLfixed:b}
&= \Pr(\OmegaAT \wedge \OmegaEV)\,
\tracedist{
  \rho^{(\lfixed)}_{K_A \allpublic E_1^n \mid \OmegaAT \wedge \OmegaEV}
  - \rho^{(\lfixed),\mathrm{ideal}}_{K_A \allpublic E_1^n \mid \OmegaAT \wedge \OmegaEV}
} \\[0.5em]
\label{eq:iidLHLfixed:c}
&\le
\Pr(\OmegaAT \wedge \OmegaEV)\,
2^{
  \frac{1-\alpha}{\alpha}
  \left(
    \Halpha(S_1^n \mid \CP_1^n \CEC \CEV \HEV E_1^n)_{\rho \mid \OmegaAT \wedge \OmegaEV}
    - \lfixed + 2
  \right)
} \\[0.5em]
\label{eq:iidLHLfixed:d}
&\le
\Pr(\OmegaAT \wedge \OmegaEV)\,
2^{
  \frac{1-\alpha}{\alpha}
  \left(
    \Halpha(S_1^n \mid \CP_1^n \HEV E_1^n)_{\rho \mid \OmegaAT \wedge \OmegaEV}
    - \leakfixed - \EVcost - \lfixed + 2
  \right)
} \\[0.5em]
\label{eq:iidLHLfixed:e}
&\le
2^{
  \frac{1-\alpha}{\alpha}
  \left(
    \Halpha(S_1^n \mid \CP_1^n \HEV E_1^n)_{\rho}
    - \leakfixed - \EVcost - \lfixed + 2
  \right)
} \\[0.5em]
\label{eq:iidLHLfixed:f}
&=
2^{
  \frac{1-\alpha}{\alpha}
  \left(
    \Halpha(S_1^n \mid \CP_1^n E_1^n)_{\rho}
    - \leakfixed - \EVcost - \lfixed + 2
  \right)
} \\[0.5em]
\label{eq:iidLHLfixed:g}
&=
2^{
  \frac{1-\alpha}{\alpha}
  \left(
    n \Halpha(S \mid \CP E)_{\QKDGmap{}[\rho_{ABE}]}
    - \leakfixed - \EVcost - \lfixed + 2
  \right)
} \\[0.5em]
\label{eq:iidLHLfixed:h}
&\le \epsPA .
\end{align}
Here $\Halpha$ denotes the Rényi entropy (see \cref{def:renyientropy}), with
$\alpha$ the Rényi parameter. In \cref{eq:iidLHLfixed:c}, we apply the Leftover
Hashing Lemma for Rényi entropy\footnote{Note that in this proof, we pretend as though the $S_1^n$ is the pre-amplification string, instead of $\PAstring$: the latter is obtained from the former via sifting. This discrepancy is address in \cref{sec:variablelengthPA} (see also \cref{remark:variablelengthinputPA}).} \cite[Theorem~8]{dupuis_privacy_2023} (restated in
\cref{lemma:LHL}). In \cref{eq:iidLHLfixed:d}, we split off the contributions from
the error-correction and error-verification registers using
\cref{lemma:EC_cost}.\footnote{Technically, Bob also sends one additional bit
indicating whether the hash values match. However, since the state is conditioned
on $\OmegaEV$, this bit takes a deterministic value and can therefore be removed
at no cost. This is argued formally in \cref{chap:MEAT}.} In \cref{eq:iidLHLfixed:e}, we remove the conditioning on the events
$\OmegaAT \wedge \OmegaEV$ using \cref{lemma:contrenyi}. In
\cref{eq:iidLHLfixed:f}, we remove the $\HEV$ register without penalty, since it
is completely uncorrelated with the remaining registers; this follows by
applying data processing inequalities (\cref{lemma:DPI}) in both directions. In
\cref{eq:iidLHLfixed:g}, we use the additivity of Rényi entropy for IID states,
together with the fact that
\[
\rho_{S_1^n \CP_1^n E_1^n}
= \QKDGmap{}^{\otimes n}[\rho_{ABE}^{\otimes n}] .
\]
Finally, \cref{eq:iidLHLfixed:h} follows from the definition of $\lfixed$ in
\cref{eq:lfixedvalue}: since $\lfixed$ is obtained by minimizing over all possible states $\rho_{ABE}$ where $\rho_{AB} \in \feasibleset$.

Thus $\max\{\epsAT,\epsPA\}$-secrecy holds. This concludes our proof. 
\end{proof}
A similar proof can be obtained using smoothed-min entropies, although in that case one has to consider states that are partial on $\OmegaAT \wedge \OmegaEV$ instead of states conditioned on that event, as we do here. However, the {\Renyi} version typically gives tighter key rates. Moreover, we require the {\Renyi} version for handling the variable-length case, as we point out in \cref{remark:technicalreason} later.

\begin{remark} \label{remark:conditioning}

 We strongly emphasize that at no point in the above argument did we make any assumption about the probability that Eve performs a particular attack. We do not know these probabilities, and Eve is free to behave adversarially in any manner she chooses. The statements we make are always of the form: “\emph{if} Eve performs this attack, then the protocol aborts,” or “\emph{if} Eve performs some other attack, then the chosen hash length ensures secrecy.” Thus, security holds \emph{regardless} of Eve’s behaviour, and without requiring any prior probability distribution over Eve’s attacks.
\end{remark}

\section{Proving Variable-length Security } \label{sec:variablelengthsecurity}
When we consider the variable-length  security requirement (\cref{eq:secdefrealidealsymmetric}), we encounter a proliferation of events corresponding to various output key lengths and, consequently, many more terms in the final trace-distance quantity that we seek to bound. A naive extension of the fixed-length analysis would look something like this.  We could first enumerate all possible acceptance sets, each of which corresponds to producing a key of a different length. Then, for each acceptance set, we could define a corresponding feasible set and  repeat the steps of the fixed-length proof. Such an approach would yield a correct proof, but it suffers from a significant drawback: the final security parameter becomes the sum of the individual $\varepsilon$-terms obtained across the different cases. If the number of possible hash lengths is large (for instance, on the order of $n$), then the overall secrecy parameter degrades by a multiplicative factor of $n$. This is clearly undesirable.\footnote{Under certain conditions—specifically, when the acceptance sets form a “nested” sequence of increasing size—this penalty can be avoided; see Ref.~\cite[Section~III]{tupkary_security_2024} for such an approach. However, this nesting condition is quite restrictive, as it does not allow the protocol to be designed to handle unpredictable channel conditions. In this thesis, we directly obtain the more general setting, and therefore do not discuss the specialized result here.}

We will now prove a general result that avoids this problem. We consider the \nameref{prot:qkdprotocol}, which directly uses the observed frequency of outcomes in $\CP_1^n$ (denoted via $\Fobs$), to determine the length of the secret key to be produced and the number of bits to be used for error-correction.  Crucially, our method involves the construction of a statistical estimator $\bstat(\Fobs)$, that with high probability is a lower bound on the {\Renyi} entropy $ \Halpha(S_1^n | \CP_1^n E_1^n )_{\rho}$ of the state $\rho_{S_1^n \CP_1^n E_1^n}$ obtained in the QKD protocol. That is, we will require an estimator $\bstat$ such that for any state $\rho_{ABE}$, it is the case that
	\begin{equation} \label{eq:bstatrequirement}
		\Pr_{\Fobs} \left( \bstat(\Fobs ) \leq \Halpha(S_1^n | \CP_1^n E_1^n )_{\rho} \right) \geq 1-\epsAT.
	\end{equation}

    We will construct this estimator for a QKD protocol later in \cref{sec:variableapplicationtoQubitbb84}, using simple concentration inequalities. A lemma that helps in this construction is obtained below.

    \begin{remark} \label{remark:fixedunknownrho}
       Note that in a QKD protocol, we deal with a \textit{fixed yet unknown} $\rho_{ABE}$, which is determined by Eve's attack, and which is \emph{not} a random variable. This $\rho_{ABE}$ then gives rise to a random variable $\Fobs$. Given that Alice and Bob observe $\Fobs$, obtained by performing measurements on $\rho_{ABE}^{\otimes n}$, we would like to construct the function $\bstat(\Fobs)$ which acts as an estimator of {\Renyi} entropy. To do so, we will need to construct a set $\variableset(\Fobs)$ with the required properties, as described in the following Lemma.
\end{remark}
\begin{lemma}\label{lemma:settobstat}
For any state $\rho_{ABE}$, let $\Fobs \in   \mathbb{P}(\mathbb{\CP})$ denote the frequency distribution of outcomes in the register $\CP$, obtained by measuring the state $\rho_{ABE}^{\otimes n}$ in the \nameref{prot:qkdprotocol}. That is, we have 
\begin{equation}
    \rho_{S_1^n Y_1^n \CP_1^n E_1^n} = \rho^{\otimes n}_{S Y \CP E}
    = \QKDGmap{}^{\otimes n}[\rho_{ABE}^{\otimes n}],
\end{equation}
and $\Fobs$ is the frequency distribution of outcomes in $\rho_{\CP_1^n}$. Suppose that we have a ``confidence set" $V(\Fobs)$\footnote{Technically, $\variableset$ is a function that takes a frequency distribution as input and outputs a set of states.} such that, for all states $\rho_{AB}$, it holds that\footnote{It is crucial to note that this lemma is making statements about a \emph{fixed but unknown} state $\rho_{AB}$, which gives rise to random variables $\Fobs$ and $\variableset(\Fobs)$. }
\begin{equation} \label{eq:variablesetproperty}
    \Pr_{\Fobs}(\rho_{AB} \in \variableset(\Fobs)) \geq 1 - \epsAT.
\end{equation}
Then any $\bstat(\Fobs)$ satisfying
\begin{equation}
\begin{aligned}
    \bstat(\Fobs)
   &\leq  n \inf_{\nu \in  \Sigma(\Fobs) }
        \Halpha(S|  \CP E)_{\nu}, \\
\Sigma(\Fobs) &\coloneq \left\{ \QKDGmap{}[\omega_{ABE} ] \; \mid \omega_{A} = \sigma_A \;, \omega_{AB} \in \variableset(\Fobs) \;  \right\},        
\end{aligned}
\end{equation}
where $E$ denotes purifying register,  satisfies the required property from \cref{eq:bstatrequirement}.
\end{lemma}
\begin{proof}
From the additivity of {\Renyi} entropy across tensor products (see \cref{lemma:additivity}), and the fact that $\rho_{S_1^n \CP_1^n E_1^n}$ is an IID state, we have that $\Halpha(S_1^n | \CP_1^n E_1^n)_{\QKDGmap{}^{\otimes n}[\rho^{\otimes n}_{ABE}]} = n\Halpha(S | \CP E)_{\QKDGmap{}[\rho_{ABE}]}  $. Thus the required statement follows from:
\begin{equation}
\begin{aligned}
    \Pr(\bstat(\Fobs) \leq \Halpha(S_1^n | \CP_1^n E_1^n)_{\rho} ) &=    \Pr(\bstat(\Fobs) \leq   n \Halpha(S | \CP E)_{\QKDGmap{}[\rho_{ABE}]} ) \\
    &\geq \Pr(\rho_{AB} \in \variableset(\Fobs)) \\
    &\geq 1-\epsAT 
    \end{aligned}
\end{equation}
where the probability is taken over $\Fobs$, where we used the fact that
\begin{equation}
\rho_{AB} \in \variableset(\Fobs) \implies \bstat(\Fobs) \leq   n \Halpha(S | \CP E)_{\QKDGmap{}[\rho_{ABE}]}
\end{equation}
for the second line, and the property of $\variableset$ (\cref{eq:variablesetproperty}) for the final line. 
\end{proof}
Thus, in order to construct the required $\bstat(\Fobs)$, we only need to construct $\variableset(\Fobs)$, which can be done using standard concentration inequalities. This then determines the length of the output key as follows:
\begin{itemize}
    \item The function $\leak(\Fobs) : \mathbb{P}(\mathcal{\CP}) \rightarrow \mathbb{N}$, which determines the number of possible transcripts of the error-correction protocol that is used, can be chosen arbitrarily. 
    \item Then, the output key length $\lkey(\Fobs)$ is given by
    \begin{equation} \label{eq:lvariablevalue}
        \lkey(\Fobs) \coloneq \max\left\{ \floor{ \bstat(\Fobs)  - \leak(\Fobs) - \theta(\epsEV,\epsPA)},0\right\}
    \end{equation}
    where for brevity, we use 
    \begin{equation}
    \theta(\epsEV,\epsPA) \coloneq \EVcost  + \frac{\alpha}{\alpha-1} \log(\frac{1}{\epsPA}) -2.
    \end{equation}
    \end{itemize}
We can now write down the security statement for variable-length protocols. 

\begin{theorem}[Variable-length security statement of $\{ \protMap{\lkey}, \sigma_A \}$ for IID collective attacks] \label{theorem:variablelengthIID}
Consider the variable-length \nameref{prot:qkdprotocol} given by $\{ \protMap{\lkey}, \sigma_A \}$, where $\lkey(\Fobs)$ is given by \cref{eq:lvariablevalue}, and $\bstat(\Fobs)$ satisfies \cref{eq:bstatrequirement}. Then, the protocol is $(\epsPA+\epsAT)$-secret against IID collective attacks, and  $(\epsPA+\epsAT+\epsEV)$-secure against IID collective attacks (see \cref{def:epsSecPromiseIIDcollective}).
\end{theorem}
\begin{proof}
Again, as in the proof of \cref{theorem:fixedlengthIID}, we only need to show that secrecy holds against IID collective attacks. Let us fix the attack Eve performs, i.e, we fix the state $\rho_{AB}$ shared between Alice and Bob,  and assume that Eve holds some purification $\rho_{ABE}$. As this protocol proceeds, we obtain the state $\rho_{S_1^n \CP_1^n \CEC \CEV \HEV E_1^n }$, where $$\rho_{S_1^n Y_1^n \CP_1^n E_1^n} = \rho^{\otimes n}_{S Y \CP E} = \QKDGmap{}^{\otimes n} [\rho_{ABE}^{\otimes n}].$$

A critical step in our proof is the division of all possible observations into two sets: one that corresponds to observations leading to a key length that is sufficiently small to guarantee secrecy (denoted by $\mathcal{T}_{\leq}$), and another that does not (denoted by $\mathcal{T}_{>}$). Security for observations in the former set follows essentially from the fact that the key length is small enough. Security for observations in the latter set follows from the fact that such cases occur only with small probability, since the key length is chosen as a suitable function of $\bstat(\Fobs)$, which itself acts as a statistical estimator of the {\Renyi} entropy.

We define the sets $\mathcal{T}_{\leq}$ as
\begin{align} \label{eq:lowersetdefined}
\mathcal{T}_{\leq} &\coloneq \left\{ \Fobs |  \lkey(\Fobs) + \leak(\Fobs) + \theta(\epsEV,\epsPA) \leq \Halpha(S_1^n | \CP_1^n E_1^n)_\rho \; \wedge \; \lkey(\Fobs) > 0 \right\}, \\
\mathcal{T}_{>} &\coloneq \left\{ \Fobs  |   \lkey(\Fobs) + \leak(\Fobs) +  \theta(\epsEV,\epsPA) > \Halpha(S_1^n | \CP_1^n E_1^n)_\rho  \; \wedge \; \lkey(\Fobs) > 0   \right\}.
\end{align}

The analysis now proceeds by writing out the sum over all possible observations $\Fobs$, and then grouping them according to the sums above.\footnote{Note that the main idea of the proof here is identical to that of published work \cite{tupkary_security_2024}; however, the presentation here is cleaner and simpler. } Let $\Omega(\Fobs)$ be the event that $\CP_1^n$ holds a value $\cP_1^n$ such that $\freq(\cP_1^n) = \Fobs$.

Let us consider the secrecy definition for variable-length protocols (\cref{def:epsSecPromiseIIDcollective}). The definition groups together terms with the same output length of the key, and the different events $\Omega(\Fobs)$ may correspond to the same output length of the key. Nevertheless,  the events $\Omega(\Fobs)$ are deterministic functions of public announcements $\CP_1^n$. Thus, the states $\rho_{K_A  \allpublic E_1^n |\Omega(\Fobs) \wedge \OmegaEV  }$  have orthogonal supports. Therefore, we have

\begin{equation}
    \begin{aligned}
  \Delta &= \tracedist{
            \left(
                \left( \Tr_{K_B} \circ \protMap{\lfixed}
                - \Tr_{K_B} \circ \protMapId{\lfixed} \right)
                \otimes \idmap_{E_1^n}
            \right)
            \left[ \rho_{ABE}^{\otimes n} \right]
        }
       \\
&\leq \sum_{ \substack{\Fobs \\ \lkey(\Fobs) > 0 } }  \Pr(\Omega(\Fobs) \wedge \OmegaEV ) \times \\
& \tracedist{ \rho_{K_A  \allpublic E_1^n |\Omega(\Fobs) \wedge \OmegaEV  } - \rho^{\mathrm{ideal}}_{K_A  \allpublic E_1^n | \Omega(\Fobs) \wedge \OmegaEV } }
\end{aligned}
        \end{equation}
where we can omit events which lead to aborts, which do not contribute to the trace distance. We can now split the sum over $\Fobs$ into two parts, one corresponding to $\Fobs \in \mathcal{T}_{>}$, and other correspoding to $\Fobs \in \mathcal{T}_{\leq}$. The first part of the above expression can be bounded as

\begin{align}
\label{eq:DeltaGreater:a}
\Delta_{>}
&\coloneq \sum_{\Fobs \in \mathcal{T}_{>} }
\Pr\!\bigl(\Omega(\Fobs) \wedge \OmegaEV\bigr) \times \nonumber \\ &
\tracedist{
  \rho_{K_A \allpublic E_1^n \mid \Omega(\Fobs) \wedge \OmegaEV}
  -
  \rho^{\mathrm{ideal}}_{K_A \allpublic E_1^n \mid \Omega(\Fobs) \wedge \OmegaEV}
} \\[0.5em]
\label{eq:DeltaGreater:b}
&\le
\sum_{\Fobs \in \mathcal{T}_{>} }
\Pr\!\bigl(\Omega(\Fobs) \wedge \OmegaEV\bigr) \\[0.5em]
\label{eq:DeltaGreater:d}
&\le
\sum_{\Fobs \in \mathcal{T}_{>}}
\Pr\!\bigl(\Omega(\Fobs)\bigr) \\[0.5em]
\label{eq:DeltaGreater:e}
&=
\Pr_{\Fobs}\!\left(
  \lkey(\Fobs)
  + \leak(\Fobs)
  + \theta(\epsEV,\epsPA)
  >
  \Halpha(S_1^n \mid \CP_1^n E_1^n) \; \wedge \; \lkey(\Fobs) > 0
\right) \\[0.5em]
\label{eq:DeltaGreater:f}
&\le
\Pr_{\Fobs}\!\left(
  \bstat(\Fobs)
  >
  \Halpha(S_1^n \mid \CP_1^n E_1^n)_\rho
\right) \\[0.5em]
\label{eq:DeltaGreater:g}
&\le \epsAT .
\end{align}

In \cref{eq:DeltaGreater:b}, we use the fact that the trace distance is upper
bounded by $1$.  In
\cref{eq:DeltaGreater:d}, we use basic properties of probability. \cref{eq:DeltaGreater:e} is a reformulation of
\cref{eq:DeltaGreater:d}, expressing the probability in terms of the defining
property of the set $\mathcal{T}_{>}$. The inequality in
\cref{eq:DeltaGreater:f} follows from the fact that the statements inside the probability in \cref{eq:DeltaGreater:e} imply $
\bstat(\Fobs)
\ge
 \Halpha(S_1^n \mid \CP_1^n E_1^n)_\rho$ (using \cref{eq:lvariablevalue}). Finally,
\cref{eq:DeltaGreater:g} follows from the defining properties of the estimator
$\bstat$, which guarantees that the probability of overestimating the Rényi
entropy is bounded by $\epsAT$.

The remaining part of $\Delta$ can be bounded via the Leftover Hashing Lemmas, via

\begin{align}
\label{eq:DeltaLeq:a}
\Delta_{\leq}
&= \sum_{\Fobs \in \mathcal{T}_{\leq} }
\Pr\!\bigl(\Omega(\Fobs) \wedge \OmegaEV\bigr) \times \nonumber\\
&\qquad
\tracedist{
  \rho_{K_A \allpublic E_1^n \mid \Omega(\Fobs) \wedge \OmegaEV}
  -
  \rho^{\mathrm{ideal}}_{K_A \allpublic E_1^n \mid \Omega(\Fobs) \wedge \OmegaEV}
} \\[0.5em]
\label{eq:DeltaLeq:b}
&\le
\sum_{\Fobs \in \mathcal{T}_{\leq} }
\Pr\!\bigl(\Omega(\Fobs) \wedge \OmegaEV\bigr)\,
2^{\frac{1-\alpha}{\alpha}
\left(
\Halpha(S_1^n \mid \CP_1^n \CEC \CEV \HEV E_1^n)_{\rho \mid \Omega(\Fobs) \wedge \OmegaEV}
- \lkey(\Fobs) + 2
\right)
} \\[0.5em]
\label{eq:DeltaLeq:c}
&\le
\sum_{\Fobs \in \mathcal{T}_{\leq} }
\Pr\!\bigl(\Omega(\Fobs) \wedge \OmegaEV\bigr)\,
2^{\frac{1-\alpha}{\alpha}
\left(
\Halpha(S_1^n \mid \CP_1^n \HEV E_1^n)_{\rho \mid \Omega(\Fobs) \wedge \OmegaEV}
- \leak(\Fobs) - \EVcost - \lkey(\Fobs) + 2
\right)
} \\[0.5em]
\label{eq:DeltaLeq:d}
&\le
\sum_{\Fobs \in \mathcal{T}_{\leq} }
\Pr\!\bigl(\Omega(\Fobs)\bigr)\,
2^{\frac{1-\alpha}{\alpha}
\left(
\Halpha(S_1^n \mid \CP_1^n \HEV E_1^n)_{\rho \mid \Omega(\Fobs)}
- \leak(\Fobs) - \EVcost - \lkey(\Fobs) + 2
\right)
} \\[0.5em]
\label{eq:DeltaLeq:e}
&=
\sum_{\Fobs \in \mathcal{T}_{\leq} }
\Pr\!\bigl(\Omega(\Fobs)\bigr)\,
2^{\frac{1-\alpha}{\alpha}
\left(
\Halpha(S_1^n \mid \CP_1^n E_1^n)_{\rho \mid \Omega(\Fobs)}
- \leak(\Fobs) - \EVcost - \lkey(\Fobs) + 2
\right)
} \\[0.5em]
\label{eq:DeltaLeq:f}
&\le
\sum_{\Fobs \in \mathcal{T}_{\leq} }
\Pr\!\bigl(\Omega(\Fobs)\bigr)\,
2^{\frac{1-\alpha}{\alpha}
\left(
\Halpha(S_1^n \mid \CP_1^n E_1^n)_{\rho \mid \Omega(\Fobs)}
-
\Halpha(S_1^n \mid \CP_1^n E_1^n)_{\rho}
+ \theta(\epsEV,\epsPA)
- \EVcost + 2
\right)
} \\[0.5em]
\label{eq:DeltaLeq:g}
&\le
2^{\frac{1-\alpha}{\alpha}
\left(
\Halpha(S_1^n \mid \CP_1^n E_1^n)_{\rho}
-
\Halpha(S_1^n \mid \CP_1^n E_1^n)_{\rho}
+ \theta(\epsEV,\epsPA)
- \EVcost + 2
\right)
} \\[0.5em]
\label{eq:DeltaLeq:h}
&\le \epsPA .
\end{align}
In \cref{eq:DeltaLeq:b}, we apply the Leftover Hashing Lemma
(\cref{lemma:LHL}).\footnote{Note that in this proof, we pretend as though the $S_1^n$ is the pre-amplification string, instead of $\PAstring$: the latter is obtained from the former via sifting. This discrepancy is address in \cref{sec:variablelengthPA} (see also \cref{remark:variablelengthinputPA}).} In \cref{eq:DeltaLeq:c}, we split off the
error correction and error verification registers using
\cref{lemma:EC_cost}.\footnote{As before, since we condition on $\OmegaEV$,
the single bit sent by Bob indicating whether the hash values match takes a
deterministic value and can therefore be ignored.}
In \cref{eq:DeltaLeq:d}, we remove the conditioning on the event $\OmegaEV$
using \cref{lemma:conditioning}. In \cref{eq:DeltaLeq:e}, we remove the
$\HEV$ register without penalty, since it is uncorrelated with the remaining
registers; this follows formally from data processing
(\cref{lemma:DPI}). In \cref{eq:DeltaLeq:f}, we use the defining property of the set
$\mathcal{T}_{\leq}$, namely that
\[
\lkey(\Fobs) + \leak(\Fobs) + \theta(\epsEV,\epsPA)
\le
\Halpha(S_1^n \mid \CP_1^n E_1^n)_\rho,
\]
to replace the $\Fobs$-dependent terms by a global bound. This step is crucial,
as it reduces the number of terms that depend on $\Fobs$. After this replacement, the expression appears to be a weighted average of {\Renyi} entropies, and one can use 
\cref{lemma:renyiweightedaverage} to combine all the terms to obtain \cref{eq:DeltaLeq:g}.
Finally, \cref{eq:DeltaLeq:h} follows from straightforward algebra.

The required statement then follows from
\begin{equation}
    \Delta = \Delta_{>}  + \Delta_{\leq} \leq \epsAT + \epsPA.
\end{equation}
This concludes the proof.
\end{proof}

\begin{remark} \label{remark:technicalreason}
		We highlight two critical steps in the proof of \cref{theorem:variablelengthIID}. The first ingredient is the appropriate construction of sets $\mathcal{T}_{\leq},\mathcal{T}_{>}$ and the relationship between these sets, $\bstat(\Fobs)$ and $\lkey(\Fobs)$. These sets depend on Eve's attack, and our argument allows us to construct them appropriately for every possible attack, and then to bound the two contributions to $\Delta$ that arise from it. 
        The second key step is the use of \cref{lemma:renyiweightedaverage}, which allows us to get rid of terms involving {\Renyi} entropies of states conditioned on events. In particular, smooth min-entropy does not straightforwardly allow a statement analogous to \cref{lemma:renyiweightedaverage}: thus, the above proof requires the use of {\Renyi} entropy.
	\end{remark}
Having proved both fixed-length and variable-length security, we will now turn to the construction of the required sets needed for key rate calculations.

\section{Constructing feasible sets confidence sets}
\label{sec:constructingsetsestimators}
In this section, we explain the construction of the sets required for using the results in this chapter ($\feasibleset$ and
$\variableset(\Fobs)$). Recall that the 
\nameref{prot:qkdprotocol} we consider produces a public announcement in each round,
recorded in the register $\CP$, which has alphabet $\mathcal{\CP}$. We partition
this alphabet of possible announcements into two disjoint subsets: one used for
key generation, denoted $\mathcal{\CP}_{\mathrm{gen}}$, and one used for testing,
denoted $\mathcal{\CP}_{\mathrm{test}}$\footnote{These labels reflect the
structure of the concrete protocols considered later, where Alice designates
each round as a $\test$ or  $\gen$ round with probabilities $\gamma$ and
$1-\gamma$, respectively. For the purpose of the security analysis, however,
this partition is simply a generic division of the announcement alphabet.}.
Thus,
\[
\mathcal{\CP} = \mathcal{\CP}_{\mathrm{test}} \cup \mathcal{\CP}_{\mathrm{gen}} .
\]

Roughly speaking, the announcements corresponding to test rounds are used to
infer properties of the underlying quantum state, whereas the announcements
corresponding to key-generation rounds are used to extract secret key bits.

Let $\{\Gamma^{(AB)}_i\}_{i \in \mathcal{\CP}}$ denote the POVM associated with the
public announcements, where each POVM element is indexed by the corresponding
announcement. In each round of the protocol, this POVM is applied and a single
outcome is obtained. Since we assume IID collective attacks, the outcomes across
rounds are independent and identically distributed. In particular, for any fixed
announcement $i \in \mathcal{\CP}$, the number of times outcome $i$ is observed
over $n$ rounds follows a binomial distribution, with an unknown success
probability determined by the underlying state.

This observation was first made and utilized in Ref.~\cite{Kamin2025}, which motivates the use of statistical concentration results for
binomial random variables, which we will employ below to construct confidence
regions for the relevant parameters. Note that our analysis here very closely mirrors the one from Ref.~\cite{Kamin2025}, but is not identical to it. In particular, Ref.~\cite{Kamin2025} obtains a slightly tighter result for fixed-length scenarios.

\subsection{Binomial tail probabilities and confidence intervals} \label{subsec:binomialexplained}

Let $X$ be a binomial random variable with parameters $(n,p)$, that is, $
X \sim \mathrm{Binomial}(n,p)$, where $p \in [0,1]$ is an unknown success probability. Suppose that $x$ successes
are observed in $n$ independent trials, and define the empirical frequency $f_{\mathrm{obs}} := \frac{x}{n}$. 
For fixed integers $n$ and $x$, the binomial tail probabilities
\[
\Pr(X \le x) =\sum_{k=0}^{x} \binom{n}{k} p^{k} (1-p)^{n-k}
\quad\text{and}\quad
\Pr(X \ge x) = \sum_{k=x}^{N} \binom{n}{k} p^{k} (1-p)^{n-k}
\]
are monotone functions of the parameter $p$. Consequently, for any
$\varepsilon \in (0,1)$, there exist unique values of $p$ for which these tail
probabilities equal $\varepsilon$, which can be expressed in terms of quantiles of the beta
distribution. Specifically, the equation
\begin{equation}
\Pr_{X \sim \mathrm{Binomial}(n,p)}(X \ge x) = \varepsilon
\label{eq:binom_lower_eq}
\end{equation}
is solved by
\begin{equation}
p = \mathcal{B}_\mathrm{beta}\left(\varepsilon;\, x,\, n-x+1 \right),
\label{eq:binom_lower_quantile}
\end{equation}
where $\mathcal{B}_\mathrm{beta}$ denotes the $q$th quantile of the beta distribution with shape
parameters $(a,b)$\footnote{Note that the exact form of the beta distribution is unimportant: there exist standard libraries in standard numerical packages that can evaluate these expressions.}. Similarly, the equation
\begin{equation}
\Pr_{X \sim \mathrm{Binomial}(n,p)}(X \le x) = \varepsilon
\label{eq:binom_upper_eq}
\end{equation}
is solved by
\begin{equation}
p = \mathcal{B}_\mathrm{beta}\left(1-\varepsilon;\, x+1,\, n-x \right).
\label{eq:binom_upper_quantile}
\end{equation}

Equivalently, we have that \cite{Rao_statistics_2000,clopper_useofconfidence_1934}
\begin{equation} \label{eq:binomialboundsweuse}
\begin{aligned}
p \le \mathcal{B}_\mathrm{beta}\left(\varepsilon;\, x,\, n-x+1 \right) \; &\implies   \;  \Pr(X \geq x) \le \varepsilon \\
p \ge \mathcal{B}_\mathrm{beta}\left(1-\varepsilon;\, x+1,\, n-x \right)  \; &\implies \; \Pr(X \leq x) \le \varepsilon 
\end{aligned}
\end{equation}

We will now use \cref{eq:binomialboundsweuse} for constructing $\feasibleset$ and $\variableset(\Fobs)$ with the desired properties.

\subsection{From binomial distribution to Set constructions}

The following lemma constructs the feasible set $\feasibleset$ from the acceptance set $\acceptanceset$ and the parameter $\epsAT$, in a way that satisfies the requirements of \cref{sec:fixedlengthsecurity}. The acceptance set (which we recall is the set of observations for which the acceptance test passes) is defined as the set of frequency vectors that lie close to a predetermined reference value $\Fbar$. This reference value is typically set to be the honest behaviour of the protocol. Our construction relies on the fact that, for each outcome $i \in \mathcal{\CP}$, the number of times outcome $i$ is observed follows a binomial distribution, and corresponds to a measurement with POVM element $\Gamma^{(AB)}_i$.

\begin{lemma}[Constructing the feasible set from \cref{sec:fixedlengthsecurity}]
\label{lemma:constructingfeasibleset}
Let $\Fbar \in \mathbb{P}(\mathcal{\CP})$ be a frequency distribution such that the acceptance set $\acceptanceset$ is defined via\footnote{One could also have $i\in \mathcal{\CP}$ in the set, instead of $i \in \mathcal{\CP}_\mathrm{test}$ if one so desired.}
\begin{equation}
    \acceptanceset \coloneq \left\{ \Fobs \mid | \Fobs_i - \Fbar_i | \leq t_i  \; \forall i \in \mathcal{\CP}_\mathrm{test}  \right\} 
 \end{equation}

Let $\Gamma^{(AB)}_{i}$ be the POVM element corresponding to $i \in \mathcal{\CP}$. Then, consider the feasible set $\feasibleset$ defined as
\begin{equation}
    \feasibleset   \coloneq \left\{ \rho \in \dop{=}(AB) \mid \Fbar_i -t_i -\kappalowerfixed{i} \leq \Tr[\Gamma^{(AB)}_{i} \rho] \leq \Fbar_i + t_i + \kappaupperfixed{i}, \; \; \forall i \in \mathcal{\CP}_\mathrm{test}   \right\},
\end{equation}
where $\kappalowerfixed{i}$ and $\kappaupperfixed{i}$ are given by\footnote{Note that, strictly speaking, in \cref{eq:kappafixeddef} (which we obtain using the  cumulative binomial distribution  in \cref{eq:binomialboundsweuse}) the second and third arguments are required to be natural numbers. This issue can be easily addressed either by introducing appropriate floor and ceiling functions, or by choosing $\Fbar_i$ and $t_i$ such that $n \Fbar_i$ and $n t_i$ are natural numbers. In this thesis we take the latter approach for simplicity in our expressions. 

Moreover, the second and third arguments also need to satisfy obvious bounds, such as positivity and being less than $n$. These constraints can likewise be enforced either via suitable use of minimum and maximum functions, or via suitable choice of $\Fbar_i$ and $t_i$. Again, we we take the latter approach for simplicity in our expressions.}
\begin{equation} \label{eq:kappafixeddef}
    \begin{aligned}
        \kappalowerfixed{i} & = \Fbar_i - t_i - \mathcal{B}_\mathrm{beta}\left( \epsAT; \; n(\Fbar_i -t_i) , \; n - n(\Fbar_i -t_i)+1  \right) \\
        \kappaupperfixed{i} &=\mathcal{B}_\mathrm{beta}\left( 1 - \epsAT; \; n(\Fbar_i+t_i)+1, n - n(\Fbar_i - t_i) \right) - \Fbar_i - t_i.
    \end{aligned}
\end{equation}
This feasible set satisfies the required property from \cref{sec:fixedlengthsecurity}, that is:

\begin{equation}
    \rho_{AB} \notin \feasibleset \implies \Pr(\OmegaAT)_{\rho} \leq \epsAT
\end{equation}

\end{lemma}
Note that in the lemma above, we make a specific choice of the acceptance set. In general, other choices are possible. For instance, see Refs.~\cite{george_numerical_2021,renner_security_2005}, where the authors use the $\ell_1$-norm  between $\Fobs$ and $\Fbar$ to define the acceptance criterion.
\begin{proof}
The proof follows from a straightforward application of the bounds on the cumulative binomial distribution from \cref{eq:binomialboundsweuse}. If $\rho \notin \feasibleset$, there exists at least one $i$ such that either
\begin{equation}
    \begin{aligned}
        \Tr[\Gamma^{(AB)}_i \rho] &\leq \Fbar_i -t_i -\kappalowerfixed{i}  \\
        &= \mathcal{B}_\mathrm{beta}\left( \epsAT; \; n(\Fbar_i -t_i) , \; n - n(\Fbar_i -t_i)+1  \right)
    \end{aligned}
\end{equation}
or \begin{equation}
    \begin{aligned}
  \Tr[\Gamma^{(AB)}_i \rho] &\geq  \Fbar_i + t_i + \kappaupperfixed{i}  \\
  &= \mathcal{B}_\mathrm{beta}\left( 1 - \epsAT; \; n(\Fbar_i+t_i)+1, n - n(\Fbar_i - t_i) \right)
  \end{aligned}
  \end{equation}
is satisfied. 

Suppose it is the former. In that case, since $\OmegaAT \implies \Fobs_i \ge \Fbar_i - t_i$,
we obtain
\begin{equation}
\Pr(\OmegaAT)
\le \Pr(\Fobs_i \ge \Fbar_i - t_i)
\le \epsAT,
\end{equation}
where the final inequality follows from \cref{eq:binomialboundsweuse}.

Similarly, for the latter case, we have $\OmegaAT \implies \Fobs_i \le \Fbar_i + t_i$, and hence
\begin{equation}
\Pr(\OmegaAT)
\le \Pr(\Fobs_i \le \Fbar_i + t_i)
\le \epsAT,
\end{equation}
again by \cref{eq:binomialboundsweuse}. This concludes the proof.
\end{proof}

Using the same approach, we have the following lemma that constructs the confidence set $\variableset(\Fobs)$ satisfying the requirements of \cref{sec:variablelengthsecurity}.

\begin{lemma}[Constructing the confidence set from \cref{sec:variablelengthsecurity}]
\label{lemma:constructingconfidenceset}

Let $\Fobs \in \mathbb{P}(\mathcal{\CP})$ be a observed frequency distribution in the public announcement registers $\CP_1^n$. Let $\Gamma^{(AB)}_{i}$ be the POVM element corresponding to $i \in \mathcal{\CP}$. Then, consider the $\variableset(\Fobs)$ defined as\footnote{The range of probabilities in the expression below is simply the Clopper-Pearson confidence interval for binomial random variables \cite{clopper_useofconfidence_1934}.}
\begin{equation}
  \variableset(\Fobs)  \coloneq \left\{ \rho \in \dop{=}(AB) \mid \Fobs_i -\kappalowervariable{i} \leq \Tr[\Gamma^{(AB)}_{i} \rho] \leq \Fobs_i + \kappauppervariable{i}   \; \forall i \in \mathcal{\CP}_\mathrm{test} \right\},
\end{equation}
where $\kappalowervariable{i}$ and $\kappaupperfixed{i}$ are given by
\begin{equation} \label{eq:kappavariabledef}
    \begin{aligned}
        \kappalowervariable{i} & = \Fobs_i - \mathcal{B}_\mathrm{beta}\left( \frac{\epsAT}{2  \abs{\mathcal{\CP}_\mathrm{test} } }; \; n(\Fbar_i -t_i) , \; n - n(\Fbar_i -t_i)+1  \right) \\
        \kappauppervariable{i} &=\mathcal{B}_\mathrm{beta}\left( 1 - \frac{\epsAT}{2  \abs{\mathcal{\CP}_\mathrm{test}} }; \; n(\Fbar_i+t_i)+1, n - n(\Fbar_i - t_i) \right) - \Fbar_i - t_i.
    \end{aligned}
\end{equation}
This confidence set satisfies the required property from \cref{sec:variablelengthsecurity}, that is, for all possible $\rho_{AB}$, we have:
\begin{equation}
 \Pr_{\Fobs}(\rho_{AB} \in \variableset(\Fobs)) > 1-\epsAT 
\end{equation}
\end{lemma}
\begin{proof}
    Consider any $i \in \mathcal{\CP}_\mathrm{test}$, and recall that $n \Fobs_i$ is a binomial random variable with probability of success given by $\Tr[\Gamma^{(AB)}_i \rho]$. Thus, using \cref{eq:binomialboundsweuse}, we obtain the following Clopper-Pearson confidence \cite{clopper_useofconfidence_1934} interval for binomial random variables, and obtain
    \begin{equation}
        \Pr_{\Fobs} ( \Fobs_i - \kappalowervariable{i}  \leq \Tr[\Gamma^{(AB)}_i \rho] \leq \Fobs_i + \kappauppervariable{i} ) \geq 1 - \frac{\epsAT} {\abs{\mathcal{\CP}_\mathrm{test} }}
    \end{equation}
    where $\kappauppervariable{i},\kappalowervariable{i}$ are defined in \cref{eq:kappavariabledef}.
    Taking the negation of this event, and taking the union bound, we obtain 
      \begin{equation}
      \begin{aligned}
       & \Pr_{\Fobs} \left(  \Tr[\Gamma^{(AB)}_i \rho]  \notin \left[\Fobs_i + \kappauppervariable{i},\Fobs_i + \kappauppervariable{i}  \right] \right) \leq  \frac{\epsAT}{\abs{\mathcal{\CP}_\mathrm{test} }}  \\
        & \Pr_{\Fobs} \left(  \bigcup_{i \in \mathcal{\CP}_\mathrm{test} } \Tr[\Gamma^{(AB)}_i \rho] \notin \left[\Fobs_i + \kappauppervariable{i},\Fobs_i + \kappauppervariable{i}  \right] \right) \leq  \epsAT \\
       & Pr_{\Fobs} \left(  \rho \in \variableset{\Fobs} \right) \geq 1- \epsAT.
       \end{aligned}
    \end{equation}
    This concludes the proof.

\end{proof}

\begin{remark}\label{remark:setcomparison}
It is instructive to compare the sets $\feasibleset$ and $\variableset(\Fobs)$,
since they appear in the key rate formulas as the domains of optimization of the
same objective function. The first notable difference is that the feasible set
$\feasibleset$ does not depend on the observed frequency vector $\Fobs$: in a
fixed-length protocol, a predetermined key length is produced whenever
$\Fobs \in \acceptanceset$, and the protocol aborts otherwise.

Consider now a situation in which honest behavior is realized, so that
$\Fobs \approx \Fbar$. In this case, there are two further notable differences.
First, the set $\feasibleset$ is larger due to the parameters $t_i$. This reflects
the fact that any realistic protocol must choose the acceptance set to have a large enough size in order to avoid aborting with overwhelming probability. We
investigate this effect further when computing expected key rates in the next
section.

Second, the parameters $\kappalowervariable{i}$ and $\kappauppervariable{i}$ are
generally larger than their fixed-length counterparts
$\kappalowerfixed{i}$ and $\kappaupperfixed{i}$. This difference arises because
the variable-length analysis places a stronger requirement on the sampling estimates: we must ensure that \emph{all} true probabilities lie within their
respective confidence intervals simultaneously. Achieving this requires the
use of a union bound in the proof. In contrast, in the construction of the
feasible set, it suffices that at least one value falls outside the relevant
range, and no union bound is utilized.
\end{remark}

\section{Application to Qubit BB84}\label{sec:variableapplicationtoQubitbb84}
We will now apply our results to Qubit BB84 protocol, specified in \cref{subsec:protocolvariations}.  At this stage, it may be useful to refer back to \cref{sec:statesandmeasurements} to recall details about the states sent, measurements performed, announcements undertaken and sifting procedure for the protocol. We first state the following corollaries concerning the security of fixed-length and variable-length Qubit BB84 protocols.

\begin{corollary}(Fixed-length Qubit BB84) \label{cor:fixedlengthqubitbb84}
Consider the fixed-length qubit BB84 variant of the \nameref{prot:qkdprotocol} as specified in \cref{sec:statesandmeasurements}, with the acceptance set and feasible set $\feasibleset$ (which depends on $\epsAT$)  as constructed in \cref{lemma:constructingfeasibleset}. Then this fixed-length Qubit BB84 protocol is $(\max(\epsAT,\epsPA) + \epsEV)$-secure against IID collective attacks, as long as the $\lfixed$ satisfies
  \begin{equation} \label{eq:lfixedvaluequbitbb84IID}
        \begin{aligned}
            \lfixed &\leq \max \Bigg\{ \Bigg \lfloor
                n \inf_{\nu \in \Sigma (\feasibleset) }
                H(S | \CP E)_{\nu} - n(\alpha-1)\log^2(1+2\dim(S))
                - \leakfixed \\
                &
                - \EVcost 
                - \frac{\alpha}{\alpha - 1}
                 \log\left( \frac{1}{ \epsPA} \right) 
                  + 2 \Bigg\rfloor, 0 \Bigg\}\\
    \Sigma(\feasibleset) &\coloneq \left\{\; \QKDGmap{}[\omega_{ABE}] \; \mid  \omega_A = \sigma_A, \omega_{AB} \in \feasibleset \right\},
        \end{aligned}
    \end{equation}
where $\alpha \in (1,1+1/\log(1+2\dim(S)))$, and $E$ is a purifying register.
\end{corollary}

\begin{proof}
The result follows directly from the fixed-length IID security statement in
\cref{theorem:fixedlengthIID}, together with the specification of the feasible
set given in \cref{lemma:constructingfeasibleset}. In addition, we use
\cref{lemma:contrenyi}, which states that
\[
\Halpha(S | \CP E)
\ge
H(S | \CP E)
- (\alpha-1)\log^2\!\bigl(1+2\dim(S)\bigr).
\]
\end{proof}

Some comments are in order. Note that we replace the {\Renyi} entropy $\Halpha(S |\CP E)$ with the von Neumann entropy $H(S | \CP E)$, since this substitution allows us to use established methods for the reliable numerical evaluation of the resulting optimization problem \cite{winick_reliable_2018,burniston_software_2024} (which is convex in $\omega_{AB}$). Directly evaluating the {\Renyi} entropy would lead to larger key lengths, see Ref.~\cite{chung2025generalized} for methods to do so.  The range of {\Renyi} parameter
$\alpha \in \bigl(1,\, 1+1/\log(1+2\dim(S))\bigr)$ is to ensure the validity of all intermediate statements used in the
proof. The key rate can be optimized over the choice of alpha $\alpha$, and it is straightforward to
verify (by differentiating the key length expression with respect to $\alpha$) that the choice
\[
\alpha
=
1 + \frac{\log(1/\epsPA)}{\sqrt{n}\,\log(2\dim(S)+1)}
\]
is optimal. This is the choice taken in the entirety of this chapter. 

We now state the corresponding statement for variable-length Qubit BB84 protocols.

\begin{corollary}(Variable-length Qubit BB84) \label{cor:variablelengthqubitbb84}
Consider the variable-length Qubit BB84 variant of the \nameref{prot:qkdprotocol} as specified in \cref{sec:statesandmeasurements}, with the confidence set $\variableset(\Fobs)$   (which depends on $\epsAT$)  as constructed in \cref{lemma:constructingconfidenceset}. Then this variable-length Qubit BB84 protocol is $(\epsAT+\epsPA + \epsEV)$-secure against IID collective attacks, as long as the $\lkey(\Fobs)$ satisfies
  \begin{equation} \label{eq:lvariablevaluequbitbb84IID}
        \begin{aligned}
            \lkey(\Fobs) &\leq \Bigg \lfloor
                n \inf_{\nu \in \Sigma (\Fobs) }
                H(S | \CP E)_{\nu} - n(\alpha-1)\log^2(1+2\dim(S))
                - \leakfixed \\
                &
                - \EVcost 
                - \frac{\alpha}{\alpha - 1}
                 \log\left( \frac{1}{ \epsPA} \right) 
                  + 2 \Bigg\rfloor,\\
    \Sigma(\Fobs) &\coloneq \left\{\; \QKDGmap{}[\omega_{ABE}] \; \mid  \omega_A = \sigma_A, \omega_{AB} \in \variableset(\Fobs) \right\},
        \end{aligned}
    \end{equation}
where $\alpha \in (1,1+1/\log(1+2\dim(S)))$
\end{corollary}
\begin{proof}
The proof follows from \cref{theorem:variablelengthIID,lemma:contrenyi,lemma:constructingconfidenceset}, analogous to the proof of \cref{cor:fixedlengthqubitbb84}.
\end{proof}

\subsection{Key Rate Plots} \label{subsec:keyrateplotsqubitbb84}
We will now apply \cref{cor:fixedlengthqubitbb84,cor:variablelengthqubitbb84} to compute
key rates for fixed-length and variable-length qubit BB84 protocols, and to
compare their performance. In addition to the usual key rates, we also compute \emph{expected key rates}. To motivate this notion, recall that a security proof specifies which key
length the protocol produces \emph{if} certain events occur. By itself, this
does not quantify how much key is produced \emph{on average} when the protocol
is executed in practice.\footnote{In fact, a protocol that always produces a key of zero length is perfectly secure.}  

To define the expected key rate, we first specify an \emph{honest behavior} of
the channel, and hence of the protocol. That is, we assume that in the absence
of an adversary the channel behaves according to a fixed model. This
assumption induces an IID state $\rho_{\mathrm{honest}}^{\otimes n}$ shared
between Alice and Bob.

We then define
\begin{equation} \label{eq:keyratesplotted}
\begin{aligned}
R_{\mathrm{fixed}}
&:= \frac{\lfixed}{n}
\qquad
\text{(fixed-length key rate \emph{if} the protocol accepts)},\\
R_{\mathrm{variable}}(\Fobs)
&\coloneq \frac{\lkey(\Fobs)}{n}
\qquad
\text{(variable-length key rate \emph{if} $\Fobs$ is observed)},\\
\widetilde{R}_{\mathrm{fixed}}
&:= \Pr(\OmegaAT \wedge \OmegaEV)_{\rho_{\mathrm{honest}}^{\otimes n}}
\cdot \frac{\lfixed}{n}
\qquad
\text{(expected fixed-length key rate)},\\
\widetilde{R}_{\mathrm{variable}}
&:= \sum_{\Fobs}
\Pr(\Omega(\Fobs) \wedge \OmegaEV)_{\rho_{\mathrm{honest}}^{\otimes n}}
\cdot \frac{\lkey(\Fobs)}{n}
\qquad
\text{(expected variable-length key rate)}.
\end{aligned}
\end{equation}

We will compute these quantities for two different scenarios of honest
behavior. Note that in our numerical evaluations, we assume that the event $\OmegaEV$
occurs with probability close to one and focus instead on the impact of variations in the observed statistics.\footnote{This assumption can be justified, for example, by using
sufficiently strong error correction.} For the qubit BB84 plots, we set $n = 10^8$, and the probability of testing to
$\gamma = 0.1$. In both test and key generation rounds, each basis is
chosen with probability $0.5$, and therefore each state is sent with
probability $0.25$. 

For fixed-length protocols, we set
$
\epssecret = \epsEV = \epsAT = \epsPA = 10^{-10},
$
resulting in an overall security parameter of
$\epssecure = 2 \times 10^{-10}$. We set
$
\leakfixed = f_{\mathrm{EC}}n H(S |Y \CP)_{\rho_{\mathrm{honest}}},
$
where $f_{\mathrm{EC}} = 1.16$ denotes the efficiency of the
error-correction protocol. 
For variable-length protocols, we set
$
\epssecret = \epsEV = 10^{-10}, \qquad
\epsAT = \epsPA = 10^{-10}/2,
$
again obtaining an overall security parameter of
$\epssecure = 2 \times 10^{-10}$. We set
$
\leak(\Fobs) = f_{\mathrm{EC}} n H(S|Y \CP)_{\Fobs}
$.\footnote{By this, we mean that $\Fobs$ induces a corresponding distribution on key generation rounds with respect to which the entropy is evaluated.} Note that many of these parameters should, in principle, be optimized to obtain the best possible key rates. However, we do not perform this optimization here, as our goal is simply to compare fixed-length and variable-length protocols.
\subsubsection{Fixed honest behaviour}

\begin{figure}[ht!]
    \centering
    \includegraphics[width=\linewidth]{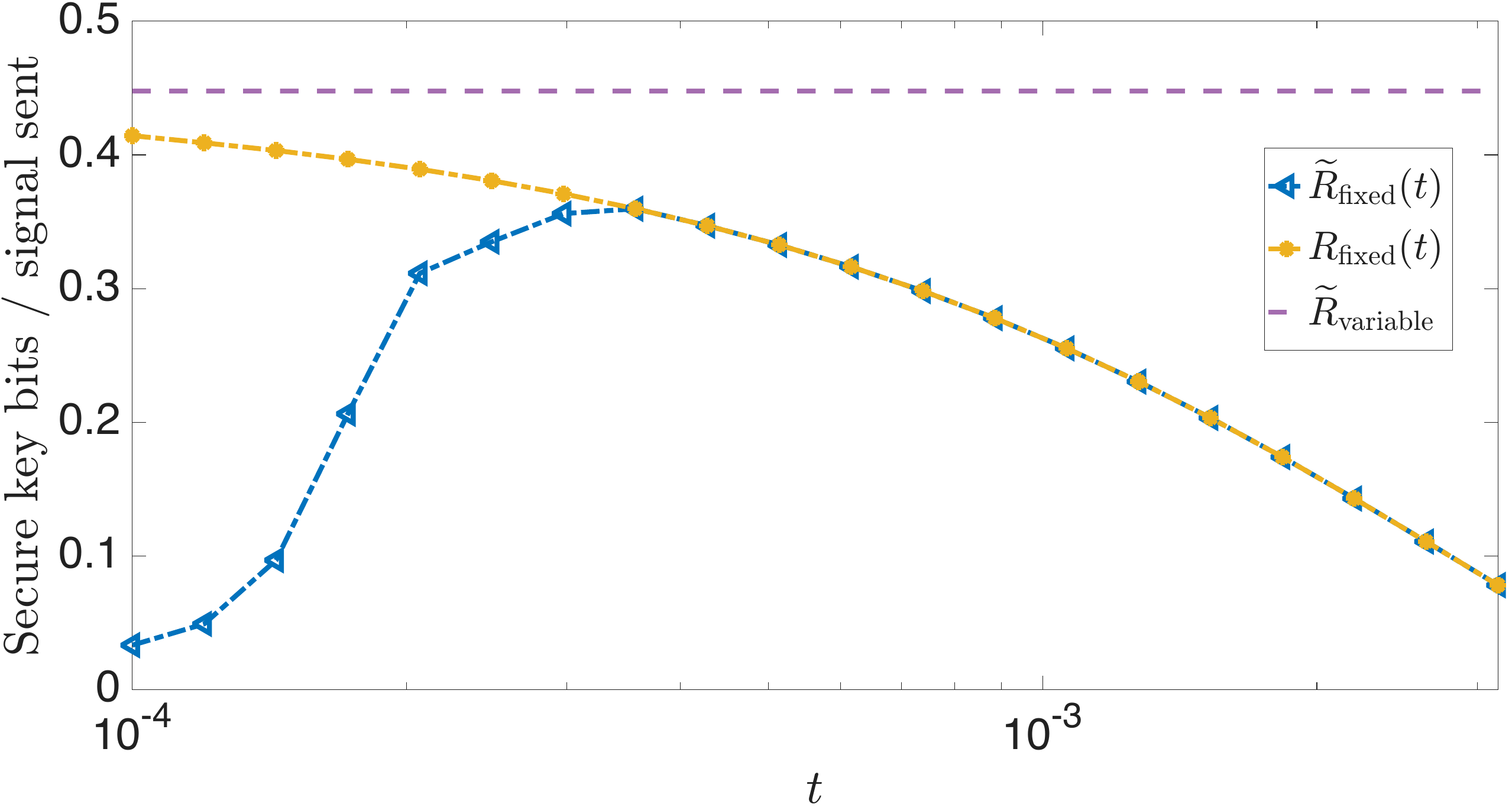}
    \caption{Expected key rate for fixed-length protocols $\widetilde{R}_\mathrm{fixed}(t)$ for various values of $t$, key rate upon acceptance for fixed-length protocols $R_\mathrm{fixed}(t)$ plotted for various values of $t$, and the expected key rate for variable-length protocol plotted $\widetilde{R}_\mathrm{variable}$, plotted for a fixed honest behaviour of the channel.}
    \label{fig:qubitBB84_honestbehaviour}
\end{figure}

We first fix the honest behaviour of the channel to have $0 \mathrm{dB}$
loss, no depolarization, and no misalignment. This determines
$\Fbar$ from the honest behaviour. We then vary the size of the acceptance
test, which we constructed in \cref{lemma:constructingfeasibleset}. To do so, we define a single parameter $t$, 
and set $t_i = t$ for all $i$ (recall that $t_i$ controlled the size of deviation from $\Fbar_i$ that is allowed for the protocol to accept). As stated earlier, one may choose other acceptance criterion. The resulting key rates are plotted in
\cref{fig:qubitBB84_honestbehaviour} and discussed below.  Note that that since the expected key rates include the probability over various observations, the value of $t$ (which controls the size of the acceptance set) is not important.

\begin{enumerate}
    \item The fixed-length key rate conditioned on acceptance is given by
    $R_{\mathrm{fixed}}$ and is a monotonically decreasing function of $t$.
    (As the size of the acceptance set increases, the worst-case observation
    within the acceptance set becomes more pessimistic, leading to a smaller
    key rate upon acceptance.)

    \item The expected fixed-length key rate is given by
    $\widetilde{R}_{\mathrm{fixed}}$. (This quantity is computed as follows.)
    \begin{enumerate}
        \item We first compute the expected honest behaviour of the channel,
        namely $\Fbar$, where
        \[
        \Fbar_i = \Tr\!\left[\Gamma^{(AB)}_i \rho_{\mathrm{honest}}\right].
        \]
        \item We then sample $n$ times from $\Fbar$, to obtain the observed frequency vector $\Fobs$, and check whether $\Fobs \in \acceptanceset$.
        \item This procedure is repeated $25$ times, and
        $\Pr(\OmegaAT)_{\rho_{\mathrm{honest}}}$ is estimated by the empirical
        fraction of samples for which $\Fobs \in \acceptanceset$.
    \end{enumerate}
    We observe that this expected key rate is small for very small values of $t$,
    since the protocol almost never accepts. It is also small for very large
    values of $t$, reflecting the fact that the key rate upon acceptance becomes
    small. The expected key rate therefore captures the trade-off between
    accepting with high probability and producing a large key upon acceptance.
    
    \item We also plot the expected variable-length key rate
    $\widetilde{R}_{\mathrm{variable}}$, which is computed analogously to
    item~(2) by sampling $n$ times from $\Fbar$ to obtain $\Fobs$, computing $\lkey(\Fobs)$, repeating this procedure  $25$ times, and averaging the
    resulting key rate.
\end{enumerate}

Crucially, we find that the variable-length protocol achieves a higher expected
key rate than the best fixed-length protocol. Since the variable-length protocol
consists of exactly the same steps as the fixed-length protocol, and differs
only in the parameters of the classical post-processing, its implementation does
not impose any additional experimental or operational difficulties. This
improvement therefore comes at no additional cost, while yielding a strictly
higher expected key rate.

\subsubsection{Unpredictable Honest Behaviour}

\begin{figure}[ht!]
    \centering
    \includegraphics[width=\linewidth]{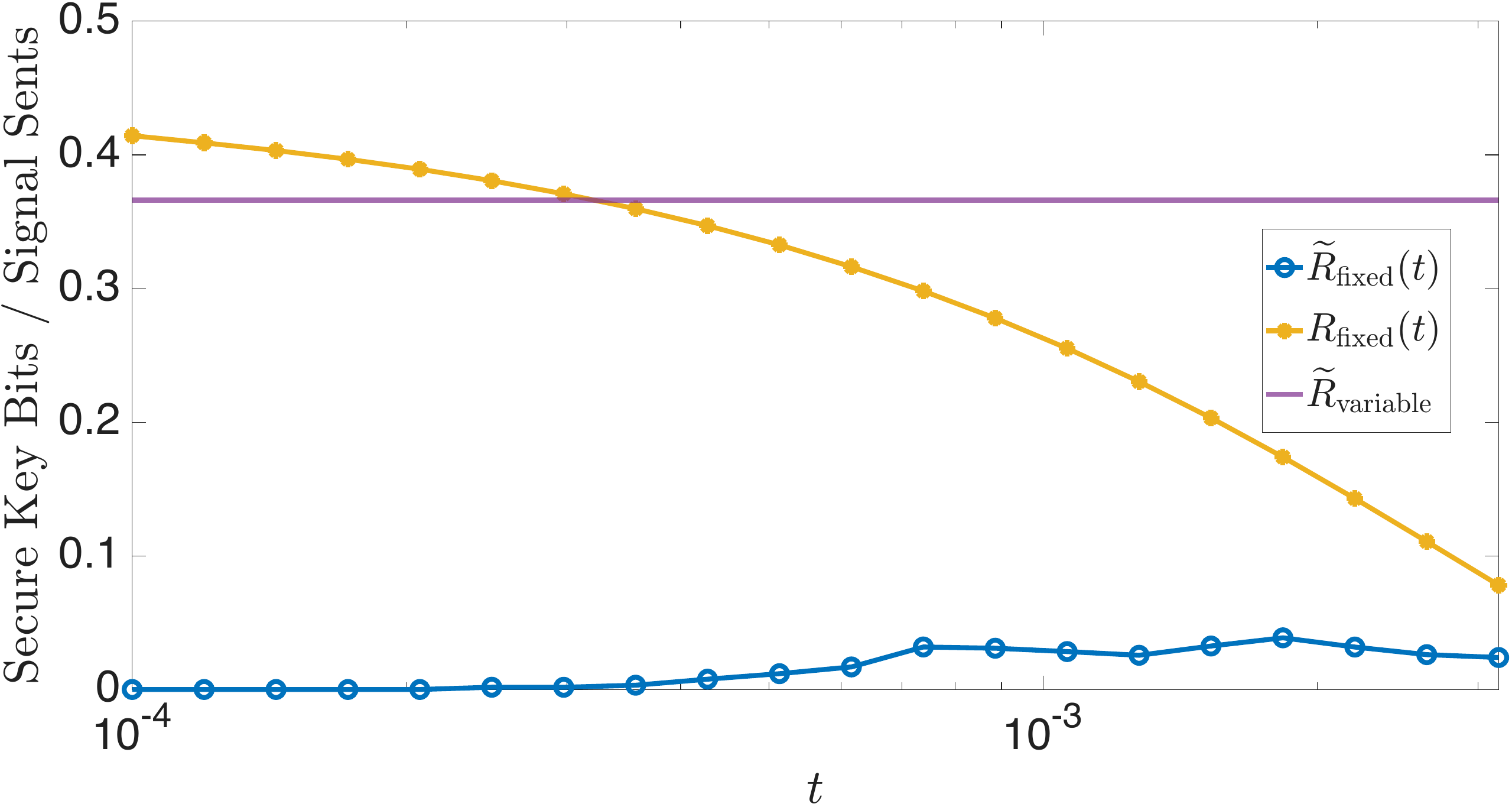}
    \caption{Expected key rate for fixed-length protocols $\widetilde{R}_\mathrm{fixed}(t)$ for various values of $t$, key rate upon acceptance for fixed-length protocols $R_\mathrm{fixed}(t)$ plotted for various values of $t$, and the expected key rate for variable-length protocol plotted $\widetilde{R}_\mathrm{variable}$, for an unpredictable honest behaviour.}
    \label{fig:qubitBB84_chaotichonestbehaviour}
\end{figure}

In this section, we compute key rates for a scenario in which the honest behaviour of the channel is unpredictable. For simplicity, we consider a channel model that takes values from a discrete set of depolarization probabilities and misalignment angles. For a given run, the depolarization probability is chosen from the set $\{0, 0.02, 0.04\}$, and the misalignment angle is chosen uniformly from the set $\{0^\circ, 4^\circ, 8^\circ\}$, with the loss fixed to $0\mathrm{dB}$. The channel therefore has $n_{\mathrm{channel}} = 9$ possible realizations, each occurring with equal probability. We denote by $\rho_{\mathrm{honest}}^{(j)}$ the state corresponding to the $j$th honest channel behaviour. The resulting key rates are plotted in \cref{fig:qubitBB84_chaotichonestbehaviour}.

\begin{enumerate}
    \item The fixed-length key rate conditioned on acceptance is given by
    $R_{\mathrm{fixed}}$ and is a monotonically decreasing function of $t$, with $\Fbar$ determined by zero misalignment and no depolarization. This is exactly the same as the key rate plotted in \cref{fig:qubitBB84_honestbehaviour}.

    \item  The expected fixed-length key rate is given by
    $\widetilde{R}_{\mathrm{fixed}}$. Since there are different realizations of the channel, we first fix the channel realization to be one out of the $n_\mathrm{channel}$ values, and compute the expected key rate for that realization by following the procedure from earlier section, repeating the sampling $25$ times. Since each channel is equally likely, we simply take the average of these computed key rates as the expected key rates. 

    Here, the expected key rate is much smaller than the key rate upon acceptance, since the protocol accepts only on a small number of channel behaviours. As $t$ grows larger, the size of the acceptance test increases, and the protocol starts acccepting on multiple possible channel realizations. However, the size of the acceptance test is already large, and key rate upon acceptance rapidly goes to zero, causing the expected key rate to also go to zero.

    \item We also plot the expected variable-length key rate
    $\widetilde{R}_{\mathrm{variable}}$, which is computed analogously to
    item~(2) by averaging the
    resulting key rate over the $9$ possible channel realizations. Crucially, we find
the expected key rate for variable-length protocols
is much higher than the expected key rate for the
fixed-length protocols.
\end{enumerate}

Note that the degree of improvement
shown by the variable-length protocol in \cref{fig:qubitBB84_chaotichonestbehaviour} depends the variability in the honest behaviour. Larger variation in the honest behaviour will lead to a bigger difference in performance between fixed-length and varibale-length protocols.

\section{Application to Decoy-State BB84}\label{sec:variableapplicationtoDecoybb84}
We now apply our variable-length security analysis to decoy-state protocols. 
We note that the components related to the decoy-state analysis were developed as part of Ref.~\cite{Kamin2025}, to which I did not contribute. 
Accordingly, these aspects are only reviewed briefly in thesis, and we refer the reader to Ref.~\cite{Kamin2025} for further details. As already emphasized in \cref{subsec:motivationfordecoy}, the central task remains unchanged: namely, to bound the relevant entropic quantity by exploiting the richer set of observed statistics available in decoy-state protocols.  We give a brief explanation of how to go about performing decoy-state analysis in Appendix~\ref{Appendix:misc}.

We plot key rates for a protocol in which both Alice and Bob choose the $\Zbasis$ basis with probability $0.9$. The decoy intensities are given by $\mu_1 = 1$, $\mu_2 = 0.01$, and $\mu_3 = 0.001$. Each round is a $\test$ round with probability $0.1$, and the decoy intensities are chosen uniformly at random. In $\gen$ rounds, only the signal intensity $\mu_1$ is chosen, with probability $1$.
Double clicks on Bob's side are randomly assigned to single clicks, which allows us to use the qubit squasher (see \cref{subsec:squashing}). We plot variable key rates for the honest behaviour of the channel as a function of loss; that is, we assume that the observed frequency vector $\Fobs$ corresponds to the honest behaviour of the channel. we set
$
\epssecret = \epsEV = 10^{-10}, \qquad
\epsAT = \epsPA = 10^{-10}/2,
$
again obtaining an overall security parameter of
$\epssecure = 2 \times 10^{-10}$. We set
$
\leak(\Fobs) = f_{\mathrm{EC}}n H(S \mid Y \CP)_{\Fobs}
$, and $f_{\mathrm{EC}}=1$. As before, many of these parameters can be optimized over for the best results, which we do not do here.

\begin{figure}[ht!]
    \centering
    \includegraphics[width=\linewidth]{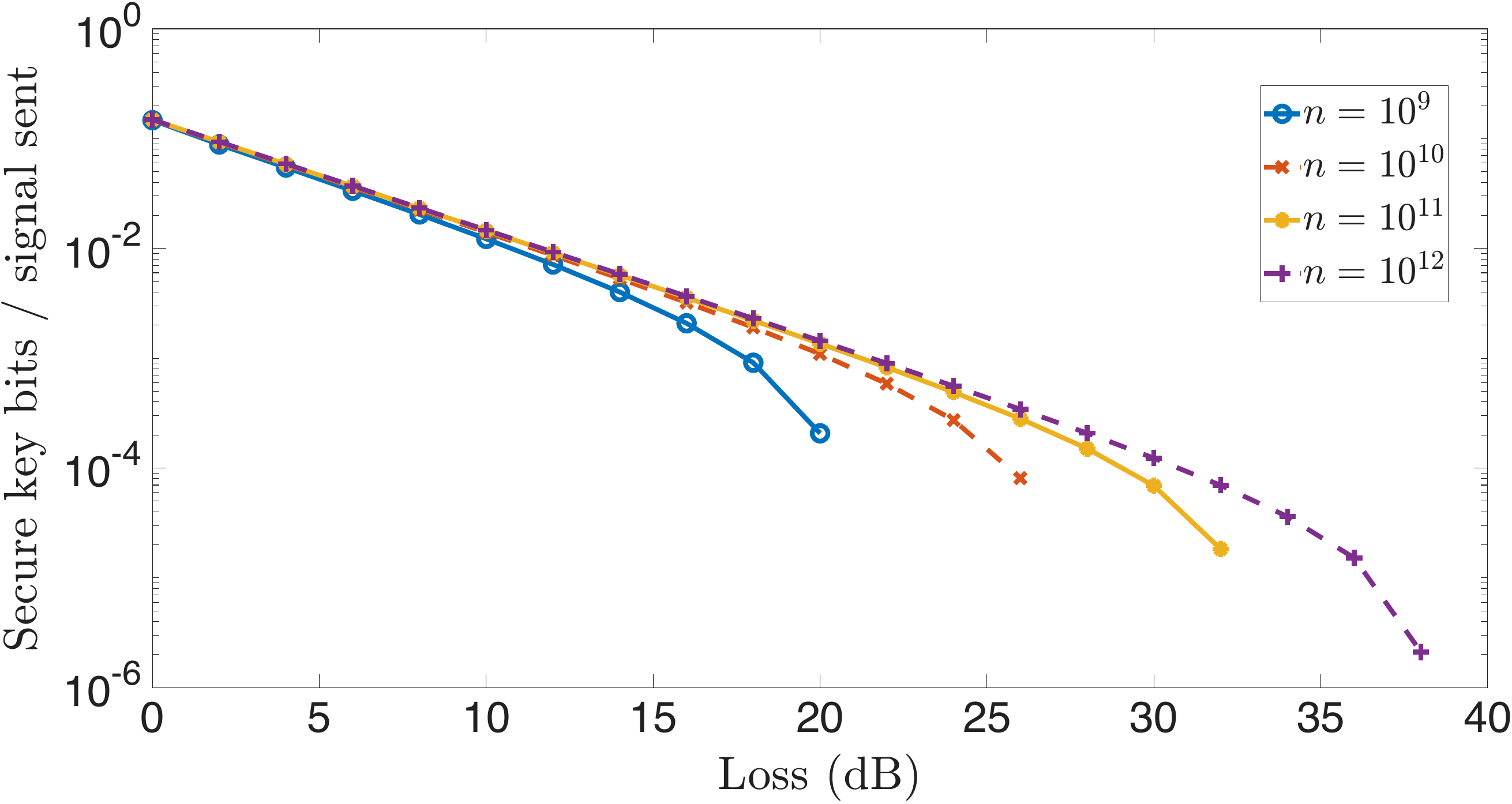}
    \caption{Key rate for variable-length decoy-state BB84 protocol plotted against loss, for  the typical observations of the protocol, against IID collective attacks. }
    \label{fig:decoyBB84_IID}
\end{figure}

\section{Variable-length input to privacy amplification} \label{sec:variablelengthPA}

So far we have studied  the variable-length aspects of the \emph{final key} that is generated after privacy amplification in QKD protocols. In this section, we will turn our attention to the variable-length aspect of the \emph{pre-amplification string} in QKD implementations, before privacy amplification. In particular, we will point out and remedy a gap (pointed out briefly in \cref{remark:variablelengthinputPA}) between the theoretical analysis of privacy amplification and its experimental implementation. For simplicity, we only consider fixed-length QKD protocols. However, our result takes the form of a modified Leftover Hashing Lemma, and can be trivially generalized to variable-length protocols. We also do make any IID attack assumption in this section, and our results hold for arbitrary attacks.

	\subsection{Sifting in QKD} \label{subsec:siftinginqkd}
 Consider the following three ways of implementing the sifting step in QKD protocols, where Alice and Bob wish to throw away certain rounds based on public announcements. Note that the \nameref{prot:qkdprotocol} implements the procedure from Case~\ref{PAcase3} below, but the procedure analyzed in this chapter corresponds to Case~\ref{PAcase1}. 
	
\begin{enumerate}
		\item\label{PAcase1} \textbf{Map the discard outcomes to $\bot$, but don't discard them:} In this case, the state prior to privacy amplification is given by $\rho_{S_1^n \CP_1^n \CEC \CEV \HEV \Eve }$\footnote{Here we ignore Bob's registers, since we are only concerned with secrecy}, where $S$ is a register that takes values in $\{0,1,\bot\}$.  In this case, we can perform privacy amplification on $S_1^n$ using universal$_2$ hashing from $S_1^n$ to $l$ bits. In particular, binary Toeplitz hashing, a widely used choice, is not possible.
		\item\label{PAcase2} \textbf{Map the discard outcomes to $0$, but don't discard them:} In this case, the state prior to privacy amplification is given by $\rho_{\hat{S}_1^N \CP_1^n \CEC \CEV \HEV \Eve }$, where $\hat{S}$ is a register that takes values in $ \{0,1\}$. In this case, we can implement privacy amplification using universal$_2$ hashing from $n$ bits to $l$ bits. In particular, binary Toeplitz hashing, a widely used choice, is possible; however, the hash matrices must always be for input strings of a \emph{fixed} length $n$.
		\item\label{PAcase3} \textbf{Actually discard the discard outcomes:}  In this case, the state prior to privacy amplification is given by $\rho_{\PAstring  \CP_1^n \CEC \CEV \HEV \Eve  }$, where $\PAstring $ is a register that takes values in the set of bitstrings of length less than or equal to $n$, which we shall denote as $\{0,1\}^{\leq n}$. In this case, one first looks at the number of bits in the register $\PAstring$, denoted by $\mathrm{len}(\PAstring)$, and chooses a universal$_2$ hashing procedure from $\mathrm{len}(\PAstring)$ bits to $\lfixed$\footnote{Here we use $\cobs$ rather than $\Fobs$, since the additional generality comes with no cost. Recall that $\Fobs = \freq(\cobs)$.} bits. This is what is commonly done in QKD experiments, and is what is specified in \nameref{prot:qkdprotocol}. Practically, one would like to use binary Toeplitz hashing in this procedure. However, we will see below that this is \emph{not a valid universal$_2$ hashing procedure} from $\{0,1\}^{\leq n}$ to $\lfixed$ bits. 
	\end{enumerate}
The theoretical analysis of Case~\ref{PAcase1} and Case~\ref{PAcase2} is straightforward, since they constitute valid universal$_2$ hashing procedures from  $\{0,1,\bot\}^n$ to $\lfixed$ bits, and $n$ bits to $l$ bits, respectively. Thus, Leftover Hashing Lemmas can be directly applied. However, Case~\ref{PAcase3} is \textit{not}  necessarily a universal$_2$ hashing procedure from $\{0,1\}^{\leq n}$ to $\lfixed$ bits, as we now explain. Thus we cannot directly apply the Leftover Hashing Lemma\footnote{Note that we ignore the nuance between ideal universal$_2$ hashing and universal$_2$ hashing (see \cref{remark:LHLequalityissue}) in this section, since our analysis can be equally applied to both variants.} in this case.

\subsection{The problem} \label{subsec:problem}
	For every $l_\mathrm{input} \in \mathbb{N}$,  let $\hashfamily{l_\mathrm{input}}{\lfixed}$ denote a universal$_2$ hash family from $l_\mathrm{input}$ bits to $\lfixed$ bits. Then, the procedure described in Case~\ref{PAcase3} above is equivalent to \textit{first}  randomly sampling $f_{l} \in \hashfamily{l}{\lfixed}$ for every $l$, followed by computing $f_{\mathrm{len}(\PAstring)}(\PAstring)$. Note that in this case, only one of the sampled $f_{l}$s is  ever applied. 
		In order for this procedure to be a valid universal$_2$ hashing procedure from $\{0,1\}^{\leq n}$ to $\lfixed$ bits, by definition it must be the case that for any two inputs $\vec{s}_1 \neq \vec{s}_2$, we have
		\begin{equation} \label{eq:twouniversalhashing}
			\Pr_{f_1,f_2,...,f_n} [f_{\mathrm{len}(\vec{s}_1)} (\vec{s}_1) = f_{\mathrm{len}(\vec{s}_2)} (\vec{s}_2)] \leq \frac{1}{2^{\lfixed}}. 
		\end{equation}
		 When $\vec{s}_1$ and $\vec{s}_2$ are of the same length, then \cref{eq:twouniversalhashing} follows from the universal$_2$ property of $\hashfamily{l}{\lfixed}$.
		 When $\vec{s}_1$ and $\vec{s}_2$ are of different length, an explicit counter example can be obtained by considering $\vec{s}_1$ and $\vec{s}_2$ to be all-zero strings of different lengths. In this case, if $\hashfamily{l}{\lfixed}$ is a universal$_2$ \textit{linear} hash family, then $f_{\mathrm{len}(\vec{s}_1)} (\vec{s}_1) = f_{\mathrm{len}(\vec{s}_2)} (\vec{s}_2) = \mathbf{0}$ with probability $1$. This contradicts the required \cref{eq:twouniversalhashing}.  Thus for binary Toeplitz hashing, Case~\ref{PAcase3} is \textit{not} a valid universal$_2$ hashing procedure. Thus we cannot directly apply the leftover hashing lemma. 
		 \begin{remark} \label{remark:hashing}
		 	We note that if every $\hashfamily{l}{\lfixed}$ is chosen such that it is universal$_2$ \textit{and} has  the following ``uniform output'' property:
		 	\begin{equation}\label{eq:uniformoutput}
		 		 \Pr_{f_l \in \hashfamily{l}{\lfixed}} [f_l(\mathbf{z}) = k] \leq 1/2^l \quad \forall \quad  \mathbf{z} \in \{0,1\}^i,k\in\{0,1\}^{\lfixed},
		 	\end{equation}
		 then it is straightforward to prove that \cref{eq:twouniversalhashing} holds and hence the described procedure is a valid universal$_2$ hashing. Furthermore, in principle any universal$_2$ hashing procedure can be modified into one that satisfies \cref{eq:uniformoutput} by increasing the amount of random seed bits required and XORing the output of the hash function with this seed. This  construction is utilized in the proof of the \cref{lemma:variableinputhashing}. However, physically implementing this conversion in an actual QKD protocol would be an undesirable additional cost, hence we instead provide a proof that shows that this is not necessary.
		 \end{remark}

	 \subsection{The solution}
		We address this issue with \cref{lemma:variableinputhashing,lemma:renyiinvariance} below. We start by proving the following modified Leftover Hashing Lemma that is applicable to Case~\ref{PAcase3}, as long as the protocol satisfies the property that the positions and values of the discarded outcomes can be determined from the public announcements $\CP_1^n$ (we return to this point after presenting the lemmas and their proofs). Our approach is to first use \cref{remark:hashing} to construct a virtual hashing procedure that is a valid universal$_2$ hashing procedure from $\{0,1\}^{\leq n} $ to $\lfixed$ bits.  We will then show that the actual output states can be obtained by performing a CPTP map on the virtual output states.  The required result then follows from data-processing inequalities. 
		
\newcommand{\rawkey}{R}
\newcommand{\suplabels}{}
\newcommand{\suplabelsbrkt}{}

\newcommand{\virthashfamily}[2]{\mathcal{F}_\mathrm{hash}^{\mathrm{virt}}(#1,#2)}

\begin{restatable}[]{lemma}{variableinputhashingfirstlemma}
 \label{lemma:variableinputhashing}
Let $\rho_{\PAstring \mathbf{C} \Eve}$ be a state classical in $\PAstring \mathbf{C}  $ (where the $\PAstring$ register takes values in $\{0,1\}^{\leq n}$), with the property that conditioned on each possible value $\bar{c}$ on the $\mathbf{C}$ register, the resulting distribution on $\PAstring$ is only supported on values in $\{0,1\}^{k_{\bar{c}}}$ for some constant $k_{\bar{c}} \in \mathbb{N}$. Note that $\mathbf{C}$ is a classical register which can be store arbitrary values for the purposes of this lemma. 
Let $\rho^{\suplabelsbrkt}_{K_A   \mathbf{C}  \HPA \Eve}$ be the state obtained from $\rho_{\PAstring  \mathbf{C} \Eve}$ by first computing the number of bits $\mathrm{len}(\PAstring)$ in the $\PAstring$ register, then implementing a universal$_2$ hashing procedure from $\mathrm{len}(\PAstring)$ bits to $\lfixed$ bits (in other words, the procedure described in Case~\ref{PAcase3} of \cref{subsec:siftinginqkd}). Then for any event $\Omega$ on the classical register $ \mathbf{C}$, we have (for $\rho^{(\suplabelsbrkt\mathrm{ideal})}_{K_A  \mathbf{C} \HPA \Eve} \coloneqq \frac{\id_{K_A}}{|K_A|} \otimes \rho^{\suplabelsbrkt}_{ \mathbf{C} \Eve}$):
\begin{equation} \label{eq:varLHL}
\begin{aligned}
&	 \Pr(\Omega) \tracedist{ \rho^{\suplabelsbrkt}_{K_A  \mathbf{C} \Eve | \Omega} - \rho^{(\suplabelsbrkt\mathrm{ideal})}_{K_A  \mathbf{C} \Eve  | \Omega}} \\
&\leq 	\Pr(\Omega)  2^{- \left( \frac{\alpha-1}{\alpha}  \right)\left( \Halpha(\PAstring |  \mathbf{C}  \Eve)_{\rho | \Omega} - \lfixed + 2 \right) } \\
&\leq 	 2^{- \left( \frac{\alpha-1}{\alpha}  \right)\left( \Halpha(\PAstring |  \mathbf{C}  \Eve)_{\rho} - \lfixed + 2 \right) } 
\end{aligned}
\end{equation}
\end{restatable}
The proof can be found in Appendix~\ref{Appendix:variable}.
		
While here we have focused on proving an analogue of the Leftover Hashing Lemma for {\Renyi} entropy, a similar result for the smooth min-entropy version can be obtained by exactly the same proof, except that when conditioning on the event $\Omega$, one should use the the subnormalized partial states.

In order to use \cref{lemma:variableinputhashing}, which justifies the use of the LHL, we have to compute bounds on the {\Renyi} entropy $ \Halpha(\PAstring | \mathbf{C} \Eve)_{\rho}$, which is computed on the state just prior to privacy amplification in Case~\ref{PAcase3}. However, we expect that if the registers that were discarded to produce $\PAstring$ from $S_1^n$ are completely determined by the register $\mathbf{C}$, then this entropy should be the same as the value before the discarding process, since the conditioning register $\mathbf{C}$ could be used to isometrically convert between the values before and after discarding some registers. We formalize this claim in the following Lemma and subsequent discussion.

\begin{restatable}[]{lemma}
  {variableinputhashingsecondlemma}  
 \label{lemma:renyiinvariance}
	Suppose $\rho_{\rawkey \mathbf{C} \Eve},\rho_{\PAstring \mathbf{C} \Eve}$ are states that are classical in $\mathbf{C}$, and related to each other as follows: letting $\rawkey_{\bar{c}}$ be a register containing the support of the conditional state $\rho_{\rawkey|\mathbf{C} = \bar{c}}$, there exist isometries $V^{(\bar{c})}_{\rawkey_{\bar{c}} \rightarrow \PAstring}$ such that\footnote{\cref{eq:unitaryaction} is a well-defined expression despite the fact that $V$ is not defined on all of $\rawkey \mathbf{C}$, because $\rho_{\rawkey \mathbf{C} \Eve}$ is only supported on the subspace on which $V$ is defined.}
	\begin{equation}  \label{eq:unitaryaction}
		\begin{gathered}
			V\rho_{\rawkey \mathbf{C} \Eve} V^\dagger =\rho_{\PAstring \mathbf{C} \Eve}, \text{ where} \\
			V \coloneqq \sum_{\bar{c}} V^{(\bar{c})}_{\rawkey_{\bar{c}} \rightarrow \PAstring} \otimes \ket{\bar{c}}\bra{\bar{c}}_{\mathbf{C}}.
		\end{gathered}
	\end{equation}
	Then we have
	\begin{equation} \label{eq:renyiinvariance}
		\Halpha(\PAstring |\mathbf{C} \Eve)_\rho = \Halpha(\rawkey | \mathbf{C} \Eve)_\rho.
	\end{equation}
\end{restatable}
The proof can be found in \cref{Appendix:variable}.

To apply \cref{lemma:renyiinvariance} in comparing Cases~\ref{PAcase1},~\ref{PAcase2} and~\ref{PAcase3} described previously, we can begin by viewing $\rawkey$ as being $S_1^n$ in Case~\ref{PAcase1} or $\hat{S}_1^n$ in Case~\ref{PAcase2}. If the protocol satisfies the condition that the positions and values of discarded outcomes are fixed by the public announcements $\mathbf{C}$, 
we can define operations $V^{(\bar{c})}_{\rawkey_{\bar{c}} \rightarrow \PAstring}$ that simply drop the discarded outcomes specified by $\bar{c}$, 
and it is not difficult to show the state $\rho_{\PAstring \mathbf{C} \Eve}$ in Case~\ref{PAcase3} has the following properties:
\begin{enumerate}
	\item These $V^{(\bar{c})}_{\rawkey_{\bar{c}} \rightarrow \PAstring}$ operations are indeed isometries, and $\rho_{\PAstring \mathbf{C} \Eve}$ is related to $\rho_{\rawkey \mathbf{C} \Eve}$ in the sense expressed in \cref{eq:unitaryaction}. 
	
	\item $\rho_{\PAstring \mathbf{C} \Eve}$ satisfies the conditions of \cref{lemma:variableinputhashing}, and hence \cref{eq:varLHL} is valid.
	
	\item $\rho_{\PAstring \mathbf{C} \Eve}$ satisfies the conditions of \cref{lemma:renyiinvariance}, and hence $H_\alpha(\PAstring |\mathbf{C} \Eve)_\rho$ in \cref{eq:varLHL} can be replaced with $H_\alpha(\rawkey |\mathbf{C} \Eve)_\rho$.
\end{enumerate}
(Basically, the above statements hold because under that protocol condition, for each value $\bar{c}$, the output length of $V^{(\bar{c})}_{\rawkey_{\bar{c}} \rightarrow \PAstring}$ is fixed, and all the discarded positions have fixed values so there are no ``collisions''.)

With this, we see that for Case~\ref{PAcase3} the bound in \cref{eq:varLHL} holds with $\Halpha(\PAstring | \mathbf{C} \Eve)_\rho$ replaced by $\Halpha(S_1^n | \mathbf{C} \Eve)_\rho$ from Case~\ref{PAcase1} or $\Halpha(\hat{S}_1^n | \mathbf{C} \Eve)_\rho$ from Case~\ref{PAcase2}; in particular, for the purposes of this thesis this means the proofs of \cref{theorem:fixedlengthIID,theorem:variablelengthIID} are valid\footnote{Note that when we further simply these expressions to von Neumann entropies using  \cref{lemma:contrenyi}, then that simplification depends  and these the bound on $\Halpha(S_1^n |\mathbf{C} \Eve)$  on the dimension of $S$ versus $\hat{S}$.} even if we apply the procedure in Case~\ref{PAcase3} rather than Case~\ref{PAcase2}.
To qualitatively summarize, under that protocol condition, the bounds obtained on the privacy amplification procedure in QKD are unaffected if the actual protocol implements Case~\ref{PAcase3} in place of Case~\ref{PAcase1} or Case~\ref{PAcase2}.

	\section{Summary and Outlook} \label{sec:variablesummary}
In this chapter we presented a security proof for variable-length QKD protocols against IID collective attacks. These protocols  are highly relevant for implementations since they do not require users to characterize their channel before running the QKD protocol. Instead, they include instructions for adjusting the length of the final key, and the amount of error-correction information, for every possible observation during the protocol. The key ingredient in this analysis was the construction of a statistical estimator that lower bounds the relevant entropic quantity, and the use the {\Renyi} entropies in the proof (see \cref{remark:technicalreason}). In fact, we will see that the MEAT framework in \cref{chap:MEAT} also uses similar ideas: however it notably makes do with a \emph{weaker} statistical estimator than the one we demanded in this chapter (see \cref{theorem:entropytovarlength}).

These results eliminate the typical trade-off between probability of the protocol accepting and the output key length it produces when it accepts that is imposed by fixed-length protocols. They also remove the need for explicit channel characterization prior to the QKD protocol run.  We exemplified our results by studying the performance of the qubit BB84 protocol implemented in this fashion. We showed that the variable-length implementation leads to a significant improvement in the expected key rate compared to fixed-length implementations, especially for scenarios where the channel is chaotic and unpredictable. We also applied our variable-length framework towards the decoy-state BB84 protocol, using the analysis from Ref.~\cite{Kamin2025}.

We highlight that  our proof approach relies on a Leftover Hashing Lemma for {\Renyi} entropies that was only recently developed in Ref.~\cite{dupuis_privacy_2023}. It does not seem entirely straightforward to construct a similar  analysis using  Leftover Hashing Lemma versions based on smooth min-entropy.

Finally, we identified and addressed a subtle gap in the analysis of sifting and privacy amplification arising from the variable length of the pre-amplification string $\PAstring$. Our fix acts as retroactively, since it does not require any changes to the protocol implementation or the output key lengths.

Our analysis so far has the significant limitation that it applies only to IID collective attacks. This is not a reasonable assumption to make in QKD, since Eve can always implement an attack that violates it (for instance by attacking odd and even round signals differently). We therefore turn to the postselection technique in the next chapter, which allows our  security analysis in this chapter to be lifted to hold against coherent attacks.

\chapter{Postselection technique lift to coherent attacks} \label{chap:postselection}

\epigraph{
Where we break free from the shackles of IID collective attacks; where we obtain a reduction to IID collective attacks using permutation
symmetries; and where we
begin to incessantly count the dimensions of Alice’s and Bob’s systems (and agonize when they grow).
}{}

Proving the security of QKD protocols against coherent attacks is a challenging task, since the security proof must hold for any arbitrary strategy employed by the adversary. This stands in contrast to the IID collective attack setting, where Eve interacts with each round in an identical and independent manner. A powerful approach to bridge these two regimes is the postselection technique \cite{christandl_postselection_2009,nahar_postselection_2024}, which provides conditions under which a security proof against IID collective attacks can be “lifted’’ to a proof that holds against general coherent attacks. In this chapter, we give a complete and self-contained treatment of the postselection technique as it applies to practical QKD protocols, which is based on Ref.~\cite{nahar_postselection_2024}. The  technique itself was originally proposed in Ref.~\cite{christandl_postselection_2009}. 

We begin with an overview of the underlying mathematical tools in \cref{sec:deFinetti}, focusing on the de Finetti reductions that lie at the heart of this method. We then explain how these reductions are used in QKD security proofs in \cref{sec:deFinettiQKD}. In \cref{sec:postselectionoptics}, we investigate how to apply the postselection technique to optical implementations of QKD, which requires additional care due to the infinite-dimensional nature of photonic systems. In \cref{sec:postselectionapplication}, we combine all these ingredients to obtain explicit key rates for the Qubit BB84 and Decoy-state BB84 protocol against coherent attacks. In \cref{sec:postselectionsummary} we summarize the results in this chapter.  

While the postselection technique is well established in principle, the existing literature did not provide a fully correct and operational proof suitable for realistic prepare-and-measure QKD protocols until Ref.~\cite{nahar_postselection_2024}, on which this chapter is based. In particular, several obstacles had to be resolved before a complete security proof could be obtained. The main issues addressed in this chapter are:
\begin{enumerate}
\item \textbf{Infinite-dimensional optical systems: }
Bob’s detection apparatus is inherently infinite-dimensional, and realistic models of his measurement process must reflect this. Although several tools exist to handle infinite-dimensional detectors \cite{fung_universal_2011, gittsovich_squashing_2014, zhang_security_2021, upadhyaya_dimension_2021}, many of them are not directly compatible with the postselection technique. We show how, and under what conditions, these tools can be utilized with the postselection technique.
Moreover, in decoy-state protocols, Alice’s emitted optical states also inhabit an infinite-dimensional Hilbert space. We rigorously apply the source maps \cite{nahar_imperfect_2023} to reduce this to the standard finite-dimensional setting of tagged states \cite{gottesman_security_2004}, allowing the postselection technique to be applied without compromising correctness.
\item \textbf{A technical flaw in prior formulations of the postselection technique: }
The original proof in Ref.~\cite{christandl_postselection_2009} contained a step that is not valid as stated \cite[Note added after publication]{renner_simplifyingps_2010}. An attempted correction in \cite[Section 3.3.2]{belzig_studying_2020} also introduced an error. In this chapter, we present the first  correct and complete version of this step.

\item \textbf{Incorporating the fixed marginal constraint required for prepare-and-measure QKD: }
Standard formulations of the postselection technique assume that the marginal on Alice’s system is unconstrained. However, in generic prepare-and-measure protocols, Alice’s marginal after the source-replacement scheme (see \cref{subsec:sourcereplacement}) is fixed by design, and the security proof must operate under this constraint. Prior versions of the postselection technique could not accommodate this, and therefore were inapplicable to prepare-and-measure protocols.\footnote{More precisely, these methods could still be applied, but would result in a reduction to an IID collective attack analysis that was required to hold for \emph{all} IID states - including those that do not satisfy the required marginal property. This would typically yield zero key rates for generic QKD protocols.} We extend the framework so that it naturally handles fixed-marginal conditions, thereby enabling its use in realistic QKD analyses.
\end{enumerate}
Beyond establishing the first fully rigorous version of the postselection technique tailored to QKD, we also obtain significant performance improvements, leading to much tighter key rates than previously achievable with postselection-based proofs (these results are primarily due to Shlok Nahar). Finally, we show how the entire analysis can be adapted to variable-length protocols from \cref{chap:variable}, which is essential for achieving optimal expected key rates in many practical settings.

\section{De Finetti Reductions} \label{sec:deFinetti}

Quantum de Finetti theorems \cite{renner_security_2005,christandl_postselection_2009,arnon-friedman_finetti_2015,fawzi_quantum_2015} are useful in reducing the analysis of various quantum information processing tasks to the IID case. In this work, we focus on quantum de Finetti reductions of following form used in Refs.~\cite{christandl_postselection_2009,arnon-friedman_finetti_2015,fawzi_quantum_2015}, where a permutation invariant state $\bar{\rho}_{A_1^n B_1^n}$ (see \cref{def:permutationInvariance}) satisfies:
\begin{equation}
    \bar{\rho}_{A_1^n B_1^n} \leq \cost\; \tau_{A_1^n B_1^n},
\end{equation}
where $\tau_{A_1^n B_1^n}=\int \sigma_{AB}^{\otimes n}\text{d}\sigma_{AB}$ is a normalized density matrix, and $g_{n,x} = 
\dimSym{x} = \binom{n+x-1}{n}$ is the dimension of the symmetric subspace $Sym{x}$, which we define in \cref{def:symmetricsubspace}. Although we will not require many detailed properties of this subspace to understand the present chapter, its dimension plays an important role. An easily computable upper bound on the dimension of the symmetric subspace is given in \cref{sec:postselectionapplication}.
We refer to a state of the form $\tau_{A_1^nB_1^n}=\int \sigma_{AB}^{\otimes n}\text{d}\sigma_{AB}$ as a ``de Finetti state".
Typically, in reducing the analysis of quantum information processing tasks to the IID case, the following factors come into play:
\begin{enumerate}
    \item The value of $\cost$ : This should be as small as possible, as it appears as a penalty in the reduction to IID states. 
    \item The integral over IID states in $\tau_{A^nB^n}$ : This should be such that the integral is only over states for which the task has been ``analyzed'', in the sense that some security property has been proven for all such IID states.
\end{enumerate}

In this section we will present several improvements to the value of $\cost$ from Ref.~\cite{nahar_postselection_2024}. These improvements are of two types.  The first improves the dimensional scaling ($\cost$) in   \cite[Lemma 3.1]{fawzi_quantum_2015} for generic states. The second improvement shows that the dimensions can be reduced for states that are invariant under certain symmetries, as an extension of Ref.~\cite{arnon-friedman_finetti_2015} to the quantum case.
Proofs of these improved statements are deferred to Ref~\cite{nahar_postselection_2024}.

\subsection{A simple de Finetti reduction}

We will now prove a simple de Finetti reduction \cite{christandl_postselection_2009}, which provides intuition for why statements of this form can be expected to be true.  The proof strategy below mirrors that of all the de Finetti reductions considered in this chapter, although later refinements are significantly more technically involved. We begin by defining the symmetric subspace, and a notion of permutation invariance for operators acting on tensor-product spaces. We use $\permset{n}$ to denote the set of all possible permutations on $n$ objects.

\begin{definition}(Symmetric subspace \cite[Chapter 7]{watrous_theory_2018}) \label{def:symmetricsubspace}
Let $\mathcal{H}$ be a finite-dimensional Hilbert space with  dimension $d_H$, that is $\mathcal{H} \cong \mathbb{C}^{d_H}$.
For each permutation $\perm \in \permset{n}$, let $\permutation{d_H}{n}$ denote the standard unitary representation of $\perm$ acting on $\mathcal{H}^{\otimes n}$.
The \term{symmetric subspace} of $\mathcal{H}^{\otimes n}$ is defined as
\[
\Sym{\mathcal{H}} \coloneqq
\left\{ \ket{\psi} \in \mathcal{H}^{\otimes n} :
\permutation{d_H}{n} \ket{\psi} = \ket{\psi}
\ \text{for all } \perm \in \permset{n} \right\}.
\]
The dimension of the $n$-fold symmetric subspace is given by
\[
\dimSym{d_H} = \binom{n+d_H-1}{n}.
\]
\end{definition}

\begin{definition}(Permutation-invariance of matrices \cite[Chapter 7]{watrous_theory_2018})
\label{def:permutationInvariance}
Given a matrix $\bar{\rho}_{H_1^n} \in \linear\!\left(H_1^n\right)$ and a permutation $\perm \in \permset{n}$ of its subsystems, we denote the action of $\perm$ on $\bar{\rho}_{H_1^n}$ as
\[
\bar{\rho}_{H_1^n}  \;\mapsto\; \permutation{d_H}{n}\, \bar{\rho}_{H_1^n} \,\permutation{d_H}{n}^{\dagger},
\]
where $\mathcal{H} \cong \mathbb{C}^{d_H}$ and $\permutation{d_H}{n}$ is the standard unitary representation of $\perm$ on $\mathcal{H}^{\otimes n}$.
We say that $\bar{\rho}_{H_1^n}$ is \term{permutation invariant} if it is invariant under the action of all permutations $\perm \in \permset{n}$.
\end{definition}

Note that permutation invariant states are \emph{not} necessarily supported on the symmetric subspace.
For example, the maximally mixed state is permutation invariant but has full support on the entire Hilbert space. Nevertheless, the following result holds.

\begin{lemma}
[Permutation-invariant states admit symmetric purifications {\cite[Lemma~4.2.2]{renner_security_2005}}] \label{lemma:permutationpurification}
Let $\bar{\rho}_{H_1^n}$ be a permutation-invariant state\footnote{This can be straightforwardly generalized to hold for matrices on $\Pos(H_1^n)$.} on $\dop{=}(H_1^{n})$, and let $\mathcal{H}$ be the finite-dimensional Hilbert space associated with $H$.
Then there exists a purification of $\bar{\rho}_{H_1^n}$ supported on the symmetric subspace
$
\Sym{\mathcal{H} \otimes \mathcal{H}}$.
\end{lemma}

Moreover, the the symmetric subspace itself can be characterized using a mixture of IID states using the following lemma.

\begin{lemma}
[Symmetric projector as a Haar mixture of IID states]
\label{lm:symprojHaar}
Let $\mathcal{H} \cong \mathbb{C}^{d_H}$ be a finite-dimensional Hilbert space, and let
$\Pi_{\Sym{\mathcal{H}}}$ denote the projector onto the symmetric subspace
$\Sym{\mathcal{H}} \subset \mathcal{H}^{\otimes n}$.
Then $\Pi_{\Sym{\mathcal{H}}}$ admits the representation
\begin{equation}
\Pi_{\Sym{\mathcal{H}}}
= \dimSym{d_H} \int \sigma_H^{\otimes n} \, \mathrm{d}\sigma_H,
\end{equation}
where the integral is taken over pure states $\sigma_H = \ketbra{\psi}{\psi}$ on $\mathcal{H}$
with respect to the unitarily invariant (Haar) probability measure.
\end{lemma}
\begin{proof}
Define
\begin{equation}
M \coloneqq \int \sigma_H^{\otimes n}\, \mathrm d\sigma_H,
\end{equation}
where the integral ranges over pure states $\sigma_H=\ketbra{\psi}{\psi}$ on $\mathcal{H}$
with respect to the unitarily invariant (Haar) probability measure. First, note that for every pure state $\ket{\psi}\in\mathcal{H}$, we have $\ket{\psi}^{\otimes n}\in \Sym{\mathcal{H}}$. Thus, we have
\begin{equation}
\Pi_{\Sym{\mathcal{H}}}\, M \,\Pi_{\Sym{\mathcal{H}}} = M,
\end{equation}
so $M$ has support contained in $\Sym{\mathcal{H}}$.

Let $U$ be any unitary on $\mathcal{H}$. Using Haar invariance of the measure,
\begin{equation}
U^{\otimes n} M (U^\dag)^{\otimes n}
= \int (U\sigma_H U^\dag)^{\otimes n}\, \mathrm d\sigma_H
= \int \sigma_H^{\otimes n}\, \mathrm d\sigma_H
= M.
\end{equation}

Now, consider the restriction of $M$ to the symmetric subspace.
Since $M$ is supported on $\Sym{\mathcal{H}}$ and commutes with $U^{\otimes n}$
for every unitary $U$, Schur's lemma \cite{watrous_theory_2018} implies that $M$ acts as a scalar multiple
of the identity on $\Sym{\mathcal{H}}$. This step relies on the fact that the action of $U^{\otimes n}$ on the symmetric subspace
$\Sym{\mathcal H}$ is irreducible, so that any operator supported on $\Sym{\mathcal H}$ and
commuting with $U^{\otimes n}$ must be proportional to the identity on $\Sym{\mathcal H}$.

Thus, there exists a constant $c\ge 0$ such that
\[
M = c\, \Pi_{\Sym{\mathcal{H}}}.
\]
Taking the trace of both sides,
\[
\Tr(M) = \int \Tr(\sigma_H^{\otimes n})\, \mathrm d\sigma_H = \int 1\, \mathrm d\sigma_H = 1,
\]
since $\sigma_H^{\otimes n}$ is a rank-one density operator and the Haar measure is normalized. Thus, $c=1/\dimSym{d_H}$, since $\Tr(\Pi_{\Sym{\mathcal{H}}}) = \dimSym{d_H}$
\end{proof}

Combining these statements, we obtain the following de Finetti reduction.

\begin{corollary}(Simple de Finetti reduction without fixed marginal \cite[Lemma 2]{christandl_postselection_2009})
\label{cor:simpleDeFinetti}
Let $\rho_{H_1^n} \in \dop{=}(H_1^n)$ be a permutation invariant state. Then, there exist a probability measure $\mathrm d\sigma_H$ on the set of states $\sigma_H \in \dop{=}(H)$ on $H$ such that
\begin{equation}
\label{eq:simpleDeFinettiNoFixedMarginal}
\rho_{H^n} \;\leq\; g_{n,d^2_H}\int \sigma_H^{\otimes n}\,\mathrm d\sigma_H,
\end{equation}
where $\cost = \dimSym{x}$.
\end{corollary}
\begin{proof}
    By \cref{lemma:permutationpurification}, there exists a purification $\rho_{H_1^n  K_1^n}$ of $\rho_{H_1^n}$, where $K \cong H$, which is supported on the symmetric subspace $\Sym{H \otimes K}$. Using \cref{lm:symprojHaar}, we obtain 
    \begin{equation}
        \rho_{H_1^n  K_1^n} \leq \Pi_{\Sym{H \otimes K}} = \dimSym{d_H d_K} \int \sigma_{H  K}^{\otimes n} d \sigma_{H  K} 
    \end{equation}
    where the integral is with respect to the unitarily invariant measure. The required statement then follows from simply taking the partial trace over $K_1^n$ in the above expression. 
\end{proof}
Notice that the dimension that shows up in $\cost$ is $d_H^2$, which is due to the fact that we consider the symmetric subspace on $H_1^n \otimes K_1^n$, where the purification lives.

\subsection{Improvements} \label{subsec:Improvement}
Notice that \cref{cor:simpleDeFinetti} involves an integral over \emph{all} IID states. In many applications, however, we would like a de Finetti reduction that preserves a fixed marginal on $A$. A statement of this form, along with a variety of improvements, is obtained below. For the purposes of this thesis, we simply state the improved de Finetti reductions below and apply them in \cref{sec:postselectionapplication} (since the author was not a primary contributor to those results).
We refer the reader to Ref.~\cite{nahar_postselection_2024} for a detailed discussion of the proof ideas, relevant techniques, extensions to group symmetries, and related works.

\begin{corollary}(\cite[Corollary 1]{nahar_postselection_2024}) \label{cor:generalMixedDeFinetti}
    Let $\promise \in \Pos(\mathbb{C}^{d_{A}})$ and $\state$ be any permutation-invariant extension of $\left(\promise\right)^{\otimes n}$. Then there exists a probability measure d$\sigma_{AB}$ on the set of non-negative extensions $\sigma_{AB}$ of $\promise$, such that
    \begin{align}
        \state \leq \cost\int \sigma_{AB}^{\otimes n} \text{d} \sigma_{AB}
    \end{align}
    holds for $x=d_{A}^2d_{B}^2$.
\end{corollary}

Note that there is a dramatic improvement\footnote{For the purposes of QKD key rates, this is indeed dramatic, as we shall soon see in \cref{sec:postselectionapplication}} over Corollary 3.2 from Ref.~\cite{fawzi_quantum_2015}, since we go from $x=d_{A}^2d_{B}^4$. Additionally, with this improvement we obtain the earlier de Finetti reduction without a fixed marginal \cite[Lemma 2]{christandl_postselection_2009} (see \cref{cor:simpleDeFinetti}) by considering $A$ to be a trivial system, thus unifying both results. 

A further improvements is available in settings where the underlying state can be assumed to be block diagonal; these are stated below.
For pedagogical reasons, when applying de Finetti reductions to QKD in \cref{sec:deFinettiQKD}, we present results based on the de Finetti reduction with fixed marginal (\cref{cor:generalMixedDeFinetti}), rather than the strengthened bounds discussed below.
Since these improvements can be incorporated in a straightforward manner, the remainder of this section may be skipped without affecting the understanding of the rest of this chapter.
Nevertheless, the results presented here are useful for improving finite-size bounds in practical key rate computations. We start with the definition of IID-block-diagonal states.

\begin{definition}[IID-block-diagonal states] \label{def:iidBLockDiagonal}
    Given a matrix $\state \in \linear((\C{d_{A} d_{B}})^{\otimes n}),$ and a set of orthogonal projections $\{\Pi_i\}_{i=1}^k \subset \linear(\C{d_{A} d_{B}})$, we say that the matrix is \textit{IID-block-diagonal} if
    \begin{align*}
        \state = \sum_{\vec{j}\in [k]^n} \Pi_{\vec{j}} \state \Pi_{\vec{j}}
    \end{align*}
    where $\Pi_{\vec{j}} \coloneqq \Pi_{j_1} \otimes \ldots \otimes \Pi_{j_n}$.
    We denote the rank of projector $\Pi_i$ as $d_i$; it corresponds to the dimension of the $i^{\text{th}}$ block. 
\end{definition}
With this definition, we state the Corollary below, which will be the main result we use in later analysis.

\begin{corollary}(de Finetti with IID block-diagonal symmetry and marginal constraint \cite[Corollary 2]{nahar_postselection_2024}) \label{cor:blockDiagDeFinetti}
    Let $\promise \in \Pos(\mathbb{C}^{d_{A}})$ and $\state$ be any permutation-invariant and IID-block-diagonal extension of $\left(\promise\right)^{\otimes n}$ with respect to projections $\{\Pi_{i}^A\otimes\Pi_{j}^B\}_{i,j=1}^{k_A,k_B}$ of dimension $\{\blockdim{i}{A}\blockdim{j}{B}\}_{i,j=1}^{k_A,k_B}$, where $\blockdim{i}{A}$ and $\blockdim{j}{B}$ are the ranks of $\Pi^A_{i}\in \linear(\C{d_A})$ and $\Pi^B_{j}\in \linear(\C{d_B})$ respectively.
    Then there exists some probability measure d$\sigma_{AB}$ over the set of block-diagonal extensions $\sigma_{AB}$ of $\promise$ such that
    \begin{align}
        \state \leq \cost \int \sigma_{AB}^{\otimes n} \ \text{d}\sigma_{AB},
    \end{align}
    where $x=\sum_{i,j=1}^{k_A,k_B} \blockdim{i}{A}^2\blockdim{j}{B}^2$.
\end{corollary}

Although the IID-block-diagonal condition might seem restrictive, we show in \cref{sec:postselectionapplication} that optical implementations often naturally result in such IID-block-diagonal structure. Thus, this would greatly tighten the analysis of optical implementations of quantum information protocols.

\section{Correct application of De Finetti reductions in QKD} \label{sec:deFinettiQKD}

   In this section, we fill in a missing gap in Ref.~\cite{christandl_postselection_2009}, in the reduction of QKD security proofs from arbitrary attacks to IID collective attacks, first noticed in \cite{belzig_studying_2020}. We also explain how the postselection technique can be applied to prepare-and-measure protocols. The approach from Ref.~\cite{christandl_postselection_2009} reduces the security proof of QKD protocols for arbitrary states to that of IID states in two steps. The first step is a reduction from the security of arbitrary states to the security of the state de Finetti state $\tau_{A_1^nB_1^n}$ (with Eve holding a purification). The second is a reduction from the security for $\tau_{A_1^nB_1^n}$ (with Eve holding a purification) to that of IID states. The second step in their analysis is argued intuitively and is not on sound mathematical grounds. Here, we present a rigorous proof of this step. For the sake of completeness, we explain the first step as well.

\subsection{Using de Finetti reductions for QKD}
\label{subsec:usingdeFinetti}

 We first recall some of our notation for QKD protocols.
Following \cref{def:epsSecPromise}, after performing source-replacement, the \nameref{prot:qkdprotocol} can be equivalently identified with $\{ \protMap{\lkey} ,\promise \}$, where $\protMap{\lkey}  \in \CPTP(A_1^nB_1^n, K_A K_B \allpublic )$ and $\lkey(\cobs)$ is the length of the output key the protocol produced, upon observing $\cobs$ in the public announcements registers $\CP_1^n$, and error-verification passing. Moreover, since $\epscorr$-correctness is already satisfied, we only need to concern ourselves with the satisfying the secrecy requirements (see \cref{lemma:securityfromcorrandsecrecy}). 
        The ideal QKD protocol $\protMapId{\lkey}\in \CPTP(A_1^nB_1^n, K_A K_B \allpublic ) $ is one which implements the actual QKD protocol $\protMap{\lkey}$, and then applies the $\idealmap$ map (see \cref{sec:securitydefinition}), which replaces Alice and Bob's key registers with the perfect key of length of appropriate length.
 The overall secrecy requirement of a QKD protocol can be described in terms of the maps $\Tr_{K_B} \circ \protMap{\lkey}$ and $\Tr_{K_B} \circ \protMap{\lkey}$ (see \cref{def:epsSecPromise}).

We now define what it means for a map to be permutation-invariant. Note that in this definition we also correct a technical error in Ref.~\cite{christandl_postselection_2009} regarding the order in which the maps are applied.
\begin{definition}[Permutation-invariance of maps] \label{def:permInvariantMaps}
    A linear map $\Delta \in \mapset(A_1^n,B)$ is \term{permutation-invariant}, if for every $\perm \in \permset{n} $, there exists a $G_{\perm} \in \CPTP(B,B)$ such that
    \begin{equation}
        G_\perm \circ \Delta \circ \mathcal{W}_{\perm} = \Delta
    \end{equation}
    where $\mathcal{W}_{\perm} (\cdot) = \permutation{d_{A}}{n} \left(\cdot\right) \permutation{d_{A}}{n}^\dagger$.
\end{definition}
Arguably, the above property might be better described as ``covariance'' rather than ``invariance'', since for instance it does not require that the map is literally ``invariant'' in the sense $\Delta \circ \mathcal{W}_{\perm} = \Delta$. However, for this work we shall follow the existing terminology in the field.

\subsubsection{Satisfying permutation invariance}

Note that typical QKD protocols are not permutation invariant. Instead, permutation invariance must be enforced explicitly, and requires the optimal random permutation step of \nameref{prot:qkdprotocol} to be implemented.  This idea is formalized in \cref{lemma:satisfyingpermutationinvariance} below and in Ref.~\cite{christandl_postselection_2009}. A proof is provided in Appendix.~\ref{Appendix:PS}.

The idea is simple: if a channel $\mathcal{F} \in \CPTP(A^n,B)$ begins by applying a uniformly random permutation to its input registers, followed by operations that do not depend on the chosen permutation, and if the applied permutation is recorded in a classical output register, then $\mathcal{F}$ is permutation invariant in the sense of the above definition (since we can undo the effect of the permutation by modifying the register in which its value is announced). This holds even though $\mathcal{F}$ need not satisfy $\mathcal{F}\circ\mathcal{W}_{\perm}=\mathcal{F}$.
Such a random permutation can be implemented using approximately $n \log (n)$ uniformly random bits. In the QKD setting, these bits may be generated locally by one party and publicly announced, and therefore do not require any pre-shared secret key \cite[Appendix B]{nahar_postselection_2024}. 

\begin{restatable}[Satisfying permutation invariance]{lemma}{satisfyingpermutationinvariance}
\label{lemma:satisfyingpermutationinvariance}
Consider the QKD protocol $\{ 
\protMap{\lkey} , \sigma_A\}$   \\
where $ \protMap{\lkey} \in \CPTP(A_1^n B_1^n , K_A \allpublic)$ is the QKD protocol map describing \nameref{prot:qkdprotocol}.
Define a new protocol map $\protMap{\lkey,\mathrm{perm}}$ by first applying a uniformly random permutation to the systems $A_1^n B_1^n$, announcing the permutation in a classical register $C_{\mathrm{perm}}$, and then applying $\protMap{\lkey}$. Explicitly,
\begin{equation}
\label{eq:perm_protocol_map}
\protMap{\lkey,\mathrm{perm}}(\cdot)
=
\protMap{\lkey}
\circ \left( 
\sum_{\pi \in \permset{n}} \frac{1}{n!}
\permutation{n}{d_A d_B}
(\cdot)
\permutation{n}{d_A d_B}^{\dagger}
\otimes
\ketbra{\pi}_{C_{\mathrm{perm}}} \right).
\end{equation}
Then the following statements hold:
\begin{enumerate}
    \item The protocol $\{ \protMap{\lkey,\mathrm{perm}}, \sigma_A \}$ describes
    \nameref{prot:qkdprotocol} with the optional permutation step applied to the
    $X_1^n , Y_1^n$ registers, where the applied permutation is announced in the (additional) register 
    $C_{\mathrm{perm}}$, which is made available to Eve.
    \item The difference between the real and ideal protocols,
    \[
  \Tr_{K_B} \circ   \protMap{\lkey,\mathrm{perm}}
    -
   \Tr_{K_B} \circ    \idealmap \circ \protMap{\lkey,\mathrm{perm}},
    \]
    is permutation invariant according to \cref{def:permutationInvariance}. 
\end{enumerate}
\end{restatable}

The following lemma (see Appendix.~\ref{Appendix:PS} for the proof) formalizes the intuitive statement that, if one is only
concerned with proving secrecy against IID \ collective attacks, then the
explicit permutation step does not affect the security analysis. We stress,
however, that this permutation step \emph{must still be implemented} in practice
in order for the analysis developed in this work to be applicable. Throughout
this chapter, we refer to $\protMap{\lkey}$ as
\nameref{prot:qkdprotocol} without the permutation step, and to
$\protMap{\lkey,\mathrm{perm}}$ as \nameref{prot:qkdprotocol} with the
permutation step.

\begin{restatable}{lemma}{permutationdoesnotmatter} \label{lemma:permutationdoesnotmatter}
Let $\{ \sigma_A, \protMap{\lkey} \}$ be a QKD protocol, and let
$\{ \sigma_A , \protMap{\lkey,\mathrm{perm}} \}$ be a related QKD protocol in
which the maps are related by a random permutation
(\cref{eq:perm_protocol_map}). Then,
\begin{equation}
\begin{aligned}
&\tracedist{
\left(
\left(
\Tr_{K_B} \circ \protMap{\lkey,\mathrm{perm}}
-
\Tr_{K_B} \circ \protMapId{\lkey,\mathrm{perm}}
\right)
\otimes \idmap_{E_1^n}
\right)
\left[
\rho_{ABE}^{\otimes n}
\right]
}
\\
&=
\tracedist{
\left(
\left(
\Tr_{K_B} \circ \protMap{\lkey}
-
\Tr_{K_B} \circ \protMapId{\lkey}
\right)
\otimes \idmap_{E_1^n}
\right)
\left[
\rho_{ABE}^{\otimes n}
\right]
}.
\end{aligned}
\end{equation}
That is, security against IID collective attacks for $\protMap{\lkey}$
implies security against IID collective attacks for
$\protMap{\lkey,\mathrm{perm}}$.
\end{restatable}

\subsubsection{Using the permutation invariance of protocols.}

We proceed in a manner similar to Ref.~\cite{christandl_postselection_2009}, and prove \cref{lemma:endOfStepOne}, which can be used to relate $\epssecret$-secrecy of arbitrary states to the $\epssecret$-secrecy of a state that is a purification of a de Finetti state. To do so, we prove \cref{lemma:psLemma,lemma:tau}. Note that \cref{lemma:psLemma} is also proved in the proof of \cite[Theorem 2]{christandl_postselection_2009}. All proofs can be found in Appendix.~\ref{Appendix:PS}

\begin{restatable}{lemma}{pslemmaone}
    \label{lemma:psLemma}
    Let $\mathcal{F}, \mathcal{F}^\prime \in \mapset(A_1^nB_1^n,K)$ be linear maps such that  $\mathcal{F}-\mathcal{F}^\prime$ is a permutation-invariant linear map. Let  $\rho_{A_1^nB_1^nR^{\prime \prime}} \in \Pos(A^nB^nR^{\prime \prime})$ be any extension of $\rho_{A_1^nB_1^n}$. Then the state $\bar{\rho}_{A_1^nB_1^n} = \frac{1}{n!} \sum_{\perm \in \permset{n}}  \mathcal{W}_\perm (\rho_{A_1^nB_1^n})$ is permutation-invariant, and for any purification $\bar{\rho}_{A_1^n B_1^n R^\prime}$ of that state, we have
    \begin{equation}
        \tracedist{ \left( \left( \mathcal{F} - \mathcal{F}^\prime \right) \otimes \idmap_{R^{\prime\prime}}\right) \left( \rho_{A_1^nB_1^nR^{\prime\prime}} \right) } \leq	\tracedist{ \left( \left(\mathcal{F} - \mathcal{F}^\prime \right) \otimes \idmap_{R^\prime}\right) \left( \bar{\rho}_{A_1^nB_1^nR^\prime} \right) }. 
    \end{equation}
\end{restatable}

\begin{restatable}{lemma}{pslemmatwo}
    \label{lemma:tau} 
    Let $\rho_{A_1^nB_1^n} \in \Pos(A_1^nB_1^n)$ and $\tau_{A_1^nB_1^n} \in \Pos(A_1^nB_1^n)$ be such that $\rho_{A_1^nB_1^n}  \leq \cost \tau_{A_1^n B_1^n}$ for some $\cost > 0 $. Let $\rho_{A_1^nB_1^nR^\prime}$ be any extension of $\rho_{A_1^nB_1^n}$, and let $\tau_{A_1^nB_1^nR}$ be any purification of $\tau_{A_1^nB_1^n}$. Then for any two maps $\mathcal{F}, \mathcal{F}^\prime \in \mapset(A_1^nB_1^n,K)$, 
    \begin{equation}
        \tracedist{ \left( \left( \mathcal{F} - \mathcal{F}^\prime \right) \otimes \idmap_{R^\prime} \right) \left( \rho_{A_1^nB_1^nR^\prime} \right) } \leq \cost \tracedist{ \left( \left(\mathcal{F} - \mathcal{F}^\prime \right) \otimes \idmap_{R} \right) \left( \tau_{A_1^nB_1^nR} \right) }	
    \end{equation}	
\end{restatable}

Combing these results, we obtain the following lemma which we utilize in the next subsection.

\begin{lemma}
 \label{lemma:endOfStepOne}
    Let $\mathcal{F}, \mathcal{F}^\prime \in \mapset(A_1^nB_1^n,K)$ be such that  $\mathcal{F}-\mathcal{F}^\prime$ is a permutation-invariant map. Let  $\rho_{A_1^nB_1^nR^{\prime\prime}} \in \Pos(A^nB^nR^{\prime\prime})$ with $\Tr_{B^n R^{\prime \prime }}\left(\rho_{A_1^nB_1^nR^{\prime\prime}}\right) = \left(\promise\right)^{\otimes n}$. Then there exists a probability measure d$\sigma_{AB}$ on the set of extensions $\sigma_{AB}$ of $\promise$ such that
    \begin{equation}
        \tracedist{ \left( \left( \mathcal{F} - \mathcal{F}^\prime \right) \otimes \idmap_{R^{\prime\prime}} \right) \left( \rho_{A_1^nB_1^nR^{\prime\prime}} \right) } \leq \cost \tracedist{ \left( \left(\mathcal{F} - \mathcal{F}^\prime \right) \otimes \idmap_{R} \right) \left( \tau_{A_1^nB_1^nR} \right) },
    \end{equation}
    where $\tau_{A_1^nB_1^nR}$ is any purification of $\tau_{A_1^nB_1^n} =  \int \text{d} \sigma_{AB}\  \sigma_{AB}^{\otimes n}$, and $x = d_A^2d_B^2$.
\end{lemma}
\begin{proof}
    
Lemma \ref{lemma:psLemma} lets us assume without loss of generality that the input state to a permutation-invariant QKD protocol is permutation-invariant on $A_1^nB_1^n$. Thus, the input states to such protocols satisfy the de Finetti reductions (\cref{cor:generalMixedDeFinetti}) described in Section \ref{sec:deFinetti}. Using \cref{lemma:tau} on this de Finetti reduction allows us to prove the following lemma after combining \cref{lemma:psLemma,lemma:tau,cor:generalMixedDeFinetti}.\end{proof}

\subsection{Reducing Security of QKD protocols to the IID case}
\cref{lemma:endOfStepOne} allows us to reduce the $\epssecret$-secrecy with prepare-and-measure QKD protocol for any arbitrary input state (satisfying the fixed marginal property on $A$), to the $\epssecret$-secrecy of the protocol when the input state is a purification $\tau_{A^nB^nR}$ of a mixture of IID states $\tau_{A^nB^n}$ with the same fixed marginal on $A$.
In this subsection, we will rigorously reduce the $\epssecret$-secrecy of a QKD protocol acting on $\tau_{A^nB^nR}$ to that of a QKD protocol against IID collective attacks. Note that Ref.~\cite{nahar_postselection_2024} presents two variants of this reduction. One is applicable to both variable-length and fixed-length protocols and is independent of the proof technique used for the IID collective attack security analysis. The other variant is slightly tighter, but applies only to fixed-length protocols and requires the IID collective attack proof to follow a specific structure. In this thesis, we present only the former, as it is the more general of the two.  

We first state the following theorem relating security against IID collective attacks to security against the purification $\tau_{A_1^n B_1^n R}$ of a mixture of IID collective attacks. It is this step which was incorrectly argued in Ref.~\cite{christandl_postselection_2009}.

\begin{restatable}[Postselection Theorem for Variable-length]{theorem}{maintheoremvar} \label{thm:maintheoremvar}
    Consider the \nameref{prot:qkdprotocol} defined via $\{ \protMap{\lkey} , \promise \}$ where  $\protMap{\lkey} \in \CPTP(A_1^nB_1^n, K_A K_B \allpublic) $ is such that the $\epssecret$-secrecy condition (\cref{def:epsSecPromise}) holds for all IID states $\rho_{A_1^nB_1^nE_1^n} = \sigma^{\otimes n}_{ABE}$ satisfying $\Tr_{BE}(\sigma_{ABE}) = \promise$. 
    Let the state $\tau_{A_1^nB_1^n}$ be given by
    \begin{align}
        \tau_{A_1^nB_1^n} = \int \sigma_{AB}^{\otimes n} \text{d} \sigma_{AB},
    \end{align}
    where d$\sigma_{AB}$ is some probability measure on the set of non-negative extensions $\sigma_{AB}$ of $\promise$ and $\tau_{A_1^nB_1^nR}$ be a purification of $\tau_{A_1^nB_1^n}$. Let $\protMap{\lkey^\prime}$ be a variable-length QKD protocol map identical to $\protMap{\lkey}$, except that it hashes to a length \begin{equation}
    \lkey^\prime (\cdot)=  \max\left\{ \floor{\lkey(\cdot)- 2 \log(\cost) - 2\log(1/2\epstilde)},0 \right\}
\end{equation}
instead of $\lkey(\cdot)$, where $x=d^2_Ad_B^2$.
     Then,
    \begin{equation}
     \tracedist{\left( \left(\Tr_{K_B} \circ  \protMap{\lkey^\prime} - \Tr_{K_B} \circ \protMapId{\lkey^\prime}  \right)\otimes \idmap_R \right) (\tau_{A_1^nB_1^nR})} \leq \sqrt{8\epssecret} + \epstilde. 
    \end{equation}
The same result also holds for the protocols that include the random permutation step on the measurement outcomes, that is, for $\protMap{\lkey,\mathrm{perm}}$.
   \end{restatable}
  Note that, while the above result is stated in terms of \nameref{prot:qkdprotocol}, it only requires that the final stage is implemented via privacy amplification. The precise details of the earlier steps do not matter (and hence the same proof works both for $\protMap{\lkey,\mathrm{perm}}$ and $\protMap{\lkey}$. Moreover although all our proofs were stated for prepare-and-measure QKD protocols, they are also applicable to entanglement-based QKD protocols by choosing the fixed marginal to be trivial. This improves upon Ref.~\cite{christandl_postselection_2009}, whose results apply only to entanglement-based QKD protocols.
\begin{proof}

Recall that in the variable-length protocol, multiple events may occur. Either the protocol aborts and does not produce any key, or it accepts and produces a key of length some length. We use $\Omega(\cobs)$ to refer to the event that $\cobs$ is observed in the public announcements. In this case, if error-verification passes, a key of length $\lkey(\cobs)$ (or $\lkey'(\cobs)$, depending on the protocol) is produced. We model aborts as producing a key of zero length, and denote that event via $\Omega_{\mathrm{len}=0}$. Thus, we write the output states for $\protMap{\lkey^\prime},\protMapId{\lkey^\prime}$, for the input state $\tau_{A_1^nB_1^nR}$, as
\begin{equation}
    \begin{aligned}
         \left(\Tr_{K_B}  \circ \protMap{\lkey^\prime} \otimes \idmap_R \right) (\tau_{A_1^nB_1^nR}) &= \sum_{\cobs : \lkey'(\cobs) > 0 } \Pr(\Omega(\cobs) \wedge \OmegaEV ) \tau^{\lkey'(\cobs)}_{K_A \allpublic R | \Omega(\cobs) \wedge \OmegaEV} \\       &+\Pr(\Omega_{\mathrm{len}=0}) \tau^{(\bot)}_{K_A \allpublic R | \Omega_{\mathrm{len}=0} }  \\
        \left(\Tr_{K_B}  \circ \protMapId{\lkey^\prime} \otimes \idmap_R \right) (\tau_{A_1^nB_1^nR}) &= \sum_{\cobs : \lkey'(\cobs) > 0} \Pr(\Omega(\cobs) \wedge \OmegaEV) \tau^{(l^\prime(\cobs), \mathrm{ideal})}_{K_A \allpublic R | \Omega(\cobs) \wedge \OmegaEV} \\
        &+ \Pr(\Omega_{\mathrm{len}=0}) \tau^{(\bot,\mathrm{ideal})}_{K_A \allpublic R | \Omega_{\mathrm{len}=0}}.
    \end{aligned}
\end{equation}
 Similarly, we write the output states for $\protMap{\lkey},\protMapId{\lkey}$ acting upon the input state $\tau_{A_1^nB_1^nE_1^n}$
 \begin{align} \label{eq:temptau}
        \tau_{A_1^nB_1^nE_1^n} = \int \sigma_{ABE}^{\otimes n} \text{d} \sigma,
    \end{align}
as
\begin{equation}
    \begin{aligned}
         \left(\Tr_{K_B} \circ \protMap{\lkey} \otimes \idmap_{E_1^n} \right) (\tau_{A_1^nB_1^nE_1^n}) &= \sum_{\cobs : \lkey(\cobs) > 0} \Pr(\Omega(\cobs) \wedge \OmegaEV) \tau^{(\lkey(\cobs))}_{K_A \allpublic E_1^n | \Omega(\cobs) \wedge \OmegaEV}\\
         &+\Pr(\Omega_{\mathrm{len}=0}) \tau^{(\bot)}_{K_A \allpublic E_1^n | \Omega_{\mathrm{len}=0} } \\
             \left(\Tr_{K_B} \circ \protMapId{\lkey} \otimes \idmap_{E_1^n} \right) (\tau_{A_1^nB_1^nE_1^n}) &= \sum_{\cobs: \lkey(\cobs)>0} \Pr(\Omega(\cobs) \wedge \OmegaEV) \tau^{(\lkey(\cobs), \mathrm{ideal})}_{K_A \allpublic E_1^n | \Omega_{\Omega(\cobs) \wedge \OmegaEV}} \\
             &+\Pr(\Omega_{\mathrm{len}=0}) \tau^{(\bot,\mathrm{ideal})}_{K_A \allpublic E_1^n | \Omega(\mathrm{len}=0)}.
    \end{aligned}
\end{equation}

For each possible observation $\cP_1^n$, we get the following term (which is multiplied by $\Pr(\Omega(\cobs) \wedge \OmegaEV )$) in the secrecy expression we wish to bound, 
\begin{equation}
	 	\zeta(\cobs) \coloneqq  \tracedist{ \tau^{(\lkey(\cobs))}_{K_A \allpublic E_1^n| \Omega(\cobs) \wedge \OmegaEV}  - \tau^{(\lkey(\cobs), \mathrm{ideal})}_{K_A  \allpublic E_1^n| \Omega(\cobs) \wedge \OmegaEV} }. 
	 	\end{equation}
Note that without loss of generality, we can assume that $\zeta(\cobs) < 1$, if equality occurs, then the required bound in \cref{eq:varproofend:b} for the proof follows trivially). The converse bound for privacy amplification \cite[Theorem 7.7]{tomamichel_quantum_2016} allows us to bound the smooth min entropy of the state prior to privacy amplification. Intuitively, we are simply using a statement that shows that if privacy amplification results in an output key that is secret, then the starting state must have atleast some amount of smooth min entropy. This allows us to obtain : 
	  \begin{equation} \label{eq:conversePA}
		  	\Hmin[\sqrt{2 \zeta(\cobs) - \zeta(\cobs)^{ 2} }] (\PAstring | \CP_1^n \CEC \CEV \HEV E_1^n)_{\tau | \Omega(\cobs) \wedge \OmegaEV } \geq \lkey(\cobs).
	 \end{equation}
     We now work with the above bound.

    Since we require $\tau_{A_1^nB_1^nR}$ to be a purification, we purify $\tau_{A_1^nB_1^nE_1^n}$ using $V$ as the purifying system. Thus, $E_1^n V$ is identified with $R$ in the theorem statement. Since each $\sigma^{\otimes n}_{ABE}$ in \cref{eq:temptau} belongs to $\Sym{\mathbb{C}^{d^2_Ad^2_B} }$,  $\tau_{A_1^nB_1^nE_1^n}$ is supported on $\Sym{\mathbb{C}^{d^2_Ad^2_B} }$. Thus, the dimension of the purifying register $V$ is bounded by  $\dim(V) \leq \dim(\Sym{\mathbb{C}^{d^2_Ad^2_B} })  = \cost$ with $x=d_A^2d_B^2$. Then, using the chain rule from \cite[Eq. (8)]{winkler_impossibility_2011} to split off this extra quantum register, we have
	\begin{equation} \label{eq:afterconversepa}
		\begin{aligned}
		\Hmin[\sqrt{2 \zeta(\cobs) - \zeta(\cobs)^{2} }] (\PAstring | \CP_1^n \CEC \CEV \HEV E_1^n V )_{\tau | \Omega(\cobs) \wedge \OmegaEV  } &\geq 
		\Hmin[\sqrt{2 \zeta(\cobs) - \zeta(\cobs)^{2} }](\PAstring | \CP_1^n \CEC \CEV \HEV E_1^n)_{\tau |  \Omega(\cobs) \wedge \OmegaEV } \\
        &- 2 \log( \dim(V))\\
		& \geq \lkey(\cobs) - 2 \log(\cost)
		\end{aligned}
	\end{equation}
	
	 Therefore, consider the modified protocol $\protMap{\lkey^\prime}$ that hashes to \begin{equation}
	     \lkey^\prime(\cdot) = \max \left\{ \floor{\lkey(\cdot) -  2\log(\cost) - 2 \log(1/ 2\epstilde) }, 0 \right\},
	 \end{equation}  
     instead of $\lkey(\cdot)$. In this case, using the Leftover Hashing Lemma (see \cref{lemma:LHL}) for smooth min entropy for the terms where $\lkey'(\cobs) > 0$, we have
	 \begin{equation} \label{eq:converseboundwithLHL}
	 	\begin{aligned}
	&	\tracedist{ \tau^{(\lkey^\prime(\cobs))}_{K_A  \allpublic E_1^nV |  \Omega(\cobs) \wedge \OmegaEV }  - \tau^{(\lkey^\prime(\cobs),\mathrm{ideal})}_{K_A  \allpublic E_1^n V| \Omega(\cobs) \wedge \OmegaEV } }\\
    &\leq \frac{1}{2} 2^{-\frac{1}{2} \left(   	\Hmin[\sqrt{2 \zeta(\cobs)- \zeta(\cobs)^{2} }] (\PAstring | \CP_1^n \CEC \CEV \HEV E_1^n V)_{\tau |  \Omega(\cobs) \wedge \OmegaEV  }  - \lkey^\prime(\cobs) \right)} + 2 \sqrt{2 \zeta(\cobs) - \zeta(\cobs)^{ 2}} \\
		&\leq  \frac{1}{2} 2^{ \log( 2 \epstilde)} + 2 \sqrt{2 \zeta(\cobs) - \zeta(\cobs)^{ 2}} \\
		& =  \epstilde +  2 \sqrt{2 \zeta(\cobs) - \zeta(\cobs)^{ 2}} \\
		\end{aligned}
	\end{equation}
  Therefore, bringing all the terms together, we obtain

\begin{align}
&\tracedist{
  \left(
    \left( \Tr_{K_B} \circ \protMap{\lkey^\prime} - \Tr_{K_B} \circ \protMapId{\lkey^\prime} \right)
    \otimes \idmap_{E_1^n V}
  \right)
  \left( \tau_{A_1^n B_1^n E_1^n V} \right)
}
\nonumber
\\
&= \sum_{\cobs : \lkey'(\cobs) > 0}
  \Pr( \Omega(\cobs) \wedge \OmegaEV )
  \tracedist{
    \tau^{(\lkey'(\cobs))}_{K_A \allpublic E_1^n V \mid \Omega(\cobs) \wedge \OmegaEV}
    -
    \tau^{(\lkey'(\cobs),\mathrm{ideal})}_{K_A \allpublic E_1^n V \mid \Omega(\cobs) \wedge \OmegaEV}
  }
\label{eq:varproofend:b}
\\
&\leq \sum_{\cobs : \lkey'(\cobs) > 0}
  \Pr( \Omega(\cobs) \wedge \OmegaEV )
  \left(
   \epstilde
    + 2 \sqrt{2 \zeta(\cobs) - \zeta(\cobs)^{ 2}}
  \right)
\label{eq:varproofend:c}
\\
&\leq \sum_{\cobs : \lkey'(\cobs) > 0}
  \Pr( \Omega(\cobs) \wedge \OmegaEV )
  \left(
    \epstilde
    + 2 \sqrt{2 \zeta(\cobs)}
  \right)
\label{eq:varproofend:d}
\\
&= \sum_{\cobs : \lkey'(\cobs) > 0}
  \Pr( \Omega(\cobs) \wedge \OmegaEV )\epstilde
  + \sqrt{8} \sum_i
  \Pr( \Omega(\cobs) \wedge \OmegaEV ) \sqrt{\zeta(\cobs)}
\label{eq:varproofend:e}
\\
&\leq \epstilde
  + \sqrt{8}
    \sqrt{
      \sum_{\cobs : \lkey'(\cobs) > 0}
      \Pr( \Omega(\cobs) \wedge \OmegaEV )
   \zeta(\cobs) }
\label{eq:varproofend:f}
\\
&\leq \epstilde
  + \sqrt{8} \sqrt{\epssecret}
\label{eq:varproofend:g}
\end{align}
where \cref{eq:varproofend:b} follows from the definition of secrecy, \cref{eq:varproofend:c} follows from \cref{eq:converseboundwithLHL}. \cref{eq:varproofend:d,eq:varproofend:e} follow from basic algebra. In \cref{eq:varproofend:f} we use the concavity of the square root function and Jensen's inequality, whereas \cref{eq:varproofend:g} follows from:
\begin{equation}
      \sum_{\cobs : \lkey'(\cobs) > 0}
      \Pr( \Omega(\cobs) \wedge \OmegaEV )
   \zeta(\cobs)  \leq   \sum_{\cobs : \lkey(\cobs) > 0}
      \Pr( \Omega(\cobs) \wedge \OmegaEV )
   \zeta(\cobs)  = \epssecret
\end{equation}
where we use the fact that   $\lkey'(\cobs) > 0 \implies \lkey(\cobs) > 0$. This concludes the proof.
\end{proof}

Through a series of lemmas and theorems, we have reduced the security proof of a QKD protocol against arbitrary attacks to the security proof of a similar QKD protocol against IID collective attacks. There are two costs to be paid for this lift. One is a cost paid to the $\epssecret$ as stated in \cref{lemma:endOfStepOne}. The other is a cost paid to the hash length that can be chosen as stated in \cref{thm:maintheoremvar}. We bring together the entire reduction formally in the following corollary.
\begin{corollary}  \label{cor:liftToCoherent}
     Consider the \nameref{prot:qkdprotocol} defined via $\{ \protMap{\lkey} , \promise \}$ where  $\protMap{\lkey} \in \CPTP(A_1^nB_1^n, K_A K_B \allpublic) $ is   a map such that
    the $\epssecret$-secrecy condition holds for all IID states $\rho_{A^nB^nE^n} = \sigma^{\otimes n}_{ABE}$ satisfying $\Tr_{BE}(\sigma_{ABE}) = \promise$ ((\cref{def:epsSecPromiseIIDcollective}) ).  That is, 
    \begin{equation}
    \begin{aligned}
        &\tracedist{
            \left(
                \left( \Tr_{K_B} \circ \protMap{\lkey}
                - \Tr_{K_B} \circ \protMapId{\lkey} \right)
                \otimes \idmap_{E_1^n}
            \right)
           \left[ \rho_{ABE}^{\otimes n} \right]
        }
        \leq \epssecret, \\
        &\forall\, \rho_{ABE} \in \dop{=}(ABE)
        \text{ such that }
        \Tr_{B E}\!\left( \rho_{A B E} \right)
        =  \sigma_A .
    \end{aligned}
\end{equation}
    Let $\protMap{\lkey^\prime,\mathrm{perm}}$ be a variable-length QKD protocol map identical to $\protMap{\lkey}$, except that \begin{enumerate}
        \item it hashes to a length $\lkey^\prime(\cdot)$ instead of $\lkey(\cdot)$,
        \item includes the random permutation step on the classical $X_1^n,Y_1^n$ (see \cref{lemma:satisfyingpermutationinvariance}), announced in the register $C_\mathrm{perm}$.
    \end{enumerate} 
    Suppose 
    $
    \lkey^\prime (\cdot)=  \max\left\{ \floor{\lkey(\cdot)- 2 \log(\cost) - 2\log(1/2\epstilde)},0 \right\}$, where  $x=d_A^2d_B^2$.  
    Then the QKD protocol $\{\protMap{\lkey^\prime,\mathrm{perm}},\sigma_{A} \}$ is $\cost \left( \sqrt{8 \epssecret} + \epstilde \right)$-secret with fixed marginal $\promise$. That is, 
\begin{equation}
    \begin{aligned}
        &\tracedist{
            \left(
                \left( \Tr_{K_B} \circ \protMap{\lkey',\mathrm{perm}}
                - \Tr_{K_B} \circ \protMapId{\lkey',\mathrm{perm}} \right)
                \otimes \idmap_{\Eve}
            \right)
           \left[ \rho_{A_1^n B_1^n \Eve} \right]
        }
        \leq\cost \left( \sqrt{8 \epssecret} + \epstilde \right), \\
        &\forall\, \rho_{A_1^n B_1^n \Eve}\in \dop{=}(A_1^n B_1^n \Eve)
        \text{ such that }
       \rho_{A_1^n} = \sigma_A^{\otimes n}.
    \end{aligned}
\end{equation}
\end{corollary}
\begin{proof} 
Since $\protMap{\lkey^\prime,\mathrm{perm}} - \protMapId{\lkey^\prime,\mathrm{perm}}$  is permutation-invariant (from \cref{lemma:satisfyingpermutationinvariance}), \cref{lemma:endOfStepOne} states that there exists a probability measure $d\sigma_{AB}$ on the set of extensions $\sigma_{AB}$ of $\promise$ such that 
 \begin{equation} \label{eq:incor2.1}
 \begin{aligned}
         &\tracedist{ \left( \left( \protMap{l^\prime,\mathrm{perm}} - \protMapId{l^\prime,\mathrm{perm}} \right) \otimes \idmap_{E_1^n} \right) \left( \rho_{A_1^nB_1^nE_1^n} \right) }  \\
         &\leq   \cost \tracedist{ \left( \left(\protMap{l^\prime,\mathrm{perm}} - \protMapId{l^\prime,\mathrm{perm}} \right) \otimes \idmap_{R} \right) \left( \tau_{A_1^nB_1^nR} \right) },
         \end{aligned}
    \end{equation}
    where $\tau_{A_1^nB_1^nR}$ is any purification of $\tau_{A_1^nB_1^n} =  \int \text{d} \sigma_{AB}\  \sigma_{AB}^{\otimes n}$, and $x=d_A^2d_B^2$. We then use \cref{thm:maintheoremvar}, which allows us to only prove security against IID states. Then, we utilize \cref{lemma:permutationdoesnotmatter} to argue that for IID states, the security of the protocol with the permutation is the same as the security of the protocol without the permutation. This concludes the proof.
\end{proof}
The above corollary can now be directly reduced the secrecy analysis of QKD protocols that satisfy the permutation invariance condition. Thus, we make rigorous a verbal argument made in Ref.~\cite{christandl_postselection_2009} in Theorem \ref{thm:maintheoremvar}. In doing so, we notice that the key secrecy parameter is worse than predicted by Ref.~\cite{christandl_postselection_2009}. In particular, Ref.~\cite{christandl_postselection_2009} obtains a secrecy parameter of $ \approx \cost \epssecret$ against arbitrary attacks, as compared to $\cost\sqrt{\epssecret}$ which we obtain, where $\epssecret$ refers to the secrecy parameter obtained against IID collective attacks.

\subsection{Improvements from block-diagonal symmetry}

In situations where the QKD protocol satisfies additional properties, one can repeat the previous analysis as shown below to obtain improved performance. We first state when a 
QKD protocol is IID-block-diagonal, in preparation for using our improved de Finetti result from \cref{cor:blockDiagDeFinetti}.
\begin{definition}[IID-block-diagonal maps] \label{def:blockDiagMaps}
    Let $\{\Pi_i\}_{i=1}^k$ be a set of orthogonal projectors on $A$. A linear map $\Delta \in \mapset(A_1^n,B_1)$ is \term{IID-block-diagonal}, if
    \begin{equation}
        \sum_{\vec{i}\in[k]^n} \Delta \circ \mathcal{P}_{\vec{i}} = \Delta
    \end{equation}
    where $\mathcal{P}_{\vec{i}} (\cdot) = \bigotimes_{j=1}^n \Pi_{i_j} \left(\cdot\right) \bigotimes_{j=1}^n \Pi_{i_j}$.
\end{definition}

  \begin{corollary}  \label{cor:liftToCoherentvarxBlockDiagonal}
     Consider the \nameref{prot:qkdprotocol} defined via $\{ \protMap{\lkey} , \promise \}$ where  $\protMap{\lkey} \in \CPTP(A_1^nB_1^n, K_A K_B \allpublic) $ is such that
 the $\epssecret$-secrecy condition (\cref{def:epsSecPromiseIIDcollective}) holds for all IID states $\rho_{A^nB^nE^n} = \sigma^{\otimes n}_{ABE}$ satisfying $\Tr_{BE}(\sigma_{ABE}) = \promise$.

 Let $\protMap{\lkey^\prime,\mathrm{perm}}$ be a variable-length QKD protocol map identical to $\protMap{\lkey,\mathrm{perm}}$, except that it 
 \begin{enumerate}
     \item  it hashes to a length $\lkey^\prime(\cdot)$ instead of $\lkey(\cdot)$, 
        \item includes the random permutation step on the classical $X_1^n,Y_1^n$ (see \cref{lemma:satisfyingpermutationinvariance}), announced in the register $C_\mathrm{perm}$.
        \item and $\protMap{\lkey^\prime,\mathrm{perm}} - \idealmap \circ \protMap{\lkey^\prime,\mathrm{perm}}$ is an IID-block-diagonal map with respect to projections $\{\Pi_i^A \otimes \Pi_j^B\}_{i,j=1}^{k_A,k_B}$ of dimension $\{d^A_id^B_j \}_{i,j=1}^{k_A,k_B}$.
 \end{enumerate}  
   Suppose 
    $
    \lkey^\prime (\cdot)=  \max\left\{ \floor{\lkey(\cdot)- 2 \log(\cost) - 2\log(1/2\epstilde)},0 \right\}$, where  $x = \sum_{i,j=1}^{k_A,k_B} (d^A_i)^2 (d^B_j)^2$.
  Then $\{\protMap{\lkey},\sigma_A\}$ is $g_{n,x} \left( \sqrt{8 \epssecret} + \epstilde \right)$-secret.
\end{corollary}
\begin{proof}
    The proof follows from a similar series of steps as \cref{cor:liftToCoherent}, but using the de Finetti reduction form \cref{cor:blockDiagDeFinetti} instead of \cref{cor:generalMixedDeFinetti}.
\end{proof}

\section{Postselection technique for optical protocols} \label{sec:postselectionoptics}
So far, we have shown how the postselection technique can be used to reduce the analysis to proving secrecy against IID collective attacks. However, our lift statements (\cref{cor:liftToCoherent,cor:liftToCoherentvarxBlockDiagonal}) depend explicitly on the dimensions of Alice’s and Bob’s systems. This creates a problem when dealing with optical protocols. In this section, we will see how to extend our analysis to handle such protocols, in particular decoy-state BB84.

If we naively apply the source-replacement scheme from \cref{chap:MEAT}, then Alice’s shield system $\Ashield$ (see \cref{subsec:sourcereplacement}) - a system that remains inside her lab - is infinite dimensional for phase-randomized weak coherent pulses. Likewise, Bob’s threshold detectors act on infinite-dimensional Hilbert spaces. Thus, neither $d_A$ nor $d_B$ is finite.

In this section, we will formally state how squashing maps and source maps can be used to reduce the protocol to one involving finite-dimensional systems. It may be helpful to revise \cref{subsec:squashing,subsec:sourcemaps} at this point. We note that squashing maps and source maps are often applied at the level of the single-round optimization problem in many works \cite{kamin_finite-size_2024,kamin_renyi_2025,zhang_improved_2017,Kamin2025}. That approach is tenable when the underlying security proof technique does not depend on the dimensions\footnote{Although we note that there are still nuances when it comes to handling truly infinite dimensions, since a statement that holds for arbitrary finite dimensional states does not necessarily hold for infinite dimensional states, see \cite[Appendix A]{inprep_BDR3}}. This is not the case for us, and therefore additional care is required. 

Thus, we will now combine the use of source maps and squashing maps with our analysis. We will first use these tools to reduce the protocol to a finite-dimensional one, after which it can be analyzed using the analysis from previous sections. For this, it is convenient to view an instance of \nameref{prot:qkdprotocol} as $\{\protMapbeforeSR{\lkey},\sigma_{XA'} \}$, where Alice first creates the classical quantum state $\sigma_{XA'}^{\otimes n}$, which is followed by Eve's attack, and $\protMapbeforeSR{\lkey}$ is the protocol map that implements Bob's measurements, and all subsequent classical postprocessing (see \cref{def:equivalentQKDprotocolbeforeSR,def:epsSecPromisechannelversion}). Note that we will include a more general version of these statements in \cref{sec:optics}.

We will first obtain the following lemma concerning source maps.
\begin{lemma}[Source maps] \label{lemmaSourceMapSecurity}
    Let $\{\sigma_{k}\}\subset\dop{=}(A')$ be the set of states prepared by Alice in a QKD protocol given by $\{\protMapbeforeSR{\lkey},\sigma_{XA'} \}$. Consider a related QKD protocol $\{\protMapbeforeSR{\lkey}, \xi_{XA''}$ that is identical to the first one in all cases except source preparation, where it prepares $\xi_k$ instead of $\sigma_k$.    
    Suppose that there exists a \term{source map} $\Psi \in\CPTP(A'',A')$ relating the real states $\{\sigma_k\}$ to a set of virtual states $\{\xi_k\}\subset\dop{=}{A''}$ such that $\sigma_k = \Psi[\xi_k]$ for all $k$. 
    Then $\epssecure$-security for the virtual protocol $\{\protMapbeforeSR{\lkey}, \xi_{XA''} \}$  implies $\epssecure$-security for the real protocol $\{\protMapbeforeSR{\lkey},\sigma_{XA'} \}$.\footnote{One can obtain the same result for secrecy instead.}
\end{lemma}
\begin{proof}
    For any attack $\attack{} \in \CPTP((A')_1^n, B_1^n \Eve)$, we have 
    \begin{equation}
\begin{aligned}
   & \tracedist{
            \left(
                \left(  \protMapbeforeSR{\lkey} 
                - \idealmap \circ  \protMapbeforeSR{\lkey} \right)
                \otimes \attack{}
            \right)
           \left[ \sigma_{X_1^n (A')_1^n } \right]} \\
           &=  \tracedist{
            \left(
                \left(  \protMapbeforeSR{\lkey} 
                - \idealmap \circ  \protMapbeforeSR{\lkey} \right)
                \otimes \attack{} \circ \Psi^{\otimes n}
            \right)
           \left[ \xi_{X_1^n (A'')_1^n } \right]} \\
           &\leq \epssecure.
\end{aligned}
\end{equation}
where we simply used the stated property of the source map for the second line, and the definition of security for the final line, where we have that  $ \attack{} \circ \Psi^{\otimes n} \in \CPTP((A'')_1^n, B_1^n \Eve)$ is a valid attack that Eve can perform on the protocol with the virtual states. .
\end{proof}

The original squashing map constructions \cite{fung_universal_2011,gittsovich_squashing_2014,zhang_security_2021,upadhyaya_dimension_2021}
 prove the applicability of squashing models by showing that their usage lower bounds the key rate under the assumption of IID collective attacks. However, this alone is \emph{insufficient} to apply the squashing model to make Bob's system finite-dimensional and apply the postselection technique.
Thus, we prove the following lemma.

\begin{restatable}[Squashing]{lemma}{lemmaSquash} \label{lemma:squash}
Consider the QKD protocol $\left\{  \protMapSquash{\lkey} \circ \left( \measChannel{\{ \Gamma^{(B)}_k \} } \right)^{\otimes n} , \sigma_{XA'} \right\}$, 
where $\measChannel{\{ \Gamma^{(B)}_k \}}$ is a channel that measures the $B$ systems and stores the outcome in $Y$, and $\protMapSquash{\lkey} \in \CPTP(X_1^n Y_1^n, \allpublic) $ implements the rest of the QKD protocol. Suppose there exists quantum channels (squashing maps, see \cref{def:squashing}) $\Lambda \in \CPTP(B, Q)$ and measurement channels $\measChannel{\left\{F_{i}^{Q}\right\}} \in \CPTP(Q,Y)$ such that
\begin{align} \label{eq:squashingConditionPS}
    \measChannel{\left\{F_{i}^{(Q)}\right\}} \circ \Lambda = \measChannel{\left\{M_{i}^{(B)}\right\}} 
\end{align}
Then, the $\epssecure$-security of the squashed QKD protocol, given by $\left\{ \protMapSquash{\lkey} \circ \left( \measChannel{\{ \Gamma^{(B)}_k \} } \right)^{\otimes n}, \sigma_{XA'} \right\}$ implies the $\epssecure$-security of the original QKD protocol, given by $ \left\{ \protMapSquash{\lkey} \circ \left( \measChannel{\{ \Gamma^{(B)}_k \} } \right)^{\otimes n} \right\}$.
\end{restatable}

\begin{proof}
We use the fact that the first step of the QKD protocol is to measure Bob's received state, i.e, the protocal map is given by $\protMapSquash{\lkey} \circ \left( \measChannel{\{ \Gamma^{(B)}_k \} } \right)^{\otimes n}$. Thus, for any attack $\attack{}\in \CPTP((A')_1^n, B_1^n \Eve) $  we have

\begin{equation}
\begin{aligned}
   & \tracedist{
            \left(
                \left(  \protMapSquash{\lkey} \circ \left( \measChannel{\{ \Gamma^{(B)}_k \} } \right)^{\otimes n }
                - \idealmap \circ  \protMapSquash{\lkey} \circ \left( \measChannel{\{ \Gamma^{(B)}_k \}  } \right)^{\otimes n}\right)
                \otimes \attack{}
            \right)
           \left[ \sigma_{X_1^n (A')_1^n } \right]} \\
           &= \tracedist{
            \left(
                \left(  \protMapSquash{\lkey} \circ \left( \measChannel{\{ \Gamma^{(Q)}_k \} } \right)^{\otimes n }
                - \idealmap \circ  \protMapSquash{\lkey} \circ \left( \measChannel{\{ \Gamma^{(Q)}_k \}  } \right)^{\otimes n}\right)
                \otimes \Lambda^{\otimes n} \circ \attack{}
            \right)
           \left[ \sigma_{X_1^n (A')_1^n } \right]} \\
           &=\tracedist{
            \left(
                \left(  \protMapSquash{\lkey} \circ \left( \measChannel{\{ \Gamma^{(Q)}_k \} } \right)^{\otimes n }
                - \idealmap \circ  \protMapSquash{\lkey} \circ \left( \measChannel{\{ \Gamma^{(Q)}_k \}  } \right)^{\otimes n}\right)
                \otimes \attackSquash{}
            \right)
           \left[ \sigma_{X_1^n (A')_1^n } \right]} \\
           &\leq \epssecure
\end{aligned}
\end{equation}
where the second line follows from \cref{eq:squashingConditionPS}, and the last line follows redefining $\attackSquash{} = \Lambda^{\otimes n} \circ \attack{} $. The required statement follows from the fact that $\attackSquash{} \in \CPTP((A')_1^n, Q_1^n \Eve)$ is a valid attack on the protocol with the squashed POVM measerements.
\end{proof}

\subsubsection{Using Squashing with postselection technique}
\label{subsec:squashingwithPS}
We now briefly discuss a subtle issue that arises when using the postselection technique with squashing arguments. The subtlety comes from having to prove the security against \textit{all} attacks in the protocol with squashed measurements. 

First, note that not all squashing models proceed via the construction of an explicit squashing map $\Lambda$. In particular, the universal squashing model \cite{fung_universal_2011} only allows the replacement of the original measurements with squashed measurements for the purposes of estimating certain statistical quantities. Moreover, the dimension-reduction method \cite{upadhyaya_dimension_2021} does not proceed via a squashing map argument. As a result, they cannot be combined with the postselection technique through \cref{lemma:squash}.\footnote{Since the dimension-reduction method is currently the only known approach for reducing CVQKD to finite dimensions, the applicability of the postselection technique to CVQKD remains an open problem.}

Existing multiphoton-to-qubit squashing maps, such as those constructed in Ref.~\cite{gittsovich_squashing_2014}, are only known to exist under highly restrictive assumptions on the detector model. In particular, they require perfectly matched (or exactly equal) efficiencies and dark count rates across all detectors, and are therefore not robust to realistic device imperfections.

Consequently, in practical settings the only viable candidate is the flag-state squasher. Moreover, recent works \cite{nahar2025imperfect,nahar_phd_2025} establish the existence of a flag-state squasher even when detector parameters are only known to lie within a characterized range. It is therefore natural to attempt to combine the flag-state squasher with the postselection technique. However, a direct application of \cref{lemma:squash} using the standard flag-state squasher leads to a trivial (zero) key rate, due to the existence of classical flags as described in \cref{subsec:squashing}.

For the postselection technique specifically, a new ``weight-preserving flag-state squasher" variant of the flag-state squasher was introduced in Ref.\cite{nahar_postselection_2024}. This construction modifies the standard flag-state squasher in a way that allows the entire squashing map to be given to Eve, while still enabling the application of \cref{lemma:squash} together with the postselection technique. We do not describe this construction in detail here, nor do we use this modified squasher in this thesis. Instead, for the numerical results presented in this chapter, we employ the qubit squasher introduced in Ref.\cite{gittsovich_squashing_2014}. The purpose of the present discussion is to highlight a subtle but crucial issue in combining squashing models with postselection, and to motivate the need for such modified squashing maps. We note that Refs.~\cite{nahar_phd_2025,nahar2025imperfect} present an alternative approach that enables the use of the flag-state squasher together with postselection, even in the presence of detector imperfections, that utilizes the weight-preserving flag-state squasher. That method is technically involved, and we refer the interested reader to the cited works for further details. 

With these clarifications in place, we now have all the necessary tools to apply our results to the qubit BB84 and decoy-state BB84 protocols, which we do in the next subsection.

\section{Application to BB84 protocols} \label{sec:postselectionapplication}

So far we have made rigorous the framework to apply the postselection technique to optical prepare-and-measure protocols. In this section, we will first outline a recipe in \cref{subsec:recipePS} that can be used to apply the postselection technique to optical protocols. Then, in \cref{subsec:QubitBB84PS,subsec:decoystateBB84PS} we apply that recipe to the Qubit BB84 and decoy-state BB84 protocols.

\subsection{Recipe}
\label{subsec:recipePS}
We use \cref{cor:liftToCoherentvarxBlockDiagonal,cor:liftToCoherent}, which state that given a variable-length security proof against IID collective attacks with key lengths $\lkey(\cobs)$ and secrecy parameter $\epssecret$, the postselection technique can be used to provide a security proof against coherent attacks with key lengths $\lkey(\cobs)-2\log(\cost)-2\log(1/2\epstilde)$ and secrecy parameter $\cost (\sqrt{8\epssecret}+\epstilde)$. Here, $\epstilde$ is a parameter that can be chosen freely and $\cost$ can be thought of as the cost of using the postselection technique that depends on the dimensions of the systems. In the presence of additional protocol structure, this cost can be reduced as detailed in the corollary statements.

In practice, this can be used as follows:
\begin{enumerate}
    \item Choose a target secrecy parameter $\epssecret$ based on the application in mind.
    \item \label{item:PScost} Determine the protocol-dependent upper bound on the cost of using the postselection technique as $\cost = \binom{n+x-1}{x-1} \leq \left(\frac{e (n+x-1)}{x-1}\right)^{x-1}$. Here, $x$ depends on the dimensions of the systems, and the structure in the protocol.
    For a generic protocol, $x = d_A^2d_B^2$ as described in \cref{cor:liftToCoherent}. In the presence of block-diagonal structure this can be improved as stated in \cref{cor:liftToCoherentvarxBlockDiagonal}.

    \item Pick value of free parameter $\epstilde$ (in principle this value can be optimised over). For our calculations, we choose this to be $\epstilde = \epssecret/2$
    \item Compute key lengths $\lkey(\cdot)$ through an IID security proof, with secrecy parameter $\frac{(\epssecret-\cost\epstilde)^2}{8\cost^2}$. For our calculations, we choose $\epsPA = \epsAT$ for the IID key length calculations. 
    \item Use key lengths $\lkey(\cobs)-2\log(\cost)-2\log(1/2\epstilde)$ as the final hash length for the protocol. This protocol is secure against coherent attacks.
\end{enumerate}

\subsubsection{Application to optical protocols using Qubit squuasher} \label{subsec:recipeForOpticalProtocols}

For detection setups with threshold detectors,
an appropriate squasher must be used to squash Bob's system down to finite-dimensions. If we use the qubit squasher from Ref.~\cite{gittsovich_squashing_2014}, Bob measures a $3$-dimensional system what is made of the vacuum and qubit space. Moreover, Bob is block-diagonal, and thus the dimensional-dependent term $x$ in \cref{item:PScost} is given by (\cref{cor:liftToCoherentvarxBlockDiagonal})
\begin{align} \label{eq:WPFSSOpticalDimension}
    x = d_A^2\left(1^2 + 2^2\right).
\end{align}
For the qubit BB84 protocol, one can also further reduce $d_A$ to $2$ instead of $4$, as argued in the next chapter. Hence, for Qubit BB84, we have $x=2^2(5)$.

For decoy-state protocols, we use the tagging source map, and therefore consider tagged states as described in \cref{eq:taggedStates}. For most encodings such as polarization, time-bin, etc. (see Ref.~\cite{nahar_postselection_2024} for formal details, and Ref.~\cite{nahar_phd_2025} for improvements), the dimensional-dependent term $x$ in Step.~\cref{item:PScost} is given by
\begin{align} \label{eq:DecoyOpticalDimension}
    x = \nint^2(\ndecoy+2)d_{\Ameas}^2(1^2+2^2),
\end{align}
where $\nint$ is the number of intensities used in the protocol, and $d_{\Ameas}$ is the dimension of Alice's subsystem that she measures (recall that $A=\Ameas \Ashield$). We set $\ndecoy=3$, and hence we obtain $x=3^2(3+2)4^2(1^2+2^2)$, for the decoy-state BB84 calculations.

\subsection{Qubit BB84}
\label{subsec:QubitBB84PS}
\begin{figure}[!ht]
    \centering
    \includegraphics[width=\linewidth]{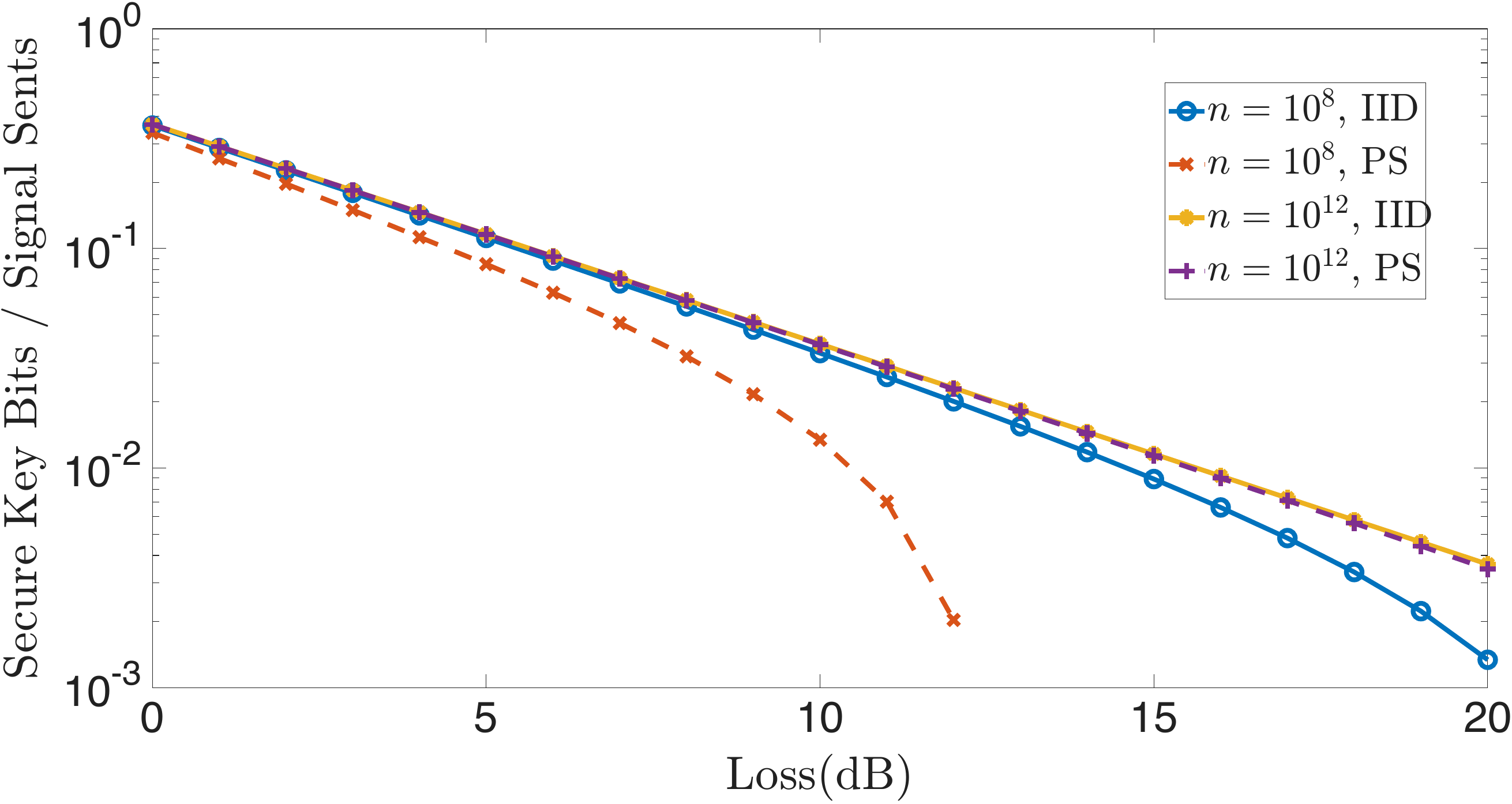}
    \caption{Key rate for variable-length qubit BB84 protocol plotted against loss, with and without the postselection lift to coherent attacks.}
    \label{fig:qubitBB84_PS}
\end{figure}

We plot key rates with and without the postselection lift for the qubit BB84 protocol in \cref{fig:qubitBB84_PS}. We plot variable-length key rates for the typical observations of the channel, i.e $\Fobs$ set to be the expected honest behaviour of the channel. Alice and Bob both choose each basis with probability $0.5$, and the probability of a $\test$ round is given by $\gamma=0.1$. We consider a channel with $2^\circ$ misalignment, and depolarization probability of $2 \%$. Key rates are plotted against loss, with the target $\epssecret = \epsEV = 10^{-10}$.
We use the same formula for key rate computations as \cref{sec:variableapplicationtoQubitbb84}, suitably modifying the parameters as required by the postselection technique. 

\subsection{Decoy-state BB84}
\label{subsec:decoystateBB84PS}
\begin{figure}[!ht]
    \centering
    \includegraphics[width=\linewidth]{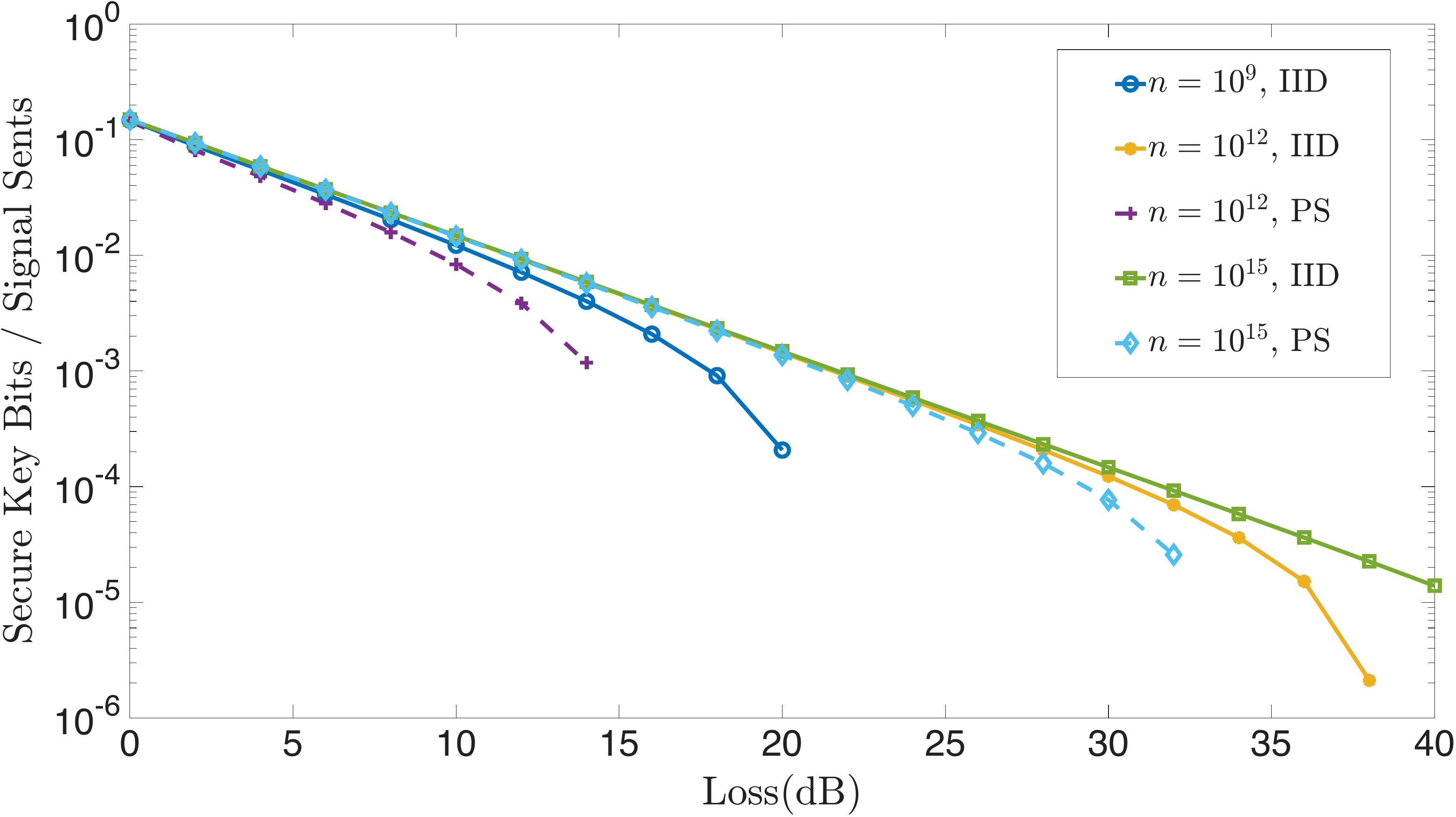}
    \caption{Key rate for variable-length decoy-state BB84 protocol plotted against loss, with and without the postselection lift to coherent attacks.}
    \label{fig:decoyBB84_PS}
\end{figure}
We plot key rates with and without the postselection lift for the decoy-state BB84 protocol in \cref{fig:qubitBB84_PS}. We plot variable-length key rates for the typical observations of the channel, i.e $\Fobs$ set to be the expected honest behaviour of the channel. Alice and Bob both choose each basis with probability $0.9$, and the probability of a $\test$ round is given by $\gamma=0.1$. We set the decoy intensities to be $\mu_1=1,\mu_2=0.01,\mu_3=0.001$, with all three intensities sent with equal probability in the $\test$ rounds, and only $\mu_1$ being used in the $\gen$ rounds. We plot key rates for a loss-only channel, with the $\epssecret = \epsEV = 10^{-10}$. 
We use the same formula for key rate computations as \cref{sec:variableapplicationtoDecoybb84} (which relies on Ref.~\cite{Kamin2025}), suitably modifying the parameters as required by the postselection technique.

We see that the usage of the postselection technique significantly reduces the key rate compared to IID collective attacks. The main impact is due to the large value of $g_{n,x}$. For a plot that shows the impact of this dimension on the key rate performance, see \cite[Fig 2]{nahar_postselection_2024}.

\section{Summary and Outlook}\label{sec:postselectionsummary}
Overall, this chapter provides the first complete, robust, and practically applicable formulation of the postselection technique%
\footnote{The author is not entirely sure why this technique was named so, as it has nothing to do with postselection of data.}
for quantum key distribution. 
It resolves several longstanding gaps in the literature and enables rigorous security proofs against coherent attacks for realistic prepare-and-measure protocols. 

The postselection technique is remarkably easy to use in practice. 
One simply needs to ensure that the protocol explicitly includes a random permutation of the classical data as one of its steps, thereby guaranteeing permutation invariance of the overall protocol. 
Once this condition is satisfied, \cref{cor:liftToCoherent} can be applied to reduce the entire security analysis to the IID collective attack setting.
We emphasize that this reduction is entirely independent of how one chooses to perform the IID security analysis.
After this reduction, one may proceed with \emph{any} IID security analysis of choice. 
This includes, for instance, approaches based on uncertainty relations or phase error correction 
\cite{Cong_sidechannelsecure_2025}.\footnote{ 
We note, however, that if one wishes to pursue a phase error correction approach at this stage, it is generally preferable to rely directly on the underlying de Finetti reductions to simplify the sampling process, as done for example in Refs.~\cite{tamaki_unconditionally_2003,matsuura_asymptoticallytight_2025,Shan_improvedfinitekey_2025}, rather than relying directly on the lifting theorems.}

Despite these strengths, the postselection technique has several fundamental drawbacks.
First, recall that the technique starts from a global state $\rho_{A_1^n B_1^n \Eve}$ describing Eve’s attack on all $n$ rounds.
In this picture, Eve’s attack is applied first, which results in the post-attack state that is to be analyzed, and only afterward are public announcements made.
As a consequence, protocols that rely on \emph{on-the-fly} announcements, where Alice and Bob announce public information before all signals have been sent and received,  cannot be handled in a straightforward manner within this framework.

Second, permutation invariance formally requires permuting the quantum systems $A_1^n B_1^n$.
In practice, one permutes only the classical measurement data obtained from these systems, and this is sufficient to justify permutation invariance if Alice and Bob perform the same measurements in each round. However, it is unclear how this technique can be extended to scenarios involving correlated measurement effects, such as detector dead times or after pulsing \cite{wang2025phase,nahar2025imperfect}, or correlated source imperfections \cite{curras_securityquantumkeydistribution_2025,yoshino_qkdcorrelated_2018}. 

Finally, the postselection technique yields highly pessimistic key rates, primarily due to the large value of the overhead factor $\cost$.
Although the improved de Finetti reductions developed in this chapter significantly mitigate this issue, the resulting key rates remain below those achievable using phase error rate based methods and the MEAT framework studied in the next two chapters.

In summary, while the postselection technique provides a conceptually clean and broadly applicable route to coherent-attack security, its limitations in handling adaptive announcements, correlated imperfections, and finite-size performance are significant drawbacks.

\chapter{Entropic Uncertainty Relations and Phase Error Estimation}
\label{chap:EUR}
\epigraph{
Where we \emph{really} break free from the shackles of IID collective attacks; where we confront Eves's global attack head-on; and where we enthusiastically consider a large number of fictitious measurements, and agonizingly obtain estimates on their measurement outcomes.
}{}
Security proofs of QKD based on entropic uncertainty relations (EUR)
\cite{tomamichel_uncertainty_2011,tomamichel_largely_2017,tomamichel_Tight_2012,tupkary2025qkdsecurityproofsdecoystate}
and on the phase error correction approach
\cite{koashi_simple_2005,koashi_simple_2009,tupkary2025qkdsecurityproofsdecoystate}
yield some of the highest key rates against coherent attacks in the finite-size regime.
Historically, these techniques were among the earliest to provide analyses of coherent
attacks, and it is only more recently that EAT-based methods (which we discuss in
\cref{chap:MEAT}) have begun to achieve comparable (and in many cases superior \cite{Kamin2025}) key
rates. Moreover, these approaches are widely used in practice, due in part to their
conceptual simplicity and the availability of closed-form or largely analytic key rate
expressions.

However, while this simplicity and elegance are most apparent in idealized BB84
scenarios, they come at a significant cost when practical imperfections are taken into
account, as we shall see in this chapter. Although source imperfections
\cite{pereira_modified_2023,tamaki_loss-tolerant_2014,zapatero2023implementationsecurityquantumkey,curras-lorenzo_security_2024}
have been extensively studied within the phase error correction framework, detector
imperfections have only recently been addressed for either proof technique 
\cite{tupkary_phase_2024}.

In particular, prior to Ref.~\cite{tupkary_phase_2024} (which is the work on which this chapter is based), these proof techniques required the probability of detection in Bob’s measurement
setup to be independent of the basis choice. This assumption is commonly referred to as
\term{basis-independent loss}, while violations of this assumption are known as
\term{basis-efficiency mismatch} or \term{detection-efficiency mismatch}. Satisfying
basis-independent loss requires the efficiencies and dark count rates of Bob’s detectors
to be \emph{exactly} identical. Consequently, justifying this assumption in practice
requires exact characterization of identical detectors, which is not achievable. Meanwhile, there have been several experimental demonstrations \cite{gerhardtFullfield2011,makarovEffects2006,lydersen_Hacking_2010,sajeedSecurity2015,qiTimeshift2007}, exploiting  basis-efficiency mismatch for attacks on QKD systems.

As a result, these proof techniques remained  inapplicable to practical
prepare-and-measure (and entanglement-based) QKD scenarios involving realistic detectors.\footnote{By contrast, measurement-device-independent QKD (MDI-QKD)
\cite{lo_Measurementdeviceindependent_2012} is able to address \emph{all} detector
imperfections and detector side channels, since it assumes the detectors to be completely
under Eve’s control. However, it is much more complex to implement and prepare-and-measure QKD protocols remain the dominant implementation.} Extending EUR and phase error correction based proof techniques
to realistic detector models has remained a well-known open problem in the field for
nearly two decades. While a variety of partial results exist \cite{lydersen_Security_2010,maroy_security_2010,trushechkin_Security_2022,bochkov_Security_2019,zhang_security_2021,Grasselli_qkdwithbasisdependent_2024,Marcomini_losstolerant_2024,fung_Security_2009,winick_reliable_2018}, none addressed the
finite-size regime against coherent attacks prior to Ref.~\cite{tupkary_phase_2024}; see
\cite[Table~1]{tupkary_phase_2024} for a detailed overview of the historical developments.

In this chapter, we prove the security of the decoy-state BB84 protocol with an active detection setup \textit{without} assuming basis-independent loss. We do so by showing that the phase error rate can be suitably bounded even without the assumption. We explicitly define metrics $\deltaone$, $\deltatwo$ that quantify the deviation from the ideal case, and bound the phase error rate in terms of these deviations. Our framework is general, and can be applied to any (IID) detector model of one's choice, as long as the relevant metrics $\deltaone$, $\deltatwo$ can be suitable bounded.
We explicitly compute these metrics for the case of detectors with basis-efficiency mismatch and unequal dark count rates. To do so, we assume the the canonical model of detectors described in \cref{subsec:detectors,sec:statesandmeasurements}. The block-diagonal structure of the detector POVMs significantly aids the computation of these metrics. Moreover, we compute these metrics directly from the experimental characterization of the detection efficiencies and dark count rates of the detectors. Our results extend the security of QKD to the following practical scenarios:
	
	\begin{enumerate}
		\item \label{caseone} Bob's detectors are \textit{not} identical, but the values of efficiency ($\eta_{b_i}$ for basis $b$ and outcome $i$) and dark count rates ($d_{b_i}$) are known. Note that while this is a useful toy model, such scenarios are impractical since they require $\eta_{b_i}, d_{b_i}$ to be known exactly.  We treat the dark count rate as a part of the POVM element, as described in \cref{subsec:detectors}.
		\item \label{casetwo} Bob's detectors are \textit{not} identical, and the values of efficiency and dark count rates are only known to be in some range $\eta_{b_i} \in   [\etadet (1 -\etachar),\etadet(1+\etachar)], d_{b_i} \in   [ \dcprob( 1-\dcchar), \dcprob (1+\dcchar)]$.  While this is again a useful toy model, a detectors response ($\eta_{b_i}, d_{b_i}$) to incoming photons typically depends on the spatio-temporal modes of incoming photons, which are in Eve's control.
		\item \label{casethree} Bob's detectors are \textit{not} identical, and the values of efficiency and dark count rates are only known to be in some range. Moreover, these values depend on the spatio-temporal modes (labelled by $\mode$) of the incoming photons, and can therefore be chosen by Eve \cite{makarovEffects2006,zhang_security_2021,sajeedSecurity2015,qiTimeshift2007}. This is expressed mathematically as $\eta_{b_i}(\mode) \in [  \etadet(\mode) (1 -\etachar), \etadet(\mode) (1+\etachar)], d_{b_i} (\mode) \in [ \dcprob (1-\dcchar), \dcprob (1+\dcchar) ]$. Note that in this model, the range of allowed values of the loss can depend on the spatio-temporal mode, whereas the dark count rates for all the modes lie in the same range. 

	\end{enumerate}
	
	Our metrics $\deltaone,\deltatwo$ involve an optimization over all possible values of $\eta_{b_i}, d_{b_i}$ in their respective ranges. Moreover, for our model of  multi-mode detectors, we find that our methods  yield the same values for Case \ref{casetwo} and Case \ref{casethree}. Thus, our methods address one practically important detector side-channel as a by-product. 

    \subsubsection{Organization of this chapter}

This chapter is organized as follows. We begin by stating a fairly abstract theorem (\cref{theorem:eurvariablegenericresult}) on the variable-length security of \nameref{prot:qkdprotocol}, which serves as basis for proving variable-length security in EUR based approaches. (This section may be skipped on a first reading, as its role will become clearer once the phase error rate bounds derived in later sections have been studied.) We will then see how suitable bounds on the phase error rate suffice for proving security, and describe some details about our notation and phase error rates in \cref{sec:protocoleurqubitbb84}.
In \cref{sec:perfectdetectors}, we show how such bounds can be obtained in the case where the basis-independent loss assumption is satisfied. In \cref{sec:imperfectdetectors}, we show how analogous bounds can be obtained when the basis-independent loss assumption is \emph{not} satisfied. Finally, in \cref{sec:decoy}, we extend the analysis to prove variable-length security for the decoy-state BB84 protocol with imperfect detectors
In \cref{sec:eurresults} we apply our results to study the effect of basis-mismatch on decoy-state BB84.  In \cref{sec:eurconclusion}
we present some concluding remars. 

\section{Variable length security in EUR} \label{sec:eurvariable}
In this subsection, we will state a theorem regarding the variable-length security of \nameref{prot:qkdprotocol}, analogous to \cite[Theorem 4]{tupkary_phase_2024}, that is suitable for use with the EUR technique. 
Recall that the \nameref{prot:qkdprotocol} makes public announcements in the register $\CP_1^n$, and we use $\cobs$ to denote the value stored in $\CP_1^n$, and $\Omega(\cobs)$ denote the event that $\cobs$ was observed in $\CP_1^n$ registers. Recall further that our goal is to lower bound the smooth min entropy of the raw key string i.e, $\Hmin[\epsbar](\PAstring| \CP_1^n \Eve)_{\rho}$. Ofcourse, from \cref{subsec:boundingentropy}, we know that we have to condition carefully on various events in order to get non-trivial results. In the EUR case, we will also have the smoothing parameter itself also depend on the probability of the event being conditioned on.  We will also need to consider various unobserved events, which we denote using $\Omega(\cnotobs)$. These refer to events that are well defined, but which we do not directly have access to (such as number of single-photon rounds in the protocol etc). 

We are now ready to state the theorem statement. Note that the proof uses the same essential idea as \cite[Supplementary Note A]{curras-lorenzo_tight_2021} and Refs.~\cite{hayashi_concise_2012,kawakami_security_nodate}, but is more general, and carefully handles conditioning on events. The proof is provided in Appendex.~\ref{Appendix:EUR}. The usage of this theorem is explained after the theorem statement. 

\begin{restatable}[Variable-length security for EUR based methods]{theorem}{eurvariabletheorem}   

			In \nameref{prot:qkdprotocol}, let $\Omega(\cnotobs)$ denote a well-defined event on the state $\rho_{\PAstring \PAstring_B \CP_1^n \CEC \CEV \HPA}$ just before privacy amplification. Therefore,  $\event{\cobs,\cnotobs}$ denotes well-defined events, where $\cobs$ denotes the value of the public announcement register $\CP_1^n$ in the protocol, and $\cnotobs$ denotes values that are not directly observed in the protocol. For every value of $\cobs$, let $\EURset(\cobs)$ be a set of possible values of $\cnotobs$.
            
            Let $\serfbound(\cobs, \cnotobs) \in [0,1)$ be such that the following bound on the smooth min entropy of the preamplification string holds,
        \begin{equation} \label{eq:genericvariablecondition1}			\Hmin[\sqrt{\serfbound(\cobs,\cnotobs)}](\PAstring |\CP_1^n \Eve)_{\rho | \Omega(\cobs,\cnotobs)} \geq \EURbeta(\cobs) \qquad \qquad  \forall \cobs, \forall \cnotobs \in \EURset(\cobs).
			\end{equation}
where the smoothing parameters $\serfbound(\cobs,\cnotobs)$ satisfy 
            \begin{equation} \label{eq:genericvariablecondition2}
				\sum_{\cobs} \sum_{\cnotobs \in \EURset(\cobs)} \Pr(\event{\cobs,\cnotobs}) \serfbound_{(\cobs,\cnotobs)}  + \sum_{\cobs} \sum_{\cnotobs\notin \EURset(\cobs)} \Pr(\event{\cobs,\cnotobs})\leq \epsAT^2.
			\end{equation}
          Let the output key length be given by \begin{equation} \label{eq:genericvariablecondition3}
			 \lkey(\cobs) \coloneqq \max\left\{ \floor{  \EURbeta(\cobs) - \leak(\cobs) - 2\log( \frac{1}{\epsPA}) - \EVcost + 2}, 0\right\}.
			\end{equation}
            Then \nameref{prot:qkdprotocol} is $(2\epsAT+ \epsPA + \epsEV)$-secure.
            \label{theorem:eurvariablegenericresult}
\end{restatable}
A typical use of this theorem is as follows.
 The bound on the smooth min entropy given by $\EURbeta(\cobs)$ will depend on the phase error rate bound we obtain, via the entropic uncertainty relations. The value $\serfbound(\cobs,\cnotobs)$ will denote the probability of our computed phase error bound failing conditioned on observing $\cobs,\cnotobs$. 
 Then,  \cref{eq:genericvariablecondition2} is simply stating that we the probability that probability of the phase error rate bound failing is small. We have to include $\cnotobs$ since, our phase error rate bounds sometimes depend on statistics obtained from decoy-analysis, which are not known exactly, but instead estimated from $\cobs$.

\section{Protocol Description and phase error rates} \label{sec:protocoleurqubitbb84}
In this section we describe the protocol we study in this chapter. It is a specific instantiation of the Qubit BB84 protocol from \cref{subsec:protocolvariations}, where we introduced specialized notation that is necessary for the EUR based analysis. This is due to the fact that EUR-based security proofs, as we shall see, are naturally suited to scenarios in which one basis (the $\Xbasis$ basis) is always used for testing, while the other basis (the $\Zbasis$ basis) is always used for key generation, with a small sample potentially reserved for estimating the error-correction parameters. This stands in contrast to the protocol variations described in \cref{subsec:protocolvariations}, which typically feature separate test and key-generation rounds, with both bases used in both types of rounds.\footnote{We emphasize, however, that \cref{theorem:eurvariablegenericresult} itself remains completely general. The above restriction is imposed only at the level of the concrete protocol specification, as we now begin deriving explicit phase error rate bounds.}

 We will first describe the protocol briefly, and then specify how it emerges as a specific instance of \nameref{prot:qkdprotocol}. We consider a protocol where Alice prepares qubit states, but Bob uses threshold detectors for measurements.

\begin{enumerate}
		\item \textbf{State Preparation: } Alice decides to send states in the basis $\Zbasis$  ($\Xbasis$) with probability $\pzA$($\pxA$). If she chooses the $\Zbasis$ basis, she sends states $\{\ket{0}_{A'},\ket{1}_{A'}\}$ with equal probability. If she chooses the $\Xbasis$ basis, she sends states $\{\ket{+}_{A'},\ket{-}_{A'}\}$ with equal probability.  Notice that this ensures 
		\begin{equation} \label{eq:sourcecondition}
			\rho_{{A'} | \Xbasis} = \frac{\ketbra{+} + \ketbra{-}}{2}  = \frac{\ketbra{0} + \ketbra{1}}{2} = \rho_{{A'} | \Zbasis} = \frac{\id_{A'}}{2}
		\end{equation}
		where $\rho_{A'| b_A}$ denotes the the state sent out from Alice's lab given that she chooses a basis $b_A$. Essentially, \cref{eq:sourcecondition} says that the Alice's signal states leak no information about the basis chosen by Alice. This can be shown rigorously as follows.
		
		Using the source-replacement scheme \cite{curty_entanglement_2004,ferenczi_symmetries_2012}, Alice's signal preparation is equivalent to her first preparing the state $\ket{\Psi_+} = \frac{\ket{00}_{AA'} + \ket{11}_{A A'}}{\sqrt{2}}$ followed by measurements on the $A$ system.  Now, if Alice prepares the states from \cref{eq:sourcecondition}, her POVM elements corresponding to the basis $b_A$ signal states sum to $p^{(A)}_{(b_A)} \id_A$. Because of this fact, one can view Alice's measurement process,
		\textit{after} using the source-replacement scheme, as equivalent to choosing the basis $\Zbasis(\Xbasis)$ with probability $\pzA(\pxA)$, followed by measuring using the POVM 
		$\{ \AlicePOVM{b_A}{0},\AlicePOVM{b_A}{1}\}$, for a given basis $b_A$. This reflects the fact that Eve has no knowledge of the basis used.
		
		The POVM elements are given by
		\begin{equation} \label{eq:alicePOVMs}
			\begin{aligned}
				\AlicePOVM{\Zbasis}{0} = \ketbra{0}, \quad  & \AlicePOVM{\Zbasis}{1} = \ketbra{1} \\
				\AlicePOVM{\Xbasis}{0} = \ketbra{+}, \quad  & \AlicePOVM{\Xbasis}{1} = \ketbra{-}.
			\end{aligned}
		\end{equation}
		
		Therefore, we now have a setup where the state $\rho_{A^nB^n}$ is shared between Alice and Bob, followed by basis choice and measurements by Alice. 
		
		\begin{remark} \label{remark:sourcereplacement}
			Without  loss of generality, one can always use the source-replacement scheme, and delay Alice's measurements until after Eve's attack has been completed, for any set of signal states. However, this process might result in POVM elements for Alice whose sum (for a specific basis) is not proportional to identity. 
            In this case, Alice's measurements are \textit{incompatible with active basis choice} after the source-replacement scheme. We utilize the fact that Alice implements active basis choice when using the EUR statement (\cref{thm:EUR}), and in bounding the phase error rate (\cref{sec:perfectdetectors,sec:imperfectdetectors}). It is precisely for this reason that \cref{eq:sourcecondition} is needed. 
            
            We note that there exist many methods to address imperfect state preparation (\cref{eq:sourcecondition} does not hold),    \cite{curras-lorenzo_security_2024,tamaki_loss-tolerant_2014,pereira_modified_2023,zapatero2023implementationsecurityquantumkey}. The approach there is fairly involved and involves a scenario where Alice prepares two different source-replaced states, depending on whether it is a $\Zbasis$ round or a $\Xbasis$ round. In this chapter, we will always consider the simpler case outlined above, and focus our attention on detector imperfections.
		\end{remark}
\item \textbf{Measurement: } Bob chooses to measure in the $\Zbasis$($\Xbasis$) basis with probability $\pzB$($\pxB$). For each basis choice, Bob has two threshhold detectors, each of which can click or not-click. Bob maps double clicks to $0/1$ randomly (this is essential, see \cref{remark:doublecountremapping}), and thus has 3 POVM elements in each basis $b$, which we denote using  $\{ \BobPOVM{b}{\bot},\BobPOVM{b}{0},\BobPOVM{b}{1}  \}$ which correspond to the inconclusive-outcome, $0$-outcome, and the $1$-outcome. In this work, we will use the following notation to write joint POVM elements, 
		\begin{equation} \label{eq:alicebobpovms} \begin{aligned} 
				\AliceBobPOVM{b_A, b_B}{i,j} &\coloneq \AlicePOVM{b_A}{i} \otimes \BobPOVM{b_B}{j}, \quad \\ 
				\AliceBobPOVM{b_A,b_B}{\neq} &\coloneq \AlicePOVM{b_A}{0} \otimes \BobPOVM{b_B}{1} +  \AlicePOVM{b_A}{1} \otimes \BobPOVM{b_B}{0} ,\\
				\AliceBobPOVM{b_A,b_B}{=} & \coloneq \AlicePOVM{b_A}{0} \otimes \BobPOVM{b_B}{0} +  \AlicePOVM{b_A}{1} \otimes \BobPOVM{b_B}{1}, \\
				\AliceBobPOVM{b_A,b_B}{\bot} & \coloneq \id_A \otimes \BobPOVM{b_B}{\bot},
			\end{aligned}
		\end{equation}
		where Alice's POVMs are defined in \cref{eq:alicePOVMs}, and Bob's in \cref{subsec:detectors}.		\begin{remark} \label{remark:doublecountremapping}
			As we will see in \cref{sec:perfectdetectors}, the mathematical assumption on Bob's detector setup needed for phase error estimation is actually given by
			\begin{equation} \label{eq:bobcondition}
				\BobPOVM{\Xbasis}{\bot}=\BobPOVM{\Zbasis}{\bot}.
			\end{equation} 
			This means that the probability of a round being inconclusive (i.e discarded) is independent of the basis for all input states. Notice that \cref{eq:bobcondition} depends on the choice of classical post-processing on Bob's side. In particular, it can be trivially satisfied by mapping no-click and double-click events to $0$ and $1$ randomly (so that $\BobPOVM{\Xbasis}{\bot}=\BobPOVM{\Zbasis}{\bot}$ is zero). However, such a protocol cannot produce a key when loss is greater than $50\%$, and is therefore impractical. In general, if one assumes the canonical model of detectors (see \cref{subsec:detectors}), and maps double-clicks to 0/1 randomly, then \cref{eq:bobcondition} requires the loss and dark count rates in each detector-arm to be equal.  This is why this condition is referred to as ``basis-independent loss", and its violation is referred to as ``detection-efficiency mismatch" in the literature. Note that even for identical detectors, one is forced remap double-click events to satisfy \cref{eq:bobcondition}.
		\end{remark}

		\item \textbf{Classical Announcements and Sifting: } For all rounds, Alice and Bob announce the basis they used. Furthermore, Bob announces whether he got a conclusive outcome ($\{ \BobPOVM{b}{0},\BobPOVM{b}{1} \}$), or inconclusive ($\{  \BobPOVM{b}{\bot} \}$). A round is said to be ``conclusive'' if Alice and Bob used the same basis, and Bob obtained a conclusive outcome. 

        All rounds where Alice sent $\Xbasis$ are used for testing, and Alice and Bob announce their measurement outcomes. These rounds are used to estimate the phase error rate. We let $\nX$ be the number of $\Xbasis$ basis conclusive rounds, and let $\eX$ be the observed error rate in these rounds. 

        All rounds where Alice sent $\Zbasis$ are used for key generation.\footnote{Note that, in practice, a very small fraction of these rounds can be used to estimate the $\Zbasis$-basis error rate, which in turn can be used to determine $\leak(\Fobs)$, the amount of error correction required. However, this sample need not be estimated accurately for the purposes of security, and therefore we may choose to use only a very small fraction of rounds for this purpose.} Rounds where no-detect events occur are discarded, and one is left with $\nK$ key generation rounds used for key generation. 
		
		All these classical announcements are stored in the register $\CP_1^n$, and 	$\event{\nX,\nK,\eX,\eZ}$ denotes the event that $\nX,\nK,\eX,\eZ$ values are observed in the protocol.

		\begin{remark}
			In this chapter, we use bold letters, such as $\bm{x}$ to denote a classical random variable, and $x$ to denote a particular value it takes. Furthermore, we will use $\event{x}$ to denote the event that $\bm{x} =x$. 
			Thus our protocol involves random  variables $\nXrv,\nKrv,\eXrv,\ephrv$, which take  values $\nX,\nK,\eX,\eph$ in any given run.
		\end{remark}        
\end{enumerate}

We use $\lkey(\params)$ to denote the key length, and $\leak(\params)$ to denote the error-correction protocol parameter in this thesis, with the understanding that $\params$ is obtained from $\cobs$.

The above protocol is a specific instantiation of \nameref{prot:qkdprotocol}, and the various probabilities can be related via simple algebraic manipulation. In particular:
\begin{itemize}
    \item $\Pr(\test)$ in \nameref{prot:qkdprotocol} is equal to $\pxA$. 
    \item $\Pr(a| \test) = \frac{1}{2}$ in \nameref{prot:qkdprotocol}  whenever $a$ corresponds to a $\Xbasis$ basis state, and $\Pr(a| \test) = 0$ whenever $a$ corresponds to $\Zbasis$ basis state.
    \item $\Pr(a| \gen) = \frac{1}{2}$ in \nameref{prot:qkdprotocol}  if $a$ corresponds to a $\Zbasis$ basis state, and $0$ otherwise. 
\end{itemize}

\subsection{Requirements on phase error estimation} \label{subsec:requirementsonphaseerror}

We now turn our attention to the phase error rate. Note that in a QKD protocol, one starts with a fixed but unknown state $\rho_{A^nB^n\Eve}$ that represents Eve's attack. As the protocol evolves, we get $\nK$ states where Alice and Bob measured in the $\Zbasis$ basis, and Bob got a detection event. The phase error rate is defined by the error rate in these rounds, if Alice and Bob instead choose to measure these rounds in the complementary $\Xbasis$ basis. Describing this formally requires viewing measurements by Alice and Bob as multi-step measurements, and is undertaken throughout this chapter. 

In this way, the state gives rise to random variables $\nXrv,\nKrv,\eXrv,\ephrv$. Here $\ephrv$ denotes the random variable corresponding to the phase error rate in the key generation rounds, when Alice and Bob measure those rounds (virtually) in the $\Xbasis$ basis. (The phase error rate is explained in greater detail in \cref{sec:perfectdetectors,sec:imperfectdetectors}). To obtain security, one must obtain a high probability upper bound on the phase error rate $\ephrv$.  We assume that one has a way to obtain the following statement (which we prove in \cref{sec:perfectdetectors,sec:imperfectdetectors}): 
	\begin{equation} \label{eq:req}
		\Pr(\ephrv \geq \Boundbasicdelta(\eXrv,\nXrv,\nKrv) ) \leq \epsAT^2.
	\end{equation}
	This states that the phase error rate is upper bounded (with high probability) by a suitable function $\Boundbasicdelta$ of the observed error rate in the $\Xbasis$ basis rounds, and the number of test and key generation rounds. 	We will obtain a suitable $\Boundbasicdelta$ satisfying \cref{eq:req} in  \cref{sec:perfectdetectors,sec:imperfectdetectors}, with and without the basis-independent loss assumption. The function $\Boundbasicdelta$ depends on the metrics $\deltaone,\deltatwo$ that quantify the deviation from ideal behavior for a given protocol description.

	\begin{remark} \label{remark:randomvariables}
		When working with random variables that are obtained via measurements on quantum states, the joint distributions of random variables can only be specified when those random variables \textit{can exist at the same time}, via some physical measurements on the state. For example, one cannot speak of the joint distribution of $\Xbasis$ and $\Zbasis$ measurement outcomes on the \textit{same} state, since such a joint distribution does not exist. 
		In the entirety of this thesis, all the random variables whose joint distribution is used in our arguments can indeed exist at the same time.
	\end{remark}
	
	Given an upper bound on the phase error rate (\cref{eq:req}), we have  the following theorem regarding the variable-length security of the QKD protocol described above, which follows from the bound \cref{eq:req} and \cref{theorem:eurvariablegenericresult} and the EUR statement \cite{tomamichel_uncertainty_2011}.

\begin{theorem} \label{theorem:eurqubitbb84}
    Consider \nameref{prot:qkdprotocol} with the details as specified in \cref{sec:protocoleurqubitbb84}. Suppose that the phase error rate bound from \cref{eq:req} holds, and the protocol produces an output key of length
\begin{equation} \label{eq:lvalueEUR}
			\begin{aligned}
				&l(\cobs) \coloneq  \max\Big(0,  \\ &\floor{ n_K\left(1- h \left( \Boundbasicdelta(\eX,\nX,\nK ) \right) \right) - \leak(\cobs)  
				- 2\log(1/\epsPA) - \EVcost + 2 }\Big).
			\end{aligned}
		\end{equation} 
    Then the QKD protocol is $(2\epsAT+\epsPA)$-secret, and $(2\epsAT + \epsPA + \epscorr)$-secure. 
\end{theorem}
\begin{proof}
   For this proof,  we will not need to consider the $\cnotobs$ from  \cref{theorem:eurvariablegenericresult}.\footnote{Formally, we can simply set $\cnotobs$ to be a dummy variable that always takes fixed value, and $\EURset(\cobs)$ to always be a set containing that fixed value.}
  We define $   \serfbound(\cobs)$ to be the probability that the phase error bound fails, \emph{conditioned} on $\Omega(\cobs)$, i.e,
    \begin{equation} \label{eq:defserfbound}
        \serfbound(\cobs) = 	\Pr(\ephrv \geq \Boundbasicdelta(\eXrv,\nXrv,\nKrv) | \Omega(\cobs) ),
    \end{equation}
    Then, the \cref{eq:genericvariablecondition2} is satisfied since the overall probability of the phase error bound failing is small (from \cref{eq:req}). Further, we set $\EURbeta(\cobs) = \nK \left(1-h(\Boundbasicdelta(\eX,\nX,\nK)) \right)$, which ensures that $\lkey(\cobs)$ form \cref{eq:lvalueEUR} matches the specification from \cref{theorem:eurvariablegenericresult} (see \cref{eq:genericvariablecondition3}). 

    Thus, we only need to prove that $\EURbeta(\cobs)$ lower bounds the smooth min entropy as as described in \cref{eq:genericvariablecondition1}.

    	To do so, focus on the state $\rho_{A_1^{\nK}B_1^{\nK} \CP_1^n \Eve | \event{\cobs} }$, which is the state on the detected key generation rounds. This state can be obtained by transforming Bob's measurement  procedure to consist of two steps, and then only implementing the first step measurement which determines the detect vs no-detect outcome. Such a state can be rigorously obtained using \cref{lemma:twostep}. For the purposes of this proof, we only need the fact that it is well defined.  We will obtain a bound on the smooth min entropy of the key generated from this state.
			
		Suppose Alice measures her $\nK$ systems in the $\Zbasis$ basis (this is what happens in the actual protocol). The post-measurement state is given by  $\rho_{\PAstring B_1^{\nK} \CP_1^n \Eve  | \event{\cobs} }$. Suppose she measures it in the $\Xbasis$ basis, and let the post-measurement state be given by $\rho^\mathrm{virt}_{\widetilde{\PAstring} B_1^{\nK}\CP_1^n \Eve  | \event{\cobs} }$. This $\Xbasis$ measurement is not actually done in the protocol, and is only required for the theoretical proof. Using the entropic uncertainty relation \cite{tomamichel_uncertainty_2011} (see \cref{thm:EUR}), we can relate the smooth min and max entropies $\left(\text{with smoothing parameter }\sqrt{\serfbound(\cobs)}\right)$ of the two states obtained via $\Zbasis$ and $\Xbasis$ measurements as 
			\begin{equation}
				 \Hmin[\sqrt{\serfbound(\cobs)}](\PAstring | \CP_1^n  \Eve)_{\rho | \event{\cobs} } + \Hmax[\sqrt{\serfbound(\cobs)}](\widetilde{\PAstring} | B_1^{\nK})_{\rho^\mathrm{virt} | \event{\cobs}} \geq \nK \qualityfactor.
				\end{equation}
			where $ \qualityfactor \coloneq \log(\frac{1}{ \max_{i,j} \norm{\AlicePOVM{\Xbasis}{i} \AlicePOVM{\Zbasis}{j}}^2_\infty})$\footnote{We need to apply the entropic uncertainty relations on the $\nK$ round measurements. However, for IID measurements, this can be then simplified easily to be $\nK$ times the computation for a single round measurement}. We have deliberately chosen an appropriate smoothing parameter in the above equation.

			\begin{remark}
				Notice that the value of $\qualityfactor$ \textit{only} depends on the POVM's used by Alice, after using the source-replacement scheme, and is equal to $1$ in this work. Thus, we set $\qualityfactor=1$ in the remainder of this work. Moreover, directly using the EUR in this context requires Alice to implement an \textit{active} basis choice measurement, which requires perfect signal state preparation. However, as stated earlier, several techniques of dealing with imperfect source preparation exist.
			\end{remark}	
	We now wish  so simply the max entropy term in the above expressions. We can make Bob measure his systems $B$ in the $\Xbasis$ basis to obtain the classical outcome $Y_1^{\nK}$. Then, using  data processing \cref{lemma:DPIsmoothmin} , we obtain
			\begin{equation} \label{eq:eurafterdpi}
				\begin{aligned}
					& \Hmin[\sqrt{\serfbound(\cobs)}](\PAstring | \CP_1^n \Eve)_{\rho | \event{\cobs} } \\
					+& \Hmax[\sqrt{\serfbound(\cobs)}](\widetilde{\PAstring} | Y_1^{\nK})_{\rho^\mathrm{virt} | \event{\cobs}} \geq \nK .
				\end{aligned}
			\end{equation}

			Recall that we have a probabilistic upper bound on $\ephrv$ (the error rate in $\widetilde{\PAstring},Y_1^{\nK}$) conditioned on the event $\event{\cobs}$. This bound fails with probability $\serfbound(\cobs)$ (see \cref{eq:defserfbound}). Thus,  using \cref{lemma:smoothmaxerror}  along with this fact, we obtain:
			\begin{equation}				\Hmax[\sqrt{\serfbound(\cobs)}](\widetilde{\PAstring} | Y_1^{\nK})_{\rho^\mathrm{virt} | \event{\cobs} } \leq \nK h \left( \Boundbasicdelta(\eX,\nX,\nK) \right),
			\end{equation}
			which along with \cref{eq:eurafterdpi} gives us
			\begin{equation} \label{eq:eurbound}
				\begin{aligned}
					\Hmin[\sqrt{ \serfbound(\cobs)}](\PAstring | \CP_1^n  \Eve)_{\rho | \event{\cobs} }  \geq \nK ( 1 - h\left( \Boundbasicdelta(\eX ,\nX,\nK)\right)).
				\end{aligned}
			\end{equation}
			This is the required bound on the smooth min entropy of the raw key, where the RHS is exactly what we have defined to be $\EURbeta(\cobs)$.  This concludes the proof. 
\end{proof}
Thus all that is left to do is obtain the required bound on the phase error rate.
\section{Phase error estimation for BB84 with perfect detectors} \label{sec:perfectdetectors}

	We will now prove \cref{eq:req} for an implementation that satisfies the basis-independent loss assumption. It is useful to refer to \cref{fig:virtualprotperfect} for this section. To prove \cref{eq:req}, we will need to modify the actual protocol to an equivalent protocol (in the sense of being the same quantum to classical channel). To do so we will use \cref{lemma:twostep} below to reformulate Alice and Bob's measurements to consist of two steps. The first step will implement a basis-\textit{independent} filtering operation that discards the inconclusive outcomes, while the second step will complete the measurement procedure. Then the required claim will follow from random sampling arguments on the second step measurements. We start by explaining the two-step protocol measurements.

	\subsection{Protocol Measurements} \label{subsec:protmeasureperfect}
	
	We will first use the following lemma to divide Alice and Bob's measurement procedure into two steps. For the proof, we refer the reader to Appendix.~\ref{Appendix:EUR}.

	\begin{restatable}[Filtering POVMs]{lemma}{twosteplemma} \label{lemma:twostep}
				Let $\{\Gamma_{k}  | k \in \mathcal{A}\}$ be a POVM on a register $Q$, and let $\{ \mathcal{A}_i\}_{i \in \mathcal{P}_\mathcal{A}}$ be a partition of $\mathcal{A}$, and let  $\rho \in \dop{=}(Q)$ be a state. The classical register storing the measurement outcomes when $\rho$ is measured using $\{\Gamma_k\}_{k \in \mathcal{A}}$ is given by
		\begin{equation}
			\rho_\mathrm{final} \coloneq	\sum_{k \in \mathcal{A}} \Tr(\Gamma_{k} \rho)  \ketbra{k}.
		\end{equation}
		This measurement procedure is equivalent (in the sense of being the same quantum to classical channel) to the following two-step measurement procedure: First doing a coarse-grained ``filtering'' measurement of $i$, using POVM $\{ \tilde{F}_i \}_{ i \in \mathcal{P}_\mathcal{A}}$, where
		\begin{equation}
			\begin{aligned}
				\tilde{F}_i  &\coloneq \sum_{j \in \mathcal{A}_i} \Gamma_{j}, \quad \quad \text{leading to the post-measurement state} \\
				\rho^\prime_\mathrm{intermediate} &= \sum_{i \in \mathcal{P}_{\mathcal{A}}}  \sqrt{\tilde{F}_i} \rho \sqrt{\tilde{F}_i}^\dagger  \otimes \ketbra{i}.
			\end{aligned}
		\end{equation}
		Upon obtaining outcome $i$ in the first step, measuring using  POVM $\{ G_{k} \}_{ k \in \mathcal{A}_i}$ where
		\begin{equation} \label{eq:twostepsecondstate}
			\begin{aligned}
				G_{k} &\coloneq \sqrt{\tilde{F}}^+_i \Gamma_{k} \sqrt{\tilde{F}}^+_i + P_{k} \quad \quad \text{leading to the post-measurement classical state} \\
				\rho^\prime_\mathrm{final} &= \sum_{i \in \mathcal{P}_{\mathcal{A}}} \sum_{k \in \mathcal{A}_i}  \Tr( G_{k }\sqrt{\tilde{F}_i} \rho \sqrt{\tilde{F}_i})  \ketbra{k},
			\end{aligned}
		\end{equation}
		where $F^+$ denotes the pseudo-inverse of $F$, and $P_{k}$ are any positive operators satisfying $\sum_{k \in \mathcal{A}_i} P_k = \id - \proj_{\tilde{F}_i}$, where $\proj_{\tilde{F}_i}$ denotes the projector onto the support of $\tilde{F}_i$.
		\end{restatable}

	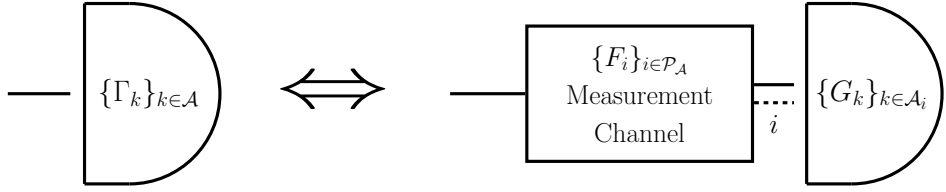
\begin{figure}[!ht]
  \centering
  \scalebox{0.6}{%
  \begin{tikzpicture}

    \begin{scope}[shift = {(-8,0)}]
      \pic {detectorLarge={det_full,  { \LARGE $\{\Gamma_k\}_{k\in\mathcal{A}}$} , black}};
      \coordinate (left_det_full) at ($(det_full) - (3,0)$);

      \draw[line width = 0.7mm] (left_det_full) -- ([xshift=0cm] det_full.west);
    \end{scope}

    \begin{scope}[shift={(8,0)}]
      \pic {detectorLarge={det_second_step, { \LARGE $\{G_k\} _{k\in\mathcal{A}_i}$ }, black}};
      \node[channel, left of = det_second_step, xshift = -4cm,
            font = {\fontsize{18}{23.76}\selectfont}](channel_first_step)
            {$\{F_i\}_{i \in \mathcal{P}_\mathcal{A}}$\\ Measurement \\Channel};
      \coordinate (left_channel) at ($(channel_first_step) - (4.2,0)$);

      \draw[line width = 0.7mm] (left_channel) -- (channel_first_step.west);
      \draw[line width = 0.7mm]
        ([yshift = 0.2cm] channel_first_step.east)
        -- ([xshift=0.15cm,yshift = 0.2cm] det_second_step.west);

      \draw[dashed, line width = 0.7mm]
        ([yshift=-0.2cm] channel_first_step.east)
        -- ([yshift=-0.2cm, xshift=0.15cm] det_second_step.west)
        node[midway, below, yshift = -0.1cm, black,
             font = {\fontsize{18}{23.76}\selectfont}]{$i$};
    \end{scope}

    \node at (-2.5,0) {\scalebox{1}{\resizebox{2.5cm}{0.6cm}{$\Leftrightarrow$}}};

  \end{tikzpicture}%
  }
  \caption{Schematic for the two-step measurement procedure from \cref{lemma:twostep}. Note that the second step measurement $\{G_k\}_{k \in \mathcal{A}_i}$ depends on the outcome of the first step measurement.}
\end{figure}

	Consider the POVMs $\{	\AliceBobPOVM{b_A,b_B}{\neq}, \AliceBobPOVM{b_A,b_B}{=}, \AliceBobPOVM{b_A,b_B}{\bot} \}$ defined in \cref{eq:alicebobpovms}, which correspond to Bob obtaining a conclusive outcome and Alice and Bob obtaining an error, Bob obtaining a conclusive outcome and Alice and Bob not obtaining an error, and Bob obtaining an inconclusive outcome respectively, for basis choices $b_A,b_B$. Without loss of generality, we can use \cref{lemma:twostep} to equivalently describe
	Alice and Bob's measurement procedure as consisting of two steps. 
	\begin{enumerate}
		\item First, they measure using POVM $\{\AliceBobPOVMFilter{b_A,b_B}{\con } ,  \AliceBobPOVMFilter{b_A,b_B}{\bot}\}$ which determines whether they obtain a conclusive and inconclusive measurement outcome. 
		\item Then, if they obtain a conclusive outcome, they measure using a second POVM \\ $\{ \AliceBobPOVMsecond{b_A,b_B}{=},\AliceBobPOVMsecond{b_A,b_B}{\neq} \}$.
	\end{enumerate}
	We use the convention that whenever an explicit basis $(\Xbasis/\Zbasis)$ is written in the subscript of these POVMs, it refers to the basis used by both Alice and Bob.	
	We refer to the first-step measurements as ``filtering" measurements, since they determine whether Bob gets a conclusive outcome (which may be kept or discarded depending on basis choice), or an inconclusive outcome (which is always discarded).
	Furthermore, due to the construction of the POVM from \cref{lemma:twostep}, we have
	\begin{equation}
		\AliceBobPOVMFilter{b_A,b_B}{\bot} =  \id_A \otimes \BobPOVM{b_B}{\bot} .
	\end{equation}

    \subsection{Constructing Equivalent Protocol} \label{subsec:equivprotperfect}

    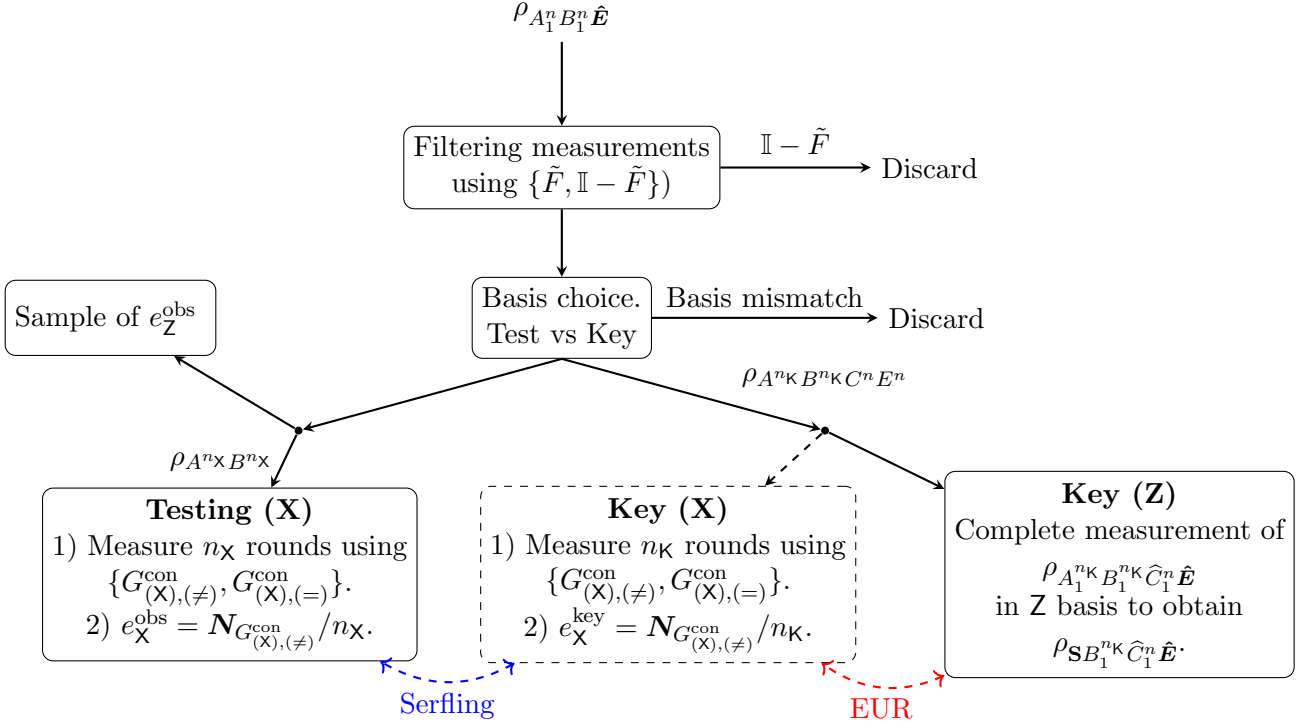
\begin{figure}[ht]
    \hspace*{-0.8cm}
    {\small \begin{tikzpicture}[node distance=2cm]
        \node (start) {$\rho_{A_1^nB_1^n \Eve}$ };
        \node (filtering) [processLarge, below of=start]
            {Filtering measurements \\ using $\{ \tilde{F} , \id - \tilde{F}\}$)};

        \node (center) [processLarge, below of=filtering]
            {Basis choice. \\ Test vs Key };

        \node (eZsample) [processLarge, left of=center, xshift = -4cm]
            { Sample of $\eZ$ };

        \node (test) [processLarge, below left of=center, xshift=-3cm, yshift=-2cm]
            { \textbf{Testing (X)} \\
              1) Measure $\nX$ rounds using \\
              $\{  \AliceBobPOVMsecond{\Xbasis}{\neq}, \AliceBobPOVMsecond{\Xbasis}{=}\}$. \\
              2) $\eX = \bm{N}_{ \AliceBobPOVMsecond{\Xbasis}{\neq}} / \nX$. };

        \node(keystate) [point, right of=center,xshift = 1.5cm,yshift = -1.5cm] {};
        \node(teststate) [point, left of=center,xshift = -1.5cm,yshift = -1.5cm] {};

        \node(keystateabove) [emptyprocess, above of=keystate,yshift = -1.3cm]
            {$\rho_{A^{\nK}B^{\nK} C^n E^n}$};

        \node (keyX) [virtualprocess, below right of=center, xshift=0cm, yshift=-2cm]
            { \textbf{Key (X)} \\
              1) Measure $\nK$ rounds using \\
              $\{  \AliceBobPOVMsecond{\Xbasis}{\neq}, \AliceBobPOVMsecond{\Xbasis}{=}\}$. \\
              2) $\eph = \bm{N}_{ \AliceBobPOVMsecond{\Xbasis}{\neq}} / \nK$. };

        \node (key) [processLarge, below right of=center, xshift=6cm, yshift=-2cm]
            { \textbf{Key (Z)} \\
              Complete measurement of \\
              $\rho_{A_1^{\nK}B_1^{\nK}\CP_1^n\Eve}$ \\
              in $\Zbasis$ basis to obtain \\
              $\rho_{\PAstring B_1^{\nK}\CP_1^n\Eve}$.};

        \draw [arrow] (start) --  ++(filtering);
        \draw [arrow] (filtering) -- ++(center);

        \draw [arrow] (center.south) -- ++ (teststate) ;
        \draw [arrow] (center.south) -- ++ (keystate) ;

        \draw [arrow] (filtering.east) -- ++(2,0)
            node[midway, above] {$\id-\tilde{F}$} node[right] {Discard};

        \draw [arrow] (center.east) -- ++(3,0)
            node[midway, above] {Basis mismatch} node[right] {Discard};

        \draw [arrow] (teststate) -- ++(test)
            node[midway,left] {$\rho_{A^{\nX} B^{\nX}}$};;

        \draw [arrow] (teststate) -- ++(eZsample);

        \draw [arrow] (keystate) -- ++(key);
        \draw [dashedarrow] (keystate) -- ++(keyX);

        \draw (key) edge [dashed, <->, thick, bend left, color = red]
            node[midway, below] {EUR} (keyX);

        \draw (keyX) edge [dashed, <->, thick, bend left, color = blue]
            node[,yshift = 0cm, midway, below] {Serfling} (test);
    \end{tikzpicture} }
    \caption{Protocol flowchart for the equivalent protocol from \cref{sec:perfectdetectors}, where basis-independent loss assumption (\cref{eq:bobcondition}) is satisfied. The dotted arrows and boxes represent virtual measurements that do not actually happen in the real protocol. Connections between different boxes are highlighted using curved arrows. We use the Serfling bound (\cref{lemma:sampling}) to obtain a bound on the phase error rate from observations. The phase error rate is then used to bound the smooth min entropy using the EUR statement. We use $\bm{N}_P$ to denote the number of $P$ measurement outcomes, where $P$ denotes a POVM element. For clarity, we have omitted the conditioning on events in the figure (but not in our proof). The basis used for measurements is indicated in each box, and refers to the basis used by \textit{both} Alice and Bob.}
    \label{fig:virtualprotperfect}
\end{figure}
	
	We will now construct an equivalent protocol that is described in \cref{fig:virtualprotperfect}. 
	\begin{enumerate}
		\item If one has $\BobPOVM{\Xbasis}{\bot} = \BobPOVM{\Zbasis}{\bot}$, then we find that the filtering measurements $\AliceBobPOVMFilter{b_A,b_B}{\con}$ is independent of the basis choices $(b_A,b_B)$. Let this basis-independent  POVM element be $\tilde{F}$. If the filtering measurement does not depend on the basis choice, then implementing the basis choice followed by filtering measurement is the same as implementing the filtering measurement followed by basis choice. Thus, we can delay basis choice until after the filtering measurements have been performed. This can also be formally argued using \cref{lemma:twostep}. This allows us to obtain the first node of \cref{fig:virtualprotperfect}, where we measure using $\{\tilde{F} , \id -\tilde{F}\}$.
		\item This is then followed by random basis choices and assignment to test vs key by Alice and Bob.  All $\Xbasis$ basis rounds are used for testing, while almost all $\Zbasis$ basis rounds are used for key generation (and a small fraction is used to estimate $\eZ$). Thus, we get the second node of \cref{fig:virtualprotperfect}.  Note that the estimate $\eZ$ of the error rate in the key bits is only used to determine the amount of error-correction required and does not affect the secrecy of the protocol. However, $\eZ$ and the choice of error-correction protocol is important to ensure that error-verification succeeds with high probability.
		
		\item The $\nX$ testing rounds are measured using $\{ \AliceBobPOVMsecond{\Xbasis}{=},\AliceBobPOVMsecond{\Xbasis}{\neq} \}$, and the error rate in these rounds is denoted by $\eX$. This is the error rate we observe. This is the \textbf{Testing (X)} node of \cref{fig:virtualprotperfect}.
		
		\item The $\nK$ key generation rounds can be measured (virtually) using the same POVM \\ $\{ \AliceBobPOVMsecond{\Xbasis}{=},\AliceBobPOVMsecond{\Xbasis}{\neq} \}$. The error rate in these rounds is denoted by $\eph$ and is the phase error rate we wish to estimate. This is the \textbf{Key (X)} node of \cref{fig:virtualprotperfect}.
		
		\item The actual $\nK$ key generation rounds are measured in the $\Zbasis$ basis to obtain the raw key. This is the \textbf{Key (Z)} node of \cref{fig:virtualprotperfect}.
	\end{enumerate}

\subsection{Sampling} \label{subsec:samplingperfect}
	We will now turn our attention to the sampling part of the argument, and obtain an estimate $\Boundbasiczero$ on the phase error rate that satisfies \cref{eq:req}. To do so, we will make use of the following Lemma, which uses the Serfling bound \cite{serfling_probability_1974}. For the proof, we refer the Appendix.~\ref{appsubsec:randomsampling}.

	\begin{restatable}[Serfling with IID sampling]{lemma}{serflinglemma} \label{lemma:sampling}
		Let $\X_1 \dots \X_{n}$ be bit-valued random variables. Suppose each position $i$ is mapped to the ``test set'' ($i \in \bm{J}_t$) with probability $p_t$, and the ``key set'' ($i \in \bm{J}_k$) with probability  $p_k$.  
		Let $\event{\nX,\nK}$ be the event that exactly $\nX$ positions are mapped to test, and exactly $\nK$ positions are mapped to key. Then, conditioned on the event $\event{\nX,\nK}$, the following statement is true:
		\begin{equation}
			\begin{aligned}
				\Pr\left(\sum_{i \in \bm{J}_k} \frac{\X_i}{\nK} \geq \sum_{i \in \bm{J}_t} \frac{\X_i}{\nX} + \gamma_\mathrm{serf} \right)_{ | \event{\nX,\nK}} &\leq e^{-2 \gamma_\mathrm{serf}^2 \fserf(\nX,\nK)} ,\\
				\fserf(\nX,\nK )&\coloneq \frac{\nK \nX^2}{(\nK+\nX)(\nX+1)}.
			\end{aligned}
		\end{equation}
	\end{restatable}
	
	To use the lemma, we will identify $X_i=1$ with error, and $X_i=0$ with the no-error  outcome, when the conclusive rounds are measured in the $\Xbasis$ basis. The test data will correspond to $\eX$, whereas the key data will correspond to $\eph$. 

	\begin{remark}
		There are two important aspects to the sampling argument. First, the Serfling bound applies in the situation where one chooses a random subset of \textit{fixed-length} for testing. However, the above procedure (and many QKD protocols) randomly assigns each round to testing vs key generation. Thus, Serfling must be applied with some care, and that is what is done here\footnote{It is also worthwhile to note that if one is interested in estimating the QBER \textit{independent} of basis, then the standard serfling argument is directly applicable (for instance in \cite{tomamichel_largely_2017}).}. This observation has been missing in many prior works. 
		Second, since we are interested in a variable-length protocol, we require slightly different statements than standard fixed-length security proofs (\cref{eq:req}). However, these can also be obtained by simple (almost trivial) modifications to existing arguments and yield the same results as before. Both these issues are addressed in the proof of \cref{lemma:sampling} in \cref{appendix:sampling}.
	\end{remark}

		Let us consider the second node in the equivalent protocol constructed in \cref{fig:virtualprotperfect}, where rounds are now randomly assigned for testing ($\Xbasis$ basis) or key generation ($\Zbasis$ basis and key generation). (The remaining rounds are used for estimating the $\Zbasis$ basis error rate or discarded and are unimportant for this discussion).  Consider the state $\rho_{ | \event{\nX,\nK}}$, where the number of rounds to be used to testing and key generation is fixed. Using \cref{lemma:sampling} on this state, we obtain
		\begin{equation} \label{eq:samplinguseperfect}
			\begin{aligned}
				\Pr(\ephrv \geq \eXrv + \gamma_\mathrm{serf})_{ | \event{\nX,\nK}} &\leq  e^{-2 \gamma_\mathrm{serf}^2 \fserf(\nX,\nK)} ,
			\end{aligned}
		\end{equation}
		Furthermore we can choose
		\begin{equation} \label{eq:gammaserfdefined}
			\gamma^{\epsAT}_\mathrm{serf}(\nX,\nK) \coloneq \sqrt{\frac{\ln(1/\epsAT^2) }{ 2 \fserf(\nX,\nK)} } \implies e^{-2 \left(		\gamma^{\epsAT}_\mathrm{serf}(\nX,\nK) \right)^2 \fserf(\nX,\nK)} = \epsAT^2.
		\end{equation}
		Thus we can choose
		\begin{equation} \label{eq:estimateperfect}
			\Boundbasiczero(\eX,\nX,\nK) = \eX + \gamma^{\epsAT}_\mathrm{serf} (\nX,\nK)
		\end{equation}
		to be our bound for the phase error rate,  where the $(0,0)$ subscript indicates that the bound is only valid when there is no deviation from basis-independent loss. Finally, since the bound is valid for any event $\Omega(\nX,\nK)$, we can get rid of this conditioning in  \cref{eq:samplinguseperfect}, to obtain \cref{eq:req} via
		\begin{equation} 
			\begin{aligned}
				\Pr(\ephrv \geq 	\Boundbasiczero(\eXrv,\nXrv,\nKrv)  )& = \sum_{\nX,\nK} \Pr(\event{\nX,\nK}) \Pr(\ephrv \geq 	\Boundbasiczero(\eXrv,\nX,\nK)  )_{ | \event{\nX,\nK}}  \\
				&\leq \sum_{\nX,\nK} \Pr(\event{\nX,\nK})  \epsAT^2 \\
				& = \epsAT^2
			\end{aligned}
		\end{equation}
	
		Thus, for the above choice of $\Boundbasiczero(\eX,\nX,\nK)$, the variable-length security of the protocol follows from \cref{theorem:eurqubitbb84}.

\section{Phase error estimation for BB84 with imperfect detectors} \label{sec:imperfectdetectors}

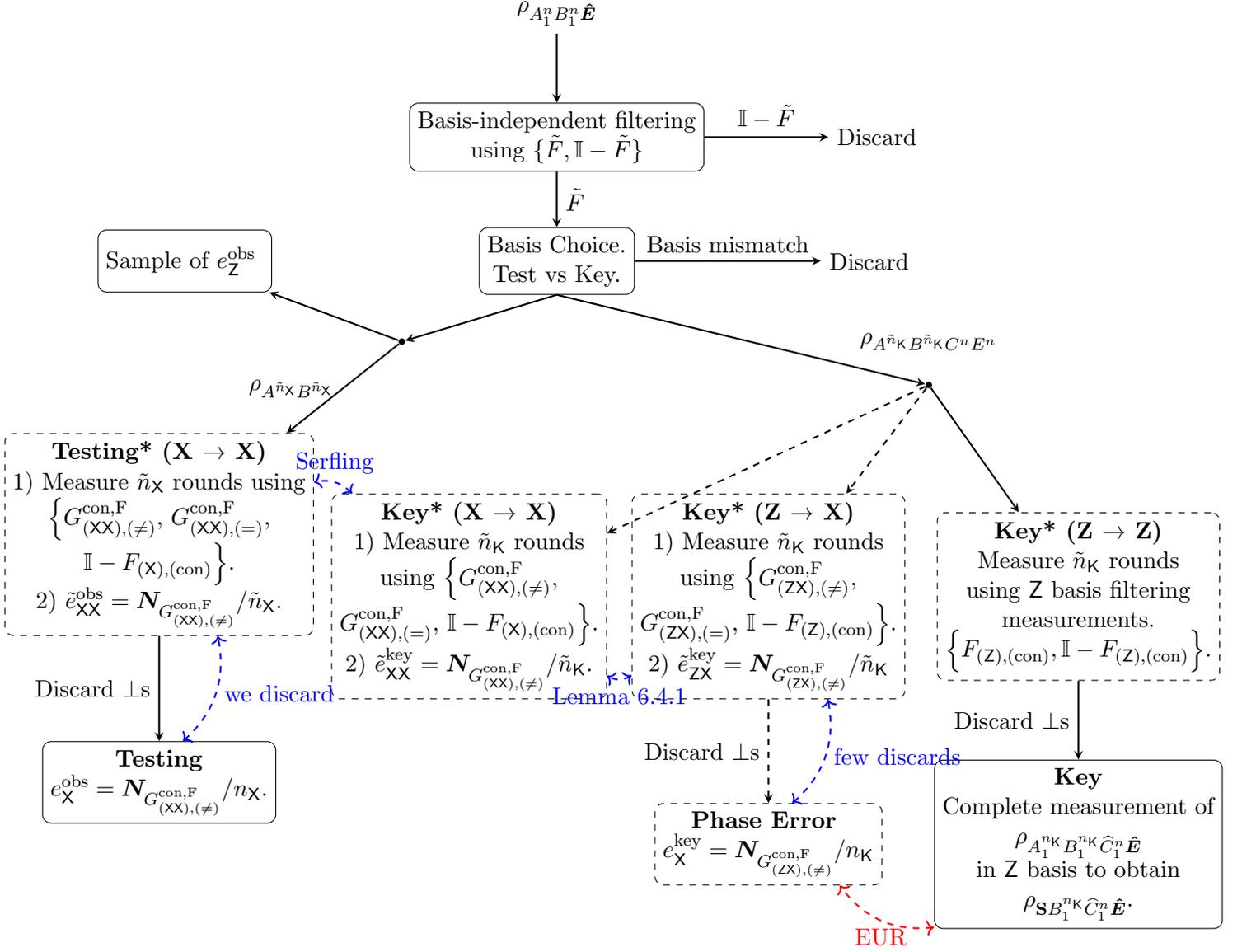
\begin{figure}[ht]
    \hspace*{-2cm} 
    {\small \begin{tikzpicture}[node distance=2cm]
        \node (start) {$ \rho_{A_1^nB_1^n\Eve}$ };
        \node (filtering) [processLarge, below of=start]
            {Basis-independent filtering \\
             using $\{ \tilde{F} , \id - \tilde{F}\}$};

        \node (center) [processLarge, below of=filtering]
            {Basis Choice. \\ Test vs Key. };

        \node (eZsample) [processLarge, left of=center, xshift = -4cm]
            { Sample of $\eZ$ };

        \node(keystate) [point, right of=center,xshift = 4cm,yshift = -2cm] {};
        \node(teststate) [point, right of=center,xshift = -4.5cm,yshift = -1.3cm] {};

        \node(keystateabove) [emptyprocess, above of=keystate,yshift = -1.3cm]
            {$\rho_{A^{\nKperf}B^{\nKperf} C^n E^n}$};

        \node (testvirt) [virtualprocess, below left of=center, xshift=-5cm, yshift=-3cm]
            { \textbf{Testing* (X $\rightarrow$ X)} \\
              1) Measure $\nXperf$ rounds using \\
              $\Big\{ \AliceBobPOVMsecondsandwich{\Xbasis \Xbasis}{\neq},$
              $  \AliceBobPOVMsecondsandwich{\Xbasis \Xbasis}{=},$ \\$
              \id - \AliceBobPOVMprimeFilter{\Xbasis}{\con}  \Big\}$. \\
              2) $\eXperf = \bm{N}_{ \AliceBobPOVMsecondsandwich{\Xbasis \Xbasis}{\neq}} / \nXperf$. };

        \node (test) [processLarge, below of=testvirt,  yshift= -2cm]
            { \textbf{Testing} \\
              $\eX = \bm{N}_{ \AliceBobPOVMsecondsandwich{\Xbasis \Xbasis}{\neq}} / \nX$. };

        \node (keyvirtXX) [virtualprocess, below left of=keystate, xshift=-6cm, yshift=-2cm]
            { \textbf{Key* (X $\rightarrow$ X)} \\
              1) Measure $\nKperf$ rounds \\
              using 
              $\Big\{ \AliceBobPOVMsecondsandwich{\Xbasis \Xbasis}{\neq},$ \\
              $  \AliceBobPOVMsecondsandwich{\Xbasis \Xbasis}{=},$  $
              \id - \AliceBobPOVMprimeFilter{\Xbasis}{\con}  \Big\}$. \\
              2) $\ephwierd = \bm{N}_{ \AliceBobPOVMsecondsandwich{\Xbasis \Xbasis}{\neq}} / \nKperf$. };

        \node (keyvirtZX) [virtualprocess, below right of=keystate, xshift=-4cm, yshift=-2cm]
            { \textbf{Key* (Z $\rightarrow$ X)} \\
              1) Measure $\nKperf$ rounds \\ using 
              $\Big\{ \AliceBobPOVMsecondsandwich{\Zbasis \Xbasis}{\neq},$ \\
              $  \AliceBobPOVMsecondsandwich{\Zbasis \Xbasis}{=},$ $
              \id - \AliceBobPOVMprimeFilter{\Zbasis }{\con} \Big\}$. \\
              2) $\ephperf = \bm{N}_{\AliceBobPOVMsecondsandwich{\Zbasis \Xbasis}{\neq}} / \nKperf$};

        \node (keyZX) [virtualprocess, below of=keyvirtZX, yshift= -2cm]
            { \textbf{Phase Error } \\
              $\eph = \bm{N}_{\AliceBobPOVMsecondsandwich{\Zbasis \Xbasis}{\neq}} / \nK$ };

        \node (keyvirt) [virtualprocess, below right of=keystate, xshift=1cm, yshift=-2cm]
            { \textbf{Key* (Z $\rightarrow$ Z)} \\
              Measure $\nKperf$ rounds   \\
              using  $\Zbasis$ basis filtering \\
              measurements.  \\
              $\Big\{\AliceBobPOVMprimeFilter{\Zbasis }{\con}, \id- \AliceBobPOVMprimeFilter{\Zbasis }{\con} \Big\}$. };

        \node (key) [processLarge, below  of=keyvirt,yshift = -2cm]
            { \textbf{Key} \\
              Complete measurement of \\
              $\rho_{A_1^{\nK} B_1^{\nK} \CP_1^n \Eve}$  \\
              in  $\Zbasis$ basis to obtain \\
              $\rho_{\PAstring B_1^{\nK} \CP_1^n \Eve}$.};

        \draw [arrow] (start) -- ++ (filtering);
        \draw [arrow] (filtering) -- ++ (center) node[midway,right] {$\tilde{F}$};

        \draw [arrow] (teststate) -- (testvirt) node[midway,left] {$\rho_{A^{\nXperf} B^{\nXperf}}$};
        \draw [arrow] (teststate) -- (eZsample);

        \draw [arrow] (center.south) -- (keystate) ;
        \draw [dashedarrow] (keystate) -- (keyvirtXX);
        \draw [dashedarrow] (keystate) -- (keyvirtZX);
        \draw [arrow] (keystate) -- (keyvirt);
        \draw [arrow] (center.south) -- (teststate);

        \draw [arrow] (testvirt) -- (test) node [midway, left] {Discard  $\bot$s};

        \draw [dashedarrow] (keyvirtZX) -- (keyZX) node [midway, left] {Discard  $\bot$s};
        \draw [arrow] (keyvirt) -- (key) node [midway, left] {Discard  $\bot$s};

        \draw [arrow] (filtering.east) -- ++(2,0)
            node[midway, above] {$\id-\tilde{F}$} node[right] {Discard};

        \draw [arrow] (center.east) -- ++(3,0)
            node[midway, above] {Basis mismatch} node[right] {Discard};

        \draw (key) edge [dashed, <->, thick, bend left, color = red]
            node[midway, below] {EUR} (keyZX);

        \draw (keyZX) edge [dashed, <->, thick, bend right, color = blue]
            node[midway, right] {few discards} (keyvirtZX);

        \draw (keyvirtZX) edge [dashed, <->, thick, bend left, color = blue]
            node[midway, below] {\cref{lemma:pulkitgeneric}} (keyvirtXX);

        \draw (keyvirtXX) edge [dashed, <->, thick, bend right, color = blue]
            node[midway, above] {Serfling} (testvirt);

        \draw (test) edge [dashed, <->, thick, bend right, color = blue]
            node[midway, right] {we discard} (testvirt);

    \end{tikzpicture} }
    \caption{Protocol flowchart for the equivalent protocol from \cref{sec:imperfectdetectors}, where basis-independent loss assumption (\cref{eq:bobcondition}) is not satisfied.  For the POVMs, the reader may refer to \cref{table:povms} or \cref{subsec:protmeasure}. Compared to \cref{fig:virtualprot}, the testing and key generation rounds go through an additional second step filtering measurement that depends on the basis used. The basis used in these measurements in indicated in each box, and indicates the basis used by \textit{both} Alice and Bob.}
    \label{fig:virtualprot}
\end{figure}

In this section, we will prove \cref{eq:req} for an implementation that \textit{does not satisfy} the basis-independent loss assumption. The argument is similar to the one presented in \cref{sec:perfectdetectors}, with important additions. It is helpful to refer to \cref{fig:virtualprot} for this section. We will first explain the idea behind the proof, before stating the proof itself.

		\subsubsection*{Proof Idea}
		We will use \cref{lemma:twostep}  in \cref{subsec:protmeasure} to construct an equivalent measurement procedure (in the sense that it is the same quantum to classical channel) for the protocol, which consists of three steps. The first step measurement is done using the POVM $\{ \tilde{F}, \id - \tilde{F}\}$ and implements basis-\textit{independent} filtering (discarding) operations. ($\tilde{F}$ here plays the same role as in \cref{sec:perfectdetectors}, but is defined differently). In particular it is the largest common filtering operation over both basis choices.
		
		Due to basis-efficiency mismatch, we will have a second step measurement that implements filtering operations that depend on the basis choice. (This will typically result in a small number of discards for a small amount of basis-dependent loss in the detectors). Once both filtering steps are done, the measurements on the remaining rounds can be completed using the third step  measurements which determines the exact measurement outcomes on the detected rounds.

		Turning our attention to \cref{fig:virtualprot}, the state first undergoes the basis-independent filtering measurement in the first node. This is then followed by random basis choice and assignment to testing and key generation at the second node. The testing rounds are further measured using second step $\Xbasis$ basis filtering POVM and third step $\Xbasis$ basis POVM at the \textbf{Testing* (X $\rightarrow$ X)} node. Similarly, the key generation rounds are measured using second step $\Zbasis$ basis filtering POVM  and third step $\Zbasis$ basis POVM. Note that we use $(b_\mathrm{2nd} \rightarrow b_\mathrm{3rd})$ to denote the basis choice $b_\mathrm{2nd}$ for the second step filtering measurement, and basis choice $b_\mathrm{3rd}$ for the third step measurement, for both Alice and Bob.

		We will consider virtual measurements on the key generation rounds corresponding to  \textbf{X $\rightarrow$ X} and \textbf{Z $\rightarrow$ X}. These are represented using dotted boxes and lines in the figure. These measurements are not performed in the protocol, but are only required in our proof. 
		We will then associate an error rate with all these choices of measurements, which corresponds to the number of rounds that resulted in an error divided by the total number of rounds on which the measurements were done.
		
		We see a variety of error rates in \cref{fig:virtualprot}. These errors are classified based on three criteria: 
        \begin{enumerate}
            \item   The basis used by Alice \textit{and} Bob in the second and third step measurements (written in the subscript),
            \item Whether the $\bot$s due to the second step measurements have been discarded from the total number of rounds or not ($e$ vs $\tilde{e}$),
            \item Whether they were done on testing rounds ($\text{obs}$ in superscript), or key generation rounds ($\text{key}$ in superscript). 
        \end{enumerate}
        The proof will follow by building a connection from our observed error rate ($\eX$), to the phase error rate  ($\eph$). These connections are highlighted using curved blue arrows in the figure. Note that we only observe the error rate $\eX$ in the protocol. 
		
		In particular, we will relate the error rates before and after discarding for the testing rounds ($\eXperf \leftrightarrow \eX$) by simply noting that we discard rounds in the second-step measurements. On the other hand, we will relate $\ephperf
		\leftrightarrow \eph$ by bounding the number of discards that can happen in the second step filtering measurements. This relation will depend on $\delta_2$, which will be the metric that quantifies the ``smallness" of the POVM element corresponding to the discard outcome.  $\eXperf$ and $\ephwierd$ correspond to error rates corresponding to exactly the same measurement, and assigned to test vs key randomly. Thus, they can be related using Serfling (\cref{lemma:sampling}), exactly as in \cref{sec:perfectdetectors}.  $\ephwierd$ and $\ephperf$ correspond to error rates on the same state, but with slightly different POVMs, and thus are expected to be similar. This can be rigorously argued using \cref{lemma:pulkitgeneric}, where we use $\deltaone$ to quantify the ``closeness" of these POVMs. Combining all these relations, we will ultimately obtain \cref{eq:estimateimperfect}.

		We will now convert the above sketch into a rigorous proof. We start by explaining the three step protocol measurements.

	\subsection{Protocol Measurements} \label{subsec:protmeasure}
		Fix the basis $b_A,b_B$ used by Alice and Bob.
		As in \cref{subsec:protmeasureperfect}, consider the POVM \\ $\{\AliceBobPOVM{b_A,b_B}{\neq},
		\AliceBobPOVM{b_A,b_B}{=}, \AliceBobPOVM{b_A,b_B}{\bot}\}$ defined in \cref{eq:alicebobpovms}, which correspond to Bob obtaining a conclusive outcome (and Alice and Bob obtaining an error), Bob obtaining a conclusive outcome (and Alice and Bob not obtaining an error), and Bob obtaining an inconclusive outcome respectively.  Since $\AliceBobPOVM{b_A,b_B}{\bot} $ now depends on the basis choices, we cannot proceed in the same way as before. This reflects the fact that the discarding is basis dependent. Thus we will reformulate the measurement process in a different way.

	To do so, consider a $\tilde{F}$ such that 
		\begin{equation}\label{eq:Ftildedefined}
			\tilde{F} \geq	\AliceBobPOVM{b_A,b_B}{=} + \AliceBobPOVM{b_A,b_B}{\neq}  \quad \forall (b_A,b_B)
		\end{equation}
		This $\tilde{F}$ will play the role of a common  ``basis-independent filtering measurement".	While any choice satisfying the above requirement will suffice, for the best results, $\tilde{F}$ must fulfil \cref{eq:Ftildedefined} as tightly as possible.
		\begin{remark}
			Since basis-mismatch rounds are discarded anyway, it is possible to argue that we only need $\tilde{F}$ to satisfy $	\AliceBobPOVM{b_A,b_B}{=}  +\AliceBobPOVM{b_A,b_B}{\neq} \leq \tilde{F}$ for $b_A = b_B$. This involves constructing a slightly different equivalent protocol where the first node decides basis match vs mismatch. The basis match events then undergo the usual filtering followed by basis choice, while the mismatch events are discarded without any filtering. If this modified requirement results in a value of $\tilde{F}$ that is ``smaller" then the original choice, then this will lead to tighter key rates. Intuitively, this is due to the fact that a smaller value of $\tilde{F}$ means that more loss is attributed to the basis-independent filtering.
		\end{remark}

		To reformulate the measurement procedure, start by considering the four-outcome POVM given by $\{\id - \tilde{F},  \tilde{F} - \AliceBobPOVM{b_A,b_B}{=}  - \AliceBobPOVM{b_A,b_B}{\neq} , \AliceBobPOVM{b_A,b_B}{=} , \AliceBobPOVM{b_A,b_B}{\neq}   \}$,  where the first two outcomes correspond to discard, the third correspond to a conclusive no-error outcome, and the fourth corresponds to a conclusive error. This  four-outcome measurement followed by classical grouping of the first two outcomes is then equivalent to the original three-outcome measurement in the protocol.
        
	Now, we can use \cref{lemma:twostep} to reformulate the four-outcome measurement as occurring in two steps. In the first step, Alice and Bob measure using POVM $\{ \tilde{F} , \id - \tilde{F} \}$ and discard the latter outcomes. If they obtain the $\tilde{F}$ outcome, they then complete the measurement using POVM
		$\{  \AliceBobPOVMFilter{b_A,b_B}{\bot}, \AliceBobPOVMFilter{b_A,b_B}{=}, \AliceBobPOVMFilter{b_A,b_B}{\neq}\}$, corresponding to discard, conclusive no-error and conclusive error outcomes respectively. 
		
		We then use \cref{lemma:twostep} again to reformulate this three-outcome measurement to consist of two steps. First, they measure using $\{\AliceBobPOVMprimeFilter{b_A,b_B}{\con} ,  \AliceBobPOVMprimeFilter{b_A,b_B}{\bot}\}$ which determines whether they obtain a conclusive or inconclusive measurement outcome. Then, if they obtain a conclusive outcome, they measure using the POVM $\{ \AliceBobPOVMsecond{b_A,b_B}{=},\AliceBobPOVMsecond{b_A,b_B}{\neq} \}$. Thus we now have a three-step measurement procedure, described in \cref{table:povms}.

		Since basis-mismatch signals are anyway discarded in the protocol, from this point onwards, we will only be concerned with POVMs that correspond to Alice and Bob choosing the same basis. As before, we will use the convention that whenever a basis is explicitly written as $\Xbasis / \Zbasis$ (or denoted using $b_1  b_2$), it represents \textit{both} Alice and Bob's basis choices.		
        
It will be convenient to recombine the second and third step measurement into a single measurement step with three outcomes. For brevity we introduce the following notation to write this POVM $\{\AliceBobPOVMsecondsandwich{b_1 b_2}{\neq},\AliceBobPOVMsecondsandwich{b_1 b_2}{=} ,\id - \AliceBobPOVMprimeFilter{b_1}{\con}  \}$ where
		\begin{equation} \label{eq:notationGsandwich}
			\begin{aligned}
				\AliceBobPOVMsecondsandwich{b_1b_2}{\neq} &= \sqrt{\AliceBobPOVMprimeFilter{b_1}{\con}}\AliceBobPOVMsecond{b_2}{\neq} \sqrt{\AliceBobPOVMprimeFilter{b_1}{\con}}, \\
				\AliceBobPOVMsecondsandwich{b_1b_2}{=} &= \sqrt{ \AliceBobPOVMprimeFilter{b_1}{\con}}\AliceBobPOVMsecond{b_2}{=} \sqrt{\AliceBobPOVMprimeFilter{b_1}{\con}}. 
			\end{aligned}
		\end{equation}
		where the subscript $b_1 b_2$ determines the basis for the second step and third step measurements by \textit{both} Alice and Bob, and the superscript $F$ indicates the merging of the two measurement steps. (Note that if $b_1 = b_2 = b$, then this simply reverses the earlier action of \cref{lemma:twostep} that split $\{  \AliceBobPOVMFilter{b,b}{\bot}, \AliceBobPOVMFilter{b,b}{=}, \AliceBobPOVMFilter{b,b}{\neq}\}$ to generate the second and third-step measurements. However, we will consider fictitious measurements where $b_1 \neq b_2$ in our proof. To describe such measurements, it is indeed necessary to split $\{  \AliceBobPOVMFilter{b_A,b_B}{\bot}, \AliceBobPOVMFilter{b_A,b_B}{=}, \AliceBobPOVMFilter{b_A,b_B}{\neq}\}$  into two separate steps.)

        \begin{table}[H]
    \centering
    \renewcommand{\arraystretch}{1.3} 
    \begin{tabularx}{\textwidth}{>{\raggedright\arraybackslash}p{4cm} X}
        \toprule
        \textbf{Symbol} & \textbf{Meaning} \\
        \midrule
				$\{ \tilde{F}, \id -\tilde{F}\}$ & First step measurement. Implements basis-independent filter. \\
			
				$\{\AliceBobPOVMprimeFilter{b_A,b_B}{\con}, \id- \AliceBobPOVMprimeFilter{b_A,b_B}{\con} \}$.  & Second step measurement. Implements filtering that is basis dependent.  \\
			
				$\{ \AliceBobPOVMsecond{b_A b_B}{=},\AliceBobPOVMsecond{b_A b_B}{\neq} \}$  & Third step measurement corresponding to no-error and error. \\
				
				$\{ \AliceBobPOVMsecondsandwich{b_1 b_2}{\neq} $ ,
				$ \AliceBobPOVMsecondsandwich{b_1 b_2}{=} ,\id - \AliceBobPOVMprimeFilter{b_1}{\con}  \}$ & Combined second and third step measurement, corresponding to no-error, error and discard. \\ 
				
				$\nXperf$ & Number of testing rounds after basis-independent filter only \\
				
				$\nKperf$ & Number of key generation rounds after basis-independent filter only \\
				
				$\nX$ & Actual number of testing rounds \\
				
				$\nK$ & Actual number of key generation rounds \\
				\bottomrule
		   \end{tabularx}
			\caption{Different symbols used in our proof. Note that $b_A,b_B$ refer to basis choice of Alice and Bob. However, $b_1, b_2$ refer to the basis used by \textit{both} Alice and Bob, for the second and third step measurements. Whenever a basis is explicitly written as $X /Z$ (or $b_1,b_2$ ) it represents \textit{both} Alice and Bob's basis choices.
			}\label{table:povms}
		\end{table}

	\subsection{Constructing an equivalent protocol} \label{subsec:equivprot}
		We will now construct the equivalent protocol from \cref{fig:virtualprot}. The construction is similar to the one from \cref{subsec:equivprotperfect}, albeit with some important modifications. 
		\begin{enumerate}
			\item As in \cref{subsec:equivprotperfect}, we observe that the first step measurement is conducted using $\{\tilde{F},\id -\tilde{F}\}$ and is independent of basis. Therefore, we can delay basis choice until after this measurement has been completed, and the $\id -\tilde{F}$ outcomes are discarded. That is the first node of \cref{fig:virtualprot}.
			\item The remaining rounds undergo random basis choice. Basis mismatch rounds are discarded, all $\Xbasis$ basis rounds are used for testing, while $\Zbasis$ basis rounds are used for key generation (with a tiny sample used for estimating $\eZ$). This allows us to obtain the second node of \cref{fig:virtualprot}.  Again, as in \cref{subsec:equivprotperfect}, the estimate we obtain on $\eZ$ does not affect the secrecy claim of the protocol, since $\eZ$ is only used to determine the amount of error-correction to be performed. 
		\end{enumerate}
		Note that unlike \cref{subsec:equivprotperfect}, we have to perform \textit{two} measurements on the testing and key generation rounds after the second node, and these rounds are \textit{not} guaranteed to result in a conclusive outcome. We describe these measurements in detail below.
		\subsubsection{Testing Rounds after basis-independent Filter}
		We will now complete the measurement steps on the test rounds (which take place in the \textbf{Testing* (X $ \rightarrow $ X)} box in \cref{fig:virtualprot}).
		Let us consider the $\Xbasis$ basis rounds used for testing at this stage. Let $\nXperf$ be the number of such rounds. Note that some of these rounds will be discarded during the remainder of the protocol, and therefore we do not know the value of $\nXperf$ in the actual protocol. However, we will see that we do not need to.
		
		These rounds must undergo the second step filtering measurement using $\{\AliceBobPOVMprimeFilter{\Xbasis}{\con}, \id- \AliceBobPOVMprimeFilter{\Xbasis}{\con} \}$, where the rounds which yield the latter outcome are discarded. Now, the remaining rounds are measured using the third step $\{ \AliceBobPOVMsecond{\Xbasis}{=},\AliceBobPOVMsecond{\Xbasis}{\neq} \}$ that determines whether Alice and Bob observe an error or no error. Recall that we use the convention that whenever a basis is explicitly written as $X/Z$, it refers to \textit{both} Alice and Bob measuring in the same basis.

		Combining the second and third measurement step, we see that measuring $\nXperf$ rounds using the above two-step procedure is equivalent to measuring directly using 
		$\Big\{ \AliceBobPOVMsecondsandwich{\Xbasis \Xbasis}{\neq} $ ,
		$ \AliceBobPOVMsecondsandwich{\Xbasis \Xbasis}{=} ,\id - \AliceBobPOVMprimeFilter{\Xbasis}{\con}  \Big\}$ (see \cref{eq:notationGsandwich}), with the outcomes corresponding conclusive and error, conclusive and no-error and inconclusive respectively.  
		We write   $\eXperf$ be the error rate in these rounds, which is the fraction of rounds that resulted in the $\AliceBobPOVMsecondsandwich{\Xbasis \Xbasis}{\neq}$-outcome. The subscript $\Xbasis \Xbasis$ reflects the fact that this is the error rate when the second step and third step measurements are in $\Xbasis$ basis. Note that we do not actually observe this error rate in the protocol. We write $\eX$ as the error rate in these rounds after discarding the $\bot$ outcomes. This is the error rate we actually observe in the protocol.      
        
 	\subsubsection{Key Generation Rounds after basis-independent Filter}
		We will now complete the virtual measurement steps on the key generation rounds, that lead to the phase error rate (which take place in the \textbf{Key* (Z $\rightarrow$ X)} box in \cref{fig:virtualprot}). Let us consider the $\Zbasis$ basis rounds selected for key generation at this stage. Let $\nKperf$ be the number of such rounds. Note that some of these rounds will be discarded during the remainder of the protocol, and therefore we do not actually know the value of $\nKperf$ in the protocol. However, as in the case of $\nXperf$, we do not need to.
		
		These rounds must undergo the second step filtering measurement using $\{\AliceBobPOVMprimeFilter{\Zbasis }{\con}, \id- \AliceBobPOVMprimeFilter{\Zbasis }{\con} \}$, where the rounds which yield the latter outcome are discarded. Now, we wish to obtain the phase error rate when the remaining rounds are measured using the third step $\{ \AliceBobPOVMsecond{\Xbasis}{\neq},\AliceBobPOVMsecond{\Xbasis}{=} \}$ that determines whether Alice and Bob observe an error or no error. 
		
		Again, the above two-step measurement procedure is equivalent to measuring directly using
		$\Big\{ \AliceBobPOVMsecondsandwich{\Zbasis \Xbasis}{\neq},$ 
		$ \AliceBobPOVMsecondsandwich{\Zbasis \Xbasis}{=},$ $\id - \AliceBobPOVMprimeFilter{\Zbasis }{\con}  \Big\}$ (see \cref{eq:notationGsandwich}), with the outcomes corresponding conclusive and error, conclusive and no-error and inconclusive respectively. 
		We let  $\ephperf$ be the error rate in these rounds, which is the fraction of rounds that resulted in the $\ \AliceBobPOVMsecondsandwich{\Zbasis \Xbasis}{\neq}$-outcome. Again, the subscripts denote the fact that this is the error rate when the second step measurement is in the $\Zbasis$ basis and the third step measurement is in the $\Xbasis$ basis. 
		The phase error rate $\eph$ is the error rate in these rounds after discarding the $\bot$ outcomes.
		\begin{remark}
			When basis-efficiency mismatch is present, one must figure out the phase error rate in the key generation rounds, which are filtered using the $\Zbasis$ basis. However the rounds for testing are filtered using the $\Xbasis$ basis. These filtering steps are not identical. Therefore it becomes very difficult to prove rigorous bounds on the phase error rate based on the observed data. The main contributions of this chapter is a rigorous derivation of such bounds, without relying on asymptotic behavior or IID assumptions.
		\end{remark}

		Since the measurements in the key generation rounds leading to $\ephperf$ are not identical to the one in the testing rounds which leads to $\eXperf$, one cannot directly use Serfling (\cref{lemma:sampling}) to relate the two, as we did in \cref{subsec:samplingperfect}.
		Therefore, we introduce another set of virtual measurements (which take place in the  \textbf{Key* (X $\rightarrow$ X)} box in \cref{fig:virtualprot}), corresponding to $\Xbasis$ basis second and third step measurements. Thus we obtain
		another error rate $\ephwierd$. This is the error rate corresponding to the case where these $\nKperf$ rounds are measured using 	$\Big\{ \AliceBobPOVMsecondsandwich{\Xbasis \Xbasis}{\neq},$ 
		$ \AliceBobPOVMsecondsandwich{\Xbasis \Xbasis}{=} , \id- \AliceBobPOVMprimeFilter{\Xbasis}{\con}\Big\}$ (the same measurement that testing rounds are subject to).

	\subsection{Cost of removing the basis-independent loss assumption} \label{subsec:assumptions}
		In removing the basis-independent loss assumption from phase error estimation, we will need to define metrics $\deltaone,\deltatwo$, which will quantify the deviation from ideal behavior. We will now explain how these metrics are defined. 
		
		Consider the POVM elements $\AliceBobPOVMsecondsandwich{\Zbasis \Xbasis}{\neq}$ and $ \AliceBobPOVMsecondsandwich{\Xbasis \Xbasis}{\neq}$ defined via \cref{eq:notationGsandwich}, which combine the second and third step measurements. In \cref{sec:perfectdetectors} they were exactly equal. 	We define $\deltaone$ to quantify the closeness of these POVM elements as
		\begin{equation} \label{eq:deltaonedefone}
			\delta_1 \coloneq 2 \norm{\AliceBobPOVMsecondsandwich{\Zbasis \Xbasis}{\neq} - \AliceBobPOVMsecondsandwich{\Xbasis \Xbasis}{\neq} }_\infty,
		\end{equation}
		and use it in \cref{lemma:pulkitgeneric} (to be discussed later) in our proof. 
		
		Consider the second step measurements, where outcomes corresponding to POVM element $ \id - \AliceBobPOVMprimeFilter{\Zbasis }{\con}$ are discarded. In \cref{sec:perfectdetectors}, there was no need of the second step filtering measurement, which is equivalent to having $\AliceBobPOVMprimeFilter{\Zbasis }{\con} = \id$. We define $\deltatwo$ to quantify the amount of deviation from this case as
		\begin{equation} \label{eq:deltatwodef}
			\deltatwo \coloneq	\norm{\id - \AliceBobPOVMprimeFilter{\Zbasis }{\con} }_\infty .
		\end{equation}
		Thus $\deltatwo$ controls the likelihood of discards in the second step filtering measurements.
		
		Having defined $\deltaone,\deltatwo$ as metrics of the deviation from the basis-independent loss assumption, we now move on to consider the relations between the error rates in the next subsection.

		\subsection{Sampling} \label{subsec:sampling}
		Let us recall the error-rates we have defined so far:
		\begin{enumerate}
			\item $\eX $ is the fraction of the $\nX$ testing rounds that resulted in the $\AliceBobPOVMsecondsandwich{\Xbasis \Xbasis}{\neq}$ outcome. We have access to $\eX$ in the protocol, since it is something we actually observe. 
			\item $\eXperf $ is the fraction of the $\nXperf$ testing rounds (after basis-independent filter only) that result in $\AliceBobPOVMsecondsandwich{\Xbasis \Xbasis}{\neq}$-outcome. $\eX$ is obtained from $\eXperf$ after some rounds are discarded in the second step measurements.
			\item $\ephwierd$ is the fraction of the $\nKperf$ key generation rounds (after basis-independent filter only) that result in $\AliceBobPOVMsecondsandwich{\Xbasis \Xbasis}{\neq}$-outcome. 
			\item $\ephperf$ is the fraction of the $\nKperf$ key generation rounds (after basis-independent filter only) that result in $\AliceBobPOVMsecondsandwich{\Zbasis \Xbasis}{\neq}$-outcome. 
			\item $\eph$ is the fraction of the $\nK$ key generation rounds that result in $\AliceBobPOVMsecondsandwich{\Zbasis \Xbasis}{\neq}$-outcome. This is the quantity we wish to estimate. $\eph$ is obtained from $\ephperf$ after some rounds are discarded in  the second step measurements.
		\end{enumerate}
We wish to prove \cref{eq:req} that relate $\eXrv$ to $\ephrv$. We do this by relating the various error-rates together as  $\eXrv \leftrightarrow \eXperfrv \leftrightarrow \ephwierdrv \leftrightarrow\ephperfrv\leftrightarrow\ephrv$. We will consider the event $\event{\nXperf,\nKperf}$, even though we do not actually observe it in the protocol. In the end, all random variables and events not directly observed in the protocol will disappear from our final expressions.
		
\begin{itemize}
			\item $\eXrv \leftrightarrow \eXperfrv$: Recall from the \textbf{Testing*(X $ \rightarrow$ X)} node in \cref{fig:virtualprot}, that $\eXrv = \num_{\AliceBobPOVMsecondsandwich{\Xbasis \Xbasis}{\neq}} / \nXrv$ and $\eXperfrv = \num_{\AliceBobPOVMsecondsandwich{\Xbasis \Xbasis}{\neq}} / \nXperfrv$. The required relation follows from the fact that we discard rounds to go from $\eXperfrv$ to $\eXrv$, i.e we have $\Pr(\nXrv \leq \nXperfrv)_{ | \event{\nXperf,\nKperf}} = 1$.  Therefore, we obtain 
			\begin{equation} \label{eq:boundone}
				\Pr( \eXperfrv \geq \eXrv )_{ | \event{\nXperf,\nKperf}} = 0.
			\end{equation}
	\item $\eXperfrv \leftrightarrow \ephwierdrv$ : These error rates correspond to measurement outcomes using the \textit{same} POVM, but with random assignment to testing vs key generation. Thus we can apply \cref{lemma:sampling} (Serfling) in  exactly the same manner as in \cref{subsec:samplingperfect}, conditioned on the event $\event{\nXperf,\nKperf}$. In doing so, we obtain
			\begin{equation} \label{eq:samplinguseimperfect}
				\begin{aligned}
					\Pr(\ephwierdrv \geq \eXperfrv + \gamma_\mathrm{serf})_{ | \event{\nXperf,\nKperf}} &\leq  e^{-2\gamma_\mathrm{serf}^2 \fserf(\nXperf,\nKperf)}.
				\end{aligned}
			\end{equation}
			Using the definition from \cref{eq:gammaserfdefined}, we have 
			\begin{equation} \label{eq:gammavalueimperfect}
				\gamma^{\epsATa}_\mathrm{serf}(\nXperf,\nKperf) = \sqrt{\frac{\ln(1/\epsATa^2) }{ 2  \fserf(\nXperf,\nKperf)} } \implies e^{ - \left( \gamma^{\epsATa }_\mathrm{serf}(\nXperf,\nKperf) \right)^2 2 \fserf(\nXperf,\nKperf)} = \epsATa^2.
			\end{equation}
			Therefore, we obtain
			\begin{equation} \label{eq:boundtwo}
				\begin{aligned}
					\Pr(\ephwierdrv \geq \eXperfrv + \gamma^{\epsATa}_\mathrm{serf}   (\nXperf,\nKperf)   )_{ | \event{\nXperf,\nKperf}} &\leq  \epsATa^2
				\end{aligned}
			\end{equation}

            	\item $ \ephwierdrv \leftrightarrow \ephperfrv$: We utilize the definition of $\deltaone$ stated in \cref{subsec:assumptions}. Since the POVM elements  generating $\ephperfrv$ ($\AliceBobPOVMsecondsandwich{\Zbasis \Xbasis}{\neq} $) and $\ephwierdrv$ ($\AliceBobPOVMsecondsandwich{\Xbasis \Xbasis}{\neq} $) are close, we expect the bounds obtained on $\ephperfrv$ and $\ephwierdrv$ to also be close. This is made precise in the following lemma proved in Appendix.~\ref{appsubsec:randomsampling}.

			\begin{restatable}[Similar measurements lead to similar observed frequencies]{lemma}{pulkitlemmageneric}\label{lemma:pulkitgeneric} Let $\rho_{Q^n} \in \dop{=}(Q^{\otimes n})$ be an arbitrary state. Let $\{P,\id - P\}$ and $\{P^\prime,\id-P^\prime\}$ be two sets of POVM elements, such that $\norm{P^\prime-P}_\infty \leq \delta$. 
				Then,
				\begin{equation} \label{eq:pulkitequationgeneric}
					\Pr(\frac{\num_{P^\prime}}{n} \geq e +  2 \delta+c) \leq \Pr(\frac{\num_P}{n} \geq e) + \binfunction{n}{2\delta}{c},
				\end{equation}
				for $e \in [0,1]$, where $\num_P$ is the number of $P$-outcomes when each subsystem of $\rho_{Q^n}$ is measured using POVM $\{P,\id - P\}$, and
				\begin{equation}
					\binfunction{n}{\delta}{c} \coloneq  \sum_{i = n (\delta+c)}^{n} {n \choose i} \delta^ i (1-\delta)^{n-i}. 
				\end{equation}
			\end{restatable}

			Thus, using  \cref{lemma:pulkitgeneric} and $\deltaone$ defined in \cref{eq:deltaonedefone}, we obtain
			\begin{equation} 
				\begin{aligned}
					\Pr(\ephperfrv \geq e + \deltaone + \cone)_{ | \event{\nXperf,\nKperf} } &\leq \Pr(\ephwierdrv \geq  e)_{| \event{\nXperf,\nKperf}}  + \binfunction{\nKperf}{ \deltaone}{ \cone}.
				\end{aligned}
			\end{equation}
			
			We would like $\binfunction{\nKperf}{ \deltaone}{ \cone}$ to be equal to a constant $\epsATb^2$ on the right hand side of the above expression. To do so, we note that $\binfunction{\nKperf}{ \deltaone}{ \cone}$ is a monotonic (and therefore invertible) function of $\cone$. Thus, we can choose  $c_1$ to be a function $	 \gamma^{\epsATb}_\mathrm{bin}(\nKperf,\deltaone)$ such that 
			\begin{equation} \label{eq:gammabindefined}
				\binfunction{\nKperf}{ \deltaone}{ \gamma^{\epsATb}_\mathrm{bin}(\nKperf,\deltaone})= \epsATb^2.
			\end{equation} 
			Using this as a definition $\gamma^{\epsATb}_\mathrm{bin}(\nKperf,\deltaone)$, we obtain
			\begin{equation}  \label{eq:boundthree}
				\Pr(\ephperfrv \geq e + \deltaone + \gamma^{\epsATb}_{\text{bin}}(\nKperf,\deltaone) )_{ | \event{\nXperf,\nKperf}}\leq \Pr(\ephwierdrv \geq e)_{| \event{\nXperf,\nKperf}}  +\epsATb^2.
			\end{equation}
			Note that $\gamma_{\text{bin}}$ can be easily computed numerically by relating $\binfunction{n}{\delta}{c}$ (and its inverse) to the cumulative binomial distribution and using root finding algorithms. 
	\item   $\ephperfrv \leftrightarrow \ephrv$:  We will use the fact that the filtering measurements result in a very small number of discards.
			
			First, note that $\ephperfrv =  \bm{N}_{\AliceBobPOVMsecondsandwich{\Zbasis \Xbasis}{\neq}} / \nKperfrv$, and $\ephrv = \bm{N}_{\AliceBobPOVMsecondsandwich{\Zbasis \Xbasis}{\neq}} / \nKrv$. Thus, we have $\ephperfrv	/ \ephrv = \nKrv /\nKperfrv$.

			Recall that $\nKrv$ is obtained by discarding rounds from $\nKperfrv$ based on $\{ \AliceBobPOVMprimeFilter{\Zbasis }{\con}, \id - \AliceBobPOVMprimeFilter{\Zbasis }{\con}\}$ measurements. We will essentially show that very few rounds are discarded in this step, using \cref{eq:deltatwodef}. To do so, we prove the following Lemma in Appendix.~\ref{appsubsec:randomsampling}.
			
			\begin{restatable}[Small POVM measurement]{lemma}{smallPOVM}
				\label{lemma:smallPOVM}
				Let $\rho_{Q^n} \in \dop{=}(Q^{\otimes n})$ be an arbitrary state. Let $\{P,I - P\}$ be a POVM such that $\norm{P}_\infty \leq \delta$. Then
				\begin{equation}
					\Pr(\frac{\num_P}{n} \geq \delta+c) \le \binfunction{n}{\delta}{c} \coloneq  \sum_{i = n (\delta+c)}^{n} {n \choose i} \delta^ i (1-\delta)^{n-i} ,
				\end{equation}
				where $\num_P$ is the number of $P$-outcomes when each subsystem of $\rho_{Q^n}$ is measured using POVM $\{P,\id - P\}$.
			\end{restatable} 
			Then, using \cref{lemma:smallPOVM} with $P=\id - \AliceBobPOVMprimeFilter{\Zbasis }{\con}$ and $\deltatwo$ defined in \cref{eq:deltatwodef}, we obtain
			\begin{equation} 
				\begin{aligned}
					\Pr( \ephperfrv \leq \ephrv (1-\deltatwo-\ctwo))_{\event{\nXperf,\nKperf}} &=	\Pr(\frac{\nKperf - \nKrv}{\nKperf} \geq \deltatwo+\ctwo)_{ | \event{\nXperf,\nKperf}} \\
					&=	\Pr(\frac{\num_{\id - \AliceBobPOVMprimeFilter{\Zbasis }{\con} }}{\nKperf} \geq \deltatwo+\ctwo)_{ | \event{\nXperf,\nKperf}} \\
					&\leq \binfunction{\nKperf}{\deltatwo}{\ctwo}. \\
				\end{aligned}
			\end{equation}

			Again, we would like $\binfunction{\nKperf}{\deltatwo}{\ctwo}$ to be a constant value $\epsATc^2$. Thus, we replace $c_2$ with $\gamma^{\epsATc}_{\text{bin}}(\nKperf,\deltatwo)$
			and obtain
			
			\begin{equation} \label{eq:boundfour} 
				\begin{aligned}
					\Pr( \ephperfrv \leq \ephrv (1-\deltatwo- \gamma^{\epsATc}_{\text{bin}}(\nKperf,\deltatwo) ))_{ | \event{\nKperf,\nKperf}} &\leq \epsATc^2
				\end{aligned}
			\end{equation}

\end{itemize}
	Thus we have relationships \cref{eq:boundone,eq:boundtwo,eq:boundthree,eq:boundfour} between all the error rates, whose complements hold with high probability. These can all be combined using straightforward but cumbersome algebra (see \cref{appendix:combining}), to obtain
		\begin{equation} \label{eq:finalbound}
			\Pr (   \ephrv \geq \frac{\eXrv+  \gamma^{\epsATa}_\mathrm{serf}   (\nXperf,\nKperf) + \deltaone + \gamma^{\epsATb}_{\text{bin}}(\nKperf,\deltaone)}{ (1-\deltatwo- \gamma^{\epsATc}_{\text{bin}}(\nKperf,\deltatwo) )}    )_{ | \event{\nXperf,\nKperf}} \leq \epsATb^2 + \epsATa^2 + \epsATc^2.
		\end{equation}
		Using the above expression requires us to know the values of $\nKperf$ and $\nXperf$ which we do  not. This problem is easily resolved by noting all the $\gamma$s are decreasing functions of $\nKperf$ and $\nXperf$, and that $\nKperf (\nXperf)$ cannot be smaller than $\nK (\nX)$ (since we discard rounds to from the former to the latter) . Thus, we can replace $\nKperf$ with $\nKrv$ and $\nXperf$ with $\nXrv$ and obtain
		
		\begin{equation} \label{eq:finalboundusable}
			\Pr (   \ephrv \geq \frac{\eXrv+  \gamma^{\epsATa}_\mathrm{serf}   (\nXrv,\nKrv) + \deltaone + \gamma^{\epsATb}_{\text{bin}}(\nKrv,\deltaone)}{ (1-\deltatwo- \gamma^{\epsATc}_{\text{bin}}(\nKrv,\deltatwo) )}    )_{ | \event{\nXperf,\nKperf}} \leq \epsATb^2 + \epsATa^2 + \epsATc^2
		\end{equation}
		We set $\epsATa^2+\epsATb^2+\epsATc^2 = \epsAT^2$, and obtain the choice of $\Boundbasicdelta$:
		\begin{equation} \label{eq:estimateimperfect}
			\Boundbasicdelta(\eX,\nX,\nK) \coloneq  \frac{\eX+  \gamma^{\epsATa}_\mathrm{serf}   (\nX,\nK) + \deltaone + \gamma^{\epsATb}_{\text{bin}}(\nK,\deltaone)}{ (1-\deltatwo- \gamma^{\epsATc}_{\text{bin}}(\nK,\deltatwo) )},
		\end{equation}
		where functions $\gamma_\mathrm{bin},\gamma_\mathrm{serf}$ are defined in \cref{eq:gammabindefined} and \cref{eq:gammaserfdefined} respectively.
		Since \cref{eq:finalboundusable} is valid for all events $\event{\nXperf,\nKperf}$, the above choice satisfies \cref{eq:req} via
		\begin{equation} 
			\begin{aligned}
				\Pr(\ephrv \geq 	\Boundbasicdelta(\eXrv,\nXrv,\nKrv)  ) &\leq \sum_{\nXperf,\nKperf} \Pr(\event{\nXperf,\nKperf}) 	\Pr(\ephrv \geq 	\Boundbasicdelta(\eXrv,\nXrv,\nKrv)  )_{ | \event{\nXperf,\nKperf}} \\
				&\leq  \sum_{\nXperf,\nKperf} \Pr(\event{\nXperf,\nKperf})  \epsAT^2  = \epsAT^2.
			\end{aligned}
		\end{equation}

	\begin{remark} Let us investigate the behavior of \cref{eq:estimateimperfect} in the limit  $\deltaone,\deltatwo \rightarrow 0$.
			Recall that $\gamma_\mathrm{bin}^{\epsAT} (n,\delta)$ was defined as the value of $c$ such that $\binfunction{n}{\delta}{c} = \sum_{i=n(\delta+c)}^{n} {n \choose k} \delta^i (1-\delta)^{n-i} \leq \epsAT^2$.
			However, notice that $\delta \rightarrow  0 \implies \binfunction{n}{\delta}{c} \rightarrow 0$ for any value of $c$. Therefore, $\delta \rightarrow 0 \implies \gamma_\mathrm{bin}(n,\delta) \rightarrow 0$. Setting these limits in \cref{eq:estimateimperfect}, we recover the result \cref{eq:estimateperfect} for the case where the basis-independent loss assumption is satisfied.
		\end{remark}

		Thus we now have a phase error estimation bound that is valid even in the presence of basis-efficiency mismatch. (A self-contained statement describing this as a sampling result can be found in \cite[Theorem 2]{tupkary_phase_2024}) 
		
\section{Application to decoy-state BB84} \label{sec:decoy}
	So far in this chapter, we have focused our attention on the BB84 protocol implemented using perfect single-photon sources for pedagogical reasons. In this section, we will extend our techniques and obtain a variable-length security proof for decoy-state BB84 \cite{Hwang_qkdiwthhighloss_2003,Lo_decoystate_2005,Ma_practicaldecoy_2005,hayashi_security_2014,wang_beating_2005} with imperfect detectors. We base our security proof approach on that of Lim et al \cite{lim_concise_2014}, while fixing some technical errors in that work (see Ref.~\cite[Section 5]{tupkary_phase_2024}).

    		\subsection{Protocol specification} \label{subsec:decoyprotocol}
		The decoy-state BB84 protocol modifies the following steps of the protocol described in \nameref{prot:qkdprotocol}.
		\begin{enumerate}
			\item \textbf{State Preparation:} Alice decides to send states in the $\Zbasis(\Xbasis)$ basis with probability $\pzA$ ($\pxA$). She additionally chooses a signal intensity $\mu_k \in \{\mu_1,\mu_2,\mu_{3}\}$ with some predetermined probability $p_{\mu_k}$ \footnote{This probability can depend on the basis used without affecting the results of this work. To incorporate this, one simply has to track the correct probability distribution through all the calculations.}. She prepares a phase-randomized weak laser pulse based on the chosen values, and sends the state to Bob.  We assume $\mu_1 > \mu_2 + \mu_3$ and $\mu_2 > \mu_3 \geq 0$. This requirement on the intensity values, as well as the total number of intensities, is not fundamental. It is used in deriving the analytical bounds in the decoy-state analysis. Note the all decoy intensities are used in both bases. 
			\item \textbf{Measurement:}  Bob chooses  the basis $\Zbasis$($\Xbasis$) with probability  $\pzB$($\pxB$) and measures the incoming state. This step of the protocol is identical to that from \cref{sec:protocoleurqubitbb84}.
			\item \textbf{Classical Announcements and Sifting:}  For all rounds, Alice and Bob announce the bases they used. Furthermore, Bob announces whether he got a conclusive outcome ($\{ \BobPOVM{b}{0},\BobPOVM{b}{1} \}$), or an inconclusive outcome ($\{  \BobPOVM{b}{\bot} \}$). A round is said to be ``conclusive'' if Alice and Bob used the same basis, and Bob obtained a conclusive outcome. As before $\Xbasis$ basis rounds are used for testing, and $\Zbasis$ basis rounds are used for key generation (with a tiny fraction used for obtaining $\eZ$). 
			
			On all $\Xbasis$ basis rounds, Alice and Bob announce their measurement outcomes and intensity choices. We let $\nXmu{k}$ be the number of $\Xbasis$ basis conclusive rounds where Alice chose intensity $\mu_k$, and let $\eXmu{k}$ be the observed error rate in these rounds. For brevity, we use the notation $\nXmu{\allk} = ( \nXmu{1} \dots \nXmu{3})$ to denote observations from all intensities. (We use similar notation for $\eXmu{\allk}$, $\nKmu{\allk}$ etc).
			
			On all $\Zbasis$ basis, Alice announces her intensity choices, and these rounds
			are used for key generation. We let $\nK$ rounds be the total number of $\Zbasis$ basis conclusive rounds used for key generation. 
			
			All announcements are stored in the register $\CP_1^n$. We use $\event{\decoyparams}$ to denote the event that $\decoyparams$ values are observed in the protocol.
			
		\end{enumerate}

The remaining steps of the protocol are the same as in \nameref{prot:qkdprotocol}. In particular, based on the observations $\decoyparams$, Alice and Bob implement one-way error-correction using  $\leak(\decoyparams)$ bits of communication, followed by error-verification, and privacy amplification to produce a key of $\lkey(\decoyparams)$ bits. Additionally note that our protocol generates key from \textit{all} intensities, instead of having a single ``signal'' intensity for key generation.

		\subsection{Required and actual phase error estimation bound} \label{subsec:decoyphase}
		In order to prove security for our decoy-state QKD protocol, we will need to bound two quantities. First, we must obtain a lower bound on the number of single-photon events that lead to key generation $\nKphrv{1}$ (since as we argued in \cref{subsec:motivationfordecoy} that multi-photon rounds leak all info to Eve). Second, we must obtain an upper bound on the phase error rate within these single-photon key generation rounds, given by $\ephphrv{1}$. This can be represented mathematically as
		\begin{equation} \label{eq:decoyreq}
			\Pr(\ephphrv{1} \geq \Bound_e(\eXmurv{\allk}, \nXmurv{\allk}, \nKmurv{\allk} ) \quad \lor \quad  \nKphrv{1} \leq \Bound_1(\nKmurv{\allk} )  ) \leq \epsAT^2,
		\end{equation}
		where	$\lor$ denotes the logical OR operator, and $\Bound_e,\Bound_1$ are functions that provide these bounds as a function of the observed values.

	This statement will be used in the proof of \cref{thm:decoyvariablesecurity} to prove the variable-length security of our protocol. We will derive the required bounds ($\Bound_e,\Bound_1$) in \cref{eq:decoyreq} in two steps. First we will use decoy analysis to convert from observations corresponding to different intensities (which we have access to) to those corresponding to different photon numbers (which we do not have access to).   We will be concerned with three outcomes $ \{\Xbasis_{\neq},\Xbasis,\Kbasis\}$, corresponding to $\Xbasis$ basis conclusive error outcome, $\Xbasis$ basis conclusive outcome, and $\Zbasis$ basis conclusive outcome used for key generation respectively.
		Thus, at the end of the first step we will obtain
		
		\begin{equation} \label{eq:decoyreqstepone}
			\begin{aligned}
				\Pr \Big(  & \eXphrv{1} \geq \frac{	\Bounddecoymax{1} (\nXneqmurv{\allk})  }{  	\Bounddecoymin{1} (\nXmurv{\allk})}
				\; \lor \;   \nXphrv{1} \leq    \Bounddecoymin{1}( \nXmurv{\allk})   \; \lor \; 
				\nKphrv{1} \leq   \Bounddecoymin{1}(\nKmurv{\allk}  ) \Big) \leq  9 \epsdecoy^2
			\end{aligned}
		\end{equation}
		where $\Bounddecoymin{m}$ and $\Bounddecoymax{m}$ are functions that compute bounds on the $m$-photon components of the input statistics. Note that we use $ \nXneqmurv{\allk} =  (\nXmurv{1} \times \eXmurv{1}, \dots, \nXmurv{\nint}  \times \eXmurv{\nint})$ to denote the number of rounds resulting both Alice and Bob using the $\Xbasis$ basis and obtaining an error,  for each intensity (and we will assume implicit conversion between these two notations). The $9$ on the RHS comes from the fact that we implement decoy analysis on $3$ different events and we have $3$ intensities. We will prove \cref{eq:decoyreqstepone} in Appendix.~\cref{appendixdecoysection}.

\begin{remark}
			Note that the only parameters actually observed in the protocol are given by $\decoyparams$. Variables like $\eXphrv{1}$ are not actually directly observed, but instead are derived from observations.
		\end{remark}
		
		In this second step, we will use $\eXphrv{1},\nXphrv{1},\nKphrv{1}$ to bound the single photon phase error rate $\ephphrv{1}$. Notice this is exactly what we showed \cref{sec:perfectdetectors,sec:imperfectdetectors}.  In particular, with $	\Boundbasicdelta$ directly obtained from \cref{eq:estimateimperfect}, we have
		\begin{equation} \label{eq:decoyreqsteptwo}
			\Pr(\ephphrv{1} \geq \Boundbasicdelta(\eXphrv{1},\nXphrv{1},\nKphrv{1}) ) \leq  \epsATsingle^2.
		\end{equation}
		where $\epsATsingle^2$ denotes the failure probability of the ``single-photon" part of our estimation.

		However, note that we do not directly observe $\eXphrv{1},\nXphrv{1},\nKphrv{1}$ in the decoy-state protocol (unlike \cref{sec:imperfectdetectors}). Thus we would like to replace these values with the bounds computed from our decoy analysis (\cref{eq:decoyreqstepone}). This is straightforward to do, since $\Boundbasicdelta$ is an increasing function of $\eXphrv{1}$, and decreasing function of $\nXphrv{1},\nKphrv{1}$.  This can be done formally by a straightforward application of the union bound for probabilities ($\Pr(\Omega_1 \lor \Omega_2) \leq \Pr(\Omega_1)+\Pr(\Omega_2)$) applied to \cref{eq:decoyreqstepone,eq:decoyreqsteptwo}. Doing so allows us to conclude that the probability of \textit{any} of the bounds in \cref{eq:decoyreqstepone,eq:decoyreqsteptwo} failing is smaller than $9 \epsdecoy^2 + \epsATsingle^2$.  Then we use the fact that if \textit{none} of the bounds inside the probabilities in \cref{eq:decoyreqstepone,eq:decoyreqsteptwo} fail, then this implies that the bounds inside the probability in \cref{eq:decoyreqcombined} below must hold. Formally, we obtain
		\begin{equation} \label{eq:decoyreqcombined}
			\begin{aligned}
				\Pr \Bigg(  & \ephphrv{1} \geq  \Boundbasicdelta \Bigg( \frac{	\Bounddecoymax{1} (\nXneqmurv{\allk})  }{  	\Bounddecoymin{1} (\nXmurv{\allk})}   ,  \Bounddecoymin{1}( \nXmurv{\allk})   ,  \Bounddecoymin{1}(\nKmurv{\allk})     \Bigg)   \quad \lor \quad \\
				& \nKphrv{1} \leq  \Bounddecoymin{1}(\nKmurv{\allk}  )\Bigg)  \leq 9\epsdecoy^2 +\epsATsingle^2 \eqcolon \epsAT^2 
			\end{aligned}
		\end{equation}
		which is the required statement.  Thus, it is now enough to prove \cref{eq:decoyreqstepone} in order to prove \cref{eq:decoyreqcombined} (equivalently \cref{eq:decoyreq}), for which we turn to decoy analysis. The decoy analysis is a standard tool, and is explained in \cref{appendixdecoysection} in \cref{Appendix:EUR}.

\subsection{Variable-length security statement for decoy-state}		 \label{subsec:decoyvariable}
		
		Having proved \cref{eq:decoyreq}, we now have the following theorem regarding variable-length security of the decoy-state BB84 protocol, which we can prove using \cref{theorem:eurvariablegenericresult}.
		
	\begin{restatable}[Variable-length security of decoy-state BB84 \cite{tupkary_phase_2024} ]{theorem}{variablelengthproofdecoy}\label{thm:decoyvariablesecurity}
			Suppose \cref{eq:decoyreq} is satisfied and let $\leak(\decoyparams)$ be a function that determines the number of bits used for error-correction in the QKD protocol. Define 	\begin{equation} \label{eq:lvaluedecoy}
				\begin{aligned}
					\lkey(\decoyparams) &\coloneq  \max\Bigg(0, \bigg\lfloor \Bound_1 \left(\nKmu{\allk} \right) \left(1- h \left( \Bound_e \left(\eXmu{\allk},\nXmu{\allk}, \nKmu{\allk}   \right) \right) \right)  \\
					&- \leak(\decoyparams)					- 2\log(1/2\epsPA) - \EVcost  \bigg \rfloor \Bigg)  
				\end{aligned}
			\end{equation}
			where $h(x)$ is the binary entropy function for $x\leq 1/2$, and $h(x) =  1$ otherwise.
			Then the variable-length decoy-state QKD protocol that produces a key of length $\lkey(\decoyparams)$ using $\leak(\decoyparams)$ bits for error-correction, 	
			upon the event $\event{\decoyparams} \wedge \OmegaEV$ is $(2 \epsAT+\epsPA+\epsEV)$-secure.
		\end{restatable}

\begin{proof}
The proof follows from a straightforward application of \cref{theorem:eurvariablegenericresult}. We identity $\cP_1^n$ with $\decoyparams$ and $\nKph{1}$ with $\cnotobs$, and consider events $\Omega(\decoyparams,\nKph{1})$, and we define 
\begin{equation}
\begin{aligned}
    \serfbound(\cP^1, \cnotobs) &= \Pr(\ephphrv{1} \geq \Bound_e(\cobs ) \quad \lor \quad  \nKphrv{1} \leq \Bound_1(\cobs )  )_{\Omega(\cobs,\nKph{1})}, \\
    \EURset(\cP_1^n)& = \{ m \in \mathbb{N} | m >\Bound_1(\nKmurv{\allk} ) \}.
\end{aligned}
\end{equation}
With this identification, the requirement from \cref{eq:genericvariablecondition2} can be shown to be satisfied. The required bound on the min entropy from \cref{eq:genericvariablecondition1} also follows from the use of the EUR relation and suitable chain rules to isolate the single-photon contribution. For more details, we refer the reader to Ref. \cite[Proof of Theorem 3]{tupkary_phase_2024}.
\end{proof}

\section{Results} \label{sec:eurresults}
	We will now apply our results to a decoy-state BB84 protocol with realistic detectors. To do so, we start by outlining a recipe for using this work to compute key rates in \cref{subsec:recipe}. We will then specify the canonical model for our detectors with efficiency mismatch in \cref{subsec:detectors}. This model is the same as that specified in \cref{sec:statesandmeasurements}, but is specified in more detail. We will apply the recipe to our model in \cref{subsec:deltabounds}. Finally, we will plot the key rate we obtain in \cref{subsec:eurplots}.

		\subsection{Recipe for computing key rates in the presence of basis-efficiency mismatch} \label{subsec:recipe} 
		In this subsection, we provide straightforward instructions for using the results of this chapter to compute key rates for decoy-state BB84 in the presence of basis-efficiency mismatch. We will start by explaining the computation of (upper bounds on) $\deltaone, \deltatwo$ for a given model of the measurement POVMs in the protocol. To do so, one has to break up the measurement process implemented by Alice and Bob into multiple steps via multiple uses of \cref{lemma:twostep}. This is done as follows:

		\begin{enumerate}
			\item Start with POVM $\{ \AliceBobPOVM{b_A,b_B}{\neq}, \AliceBobPOVM{b_A,b_B}{=}, \AliceBobPOVM{b_A,b_B}{\bot} \}$  which describe Alice and Bob measuring in the $(b_A,b_B)$ basis, and obtaining a conclusive error, a conclusive no-error, and an inconclusive outcome respectively. (In this work, we apply this recipe on the POVMs defined in \cref{eq:alicebobpovmmodel}.)
			
			\item Pick a $\tilde{F} \geq \AliceBobPOVM{b_A,b_B}{\neq}+\AliceBobPOVM{b_A,b_B}{=}$ for all $(b_A,b_B)$. Consider the four-outcome POVM \\
			$\{ \id - \tilde{F} , \tilde{F} - \AliceBobPOVM{b_A,b_B}{=} -\AliceBobPOVM{b_A,b_B}{\neq}, \AliceBobPOVM{b_A,b_B}{=},\AliceBobPOVM{b_A,b_B}{\neq} \}$. Group the last three outcomes together, and use \cref{lemma:twostep} to divide this measurement into two steps. In the first step, $\{ \tilde{F}, \id - \tilde{F}  \}$ is measured and latter outcomes discarded. The remaining rounds are measured using $\{ \AliceBobPOVMFilter{b_A,b_B}{\bot}, \AliceBobPOVMFilter{b_A,b_B}{=}, \AliceBobPOVMFilter{b_A,b_B}{\neq}\}$ where 
			\begin{equation} \label{eq:filteredFilters}
				\begin{aligned}
					\AliceBobPOVMFilter{b_A,b_B}{\bot}&=  \sqrt{\tilde{F}}^+ (\tilde{F} -  \AliceBobPOVM{b_A,b_B}{\neq} -  \AliceBobPOVM{b_A,b_B}{\neq})   \sqrt{\tilde{F}}^+  + \id - \proj_{\tilde{F}} \\
					\AliceBobPOVMFilter{b_A,b_B}{\neq} &= \sqrt{\tilde{F}}^+  \AliceBobPOVM{b_A,b_B}{\neq} \sqrt{\tilde{F}}^+ \\
					\AliceBobPOVMFilter{b_A,b_B}{=} &= \sqrt{\tilde{F}}^+  \AliceBobPOVM{b_A,b_B}{=}  \sqrt{\tilde{F}}^+
				\end{aligned}
			\end{equation}
			where $\proj_{\tilde{F}}$ denotes the projector onto the support of $\tilde{F}$.
			
			\item Consider the new POVM $\{ \AliceBobPOVMFilter{b_A,b_B}{\bot}, \AliceBobPOVMFilter{b_A,b_B}{=}, \AliceBobPOVMFilter{b_A,b_B}{\neq}\}$ .  Using \cref{lemma:twostep} again, divide this POVM measurement into two steps. The first step is implemented using $\{	\AliceBobPOVMprimeFilter{b_A,b_B}{\con}, 	\AliceBobPOVMprimeFilter{b_A,b_B}{\bot}\}$ and decides whether the outcome is conclusive or inconclusive. The conclusive outcomes are further measured using  $\{	\AliceBobPOVMsecond{b}{\neq} ,	\AliceBobPOVMsecond{b}{=} \}$. These POVM elements are given by
			\begin{equation} \label{eq:recipeone}
				\begin{aligned}
					\AliceBobPOVMprimeFilter{b_A,b_B}{\con} &= \AliceBobPOVMFilter{b_A,b_B}{\neq} + \AliceBobPOVMFilter{b_A,b_B}{=}  \\
					\AliceBobPOVMprimeFilter{b_A,b_B}{\bot} &= \AliceBobPOVMFilter{b_A,b_B}{\bot}  \\
					\AliceBobPOVMsecond{b_A,b_B}{\neq} &= \sqrt{\AliceBobPOVMprimeFilter{b_A,b_B}{\con}}^+ \AliceBobPOVMFilter{b_A,b_B}{\neq}  \sqrt{\AliceBobPOVMprimeFilter{b_A,b_B}{\con}}^+ \\
					\AliceBobPOVMsecond{b_A,b_B}{=} &= \sqrt{\AliceBobPOVMprimeFilter{b_A,b_B}{\con}}^+ \AliceBobPOVMFilter{b_A,b_B}{=}  \sqrt{\AliceBobPOVMprimeFilter{b_A,b_B}{\con}}^+ + \id - \proj_{\AliceBobPOVMprimeFilter{b_A,b_B}{\con}} \\
                    & = \id - 	\AliceBobPOVMsecond{b_A,b_B}{\neq}
				\end{aligned}
			\end{equation}
			where $\proj_{\AliceBobPOVMprimeFilter{b_A,b_B}{\con}} $ is the projector onto the support of $\AliceBobPOVMprimeFilter{b_A,b_B}{\con}$. This projector plays a trivial rule in the measurement itself, and is only included to ensure that we obtain a valid POVM.
			
			\item Compute
			\begin{equation} \label{eq:deltaCostsRecipe}
				\begin{aligned}
					\deltaone =& 2 \norm{\sqrt{\AliceBobPOVMprimeFilter{\Zbasis }{\con}}\AliceBobPOVMsecond{\Xbasis}{\neq} \sqrt{\AliceBobPOVMprimeFilter{\Zbasis }{\con}} -  \sqrt{\AliceBobPOVMprimeFilter{\Xbasis}{\con}}\AliceBobPOVMsecond{\Xbasis}{\neq} \sqrt{\AliceBobPOVMprimeFilter{\Xbasis}{\con}} }_\infty \\
					\deltatwo =&	\norm{\id - \AliceBobPOVMprimeFilter{\Zbasis }{\con} }_\infty 
				\end{aligned}
			\end{equation}
			where we recall that whenever the basis is explicitly written as $\Xbasis/\Zbasis$, it represents both Alice and Bob’s basis choices.
			\item For the analysis of practical scenarios, where the detector parameters $\eta_{b_i}, d_{b_i}$ are not known exactly but are instead known to be in some range, one must also additionally maximize \cref{eq:deltaCostsRecipe} over all possible choices of parameters $\eta_{b_i}, d_{b_i}$. 
		\end{enumerate}

		 Once $\deltaone,\deltatwo$ are computed via the procedure above, we can compute key rates as follows. The key rate expression for the decoy-state BB84 protocol is given by \cref{eq:lvaluedecoy}. To use this expression,  refer to  \cref{eq:decoyreqcombined,eq:decoyreq} (which are notationally equivalent). The bounds for the decoy analysis in \cref{eq:decoyreqcombined} are in turn found in \cref{eq:decoyboundone,eq:decoyboundtwo,eq:decoyboundthree}, whereas the bound for the phase error estimation is found in \cref{eq:estimateimperfect}.  For the BB84 protocol where Alice sends single photons, the key rate is given by \cref{eq:lvalueEUR,eq:estimateimperfect}.

\subsection{Detector Model} \label{subsec:detectors}

In this section, we specify the canonical model of Bob's detectors (for active BB84) we use in this work. Let $\eta_{b_i}, d_{b_i}$ denote the efficiency and dark count rate of Bob's POVM corresponding to basis $b$, and bit $i$. We first define Bob's double click POVM for basis $b\in\{\Zbasis, \Xbasis\}$ to be $\BobPOVM{b}{\dc} = \sum_{N_0, N_1 =0}^\infty (1-(1-d_{b_0})(1-\eta_{b_0})^{N_0}) (1-(1-d_{b_1})(1-\eta_{b_1})^{N_1}) \ketbra{N_0,N_1}_b$ ,	where $\ketbra{N_0, N_1}_b$ is the state with $N_0$ photons in the mode 1, and $N_1$ photons in mode 2, where the modes are defined with respect to basis $b$. For example, for polarization-encoded BB84, $\ketbra{2,1}_{\Zbasis }$ would signify the state with $2$ horizontally-polarised photons and $1$ vertically polarised photon. Recall that double clicks are mapped to single clicks randomly in our protocol. 
		Thus, we can write Bob's POVM elements as
		\begin{equation} \label{eq:bobpovmmodel}
			\begin{aligned}
				\BobPOVM{b}{\bot} &= \sum_{N_0, N_1=0}^\infty (1-d_{b_0})(1-d_{b_1})(1-\eta_{b_0})^{N_0} (1-\eta_{b_1})^{N_1} \ketbra{N_0,N_1}_b\\
				\BobPOVM{b}{0} &= (1-d_{b_1})\sum_{N_0, N_1 =0}^\infty (1-(1-d_{b_0})(1-\eta_{b_0})^{N_0}) (1-\eta_{b_1})^{N_1} \ketbra{N_0,N_1}_b + \frac{1}{2}\BobPOVM{b}{\dc}\\
				\BobPOVM{b}{1} &= (1-d_{b_0})\sum_{N_0, N_1=0}^\infty (1-\eta_{b_0})^{N_0} (1-(1-d_{b_1})(1-\eta_{b_1})^{N_1}) \ketbra{N_0,N_1}_b+\frac{1}{2}\BobPOVM{b}{\dc}.
			\end{aligned}
		\end{equation}
		Decoy methods allow us to restrict out attention to rounds where Alice sent single photons. Thus her Hilbert space is qubit while Bob holds two optical modes. The joint Alice-Bob POVM elements for the basis $b$ can be constructed via \cref{eq:alicePOVMs,eq:alicebobpovms,eq:bobpovmmodel} and are given by
		\begin{equation} \label{eq:alicebobpovmmodel}
			\begin{aligned}
				\AliceBobPOVM{b,b}{\bot} =& \id_A\otimes \sum_{N_0, N_1=0}^\infty (1-d_{b_0})(1-d_{b_1})(1-\eta_{b_0})^{N_0} (1-\eta_{b_1})^{N_1} \ketbra{N_0,N_1}_b\\
				\AliceBobPOVM{b,b}{\neq} =& \ketbra{0}_b \otimes (1-d_{b_0})\sum_{N_0, N_1=0}^\infty (1-\eta_{b_0})^{N_0} (1-(1-d_{b_1})(1-\eta_{b_1})^{N_1}) \ketbra{N_0,N_1}_b\\
				&+ \ketbra{1}_b \otimes (1-d_{b_1})\sum_{N_0, N_1 =0}^\infty (1-(1-d_{b_0})(1-\eta_{b_0})^{N_0}) (1-\eta_{b_1})^{N_1} \ketbra{N_0,N_1}_b\\
				&+ \id_A \otimes \frac{1}{2}\BobPOVM{b}{\dc}\\
				\AliceBobPOVM{b,b}{=} =& \ketbra{0}_b \otimes (1-d_{b_1})\sum_{N_0, N_1 =0}^\infty (1-(1-d_{b_0})(1-\eta_{b_0})^{N_0}) (1-\eta_{b_1})^{N_1} \ketbra{N_0,N_1}_b\\
				&+ \ketbra{1}_b \otimes (1-d_{b_0})\sum_{N_0, N_1=0}^\infty (1-\eta_{b_0})^{N_0} (1-(1-d_{b_1})(1-\eta_{b_1})^{N_1}) \ketbra{N_0,N_1}_b\\
				&+ \id_A \otimes \frac{1}{2}\BobPOVM{b}{\dc},
			\end{aligned}
		\end{equation}
	where $\ketbra{0}_b$ on Alice's system is the $\ket{0}$ state encoded in basis $b$. Note that this is different from the vacuum state $\ketbra{0,0}_b$ on Bob's system, the state with 0 photons in all modes.
		
		In any practical protocol, the detection efficiencies $\eta_{b_i}$ and dark count rates $d_{b_i}$ cannot be characterized exactly. Therefore, instead of assuming exact knowledge of these parameters, we assume that they are characterized upto some tolerances $\etachar, \dcchar$ given by
		\begin{equation}\label{eq:etadcmodel}
			\begin{aligned}
				\eta_{b_i} &\in [\etadet(1-\etachar ), \etadet(1+\etachar )], \\
				d_{b_i} &\in [\dcprob(1-\dcchar ), \dcprob(1+\dcchar )].
			\end{aligned}
		\end{equation}

	\subsection{Computing bounds on \texorpdfstring{$\deltaone,\deltatwo$}{delta1, delta2}.}
    \label{subsec:deltabounds}
        The upper bounds on $\deltaone,\deltatwo$ are straightforward cumborsome algebra, and we refer the reader to \cite[Appendix G]{tupkary_phase_2024}. The main idea is to utilize the block-diagonal structure in photon number, and compute the bounds separately for each block.
			
		\subsubsection{Active BB84 detection setup without any hardware modification}
In this case, we obtain
		\begin{equation} \label{eq:finaldeltas}
			\begin{aligned}
				\deltaone  &\leq  \max \left\{  \left(1-\frac{1-(1-\dMin)^2}{1-(1-\dMax)^2}\right)\frac{\dMax(2-\dMin)}{1-(1-\dMin)^2},  4  \abs{ 1- \sqrt{ 1-(1-\dMin)^2(1-\etaRenorm)}} \right\}, \\
				\deltatwo &\leq \max \left\{1-\frac{1-(1-\dMin)^2}{1-(1-\dMax)^2},(1-\dMin)^2 (1-\etaRenorm) \right\}, \\
			\end{aligned}
		\end{equation}
		where
		\begin{equation} \label{eq:boundsetadc}
			\begin{aligned}
				\etaRenorm &= \etaMin / \etaMax \\
				\dMax &= \max\{d_{X_0},d_{X_1},d_{Z_0},d_{Z_1}\} \leq \dcprob (1+\dcchar), \\
                \dMin &= \min \{d_{X_0},d_{X_1},d_{Z_0},d_{Z_1}\} \geq \dcprob(1-\dcchar), \\ 
				\etaMax &= \max\{\eta_{X_0},\eta_{X_1},\eta_{Z_0},\eta_{Z_1}\} \leq  \etadet(1+\etachar),\\
                \etaMin &= \min \{\eta_{X_0},\eta_{X_1},\eta_{Z_0},\eta_{Z_1}\} \geq \etadet(1-\etachar). 
			\end{aligned}
		\end{equation}
		Thus, upper bounds on $\deltaone,\deltatwo$ can be computed using \cref{eq:finaldeltas} and the bounds in \cref{eq:boundsetadc}. It is these bounds that we use to compute key rates.

		\subsubsection{Random Swapping of 0 and 1 Detectors}
		
		In Ref.~\cite{fung_Security_2009} it was argued that random swapping of the 0 and the 1 detector can be used to remove basis-efficiency mismatch for single-photon pulses entering Bob's detectors. Note that this trick \textit{only works for the single-photon subspace}. We will now adapt our analysis to the case where Bob randomly swaps the $0$ and the $1$ detector. 
		
		In the scenario where we randomly swap the $0$ and the $1$ detectors, we make certain physically motivated assumptions (\cref{eq:etaXIsEtaZ}) about the detector setup. In particular, we assume that the dark count rate is a property of the detector only. Furthermore, we assume that the basis choice setting does not change the detector parameters. This means that  the dark count rate and detection efficiency in both bases is the same (though these can be different for each detector).  Thus, we have
		\begin{equation} \label{eq:etaXIsEtaZ}
			\begin{aligned}
				\eta_{X_0} &= \eta_{Z_0}\eqqcolon \eta_0\\
				\eta_{X_1} &= \eta_{Z_1}\eqqcolon \eta_1\\
				d_{X_0} &= d_{Z_0}\eqqcolon d_0\\
				d_{X_1} &= d_{Z_1}\eqqcolon d_1.
			\end{aligned}
		\end{equation}

		We will see that this indeed allows us to obtain improved results, even though it does not completely remove efficiency mismatch. In particular, the leading order terms in $\deltaone, \deltatwo$ are improved in the new bounds obtained  in \cref{eq:deltacalcsnew,eq:deltacalcsnewbound}.
		 Note that our metrics $\deltaone, \deltatwo$ do not improve unless we make these assumptions. These assumptions are also implicit in the claims presented in Ref.~\cite{fung_Security_2009}.
		
		If the random swapping is implemented, the Bob's POVM elements also need to be suitably modified, as described in \cite[Section 6.4.2]{tupkary_phase_2024}, and the $\deltaone,\deltatwo$ computations are undertake for these modified POVMs.

		\begin{equation} \label{eq:deltacalcsnew}
			\begin{aligned}
			\deltaone &\leq 4 \left(1-\sqrt{1-(1-\dMultAvg)^2\frac{(1-\etaRenorm)^2}{2}}\right), \\
				\deltatwo &\leq (1-\dMultAvg)^2\frac{(1-\etaRenorm)^2}{2},
				\end{aligned}
		\end{equation}
		where
		\begin{equation} \label{eq:deltacalcsnewbound}
			\begin{aligned}
			 \dMultAvg &= 1-\sqrt{(1-d_0)(1-d_1)} \geq \dMin ,\\
			 \etaRenorm &= \frac{\etaMin}{ \etaMax} \geq \frac{1-\etachar}{1+\etachar}.
			\end{aligned}
		\end{equation}
			Thus, upper bounds on $\deltaone,\deltatwo$ in case of random swapping of detectors can be computed using \cref{eq:deltacalcsnew} and the bounds in \cref{eq:deltacalcsnewbound}. 
		We see that these bounds are better than the earlier bounds from \cref{eq:finaldeltas}. On inspecting our calculations, we find that the zero-photon component of $\deltaone,\deltatwo$ goes to zero due to  $d_X = d_Z$. Furthermore, random swapping in addition to the assumption of $\eta_X = \eta_Z$ leads to the single-photon contribution also being zero. Thus, we are left with the two-photon contribution.

    \subsection{Plots} \label{subsec:eurplots}

				\begin{figure}[!ht] 
                \centering
			\includegraphics[width=\linewidth]{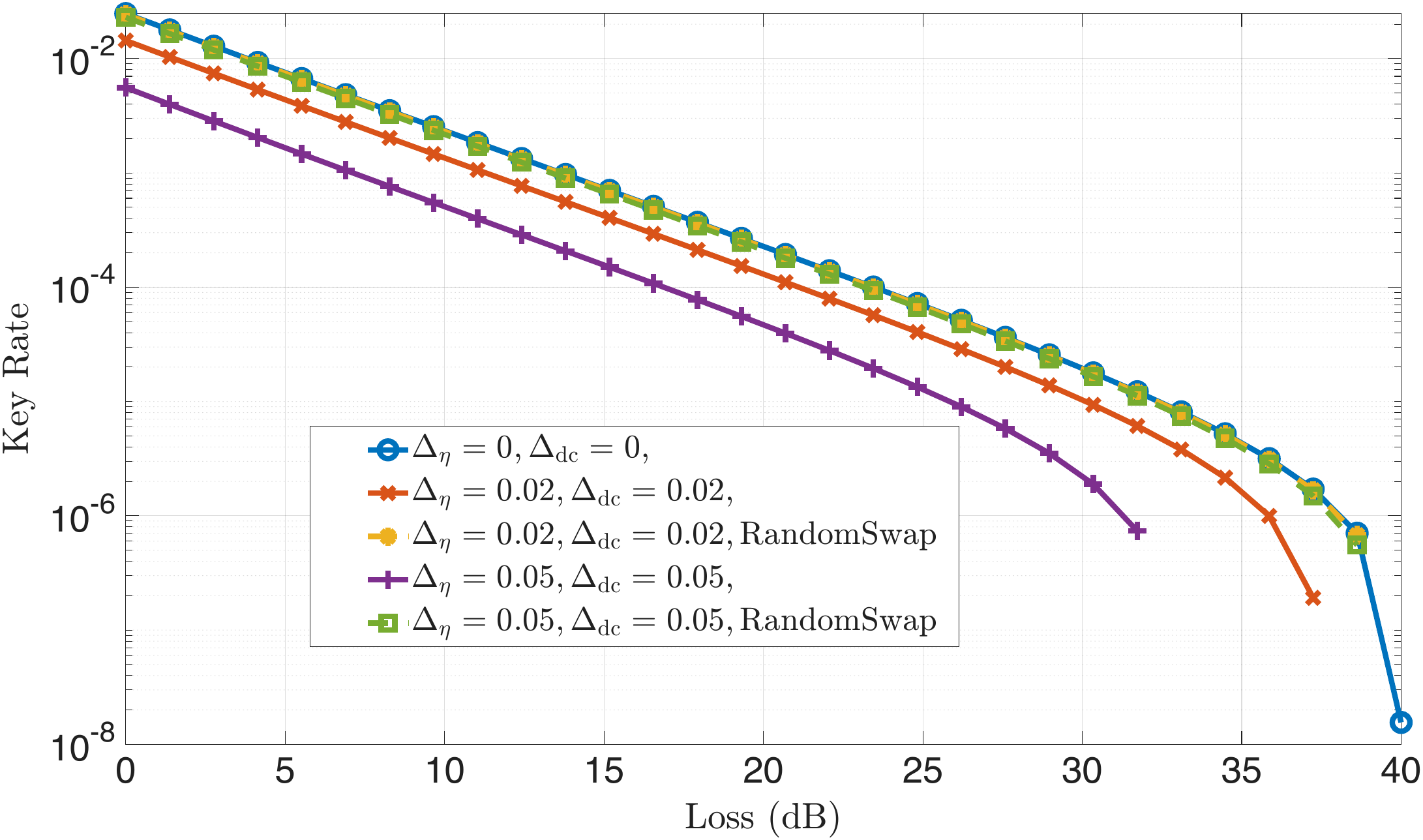}
			\caption{Finite-size key rates in the presence of basis-efficiency mismatch, for the decoy-state BB84 protocol, against loss. We plot key rates for $n = 10^{12}$ number of total signals sent, for various values of $\etachar,\dcchar$. We find that random swapping of the $0$ and $1$ detectors drastically improves the key rates obtained.}  \label{fig:figOne}
		\end{figure}
	
	\begin{figure} [!ht]
    \centering
			\includegraphics[width=\linewidth]{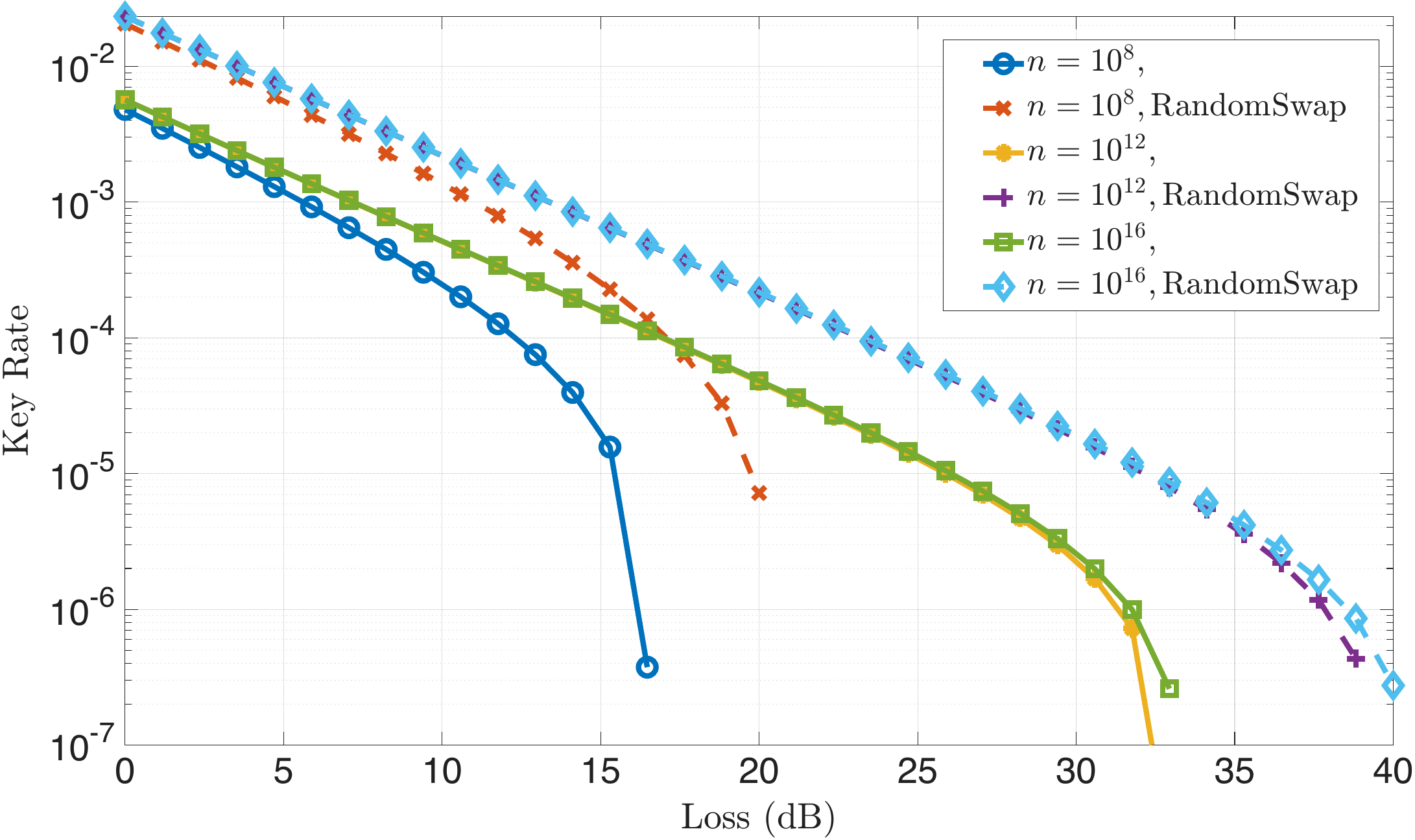}
			\caption{Finite-size key rates in the presence of basis-efficiency mismatch, for the decoy-state BB84 protocol against loss. We plot key rates for various values of  total number of signals sent ($n$), for $\etachar=\dcchar=0.05$.}  \label{fig:figTwo}
		\end{figure}

\begin{figure}[!ht]
\centering
	\includegraphics[width=\linewidth]{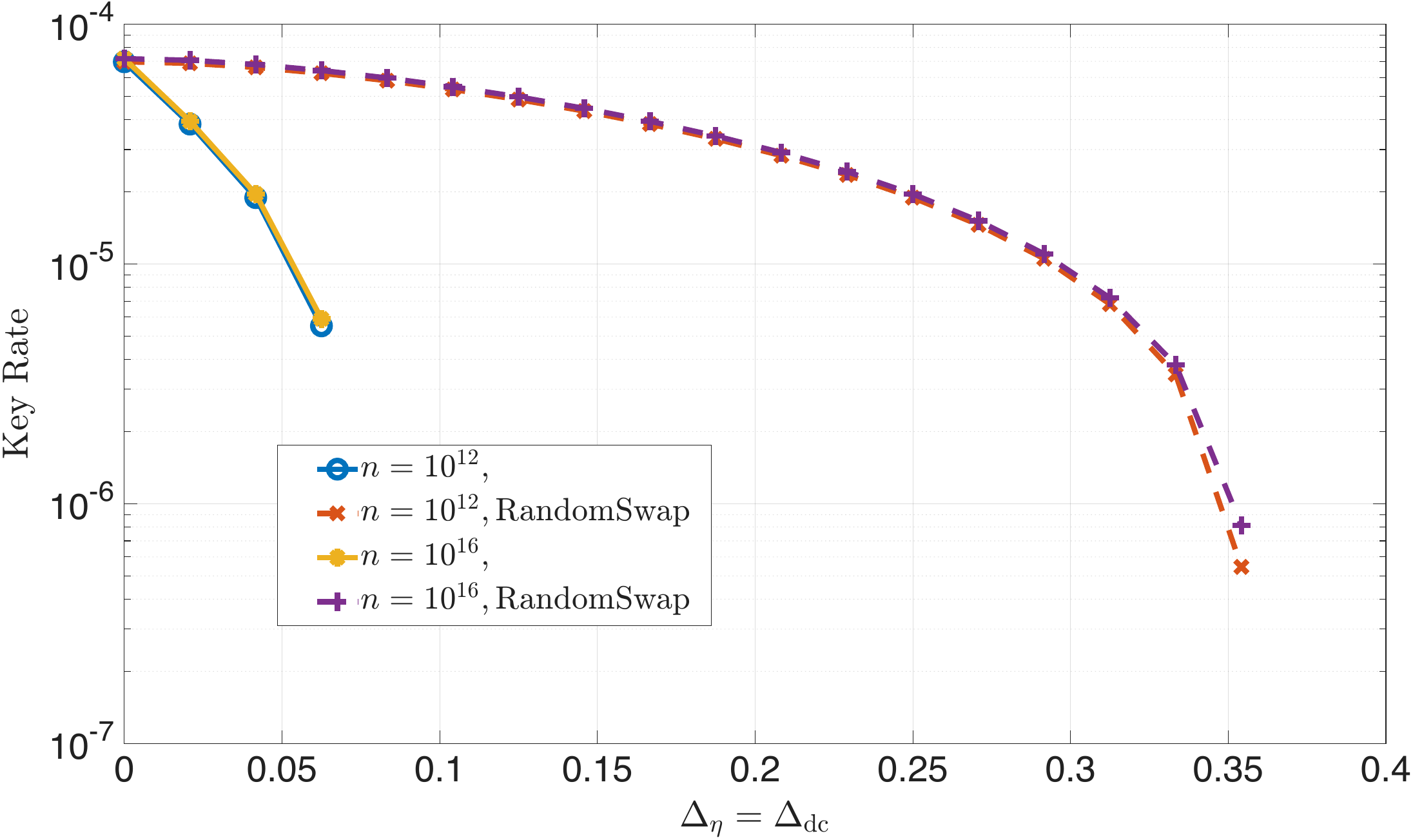}
	\caption{Finite-size key rates in the presence of basis-efficiency mismatch, for the decoy-state BB84 protocol against detector characterization parameters $\etachar,\dcchar$. We plot key rates for a channel with $25$dB loss.}  \label{fig:figThree}
\end{figure}

We plot finite size key rates for the decoy-state BB84 protocol described in \cref{sec:decoy}. We choose typical protocol parameters and plot the key rate for the expected observations for the given channel model.  For all plots, we set the basis choice probabilities to be $\pzA = \pzB = 0.5$ and $\pxA = \pxB = 0.5$, 
We set the detector parameters to be $\etadet = 0.7$ and $\dcprob = 10^{-6}$. 
	We set the misalignment angle to be $2^\circ$. We set the number of bits used for error-correction to be $ \leak(\decoyparams) = f_{\text{EC}} \nK h(\eZ)$, where $f_\text{EC} = 1.16$ is the error-correction efficiency.  The decoy intensities are chosen to be $\mu_1 = 1$, $\mu_2 = 0.1$, and $\mu_3=0.01$. Each intensity is chosen with equal probability.
    We set the overall security parameter to be $\epssecure = 2 \times 10^{-10}$, with $\epsEV = 10^{-10}$, and $\epssecret = 10^{-10}$. Then, the various $\varepsilon$ values can be further distributed as : $\epsPA = 10^{-10} / 2$, $\epsAT = 10^{-10} / 4$, and $\epsATa = \epsATb= \epsATc = \epsdecoy = 10^{-10} / 4\sqrt{12}$.
 Due to machine precision issues arising from small values of $\epsAT^2$, we use Hoeffdings inequality to bound $\gamma_{\text{bin}}$ (\cref{eq:gammabindefined}) instead of using the cumulative binomial distribution (which is tighter). As before, note that several of these parameters can be optimized over for the best key rate results. We do not do that here, since we wish to simply study the impact of efficiency mismatch on the key rates.

	\begin{enumerate}
			\item 	In \cref{fig:figOne}, we plot the finite size key rate against loss for various values of detector characterizations $\etachar, \dcchar$ for $n_\text{total} = 10^{12}$ number of total signals. 
			For $\etachar = \dcchar = 0$, we have $\deltaone = \deltatwo = 0$. Therefore the phase error rate bound from \cref{eq:estimateimperfect}	reduces to the scenario where the basis-independent loss assumption is satisfied (\cref{eq:estimateperfect}). For non-zero values of $\etachar, \dcchar$, the key rate is reduced. This is mostly due to the increase in the bound for the phase error rate from \cref{eq:estimateimperfect} from $\deltaone$. We find that random swapping leads to a dramatic improvement in performance.
			\item 
			In \cref{fig:figTwo}, we plot the finite size key rate against loss for various values of total signals sent. We set  $\etachar = \dcchar = 0.05$. 
			We find that we get close to asymptotic key rates already at $n =10^{12}$ signals sent. 
			\item 
			In \cref{fig:figThree}, we plot the finite size key rate against detector characterization parameters $\etachar,\dcchar$. We find that our methods can tolerate a significant amount of error in detector characterization. In fact, with random swapping of detectors, we get positive key rate for $n=10^{12}$ signals sent for $\etachar,\dcchar$  upto $0.35$.

    \end{enumerate}
We also refer the reader to Ref.~\cite{nahar_phd_2025} for an improvement over these results.

\subsection{Extension to multi-mode case} \label{subsec:multimodeeur}
The analysis presented so far assumes that the detector behavior, while possibly varying between rounds or exhibiting correlations, is described by a single mode characterized by bounded loss and dark count rates $\eta_{b_i}$ and $d_{b_i}$, as specified in \cref{eq:etadcmodel}.
Thus, we have presented an analysis of Case~\ref{casetwo} from the introduction to this chapter. This analysis allows us to drastically reduce the requirements on device characterization: the proof technique is now robust to imperfect characterization.
However, physical implementations of QKD protocols are vulnerable to side-channel attacks where Eve can control, to a limited extent, the POVMs used. For example, by controlling the frequency, spatial mode \cite{rau2014spatial,sajeedSecurity2015} or arrival time \cite{qiTimeshift2007} of the light, Eve can partially choose the detector efficiencies and induce a suitable basis-efficiency mismatch. This is the scenario described by Case~\ref{casethree}.

While our proof technique advances the theory to the point where this case can be handled in principle, a complete analysis first requires the physical modeling of multi-mode detectors, which remains an open problem. In this section, we outline how the results of this work can be applied to a simple multi-mode model.
 
We expand our detector model (and Bob's Hilbert space) to account for spatio-temporal modes \cite{zhang_security_2021} as 
\begin{equation} \label{eq:povmmodes}
    \Gamma^{\text{multi}}_{(b_A,b_B),(k)} = \bigoplus_{\mode} \Gamma_{(b_A,b_B),(k)} (\{{\eta_{b_i}(\mode),  d_{b_i}(\mode)}\}),
\end{equation}
where $\mode$ denotes the spatio-temporal mode, and $ \Gamma_{(b_A,b_B),(k)} (\{{\eta_{b_i}(\mode),  d_{b_i}(\mode)}\})$ denotes the single-mode POVM element corresponding to that mode, and is given by \cref{eq:alicebobpovmmodel}. The multi-mode detector has loss $\eta_{b_i}(\mode)$ for this mode, and a dark count rate of $d_{b_i}(\mode)$. 

The block-diagonal structure with respect to $\mode$ in the above equation reflects the fact that our model assumes no interference between any pair of spatio-temporal modes during the measurement process.
In particular, it captures the possibility that an adversary may exploit different times-of-arrival, frequencies, or angles of incidence to attack the system, provided that each instance corresponds to a definite spatio-temporal mode and no coherent superpositions across modes, or multi-excitation states that simultaneously occupy several modes are used. Even with these limitations, the model protects against a wide range of known classical side-channel attacks.
For instance, the time-shift attack \cite{makarovEffects2006,qiTimeshift2007} is fully captured within this model, as it simply corresponds to Eve selecting different times-of-arrival to exploit the time-dependent efficiency mismatch of the gated detectors.
Thus, the block-diagonal model represents a first step toward a more complete analysis of realistic side-channels.
This perspective also captures other potential attack strategies, such as modifying the temperature of the detection setup.
   
\begin{remark} \label{remark:sidechannelmodel}
   We stress that our results in this subsection should be interpreted within the context of this model, and may not accurately describe the physical reality of multi-mode detectors. Nevertheless, while we only consider models of the above form in this work, our proof provides a framework to accommodate more complicated models of multi-mode detectors with off-diagonal blocks, as long as one can suitably bound $\deltaone,\deltatwo$.  In general, this would require a model of the detectors, and characterization of the detectors over all the modes. For examples of such attempts to experimentally characterize all the modes, see Ref.~\cite{rau2014spatial,sajeedSecurity2015}. 
\end{remark}

Due to the block-diagonal structure of the above POVM element \cref{eq:povmmodes}, and the fact that $\deltaone, \deltatwo$ are $\infty$-norms which can be computed on each block-diagonal part separately, it is straightforward to see that our computation of $\deltaone,\deltatwo$  is directly applicable to the above scenario.  To see this,  note that our metrics are obtained by first constructing POVMs corresponding to a  multi-step measurement process, as outlined in \cref{subsec:recipe}. This construction preserves the block-diagonal structure of \cref{eq:povmmodes}.  Thus, if $\deltaone(\{{\eta_{b_i}(\mode),  d_{b_i}(\mode)}\}), \deltatwo(\{{\eta_{b_i}(\mode),  d_{b_i}(\mode)}\}) $ are the values of these metrics computed according to \cref{eq:deltaCostsRecipe}, for the appropriate single-mode POVMs,  then the metrics for the multi-mode case are given by 
    \begin{equation} \label{eq:deltamultimode}
        \begin{aligned}
        \deltaone^\text{multi} =& \max_{\mode}  \deltaone(\{{\eta_{b_i}(\mode),  d_{b_i}(\mode)}\}), \\
            \deltatwo^\text{multi} =& \max_{\mode}  \deltatwo(\{{\eta_{b_i}(\mode),  d_{b_i}(\mode)}\}).
        \end{aligned}
    \end{equation}
    If the values of $\{{\eta_{b_i}(\mode),  d_{b_i}(\mode)}\}$ are characterized and satisfy  \cref{eq:etadcmodel} for all $\mode$, then 	\cref{eq:deltamultimode} is exactly the same as the computation as in Step (5) of \cref{subsec:recipe} (which corresponds to computing $\deltaone,\deltatwo$ for  Case~\ref{casetwo} (see start of this chapter).		

This means that the recipe from \cref{subsec:recipe}, and the computed key rates from \cref{subsec:eurplots} are valid for the scenario where Eve can choose the value of $\eta_{b_i}, d_{b_i}$ in the specified ranges (\cref{eq:etadcmodel}), via some extra spatio-temporal modes. Most importantly, our analysis does not depend on the \textit{number} of such spatio-temporal modes. Thus, we are able to address scenarios where Eve has an arbitrary number of spatio-temporal modes, to induce (a bounded amount of) basis-efficiency mismatch in the detector.

			\begin{remark} \label{remark:allowingevedarkcounts}
				As discussed above, our methods are such that allowing Eve to choose the detector parameters within the characterized range yields the same key rate as having fixed detector parameters that are characterized within the same range. However, this observation need not be fundamental, and may be a consequence of the proof technique used in this work. This is because intuitively, we expect scenarios where Eve cannot choose the detector parameters (from within their respective ranges), to lead to higher key rates than scenarios where she can, since she is strictly stronger in the latter scenario.
				Moreover, while we do not know of a physical mechanism by which Eve can choose dark count rates, we allow Eve to choose them along with the detection efficiency. 
			\end{remark}

We have picked this model for its theoretical simplicity. However, more realistic models such as the one introduced in \cite[Section 3]{fung_Security_2009} can also be analysed with the results in this work. 
Finally we note that this work does not apply to \textit{all} detector side-channels. For instance, our model does not fit Trojan horse attacks \cite{Gisin_2006}. Moreover, some blinding attacks on detectors \cite{Gerhardt_2011} lead to complete  knowledge of Bob's detection events to Eve. In this case, our methods naturally lead to trivial key rates, since no key generation is possible.

\section{Summary and Outlook}
\label{sec:eurconclusion}

In this chapter, we presented a finite-size security proof of the decoy-state BB84 protocol in the presence of imperfectly characterized and bounded adversarial control over basis-efficiency mismatch, using a phase error rate based approach.
In doing so, we addressed a longstanding and well-known assumption underlying security proofs within the entropic uncertainty relation (EUR) and phase error correction frameworks.
Prior to this work, security proofs within these frameworks were fundamentally unstable: even infinitesimal amounts of basis-efficiency mismatch were sufficient to invalidate the analysis.
(This issue does not arise for measurement-device-independent QKD, and is not addressed in this chapter for entanglement-based protocols.) By allowing for bounded adversarial control over the efficiency mismatch, our methods also provide a framework for addressing an important class of detector side channels that had remained unresolved in existing security proofs for standard prepare-and-measure QKD with trusted detection setups.

We applied our framework to the decoy-state BB84 protocol and demonstrated that practical key rates can still be achieved in the finite-size regime, even in the presence of basis-efficiency mismatch.
In addition, we quantitatively investigated mitigation strategies such as random swapping of detectors, which reduce the effective mismatch and thereby improve performance.
Taken together, these results represent a significant step toward both the protocol-level security of EUR-based proof techniques and the implementation-level security of QKD systems employing trusted detectors.

\paragraph{Detector Models: } The rigorous results obtained in this chapter for multi-mode detectors rely on a specific detector model.
While this model already captures several realistic features, we expect the framework to be adaptable to more refined and experimentally accurate models in future work.
In particular, computing key rates within our framework requires bounds on the parameters $\deltaone$ and $\deltatwo$.
Although our current analysis relies on simplifying assumptions to obtain these bounds, there is clear scope for improvement.
More refined detector characterizations could lead to tighter bounds and improved key rates.
More broadly, extending the framework to encompass a wider range of detector imperfections—well characterized by state-of-the-art experimental techniques—would significantly broaden the applicability of the results. 

\paragraph{Connections to subsequent work: } We note that the results of this chapter have already informed subsequent work on QKD security analysis.
For example, basis-efficiency mismatch becomes more subtle in passive detection setups, where the beam splitter implementing the basis choice exhibits photon-number-dependent behavior.
Ref.~\cite{wang2025phase} builds directly on the framework developed here and extends it to passive protocols.  Furthermore, certain classes of correlated detector imperfections can be handled using the methods outlined in Ref.~\cite{tupkary_phase_2024} (not presented in this thesis), which were later applied rigorously in Ref.~\cite{wang2025phase}.

Another natural extension is the integration of our methods with established techniques for addressing source imperfections \cite{pereira_modified_2023,tamaki_loss-tolerant_2014,curras-lorenzo_security_2024}.
Such a combination would yield a security proof robust to both source and detector imperfections.
A result of this form has recently been obtained in Ref.~\cite{curras_securityquantumkeydistribution_2025}. Thus, these results form a foundation for a growing body of results addressing detector imperfections in QKD security proofs.

Relatedly, Ref.~\cite{nahar2025imperfect} presents a complementary approach to detector imperfections based on the flag-state squasher.
That work demonstrates that the flag-state squasher can be used to replace imperfect detectors with ideal (squashed) ones even in the presence of imperfectly characterized loss and dark count rates, and outlines a pathway toward handling correlated detector effects. Moreover, elements from all these works \cite{nahar2025imperfect,tupkary_phase_2024,wang2025phase} can be combined to yield tighter versions of the results in this chapter, as shown in Ref.~\cite{nahar_phd_2025}. A reader interested in detector imperfections is encouraged to refer to Ref.~\cite{nahar_phd_2025}.

\paragraph{Limitations of Phase Error Based Methods:} Notice that the problem of efficiency mismatch is, in fact, specific to phase error based proof techniques.
Other approaches, such as postselection or entropy accumulation–based methods, do not rely on symmetry assumptions\footnote{The postselection requirement on permutation invariance can be satisfied via protocol design.} about the underlying POVMs and therefore do not suffer from this issue.
This highlights a fundamental limitation of phase error based approaches.

While phase error based methods yield rigorous and often analytically tight bounds on the key rate, they rely heavily on symmetries that are specific to idealized BB84 implementations.
Any deviation from perfect symmetry - whether originating from source imperfections or detector imperfections - can invalidate the original analysis.

Although such deviations from symmetry can, in principle, be handled through careful and technically involved arguments, the resulting security statements are rarely modular.
Instead, they depend intricately on the precise protocol details and on the specific physical model under consideration, with even small modifications often requiring a substantial reworking of the entire proof.\footnote{This observation is supported by the length and complexity of Refs.~\cite{curras-lorenzo_security_2024,wang2025phase}, which, despite their technical sophistication, ultimately remain tailored to BB84-like protocols.}
From the perspective of this thesis, this lack of modularity constitutes the primary drawback of phase error based security proofs. At the same time, these methods have been studied extensively and are widely used, largely because they are relatively simple to apply and because many relevant source imperfections have been thoroughly analyzed in the literature \cite{curras-lorenzo_security_2024,zapatero2023implementationsecurityquantumkey,tamaki_loss-tolerant_2014,curraslorenzo_rigorousphaseerrorestimationsecurityframework_2026}, and which can potentially be combined with the analysis of this chapter.

In the next chapter, we turn to MEAT, which provides with a security proof approach that is tight, modular, and significantly more general than any we have seen so far. 

\chapter{Security using Marginal-constrained Entropy Accumulation Theorem} \label{chap:MEAT}
\epigraph{Where we invoke the MEAT and obtain a solution to a surprisingly large number of our problems; and where we present a complete and rigorous security analysis framework that accommodates almost all features a reasonable person could hope for in QKD. 
}{}

In this chapter, we present a rigorous and complete security proof for a generic QKD protocol using the marginal-constrained entropy accumulation theorem (MEAT) \cite{arqand_marginal_2025}, which can be applied to the decoy-state BB84 protocol \cite{Lo_decoystate_2005,Hwang_qkdiwthhighloss_2003,Ma_practicaldecoy_2005,Bennett_2014}. To support this goal, our approach extends far beyond this specific case: we will consolidate several key elements required for a modern QKD security analysis into a unified and modular framework. Our treatment will encompasses all the major components essential for the analysis of realistic QKD protocols, including classical authentication and postprocessing \cite{portmann_key_2014,fung_practical_2010}, finite-size effects \cite{renner_security_2005,scarani_security_2008,hayashi_upper_2007,cai_finitekey_2009,curty_finitekey_2014,hayashi_concise_2012,hayashi_security_2014,lim_concise_2014,tomamichel_largely_2017,george_numerical_2021,metger_security_2023,dupuis_entropy_2020}, source-replacement schemes \cite{bennett_quantum_1992,curty_entanglement_2004}, source maps \cite{gottesman_security_2004}, squashing maps \cite{tsurumaru_security_2008,beaudry_squashing_2008,tsurumaru_squash_2010,gittsovich_squashing_2014,zhang_security_2021} and decoy-state methods. Consequently, the framework we develop can be readily adapted to obtain security proofs for other QKD protocols, as we illustrate throughout this chapter. In particular, \cref{subsubsec:recipe} provides a concrete recipe that can be followed to derive security proofs for a broad class of QKD protocols.  We emphasize that this analysis relies heavily on a body of prior work in which many of the individual ingredients required for QKD security proofs were developed (and often combined) over the last few decades. The primary contribution of this chapter is the clean and fully explicit integration of these ingredients into a single, coherent, rigorous and complete security analysis.  This chapter is based on Ref.~\cite{inprep_BDR3}.

\paragraph{Need for a self-contained, rigorous security proof for QKD.} Over the past few decades, quantum key distribution (QKD) has evolved from a theoretical concept \cite{Bennett_2014} into a mature technology poised for deployment and commercial use. A critical milestone that must be achieved before QKD can be widely adopted is the certification of QKD devices by a relevant certification authority. A key ingredient of this step is the existence of a rigorous and complete security proof for the underlying QKD protocol, which can be scrutinized and vetted by the wider community \cite{bundPositionPaper}. Such a proof must satisfy several key criteria, as outlined in Refs.~\cite{bundPositionPaper,tupkary2025qkdsecurityproofsdecoystate}:
\begin{itemize}
    \item It must specify a complete protocol relevant to practical implementations.
    \item It must clearly state all assumptions about the devices used. 
    \item It must precisely state the security criterion to be satisfied.
    \item It must provide a rigorous mathematical proof that, under the stated assumptions, the protocol achieves the desired level of security with the specified parameters.
\end{itemize}
As discussed in Ref.~\cite{tupkary2025qkdsecurityproofsdecoystate}, no existing security analysis fully meets all of the above requirements simultaneously. A recent attempt to address this gap is Ref.~\cite{mizutani2025protocolleveldescriptionselfcontainedsecurity}, which is currently under active scrutiny by the community. Ref.~\cite{tomamichel_largely_2017} presents another exemplary and fully rigorous security proof; however, it applies to an idealized qubit-based BB84 protocol and does not address the practical decoy-state setting. In this chapter, we will present our attempt at filling this gap.

\paragraph{Reasons for using MEAT.} We employ the recently developed marginal-constrained entropy accumulation theorem (MEAT) \cite{arqand_marginal_2025} for our analysis. This approach is motivated by its ability to yield tight key rates \cite{kamin_renyi_2025}, accommodate a broad range of protocol variations, and maintain a modular structure (see \cref{subsec:discussionabstractproof}).   Moreover, it enables the framework developed here to be readily extended to incorporate device imperfections, providing a systematic foundation for future implementation-level security analyses.
The main drawbacks of this approach are its relatively recent development (limited external scrutiny) and its reliance on numerical optimization to extract key rates. We believe, however, that the former limitation will naturally diminish with time and further community validation.   Furthermore, the required numerical optimizations are performed only once during the security analysis phase, and not during the execution of the QKD protocol itself.  Overall, the use of the MEAT enables us to incorporate 
a number of important features (explained in more detail later in this chapter) such as 
on-the-fly announcements, fully adaptive key-rates, ability to handle channel variability, and robustness to source and detector imperfections, while maintaining compatibility with a wide range of state preparation, measurement, and classical post-processing choices (which are also allowed to vary across rounds).

\paragraph{Prior work.} We note that the MEAT \cite{arqand_marginal_2025} is a recent variant of the entropy accumulation theorem \cite{dupuis_entropy_2020,metger_generalised_2022,arqand_generalized_2024}. Moreover, the MEAT was first applied to the security analysis of decoy-state BB84 in Ref.~\cite{kamin_renyi_2025}, which also developed numerical techniques for the evaluation of finite-size key rates (see also Ref.~\cite{navarro_finite_2025}), and which we leverage in this chapter. The key rates obtained in the present work match those reported in Ref.~\cite{kamin_renyi_2025}. Ref.~\cite{kamin_renyi_2025} treats both variable-length and fixed-length protocols, and explicitly analyzes phase and intensity imperfections. In comparison, the analysis presented here does not explicitly model device imperfections (although it can be extended to do so). However, it is able to handle on-the-fly announcements (which arises from the way the MEAT is invoked in the security analysis) and is more general in its treatment of several structural aspects of the security proof, such as squashing maps and source maps. Moreover, it addresses certain technical issues related to classical authentication, timing of various operations, and the treatment of infinite-dimensional systems and makes explicit a few additional formal steps in the analysis.\footnote{The manuscript Ref.~\cite{inprep_BDR3} on which this chapter is based presents an even more detailed analysis.} 
The emphasis of the present work is therefore complementary: we aim to provide a fully specified protocol together with a rigorous, self-contained security analysis in which all assumptions, protocol steps, and proof components are made explicit and assembled into a single coherent framework, which is not the goal of Ref.~\cite{kamin_renyi_2025}.

\subsubsection*{Organization of this chapter}
This chapter is organized as follows. In \cref{sec:protocoldesc}, we describe the QKD protocol to be analyzed. We do not attempt to specify a complete protocol here, but specify it in as much detail as required for the security analysis. In particular, we present a general formulation of a QKD protocol without fixing the specific states, measurements, or classical post-processing steps --- leaving these components abstract. Note that a fully specified protocol can be obtain via suitable instantiation of various parameters described in \cref{table:abstractparameters} (see \cite[Section 4.2]{inprep_BDR3}). 
In \cref{sec:backgroundinfotheory}, we summarize the relevant tools and definitions from quantum information theory that are used in our security analysis of the QKD protocol.  In  \cref{sec:proof}, we rigorously establish the security of the generic QKD protocol (as defined in \cref{sec:protocoldesc}) under the assumption that Alice sends finite-dimensional states and Bob performs finite-dimensional measurements. This culminates in \cref{theorem:abstractsecuritystatement}, which states the central security result.
In \cref{sec:optics}, we address the infinite-dimensional nature of practical implementations and introduce the notions of source maps and squashing maps to reduce the analysis to the finite-dimensional case. This enables the analysis to be extended to optical protocols such as decoy-state BB84.
To extract concrete key rates, our results require solving a finite-dimensional convex 
optimization problem. In practice, any suitable numerical method may be employed, provided 
it satisfies certain requirements - namely, that it guarantees a reliable lower bound for the 
relevant minimization problem. However, such methods are quite involved. In this chapter, we directly utilize the work of Kamin et al Ref.~\cite{kamin_renyi_2025}, which computes key rates for decoy-state BB84, in \cref{sec:plotsMEAT}.  Finally, in \cref{sec:summaryMEAT}, we present our concluding remarks.  

\section{Generic Protocol Specification} \label{sec:protocoldesc}
In this section, we describe a generic QKD protocol whose security will be proven later in \cref{sec:proof}. When supplemented with detailed specifications, and after fixing the values of various free parameters, this generic protocol turns into a particular instance of the decoy-state BB84 protocol. As usual, we assume that the protocol begins with Alice and Bob having access to  local randomness stored in classical registers, which they use to implement the steps of the QKD protocol. The adversary is uncorrelated to this randomness, which is only accessible to the honest parties, and is never leaked to the adversary (even after the termination of the QKD protocol).\footnote{While this randomness is sometimes used to generate states or classical messages that are publicly released at some point (when explicitly specified in the protocol), what we mean by this requirement is that the ``raw'' values of the randomness are never made accessible to the adversary.}  
We do not explicitly describe the generation and use of these local random numbers in the QKD protocol execution; they are implicitly utilized in choosing the signal states sent, basis choices, and choosing seeds for hashing. Thus, the assumption that Alice and Bob correctly implement the prescribed state preparation and measurements (and later, hash choices) implicitly assumes that they have access to adequate randomness for doing so. We note that the assumption of perfect random numbers can be relaxed by utilizing random number generators that are $\eps_\text{rng}$-close  to perfect, in the composable security framework.\footnote{Random number generators are often designed to output uniformly random numbers. 
However, the QKD protocol may need to transform this output into nonuniform random values, as some steps such as signal preparation or basis choice are often based on nonuniform probability distributions. This can be done at essentially small cost, using the ``interval algorithm'' described in~\cite{han1997interval} or a variant thereof in~\cite{brown2020framework}.}

In what follows, when specifying the protocol steps and during the security analysis in \cref{sec:proof}, we restrict attention to the idealized setting in which the authentication mechanism behaves honestly, meaning that every authenticated classical message sent by one party is received correctly by the other party at some later time. This is the standard assumption adopted in QKD security analyses. Under this assumption, Alice and Bob agree on all public announcements. Consequently, the public communication can be represented by a single classical register that is accessible to all parties, including Eve, rather than by separate registers corresponding to each party’s local copy of the public information.
This assumption can be relaxed to more realistic authentication models; we discuss such extensions in \cref{chap:classicalauthentication}.

 During the execution of the protocol, the parties exchange the following  classical communication which is described in the protocol steps below:
\begin{equation}
\CP^n_1, \flagkey,\flagEC, \CEC, \HEV, \CEV, \flagEV, \HPA.
\end{equation}
The final output of the protocol consists of the secret keys \( K_A \) and \( K_B \), generated by Alice and Bob, respectively. All parameters required to instantiate an instance of this \nameref{prot:abstractqkdprotocol} are given in \cref{table:abstractparameters}. Note that the \nameref{prot:abstractqkdprotocol} considered here is a more general version of the \nameref{prot:qkdprotocol} from \cref{chap:qkdbackground}. It uses similar notation, but allows the state preparation, measurements, and classical processing to vary across rounds, as well as on-the-fly announcements, unlike \nameref{prot:qkdprotocol}. This is because earlier proof techniques cannot handle these features, whereas the MEAT can. We also specify the protocol in greater detail (such as those pertaining to the timing of operations), and consequently introduce additional registers and explicitly defined functions.

\begin{table}[h!]
    \centering
    \renewcommand{\arraystretch}{1.3} 
    \begin{tabularx}{\textwidth}{>{\raggedright\arraybackslash}p{4cm} X}
        \toprule
        \textbf{Symbol} & \textbf{Meaning} \\
        \midrule
      $n \in \mathbb{N}$& Total number of rounds of the QKD protocol. \\
      $\sigma^{(j)}_k \in \dop{=}(A')$ & The $k$th signal state sent by Alice, in round $j$.\\
      $p^{(j)}_k \in [0,1]$ & Probability with which Alice sends $k$th signal state in round $j$. \\
      $M^{(j)}_k \in \Pos(B_j)$  & The POVM element corresponding to for outcome $k$ of Bob's measurement in the $j$th round.  \\
$\announcementfunction^{(j)} : \mathcal{X} \otimes \mathcal{Y} \rightarrow \mathcal{\CP}$ & Function mapping Alice and Bob's local data ($X_j, Y_j)$ to public announcements, for the $j$th round.  \\
    $\keymapfunction^{(j)}(\cdot,\cdot) : \mathcal{X} \times \mathcal{\CP} \rightarrow \mathcal{S} $ & Function implementing the mapping Alice's local data stored in $X_j$ to $S_j$, based on public announcements stored in $\CP_j$, for round $j$.\\
    $\leak(\cdot) : \mathcal{\CP}^n \rightarrow \mathbb{N}$ & Function that determines the number of transcripts of error-correction protocol, as a function of public announcements $\CP_1^n$. \\
      $\lkey(\cdot) : \mathcal{\CP}^n \rightarrow \mathbb{N}$ & Function that determines the number of bits of output key, as a function of public announcements $\CP_1^n$. \\
      $\epsEV \in [0,1]$ & Epsilon for error-verification. Determines output length of hash family used in error-verification, and the final correctness parameter. \\
      $\epsPA \in [0,1]$ & Epsilon for privacy amplification. The exact input and output lengths are determined by the protocol during runtime. \\ 
      $\hashfamily{l_\mathrm{in}}{l_\mathrm{out}}$ & Universal$_2$ hash family from $l_\mathrm{in}$ bits to $l_\mathrm{out}$ bits.  Used for error-verification. The exact input and output lengths are determined by the protocol during runtime. \\
$\idealhashfamily{l_\mathrm{in}}{l_\mathrm{out}}$  & Ideal universal$_2$ hash family from $l_\mathrm{in}$ bits to $l_\mathrm{out}$ bits. Used for privacy-amplification. \\
        \bottomrule
    \end{tabularx}
    \caption{ Parameters required to define an instance of \nameref{prot:abstractqkdprotocol}.}
\label{table:abstractparameters}
\end{table}

Note that we do \emph{not} assume that Alice and Bob have synchronized clocks for our security analysis. All timings that we refer to in this description refer to the global or true time in which these events occur, and not to Alice's and Bob's local time. Thus, we do not require Alice or Bob to have access to these global time values. They are simply introduced here since they are required in our theoretical proof. 

\begin{prot}[Generic QKD Protocol] \label{prot:abstractqkdprotocol} 
\leavevmode \\
    \begin{enumerate}
 \item For rounds $j$ from $1$ to $n$, Alice and Bob perform the following operations:
   
	\begin{enumerate}
		\item  \textbf{State preparation and transmission:} In round $j$, Alice performs the following steps. 
        \begin{enumerate}
            \item  Alice prepares one of \(\numstates\) possible signal states \(\{\sigma^{(j)}_k\}_{k}\), according to the probability distribution $p^{(j)}_{k}$, where $\sigma^{(j)}_k\in \dop{=} (A^\prime_j)$.\footnote{In this protocol specification, and in our proof, Alice is allowed send different states with different probabilities in the different rounds $j$, as long as the distributions are independent across rounds.} This step requires the use of local randomness. (Note that we assume that the adversary is uncorrelated with the state preparation; in particular, they do not hold any purification of the signal state, nor is any such purification made available to the adversary at any point during or after the protocol.)  
            \item She stores the label $k$ for her choice of the signal state in a classical register \(X_j\). The register $X_j$ has the alphabet $\mathcal{X}$.
            \item Alice sends the signal state to Bob via an insecure quantum channel. \item We let $t^\mathrm{A}_j$ be the (global) time this state leaves Alice's lab.  
        \end{enumerate} 
  
		\item \textbf{Measurements:} In round $j$, Bob performs the following steps:
        \begin{enumerate}
            \item  Bob measures his received state using a POVM  \(\{M_k^{(B_j)}\}_{k=1 \dots \nummeas}\), obtains one of $d_B$ possible outcomes, and stores his results in a classical register \(Y_j\) (which has alphabet $\mathcal{Y}$). Depending on the exact detection setup used, this step may require the use of local randomness.\footnote{Note that we also allow Bob's measurement POVMs to depend on the rounds $j$.}
            \item We let $t^\mathrm{B}_j$ be the (global) time this measurement is completed. 
        \end{enumerate}

     \item    \textbf{Public announcement:} Alice and Bob engage in public announcements in parallel to their signal transmission and measurement steps described above. 
        Alice and Bob perform the following steps for each round $j$:
        \begin{enumerate}

    \item Bob sends a fixed message to Alice indicating that he has completed the measurement for the $j$th round, after he has done so.\footnote{Note that  this is a fixed message which is already known to Eve, since she knows the time at which Bob measures. Thus, this message leaks no information to the adversary, and we can exclude it from the public announcement registers. } 
    \item Alice and Bob engage in further interactive public communication, which depends on the classical registers $X_j, Y_j$ as well as the values announced so far. The first message in this exchange is sent by Alice at time $t^\mathrm{ann}_j$, but only after two conditions are met: (i) she has already transmitted the $j$th quantum state to Bob, and (ii) she has received Bob’s message confirming that his measurement for that round is complete. This ordering guarantees that announcements never precede the corresponding quantum transmission or measurement. Consequently, under the honest authentication setting, the relation $t^\mathrm{ann}_j > t^\mathrm{A}_j, t^\mathrm{B}_j$ is always satisfied.
    \item This classical communication can be two-way and may differ from round to round. We use $\CP_j$ (which has alphabet $\mathcal{\CP}$) to denote the classical register storing all public announcements in the $j$th round.  For clarity, we stress that $\CP_j$ is determined solely by the Alice's and Bob's classical data of that round, namely the registers $X_j$ and $Y_j$, and can be computed by some function $\announcementfunction^{(j)}$ acting on $X_j,Y_j$.\footnote{For the security analysis, we only require that $\CP_j$ should be obtained by applying a CPTP map on $X_j,Y_j$. Thus, we can also allow $\announcementfunction^{(j)}$ to be a stochastic map. Of course, since these public announcements are carried out by two separate parties, each of which starts with access to $X_j$ and $Y_j$ respectively, this imposes constraints on the types of announcement functions $\announcementfunction^{(j)}$ that can be actually implemented in the protocol. } 

 \begin{remark} \label{remark:groupingannouncements} We stress that we do \emph{not} require Alice and Bob to complete the announcements for round $j$, before proceeding to signal preparation and measurement for the next round. We only require that the announcements for round $j$ occur at some time after signal preparation and measurement for that round. This feature of allowing Alice and Bob to perform public announcements \emph{before} all signals are sent are received, is referred to as \term{on-the-fly} announcements.  In particular,  Alice and Bob may announce their classical registers either immediately after each round or in grouped form, where several rounds are announced together in a block of size $\block$ rounds. In the latter, $\CP_j$ for all rounds $(k_\mathrm{block}-1)\block < j \leq k_\mathrm{block}\block$ is announced  simultaneously, at some time after the signal preparation and measurement of the $k_\mathrm{block}\block$th round. Grouping announcements in this way can reduce the overall consumption of authentication keys (see \cref{sec:delayedauthentication}).   As shown in \cref{sec:proof}, the security analysis remains valid under either choice. 
    \end{remark}
   
        \end{enumerate}
       	\item  \textbf{Sifting and key map:} Alice uses knowledge of previous public announcements (and local randomness, if needed), to map her  register $X_1^n$ to the \term{pre-amplification string} $\PAstring$. 
        \begin{enumerate}
            \item Alice maps her local data $X_j$ to the classical registers $S_j$ (which has alphabet $\mathcal{S}$), using public announcements $\CP_j$. This is described by a function $\keymapfunction^{(j)}$ acting on $X_j,\CP_j$.\footnote{As with $\announcementfunction$, we only require this mapping to be CPTP. Thus we can allow $\keymapfunction$ to be a stochastic map.  }
           \item She then applies a deterministic rule, based solely on $\CP_j$, to discard certain rounds from $S_j$. This produces the (potentially shortened) pre-amplification string, which is stored in the register $\PAstring$.\footnote{Thus, the register $\PAstring$ takes values from the set of all possible strings of length less than or equal to $n$, and composed of symbols from the alphabet of the register $S_j$ that are \textit{not} discarded. In most common scenarios, $S_j$ takes values in $\{0,1,\singleRoundBot\}$ (based on $\CP_j$,$X_j$), and the last outcome refers to the outcome that is discarded. Thus, $\PAstring$ stores a binary string of length up to $n$ bits.} In this work, we consider protocols where  Alice’s remapping and discarding are defined so that whenever a value of $S_j$ will be discarded (based on $\CP_j$), $S_j$ is set to a fixed placeholder symbol $\singleRoundBot$. Since Alice determines the mapping to $S_j$, she can directly encode the rounds to be discarded in this way.\footnote{In principle, this means that Alice could implement the discarding step by inspecting the sequence $S_1^n$ alone, without separately referring to the announcements $\CP_1^n$.}
           \item  This step can be performed at any time, as long as Alice has the necessary registers $(X_j, \CP_j)$ to perform the required operations.  
        \end{enumerate}

  After all the announcements and sifting and key map operations are completed, the state of the protocol is given by 
\begin{equation}
    \rho_{ \PAstring S_1^n X_1^n Y_1^n  \CP_1^n   E_n  }.
    \end{equation} 
 where $E_n$ denotes Eve's quantum side information at the end of her attack on the $n$ rounds (which may contain a copy of the announcements $\CP_1^n$). Alice moves on to the next step in the protocol only after all public announcements are completed, i.e, after she has sent her last message and received the last message for the public announcements corresponding to round~$n$. Note that since Alice and Bob only perform announcements after they signal preparation and measurement for that round, this implies that Alice and only moves on to the next step after all signal preparation and measurements are completed.\footnote{In our protocol, it is Alice who sends the next message containing (i) the length of the output key to be produced and (ii) a parameter related to error correction, to Bob. If desired, this step could instead be initiated by Bob via a message to Alice; in that case, Bob would analogously proceed only after sending (and receiving) the final message associated with the public announcements.} 

\begin{remark} \label{rem:diffclassicalregisters}
We make the following clarification regarding the treatment of classical registers in this analysis.
During the state preparation, transmission, and measurement phases, all public announcements are recorded in the registers $\CP_j$. For each such announcement, we provide Eve with an explicit copy, denoted $\CPhat_j$ (introduced later in our security analysis), which she may choose to store, process, or discard. Consequently, all public announcements made during these phases are already included in Eve’s final quantum system $E_n$ at the end of the quantum communication stage. Thus, we do not necessarily need to explicitly specify that Eve has access to $\CP_1^n$ (although doing so does not cause any difference to the security analysis, since Eve is allowed to copy all announcements and keep them in $E_n$ anyway). 

In the remaining steps of the protocol, only classical communication takes place. We model these by introducing  classical registers that are also accessible to Eve. If Eve wishes, she may use these classical registers together with her existing quantum system $E_n$ to generate an updated system $E'_n$. However, all
operations performed by Alice and Bob during this stage act solely on their own classical
registers and thus commute with any operation Eve might apply to her systems. Consequently, we may, without loss of generality, postpone all of Eve’s remaining operations
until after the protocol has completed. Moreover, since the trace distance is non-increasing
under CPTP maps, it follows that if the required security criterion holds for the state
\emph{before} Eve’s postponed operations, then it also holds for the state \emph{after} she
applies them (which corresponds to the actual state she obtains as a result of her attack). Hence it suffices to focus solely on analyzing the former state; in other words, we can suppose without loss of generality that Eve does not act on these classical registers at all (other than storing them). 
\end{remark}
      
	\end{enumerate}

\item  \textbf{Variable length decision:} Alice and Bob use predetermined functions of the public announcements $\CP_1^n$ to determine parameters for error correction and key generation. We use $\cobs$ to denote the value stored in the classical register $\CP_1^n$.
\begin{enumerate}
    \item Alice computes the value $\leak(\cobs) \in \mathbb{N}$ (a parameter that will be used during error correction) and stores it in $\flagEC$.
    \item Alice computes the length of the output key to be produced, given by $\lkey(\cobs) \in \mathbb{N}$, using a predetermined function $\lkey(\cdot)$, and stores this in the register $\flagkey$.
    \item Alice sends $\flagkey, \flagEC$ to Bob using the authenticated classical channel.

\end{enumerate}

The state in the protocol, conditioned on observing a specific value in the $\CP_1^n$ registers, is given by
\begin{equation}
    \rho_{ \PAstring S_1^n X_1^n Y_1^n  \CP_1^n  \flagkey \flagEC E_n | \Omega(\cobs) },
    \end{equation}
    where $\Omega(\cobs)$ denotes the event that  the value $\cobs$ is observed in the $\CP_1^n$ registers. Note that the functions $\lkey(\cdot)$  and $\leak(\cdot)$ are related, since $\lkey(\cdot)$ must account for the amount of information leaked during error correction, which is quantified by $\leak(\cdot)$. The relationship is established by the security proof. 

\item[]  \textit{ \textbf{Abort Condition: } If $\flagkey$ stores $\lkey(\cobs)=0$, Alice and Bob abort the protocol. That is, they replace their key registers with $\bot$s and send $\abort$ for all future communication.}\footnote{In this case one would want Alice and Bob to avoid wasting authentication keys on messages since they intend to abort in the end anyway and instead simply jump to the end of the protocol. While this should not pose any real issue, we do not attempt to formalize this detail here.}

		\item \textbf{Error correction:} Alice and Bob perform an error-correction subprotocol that results in Bob outputting a guess for Alice's pre-amplification string, by performing the following steps:
        \begin{enumerate}
            \item Alice and Bob read the value of $\leak(\cobs)$ from the classical register $\flagEC$.
            \item They run an error-correction subprotocol that  is designed such that the total number of possible communication transcripts is at most  $2^{\leak(\cobs)}$.\footnote{For one-way error-correction protocols that leak a fixed number of bits $\leakfixed$, the number of possible transcripts is $2^{\leakfixed}$.} 
            \item The error correction procedure can have any number of communication rounds. Alice's steps can depend on her local information $\PAstring,S_1^n, X_1^n$, prior public announcements $\CP_1^n$, and announcements during the error-correction procedure itself. Bob's steps can depend on his local information $Y_1^n$, public announcements $\CP_1^n$, and announcements during the error-correction procedure.
            \item The communication transcript is stored in the classical register $\CEC$. 
            \item At the end of this procedure, Bob outputs a classical string stored in \(\PAstring_B\), intended as his estimate of Alice’s pre-amplification string stored in \(\PAstring\). The protocol ensures that $\PAstring$ and $\PAstring_B$ are of the same length.\footnote{Since Alice's procedure of discarding rounds from $S_1^n$ is based on public announcements $\CP^n_i$, Bob can compute this length himself, and ensure that his guess is of the same length.}  
            \item From this point on, we ignore the registers $S_1^n, X_1^n, Y_1^n$. This is justified because these registers are either explicitly deleted by Alice and Bob after their use, or they no longer play any role in the subsequent steps of the protocol. In either case, they are never leaked to the adversary.
        \end{enumerate}
    		
The state in the protocol, conditioned on observing a specific value in the $\CP_1^n$ registers, is given by
\begin{equation}
    \rho_{ \PAstring \PAstring_B \CP_1^n  \CEC \flagkey \flagEC  E_n | \Omega(\cobs) }.
    \end{equation}

         \item  \textbf{Error verification:}  Alice and Bob perform universal$_2$ hashing and compare the hash values, by performing the following: 
        \begin{enumerate}
            \item Alice and Bob look at the length $\lprePA$ of  the string in $\PAstring$ ($\PAstring_B$ for Bob). This determines the universal$_2$ hash family from the set of $\lprePA$-length strings to $\ceil{\log(1/\epsEV)}$ bits, which we denote using $\hashfamily{\lprePA}{\ceil{\log(1/\epsEV)}}$.\footnote{More generally, one could also use a $\delta$-almost-universal family. In this case, $\epsEV = \delta$, and $\ceil{\log(1/\epsEV)}$ is replaced by the number of the bits in the output of the chosen hash family.} 
            \item Alice uses her local randomness to choose a hash function from the chosen universal$_2$ hash family. She announces the choice of the function chosen in the classical register $\HEV$.
            \item Alice announces the result of the hash function applied to $\PAstring$, in the classical register $\CEV$. 
            \item Bob uses the announcement $\HEV$ to choose the same hash function as Alice, and applies it to his guess $\PAstring_\mathrm{B}$. Bob compares his computed hash value with Alice's announced value in $\CEV$, and outputs the binary result of the match in the register $\flagEV$. Bob sends $1$ in $\flagEV$ to indicate a match, and $0$ for a mismatch. 
        \end{enumerate}
       
\item[]        \textit{\textbf{Abort Condition:} If \(\flagEV\) indicates a mismatch, Alice and Bob  abort the protocol. That is, they replace their key registers with $\bot$s and send $\abort$ for all future communication.}
The state in the protocol, conditioned on $\Omega(\cobs)$ and error-verification passing, is given by
\begin{equation}
    \rho_{ \PAstring \PAstring_B \CP_1^n  \CEC \CEV \flagkey \flagEC \flagEV \HEV E_n | \Omega(\cobs) \wedge \OmegaEV }.
    \end{equation}
where $\wedge$ denotes the logical `and' operator and $\OmegaEV$ denotes the event that error-verification passes, i.e $\flagEV$ indicates a match.
        
		\item \textbf{Privacy amplification:}  Alice and Bob perform \emph{ideal}  universal$_2$\footnote{Recall that ideal universal$_2$ (see \cref{def:2universal}) corresponds to the case where the collision probability of the hash is \emph{equal} to $1/|D'|$, where $D'$ is the codomain of the hash family. Note that this requirement can be relaxed to universal$_2$ by sacrificing one additional bit of output key, see \cref{remark:LHLequalityissue} for a discussion.} hashing on their pre-amplification strings to generate the output key, as follows:
        \begin{enumerate}
           \item Alice looks at the length $\lprePA$ of the string in $\PAstring$, and the register $\flagkey$ to determine the length $\lkey(\cobs)$  of the output key. This determines the ideal universal$_2$ hash family from the set of $\lprePA$-length strings to $\lkey(\cobs)$ bits, which we denote using $\idealhashfamily{\lprePA}{\lkey(\cobs)}$.  
           \item Alice uses her local randomness to choose a hash function from the chosen universal$_2$ hash family. She announces the choice of the function chosen in the classical register $\HPA$. Bob uses the announcement $\HPA$ to choose the same function as Alice.
           \item Alice (Bob) applies the hash function to the $\PAstring$ ($\PAstring_\mathrm{B}$) register, and stores the result in the $K_A$ ($K_B$) registers.
           \item From this point on in the protocol, we ignore the registers $\PAstring,\PAstring_B$, since they are longer play any role in the subsequent steps of the protocol and are never leaked to the adversary. 
        \end{enumerate}
  The state in the protocol, conditioned on $\Omega(\cobs) \wedge \OmegaEV$, is given by 
\begin{equation}
    \rho_{K_A K_B  \CP_1^n  \CEC \CEV \flagkey \flagEC \flagEV \HEV \HPA E_n  | \Omega(\cobs) \wedge \OmegaEV }.
    \end{equation}
\end{enumerate}
\end{prot}

We prove the security of \nameref{prot:abstractqkdprotocol}, in terms of the parameters used to describe it (see \cref{table:abstractparameters}),  in \cref{sec:proof}.  Note that with the details specified in \cref{sec:statesandmeasurements}, one can now 
instantiate the \nameref{prot:abstractqkdprotocol} as a polarization-encoded, decoy-state BB84 protocol.

\section{Additional Background} \label{sec:backgroundinfotheory}

In this section, we collect some additional definitions and background from information theory that will be used in our proof. We start with defining $f$-weighted {\Renyi} entropies.

\begin{definition}($f$-weighted {\Renyi} entropies \cite[Definition~4.1]{arqand_marginal_2025}  \cite[Definition~4.1]{arqand_generalized_2024})
	\label{def:QES}
	Let $\rho \in \dop{=}(\CP Q Q')$ be a state where $\CP$ is classical with alphabets $\widehat{\mathcal{C}}$. A \term{tradeoff function on $\CP$} is simply a function $f:\widehat{\mathcal{C}} \to \mathbb{R}$; equivalently, we may denote it as a real-valued tuple $\mbf{f}
	\in \mathbb{R}^{|\widehat{\mathcal{C}}|}$ where each term in the tuple specifies the value $f(\cP)$. Given a tradeoff function $f$ and a value $\alpha\in(0,1)\cup (1,\infty)$, we define the  \term{$f$-weighted entropy of order $\alpha$} for $\rho$ as
    \begin{align}
	\label{eq:fweightedonlyH}
	\Halpha[f](Q|\CP Q')_{\rho} &\coloneq
	\frac{\alpha}{1-\alpha} \log \left( \sum_{\cP} \rho(\cP) \, 2^{\left(\frac{1-\alpha}{\alpha}\right) \left(\Halpha(Q|Q')_{\rho_{|\cP}} - f(\cP) \right) } \right),
\end{align}
\end{definition}
Note that the $f$-weighted {\Renyi} entropy reduces to the usual {\Renyi} entropy when the function $f$ is set to be zero for all inputs \cite[Proposition 5.1]{tomamichel_quantum_2016}.  To gain some intuition of this quantity,
we can define the notion of ``log-mean-exponential'' of a random variable with respect to a base $b \in (0, \infty)$ as:
\begin{align}\label{eq:lme}
\underset{P_X}{\operatorname{lme}_b} \left[g(X)\right] \coloneq \log_b \left( \sum_{x} P_X(x) \, b^{g(x)} \right).
\end{align}
With this, Eq.~\eqref{eq:fweightedonlyH} can be written as
\begin{align}\label{eq:interpretaslogmeanexp}
\Halpha[f](Q|\CP Q')_{\rho} = \underset{\boldsymbol{\rho}_{\CP}}{\operatorname{lme}_b} \left[\Halpha[f](Q|Q')_{\rho_{|\CP}} - f(\CP) \right] , \quad\text{where } b = 2^{1-\alpha} \in (0,\infty).
\end{align}
This means non-negativity of  $\Halpha[f](Q|\CP Q')_\rho\geq 0$ is equivalent to having $f(\CP)$ lower bound $\Halpha(Q|Q')_{\rho_{|\CP}}$ in a log-mean-exponential sense.

\begin{lemma}\label{lemma:DPIfweighted}
	(Data-processing \cite[Lemma~4.4]{arqand_marginal_2025}) Let $\rho \in \dop{=}(\CP Q Q')$ be classical on $\CP$, let $f$ be a tradeoff function on $\CP$, and take any $\alpha\in [\frac{1}{2},1)\cup(1,\infty]$. Then for any channel $\mathcal{E} \in \CPTP(Q',Q'')$, 
	\begin{align}
		\Halpha[f](Q|\CP Q'')_{\mathcal{E}[\rho]} \geq \Halpha[f](Q|\CP Q')_{\rho}.
	\end{align}
	If $\mathcal{E}$ is an isometry, then we have equality in the above bound.
\end{lemma}
\begin{lemma} \label{lemma:dpinonconditioningregister}
    (Data-processing on non-conditioning register) Let $\rho \in \dop{=}(\CP Q Q')$ be classical on $\CP$, let $f$ be a tradeoff function on $\CP$, and take any $\alpha\in [\frac{1}{2},1)\cup(1,\infty]$. Then for any unital channel $\mathcal{F} \in \CPTP(Q,Q'')$, i.e, $\mathcal{F}$ maps the identity operator $\id_Q$ to $\id_{Q'}$, we have 
    \begin{align}
		\Halpha[f](Q''|\CP Q')_{\mathcal{F}[\rho]} \geq \Halpha[f](Q|\CP Q')_{\rho}.
	\end{align}
	If $\mathcal{F}$ is an isometry, then we have equality in the above bound.
\begin{proof}
    Let $\alpha\in\left(\frac{1}{2},1\right]$, then we have
    \begin{align}
        &\Halpha(Q|Q')_{\rho_{|\cP}}\leq\Halpha(Q''|Q')_{\mathcal{F}[\rho_{|\cP}]}\qquad ;\forall\cP\in\CP\nonumber\\
        &\Longrightarrow \rho(\cP)2^{\left(\frac{1-\alpha}{\alpha}\right)\left(\Halpha(Q|Q')_{\rho_{|\cP}}-f(\cP)\right)} \leq \rho(\cP)2^{\left(\frac{1-\alpha}{\alpha}\right)\left(\Halpha(Q''|Q')_{\mathcal{F}[\rho_{|\cP}]}-f(\cP)\right)},
    \end{align}
    where the first line is an application of data processing for unital channels~\cite[Corollary~5.1]{tomamichel_quantum_2016}. Note that the inequality is saturated if $\mathcal{F}$ is an isometry. Summing over all values of $\cP$, followed by taking log and multiplying both sides by $\frac{\alpha}{1-\alpha}$, yields the desired results. The proof for $\alpha>1$ follows similarly.
\end{proof}
\end{lemma}

\begin{lemma}(Monotonicity in tradeoff function \cite[Lemma 6.4]{inprep_authentication})\label{lemma:Monotonicity_f}
	(Monotonicity in $f$) Let $\rho \in \dop{=}(\CP Q Q')$ be classical on $\CP$, let $f,g$ be tradeoff functions on $\CP$ such that for all $\cP\in\CP$ we have $f(\cP)\leq g(\cP)$. Then, for any $\alpha\in [\frac{1}{2},1)\cup(1,\infty]$, we have
	\begin{align}
		\Halpha[g](Q|\CP Q')_{\rho} \leq \Halpha[f](Q|\CP Q')_{\rho}.
	\end{align}
\end{lemma}

We also reproduce below the core result of the MEAT framework \cite{arqand_marginal_2025}. The $A$ registers are renumbered to reflect their role in this work, and we include only the statements that are required for our analysis.

\begin{theorem}(MEAT \cite[Theorem 4.1a]{arqand_marginal_2025}) \label{theorem:MEAT}
	For each $j\in\{1,2,\cdots,n\}$, take a state $\sigma^{(j)}\in\dop{=}(A_{j})$, and a channel $\mathcal{M}_j\in\CPTP(A_{j}E_{j-1},S_j\CP_jE_j)$, such that $\CP_j$ are classical. Let $\rho$ be a state of the form $\rho_{S_1^n\CP_1^nE_n}=\mathcal{M}_n\circ\cdots\circ\mathcal{M}_1[\omega_{A_1^{n}E_0}]$ for some $\omega\in\dop{=}(A_1^{n}E_0)$, such that $\omega_{A_1^{n}}=\sigma_{A_1}^{(1)}\otimes\cdots\otimes\sigma_{A_{n}}^{(n)}$. For each $j$, suppose that for every value $\cP_1^{j-1}$, we have a tradeoff function $f_{|\cP_1^{j-1}}$ on registers $\CP_j$. Define
    \begin{equation}
        \begin{aligned}
            \kappa_{\cP_1^{j-1}} &\coloneq \inf_{\nu\in\Sigma_j} \Halpha[ f_{|\cP_1^{j-1}}](S_j| \CP_j E_j \widetilde{E})_{\nu}, \\
             \Sigma_j &\coloneq \left\{
\mathcal{M}_j\left[\omega_{A_{j}E_{j-1}\widetilde{E}}\right] 
\;\middle|\;
\omega \in \dop{=}(A_{j}E_{j-1}\widetilde{E}) \;\text{s.t.}\; \omega_{A_{j}}=\sigma_{A_{j}}^{(j)}
\right\},
        \end{aligned}
    \end{equation}
  with $\widetilde{E}$ being a register of large enough dimension to serve as a purifying register for any of the $A_{j}E_{j}$ registers.   Define the following ``normalized'' tradeoff function on $\CP_1^n$:
	\begin{align}\label{eq:fullQESnorm_simp}
		\hat{f}_\mathrm{full}( \cP_1^n) \coloneq \sum_{j=1}^n \hat{f}_{|\cP_1^{j-1}}(\cP_j), \quad\text{where}\quad \hat{f}_{|\cP_1^{j-1}}(\cP_j) \coloneq f_{|\cP_1^{j-1}}(\cP_j) + \kappa_{\cP_1^{j-1}}.
	\end{align}
	Then for any $\alpha\in (1,\infty]$ we have
	\begin{align}\label{eq:chainQESnorm_simp}
	H[\hat{f}_\mathrm{full}](S_1^n | \CP_1^n E_n)_\rho \geq 0.
	\end{align}  

\end{theorem}

\section{Proving security of QKD Protocol using MEAT} \label{sec:proof}

In this section, we embark upon proving the security of the \nameref{prot:abstractqkdprotocol}. We will first begin by expressing the \nameref{prot:abstractqkdprotocol} as a sequence of CPTP maps, before analyzing the sequence of maps using MEAT \cite[Theorem 4.1a]{arqand_marginal_2025}.

\subsection{Expressing the protocol as a sequence of channels} \label{subsec:expressingprotocolassequence}
The analysis in this section is identical to the analysis found in \cite[Section 5]{arqand_marginal_2025}. The goal of this subsection is to reduce the QKD protocol we are interested in analyzing to one which can be expressed as a sequence of channels as described in \cref{lemma:stateevolution,fig:MEAT_Attack}, in preparation for the application of MEAT \cite[Theorem 4.1a]{arqand_marginal_2025}.

\subsubsection{Modifying timings of the QKD protocol}

Let us consider the \nameref{prot:abstractqkdprotocol} described in \cref{sec:protocoldesc}. 
In this case, the protocol can be equivalently expressed as the following \nameref{prot:pmprotocoll}, which will be the object of our subsequent analysis. Note that since we will only be modifying certain timings and signal-preparation steps of this protocol, only the relevant parameters are listed explicitly below.

\begin{prot}[Prepare-and-measure protocol]\label{prot:pmprotocoll} 

\leavevmode \\

\noindent \textbf{Parameters:}

\begin{tabularx}{0.9\linewidth}{r c X}
\(n \in \mathbb{N}_0\) 			    &:& 	Total number of rounds \\
\(\{\sigma^{(j)}_{X_jA'_j}\}_{j=1}^n\) 	&:& 	Classical-quantum states prepared by Alice \\
\(\{t_j^A,t_j^B,t^{\mathrm{ann}}_j\}_{j=1}^n\) 						&:& 	Timings of various steps (see below)
\end{tabularx}

\noindent We assume that the times listed above satisfy the following conditions (see \cref{remark:justifytiming}) :
\begin{itemize}
\item For each $j \in \{1,2,\dots,n\}$, we have $t^A_j,t^B_j<t^{\mathrm{ann}}_j$. 
\item Each of the sequences $\{t^A_j\}_{j=1}^n$, $\{t^B_j\}_{j=1}^n$, and $\{t^{\mathrm{ann}}_j\}_{j=1}^n$ is monotonically increasing.
\end{itemize}
We do not impose \emph{any} other restrictions on those timings; for instance, the sequences $\{t^A_j\}_{j=1}^n$, $\{t^B_j\}_{j=1}^n$, and $\{t^{\mathrm{ann}}_j\}_{j=1}^n$ can be ``interleaved'' with each other in an arbitrary fashion as long as the first condition is satisfied. In particular, we do \emph{not} require the public announcements to have  blocksize $\block=1$; larger blocksizes are permitted precisely because such interleaving is allowed.

\noindent \textbf{Protocol steps:}
\begin{enumerate}
\item Alice and Bob perform the following steps for every $j \in \{1,2,\dots,n\}$. \label{step:singlerounds}
\begin{enumerate}
\item \textbf{State preparation and transmission:} At time \(t_j^A\), independently for each round, Alice prepares some state \(\sigma^{(j)}_{k} \in S_{=} (A'_j) \) with probability $p^{(j)}_k$, and stores $k$ in the classical register \(X_j\).
She then sends out \({A'_j}\) via a public quantum channel, after which Eve may interact freely with it. This is represented via the preparation of the state $\sigma^{(j)}_{X_j A'_j} = \sum_{k=1}^{d_A} p^{(j)}_k \ketbra{k}_{X_j} \otimes \sigma^{(j)}_k$.

\item \textbf{Measurements:} 
At time $t^B_j$, Bob receives some quantum register from Eve and measures it using some POVM, storing the outcome in a register \(Y_j\). 

\item \label{step:announce} \textbf{Public announcement:} At time \(t_j^\mathrm{ann}\), Alice and Bob commence announcements for round $j$, based only on the values $X_j Y_j$. We denote all such public information as a classical register $\CP_j$.  Formally, $\CP_j$ can be viewed as the output of a CPTP map acting on the registers $X_j Y_j$.

\item \label{step:sift} \textbf{Sifting and key map:} 
At any time after \(t_j^\mathrm{ann}\), Alice computes a classical value \(S_j\) based only on her raw data \(X_j\) and the public announcements \(\CP_j\). This value \(S_j\) will later be used to generate her final key via privacy amplification. 

\end{enumerate}
\item  
Alice and Bob apply various further classical procedures such as discarding values from $S_1^n$, variable-length decision, error correction, error verification, and privacy amplification as described in \nameref{prot:abstractqkdprotocol}.
\end{enumerate}
\end{prot}

\begin{remark} \label{remark:justifytiming}
Note that these conditions on the timings are extremely minimal. The first condition states that the announcements in round $j$ only happen after Alice has sent the state in that round and Bob has measured it. This is enforced in our protocol by design: Alice only starts announcing after she has received Bob's announcement, which in turn only occurs after Bob has finished measuring.  Moreover, under the honest authentication assumption, Eve cannot cause Alice to announce prematurely by impersonating Bob. The second condition is simply a basic time-ordering condition on the sequences of preparations, measurements, and announcements \emph{individually}, without otherwise constraining them with respect to each other. These can be enforced by each party locally. In particular, these conditions do \emph{not} limit the repetition rate as in the existing generalized EAT-based proofs \cite{metger_security_2023}, which require the protocol to ensure that Eve has access to only one signal at a time (i.e, Alice can only send the state for the next round after Bob has measured the current round).
\end{remark}

For the security proof, as we will see later, we are interested in analyzing the state at the end of Step 1 in the \nameref{prot:pmprotocoll}. However, the structure of the above protocol is not yet compatible with the MEAT. Hence we first perform some modifications to the above protocol, which modify the timings $t^\mathrm{A}_j,t^\mathrm{B}_j, t^\mathrm{ann}_j$ of various steps to new values  $\tilde{t}^\mathrm{A}_j,\tilde{t}^\mathrm{B}_j, \tilde{t}^\mathrm{ann}_j$ satisfying:
\begin{equation} \label{eq:newtimes}
    \tilde{t}_1^\mathrm{A} = \dots = \tilde{t}^\mathrm{A}_n < \tilde{t}^\mathrm{B}_1 < \tilde{t}^\mathrm{ann}_1 < \tilde{t}^\mathrm{B}_2 < \tilde{t}^\mathrm{ann}_2 < \dots < \tilde{t}^\mathrm{ann}_n.
\end{equation}

To perform the above modification, we 
\begin{itemize}
\item Do not modify Bob's measurement timings, i.e, we set $\tilde{t}_j^\mathrm{B} = t^\mathrm{B}_j$.
    \item Move forward all of Alice's preparations such that they happen before Bob's first measurement.
    \item Move forward all announcements for round $j$ such that they happen right after Bob's measurement for that round, i.e, $\tilde{t}^\mathrm{B}_j < \tilde{t}^\mathrm{ann}_j < \tilde{t}^\mathrm{B}_{j+1}$. In this step, we ensure that \textit{all} announcements for round $j$ happen at $\tilde{t}^\mathrm{ann}_j$, 
\end{itemize}
Thus, in the modified protocol, Alice prepares and sends out all states at time $\tilde{t}^\mathrm{A}_1=\tilde{t}^\mathrm{A}_n$. Eve is allowed to perform any operation she wants on these states. After all states have been sent, Bob begins receiving and measuring states. After he measures the first state, Alice and Bob perform all announcements for round $j$. Bob then measures the second state and so on.

We stress that we have not lost any generality in considering the modified protocol. This is because moving the timings of Alice's preparations and public announcements to an earlier time does not reduce Eve's capabilities in any way: any action that Eve could have done in \nameref{prot:pmprotocoll} is also possible in the modified protocol, since at any point she always has access to more registers in the modified protocol,  and she is free to ignore them until the time she would have interacted with them in the original protocol. This argument lets us write the following lemma.

\begin{lemma}[Modified Protocol] \label{lemma:modifiedprotocol}
Consider \nameref{prot:abstractqkdprotocol} with timings $t^\mathrm{A}_j,t^\mathrm{B}_j,t^\mathrm{ann}_j$ such that $t^\mathrm{A}_j,t^\mathrm{B}_j < t^\mathrm{ann}_j$, and  each of the sequences $\{t^A_j\}_{j=1}^n$, $\{t^B_j\}_{j=1}^n$, and $\{t^{\mathrm{ann}}_j\}_{j=1}^n$ is monotonically increasing. Consider the modified protocol that is identical to \nameref{prot:abstractqkdprotocol} except that the timings $\tilde{t}^\mathrm{A}_j,\tilde{t}^\mathrm{B}_j,\tilde{t}^\mathrm{ann}_j$ satisfy \cref{eq:newtimes}. Any Alice-Bob-Eve state that can be obtained in the original protocol, can also be obtained in the modified protocol.
\end{lemma}
\begin{proof}
    As described above.
\end{proof}
Therefore, for the purposes of proving security, we can now focus on analyzing the modified version. In the modified version, Alice prepares and sends out all states before Bob's first measurement, i.e, she prepares the global state 
\begin{equation} \label{eq:aliceprepearesstatesflobal}
\begin{aligned}
 \sigma_{X_1^n (A')_1^n} &=   \bigotimes_{j=1}^n \sigma^{(j)}_{X_j A'_j}, \\
   \sigma^{(j)}_{X_j A'_j} &= \sum_{k=1}^{d_A} p^{(j)}_k \ketbra{k}_{X_j} \otimes (\sigma^{(j)}_{k})_{A'_j},
\end{aligned}
\end{equation}
 before sending the $(A')_1^n$ systems to Eve. The evolution of the state in the modified protocol can now be described using the following lemma. Note that it is this protocol with modified timings for which for we prove certain results in later sections. For this reason, we also formalize the notion of this protocol in terms of its internal maps and states (\cref{def:PMQKDMEAT}), and we express the relevant security requirement directly in terms of the maps that arise in the state evolution (\cref{def:qkdsecurityChannelVersionMEAT}).

 \begin{lemma}[State evolution for modified protocol] \label{lemma:stateevolutionmodifiedprotocol}
Consider the \nameref{prot:abstractqkdprotocol} with timings $t^\mathrm{A}_j,t^\mathrm{B}_j,t^\mathrm{ann}_j$ such that $t^\mathrm{A}_j,t^\mathrm{B}_j < t^\mathrm{ann}_j$, and  each of the sequences $\{t^A_j\}_{j=1}^n$, $\{t^B_j\}_{j=1}^n$, and $\{t^{\mathrm{ann}}_j\}_{j=1}^n$ is monotonically increasing. Consider the modified protocol that is identical to the \nameref{prot:abstractqkdprotocol} except that the timings $\tilde{t}^\mathrm{A}_j,\tilde{t}^\mathrm{B}_j,\tilde{t}^\mathrm{ann}_j$ satisfy \cref{eq:newtimes}. Then, the evolution of the state through the modified protocol can be described as follows:
\begin{enumerate}
\item Alice prepares the global state $ \sigma_{X_1^n (A')_1^n}$  as in \cref{eq:aliceprepearesstatesflobal}. She sends $(A')_1^n$ to Eve. Eve thus starts with $E_0 = (A')_1^n$.\footnote{Of course, Eve may always start with additional auxiliary systems. However, since these systems are completely uncorrelated (i.e., in a trivial tensor-product form) with all the other registers in the QKD protocol, they can simply be added by Eve whenever needed, for instance when she implements her first attack channel $\attack{1}$.}

She keeps $X_1^n$ to herself, and sends the $(A')_1^n$ systems to Eve, which we relabel to be $E_0$.   
\item For each round $j \in \{1,\dots,n\}$, the following maps are applied:
\begin{enumerate}
    \item In each round $j$, Eve implements her attack $\attack{j} \in \CPTP(E_{j-1},B_j E^\prime_j)$.  \item The $B_j$ system is forwarded to Bob. Bob performs measurements on the $B_j$ system, and stores the outcome in the $Y_j$ register (Alice already has the $X_j$ register). They perform public announcements for the round $j$, which are stored in the register $\CP_j$. A copy  of the public announcements is given to Eve in the register $\CPhat_j$. Alice and Bob perform sifting to generate the $S_j$ registers. This entire process can be described by a map $\QKDGmapfullbeforeSR{j} \in \CPTP(X_j B_j , S_j  X_j Y_j \CP_j \CPhat_j)$.
    \item The $E'_j$ register and the $\CPhat_j$ register is combined into a unified $E_j$ register, which is passed into Eve's attack channel for the next round $\attack{j+1}$.\footnote{When $j=n$ is the last round, there is no $\attack{j+1}$ that is applied.}
\end{enumerate}
Thus, the state after the application of these maps is given by
\begin{equation}
      \rho_{S_1^n X_1^n Y_1^n \CP_1^n E_n } = \left( \QKDGmapfullbeforeSR{n} \circ \attack{n} \dots \QKDGmapfullbeforeSR{1} \circ \attack{1} \right) \left[ \sigma_{X_1^n (A')_1^n} \right] 
\end{equation}
\item Alice performs the final sifting step, generating the register $\PAstring$ from $S_1^n$ and $\CP_1^n$.
\item Alice and Bob perform steps such as the variable-length decision, error-correction, error-verification and privacy amplification. The final sifting operation, and the above procedures are all classical operations that are described by a map $\QKDpostprocessingmap\in \CPTP(X_1^n Y_1^n S_1^n \CP_1^n, K_A K_B \flagkey \flagEC \CEC \HEV \flagEV \HPA)$.
\end{enumerate}
\end{lemma}
\begin{proof}
    The proof follows from the description of the \nameref{prot:abstractqkdprotocol} with the modified timings. 
\end{proof}
 At this stage, it is convenient to use work with the following equivalent definitions of QKD protocols and security. These are generalized versions of \cref{def:epsSecPromisechannelversion,def:equivalentQKDprotocolbeforeSR}, since they allow the protocol to have on-the-fly announcements.

\begin{definition}\label{def:PMQKDMEAT}
   We use $\left\{\{\QKDGmapfullbeforeSR{j}\}_{j=1}^n,\QKDpostprocessingmap, \sigma_{X_1^n (A')_1^n} \right\}$ to define an instance of the time-modified \nameref{prot:abstractqkdprotocol}, where $\QKDGmapfullbeforeSR{j},\QKDpostprocessingmap,\sigma_{X_1^n (A')_1^n}$ are defined in \cref{lemma:stateevolutionmodifiedprotocol}.
\end{definition} 
We can also restate the security definition \cref{def:qkdsecuritysymmetric} in terms of this sequence of channels.
\begin{definition}[QKD security as sequence of channels] \label{def:qkdsecurityChannelVersionMEAT}
Let $\left\{\{\QKDGmapfullbeforeSR{j}\}_{j=1}^n,\QKDpostprocessingmap, \sigma_{X_1^n (A')_1^n} \right\}$ define a QKD protocol that is described through the series of maps from \cref{lemma:stateevolutionmodifiedprotocol}. Let $\idealmap$ be the map that produces the ideal output state as defined in \cref{sec:securitydefinition}.
Then, the QKD protocol is $\epssecure$-secure if, for all Eve's attack channels $\attack{j}\in \CPTP(E_{j-1},B_jE'_j)$, the following inequality is satisfied:
\begin{equation}
\begin{aligned}
  &   \left\|
        \left(
            \QKDpostprocessingmap \circ \QKDGmapfullbeforeSR{n} \circ \attack{n} \circ \cdots \circ \QKDGmapfullbeforeSR{1}\circ\attack{1}
            - \idealmap \circ \QKDpostprocessingmap \circ \QKDGmapfullbeforeSR{n} \circ \attack{n} \circ \cdots \circ \QKDGmapfullbeforeSR{1}\circ\attack{1}
        \right)
        \left[\sigma_{X_1^n (A'')_1^n}\right]
    \right\|_1 \\
    &\leq \epssecure.
\end{aligned}
\end{equation}
Here, without loss of generality, we can assume that $E_j,E'_j = \mathcal{H}_{\infty}$ is infinite-dimensional; embedding any smaller spaces in $\mathcal{H}_{\infty}$ if necessary.
\end{definition} 

\subsubsection{Applying the source-replacement scheme}

 In the above description, the starting state is one where Alice holds the classical register $X_1^n$  storing the labels for the states she sent in the protocol. This state then goes through the sequence of channels described in the above lemma. While the MEAT can be be applied to this sequence as well, doing so yields trivial results. Intuitively, this is because the application of the MEAT results in a single-round optimization that involves granting Eve access to an arbitrary purification of the system under Alice's control, which is a common feature in QKD security analysis (which we saw in earlier chapters). In this scenario, that would allow Eve to hold a purifying system for $X_j$ which is simply a copy of $X_j$. This issue is straightforwardly remedied by the use of the source-replacement scheme \cite{bennett_quantum_1992,curty_entanglement_2004}, which equivalently describes Alice's initial operation as preparing an entangled states followed by performing measurements on it. This is obtained in the following lemma. Note that this is again a generalized version of \cref{lemma:backgroundsourcereplacementandshield} that accounts for on-the-fly announcements.

\begin{lemma}[Shield System and Source Replacement] \label{lemma:shieldandsourcereplacement}
Consider a protocol where Alice first prepares the global state
\begin{equation}\label{eq:AlicePreMeasuredClassicalState}
\begin{aligned}
 \sigma_{X_1^n (A')_1^n} &=   \bigotimes_{j=1}^n \sigma^{(j)}_{X_j A'_j} \\
   \sigma^{(j)}_{X_j A'_j} &= \sum_{k=1}^{d_A} p^{(j)}_k \ketbra{k}_{X_j} \otimes (\sigma^{(j)}_{k})_{A'_j} &
\end{aligned}
\end{equation}
where $\{\ket{k}\}_{X_j}$ denotes the 
classical basis of 
$X_j$. She keeps $X_1^n$ to herself, and sends the $(A')_1^n$ systems to Eve.  
Then, the state $\sigma_{X_1^n (A')_1^n}$ can equivalently be obtained by
\begin{enumerate}
    \item Preparing the global state
   \begin{equation} \label{eq:globalsourcereplacedstate}
       \begin{aligned}
      \sourcesymbol_{\Ameas_1^n \Ashield_1^n (A')_1^n} &=\bigotimes_{j=1}^n \sourcesymbol^{(j)}_{\Ameas_j \Ashield_j A'_j}\\
          \sourcesymbol^{(j)}_{\Ameas_j \Ashield_j A'_j} &=\sum_ {k,i=1}^{d_A}\sqrt{p^{(j)}_ip^{(j)}_k} \ketbra{i}{k}_{\Ameas} \otimes \ketbra{\sigma^{(j)}_{i}}{\sigma^{(j)}_k}_{\Ashield_j A'_j}
       \end{aligned}
   \end{equation}
  where $\ket{\sigma_k^{(j)}}_{\Ashield_j A'_j}$ denotes a purification of the state $(\sourcesymbol_k^{(j)})_{A'_j}$, $\{\ket{k}\}_{k =1,\dots,d_A}$ is an orthornomal basis for $\Ameas$, and $\Ashield_j$ is referred to as the shield system.
   \item Sending the state $(A')_1^n$ to Eve,
   \item For each $j \in \{1,\dots, n\}$, measuring the $\Ameas_j \Ashield_j$ systems using the POVM $\{\ketbra{k}_{\Ameas_j}\otimes\mathbb{I}_{\Ashield_j}\}_{k\in\{1,\dots,d_A\}}$, and storing the result in $X_j$.
\end{enumerate}
\end{lemma}
\begin{proof}
For any round $j$, first consider the single round states $\sigma_{X_j A'_j}$ and $\sourcesymbol_{\Ameas_j \Ashield_j A'_j}$ as defined in \cref{eq:AlicePreMeasuredClassicalState,eq:globalsourcereplacedstate}. Simple algebra tell us that measuring $\sourcesymbol_{\Ameas_j \Ashield_j A'_j}$ with the POVM $\{\ketbra{k}_{\Ameas_j}\otimes\mathbb{I}_{\Ashield_j}\}_{k\in\{1,\dots,d_A\}}$, and storing the result in $X_j$ would result exactly in $\sigma_{X_j A'_j}$.
The $n$-round case follows similarly by considering the $n$-round states and $n$-round IID POVM instead.
\end{proof}

Note that  $\ket{\sigma_k^{(j)}}_{\Ashield_j A'_j}$ can be \emph{any} purification of $(\sigma_k^{(j)})_{A'_j }$. Crucially, we now observe that after the usage of the source replacement scheme to describe signal preparation using \cref{lemma:shieldandsourcereplacement}, Alice's operations are now equivalent to her preparing a global pure state $\sourcesymbol_{\Ameas_1^n \Ashield_1^n (A')_1^n}$, and then performing a suitable measurement on the $\Ameas_j$ registers to produce classical data in the $X_j$ systems. This has the critical property that Alice's measurements on the $\Ameas_j$ systems commute with all other operations that do not depend on the $X_j$ registers. In particular, Alice's measurements commutes with Eve's attack, and thus we can delay her measurements and assume that she does her measurements at the same time as Bob. This allows us to focus our analysis to that of the following virtual entanglement-based protocol, which produces the \textit{same exact} final state as the modified protocol from \cref{lemma:modifiedprotocol} (or the evolution of the state described in \cref{lemma:stateevolutionmodifiedprotocol}). This will be the protocol we analyze in subsequent sections. A security proof for this protocol implies the security proof of \nameref{prot:abstractqkdprotocol} via \cref{lemma:shieldandsourcereplacement,lemma:modifiedprotocol,lemma:stateevolutionmodifiedprotocol}.

\begin{remark} \label{remark:timecommutealice}
    Note that in the arguments above, we modify the timings of Alice's operations \emph{twice} - first by moving them earlier in time, and then by moving some of them later in time. Specifically, we initially argue that without loss of generality, one can assume that all state preparations are performed at the start of the protocol. This modification gives more power to Eve and thus only strengthens our security claims. Next, we apply the source-replacement technique, replacing Alice's state sending operations with the preparation of a global entangled state, part of which is sent to Eve ($(A')_1^n$), part of which is kept with Alice but not measured  ($\Ashield_1^n$), and part of which is kept with Alice and measured ($\Ameas_1^n$). Finally, we argue that since Alice's measurements commute with all operations except the public announcements (which depend on $X_1^n$), those measurements can be postponed to  after Eve's attack and just before public announcements for that round, without affecting Eve's strategy. Note that Alice's global state preparation still occurs at the initial time.
\end{remark}


\newcommand{\MyTikzFigure}{

\begin{tikzpicture}[>=Stealth,auto,
  box/.style  ={draw,minimum width=1cm,minimum height=1cm,
                align=center,fill=red!20},
  cpbox/.style ={draw,minimum width=1.4cm,minimum height=0.8cm,
                align=center,fill=gray!20},
  sig/.style  ={->,rounded corners=4pt},
  merge/.style={circle,fill,inner sep=1.2pt}
]


\node[box]   (N1) {$\attack{1}$};
\node[cpbox, below right=\offsetY and \offsetX of N1] (L1) {$\QKDGmap{1}$};

\node[box,   right=\colsep of N1] (N2) {$\attack{2}$};
\node[cpbox, below right=\offsetY and \offsetX of N2] (L2) {$\QKDGmap{2}$};

\node[right=\colsep of N2] (dots) {$\cdots$};
\node[box,   right= of dots] (Nn) {$\attack{n}$};
\node[cpbox, below right=\offsetY and \offsetX of Nn] (Ln) {$\QKDGmap{n}$};

\node[left =1.4cm of N1] (E0) {};
\node[right=4cm   of Nn] (En) {};

\coordinate (join1) at ($(L1.east |- N1) + (1.0,0)$);
\coordinate (join2) at ($(L2.east |- N2) + (1.0,0)$);
\coordinate (joinn) at ($(Ln.east |- Nn) + (1.0,0)$);

\draw[sig] (E0) node[above=2pt, xshift=10pt]{$E_{0}=(A')_1^n$} -- (N1.west);

\draw[sig] (N1.east) node[above=2pt, xshift=10pt] {$E'_1$} -- (join1);
\draw[sig] (L1.east) node[above=2pt, xshift=10pt]   {$\CPhat_{1}$}
           -- ++(0.6,0) |- (join1);
\node[merge] at (join1) {};
\draw[sig] (join1) -- node[above=2pt,xshift=1pt] {$E_{1}$} (N2.west);

\draw[sig] (N2.east) node[above=2pt, xshift=10pt] {$E'_2$} -- (join2);
\draw[sig] (L2.east) node[above=2pt, xshift=10pt]   {$\CPhat_{2}$}
           -- ++(0.6,0) |- (join2);
\node[merge] at (join2) {};
\draw[sig] (join2) -- node[above=2pt,xshift=1pt] {$E_{2}$} (dots.west);


\draw[sig] (Nn.east) node[above=2pt, xshift=10pt] {$E'_n$} -- (joinn);
\draw[sig] (Ln.east) node[above=2pt, xshift=10pt]   {$\CPhat_{n}$}
           -- ++(0.6,0) |- (joinn);
\node[merge] at (joinn) {};

\draw[sig] (dots.east) node[above=2pt,xshift = 3pt] {$E_{n-1}$} -- (Nn.west);
\draw[sig] (joinn) -- node[above=2pt,xshift=1pt] {$E_{n}$} (En);

\foreach \i/\N/\L in {1/N1/L1, 2/N2/L2, n/Nn/Ln}{
  \draw[sig] (\N.south) -- ++(0,-0.6)
             |- node[above right=3pt] {$B_{\i}$} (\L.west);
}

\foreach \i/\L in {1/L1, 2/L2, n/Ln}{
  \node[below=1.5cm of \L] (S\i) {$S_{\i}\widehat{C}_{\i}$};
  \draw[sig] (\L.south) -- (S\i);

  \node[above=2.5cm of \L] (A\i) {$A_{\i}$};
  \draw[sig,<-] (\L.north) -- (A\i);
}

\draw[red,thick,rounded corners]
  ($ (N1.north west) + (-3pt,3pt)$)  
  rectangle
  ($ (join1 |- L1.south) + (3pt,-6pt)$);   
\node[text=red] at ($($(N1.north west) + (0pt, 10pt)$)$) {\(\mathcal{M}_1\)};
\draw[red,thick,rounded corners]
  ($ (N2.north west) + (-3pt,3pt)$)
  rectangle
  ($ (join2 |- L2.south) + (3pt,-6pt)$);
\node[text=red] at ($($(N2.north west) + (0pt, 10pt)$)$) {\(\mathcal{M}_2\)};
\draw[red,thick,rounded corners]
  ($ (Nn.north west) + (-3pt,3pt)$)
  rectangle
  ($ (joinn |- Ln.south) + (3pt,-6pt)$);
\node[text=red] at ($($(Nn.north west) + (0pt, 10pt)$)$) {\(\mathcal{M}_n\)};
\end{tikzpicture}

}

\begin{figure}[ht]
    \centering
    \hspace*{-4cm}
 \makebox[\linewidth][l]{%
        \scalebox{1}{\MyTikzFigure}%
    }
    \caption{The evolution of the state through Eve's attack channels $\{ \attack{j} \}$, and Alice and Bob's operations $\{\QKDGmap{j}\}$ for the \nameref{prot:virtualprotocoll}. The evolution of states is also described in \cref{lemma:stateevolution}. Note that the announcements $\CP_j$ are made available to Eve through an explicit copy $\CPhat_j$, which gets merged with $E'_j$ to form $E_j$.}
    \label{fig:MEAT_Attack}
\end{figure}

\begin{prot}[Virtual entanglement-based protocol].\label{prot:virtualprotocoll} 

\noindent \textbf{Parameters:}

\begin{tabularx}{0.9\linewidth}{r c X}
\(n \in \mathbb{N}_0\) 			    &:& 	Total number of rounds \\
\(\{\sourcesymbol^{(j)}_{\Ameas_j \Ashield_jA'_j}\}_{j=1}^n\) 	&:& 	Pure quantum states prepared by Alice \\
\(\{\tilde{t}_j^A=\tilde{t}_j^B,\tilde{t}^{\mathrm{ann}}_j\}_{j=1}^n\) 						&:& 	Timings of various steps (see below), satisfying $\tilde{t}^\mathrm{B}_1 < \tilde{t}^\mathrm{ann}_1 < \tilde{t}^\mathrm{B}_2 < \tilde{t}^\mathrm{ann}_2 < \dots < \tilde{t}^\mathrm{ann}_n.$
\end{tabularx}

\noindent \textbf{Protocol steps:}
\begin{enumerate}

\item \textbf{State preparation and transmission:} Alice prepares a global pure state $\sourcesymbol_{\Ameas_1^n \Ashield_1^n (A')_1^n} = \bigotimes_{j=1}^n \sourcesymbol^{(j)}_{\Ameas_j \Ashield_j A'_j}$. She sends out all of the $(A')_1^n$ registers to Eve. She keeps $\Ameas_1^n \Ashield_1^n$ with herself. 
\item Alice and Bob perform the following steps for every $j \in \{1,2,\dots,n\}$. \label{step:singleroundsvirtual}
\begin{enumerate}
\item \textbf{Measurements:} \label{step:announcevirtual} At time $\tilde{t}_j^\mathrm{B}$, Bob receives quantum register $B_j$ from Eve, and performs a measurement on it, storing the outcome in the register $Y_j$. At the same time\footnote{Since this is a purely virtual protocol used only within the security proof, there is no need for Alice and Bob to physically implement these operations simultaneously.}
, Alice performs a measurement in the computational basis on the $\Ameas_j$ system, and stores the measurement outcome in the $X_j$ register. 
\item \textbf{Public Announcements:} At time $\tilde{t}_j^\mathrm{ann}$, Alice and Bob perform some public announcements which may be interactive. We denote all public announcements for round $j$ with the register $\CP_j$. A copy of these announcements in the register $\CPhat_{j}$ is made available to Eve. 
\item \textbf{Sifting and key map:}\label{step:siftvirtual} At time $\tilde{t}_j^\mathrm{ann}$, Alice maps her private data $X_j$ and public announcements $\CP_j$ to a classical value $S_j$.
\end{enumerate}

\item 
Alice and Bob apply various further classical procedures such as variable-length decision, error correction, error certification, and privacy amplification as described in \nameref{prot:abstractqkdprotocol}.
\end{enumerate}
\end{prot}	

It is convenient to view the above \nameref{prot:virtualprotocoll} as a sequence of maps, depicted in \cref{fig:MEAT_Attack} and described in the following \cref{lemma:stateevolution}.

\begin{lemma}[Describing evolution of states for QKD protocol from \nameref{prot:virtualprotocoll}] \label{lemma:stateevolution}
The evolution of the state in the QKD protocol can be treated as occurring as follows:
\begin{enumerate}
    \item Alice prepares the state $\sourcesymbol_{\Ameas_1^n \Ashield_1^n (A')_1^n}$, and sends ${A'}_1^n$ to Eve. We treat this as $E_0$.  
    \item For each round $j$ from $1$ to $n$, the following maps are applied:
    \begin{enumerate}
    \item In each round $j$, Eve implements her attack $\attack{j} \in \CPTP(E_{j-1}, B_j E^\prime_j)$.
    \item The $B_j$ system is forwarded to Bob. Alice and Bob then perform measurements on the $\Ameas_j$ and $B_j$ systems, and store them in registers $X_j$ and $Y_j$, and perform public announcements $\CP_j$. A copy $\CPhat_j$ of the announcements $\CP_j$ is given to Eve. They perform sifting to generate the $S_j$ registers. This entire process can be described by a map $\QKDGmapfull{j} \in \CPTP(A_j B_j , S_j X_j Y_j \CP_j \CPhat_j)$, where $A_j = \Ameas_j \Ashield_j$ denotes the registers that do not leave Alice's lab. We let $\QKDGmap{j} \in \CPTP(A_j B_j , S_j \CP_j \CPhat{j})$ denote the map $\QKDGmap{j} \coloneq \Tr_{X_j Y_j} \circ \QKDGmapfull{j}$, i.e.~$\QKDGmapfull{j}$ with its output registers restricted to the secret register and public announcements. 
    \item The $E^\prime_j$ register and the $\CPhat_j$ register is combined into a unified $E_j$ register, which is passed into Eve's attack channel for the next round $\attack{j+1}$.\footnote{When $j=n$, there is no $\attack{n+1}$ that is applied.}
    \end{enumerate}
    Thus, the state after the application of these maps is given by
    \begin{equation}\label{eq:stateevolved}
    \begin{aligned}
        \rho_{S_1^n X_1^n Y_1^n \CP_1^n E_n } &= \left( \QKDGmapfull{n} \circ \attack{n} \dots \QKDGmapfull{1} \circ \attack{1} \right) \left[  \sourcesymbol_{\Ameas_1^n (\Ashield)_1^n (A')_1^n} \right] \\
         \rho_{S_1^n \CP_1^n E_n } &= \left( \QKDGmap{n} \circ \attack{n} \dots \QKDGmap{1} \circ \attack{1} \right) \left[  \sourcesymbol_{\Ameas_1^n (\Ashield)_1^n (A')_1^n} \right].
         \end{aligned}       
    \end{equation}
    Furthermore, we have $\QKDGmapfull{j} = \QKDGmapfullbeforeSR{j} \circ \measChannel{\{\ketbra{k}_{\Ameas_j}  \otimes \id_{\Ashield} \}}$, where $\measChannel{\{\ketbra{k}_{\Ameas_j}\otimes \id_{\Ashield}\}}  \in \CPTP(\Ameas_j \Ashield_j, X_j)$ is a channel that measures the $\Ameas_j$ system using POVM $\{\ketbra{k}_{\Ameas_j} \otimes \id_{\Ashield} \}_k$ and stores the outcome in the $X_j$ system (see \cref{def:measurementchannels}), and  
    $\QKDGmapfullbeforeSR{j}$ does all remaining operations except the initial measurement by Alice, and is the same as in \cref{lemma:stateevolutionmodifiedprotocol}.
    \item Alice performs the final sifting step, generating the register $\PAstring$ from $S_1^n$ and $\CP_1^n$.
    \item Alice and Bob perform steps such as the variable-length decision, error-correction, error-verification and privacy amplification. The final sifting operation, and the above procedures are together described by a map \\ $\QKDpostprocessingmap \in \CPTP(X_1^n Y_1^n S_1^n \CP_1^n, K_A K_B \flagkey \flagEC \CEC \HEV \flagEV \HPA)$.
\end{enumerate}
\end{lemma}
\begin{proof}
The proof follows directly from the description of \nameref{prot:virtualprotocoll}. Recall that this protocol is obtained from \nameref{prot:pmprotocoll} (with the modified timings specified in \cref{lemma:modifiedprotocol}) by replacing the state preparation step through the source-replacement scheme (\cref{lemma:shieldandsourcereplacement}). The state evolution of \nameref{prot:pmprotocoll} with the modified timings is given in \cref{lemma:stateevolutionmodifiedprotocol}. The required state evolution for \nameref{prot:virtualprotocoll} is then obtained by applying the same modification (arising from the source-replacement scheme) to the state evolution from \cref{lemma:stateevolutionmodifiedprotocol}.
    \end{proof}

    \subsection{Dimensions of Eve's systems} \label{subsec:evedimensions}
Our goal is to prove that the security statement holds for \emph{all} possible attacks by Eve. Concretely, this requires showing that the security definition (\cref{def:qkdsecurityChannelVersionMEAT}) is satisfied for \textit{every} possible attack channel $\attack{j}$. A subtlety arises here: since  $\attack{j} \in \CPTP(E_{j-1}, B_j E'_j)$, our analysis must also account for the \emph{dimensions} of the auxiliary spaces $E'_j$ that Eve may introduce. In principle, these spaces could even be infinite-dimensional. Moreover, Alice’s state preparation ($\Ashield_j, A'_j$) and Bob’s measurements ($B_j$) may themselves involve infinite-dimensional systems. However, the MEAT theorem is formulated only under the assumption that all registers are finite-dimensional. To reconcile this mismatch, we proceed as follows:

\begin{enumerate}

\item \textbf{Dimensions of Alice and Bob:} We first assume that Alice prepares finite-dimensional states and Bob performs finite-dimensional measurements. The proof in this section is restricted to this finite-dimensional Alice and Bob setting. Later, in \cref{sec:optics}, we show how these restrictions can be lifted by introducing appropriate source maps and squashing maps. With this, the only remaining issue is the dimensionality of Eve’s systems.

  \item \textbf{Security against arbitrary but finite-dimensional Eve:} Next, we fix an arbitrary finite dimension $d_j$ for each $E'_j$. For this setting, we consider a sequence of attack channels $\{\attack{j}\}_j$ with $\attack{j} \in \CPTP(E'_{j-1} \CPhat_{j-1}, E'_j B_j)$ and establish security in a way that does not depend on the particular choice of $\{\attack{j}\}_j$. Critically, the resulting security statement we obtain is independent of the values of $d_j$. Hence, security holds against all attacks where Eve’s systems are finite-dimensional.

\item \textbf{Infinite-dimensional Eve: } Finally, we extend the argument to infinite-dimensional systems. Intuitively, since our security statement holds for arbitrary finite dimensions $d_j$ of Eve's systems, one can take the limit $d_j \rightarrow \infty$ to recover the result when Eve uses infinite-dimensional systems. This limiting argument is made rigorous in \cite[Appendix A]{inprep_BDR3}, and is not included in this thesis.
\end{enumerate}
 
Thus, in the following subsections, we fix a finite dimension $d_j$ for each $E'_j$. Note that this issue of specifying Eve’s dimension is primarily a technical nuisance rather than a conceptual obstacle. Intuitively, in a protocol where state preparation, measurements, and announcements are all finite-dimensional, Eve cannot do better than implement some unitary operation in each round as part of her attack. In such a setting, she only ever requires finite-dimensional registers. For a fully rigorous analysis, one must formalize a suitable version of this intuition \cite[Appendix A]{inprep_BDR3}.

\begin{remark} \label{remark:infinitedimensions}
Let us focus on the infinite-dimensional aspects of Eve. Note that this issue does not arise when using the post-selection technique (\cref{chap:postselection}): because that approach has an explicit dimension dependence, one must first reduce the protocol itself to a finite-dimensional Alice-Bob setting. After this reduction, Eve’s attack can be taken to consist of holding a purification of a finite-dimensional state, which can (without loss of generality) be assumed to be finite-dimensional as well.

In contrast, the EUR approach in \cref{chap:EUR} has no explicit dimension dependence. In that setting, the finite dimensional Alice-Bob setting results in Eve holding some purification of this state, and security is straightforward. However, since there is no dimension dependence, our analysis in \cref{chap:EUR} is applied without any squashing operation, where Bob measures infinite-dimensional systems. Thus, it invokes several entropy statements that are technically stated only for finite-dimensional systems, but applied them to infinite-dimensional systems. We note that this is a common technical gap in many analyses, where statements  proved for arbitrary finite-dimensional systems without explicit dimensional dependence are applied to infinite-dimensional systems. This discrepancy can be resolved using the approach from \cite[Appendix~A]{inprep_BDR3}
\end{remark}

\subsection{Security proof for \nameref{prot:abstractqkdprotocol} (finite-dimensional case)}
\label{subsec:proofgenericstuff} 
Recall that we are in the setting where Alice prepares finite-dimensional states, Bob performs finite-dimensional measurements and each $E'_j$ has dimension $d_j$. We start by fixing the set of attack channels $\{\attack{j}\}_j$, and consider the fixed state obtained corresponding to this attack. As usual, we first use \cref{lemma:securityfromcorrandsecrecy} to break up the security requirement into secrecy and correctness. The fact that the protocol is $\epscorr$-correct follows in exactly the same manner as the proof of \cref{lemma:correctnessissatisfied}. Thus, we focus only on proving secrecy. 

Let us first consider the event $\Omega(\cobs)$, defined as the event that the public announcements $\CP_1^n$ took the value $\cobs$. Observe that in the protocol, the privacy amplification step is performed conditioned on the value $\cobs$ (and conditioned on error verification accepting), i.e.~it is applied on the state $\rho_{\PAstring  \CEC \CEV \flagkey \flagEC \flagEV \HEV E_n | \Omega(\cobs) \wedge \OmegaEV}$. Hence we would be interested in lower bounding the {\Renyi} entropy of register $\PAstring$ in this state, which will be used in the application of the Leftover Hashing Lemma (\cref{lemma:LHL}). However, the formulation of MEAT \cite[Theorem 4.1a]{arqand_marginal_2025} that we use later in \cref{theorem:MEATQKDfirst} is best suited in analyzing \textit{sequential} processes, and thus allows us to lower bound the {\Renyi} entropy of register $S_1^n$ of the state   $\rho_{S_1^n  \CEC \CEV \flagkey \flagEC \flagEV \HEV E_n | \Omega(\cobs) \wedge \OmegaEV}$. Since $S_1^n$ and $\PAstring$ can be transformed into one another by using the public announcements $\CP_1^n$, we expect the two entropies to be equal. This can be argued using the following lemma.

\begin{lemma}[Equality of entropies of $S_1^n$ and $\PAstring$] 
\label{lemma:equalityentropy}
For any event $\Omega(\cobs)$, consider the state just after the key map and sifting steps in the QKD protocol (see \cref{lemma:stateevolution}), given by $\rho_{\PAstring S_1^n X_1^n Y_1^n \CP_1^n E_n}$. Then, the following statement holds:
\begin{equation}
\Halpha(\PAstring| E_n)_{\rho_{ | \Omega(\cobs) }} = \Halpha(S_1^n| E_n)_{\rho_{ | \Omega(\cobs) }}.
\end{equation}
Moreover, this equality continues to hold at any later point in the protocol prior to privacy amplification. In particular, 
\begin{equation}
\Halpha(\PAstring|\CEC \CEV \flagkey \flagEC \flagEV  \HEV E_n)_{\rho_{ | \Omega(\cobs) \wedge \OmegaEV}} = \Halpha(S_1^n|\CEC \CEV \flagkey \flagEC \flagEV \HEV E_n)_{\rho_{ | \Omega(\cobs) \wedge \OmegaEV}},
\end{equation}
where we consider a version of the protocol in which the register $S_1^n$ is not deleted after use in the above equation. 
\end{lemma}
\begin{proof}
Recall that the protocol generates $\PAstring$ from $S_1^n$ by applying a deterministic discard rule based on $\CP_1^n$. In particular, $S_1^n$ can be transformed into $\PAstring$ by discarding certain positions (as specified by $\CP_1^n$), and $\PAstring$ can be transformed back into $S_1^n$ by inserting $\singleRoundBot$ symbols at the appropriate locations (again determined by $\CP_1^n$).
Since we are considering states conditioned on specific values of $\CP_1^n$, the two states are related by isometries $V_{\PAstring \rightarrow S_1^n}$ and $V_{S_1^n \rightarrow \PAstring}$ (both of which depend on $\cobs$), such that
   \begin{equation}
\begin{aligned}
   V_{\PAstring \rightarrow S_1^n} \rho_{\PAstring  E_n | \Omega(\cobs) \wedge \OmegaEV}V^\dagger_{\PAstring \rightarrow S_1^n} &= \rho_{S_1^n  E_n | \Omega(\cobs) \wedge \OmegaEV} \\
 V_{S_1^n \rightarrow \PAstring}\rho_{S_1^n  E_n | \Omega(\cobs) \wedge \OmegaEV} V^\dagger_{S_1^n \rightarrow \PAstring} &=\rho_{\PAstring  E_n | \Omega(\cobs) \wedge \OmegaEV}.
    \end{aligned}
\end{equation}
Then, the required statement follows from the fact that {\Renyi} entropies are invariant under isometries on the first system (\cref{lemma:dpinonconditioningregister}).\footnote{Note that we only require the isometry to hold in one direction for this proof to work. } Note that the statement holds throughout the protocol, since the same isometries exist at every stage. 
\end{proof}

We will now state the following theorem, which states that if the $f$-weighted {\Renyi} entropy of the private string $S_1^n$ is positive, then the secrecy requirement can be satisfied by a suitable choice of the $l(\cdot)$ and $\leak(\cdot)$ functions. This suggests that the $f$-weighted {\Renyi} entropy is a particularly useful object of study for the purposes of QKD security analysis. The proof of the following theorem is essentially the same as the one developed in Ref.~\cite{inprep_vanhimbeeck_tight_2024} (reproduced with permission in Ref.~\cite{kamin_renyi_2025}), and that work should be cited as the source whenever possible.

\begin{theorem}[$f$-weighted entropy to variable-length secrecy] \label{theorem:entropytovarlength}
Consider the state just after the key map step in \nameref{prot:virtualprotocoll} that is obtained from the \nameref{prot:abstractqkdprotocol}. 
Consider any $\alpha \in (1,2)$, and let $\fhatfull$ be any tradeoff function on the classical registers $\CP_1^n$ such that
\begin{equation} \label{eq:conditiononentropy}
    \Halpha[\fhatfull](S_1^n | \CP_1^n E_n)_{\rho} \geq 0,
\end{equation}
and let $l(\cobs)$ be given by
\begin{equation} \label{eq:lexpression} 
     l(\cobs) = \max\left\{ 0 , \floor{ \fhatfullQKD(\cobs)  - \leak(\cobs) - \ceil{ \log(\frac{1}{\epsEV})} -  \frac{\alpha}{\alpha-1} \log(\frac{1}{\epsPA})  + 2 }   \right\},
\end{equation}
where $\fhatfullQKD$ ``lower-bounds'' $\fhatfull$, i.e, satisfies
\begin{equation} \label{eq:fqkdcondition} 
    \fhatfullQKD(\cobs) \leq \fhatfull(\cobs) \qquad \forall \cobs.
\end{equation}
Then, the state obtained by performing sifting, error-correction and privacy amplification on this state as described in \nameref{prot:abstractqkdprotocol} is $\epsPA$-secret. 
\end{theorem}
\begin{proof}
Recall that for proving $\epsPA$-secrecy, one has to show that that the distance between the real and the ideal Alice-Eve states is smaller than $\epsPA$, i.e,
\begin{equation}
    \sum_{l} \Pr(\Omega_{\mathrm{len}=l}) \tracedist{\rho_{K_A \CEC \CEV \flagkey \flagEC \flagEV \HPA \HEV E_n | \Omega_{\mathrm{len}=l} }  - \tau^{(l)}_{K_A} \otimes \rho_{ \CEC \CEV \flagkey \flagEC \flagEV \HPA \HEV E_n | \Omega_{\mathrm{len}=l}} } \leq \epsPA
\end{equation}
where we recall that $\tau^{(l_A)}_{K_A} = \sum_{k \in\{0,1\}^{l_A}} \frac{1}{2^{l_A}} \ketbra{k}$ was defined in \cref{eq:taukAkB} to be the ideal output key state. 
The required statement follows from the following series of inequalities (explained in the paragraph later):

\begin{align}
  & \sum_{l=0}^\infty \Pr(\Omega_{\mathrm{len}=l})
  \tracedist{
    \rho_{K_A \CEC \CEV \flagkey \flagEC \flagEV \HPA \HEV E_n |\Omega_{\mathrm{len}=l}}
    - \tau^{(l)}_{K_A} \otimes
    \rho_{\CEC \CEV \flagkey \flagEC \flagEV \HPA \HEV E_n | \Omega_{\mathrm{len}=l}}
  }
  \label{eq:conditiononkeylength:a} \\[0.5em]
  &\leq \sum_{\{\cobs \mid \lkey(\cobs) > 0\}}
  \Pr(\Omega(\cobs) \wedge \OmegaEV) \times   \label{eq:conditiononkeylength:b} \\ 
  & 
  \tracedist{
    \rho_{K_A \CEC \CEV \flagkey \flagEC \flagEV \HPA \HEV E_n | \Omega(\cobs) \wedge \OmegaEV}
    - \tau_{K_A} \otimes
    \rho_{\CEC \CEV \flagkey \flagEC \flagEV \HPA \HEV E_n | \Omega(\cobs) \wedge \OmegaEV}
  }
 \nonumber  \\[0.5em]
  &\leq \sum_{\{\cobs \mid \lkey(\cobs) > 0\}}
  \Pr(\Omega(\cobs) \wedge \OmegaEV)
  2^{\frac{1-\alpha}{\alpha}\left(
    \Halpha(\PAstring | \CEC \CEV \flagEC \flagEV \flagkey \HEV E_n)_{\rho| \Omega(\cobs) \wedge \OmegaEV}
    - \lkey(\cobs) + 2
  \right)}
  \label{eq:conditiononkeylength:c} \\[0.5em]
  &= \sum_{\{\cobs \mid \lkey(\cobs) > 0\}}
  \Pr(\Omega(\cobs) \wedge \OmegaEV)
  2^{\frac{1-\alpha}{\alpha}\left(
    \Halpha(\PAstring | \CEC \CEV \HEV E_n)_{\rho| \Omega(\cobs) \wedge \OmegaEV}
    - \lkey(\cobs) + 2
  \right)}
  \label{eq:conditiononkeylength:d} \\[0.5em]
  &\leq \sum_{\{\cobs \mid \lkey(\cobs) > 0\}}
  \Pr(\Omega(\cobs))
  2^{\frac{1-\alpha}{\alpha}\left(
    \Halpha(\PAstring | \CEC \CEV \HEV E_n)_{\rho|\Omega(\cobs)}
    - \lkey(\cobs) + 2
  \right)}
  \label{eq:conditiononkeylength:e}
\end{align}

Here, \cref{eq:conditiononkeylength:a} just restates the quantity we wish to bound. In \cref{eq:conditiononkeylength:b}, we rewrite the sum over all possible values of  $\cobs$, and use the triangle inequality and the fact that the distance between the real and ideal states is zero when the protocol does not produce a key. Hence, we only need to keep events corresponding to a key of non-zero length being generated in the sum over $\cobs$. 
In \cref{eq:conditiononkeylength:c}, we used the Leftover Hashing Lemma (\cref{lemma:LHL}). In \cref{eq:conditiononkeylength:d}, we used the fact that values of $\flagEC,\flagEV,\flagkey$ are uniquely determined by the event $\Omega(\cobs)$, and therefore can be removed from the conditioning registers without penalty.\footnote{This can be seen formally by observing that when we condition on $\Omega(\cobs) \wedge \OmegaEV$, one can transform between the state with the $\flagEC,\flagEV,\flagkey$ registers and the state without them using a CPTP map in either direction. In both directions, the CPTP map acts only on the conditioning registers. Therefore, \cref{lemma:DPI} applies both ways, yielding the desired equality.} \cref{eq:conditiononkeylength:e} follows by using $\rho_{| \Omega(\cobs) \wedge \OmegaEV} = \rho_{( | \Omega(\cobs)) | \OmegaEV}$, and a direct application of \cref{lemma:conditioning} to get rid of the conditioning on $\OmegaEV$, where we absorb the correction term into the probability before the exponential.  Continuing this chain of inequalities, we get

\begin{align}
&\leq \sum_{\{\cobs \mid \lkey(\cobs) > 0\}}
  \Pr(\Omega(\cobs))
  2^{\frac{1-\alpha}{\alpha}\left(
    \Halpha(\PAstring | \HEV E_n)_{\rho|\Omega(\cobs)}
    - \leak(\cobs)
    - \ceil{\log\!\left(\frac{1}{\epsEV}\right)}
    - \lkey(\cobs) + 2
  \right)}
  \label{eq:conditiononkeylength:f} \\[0.5em]
  &\leq \sum_{\{\cobs \mid \lkey(\cobs) > 0\}}
  \Pr(\Omega(\cobs))
  2^{\frac{1-\alpha}{\alpha}\left(
    \Halpha(\PAstring | E_n)_{\rho|\Omega(\cobs)}
    - \leak(\cobs)
    - \ceil{\log\!\left(\frac{1}{\epsEV}\right)}
    - \lkey(\cobs) + 2
  \right)}
  \label{eq:conditiononkeylength:g} \\[0.5em]
  &\leq \sum_{\{\cobs \mid \lkey(\cobs) > 0\}}
  \Pr(\Omega(\cobs))
  2^{\frac{1-\alpha}{\alpha}\left(
    \Halpha(\PAstring | E_n)_{\rho|\Omega(\cobs)}
    - \fhatfull(\cobs)
    + \frac{\alpha}{\alpha-1}\log\!\left(\frac{1}{\epsPA}\right)
  \right)}
  \label{eq:conditiononkeylength:h} \\[0.5em]
  &\leq \epsPA
  \sum_{\{\cobs \mid \lkey(\cobs) > 0\}}
  \Pr(\Omega(\cobs))
  2^{\frac{1-\alpha}{\alpha}\left(
    \Halpha(S_1^n | E_n)_{\rho|\Omega(\cobs)}
    - \fhatfull(\cobs)
  \right)}
  \label{eq:conditiononkeylength:i} \\[0.5em]
  &\leq \epsPA
  \sum_{\cobs}
  \Pr(\Omega(\cobs))
  2^{\frac{1-\alpha}{\alpha}\left(
    \Halpha(S_1^n | E_n)_{\rho|\Omega(\cobs)}
    - \fhatfull(\cobs)
  \right)}
  \label{eq:conditiononkeylength:j} \\[0.5em]
  &= \epsPA\,
  2^{\frac{1-\alpha}{\alpha}\left(
    \Halpha[\fhatfull](S_1^n | \CP_1^n E_n)_\rho
  \right)}
  \label{eq:conditiononkeylength:k} \\[0.5em]
  &\leq \epsPA.
  \label{eq:conditiononkeylength:l}  
\end{align}
\cref{eq:conditiononkeylength:f} follows by using \cref{lemma:EC_cost} to split off the error-correction and error-verification registers. \cref{eq:conditiononkeylength:g} then follows since we can now remove the $\HEV$ register which stored the hash-choice for error verification, since it is independent of the rest of the registers, without penalty, by using data processing in both directions. \cref{eq:conditiononkeylength:h} follows from the fact that the output key length $\lkey(\cobs)$ satisfies \cref{eq:lexpression,eq:fqkdcondition}. \cref{eq:conditiononkeylength:i} is simple algebra. For \cref{eq:conditiononkeylength:j}, observe that we add more (positive) terms to the sum over $\cobs$, and thus the resulting quantity must be larger. \cref{eq:conditiononkeylength:k} follows from the definition of $f$-weighted {\Renyi} entropy from \cref{def:QES}. \cref{eq:conditiononkeylength:l} follows from $\alpha>1$ and the fact that the $\fhatfull$-weighted {\Renyi} entropy is positive, i.e, \cref{eq:conditiononentropy}. This concludes our proof.    
\end{proof}

Thus, the task reduces to finding a tradeoff function $\fhatfull$ such that
$$
\Halpha[\fhatfull](S_1^n \mid \CP_1^n E_n)_{\rho} \geq 0 .
$$
This is precisely the setting in which the MEAT can be applied, as we show in the next subsection. We will first carry out this analysis for a fixed attack by the adversary, and subsequently remove this restriction by performing a worst-case analysis over all possible attacks.

\begin{remark}
Recall that the above requirement is demanding that
$\fhatfull(\cP_1^n)$ lower bounds $\Halpha(S_1^n | E_n)$ in the
log-mean-exponential sense (see \cref{sec:backgroundinfotheory}).
This is a somewhat weaker requirement than the one imposed in the
variable length security proofs of \cref{chap:variable}, where a statistical estimator
$\bstat(\cP_1^n)$ was required to lower bound
$\Halpha(S_1^n | E_n)$ with high probability.
\end{remark}

\subsection{Security via application of MEAT} \label{subsec:securityfromMEAT}

\begin{figure}[ht]
  \centering
    \hspace*{-4cm}
 \makebox[\linewidth][l]{%
        \scalebox{1}{\MyTikzFigure}%
    }
  \caption{(Same as \cref{fig:MEAT_Attack}). The evolution of the state through Eve's attack channels $\{ \attack{j} \}$, and Alice and Bob's operations $\{\QKDGmap{j}\}$ for the \nameref{prot:virtualprotocoll}. The evolution of states is also described in \cref{lemma:stateevolution}. Note that the announcements $\CP_j$ are made available to Eve through an explicit copy $\CPhat_j$, which gets merged with $E'_j$ to form $E_j$. The MEAT \cite[Theorem 4.1a]{arqand_marginal_2025} is applied for the sequence of channels $\{\mathcal{M}_j\}$.}
  \label{fig:MEAT_Attack_repeated}
\end{figure}

\begin{theorem}[Obtaining $\fhatfull$ satisfying \cref{eq:conditiononentropy}] \label{theorem:MEATQKDfirst}
Let $\rho_{S_1^n \CP_1^n E_n}$ be the state obtained in the QKD protocol, i.e, via \cref{eq:stateevolved} from \cref{lemma:stateevolution} or \cref{fig:MEAT_Attack_repeated}. For each $j$, and every value of $\cP_1^{j-1}$, 
let $f_{|\cP_1^{j-1}}$ be a tradeoff function on the register $\CP_j$, and define $\kappafuncgeneric{f_{|\cP_1^{j-1}} }{ \sigma^{(j)}_{A_j} } { \QKDGmap{j} }{ \attack{j} }$ as:
\begin{equation} \label{eq:kappadefined}
   \kappafuncgeneric{f_{|\cP_1^{j-1}} }{ \sigma^{(j)}_{A_j} } { \QKDGmap{j} }{ \attack{j} } \coloneq \inf_{\nu \in \Sigma_j(\attack{j})} \Halpha[f_{|\cP_1^{j-1}}] (S_j | \CP_j E_j \widetilde{E} )_\nu
\end{equation}
where
\begin{equation} \label{eq:Sigmajdefined}
    \Sigma_j (\attack{j})\coloneq \left\{ \QKDGmap{j} \circ \attack{j} \left( \omega_{A_j E_{j-1} \widetilde{E}} \right) \;\middle|\; \omega_{A_j} = \sigmaconstraint{j}  \right\},
\end{equation}
where $\widetilde{E}$ is a purifying register any of the $A_j E_{j-1}$ registers. 
Then the following ``normalized" tradeoff function on $\hat{C}_1^n$:
\begin{equation}
    \fhatfull(\cobs) \coloneq \sum_{j=1}^n \left( f_{| \cP_1^{j-1}} (\cP_j) +    \kappafuncgeneric{f_{|\cP_1^{j-1}} }{ \sigma^{(j)}_{A_j} } { \QKDGmap{j} }{\attack{j} } \right)
\end{equation}
satisfies
\begin{equation}
    \Halpha[\fhatfull](S_1^n | \CP_1^n E_n) \geq 0.
\end{equation}
\end{theorem}
\begin{proof}
   The statement follows from a direct application of \cite[Theorem 4.1a]{arqand_marginal_2025}, and we simply describe how that theorem can be applied here. The final state $\rho_{S_1^n \CP_1^n E_c}$  is of the form 
   \begin{equation}
       \rho_{S_1^n \CP_1^n E_c} = \mathcal{M}_n \circ \dots \circ \mathcal{M}_1 \left( \rho_{A_1^n E_0} \right)
   \end{equation}
   where $\rho_{A_1^n E_0}$ is the global source-replaced state defined in \cref{eq:globalsourcereplacedstate}, where we identify $A_j = \Ameas_j \Ashield_j$ as the register that does not leave Alice's lab, and $E_0=(A')_1^n$ as the set of states that leave Alice's lab. Furthermore, $\rho_{A_1^n} = \bigotimes_{j=1}^n \sigma^{(j)}_{A_j}$, and $\mathcal{M}_j   
   = \QKDGmap{j} \circ \attack{j} \in \CPTP(E_{j-1} A_j , S_j \CP_j E_j)
   $. The statement thus follows directly from \cite[Theorem 4.1a]{arqand_marginal_2025}, where the symbols $\mathcal{M}_j,E_j, \CP_j,\widetilde{E},\fhatfull,f_{|\cP_1^{j-1}}$ correspond to the same symbols used in that theorem. The only difference is that $A_j$ here should be identified with $A_{j-1}$ in Ref.~\cite{arqand_marginal_2025}, and we explicitly write $\mathcal{M}_j$ as a concatenation of two maps, $\QKDGmap{j}\circ \attack{j} $.
\end{proof}

Thus, we are now at a stage where, for a given sequence of attack channels $\{ \attack{j} \}$ and operations by Alice and Bob in each round $\{\QKDGmap{j}\}$, an appropriate ``global" tradeoff function $\fhatfull$ satisfying $\Halpha[\fhatfull](S_1^n | \CP_1^n E_n)_{\rho} \geq 0$ can be obtained via \cref{theorem:MEATQKDfirst}. 

\begin{remark} \label{remark:fintuition}
Recall that the statement $\Halpha[\fhatfull](S_1^n | \CP_1^n E_n)_{\rho} \geq 0$ can be understood as the function \(\fhatfull\) lower bounding the entropy \(\Halpha(S_1^n \mid \CP_1^n E_n)_\rho\) in the log-mean-exponential sense (see \cref{eq:lme}). In this light, it is helpful to think of \(f_{|\cP_1^{j-1}}\) as assigning a preliminary “score” to the public announcement in each round. Intuitively, a good choice assigns larger values to test-round announcements that indicate little or no adversarial interference (as opposed to announcements indicating errors), and to generation-round announcements indicating that the round is kept (as opposed to discarded). One is free to start with any such preliminary score that one desires (which may or may not be compatible with the above intuition), and such a score need not have the desired property of lower bounding the relevant {\Renyi} entropy.  This preliminary score must then be reduced by the amount \(\kappafuncgeneric{f_{|\cP_1^{j-1}}}{\sigma^{(j)}_{A_j}}{\QKDGmap{j}}{\attack{j}}\). The resulting per-round contributions can subsequently be combined to yield the global tradeoff function \(\fhatfull\), which indeed satisfies the required property of providing a lower bound on \(\Halpha(S_1^n \mid \CP_1^n E_n)_\rho\) in the log-mean-exponential sense.
\end{remark}

However, the function $\fhatfull$ obtained from \cref{theorem:MEATQKDfirst} depends on the attack performed, and therefore 
this result is still not yet sufficient to yield 
security of the QKD protocol. To circumvent this issue, we require the following definition, that constructs a lower bound on 
$\kappafuncgeneric{f_{|\cP_1^{j-1}} }{ \sigma^{(j)}_{A_j} } { \QKDGmap{j} }{ \attack{j}  }$ by minimizing over all possible attacks. 

\begin{remark} \label{remark:attackset}
Since our goal is to prove security against arbitrary attacks by Eve (subject to $E'j$ having dimension $d_j$), we must in principle optimize
$\kappafuncgeneric{f_{|\cP_1^{j-1}}}{\sigma^{(j)}_{A_j}}{\QKDGmap{j}}{\attack{j}}$
over all channels $\attack{j} \in \CPTP(E_{j-1}, B_j E'_j)$. This is indeed the fundamental task we carry out.

That said, in many scenarios it is possible to restrict attention to a subset of CPTP maps without loss of generality. For example, when squashing maps are used to reduce a protocol with an infinite-dimensional measurement to an equivalent protocol with a finite-dimensional one (see \cref{sec:optics}), Eve’s effective attack channel is given by composing her original channel with the squashing map. Depending on the details of the squashing map, these composed channels may lie in a subset
\[
\attackset{j}(E_{j-1}, B_j E'_j) \subseteq \CPTP(E_{j-1}, B_j E'_j).
\]
This motivates the general definition adopted here.

Note to readers: For a first reading, it is perfectly fine to regard $\attackset{j}(E_{j-1},B_j E'_j)$ as the full set of CPTP maps. The subtleties introduced by squashing-based reductions, which motivate restricting to smaller attack sets, are explained later in \cref{sec:optics}. Readers can return to this point after that discussion to fully appreciate these nuances.
\end{remark}

\begin{definition} \label{def:kappalowergeneric}
    Let $j \in \{1,\dots,n\}$, let  $f_{|\cP_1^{j-1}}$ be a tradeoff function on $\CP_j$, let $\sigma^{(j)}_{A_j} \in \dop{=}(A_j)$, and let $\QKDGmap{j} \in \CPTP(A_j B_j, S_j \CP_j \CPhat{j})$. Let $\attackset{j}(E_{j-1},B_j E_j) \subseteq \CPTP(E_{j-1}, B_j E'_j) $ be a subset of all possible attack channels. Then, we define a value for $  \kappafuncgeneric{f_{|\cP_1^{j-1}} }{ \sigma^{(j)}_{A_j} } { \QKDGmap{j}  }{\attackset{j}}$\footnote{In this work, each $\attackset{j}$ is always associated with specific registers 
$E_{j-1}, B_j, E'_j$, and should formally be written as 
$\attackset{j}(E_{j-1}, B_j E'_j)$.  
While we typically include this explicitly, we occasionally omit the register 
labels to avoid notational clutter.} over the set of attack channels via:
    \begin{equation}\label{eq:kappalowergeneric}
    \begin{aligned}
        \kappafuncgeneric{f_{|\cP_1^{j-1}} }{ \sigma^{(j)}_{A_j} } { \QKDGmap{j}  }{\attackset{j}(E_{j-1},B_j E_j)} &\coloneq \inf_{\attack{j} \in \attackset{j}} \kappafuncgeneric{f_{|\cP_1^{j-1}} }{ \sigma^{(j)}_{A_j} } { \QKDGmap{j} }{\attack{j}  } \\
        &\coloneq \inf_{\nu \in \Sigma_j (\attackset{j})} \Halpha[f_{|\cP_1^{j-1}}] (S_j | \CP_j E_j \widetilde{E} )_\nu
        \end{aligned}
    \end{equation}
    where
\begin{equation} 
    \Sigma_j (\attackset{j}) \coloneq \left\{ \QKDGmap{j} \circ \attack{j} \left( \omega_{A_j E_{j-1} \widetilde{E}} \right) \;\middle|\; \omega_{A_j} = \sigmaconstraint{j}, \; \attack{j} \in \attackset{j}  \right\},
\end{equation}
where $\widetilde{E}$ is a purifying register for any of the $A_j E_{j-1}$ registers. 
\end{definition}

Since
\[
\kappafuncgeneric{f_{|\cP_1^{j-1}}}{\sigma^{(j)}_{A_j}}{\QKDGmap{j}}{\attackset{j}(E_{j-1}, B_j E'_j)}
\;\le\;
\kappafuncgeneric{f_{|\cP_1^{j-1}}}{\sigma^{(j)}_{A_j}}{\QKDGmap{j}}{\attack{j}}, \qquad  \forall \; \attack{j}\in \attackset{j}(E_{j-1}, B_j E'_j)
\]
one might be tempted to use the left-hand side to define $\fhatfull$ and thereby the key length (via \cref{eq:lexpression}), since that value would be the worst case value over all possible attacks. However, this formulation appears to depend on the dimensions assigned to Eve’s side-information registers, and thus we would  require an additional optimization over all possible choices of those dimensions as well. The goal of the next subsection is to present a reformulation of $\kappafuncgeneric{f_{|\cP_1^{j-1}}}{\sigma^{(j)}_{A_j}}{\QKDGmap{j}}{\attackset{j}(E_{j-1}, B_j E'_j)}$ that avoids these issues by getting rid of the dependence on the dimensions of Eve's registers.

\subsection{Reformulating the minimization over attack channels} \label{subsec:reformulatingattackmin}

In order to reformulate $\kappafuncgeneric{f_{|\cP_1^{j-1}}}{\sigma^{(j)}_{A_j}}{\QKDGmap{j}}{\attackset{j}(E_{j-1}, B_j E'_j)}$ in a manner where the dependence on the Eve's dimensions disappears, it is convenient to consider two related notions of attack channels, as follows:
\begin{itemize}
	\item	$\attack{j} \in \attackset{j} (E_{j-1}, B_j E'_j) \subseteq \CPTP(E_{j-1}, B_j E'_j)$, which describe Eve’s action from her side-information register $E_{j-1}$ to Bob’s system $B_j$ together with a new side register $E'_j$. In particular, the set $\attackset{j}$ depends on the dimension we allow for Eve’s side-information registers $E'_j$. 
	\item	$\Qattack{j} \in \Qset{j} \subseteq \CPTP(A'_j, B_j)$, which describe maps directly from Alice’s emitted signal $A'_j$ to Bob’s register $B_j$. Intuitively, Eve’s attack in this perspective can be seen as first applying such a channel to the source-replacement state $\sourcesymbol^{(j)}_{A_j A'_j}$, after which we purify the post-attack output state, and give the purifying register to Eve. Notice that $\Qset{j}$ does not depend on the dimensions we allow for Eve's side-information registers $E'_j$.
    \end{itemize}

Instead of considering all attack channels in $\attackset{j} (E_{j-1}, B_j E'_j)$ for the security analysis, we would instead like to consider all maps in $\Qset{j}$. In order to do so, we require $\Qset{j}$ to satisfy a specific property stated in  \cref{def:Qjcondition}. (Informally, we require that applying maps in $\Qset{j}$ and then giving a purification to Eve is at least as powerful as applying maps in $\attackset{j}$).  That is, for every attack in $\attackset{j} (E_{j-1}, B_j E'_j)$, there exists a channel in $\Qset{j}$ such that the outputs on the $A_jB_j$ system is the same. This is formalized in the following definition.

\begin{definition}[Marginal of $\attackset{j}(E'_{j-1} \CPhat_{j},B_j E'_j)$] \label{def:Qjcondition}
Let $\attackset{j}(E'_{j-1} \CPhat_{j},B_j E'_j)  \subseteq \CPTP(E'_{j-1} \CPhat_{j}, B_j E'_j) $ and $\Qset{j} \subseteq \CPTP(A'_j, B_j)$ be attack sets. 
We say that $\Qset{j}$  is a \term{marginal} of $\attackset{j}(E'_{j-1} \CPhat_{j},B_j E'_j)$ if the following \emph{marginalization property} holds: for every $\omega_{A_j E_{j-1} \widetilde{E}}$ such that $\omega_{A_j} = \sourcesymbol^{(j)}_{A_j}$ and $\widetilde{E}$ is a purifying register for the $A_jE_{j-1}$ registers, and every $\attack{j} \in \attackset{j}$, there exists a $\Qattack{j} \in \Qset{j}$ such that 
\begin{equation} \label{eq:Qjcondition}
    \Qattack{j}\left[ \sourcesymbol^{(j)}_{A_j A'_j} \right] 
   = \tr_{E_j \widetilde{E}} \circ \attack{j} \left[ \omega_{A_j E_{j-1} \widetilde{E}} \right].
\end{equation}
where $\sourcesymbol^{(j)}_{A_j A'_j}$ is the source-replacement state (see \cref{eq:globalsourcereplacedstate}). We say that the collection $\left\{\attackset{j}(E'_{j-1} \CPhat_{j},B_j E'_j)\right\}_{E'_{j-1},E'_j}$ satisfies the \term{dimension-independent marginal property} with respect to $\Qset{j}$ if, for all possible choices of the auxiliary dimension of Eve’s registers $E'_j,E'_{j-1}$, the collection $\left\{\attackset{j}(E'_{j-1} \CPhat_{j},B_j E'_j)\right\}_{E'_{j-1},E'_j}$ all possess the same marginal $\Qset{j}$.
\end{definition}

In this way, $\Qset{j}$ is dimension-independent, whereas $\attackset{j}(E'_{j-1} \CPhat_{j}, B_j E'_j) $ depends explicitly on the dimension assigned to $E'_j$. Notice that any state $ \attack{j} \left[ \omega_{A_j E_{j-1} \widetilde{E}} \right]$ on the RHS of \cref{eq:Qjcondition}, can be obtained from some purification of $  \Qattack{j}\left[ \sourcesymbol^{(j)}_{A_j A'_j} \right] $, which is the LHS of \cref{eq:Qjcondition}, by a map acting only on the purifying system. Thus, due to data-processing (\cref{lemma:DPIfweighted}), it suffices to consider purifications of $  \Qattack{j}\left[ \sourcesymbol^{(j)}_{A_j A'_j} \right] $ when performing the infimum over all attacks that appears in $\kappa$ from \cref{eq:kappalowergeneric}. We formalize this in the following lemma.

\begin{lemma}
    \label{lemma:MEATuniform_lowerbound}
    Let $\kappafuncgeneric{f_{|\cP_1^{j-1}} }{ \sigma^{(j)}_{A_j} } { \QKDGmap{j}  }{\attackset{j}(E_{j-1},B_j E'_j)}$ be as defined in \cref{def:kappalowergeneric}, and let $\Qset{j}$ be a marginal of $\attackset{j}(E_{j-1},B_j E'_j)$ in the sense of Definition~\ref{def:Qjcondition}. Then, for any $\alpha\geq 1$ we have:
   \begin{equation}
    \begin{aligned}
        \label{eq:kappalower_oneside}
    \kappafuncgeneric{f_{|\cP_1^{j-1}} }{ \sigma^{(j)}_{A_j} } { \QKDGmap{j}  }{\attackset{j}(E_{j-1},B_j E_j)}&\geq\inf_{\Qattack{j}\in\Qset{j}}  \Halpha[f_{|\cP_1^{j-1}}](S_j| \CP_j \CPhat_j \widehat{E})_{\QKDGmap{j}\left[    \pf\left(\Qattack{j}\left[
\sourcesymbol^{(j)}_{A_j A'_j}\right]\right)\right]}  \\  &=\inf_{\overline{\omega}\in{\overline{\Sigma}}_j\left(\Qset{j}\right)}  \Halpha[f_{|\cP_1^{j-1}}](S_j| \CP_j \CPhat_j \widehat{E})_{\QKDGmap{j}\left[\pf\left(\overline{\omega}_{A_jB_j}\right)\right]}
    \end{aligned}
    \end{equation}
   where $\pf$ is a purifying function of $A_jB_j$ onto $\widehat{E}$, and
    \begin{align}
        &{\overline{\Sigma}}_j\left(\Qset{j}\right) \coloneq \left\{\Qattack{j}\left[{\sigma^{(j)}_{A_jA'_j}}\right]
            \;\middle|\;
            \Qattack{j}\in\Qset{j} \right\},
    \end{align}
    where $\sourcesymbol^{(j)}_{A_jA'_j}$ is the source-replaced state from \cref{eq:globalsourcereplacedstate}.
    If $\Qset{j}$ satisfies the dimension-independent marginal property, i.e is the marginal of all $\left\{\attackset{j}(E'_{j-1} \CPhat_{j},B_j E'_j)\right\}_{E'_{j-1},E'_j}$, then \cref{eq:kappalower_oneside} is obtained for any dimensions of Eve's side-information registers.
    Furthermore, the objective function in the RHS of the first line of \cref{eq:kappalower_oneside} is convex in $\Qattack{j}$, and in the second line of \cref{eq:kappalower_oneside} is convex in $\overline{\omega}_{A_jB_j}$.\footnote{If $\Qset{j}$ is convex, then the RHS of \cref{eq:kappalower_oneside} is a convex optimization problem.}
    \end{lemma}

     \begin{proof}
    The first line of \cref{eq:kappalower_oneside} is merely an equivalent reformulation of the second line, obtained by an affine transformation of the input variable $\bar{\omega}$ (which is the post-attack state) to $\Qattack{j}$ (which is the attack channel). Since affine transformations preserve convexity, the convexity of the  objective function in the first line follows. Thus, we focus on proving the inequality in \cref{eq:kappalower_oneside}. In other words, we show that
\begin{equation} \label{eq:kappasameproof}
    \inf_{\attack{j}\in\attackset{j}} \inf_{\omega\in\Sigma_j(\attack{j})}\Halpha[f_{|\cP_1^{j-1}}](S_j| \CP_j E_j \widetilde{E})_{ \QKDGmap{j} \circ \attack{j} \left[ \omega_{A_{j} E_{j-1}\widetilde{E}} \right]} \geq\inf_{\Qattack{j}\in\Qset{j}}  \Halpha[f_{|\cP_1^{j-1}}](S_j| \CP_j \CPhat_j \widehat{E})_{\QKDGmap{j}\left[    \pf\left(\Qattack{j}\left[
    \sourcesymbol^{(j)}_{A_j A'_j}\right]\right)\right]},
\end{equation}
where $\widetilde{E}$ is a register of large enough dimension to accommodate for purification of all $A_jE_{j-1}$ registers.
Let $\left(\omega^*_{A_jE_{j-1}\widetilde{E}},\attack{j}^*\right)$ be a feasible point on the LHS of \cref{eq:kappasameproof}, and let $\omega^*_{A_j B_j E'_j \widetilde{E}} = \attack{j}^* \left[ \omega^*_{A_{j} E_{j-1}\widetilde{E}}\right] $; noting that it satisfies $\omega^*_{A_j} = \sigma^{(j)}_{A_j}$. Since $\Qset{j}$ is a marginal of $\attackset{j}$ in the sense of Definition~\ref{def:Qjcondition}, then \cref{eq:Qjcondition} states that there exists a channel $\Qattack{j}^*\in\Qset{j}$ such that
\begin{align}
    \Qattack{j}^*\left[ \sourcesymbol^{(j)}_{A_j A'_j} \right] 
   = \tr_{E'_j \widetilde{E}} \circ \attack{j} \left[ \omega^*_{A_j E_{j-1} \widetilde{E}} \right]=\omega^*_{A_jB_j}.
\end{align}
 Let $\pf$ be a purifying function of $A_jB_j$ onto $\widehat{E}$. Then, there exists a channel $\mathcal{N}_j \in \CPTP(\hat{E},E'_{j} \widetilde{E})$ such that $\omega^*_{A_j B_j E'_j \widetilde{E}} = \mathcal{N}_j \left[ \pf\left(\Qattack{j}^*\left[\sigma^{(j)}_{A_jA'_j}\right]\right) \right]$ (see \cref{lemma:pur_to_ext}).
Thus, we have:
\begin{align}
           \Halpha[f_{|\cP_1^{j-1}}](S_j| \CP_j E_j \widetilde{E})_{\QKDGmap{j}\circ\attack{j}\left[\omega^*_{A_{j}E_{j-1}\widetilde{E}}\right]}&= \Halpha[f_{|\cP_1^{j-1}}](S_j| \CP_j \CPhat_j E'_j \widetilde{E})_{\QKDGmap{j}\circ\attack{j}\left[\omega^*_{A_{j}E_{j-1}\widetilde{E}}\right]}\nonumber\\
            &=  \Halpha[f_{|\cP_1^{j-1}}](S_j| \CP_j \CPhat_j E'_j \widetilde{E})_{\QKDGmap{j}\circ\mathcal{N}_j\left[ \pf\left(\Qattack{j}^*\left[\sigma^{(j)}_{A_jA'_j}\right]\right) \right]}\nonumber\\
            &=\Halpha[f_{|\cP_1^{j-1}}](S_j| \CP_j \CPhat_j E'_j \widetilde{E})_{\mathcal{N}_j\circ\QKDGmap{j}\left[\pf\left(\Qattack{j}^*\left[\sigma^{(j)}_{A_jA'_j}\right]\right)\right]}\nonumber\\
            &\geq \Halpha[f_{|\cP_1^{j-1}}](S_j| \CP_j \CPhat_j \widehat{E})_{\QKDGmap{j}\left[\pf\left(\Qattack{j}^*\left[\sigma^{(j)}_{A_jA'_j}\right]\right)\right]}.
        \end{align}
        The first line simply rewrites $E_j$ as $\CPhat{j} E'_j$, and the second line replaces $\attack{j} [\omega^*_{A_j E_{j-1} \widetilde{E} }]$ with $ \mathcal{N}_j \left[ \pf\left(\Qattack{j}^*\left[\sigma^{(j)}_{A_jA'_j}\right]\right) \right]$ $\attackset{j} $ since they are equal. The third line holds due to the fact that the maps $\mathcal{N}_j$ and $\QKDGmap{j}$ commute (since they 
        act on different registers), and the last line follows from data processing inequality (\cref{lemma:DPIfweighted}). Thus, for any feasible point on the LHS, we can find a feasible point on the RHS such that the LHS is greater than or equal to the RHS. This proves the first inequality in~\cref{eq:kappalower_oneside}. 
        
     The convexity of the objective function in the second line of \cref{eq:kappalower_oneside} was established in~\cite[Lemma~4.10]{arqand_marginal_2025}. 
    \end{proof}

Note that \cite[Corollary 8.9]{inprep_BDR3}
shows that regardless of the choice of $\attackset{j}(E_{j-1}, B_j E'_j)$, the set $\Qset{j}$ can always be taken to be $\CPTP(A'_j, B_j)$  (although doing so when using the flag-state squasher leads to trivial key rates; there a non-trivial restriction needs to be imposed on $\Qset{j}$ ).  Furthermore, 
we note that although \cref{lemma:MEATuniform_lowerbound} only establishes an inequality, this inequality is typically saturated for all scenarios considered in chapter. Thus, we can instead focus on computing the objective function formulated in terms of $\Qset{j}$, where the dimension of Eve’s side-information registers plays no role. We are now ready to bring all the pieces together and present the final security statement for the protocol under study.

\subsection{Final security statement} \label{subsec:finalsecuritystatem}
We now state the final security statement concerning the security of \nameref{prot:abstractqkdprotocol}.
\begin{theorem}[Security statement for \nameref{prot:abstractqkdprotocol} under finite-dimensional Alice–Bob setting]
\label{theorem:abstractsecuritystatement}
Consider the \nameref{prot:abstractqkdprotocol} where Alice’s transmitted systems ($A'_j$) are finite-dimensional and Bob performs finite-dimensional measurements ($B_j$ is finite-dimensional). For this protocol, after applying the source-replacement scheme (\cref{lemma:shieldandsourcereplacement}), let $\sourcesymbol^{(j)}_{A_jA'_j}$ denote the resulting source states in round~$j$, and let $\QKDGmap{j}$ denote the operations Alice and Bob perform in round~$j$.
 (These are the same as those that appear in the corresponding \nameref{prot:virtualprotocoll}, with state evolution described in \cref{lemma:stateevolution}). For each $j$ and every value of $\cP_1^{j-1}$, let $f_{|\cP_1^{j-1}}$ denote a tradeoff function\footnote{See \cref{remark:fintuition} for an intuitive interpretation of this function.} on the register $\CP_j$. Let $\kappaQKDfunc{f_{|\cP_1^{j-1}}}{\sigma^{(j)}_{A_j}}{\QKDGmap{j}}$ be any value satisfying
\begin{equation}
\begin{aligned}
\kappaQKDfunc{f_{|\cP_1^{j-1}}}{\sigma^{(j)}_{A_j}}{\QKDGmap{j}}
&\leq
\inf_{\Qattack{j}\in\Qset{j}}
\Halpha[f_{|\cP_1^{j-1}}](S_j| \CP_j \CPhat_j \widehat{E})_{\QKDGmap{j}\left[\pf\left(\Qattack{j}\left[\sourcesymbol^{(j)}_{A_j A'_j}\right]\right)\right]}, \\
\Qset{j} &= \CPTP(A'_j, B_j),
\end{aligned}
\end{equation}
where $\pf$ is a purifying function of $A_j B_j$ onto $\widehat{E}$, and define 
\begin{equation} \label{eq:fhatfullQKDdefinition}
\fhatfullQKD(\cobs) \coloneq \sum_{j=1}^n \left( f_{|\cP_1^{j-1}}(\cP_j) + \kappaQKDfunc{f_{|\cP_1^{j-1}}}{\sigma^{(j)}_{A_j}}{\QKDGmap{j}} \right).
\end{equation}
Then, the protocol is $(\epsPA + \epsEV)$-secure (according to \cref{def:qkdsecurityasymmetric}) if the key length $\lkey(\cobs)$ is chosen as
\begin{equation}
 \lkey(\cobs) = \max\left\{ 0 , \floor{ \fhatfullQKD(\cobs)  - \leak(\cobs) - \ceil{ \log(\frac{1}{\epsEV})} -  \frac{\alpha}{\alpha-1} \log(\frac{1}{\epsPA}) + 2 } \right\}.
\end{equation}
\end{theorem}

\begin{proof}
    The required statement follows from an appropriate combination of all the statements we have shown previously, which we recall as follows. Recall that we only need to show the  $\epsPA$-secrecy of the protocol, since $\epsEV$-correctness has already been shown in \cref{lemma:correctnessissatisfied}.

       \paragraph*{Reducing to  \nameref{prot:virtualprotocoll}.}

    We begin by applying \cref{lemma:modifiedprotocol}, which allows us to construct a modified version of \nameref{prot:abstractqkdprotocol} in which Alice's state preparation is moved earlier in time, such that it occurs before Bob's first measurement. As argued in \cref{lemma:modifiedprotocol}, this can be done without loss of generality, and ensures that the security of this modified protocol implies the security of the original protocol. Next, we invoke \cref{lemma:shieldandsourcereplacement}, which shows that Alice's operations can be viewed as preparing a gobal source-replaced state $ \sourcesymbol_{\Ameas_1^n (\Ashield)_1^n (A')_1^n}$ (that is tensor-product across rounds) and then measuring the $\Ameas_1^n$ systems. Since Alice’s measurements commute with all operations except the public announcements (which, by design, occur only after Bob’s measurement in each round), we can postpone Alice’s measurements to occur simultaneously with Bob’s. 

    After these modifications, the protocol for which we must prove security is given by  \nameref{prot:virtualprotocoll}. In particular, it admits the structure of sequential channels described in \cref{lemma:stateevolution,fig:MEAT_Attack}, where $\{\attack{j}\}$ denotes the sequence of Eve's attack channels, and the maps $\{\QKDGmap{j}\}$ implement Alice and Bob's measurements, postprocessing and public announcements. We note that the arguments so far do not require Alice, Bob or Eve to be finite-dimensional. 
    \paragraph*{Security analysis for state obtained in \nameref{prot:virtualprotocoll}.} At this stage, we fix a particular sequence of attack channels, and analyze the output state corresponding to it. In particular, we also fix the dimensions of all of Eve's side-information registers ($E'_j$) to be some finite value $d_j$. We will later obtain a result that does not depend on any particular sequence of attack channels or dimensions, and holds for any sequence of attack channels that Eve may perform. 

    From \cref{lemma:correctnessissatisfied} (which also applies to this protocol), we have that $\epsEV$-correctness is satisfies. Thus, we are now concerned with proving $\epsPA$-secrecy of the state. In \cref{theorem:entropytovarlength}, we show that if 
    \begin{equation} \label{eq:fullprooftempgreater0}
    \Halpha[\fhatfull](S_1^n | \CP_1^n E_n) \geq 0,
    \end{equation} 
    then the function $\lkey(\cobs)$ (see \cref{eq:lexpression}) that determines the length of the output key given by
    \begin{equation}
             l(c_1^n) = \max\left\{ 0 , \floor{ \fhatfullQKD(\cobs)  - \leak(\cobs) - \ceil{ \log(\frac{1}{\epsEV})} -  \frac{\alpha}{\alpha-1} \log(\frac{1}{\epsPA}) + 2  } \right\}
    \end{equation}
    guarantees $\epsPA$-secrecy, provided that  
    \begin{equation} \label{eq:fullprooftempfcondition}
    \fhatfullQKD(\cobs) \leq \fhatfull(\cobs) \quad \forall \cobs.
    \end{equation}  
    Thus, our task reduces to constructing a global tradeoff function $\fhatfullQKD$ satisfying above property.
    
    \paragraph*{Applying MEAT.} To do so, we turn to \cref{theorem:MEATQKDfirst}, where we apply MEAT to the sequence of channels described in \cref{lemma:stateevolution,fig:MEAT_Attack}. This yields a function $\fhatfull$ satisfying $\Halpha[\fhatfull](S_1^n | \CP_1^n E_n) \geq 0$, given by
    \begin{equation}
    \fhatfull(\cobs) \coloneq \sum_{j=1}^n \left( f_{| \cP_1^{j-1}} (\cP_j) +    \kappafuncgeneric{f_{|\cP_1^{j-1}} }{ \sigma^{(j)}_{A_j} } { \QKDGmap{j}}{ \attack{j} } \right), 
\end{equation}
where $f_{| \cP_1^{j-1}}$ are arbitrary tradeoff functions on $\CP_j$ (that we choose), and the constants $\kappafuncgeneric{f_{|\cP_1^{j-1}} }{ \sigma^{(j)}_{A_j} } { \QKDGmap{j}}{ \attack{j} }$ are defined in \cref{eq:kappadefined}. However, both $\kappafuncgeneric{f_{|\cP_1^{j-1}} }{ \sigma^{(j)}_{A_j} } { \QKDGmap{j} }{ \attack{j} }$ and the resulting $\fhatfull$ constructed in this fashion depend on Eve's attack channels, and the dimensions of her registers. We would like to remove this dependence by performing a worst-case analysis over all possible attack channels $\attackset{j}(E_{j-1}, B_j E'_j) = \CPTP(E_{j-1}, B_j E'_j)$ and all possible dimensions of Eve's registers.
\paragraph*{Reformulations to remove dependence on Eve's dimensions and attack channels.} 
To address this, we construct a different kind of attack channels, and denote this set via \(\Qset{j}\). Here, we represent any attack as the application of one such channel to the source-replaced state, and then assume that Eve holds a purification of the resulting output state. We
set $\Qset{j} = \CPTP(A'_j, B_j)$, and apply \cref{lemma:MEATuniform_lowerbound}. This, along with the definition of $\kappaQKDfunc{f_{|\cP_1^{j-1}}}{\sigma^{(j)}_{A_j}}{\QKDGmap{j}}$ in the theorem statement, allows us to obtain
\begin{equation}
\kappafuncgeneric{f_{|\cP_1^{j-1}}}{\sigma^{(j)}_{A_j}}{\QKDGmap{j}}{\attack{j}}
\geq
\inf_{\Qattack{j}\in\Qset{j}}
\Halpha[f_{|\cP_1^{j-1}}](S_j| \CP_j \CPhat_j \widehat{E})_{\QKDGmap{j}\left[\pf\left(\Qattack{j}\left[\sourcesymbol^{(j)}_{A_j A'_j}\right]\right)\right]} \geq \kappaQKDfunc{f_{|\cP_1^{j-1}}}{\sigma^{(j)}_{A_j}}{\QKDGmap{j}},
\end{equation}
 for any attack $\attack{j} \in \CPTP(E_{j-1} , B_j E'_{j-1})$. Thus, if we use $\kappaQKDfunc{f_{|\cP_1^{j-1}}}{\sigma^{(j)}_{A_j}}{\QKDGmap{j}}$  to define 
\begin{equation} 
\fhatfullQKD(\cobs) \coloneq \sum_{j=1}^n \left( f_{|\cP_1^{j-1}}(\cP_j) + \kappaQKDfunc{f_{|\cP_1^{j-1}}}{\sigma^{(j)}_{A_j}}{\QKDGmap{j}} \right),
\end{equation}
then the resultant global tradeoff function $\fhatfullQKD(\cobs)$ satisfies the required property \cref{eq:fullprooftempfcondition}. Moreover,  neither $\kappaQKDfunc{f_{|\cP_1^{j-1}}}{\sigma^{(j)}_{A_j}}{\QKDGmap{j}}$ nor $\fhatfullQKD(\cobs)$ has any dependence on the dimensions of Eve's systems, or her choice of attack. Additionally,  as pointed out in \cref{lemma:MEATuniform_lowerbound}, the resulting optimization problem is finite-dimensional and convex. Thus, $\lkey(\cobs)$, as defined in the theorem statement, is a valid choice for obtaining a $(\epsPA + \epsEV)$-secure protocol when considering any attack in which Eve employs finite-dimensional side registers.

\paragraph*{Relaxing the finite-dimensional Eve assumption.}
We now address the technicality that Eve could, in principle, use infinite-dimensional side registers in her attack. So far, we have proven security for all attacks where Eve’s registers are finite-dimensional. However, since our security statement yields the same security parameter $(\epsPA + \epsEV)$ regardless of the dimensions assigned to Eve’s side registers, we can invoke the results of \cite[Appendix A]{inprep_BDR3} to argue that the same security guarantee extends to the infinite-dimensional case. This concludes our proof. 
\end{proof}

Having completed the security analysis for \nameref{prot:abstractqkdprotocol}, we now turn to the next subsection to elaborate on certain aspects, highlight subtleties, and provide additional clarification.

\subsection{Discussion} \label{subsec:discussionabstractproof}

We have now proved \cref{theorem:abstractsecuritystatement}, which guarantees the security of the \nameref{prot:abstractqkdprotocol} against arbitrary attacks by Eve, as long as one can compute a lower bound on a finite-dimensional convex optimization. Our theorem holds for the case where Alice sends finite-dimensional states, and Bob measures finite-dimensional systems. Naturally, this security statement depends on the parameters specified in the protocol, and the secure output key length depends on these parameters. Before proceeding further, let us take stock of what has been accomplished so far and identify the remaining steps needed to complete the analysis and obtain a security proof for a practical decoy-state BB84 protocol. We have two main aspects that need to be addressed.

\begin{itemize}
    \item \textbf{Infinite dimensions of Alice and Bob:} First, recall our earlier caveat: the MEAT is formally applicable only to finite-dimensional systems, even though its statement does not dependent on the dimensions of the underlying systems. However, QKD protocols are implemented using optical systems, which are naturally described by infinite-dimensional Hilbert spaces. In particular, Alice sends infinite-dimensional states, and Bob POVM lives in an infinite-dimensional Hilbert space. 
Ref.~\cite{nahar_postselection_2024} (\cref{sec:postselectionoptics}) presented an argument to reduce to finite dimensions at the level of the security definition, that is, however it does not apply here due to on-the-fly announcements. Thus we generalize the approach in Ref.~\cite{nahar_postselection_2024} to include on-the-fly announcements. This method is general and potentially compatible with a broad class of squashing maps and source maps. This is undertaken in \cref{sec:optics}. Once this reduction is complete, the resulting finite-dimensional protocol can be analyzed using MEAT, following the procedure outlined \cref{subsec:expressingprotocolassequence,subsec:proofgenericstuff,subsec:securityfromMEAT,subsec:reformulatingattackmin,subsec:finalsecuritystatem}.

\item \textbf{Numerics:} Second, even after reducing to finite dimensions, computing the key rate still requires us to solve a highly non-trivial convex optimization problem. In essence, the difficulty arises from the fact that we aim to obtain a value that is provably \emph{lower} than the infimum --- i.e, a guaranteed lower bound. We state the numerical computations needed to compute key rates in  \cref{sec:plotsMEAT}, and refer the reader to \cite{kamin_renyi_2025,navarro_finite_2025} for some recent work on this topic. We use the code from Ref.~\cite{kamin_renyi_2025} to plot key rates. 
\end{itemize}

Having discussed the above issues - which are addressed in more detail in subsequent sections - we now outline a general recipe for applying the results of the previous subsection to compute QKD key rates. 

\subsubsection{Recipe} \label{subsubsec:recipe}
Since the \nameref{prot:abstractqkdprotocol} is highly general, the security of a wide variety of QKD protocols can be established using the results developed so far. To aid in this task, we  include a practical “recipe’’ for applying our results to compute key rates for a protocol of interest.
We present this recipe in a way that also accommodates squashing maps, source maps, and the optical modeling introduced in \cref{sec:optics}. For a first reading, the reader may safely skip the optical details and return to them after going through \cref{sec:optics}.
\begin{enumerate} 
\item[] \textbf{Recipe}
    \item Specify all details of the QKD protocol, and ensure that it fits the structure described in \nameref{prot:abstractqkdprotocol}.
    \item If required, apply the  source map \cref{lemma:sourcemapsecurityMEAT} to reduce to a protocol with a different state preparation by Alice.
    \item If required, apply the  squashing map \cref{lemma:SquashMapSecurityMEAT}  to reduce to a protocol where Bob performs a different measurement. 
    \item Consider the \nameref{prot:virtualprotocoll}, which is obtained after source-replacement and timing modifications as specified  \cref{subsec:expressingprotocolassequence}.  In particular, this fixes $\sourcesymbol^{(j)}_{A_j}$, $\QKDGmap{j}$ for all $j \in \{1,\dots,n\}$, which describe Alice's (source-replaced) state preparation, and the round-by-round operations undertaken by Alice and Bob in round $j$.. Note that these quantities are obtained by considering the \nameref{prot:virtualprotocoll} version of the protocol obtained \textit{after} the application of the source-map and squashing transformations.
    \item For all $j\in \{1,\dots,n\}$ and $\cP_1^{j-1}$, specify the tradeoff functions $f_{|\cP_1^{j-1}}$ (or a deterministic procedure for specifying them, see \cref{remark:specifyfatstart}).
    \item  Determine $\attackset{j}(E_{j-1}, B_j E'_j)$ (which depends the details of the squashing map used). From it, determine the corresponding $\Qset{j}$ that satisfies the dimension-independent extension property (\cref{def:Qjcondition}). If we do not use squashing maps, this reduces to $
\Qset{j} = \CPTP(A'_j, B_j)$.
    \item  Compute a lower bound on $\kappafuncgeneric{f_{|\cP_1^{j-1}} }{ \sigma^{(j)}_{A_j} } { \QKDGmap{j}  } {\attackset{j}}$, given by 
    \begin{equation} \label{eq:kappacompute}
        \inf_{\Qattack{j}\in\Qset{j}}
\Halpha[f_{|\cP_1^{j-1}}](S_j| \CP_j \CPhat_j \widehat{E})_{\QKDGmap{j}\left[\pf\left(\Qattack{j}\left[\sourcesymbol^{(j)}_{A_j A'_j}\right]\right)\right]}
    \end{equation}
    where $\pf$ is a purifying function from $A_j B_j$ onto $\widehat{E}$. Denote this lower bound by $\kappaQKDfunc{f_{|\cP_1^{j-1}}}{\sigma^{(j)}_{A_j}}{\QKDGmap{j}}$. 
    \item Define $\fhatfullQKD =  \sum_{j=1}^n \left( f_{| \cP_1^{j-1}} (\cP_j) +    \kappaQKDfunc{f_{|\cP_1^{j-1}} }{ \sigma^{(j)}_{A_j} } { \QKDGmap{j}  } \right)$. 
    \item Set the variable-length decision to be 
    \begin{equation}
        \lkey(\cobs) = \max\left\{ 0 , \floor{ \fhatfullQKD(\cobs)  - \leak(\cobs) - \ceil{ \log(\frac{1}{\epsEV})} -  \frac{\alpha}{\alpha-1} \log(\frac{1}{\epsPA}) + 2  } \right\}.
    \end{equation} The resulting protocol is $(\epsPA+\epsEV)$-secure.
    \end{enumerate}
We now highlight some subtleties and provide additional clarification in the following remarks.

\begin{remark} \label{remark:generality} 
Notice that \cref{theorem:abstractsecuritystatement} is fairly general, since the \nameref{prot:abstractqkdprotocol} is itself fairly general. For example, we can straightforwardly accommodate scenarios where Alice's state preparation is different in different rounds, since we allow for the $\sigma^{(j)}_{A_j}$ to vary from round to round, although we do require ``ìndependent"' state preparation across rounds, i.e, $\sigma_{A_1^n} = \sigma^{(1)}_{A_1} \otimes \dots \otimes \sigma^{(n)}_{A_n}$. Interestingly, we can also accommodate fairly contrived protocols where Alice and Bob perform \emph{different} protocol operations in each round - for example, a protocol that alternates between BB84 signal preparation and measurement in odd-numbered rounds and six-state signal preparation and measurement in even-numbered rounds.  This 
manifests in the analysis as $\QKDGmap{j}$ now being different across rounds. This flexibility makes our analysis, which inherits these features from the abstract MEAT statement itself, extremely general. The only additional cost incurred is the need to perform more computations.
\end{remark}

\begin{remark} \label{remark:anyfworks}
Note that when applying the above recipe, one may start with \emph{any} choice of tradeoff functions $f_{| \cP_1^{j-1}}$ that one desires --- there are no constraints whatsoever on the original selection of these functions. This choice determines the the values of $\kappaQKDfunc{f_{|\cP_1^{j-1}} }{ \sigma^{(j)}_{A_j} } { \QKDGmap{j}  }$, which in turn influences the normalized tradeoff function $\fhatfullQKD$. While the theorem permits arbitrary choices of the tradeoff $f_{| \cP_1^{j-1}}$, certain choices lead to superior key rates compared to other choices (in fact, a careful choice is typically important to achieve positive key rates). In particular, the \emph{optimal} $f_{| \cP_1^{j-1}}$, in the sense that it generates the best performance for a  fixed honest behaviour of the channel, can be chosen following a procedure described in \cite[Sec.~VI]{kamin_renyi_2025} or~\cite[Sec.~5.3]{arqand_generalized_2024}. Since the initial choice of  $f_{| \cP_1^{j-1}}$ is not relevant for the security claim to hold, and only necessary for ensuring the best performance, we only briefly discuss the procedure to choose   $f_{| \cP_1^{j-1}}$ in \cref{sec:plotsMEAT}.
\end{remark}

\begin{remark} \label{remark:adaptivef}
Note that the above recipe, and \cref{theorem:MEATQKDfirst,theorem:abstractsecuritystatement}, allows us to choose \emph{different} tradeoff functions $f_{| \cP_1^{j-1}}$ in different rounds, which can also depend on the observed public announcements of preceding rounds (although one has to determine how this choice is made beforehand). This allows for ``fully adaptive" QKD protocols \cite{zhang_knill_qpe,arqand_generalized_2024}, where the tradeoff functions $f_{| \cP_1^{j-1}}$ can be updated over the the course of the protocol, a feature that is important when performing QKD over noisy, or unpredictable channels. This too, is a feature inherited from the abstract MEAT statement itself.  However note that updating  $f_{| \cP_1^{j-1}}$ frequently leads to additional computation cost. The simplest scenario is the one where all $\sigma^{(j)}_{A_j}$s are identical, all $f_{| \cP_1^{j-1}}$s are identical and have no dependence on $\cP_1^{j-1}$, and all $\QKDGmap{j}$s are identical.\footnote{Strictly speaking, these objects act on different registers and are therefore not equal as mathematical objects. Rather, we mean that each $\sigma^{(j)}_{A_j}$ and $\QKDGmap{j}$ is defined on a distinct but isomorphic copy of the same underlying Hilbert space, and coincides with a fixed reference state and map under the corresponding canonical identification. Similarly, we mean that the functions $f_{|\cP_1^{j-1}}$ coincide under the appropriate relabeling of the classical registers. } 
\end{remark}

\begin{remark} \label{remark:specifyfatstart}
Since the function $\fhatfullQKD$ depends on all the $f_{| \cP_1^{j-1}}$ and $\kappaQKDfunc{f_{|\cP_1^{j-1}} }{ \sigma^{(j)}_{A_j} } { \QKDGmap{j}  }$, it might seem as though the computation of $\fhatfullQKD(\cobs)$ requires the computation of all $\kappaQKDfunc{f_{|\cP_1^{j-1}} }{ \sigma^{(j)}_{A_j} } { \QKDGmap{j}  }$ for all the possible inputs, which is prohibitively expensive. However, as argued in Ref.~\cite{zhang_knill_qpe}, when applying the above result in practice, one only needs to compute $\fhatfull(\cobs)$ on the specific sequence $\cobs$ observed in the protocol. This allows for an efficient, iterative procedure: for each round $j$, one examines the past announcements $\cP_1^{j-1}$, makes some choice of $f_{| \cP_1^{j-1}}$ (which depends on those values) and computes $\kappaQKDfunc{f_{|\cP_1^{j-1}} }{ \sigma^{(j)}_{A_j} } { \QKDGmap{j}  }$, then considers a similar computation for the $(j+1)^\text{th}$ round and so on. Note that there is no requirement to finish each of these computations before the next round occurs; in other words, they can be computed at any convenient time between generating the relevant data and the final privacy amplification step --- this is because the computed values are only used to determine the key length for privacy amplification, and not for any other aspect of the protocol.
\end{remark}

\begin{remark} \label{remark:fcanbelarge}
We remark that, when running the numerical routines from  Ref.~\cite{kamin_renyi_2025} for the decoy-state BB84 protocol, one observes that the quantity
$$
f_{|\cP_1^{j-1}}(\cdot)\;+\;
\kappaQKDfunc{f_{|\cP_1^{j-1}}}{\sigma^{(j)}_{A_j}}{\QKDGmap{j}}
$$
can take fairly large positive or negative values. As a consequence, for some realizations of the public announcements \(\cobs\), the globally normalized tradeoff function \(\fhatfullQKD(\cobs)\), and the key length $\lkey(\cobs)$ may exceed the total number of rounds $n$.
This behaviour may appear counterintuitive: the Rényi entropy of the pre-amplification string is always upper bounded by $n$, whereas \(\fhatfullQKD(\cobs)\) has a qualitative interpretation as a lower bound on this quantity ``on average''. Nevertheless, this does not invalidate the security proof. The framework ensures that such extreme values arise only with sufficiently small probability, and our obtained security statement is still valid --- recall for instance that the security definition (\cref{def:qkdsecuritysymmetric}) is averaged over the possible key lengths, and thus key lengths that occur with extremely small probability do not affect it much. (Similar properties hold for the ``log-mean-exponential'' interpretation (\cref{eq:interpretaslogmeanexp}) of the $f$-weighted entropy bounds, as these are also ``averaged'' quantities.) That said, for practical implementations, if one encounters an observation \(\cobs\) for which
\[
\fhatfullQKD(\cobs) > n,
\]
it is natural to propose that its value should be reduced to $n$ (i.e.~the maximal possible {\Renyi} entropy); the protocol still straightforwardly remains secure in this case.\footnote{This can be argued using the monotonicity of $f$-weighted {\Renyi} entropies wrt $f$, see \cref{lemma:Monotonicity_f}.} Importantly, however, even without such corrective adjustments, the formal security guarantees from \cref{theorem:abstractsecuritystatement} are fully valid; the adjustment is purely for practical considerations such as implementation difficulties in producing an extremely long output key. 
\end{remark}

We now turn our attention to the remaining steps outlined at the start of this subsection. In the next section, we  consider the quantum optical nature of realistic implementations, which causes Alice’s signal states and Bob’s POVMs to reside in infinite-dimensional Hilbert spaces.

\section{Extending Security to optical protocols} \label{sec:optics}
We will now outline the additional steps required to adapt our analysis to optical QKD implementations. We will do so by reducing the security analysis of the original protocol, where Alice's signal states and Bob's POVMs belong to infinite-dimensional Hilbert spaces, to a protocol where they belong to finite-dimensional Hilbert spaces. We start by focusing on Alice's signal states. This section will essentially generalize the analysis from \cref{chap:postselection} pertaining to the use of source maps (\cref{subsec:sourcemaps}) and squashing maps (\cref{subsec:squashing}).

\begin{lemma}[Source maps]\label{lemma:sourcemapsecurityMEAT}
Let $\left\{\{\QKDGmapfullbeforeSR{j}\}_{j=1}^n,\QKDpostprocessingmap, \sigma_{X_1^n (A')_1^n} \right\}$ determine a QKD protocol (see \cref{def:PMQKDMEAT}) where Alice prepares the global state
\begin{equation}
    \begin{aligned}\nonumber
        \sigma_{X_1^n (A')_1^n} &=   \sum_{x_1^n} p(x_1^n) \ketbra{x_1^n}_{X_1^n} \otimes \sigma^{(x_1^n)}_{(A')_1^n}.
    \end{aligned}
\end{equation}
Suppose there exists a quantum channel (source map) $\Psi \in  \CPTP((A'')_1^n, (A')_1^n)$ and a set of \emph{virtual states} $\{\xi^{(x_1^n)}_{(A'')_1^n}\}_{x_1^n} \subset \dop{=}( (A'')_1^n )$ such that
\begin{align}
    \sigma_{X_1^n (A')_1^n} &= (\idmap \otimes \Psi) \left[ \xi_{X_1^n (A'')_1^n}\right] \\
    \text{where } \quad \xi_{X_1^n (A'')_1^n} &= \sum_{x_1^n} p(x_1^n) \ketbra{x_1^n}_{X_1^n} \otimes \xi^{(x_1^n)}_{(A'')_1^n}. 
\end{align}
In other words, $\sigma^{(x_1^n)}_{(A')_1^n} = \Psi\left[\xi^{(x_1^n)}_{(A'')_1^n}\right]$ for all  $x_1^n$. Then if the protocol $\left\{\{\QKDGmapfullbeforeSR{j}\}_{j=1}^n,\QKDpostprocessingmap, \xi_{X_1^n (A'')_1^n} \right\}$
using the virtual states is $\epssecure$‑secure, the protocol $\left\{\{\QKDGmapfullbeforeSR{j}\}_{j=1}^n,\QKDpostprocessingmap, \sigma_{X_1^n(A')_1^n} \right\}$ using the real states is also $\epssecure$‑secure.
\end{lemma}
\begin{proof}
If the virtual protocol $\left\{\{\QKDGmapfullbeforeSR{j}\}_{j=1}^n,\QKDpostprocessingmap, \xi_{X_1^n (A'')_1^n} \right\}$ is $\epssecure$-secure, then for all attack channels $\{\attack{j}\}_{j=2}^n$ and $\attack{1}' \in \CPTP((A'')_1^n, B_1 E_1)$, we have (as in \cref{def:qkdsecurityChannelVersionMEAT})
\begin{align} \label{eq:EpsilonSecurityTau}
\hspace{-2em}
   & \nonumber \Big\|
        \Big(
            \QKDpostprocessingmap \circ \QKDGmapfullbeforeSR{n} \circ \attack{n} \circ \cdots \circ \QKDGmapfullbeforeSR{1} 
            - \idealmap \circ \QKDpostprocessingmap \circ \QKDGmapfullbeforeSR{n} \circ \attack{n} \circ \cdots \circ \QKDGmapfullbeforeSR{1}
        \Big) \\& \quad
        \circ \left(\idmap_{X_1^n} \otimes \attack{1}'\right)\left[\xi_{X_1^n (A'')_1^n}\right]
    \Big\|_1  \leq \epssecure,
\end{align}
where we used \cref{lemma:stateevolutionmodifiedprotocol} to represent the output state obtained in the protocol. We treat the first attack channel $\attack{1}'$ separately from the rest of the attack channels in our proof, since this attack channel is the one that can implement the source map. Consider the restricted subset of channels $  \mathcal{C}_\Psi $, where we first apply the source map $\Psi$ before applying an attack channel $\attack{1} \in \CPTP((A')_1^n, B_1 E_1)$
\begin{align} \label{eq:SourceMapChannelSubset}
    \mathcal{C}_\Psi \coloneqq \left\{\attack{1} \circ \Psi \,\vert\, \attack{1} \in \CPTP((A')_1^n, B_1 E_1) \right\}.
\end{align}
Since \cref{eq:EpsilonSecurityTau} holds for all $\attack{1}'$, it also holds for any $\attack{1}' \in \mathcal{C}_\Psi$. Thus,  for any $\attack{1} \in \CPTP({A'}_1^n, B_1 E_1)$, we set $\attack{1}' = \attack{1} \circ \Psi$. With this identification, we have
\begin{align} \label{eq:tempsourceproof}
  \left(\idmap_{X_1^n} \otimes \attack{1}'\right)[\xi_{X_1^n (A'')_1^n}]
    = \left(\idmap_{X_1^n} \otimes \attack{1} \circ \Psi\right)[\xi_{X_1^n (A'')_1^n}]
    = \left(\idmap_{X_1^n} \otimes \attack{1}\right)[\sigma_{X_1^n (A')_1^n}].
\end{align}
Now, using \cref{eq:tempsourceproof,eq:EpsilonSecurityTau} we obtain
\begin{align}
\hspace{-2em}
   & \nonumber
    \Big\|
        \Big(
            \QKDpostprocessingmap \circ \QKDGmapfullbeforeSR{n} \circ \attack{n} \circ \cdots \circ \QKDGmapfullbeforeSR{1}
            - \idealmap \circ \QKDpostprocessingmap \circ \QKDGmapfullbeforeSR{n} \circ \attack{n} \circ \cdots \circ \QKDGmapfullbeforeSR{1}
        \Big) \\
        & \quad 
        \circ \left(\idmap_{X_1^n} \otimes \attack{1}\right)\left[\sigma_{X_1^n (A')_1^n}\right]
    \Big\|_1 \leq \epssecure,
\end{align}
for all attack channels $\attack{1} \in \CPTP((A')_1^n, E_1 B_1)$, and for all $\{\attack{j}\}_{j=2}^n$.
This proves that the real protocol is $\epssecure$-secure.
\end{proof}

For squashing maps, we obtain a similar result.

\begin{restatable}[Squashing maps]{lemma}{lemma:squashingmapsecurity} \label{lemma:SquashMapSecurityMEAT}
Let $\left\{\{\QKDGmapfullbeforeSR{j}\}_{j=1}^n,\QKDpostprocessingmap, \xi_{X_1^n (A'')_1^n} \right\}$ be a QKD protocol (see \cref{def:PMQKDMEAT}) where Bob measures his received state in round $j$ with the POVM $\left\{M_{i}^{(B_j)}\right\}$. That is, we have, $\QKDGmapfullbeforeSR{j} = \QKDmapfullwithoutBobMeas{j}\circ\measChannel{\left\{M_{i}^{(B_j)}\right\}}$, where $\measChannel{\left\{M_{i}^{(B_j)}\right\}} \in \CPTP(B_j, Y_j)$ and $\QKDmapfullwithoutBobMeas{j} \in \CPTP(X_j Y_j, S_j X_j Y_j \CP_j \CPhat{j})$. Suppose there exists quantum channels (squashing maps) $\Lambda_j \in \CPTP(B_j, Q_j)$ and measurement channels $\measChannel{\left\{F_{i}^{Q_j}\right\}} \in \CPTP(Q_j,Y_j)$ such that
\begin{align} \label{eq:squashingCondition}
    \measChannel{\left\{F_{i}^{(Q_j)}\right\}} \circ \Lambda_j = \measChannel{\left\{M_{i}^{(B_j)}\right\}} \quad \text{for all } j.
\end{align}
Moreover, suppose that the squashed protocol $\left\{\left\{\QKDmapfullwithoutBobMeas{j}\circ\measChannel{\left\{F_{i}^{(Q_j)}\right\}}\right\}_{j=1}^n,\QKDpostprocessingmap, \xi_{X_1^n (A'')_1^n} \right\}$ using the measurement channel $\measChannel{\left\{F_{i}^{(Q_j)}\right\}}$ is $\epssecure$-secure against all attacks $\attackSquash{j} \in \attackset{j}$, where 
\begin{align} \label{eq:attackSetConstraint}
    \attackset{j} \supset \Lambda_j \circ \CPTP(E_{j-1},B_jE_j) \quad \text{for all } j.
\end{align}
Then the real protocol $\left\{\left\{\QKDmapfullwithoutBobMeas{j}\circ\measChannel{\left\{M_{i}^{(B_j)}\right\}}\right\}_{j=1}^n,\QKDpostprocessingmap, \xi_{X_1^n (A'')_1^n} \right\}$ using the measurement channel $\measChannel{\left\{M_{i}^{(B_j)}\right\}}$ is $\epssecure$-secure against all attacks $\attack{j} \in \CPTP(E_{j-1},B_jE_j)$ in round $j$.
\end{restatable}

\begin{proof}
    We start with Eve's attack  $\attack{j}\in\CPTP(E_{j-1},B_jE_j)$ for the real protocol\newline $\left\{\left\{\QKDmapfullwithoutBobMeas{j}\circ\measChannel{\left\{M_{i}^{(B_j)}\right\}}\right\}_{j=1}^n,\QKDpostprocessingmap, \xi_{X_1^n (A'')_1^n} \right\}$ and reduce its security to that of the squashed protocol as follows
    \begin{align}
        \nonumber&\Big\|
        \Big(
            \QKDpostprocessingmap \circ \QKDmapfullwithoutBobMeas{n}\circ\measChannel{\left\{M_{i}^{B_n}\right\}} \circ \attack{n} \circ \cdots \circ \QKDmapfullwithoutBobMeas{1}\circ\measChannel{\left\{M_{i}^{B_1}\right\}}\circ\attack{1}\\
            \nonumber & 
            \qquad
            - \idealmap \circ \QKDpostprocessingmap \circ \QKDmapfullwithoutBobMeas{n}\circ\measChannel{\left\{M_{i}^{B_n}\right\}} \circ \attack{n} \circ \cdots \circ \QKDmapfullwithoutBobMeas{1}\circ\measChannel{\left\{M_{i}^{B_1}\right\}}\circ\attack{1}
            \Big)
            \left[\xi_{X_1^n (A'')_1^n}\right]
        \Big\|_1\\
        =& \nonumber\Big\|
        \Big(
            \QKDpostprocessingmap \circ \QKDmapfullwithoutBobMeas{n}\circ\measChannel{\left\{F_{i}^{Q_n}\right\}} \circ \Lambda_n \circ \attack{n} \circ \cdots \circ \QKDmapfullwithoutBobMeas{1}\circ\measChannel{\left\{F_{i}^{Q_1}\right\}} \circ \Lambda_1\circ\attack{1}\\
            \nonumber & 
           \qquad 
            - \idealmap \circ \QKDpostprocessingmap \circ \QKDmapfullwithoutBobMeas{n}\circ\measChannel{\left\{F_{i}^{Q_n}\right\}} \circ \Lambda_n \circ \attack{n} \circ \cdots \circ \QKDmapfullwithoutBobMeas{1}\circ\measChannel{\left\{F_{i}^{Q_1}\right\}} \circ \Lambda_1\circ\attack{1}
            \Big)
            \left[\xi_{X_1^n (A'')_1^n}\right]
        \Big\|_1\\
        =& \nonumber\Big\|
        \Big(
            \QKDpostprocessingmap \circ \QKDmapfullwithoutBobMeas{n}\circ\measChannel{\left\{F_{i}^{Q_n}\right\}} \circ \attackSquash{n} \circ \cdots \circ \QKDmapfullwithoutBobMeas{1}\circ\measChannel{\left\{F_{i}^{Q_1}\right\}} \circ\attackSquash{1}\\
            \label{eq:squashedProtSecurityReduction} & 
            \qquad 
            - \idealmap \circ \QKDpostprocessingmap \circ \QKDmapfullwithoutBobMeas{n}\circ\measChannel{\left\{F_{i}^{Q_n}\right\}} \circ \attackSquash{n} \circ \cdots \circ \QKDmapfullwithoutBobMeas{1}\circ\measChannel{\left\{F_{i}^{(Q_j)}\right\}} \circ \circ\attackSquash{1}
            \Big)
            \left[\xi_{X_1^n (A'')_1^n}\right]
        \Big\|_1,
    \end{align}
    where the first equation follows from \cref{eq:squashingCondition}, and the second equation follows by relabelling $\attackSquash{j}\coloneqq \Lambda_j \circ \attack{j}$. Note that $\attackSquash{j}\in \attackset{j}$ follows from \cref{eq:attackSetConstraint}. Since this holds for every attack $\attack{j} \in \CPTP(E_{j-1}, B_j E_j)$ for the real protocol, minimizing the final expression in \cref{eq:squashedProtSecurityReduction} over all
$\attackSquash{j} \in \attackset{j} \subseteq \CPTP(E_{j-1}, B_j E_j)$
is guaranteed to yield a value that is less than or equal to the result of minimizing the first expression in \cref{eq:squashedProtSecurityReduction} over all
$\attack{j} \in \CPTP(E_{j-1}, B_j E_j)$.
This completes the proof, by identifying these minimizations with the corresponding definitions of security given in \cref{def:qkdsecurityChannelVersionMEAT}.  
\end{proof}

Note that if one employs the simple qubit squasher to map the measurement
register from $B$ to $Q$, then one may simply take
$\attackSquash{j} \in \attackset{j} = \CPTP(E_{j-1}, Q_j E_j)$.
We adopt this choice when computing key rates in \cref{sec:plotsMEAT}.

When using the flag-state squasher, however, an additional restriction is
required to prevent Eve from exploiting the classical flag registers alone (see
\cref{subsec:squashing}). This can be implemented by restricting the
allowed attack maps $\attackset{j}$ to a proper subset of
$\CPTP(E_{j-1}, Q_j E_j)$.
Such a restriction is described in Ref.~\cite[Section 9]{inprep_BDR3}.
While we do not formally incorporate this restriction here, our methods
are fully compatible with the use of the flag-state squasher. For comparison, Ref.~\cite{kamin_renyi_2025} also makes use of the
flag-state squasher, but does so only after applying MEAT in the
infinite-dimensional setting, at the level of a single-round
optimization, which we avoid here.

\section{Plots} \label{sec:plotsMEAT}

To generate key rates from \cref{theorem:abstractsecuritystatement}, we must compute the quantity $\kappaQKDfunc{f_{|\cP_1^{j-1}}}{\sigma^{(j)}_{A_j}}{\QKDGmap{j}}$ defined in \cref{eq:kappacompute}. We have shown in \cref{lemma:MEATuniform_lowerbound} that this quantity can be expressed as a finite-dimensional convex optimization problem, making it suitable for numerical computation. If the protocol of interest is such that Alice and Bob perform the same operation in each round, then all the maps $\QKDGmap{j}$ are identical and states $\sigma^{(j)}_{A_j}$ are identical. Furthermore, one can choose the same tradeoff function $f_{|\cP_1^{j-1}}$ for all $j, \cP_1^{j-1}$. Thus the dependence on $j$ disappears and in what follows
we drop the subscript $j$ and write it more compactly as $\kappaQKDfunc{f}{\sigma_A}{\QKDGmap{}}$:
\begin{equation} \label{eq:decoykappacomputenumerics}
    \begin{aligned}
        \kappaQKDfunc{f}{\sigma_A}{\QKDGmap{}}
        &\leq
        \inf_{\Qattack{}\in\Qset{}}
        \Halpha[f](S| \CP \CPhat \widehat{E})_{\QKDGmap{}\left[\pf\left(\Qattack{}\left[\sourcesymbol_{A A''}\right]\right)\right]}, 
    \end{aligned}
\end{equation}
The above optimization problem is exactly the one solved in Ref.~\cite{araujo2023quantum,kamin_renyi_2025}.\footnote{Note that these works impose certain conditions in order to render the
associated optimization problems more tractable. Informally, they require
the probability of $\test$ rounds to be a fixed value, and they assign the same
tradeoff score to all public announcements occurring in $\gen$ rounds.
These assumptions are not severely restrictive in practice. We do not go
into the details of these conditions here, except to note that the
protocols considered in this thesis satisfy them.}  Furthermore, one also has a choice in choosing the tradeoff functions $f_{\cP_1^{j-1}}$. In fact, a suitable choice of these functions is critically important to obtain good performance, and a procedure for choosing optimal functions is provided in Ref.~\cite{kamin_renyi_2025}.

With these details in place, we are now ready to compute key rates using
the machinery and codebase provided in Ref.~\cite{kamin_renyi_2025}.
We emphasize that all credit for the numerical methods and implementation
belongs to the authors of that work. The plots presented here are intended
purely for illustrative purposes and are generated using their publicly
available code, with only minimal modifications on our part. Furthermore, Ref.~\cite{kamin_renyi_2025} computes key rates for a larger class of protocols, including passive setups using the flag-state squasher, includes imperfections, and also optimized over many of the free parameters. 

\subsection{Qubit BB84}

\begin{figure}[h!]
    \centering
    \includegraphics[width=\linewidth]{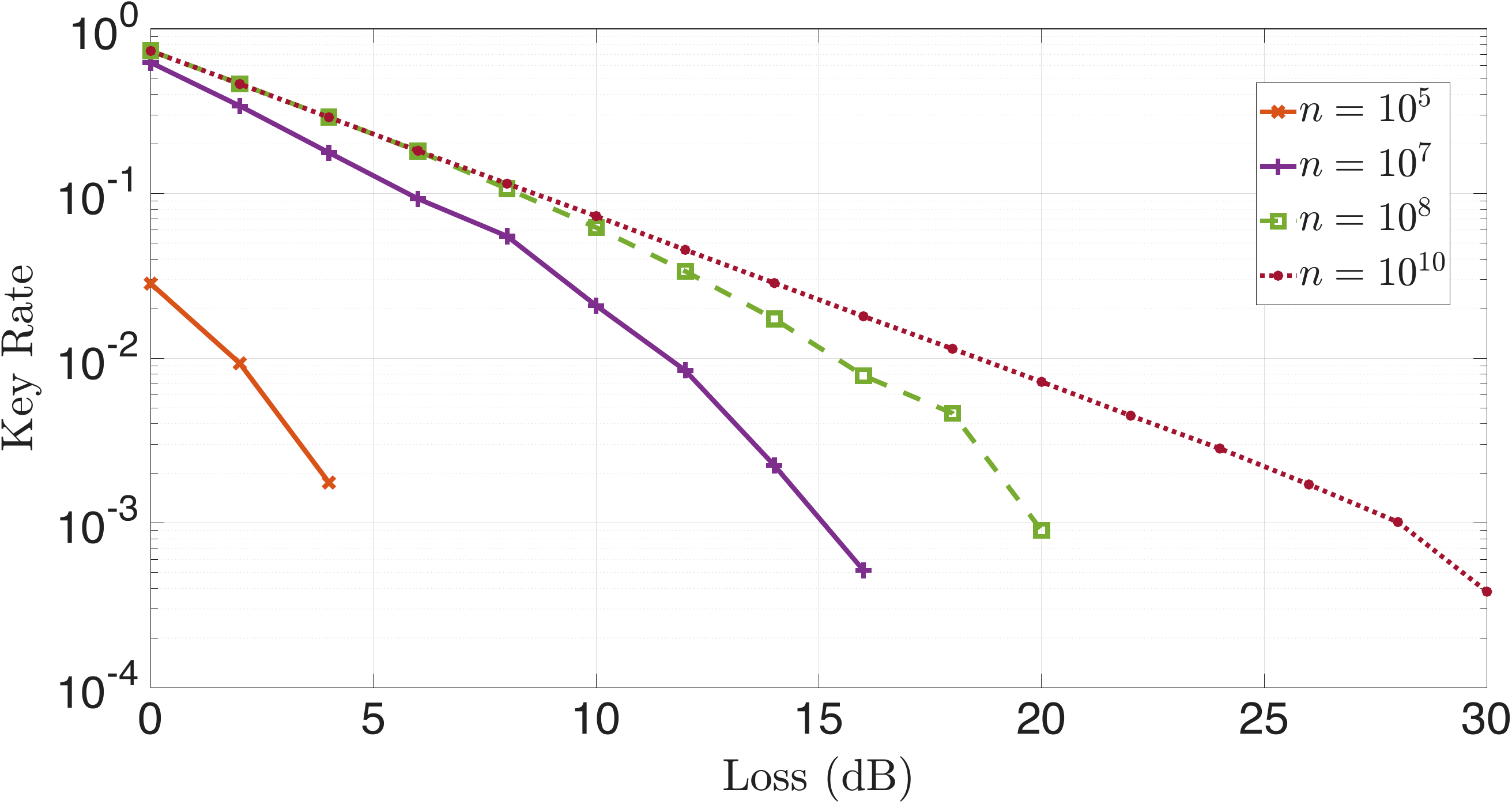}
    \caption{Key rate for variable-length qubit BB84 protocol plotted against loss, using MEAT, computed using open source code from Ref.~\cite{kamin_renyi_2025}.}
    \label{fig:qubitBB84_meat}
\end{figure}

We plot key rates for a loss-only channel in \cref{fig:qubitBB84_meat}, for the qubit BB84 protocol. We set $\epsPA = \epsAT = 10^{-10}$ leading to an overall security parameter of $\epssecure = 2 \times 10^{-10}$. A round is chosen to be a $\test$ round with probability $\gamma=0.1$, and a $\gen$ round with probability $1-\gamma$. For both $\test$ and $\gen$ rounds, Alice (and Bob) chooses the $\Zbasis$ basis with probability $0.9$ ($\Xbasis$ with probability $0.1$), with each basis state sent with equal probability. We set $\leak(\cP_1^n) = f_\mathrm{EC} n H(S | Y \CP)_{\freq(\cP_1^n)}$ (meaning that the entropy is computed for the distribution implied by the announcements), with $f_\mathrm{EC} = 1.16$. We use the optimized values of $\alpha$ for each loss value, computed in Ref.~\cite{kamin_renyi_2025}.\footnote{Note that the protocol for which these optimal values were computed is a slightly different one from the one we consider here. This does not pose a problem, since every value of $\alpha$ yields a valid key rate.}. We plot key rates for the typical observations expected  for the given channel parameters.

\subsection{Decoy-state BB84}
\begin{figure}[h!]
    \centering
    \includegraphics[width=\linewidth]{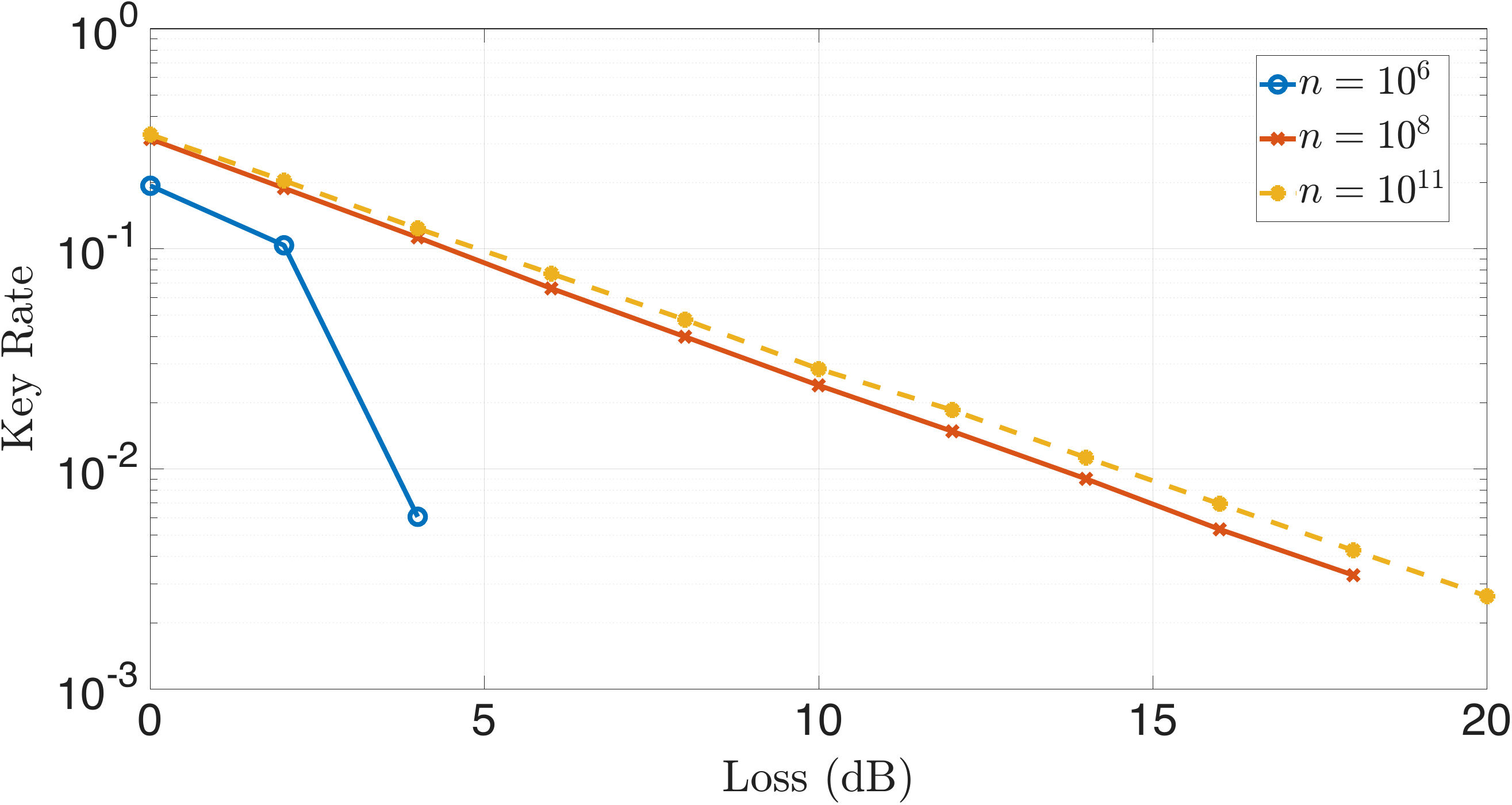}
    \caption{Key rate for the variable-length decoy-state BB84 protocol plotted as a function of loss, computed using MEAT. While key rates can, in principle, be evaluated at higher loss values, this becomes numerically challenging due to the optimization over the {\Renyi} parameter. The results shown here are obtained using open-source code from Ref.~\cite{kamin_renyi_2025}; we refer the reader to that work for additional plots covering a broader range of loss values.}
\label{fig:decoyBB84_meat}
\end{figure}

We plot key rates for a loss-only channel in \cref{fig:decoyBB84_meat} for the decoy-state BB84 protocol. We set $\epsPA = \epsAT = 10^{-10}$ leading to an overall security parameter of $\epssecure = 2 \times 10^{-10}$. The decoy intensities are given by $\mu_1=1,\mu_2=0.1,\mu_3=0.001$, and the probability of $\test$ rounds is set to $\gamma=0.1$.
We consider a protocol where, in $\test$ rounds, Alice sends $\Xbasis$ states with equal probability, using all three intensities (also with equal probabilities). In the $\gen$ rounds, Alice uses only $m_1$ intensity, and sends both $\Zbasis$ states with equal probability.  We set $\leak(\cP_1^n) = f_\mathrm{EC} n H(S | Y \CP)_{\freq(\cP_1^n)}$ (meaning that the entropy is computed for the distribution implied by the announcements), with $f_\mathrm{EC} = 1.16$. 
We plot key rates for the typical observations expected  for the given channel parameters. For every key rate point, we optimize over $\alpha$. We consider active detection setup by Bob, and use the qubit squasher, with double-clicks in each basis randomly mapped to single-clicks.

\section{Summary and Outlook} \label{sec:summaryMEAT}

The analysis presented in this chapter constitutes a very general and flexible application of the MEAT technique to QKD security analysis. It begins by considering an abstract formulation of a generic QKD protocol (\nameref{prot:abstractqkdprotocol}) and establishes security under the assumption of finite-dimensional state preparation and measurements. This framework is then extended to practical optical implementations (which inherently involve infinite-dimensional systems) through the use of standard techniques such as squashing (\cref{lemma:SquashMapSecurityMEAT}) and source maps (\cref{lemma:sourcemapsecurityMEAT}). The framework is general and can, in principle, accommodate new classes of source and squashing maps that have not yet been formalized in the literature - for instance, those designed to model correlated imperfections. The entire analysis supports on-the-fly announcements, a feature that is important to efficiently utilize classical memory in implementations, and fully adaptive key rates (\cref{remark:adaptivef}), an aspect especially relevant for satellite QKD. Moreover, much of our analysis—such as the use of squashing, source maps, and the treatment of an
infinite-dimensional Eve—is not tied to the specific formulation of MEAT used here. These
components are likely to remain valid for future variations in QKD proof techniques, such as the
new EAT variants, for example.

As an example of the modularity and flexibility of our approach, consider the security proof of
the six-state protocol \cite{bruss_optimal_1998} implemented using, for instance, a single-photon
source. This can be obtained straightforwardly by (1) expressing the protocol as an instance of
our \nameref{prot:abstractqkdprotocol}, (2) applying the appropriate squashing and source map\footnote{Source maps are not required if the single-photon source is assumed to have ideal properties.},
and (3) evaluating the resulting optimization problem using Ref.~\cite{kamin_renyi_2025}. This procedure is outlined as a recipe in
\cref{subsubsec:recipe}. In fact, the present framework can be applied to essentially the same protocol considered in Ref.~\cite{mizutani2025protocolleveldescriptionselfcontainedsecurity}, which likewise aims to provide a self-contained and rigorous security analysis of decoy-state BB84, by following the steps outlined above.
The only minor distinction is that Ref.~\cite{mizutani2025protocolleveldescriptionselfcontainedsecurity} employs dual-universal$_2$
  hash families for privacy amplification, whereas we require  universal$_2$
  hash families.
  
More generally, as discussed in \cref{remark:adaptivef}, our results also extend to fully adaptive protocols - a feature that is not currently available in any other proof technique. Moreover, it may prove particularly valuable for scenarios such as satellite QKD, where the channel behaviour changes with time. While this work provides the complete theoretical framework required for such analyses, a detailed study of various ways to update the optimal trade-off function and its impact on achievable key rates in time-varying channels is left for future work.

Another important direction for future work is the detailed treatment of device imperfections. A  wide class of imperfections can, in principle, be incorporated within our framework, as outlined in \cite[Section 11]{inprep_BDR3}. Integrating them rigorously with the above framework is an open and formidable task. Nonetheless,  the framework and results presented in this work provide the essential tools and foundations for such an endeavor.

In summary, this chapter establishes a rigorous, extensible, and implementation-oriented foundation for the security analysis of practical QKD protocols. The key ingredient in this analysis is the generality and the MEAT statement itself, which then lends itself to a very general analysis of QKD protocols.

\chapter{Classical Authentication in QKD security proofs} \label{chap:classicalauthentication}
\epigraph{Where we consider classical communication and find that the security definition of QKD is wrong, and fix it; where we find that the corresponding security analysis of QKD is flawed too, and fix \emph{that} as well; and where our proposed solution fixes both these problems retroactively. 
}{}

QKD protocols rely on both quantum and classical communication. In particular, they require the use of authenticated classical channels to ensure the integrity of exchanged messages. A substantial body of work addresses this requirement, including information-theoretically secure authentication schemes~\cite{wegman_new_1981}, performance-optimized constructions ~\cite{krawzyck_LFSR-based_1994,kiktenko_lightweight_2020}, and formulations within composable security frameworks~\cite{portmann_key_2014}. Typically, in most existing QKD security proofs (see for instance \cite{tomamichel_largely_2017,mizutani2025protocolleveldescriptionselfcontainedsecurity}), the classical authentication is assumed to behave “honestly”: the authenticated channel never aborts, and all classical messages are delivered faithfully with their original timing preserved. All our security analysis in thesis so far has been under this assumption. We refer to this as the honest authentication setting.
While this assumption simplifies security analyses, such an idealized channel cannot be realized in practice.

What can be constructed in practice is an authenticated channel that is close in functionality to one which either transmits each message faithfully or delivers a special symbol $\authabort$ to the receiving party, indicating that authentication has failed \cite{portmann_security_2022,portmann_key_2014}.
An authentication abort occurs, for example,  when the authentication tag attached to a message does not match the expected value, which can happen if the message has been modified or if an adversary attempts to inject a new message without the correct tag.
In either case, the receiver discards the message and registers $\authabort$.
Furthermore, an adversary (Eve) is permitted to delay, block, or reorder messages, or perform other timing-related manipulations—some of which may themselves result in an $\authabort$ being delivered to the receiver. We refer to this as the {\realistic}\footnote{By ``\realistic'' we mean a setting that is not itself perfectly achievable, but for which practical constructions can approximate the idealized functionality in the composable sense. See \cref{subsec:classicalcommunicationsmodel} for more discussion.} authentication setting.

This discrepancy leads to the following issues in the QKD analysis:
\begin{enumerate}
    \item First, the QKD protocol must now specify what happens when the authentication aborts. The natural choice here is to abort the protocol whenever authentication aborts (and attempt to communicate to the other party that one has aborted). 
    \item Second, only the receiving party is informed of the authentication aborts. Thus, Eve can 
    generally
    force \textit{one} party to abort in the QKD protocol while the other accepts (for instance by only interfering with the final message sent between the two parties). 
    \item Third, since the timing of messages may be modified, one can no longer assume a fixed ordering of actions performed by Alice and Bob. For instance, in the absence of synchronized clock assumptions in the security proof, Alice and Bob typically use messages to inform each other of completion of various operations in the protocol. If the timing of these messages is affected, then the ordering of actions performed by Alice and Bob is also affected.
\end{enumerate}  

These challenges arise in any setting where the authenticated classical channel may abort asymmetrically and where message timing is not rigidly preserved. Moreover, they require significant modifications to the security analysis:
\begin{itemize}
\item Due to Eve's ability to force asymmetric aborts in the realistic authentication setting, the usual security definition of QKD as specified in \cite{portmann_security_2022,ferradini2025definingsecurityquantumkey,tupkary2025qkdsecurityproofsdecoystate,ben-or_universal_2004,renner_security_2005}, which only covers symmetric aborts, cannot be satisfied. This observation has been noted in prior works, see for instance, \cite[Section VII]{portmann_security_2022} 
\cite[Section 5.2.1]{tupkary2025qkdsecurityproofsdecoystate} \cite{ferradini2025definingsecurityquantumkey}. 
\item  Even if one chooses to disregard the problem of one-sided aborts, it remains important to recognize that existing QKD security analyses rely (often implicitly) on a fixed time ordering of classical communications, which is not guaranteed in practice.
\end{itemize}

In this chapter, we address this gap as follows.
In \cref{chapauth:secdef}, we introduce the modified security definition for QKD protocols from Ref.~\cite{ferradini2025definingsecurityquantumkey}, which remains valid even when authentication can lead to receiver-side aborts.
This definition generalizes the usual trace-distance criterion (\cref{def:qkdsecuritysymmetric}) by explicitly incorporating asymmetric abort events.
In \cref{subsec:classicalcommunicationsmodel}, we specify a detailed model of interactive classical communication where authentication can result in one-sided aborts and where the adversary may modify the timing of classical messages (possibly resulting in authentication aborts).
We also briefly discuss how such a model can be implemented in practice. We then consider the scenario where we have an arbitrary ``core" QKD protocol, which is followed by a short authentication post-processing (APP) step, described in \cref{subsec:APPprotocol}.
Our goal is to analyze the security of the combined core QKD + APP protocol in the {\realistic} setting where authentication can lead to asymmetric aborts and where message timing may be influenced by the adversary.
In \cref{subsec:reductionstatement}, we state our main result: a reduction theorem showing that the security analysis of this combined protocol can be reduced to that of the core QKD protocol alone, under the assumption of honest authentication. This provides a clean separation between authentication and QKD security analysis, since one need not be concerned with authentication aborting or the timing of messages during classical communication while studying the security of the core QKD protocol. Moreover, it also retroactively lifts all prior QKD security proofs that were undertaken in the regime where the authentication was assumed to be honest to the more {\realistic} scenario, with the caveat that the protocol must now include the additional authentication post-processing step. The proof of this reduction is presented in \cref{sec:provingthereduction}.
Finally, in \cref{sec:delayedauthentication}, motivated by practical considerations of authentication key usage, we extend our analysis to the scenario of delayed authentication, where all classical communication during the core QKD protocol is undertaken using unauthenticated classical communication, and the entire communication transcript is authenticated at the end of the protocol. We  also discuss the trade-offs associated with this choice.
Concluding remarks are presented in \cref{sec:authenticationsummary}. 

We note that some prior works have analyzed the security of QKD protocols in conjunction with a realistic model of authentication, see, for example Ref.~\cite{kon_quantumauthenticated_2024}. However, the analysis in that work is tailored to a specific QKD protocol combined with a specific authentication protocol. In contrast, we establish a general result that applies to generic QKD protocols.

\section{Security Definition with asymmetric aborts} \label{chapauth:secdef}
Let us recall some notation we used for stating the security definition in \cref{chap:qkdbackground}. 
Let us focus on the output state of a generic QKD protocol, defined on the registers $K_A K_B \Esecdef$, in the setting where asymmetric aborts are possible.  Alice and Bob possess classical registers $K_A$ and $K_B$. 
The classical registers $K_A$, $K_B$ encode keys of arbitrary length for Alice and Bob --- this is formalized by having $K_A$ consist of a direct sum $K_A = \bigoplus_{l_A} K_A^{l_A}$, where $K_A^{l_A}$ is a classical register holding keys of length $l_A$, and analogously for $K_B$. 
We treat any party aborting as them storing a key of length $0$ in their registers, which we denote with a special symbol $\bot$. The $\Esecdef$ register denotes all of Eve's information at the end of the QKD protocol, and may include a copy of the classical communications that occurred in the protocol. The precise modeling of the authenticated classical communication, message timing, and one-sided authentication aborts is not essential for the security definition of QKD, which is only concerned with the output state of the QKD protocol, and will therefore be deferred to \cref{subsec:classicalcommunicationsmodel}. What matters for now is simply that the QKD protocol may output a state in which one party aborts while the other does not.

The output state of a generic QKD protocol can be written as \cite{ferradini2025definingsecurityquantumkey}:
\begin{equation} \label{cref:realoutputstate}
\rho^\text{real}_{K_A K_B  \Esecdef} \coloneq \bigoplus_{l_A, l_B \in \keyspace} \Pr(\OlAlB) \rho^\mathrm{real}_{K_A^{l_A} K_B^{l_B}  \Esecdef | \OlAlB},
\end{equation}
where $\OlAlB
$ denotes the event that Alice and Bob produce keys of lengths $l_A$ and $l_B$, respectively, and $\keyspace$ denotes the set of possible output key length combinations. The subregisters $K_A^{l_A}$ and $K_B^{l_B}$ store the keys of those specific lengths. As argued in Ref.~\cite{ferradini2025definingsecurityquantumkey}, 
we restrict the set of possible output key length combinations to the following: 
\begin{equation}
    \keyspace = \{  (l_A, l_B) \mid l_A = l_B \;\lor\; l_A = 0 \;\lor\; l_B = 0 \}.
\end{equation}
That is, Alice and Bob either share keys of the same length, or at least one of them aborts the protocol.

The ideal output state $\rho^\mathrm{ideal}_{K_A K_B  \Esecdef}$ is defined to be the one obtained by acting a map $\idealmap \in \CPTP( K_A K_B, K_A K_B)$ acting on the real output state $\rho^\mathrm{real}_{K_A K_B  \Esecdef}$ as
\begin{equation} \label{eq:realidealstates}
       \rho^\mathrm{ideal}_{K_A K_B  \Esecdef} \coloneq \idealmap\left[\rho^\mathrm{real}_{K_A K_B  \Esecdef} \right]. 
\end{equation}
Throughout this chapter, we adopt the convention that the ideal and real states of various kinds are related analogously to \cref{eq:realidealstates}; that is, the ideal state is obtained by applying the map $\idealmap$ to the corresponding real state. The action of the map $\idealmap$ was defined in \cref{sec:securitydefinition} for scenarios where key lengths are the same. We simply extend that definition to the scenario where the key lengths are different. The map $\idealmap$ performs the following operations:
\begin{itemize}
    \item It looks at the length of the keys stored in registers $K_A, K_B$ to compute $l_A, l_B$. 
    \item It replaces the $K_A,K_B$ registers with the state 
  \begin{equation}\label{eq:taukAkB}
\begin{aligned}
\tau^{l_A, l_B}_{K_A K_B} &\coloneq
\begin{cases}
\displaystyle \frac{1}{2^{l_A}} \sum_{k \in \{0,1\}^{l_A}} \ketbra{kk}_{K_A K_B}, 
& \text{if } l_A = l_B, \\[1.2em]
\tau^{l_A}_{K_A} \otimes \tau^{l_B}_{K_B}
& \text{if } l_A \neq l_B,
\end{cases} \\[1.2em]
\tau^{l_A}_{K_A} &\coloneq \frac{1}{2^{l_A}} \sum_{k \in \{0,1\}^{l_A}} \ketbra{k}_{K_A}, \qquad 
\tau^{l_B}_{K_B} \coloneq \frac{1}{2^{l_B}} \sum_{k \in \{0,1\}^{l_B}} \ketbra{k}_{K_B} .
\end{aligned}
\end{equation}
    That is, if the output lengths are the same, the key registers are replaced with perfectly uniform identical keys of that length, independent of all other registers (which the same action as before). If the output key lengths are not the same, the key registers are \emph{individually} replaced with perfectly uniform keys of the corresponding lengths, independent of all other registers.
\end{itemize}

Thus intuitively, any key obtained from the ideal state is safe to use, regardless of symmetric or asymmetric aborts, since the key is always  independent of Eve's side-information registers. Note that we have
\begin{equation}
    \begin{aligned}
          \rho^\text{ideal}_{K_A K_B  \Esecdef}  &= \idealmap \left[ \rho^\text{real}_{K_A K_B  \Esecdef}   \right] \\
        &= \sum_{ \substack{l_A = l_B = l \\ l_A,l_B \in \keyspace}}  \Pr(\Omega_{l,l})  \tau^{l,l}_{K_A K_B} \otimes \rho^\text{real}_{  \Esecdef | \Omega_{l,l}} + \sum_{\substack{l_A \neq l_B \\ l_A,l_B \in \keyspace}} \Pr(\OlAlB) \tau^{l_A}_{K_B} \otimes \tau^{l_B}_{K_B} \otimes \rho^\text{real}_{  \Esecdef | \OlAlB}.
    \end{aligned}
\end{equation}
Moreover, $\idealmap$ acts independently on each combination of key length subregisters, i.e, it can be written as 
\begin{equation} \label{eq:idealmapdecomp}
    \idealmap = \bigoplus_{l_A, l_B} \idealmap^{(l_A,l_B)},  \qquad \text{ where $\idealmap^{(l_A,l_B)} \in \CPTP(K_A^{l_A} K_B^{l_B} , K_A^{l_A} K_B^{l_B} )$.}
\end{equation}

We now state the security definition, which will require us to talk about protocols and their corresponding output states. We therefore set up some notation first.
Let us consider a protocol \(\mathcal{P}\). What we will typically be concerned with is the set of output states produced by the protocol, which we denote by \(\mathcal{W}(\mathcal{P})\). This set of possible output states  depends on various assumptions ($\mathcal{W}$) under which we analyze the protocol. In this chapter, these assumptions will be related to the authenticated channel that is used during communication, and are explained in \cref{subsec:classicalcommunicationsmodel}.\footnote{For example, an assumption may be that Alice and Bob use unauthenticated classical communication, and Eve is allowed to tamper with classical messages. A different assumption may be that they use authenticated classical communication, and no tampering of the classical communication is allowed. These give rise to different sets of possible output states.}  Thus, these assumptions fix the set of possible attacks Eve can perform, and the protocol $\mathcal{P}$ along with Eve's attack together determine a channel mapping the relevant input states to output states. When we write \(\mathcal{W}(\mathcal{P})\), we refer to the set of all output states that can arise under all possible choices of Eve’s attack. Moreover, when we compose two protocols via \(\mathcal{P}_2 \circ \mathcal{P}_1\), the composed protocol is understood in the sense of channel composition, where the resulting channel is fixed by the protocol descriptions and Eve's attack on both protocols.  This level of formalism is sufficient for our purposes. 
 We can now state the QKD security definition.

\begin{definition}[QKD Security with asymmetric aborts \cite{ferradini2025definingsecurityquantumkey}] \label{def:qkdsecurityasymmetric}
Let $\QKDprotocol$ be a QKD protocol, and let $\rho^\text{real}_{K_A K_B \Esecdef}$ be the output state of the QKD protocol, and let $\mathcal{W}\left( \QKDprotocol \right)$ denote the set of possible output states of the QKD protocol.  Let $\rho^\text{ideal}_{K_A K_B  \Esecdef}$ be the ideal output state, obtained by acting the map $\idealmap$ on the actual output state. That is
\begin{equation} \label{eq:secdefrealidealgeneric}
    \begin{aligned}
        \rho^\text{real}_{K_A K_B  \Esecdef} &\coloneq \bigoplus_{l_A, l_B} \Pr(\OlAlB) \rho^\text{real}_{K_A^{l_A} K_B^{l_B} \Esecdef| \OlAlB} \\
        \rho^\text{ideal}_{K_A K_B \Esecdef}  &\coloneq \idealmap \left[ \rho^\text{real}_{K_A K_B  \Esecdef}   \right].
    \end{aligned}
\end{equation}
Then, the (variable-length) QKD protocol is $\epssecure$-secure if, for all output state $\rho^\text{real}_{K_A K_B \Esecdef} \in \mathcal{W}(\QKDprotocol)$, the following inequality is satisfied\footnote{As noted earlier, in Ref.~\cite{ferradini2025definingsecurityquantumkey}, the trace norm appearing in the security definition is not divided by $2$. In contrast, the typical security definition \cite{ben-or_universal_2004,portmann_security_2022} includes the explicit factor of $1/2$.  The definition used in Ref.~\cite{ferradini2025definingsecurityquantumkey} is deliberate and well motivated within that work; we stress that the difference amounts only to an overall factor of $2$ in the  security parameter.}:
\begin{equation}
    \tracedist{ \rho^\text{real}_{K_A K_B  \Esecdef} -  \rho^\text{ideal}_{K_A K_B  \Esecdef}} \leq \epssecure.
    \end{equation}
\end{definition}

Note that \cref{def:qkdsecurityasymmetric} generalizes the standard definition of QKD security (\cref{def:qkdsecuritysymmetric}) \cite{ben-or_universal_2004,portmann_security_2022,ferradini2025definingsecurityquantumkey} (which corresponds to the case where $\Pr(\OlAlB ) = 0$ whenever $l_A \neq l_B$ in \cref{eq:secdefrealidealgeneric})
  to the case where the output state may have asymmetric aborts. Furthermore, observe that the definition depends on the set of possible output states $\mathcal{W}(\QKDprotocol)$, which in turn depends on the authentication setting under consideration. Under the assumption of honest authentication (and appropriate protocol design), asymmetric aborts cannot occur. Throughout this chapter, we adopt the convention that the ideal and real states of various kinds are related analogously to \cref{eq:secdefrealidealgeneric}; that is, the ideal state is obtained by applying the map $\idealmap$ to the corresponding real state.

\section{Model and Reduction Statement} \label{sec:modelandreductionstatement}
We will now specify various protocols, various assumptions on authentication and the classical communication model, and state our reduction theorem.

\subsection{Authenticated Classical Communication Model} \label{subsec:classicalcommunicationsmodel}
We will now explain the authenticated classical communications model we assume for the {\realistic} authentication setting. We stress that this model still contains idealized properties, in the sense that Eve has zero probability of faking messages without the authentication aborting. However, it is {\realistic} in the sense that it accounts for timing tampering and one-sided aborts. Moreover, we believe that an $\varepsilon_\mathrm{auth}$-close construction of such a channel is achievable within a composable framework (see \cref{subsubsec:implementingclassicalcomms} later for further discussion).

We begin by introducing notation to describe the sending and receiving of classical messages between Alice and Bob, which are mediated by Eve. For each message, we associate a register label from both the sender’s and receiver’s perspectives, and a global time (see \cref{fig:classicalcommmodel}). We emphasize that no assumptions are made about the alignment between the sender’s and receiver’s message ordering; these may differ in the presence of an active adversary. Moreover, we do not assume that Alice and Bob share synchronized clocks, nor do they need to record the timing of messages during the protocol. The times introduced here are used only for the purpose of theoretical analysis and refer to the global time when these events occur.

\begin{itemize}
    \item \textbf{Alice sending: } Alice sends messages  in registers $C^{(i)}_{A \rightarrow E}$, where the index $i$ denotes the ordering of messages from her perspective, i.e, in the order she sends them. We let $t^{(i)}_{A \rightarrow E}$ denote the time at which this message leaves Alice. 
     \item \textbf{Alice receiving: } Alice receives messages  $C^{(i)}_{E \rightarrow A}$, where the index $i$ denotes the ordering of messages from her perspective, i.e, in the order she receives them. We let $t^{(i)}_{E \rightarrow A}$ denote the time at which this message is received by Alice. 
    \item \textbf{Bob sending:} Bob sends messages in registers $C^{(i)}_{B \rightarrow E}$, where the index $i$ denotes the ordering of messages from Bob’s perspective, i.e, in the order he sends them. We let $t^{(i)}_{B \rightarrow E}$ denote the time at which this message leaves Bob. 

    \item \textbf{Bob receiving: } Bob receives messages  $C^{(i)}_{E \rightarrow B}$, where the index  $i$ denotes the ordering of messages from his perspective, i.e, in the order he receives them. We let $t^{(i)}_{E \rightarrow B}$ denote the time at which this message is received by Bob. 
\end{itemize}
Thus, each sent (received) message is indexed from the sender’s (receiver’s) point of view, and the time values reflect the true time at which these events occur (which are not known to Alice and Bob). We will modify this setting later in \cref{sec:delayedauthentication}.

\newcommand{\rowsep}{1.3}            
\tikzset{
  participant/.style={font=\small\bfseries},
  lifeline/.style={gray,dashed},
  message/.style={-Latex,thick},    
  attackbox/.style={draw,rounded corners,fill=blue!8,inner sep=4pt,font=\footnotesize}
}

\newcommand{\nextrow}[2]{\coordinate (#1) at ($(#2)+(0,-\rowsep)$);}
\begin{figure}[ht!]
  \centering
  \begin{tikzpicture}[every node/.style={inner sep=1pt}]
    \node[participant] (Alice) {Alice};
    \node[participant] (Eve)   [right=5cm of Alice] {Eve};
    \node[participant] (Bob)   [right=5cm of Eve]   {Bob};

    \draw[lifeline] (Alice) -- coordinate (AliceEnd) ++(0,-8);
    \draw[lifeline] (Eve)   -- ++(0,-8);
    \draw[lifeline] (Bob)   -- coordinate (BobEnd) ++(0,-8);

    \coordinate (A0) at ($(Alice)+(0,-\rowsep)$);
    \coordinate (B0) at ($(Bob)+(0,-\rowsep)$);

    \node[attackbox,align=center,minimum height=7.2cm,anchor=north] (Ebox)
      at ($(Eve)+(0,-0.7)$) {Eve
      may delay, drop, \\
      or replace messages \\
      (subject to the specified details \\
      of the communication model)};

\node at ($ ($(Alice)!0.5!(A0)$) + ( 2,-0.5)$) {$\vdots$};
\node at ($ ($(Bob)!0.5!(B0)$)   + (-2,-0.5)$) {$\vdots$};

    \nextrow{A1}{A0}
    \nextrow{B1}{B0}

    \draw[message] (A1) -- node[midway,above] {$C_{A\to E}^{(5)}$} (Ebox.west |- A1);
    \draw[message] (Ebox.east |- A1) -- node[midway,above] {$C_{E\to B}^{(5)}$} (B1);

    \nextrow{A2}{A1}
    \nextrow{B2}{B1}

    \draw[message] (B2) -- node[midway,above] {$C_{B\to E}^{(14)}$} (Ebox.east |- B2);

    \nextrow{A3}{A2}
    \nextrow{B3}{B2}

    \draw[message] (Ebox.east |- A3) -- node[midway,above] {$C_{E\to B}^{(6)}$} (B3);

    \nextrow{A4}{A3}
    \nextrow{B4}{B3}

    \draw[message] (B4) -- node[midway,above] {$C_{B\to E}^{(15)}$} (Ebox.east |- B4);
    \draw[message] (Ebox.west |- B4) -- node[midway,above] {$C_{E\to A}^{(14)}$} (A4);

    \node at ($(A4)+(2,-0.5*\rowsep)$) {$\vdots$};
    \node at ($(B4)+(-2,-0.5*\rowsep)$) {$\vdots$};

\draw[->, dashed]
  ($(Alice)+(-2,0)$) -- ($(AliceEnd)+(-2,-4)$)
  node[midway, left] {time};

  \end{tikzpicture}
\caption{The {\realistic} authenticated  classical communication model used in this chapter. Messages pass through Eve, who may delay, drop, or substitute them with $\authabort$, subject to the constraints described in \cref{subsec:classicalcommunicationsmodel}. Time flows from top to bottom in the figure, which illustrates an example scenario: in earlier parts of the protocol (not shown in the figure), 4 messages have been sent from Alice to Bob, and 13 messages from Bob to Alice.
Eve does not interfere with Alice’s 5th message to Bob. However, she chooses to delay Bob’s 14th message. (Presumably, Alice does not send a new message during this period because she is waiting to receive one.) During the delay, Eve receives Bob’s 15th message and also delivers the 6th message to Bob. According to our communication model, this implies that $C^{(6)}_{E \rightarrow B}$ must be $\authabort$, since it was received before Alice sent her $6$th message.}
\label{fig:classicalcommmodel}
\end{figure}
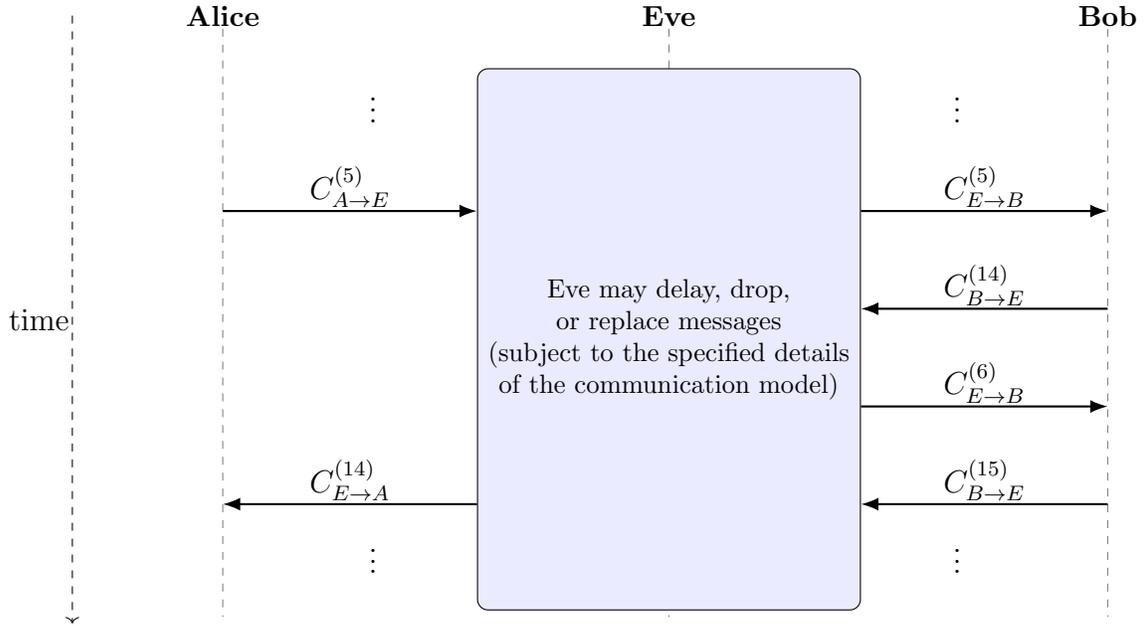

We assume that the classical authenticated channel  between the two honest parties has the following properties (see \cref{fig:classicalcommmodel}):
\begin{enumerate}
    \item \textbf{Timing:} If the $i$th message is received \emph{before} the $i$th message was sent, then the received message is the special symbol $\authabort$. Formally,
    \begin{equation}
        \begin{aligned}
        &t^{(i)}_{A\rightarrow E} > t^{(i)}_{E \rightarrow B} \quad \implies \quad C^{(i)}_{E\rightarrow B} \text{ stores } \authabort, \quad \forall i.  \\
        &t^{(i)}_{B\rightarrow E} > t^{(i)}_{E \rightarrow A} \quad \implies \quad C^{(i)}_{E\rightarrow A} \text{ stores }\authabort, \quad \forall i.  
        \end{aligned}
    \end{equation}
We refer to the case where $t^{(i)}_{B\rightarrow E} \leq  t^{(i)}_{E \rightarrow A}$ and  $t^{(i)}_{A\rightarrow E} \leq t^{(i)}_{E \rightarrow B} $ holds  for all $i$ as ``relative time ordering being preserved".

    \item \textbf{Modifying messages:} If $i$th message is received \textit{after} the $i$th message was sent, then the received message is either a copy of the sent message or it is an $\authabort$. Formally,
     \begin{equation}
        \begin{aligned}
        t^{(i)}_{A\rightarrow E} \leq t^{(i)}_{E \rightarrow B} \quad \implies \quad &\text{Either $C^{(i)}_{E\rightarrow B}$ and $C^{(i)}_{A\rightarrow E}$ store identical messages,}\\
        &\text{or $C^{(i)}_{E\rightarrow B}$ stores  \authabort}, \quad \forall i.  \\
        t^{(i)}_{B\rightarrow E} \leq t^{(i)}_{E \rightarrow A} \quad \implies \quad &\text{Either $C^{(i)}_{E\rightarrow A}$ and $C^{(i)}_{B\rightarrow E}$ store identical messages,} \\
        &\text{or $C^{(i)}_{E\rightarrow A}$ stores  \authabort}, \quad \forall i.  \\
        \end{aligned}
    \end{equation}
We assume if Eve attempts to block messages 
for longer than some preselected finite duration, then this results in the ``received message'' being an \authabort (which can be implemented simply by having the receiving party record \authabort after that duration has elapsed).
\end{enumerate}

For a given QKD protocol $\coreQKDprotocol$, we denote the set of output states possible in the above model for authenticated classical communication via $\worldreal(\coreQKDprotocol)$. In the honest authentication setting, we assume that Eve is not allowed to perform any operation that can results in an $\authabort$. That is, we are guaranteed to have:
\begin{equation}
    \begin{aligned}
        t^{(i)}_{A\rightarrow E} &\leq t^{(i)}_{E \rightarrow B} \qquad \qquad \forall i \\
        t^{(i)}_{B\rightarrow E} &\leq t^{(i)}_{E \rightarrow A} \qquad \qquad \forall i \\
        C^{(i)}_{E\rightarrow B} & \text{ and } C^{(i)}_{A\rightarrow E} \text{ store identical messages } \qquad \qquad \forall i \\  C^{(i)}_{E\rightarrow A} & \text{ and } C^{(i)}_{B\rightarrow E} \text{ store identical messages } \qquad \qquad \forall i
        \end{aligned}
\end{equation}
We denote the set of output states possible under this setting via $\worldhonest(\coreQKDprotocol)$.

\subsection{Implementing the authenticated communication model} \label{subsubsec:implementingclassicalcomms}
We note that there exist implementations of authenticated classical channels that closely approximate the functionality described by our model. One such implementation involves Alice and Bob sharing a pool of pre-distributed keys, which are used to authenticate each classical message using a message authentication scheme. As messages are sent and received, the parties iterate through their sending and receiving key pools. For example, they may employ Wegman–Carter authentication \cite{wegman_new_1981}, potentially with key recycling, as discussed in Ref.~\cite{portmann_key_2014}.

In such a setup, assuming the authentication keys remain secret, any attempt by Eve to modify a message will result in an invalid authentication tag with high probability, which causes the receiver to interpret the message as an $\authabort$. Moreover, if Eve attempts to deliver a message to the receiver \emph{before} the corresponding message has been sent, she will not have access to a valid message–tag pair. In this case as well, the receiver will reject the message as invalid (with high probability) and interpret it as $\authabort$. The same holds if e.g.~a pair of messages are swapped, since both messages would then result in $\authabort$s (with high probability). Note that in this description, there is no requirement for Alice and Bob to include the message index as a part of the message.

In practice, however, no implementation can perfectly realize the idealized model. There is always a small probability that Eve successfully forges a valid tag for a modified message.
To account for this, we could proceed in either of two ways. First, we could track this probability explicitly throughout our entire analysis, i.e.~noting at every step that there is some small probability of Eve forging the message without being detected, and writing the proof such that this event is explicitly tracked and accounted for. Alternatively, we could just perform our analysis entirely under the idealized model, and then rely on a separate proof that the implemented authentication protocol is a composably $\varepsilon_{\mathrm{auth}}$-secure construction (in some composable security framework) of the  authenticated channel described above - given these, one can invoke composability to lift the security claims based entirely on the ideal case to the real implemented scenario. In this chapter, our analysis will be based on the latter approach, i.e.~we assume the parties have access to the authenticated channel with the functionality described earlier as a starting resource.

We acknowledge however that to our knowledge,  the existing literature on composable security currently does not contain an explicit construction of that resource \cite{portmann_key_2014,portmann_security_2022,broadbent_2023}; still, we believe such a construction to be achievable via similar arguments as in Ref.~\cite{portmann_key_2014}, and leave it as a point to be resolved in future work. In fact, if one disregards the technicalities introduced by message timing, the construction of a multi-use authenticated channel with one-sided aborts has already been demonstrated in Ref.~\cite{portmann_security_2022}.  

\subsection{Authentication Post-Processing Protocol (APP)} \label{subsec:APPprotocol}

Recall our setting, where we have a core QKD protocol $\coreQKDprotocol$, followed by an Authentication Post-Processing protocol $\APPprotocol$. We refer to the resulting combined protocol as $\QKDprotocol = \APPprotocol \circ \coreQKDprotocol$. 
Our goal is to reduce the security of the combined protocol $\QKDprotocol$  (in the {\realistic} authentication setting) to the security of the core QKD protocol $\coreQKDprotocol$ alone (in the honest authentication setting). To achieve this, we design the Authentication Post-Processing protocol $\APPprotocol$ such that
\begin{itemize}
    \item both parties abort whenever any message during the core QKD protocol $\coreQKDprotocol$ results in an $\authabort$, 
    \item it commutes with the map $\idealmap$.  
\end{itemize}
These properties are crucial in our proof of the reduction in \cref{sec:provingthereduction}. We will now specify the Authentication Post-Processing protocol.

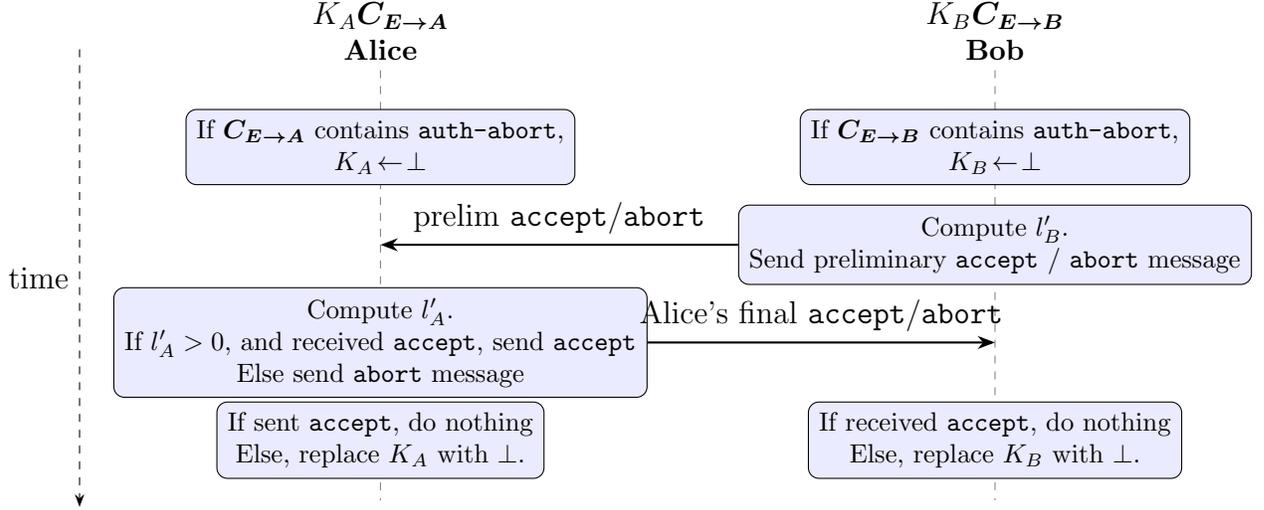
\begin{figure}[ht!] 
\begin{tikzpicture}[
    >=Stealth,
    participant/.style={font=\small\bfseries},
    lifeline/.style  ={gray,dashed},
    message/.style   ={thick,-Stealth},
    localop/.style   ={draw,rectangle,rounded corners,
                       fill=blue!8,inner sep=4pt,
                       font=\footnotesize,align=center},
]

\def\rowsep{1.3} 

\node[participant] (Alice) {Alice};
\node[participant] (Bob)   [right=7cm of Alice] {Bob};

\node at ($(Alice)+(0,0.45)$) {$K_A \CAr $};
\node at ($(Bob)  +(0,0.45)$) {$K_B \CBr $};

\draw[lifeline] (Alice) -- ++(0,-6);
\draw[lifeline] (Bob)   -- ++(0,-6);

\coordinate (A0) at ($(Alice)+(0,-\rowsep)$);
\coordinate (B0) at ($(Bob)  +(0,-\rowsep)$);

\node[localop] (A1) at (A0) 
{If  $\CAr$ contains $\authabort$,\\
$K_A\!\leftarrow\!\bot$};
\node[localop] (B1) at (B0) 
{If  $\CBr$ contains $\authabort$,\\
$K_B\!\leftarrow\!\bot$};

\nextrow{B2}{B1}
\nextrow{A2}{A1}

\node[localop] (B2op) at (B2) {Compute $l_B'$. \\
Send preliminary $\accept$ / $\abort$ message};

\draw[message] (B2op) -- node[midway,above] {prelim \texttt{accept}/\texttt{abort}} (A2);


\nextrow{A3}{A2}
\nextrow{B3}{B2}

\node[localop] (A3op) at (A3) {Compute $l_A'$. \\
If $l_A' >0$, and received $\accept$, send $\accept$ \\
Else send $\abort$ message};

\draw[message] (A3op) -- node[midway,above] {Alice's final \texttt{accept}/\texttt{abort}} (B3);
\nextrow{A4}{A3}
\nextrow{B4}{B3}

\node[localop] (A4op) at (A4) {If sent $\accept$, do nothing \\
Else, replace $K_A$ with $\bot$.};

\node[localop] (B4op) at (B4) {If received $\accept$, do nothing \\
Else, replace $K_B$ with $\bot$.};

\draw[->, dashed]
  ($(Alice)+(-4,0)$) -- ($(AliceEnd)+(-4,-2)$)
  node[midway, left] {time};

\end{tikzpicture}
\caption{Schematic of \nameref{prot:authpp}  described in \cref{subsec:APPprotocol}. Alice and Bob first update their key registers based on whether they received an $\authabort$ in any of the prior communication. They then communicate their tentatively $\accept$ / $\abort$ decisions. They then perform a final update operation on their key registers depending on their final $\accept$ / $\abort$ decision.}
\label{fig:appprotocol}
\end{figure}

Let $\rho^\mathrm{real}_{K_A K_B \CfinalQKD \EfinalQKD}$ denote the final state obtained after the execution of the core QKD protocol $\coreQKDprotocol$, where $\CfinalQKD = \CAs \CBr \CBs \CAr$ collects all classical communication that occurred during the core protocol, and $\EfinalQKD$ denotes all of Eve’s side information, which may include a copy of the classical communication. During the authentication post-processing phase, Alice and Bob start with the state  $\rho^\mathrm{real}_{K_A K_B \CfinalQKD \EfinalQKD}$, perform some classical operations and communicate using registers $\CAsauth,\CBrauth,\CBsauth,\CArauth$ (see \cref{fig:appprotocol}). 
\vspace{2em}
\begin{prot}[AuthPP Protocol] \label{prot:authpp} 
Starts with state 
$\rho^\mathrm{real}_{K_A K_B \CfinalQKD \EfinalQKD}$. Classical communication is undertaken in the registers
$\CAsauth, \CBrauth, \CBsauth, \CArauth$.
\begin{enumerate}[label=\textbf{APP~\arabic*}, ref=APP~\arabic*] 

\item \label{prot:authpp-alice} Alice checks whether any of her received messages in $\CAr$ is a \authabort. 
If she finds one, she replaces $K_A$ with $\bot$. 

\item \label{prot:authpp-bob} Bob checks whether any of his received messages in $\CBr$ is an \authabort. 
If he finds one, he replaces $K_B$ with $\bot$.

    \item \label{prot:authpp-bob2} Bob computes $l^\prime_B$ from $K_B$. If it is non-zero, he sends a preliminary \accept message to Alice. Otherwise, he sends an \abort message to Alice.
    
    \item \label{prot:authpp-alice2}Alice computes $l^\prime_A$ from $K_A$. If it is non-zero, \textit{and} she receives an \accept message from Bob, she sends her final \accept message to Bob. Else she sends \abort message to Bob.

       \item \label{prot:authpp-alice3}If Alice sent \accept message, she does nothing. If she sent \abort message, she replaces her key registers with $\bot$.
    
    \item \label{prot:authpp-bob3} If Bob receives an \accept message from Alice, he does nothing. If Bob receives either a \authabort\ or an \abort message, he replaces his key register $K_B$ with $\bot$. We let $l_B$ denote the final key length stored in $K_B$.
    
\end{enumerate}
\end{prot}

%

Mathematically, the above protocols can be described as follows:
\begin{enumerate}
    \item In \cref{prot:authpp-alice} and \cref{prot:authpp-bob}, Alice and Bob simply replace their key registers with $\bot$s if they received an $\authabort$ during the prior protocol $\coreQKDprotocol$. This is described by  a map $\authmap \in \CPTP(\Cfinal K_A K_B,\Cfinal K_A K_B)$ (see \cref{remark:authprotsteps}).
    \item In \cref{prot:authpp-alice2,prot:authpp-bob2}, Alice and Bob first compute the key lengths $l^\prime_A, l^\prime_B$ stored in their key registers $K_A,K_B$ respectively at that point in time. Based on these values, they engage in two rounds of communications. For a given value of $l^\prime_A, l^\prime_B$, this can be described as a map (influenced by Eve) $\authcommmap^{(l^\prime_A,l^\prime_B)} \in \CPTP( \EfinalQKD, \Efinal \Cauth )$. The overall map is given by $\authcommmap \in \CPTP( K_A K_B \EfinalQKD, K_A K_B \Efinal \Cauth )$. Note that $\Cauth$ here contains separate messages that were sent and received, since Eve may attack this communication in any manner she desires, potentially modifying her quantum side-information as well \footnote{Some messages are still yet to be sent or received; however, for notational simplicity we let the map act on the entirety of $\Cauth$, with the understanding that it leaves registers meant to store future messages unchanged.}.
    \item In \cref{prot:authpp-alice3,prot:authpp-bob3}, Alice and Bob  use the result of the communication in the previous step $\Cauth$ to determine their final $\accept$ / $\abort$ status. This is described as a map 
    $\authupdatemap \in \CPTP( K_A K_B \Cauth, K_A K_B \Cauth )$
\end{enumerate}
The final output state is denoted by  $\rho^\mathrm{real,final}_{K_A K_B \Cfinal \Efinal} = \authupdatemap \circ \authcommmap \circ  \authmap  \left(\rho^\mathrm{real}_{K_A K_B \CfinalQKD \EfinalQKD}\right)$, where $\Cfinal = \CfinalQKD \Cauth$, where add a superscript `final' to the state to denote that this is the final output.

\begin{remark} \label{remark:authprotsteps}
Note that the first two steps in the above protocol serve to ensure that if any classical communication resulted in an \authabort, then the receiving party stores a $\bot$  in the key register. In practice, one can simply enforce this property by appropriate design of the QKD protocol, i.e, by ensuring that the QKD protocol itself stores $\bot$ in the key register whenever \authabort is received. In such cases, we can omit \cref{prot:authpp-alice,prot:authpp-bob} from the authentication post-processing steps, as is done in Ref.~\cite{inprep_BDR3}. Here  we wish to prove a statement that is agnostic to the nature of the QKD protocol implemented. Hence we include \cref{prot:authpp-alice,prot:authpp-bob} as explicit steps. 
\end{remark}

\subsection{Reduction Statement} \label{subsec:reductionstatement}
We are now ready to state the theorem that reduces the security analysis of QKD protocols to the setting in which authentication behaves honestly.

\begin{restatable}[Reduction of QKD security analysis to the honest authentication setting]{theorem}{reductionstatement}
\label{theorem:reductionstatement}

Let $\coreQKDprotocol$ be an arbitrary QKD protocol. Let $\APPprotocol$ be the \nameref{prot:authpp} described in \cref{subsec:APPprotocol}, executed after the core QKD protocol $\coreQKDprotocol$. Let $\QKDprotocol = \APPprotocol \circ \coreQKDprotocol$ denote the resulting QKD protocol. Let $\worldhonest(\coreQKDprotocol)$ denote the set of possible output states of  $\coreQKDprotocol$ in the honest authentication setting (see \cref{subsec:classicalcommunicationsmodel}). Let $\worldreal(\QKDprotocol)$ denote the set of possible output states of $\QKDprotocol$ in the {\realistic}  authentication setting (see \cref{subsec:classicalcommunicationsmodel}). 
Then, the $\epssecure$-security for all output states in $\worldhonest(\coreQKDprotocol)$ implies $\epssecure$-security for all output states\footnote{Note that any scenario in which Eve chooses to ignore or forget part of the public classical communication can be treated as one where she first records all communication and then traces out whatever she wishes at the end of the QKD protocol. Moreover, since QKD security analysis already allows Eve to retain all public communication, keeping the public communication register explicit and accessible to her is without loss of generality.} in $\worldreal(\QKDprotocol)$. That is,
\begin{equation}
    \begin{aligned}
         \tracedist{
\rho^\mathrm{real,hon}_{K_A K_B \CfinalQKD \EfinalQKD}
-
\rho^\mathrm{ideal,hon}_{K_A K_B \CfinalQKD \EfinalQKD}
} &\leq \epssecure \qquad \forall \rho^\mathrm{real,hon}_{K_A K_B \CfinalQKD \EfinalQKD} \in \worldhonest(\coreQKDprotocol) \\
          &\Downarrow \\
     \tracedist{ \rho^\mathrm{real}_{K_A K_B \Cfinal \Efinal} -  \rho^\mathrm{ideal}_{K_A K_B \Cfinal \Efinal}} &\leq \epssecure \qquad \forall \rho^\mathrm{real}_{K_A K_B \Cfinal \Efinal} \in \worldreal(\QKDprotocol).
    \end{aligned}
\end{equation}
\end{restatable}

\textit{Proof Idea.} The detailed proof of this theorem is presented in \cref{sec:provingthereduction}.
We define an event $\Onice$, corresponding to the case where no $\authabort$s are received during the execution of the core protocol $\coreQKDprotocol$. Under the communication model from \cref{subsec:classicalcommunicationsmodel}, any attempt by Eve to tamper with the message content or disturb the relative timing of messages leads to an $\authabort$. Thus, if we consider states partial on   $\Onice$, we may assume that the authentication behaves honestly. (Recall that the state partial on $\Onice$ refers to the state that is conditioned on the event $\Onice$, but which is not re-normalized after the conditioning).

Since $\APPprotocol$ ensures that both parties abort whenever an $\authabort$ is received, it suffices to upper bound the trace distance from \cref{def:qkdsecurityasymmetric} for the output state of $\worldreal(\QKDprotocol)$ partial on $\Onice$. This is because if $\Onice$ does not occur, then both parties abort and the trace distance is zero by definition (see \cref{lemma:bothabort}).

Next, we observe that $\APPprotocol$ commutes with the ideal map $\idealmap$. We use this to show that, it is sufficient to prove security for all states in $\worldreal(\coreQKDprotocol)$ partial on the event $\Onice$ (see \cref{lemma:commutationidealauth}).  The final step is showing that it then suffices to prove the security of the core QKD protocol $\coreQKDprotocol$ in the honest authentication setting. This is shown in \cref{lemma:reductionone,lemma:reductiontwo,lemma:reductionthree}.

\begin{remark}\label{remark:authnotetoreader}
\cref{theorem:reductionstatement} is quite general and serves as a bridge between the standard QKD security analyses performed under the assumption of honest authentication and the more realistic setting where the authentication channel may be actively attacked—both in terms of message content and timing. Furthermore, we emphasize that it is entirely independent of the specific details of the QKD protocol being implemented, whether it is device-dependent or device-independent, prepare-and-measure or entanglement-based, etc. 
\end{remark}

\section{Proof of the Reduction \texorpdfstring{\cref{theorem:reductionstatement}}{Theorem}}\label{sec:provingthereduction}
In this section, we will provide a rigorous proof of \cref{theorem:reductionstatement}.

\subsection{Proving that \authabort in the core QKD protocol results in both parties aborting}
We start by proving the following lemma, which states that if either Alice or Bob received an $\authabort$ in $\CfinalQKD$ during the core QKD protocol $\coreQKDprotocol$, then both parties will abort during authentication postprocessing. We use $\Onice$ to denote the event that neither Alice nor Bob receive an \authabort in $\CfinalQKD$ during $\coreQKDprotocol$, and write $\Onice^\complement$ for its complement.

\begin{lemma} \label{lemma:bothabort}
Let $\Onice$ be the event where neither Alice nor Bob receive an \authabort in $\CfinalQKD$ during $\coreQKDprotocol$, and let $\rho^\mathrm{real,final}_{K_A K_B \Cfinal \Efinal}$ denote the final output state at the end of the full QKD protocol $\QKDprotocol$. Then the following equality holds
\begin{equation} \label{eq:niceeventequality}
 \idealmap \left[ \rho^\mathrm{real,final}_{K_A K_B \Cfinal \Efinal | \Onice^\complement} \right] \eqqcolon \rho^\mathrm{ideal,final}_{K_A K_B \Cfinal \Efinal | \Onice^\complement} = \rho^\mathrm{real,final}_{K_A K_B \Cfinal \Efinal | \Onice^\complement}.
\end{equation}
Therefore, 
\begin{equation} \label{eq:niceeventinequality}
\tracedist{\rho^\mathrm{real,final}_{K_A K_B \Cfinal \Efinal } - \rho^\mathrm{ideal,final}_{K_A K_B \Cfinal \Efinal }} =
    \tracedist{\rho^\mathrm{real,final}_{K_A K_B \Cfinal \Efinal \wedge \Onice} - \rho^\mathrm{ideal,final}_{K_A K_B \Cfinal \Efinal \wedge \Onice}} 
\end{equation}
\end{lemma}
\begin{proof}
    The proof of \cref{eq:niceeventequality} follows straightforwardly from the structure of the \nameref{prot:authpp} by considering the state  $\rho^\mathrm{real}_{K_A K_B \CfinalQKD \EfinalQKD | \Onice^\complement}$ and tracking it throughout the protocol.
    
    Let us suppose that Alice receives at least one $\authabort$. Then, \cref{prot:authpp-alice} replaces $K_A$ with $\bot$, and Alice announces \abort to Bob in \ref{prot:authpp-alice2}. Bob either  receives an $\abort$ or $\authabort$, and in either case, replaces his $K_B$ register with $\bot$ in \cref{prot:authpp-bob3}. The final output key length for both parties is $0$. 

Similarly, let us suppose that Bob receives at least one $\authabort$. Then, \cref{prot:authpp-bob} replaces $K_B$ with $\bot$, and Bob sends an $\abort$ message to Alice in \cref{prot:authpp-bob2}. This  leads to Alice sending an $\abort$ message to Bob in \cref{prot:authpp-alice2} and replacing her key register $K_A$ with $\bot$ in \cref{prot:authpp-alice3}. Thus, the final output key length for both parties is $0$.

The required \cref{eq:niceeventequality} follows by noting that $\idealmap$ acts as identity  when $l_A = l_B = 0$. Finally, \cref{eq:niceeventinequality} follows from \cref{eq:niceeventequality} via

\begin{equation} \begin{aligned}
\tracedist{\rho^\mathrm{real,final}_{K_A K_B \Cfinal \Efinal } - \rho^\mathrm{ideal,final}_{K_A K_B \Cfinal \Efinal }} 
&=   \tracedist{\rho^\mathrm{real,final}_{K_A K_B \Cfinal \Efinal \wedge \Onice} - \rho^\mathrm{ideal,final}_{K_A K_B \Cfinal \Efinal \wedge \Onice}}  \\
&+\tracedist{\rho^\mathrm{real,final}_{K_A K_B \Cfinal \Efinal \wedge \Onice^\complement} - \rho^\mathrm{ideal,final}_{K_A K_B \Cfinal \Efinal \wedge \Onice^\complement}}  \\
&=\tracedist{\rho^\mathrm{real,final}_{K_A K_B \Cfinal \Efinal \wedge \Onice} - \rho^\mathrm{ideal,final}_{K_A K_B \Cfinal \Efinal \wedge \Onice}}
    \end{aligned}
\end{equation}
where the first equality follows from the fact that the states conditioned on $\Onice$ and $\Onice^\complement$ live on orthogonal spaces, and the final equality follows from \cref{eq:niceeventequality}.
\end{proof}

\subsection{Reducing to security before authentication post-processing} \label{subsec:reducingtobeforeAPP}

From \cref{eq:niceeventequality} we see that we only need to prove security for output states of $\QKDprotocol$ partial on $\Onice$. We would like to reduce the analysis to output states of $\coreQKDprotocol$ partial on $\Onice$.
We will show in \cref{lemma:commutationidealauth} (see also \cref{fig:comm-diagram}) that, conditioned on $\Onice$, the final real and ideal output states (at the end of $\QKDprotocol$) can be obtained by the action of $\authupdatemap \circ \authcommmap \circ \authmap $ on the real and ideal states at the end of the core QKD protocol $\coreQKDprotocol$.  Using the fact that the one-norm is non-increasing under CPTP maps, we can thus instead focus on the distance between the real and ideal output states \emph{before} the authentication post-processing. This distance will then be related to the usual security guarantee obtained under the assumption that authentication behaves honestly.

\begin{lemma}[Commutation of $\idealmap$ and \nameref{prot:authpp}] \label{lemma:commutationidealauth}
Let $\Onice$ denote the event where neither Alice nor Bob received an $\authabort$ in $\CfinalQKD$ during the core QKD protocol $\coreQKDprotocol$. Let $\rho^\mathrm{real}_{K_A K_B \CfinalQKD \EfinalQKD | \Onice}$ denote the real state at the end of the core QKD protocol  conditioned on $\Onice$. Let the following states denote its evolution through  \nameref{prot:authpp}, 
\begin{equation}
\begin{aligned}
   \rho^\mathrm{real,repl}_{K_A K_B \CfinalQKD \EfinalQKD| \Onice} &\coloneq \authmap \left[ \rho^\mathrm{real}_{K_A K_B \CfinalQKD \EfinalQKD| \Onice} \right], \\
    \rho^\mathrm{real,comm}_{K_A K_B \CfinalQKD \Cauth  \Efinal| \Onice} &\coloneq \authcommmap \left[ \rho^\mathrm{real,repl}_{K_A K_B \CfinalQKD \EfinalQKD| \Onice} \right], \\
     \rho^\mathrm{real,final}_{K_A K_B \Cfinal\Efinal| \Onice} &\coloneq \authupdatemap \left[ \rho^\mathrm{real,comm}_{K_A K_B \CfinalQKD \Cauth  \Efinal| \Onice} \right].
   \end{aligned}
\end{equation}
Define the corresponding ideal states by the action of the map $\idealmap$ on the real states, i.e 
\begin{equation}
\begin{aligned}
   \rho^\mathrm{ideal}_{K_A K_B \CfinalQKD \EfinalQKD| \Onice} &\coloneq \idealmap \left[   \rho^\mathrm{real}_{K_A K_B \CfinalQKD \EfinalQKD| \Onice} \right], \\
   \rho^\mathrm{ideal,repl}_{K_A K_B \CfinalQKD \EfinalQKD| \Onice} &\coloneq \idealmap \left[   \rho^\mathrm{real,repl}_{K_A K_B \CfinalQKD \EfinalQKD| \Onice} \right],\\
   \rho^\mathrm{ideal,comm}_{K_A K_B \CfinalQKD \Cauth \Efinal| \Onice} &\coloneq \idealmap \left[   \rho^\mathrm{real,comm}_{K_A K_B \CfinalQKD \Cauth\Efinal| \Onice} \right], \\
     \rho^\mathrm{ideal,final}_{K_A K_B \Cfinal\Efinal| \Onice} &\coloneq \idealmap \left[  \rho^\mathrm{real,final}_{K_A K_B \Cfinal\Efinal| \Onice}  \right].
   \end{aligned}
\end{equation}
Then, the ideal states defined above are the same as those obtained by evolving  $ \rho^\mathrm{ideal}_{K_A K_B \CfinalQKD \EfinalQKD| \Onice}  $ through  the \nameref{prot:authpp}, i.e,
\begin{equation}
\begin{aligned}
      \rho^\mathrm{ideal,repl}_{K_A K_B \CfinalQKD \EfinalQKD| \Onice} &= \authmap \left[ \rho^\mathrm{ideal}_{K_A K_B \CfinalQKD \EfinalQKD| \Onice} \right], \\
    \rho^\mathrm{ideal,comm}_{K_A K_B \Cfinal \Cauth  \Efinal| \Onice} &= \authcommmap \left[ \rho^\mathrm{ideal,repl}_{K_A K_B \CfinalQKD \Efinal| \Onice} \right], \\
     \rho^\mathrm{ideal,final}_{K_A K_B \Cfinal\Efinal| \Onice} &= \authupdatemap \left[\rho^\mathrm{ideal,comm}_{K_A K_B \CfinalQKD \Cauth  \EfinalQKD| \Onice} \right].
     \end{aligned}
\end{equation} 
\end{lemma}

\begin{figure}[t!] 
\centering
\[
\begin{tikzcd}[row sep=5em, column sep=6em]
\rho^\text{real}_{K_A K_B \CfinalQKD \EfinalQKD | \Onice} \arrow[r, "\idealmap"] \arrow[d, "\authmap"'] 
      & \rho^\text{ideal}_{K_A K_B \Cfinal \Efinal| \Onice} \arrow[d, "\authmap \quad {\color{red} ?}"] \\
\rho^\text{real,repl}_{K_A K_B \CfinalQKD \EfinalQKD| \Onice} \arrow[r, "\idealmap"'] \arrow[d, "\authcommmap"'] 
  &  \rho^\text{ideal,repl}_{K_A K_B \CfinalQKD \EfinalQKD| \Onice}  \arrow[d, "\authcommmap\quad {\color{red} ?}"] \\
\rho^\text{real,comm}_{K_A K_B \Cfinal \Efinal| \Onice}  \arrow[r, "\idealmap"'] \arrow[d,"\authupdatemap  "]
  & \rho^\text{ideal,comm}_{K_A K_B \Cfinal \Efinal| \Onice} \arrow[d,"\authupdatemap \quad {\color{red} ?}"] \\
  \rho^\text{real,final}_{K_A K_B \Cfinal \Efinal| \Onice}  \arrow[r, "\idealmap"'] 
  & \rho^\text{ideal,final}_{K_A K_B \Cfinal \Efinal| \Onice}
\end{tikzcd}
\]
\caption{A diagram illustrating transformations between real and ideal states evolving through the \nameref{prot:authpp}. The `{\color{red} $?$}' indicates the transformations that must be shown to be true in \cref{lemma:commutationidealauth}. The states go through the map $\authmap \in \CPTP(K_A K_B \CfinalQKD, K_A K_B \CfinalQKD)$ that replaces the key registers depending on $\CfinalQKD$. They then go through some communication steps given by $\authcommmap \in \CPTP(K_A K_B \EfinalQKD, K_A K_B \Efinal \Cauth) $ (influenced by Eve). Finally, Alice and Bob perform the final updates to their key registers described by $\authupdatemap \in \CPTP(K_A K_B \Cauth, K_A K_B \Cauth)$.}
\label{fig:comm-diagram}
\end{figure}

Note that this lemma also holds when conditioning on $\Onice^\complement$ instead of $\Onice$. However, the proof in that case is considerably more cumbersome, and we do not require it here. Recall that $\Cfinal = \CfinalQKD \Cauth$.
\begin{proof}
The proof consists of some straightforward though fairly cumbersome algebra. 
We wish to show that the three vertical arrows on the right hand side of \cref{fig:comm-diagram} are satisfied. We also note that none of the maps in $\APPprotocol$ ever affect the register $\CfinalQKD$ (they only read from the register), thus the event $\Onice$ can be defined at the start, and is ``preserved'' throughout the evolution of the real or ideal state. 

\paragraph*{First arrow:} For the first vertical arrow, we simply note that if $\Onice$ occurs, then the map $\authmap$ acts identically on the input state. Thus, 
\begin{equation} \label{eq:firstarrow}
  \authmap \circ \idealmap \left[  \rho^\text{real}_{K_A K_B \CfinalQKD \EfinalQKD | \Onice} \right] = \idealmap \circ \authmap  \left[  \rho^\text{real}_{K_A K_B \CfinalQKD \EfinalQKD | \Onice} \right]
\end{equation}
is trivially satisfied. 

\paragraph*{Second arrow:} For the second vertical arrow, we note that the $\authcommmap$ first looks at the key lengths stored in Alice and Bob's registers, and then simply makes some announcements (and implements Eve's attacks). The $\idealmap$ map also simply replaces the key registers based on the key lengths. Neither maps affects the key lengths stored in $K_A, K_B$. 
Let $\projkeymap \in \CPTP(K_A K_B, K_A K_B) $ denote the map that projects onto the subspace where the key registers store keys of lengths $l_A^\prime, l_B^\prime$. Then, 
since Alice and Bob's announcements only depend on the length of their key registers $l_A^\prime$, $l_B^\prime$,  $\authcommmap$ has the following structure:
\begin{equation} \label{eq:authproofeq1}
    \authcommmap = \sum_{l^\prime_A , l^\prime_B}   \projkeymap \otimes \authcommmap^{(l_A^\prime, l_B^\prime)},
\end{equation}
 where $\authcommmap^{(l_A^\prime, l_B^\prime)} \in \CPTP(\CfinalQKD \EfinalQKD, \CfinalQKD \Cauth \Efinal)$ \footnote{It actually leaves the $\CfinalQKD$ register untouched.}, and does not act on the $K_A K_B$ registers. Moreover, the $\idealmap$ map also has a similar structure (\cref{eq:idealmapdecomp}), namely 
 \begin{equation} \label{eq:authproofeq2}
    \idealmap = \bigoplus_{l^\prime_A , l^\prime_B}  \idealmap^{(l_A^\prime, l_B^\prime)},  \qquad \text{ where $\idealmap^{(l^\prime_A,l^\prime_B)} \in \CPTP(K_A^{l^\prime_A} K_B^{l^\prime_B} , K_A^{l^\prime_A} K_B^{l^\prime_B} )$.}
\end{equation}
Then, we note that
\begin{equation} 
\begin{aligned}
\label{eq:authproofeq3}
   \authcommmap \circ \idealmap &= \sum_{l_A^\prime, l_B^\prime}  \projkeymap \circ \idealmap^{(l^\prime_A,l^\prime_B)}  \otimes \authcommmap^{(l_A^\prime, l_B^\prime)} \\
   &= \sum_{l_A^\prime, l_B^\prime}  \idealmap^{(l^\prime_A,l^\prime_B)} \circ  \projkeymap   \otimes \authcommmap^{(l_A^\prime, l_B^\prime)}  \\
   &= \idealmap \circ \authcommmap,
    \end{aligned}
\end{equation}
where all the equalities follow from the definition of the maps.  The required claim then follows from \cref{eq:authproofeq3}.

\paragraph*{Third arrow:} For the third vertical arrow, we use the fact that $\authupdatemap$ simply reads from the $\Cauth$ register and performs operations on the $K_A,K_B$ registers. Its operations depend on the values observed in the $\Cauth$ register. To proceed we will require some additional notation.

Let us use $\finalacceptabortevent{i}{j}$, where $i,j$ can either be $\top$ or $\bot$ to denote the final accept or abort decisions (and communication) undertaken by Alice and Bob as follows: 
\begin{itemize}
    \item If $i=\top$, Alice  does nothing in \cref{prot:authpp-alice3}. This corresponds to Alice sending a final \accept message in \cref{prot:authpp-alice2}.
    \item If $i=\bot$, Alice   replaces the key register with the $\bot$ value in \cref{prot:authpp-alice3}. This corresponds to Alice sending a final \abort message in \cref{prot:authpp-alice2}.
    \item If $j = \top$, Bob does nothing in \cref{prot:authpp-bob3}. This corresponds to him receiving an \accept message from Alice in \cref{prot:authpp-bob2}.
    \item If $j = \bot$, Bob replaces his key register $\bot$ in \cref{prot:authpp-bob3}. This corresponds to him receiving either an \abort or \authabort message from Alice in \cref{prot:authpp-bob2}. 
\end{itemize}   
The $\idealmap$ map also performs operations on the $K_A K_B$ register, and its operations only depend on the lengths of the strings stored in $K_A,K_B$. Recall from \cref{subsec:APPprotocol} that we use $l_A^\prime, l_B^\prime$ to denote these lengths before the update operation, and $l_A, l_B$ to denote these lengths after the update operation.

Let us first consider the scenario where we first apply $\authupdatemap$ and then $\idealmap$. Thus, we will focus first on the state $ \rho^\mathrm{real,comm}_{K_A K_B \Cfinal \Efinal | \Onice \wedge \OlAlBprime}$. This state is a mixture of the following states which correspond to various final accept or abort decisions:
\begin{equation} \label{eq:commproofmixture}
\begin{aligned}
    \Big\{ &\rho^\mathrm{real,comm}_{K_A K_B \Cfinal \Efinal | \Onice \wedge \OlAlBprime\wedge \finalacceptabortevent{\top}{\top} }, \\
    &\rho^\mathrm{real,comm}_{K_A K_B \Cfinal \Efinal | \Onice \wedge \OlAlBprime\wedge\finalacceptabortevent{\bot}{\bot} }, \\
    &\rho^\mathrm{real,comm}_{K_A K_B \Cfinal \Efinal | \Onice \wedge \OlAlBprime\wedge\finalacceptabortevent{\top}{\bot} }, \\
    &\rho^\mathrm{real,comm}_{K_A K_B \Cfinal \Efinal | \Onice \wedge \OlAlBprime\wedge\finalacceptabortevent{\bot}{\top} } \Big\}.
    \end{aligned}
\end{equation}
(Note that for the \nameref{prot:authpp} considered in this chapter, we can never have the $\finalacceptabortevent{\bot}{\top}$ occur. However, we leave it in our analysis to enable this proof to be easily adapted to variations of \nameref{prot:authpp}).
From the actions of the update map (see \cref{prot:authpp-alice3,prot:authpp-bob3}), it is straightforward to verify that  
\begin{equation}
    \authupdatemap \circ \idealmap \left[ \cdot \right] = \idealmap \circ \authupdatemap \left[ \cdot \right] \qquad \text{where $\cdot$ is any state from \cref{eq:commproofmixture}}.
\end{equation}
That is, the required commutation holds conditioned on any combination of final \accept / \abort outcomes. This implies 

\begin{equation}
    \authupdatemap \circ \idealmap \left[ \rho^\mathrm{real,comm}_{K_A K_B \Cfinal \Efinal | \Onice \wedge \OlAlBprime}\right] = \idealmap \circ \authupdatemap \left[ \rho^\mathrm{real,comm}_{K_A K_B \Cfinal \Efinal | \Onice \wedge \OlAlBprime} \right]  
\end{equation}
holds $\forall l_A^\prime ,l_B^\prime$. Since the above equation holds for all values of the intermediate key lengths $l_A^\prime, l_B^\prime$, we have the required claim: 

\begin{equation} \label{eq:arrow3}
    \authupdatemap \circ \idealmap \left[ \rho^\mathrm{real,comm}_{K_A K_B \Cfinal \Efinal | \Onice }\right] = \idealmap \circ \authupdatemap \left[ \rho^\mathrm{real,comm}_{K_A K_B \Cfinal \Efinal | \Onice } \right].
\end{equation}

Thus, the stated Lemma follows from \cref{eq:firstarrow,eq:arrow3,eq:authproofeq3}.
This concludes the proof.
\end{proof}

This commuting property allows us to focus on the distance between the real and ideal states \emph{before} \nameref{prot:authpp}, rather than after. This can be formalized using the following corollary.

\begin{corollary} \label{corr:reduction}
Let $\coreQKDprotocol$, $\APPprotocol$, and $\QKDprotocol$ be protocols such that the overall QKD protocol is given by $\QKDprotocol = \APPprotocol \circ \coreQKDprotocol$, where $\APPprotocol$ denotes a post-processing routine executed after the core QKD protocol $\coreQKDprotocol$. Assume that classical communication in the {\realistic} and honest setting behaves as specified in \cref{subsec:classicalcommunicationsmodel}, and suppose that $\APPprotocol$ is as described in \nameref{prot:authpp}. That is, consider the same setup as in \cref{theorem:reductionstatement}. Then, the $\epssecure$-security for all output states in $\worldreal(\coreQKDprotocol)$ partial on the event $\Onice$, implies $\epssecure$-security for all output states in $\worldreal(\QKDprotocol)$. That is 
  \begin{equation}
      \begin{aligned}
    \tracedist{\rho^\mathrm{real}_{K_A K_B \CfinalQKD \EfinalQKD \wedge \Onice }  - \rho^\mathrm{ideal}_{K_A K_B \CfinalQKD \EfinalQKD \wedge \Onice} }  &\leq \epssecure \qquad \forall \rho^\mathrm{real}_{K_A K_B \CfinalQKD \EfinalQKD } \in  \worldreal(\coreQKDprotocol) \\
    &\Downarrow \\
     \tracedist{\rho^\mathrm{real,final}_{K_A K_B \Cfinal \Efinal }  - \rho^\mathrm{ideal,final}_{K_A K_B \Cfinal \Efinal } }  &\leq \epssecure \qquad \forall \rho^\mathrm{real,final}_{K_A K_B \Cfinal \Efinal } \in  \worldreal(\QKDprotocol) 
      \end{aligned}
  \end{equation}
\end{corollary}
\begin{proof}
   For any output state $\rho^\mathrm{real,final}_{K_A K_B \Cfinal \Efinal } \in \worldreal(\QKDprotocol)$ of the full QKD protocol, consider the corresponding state $\rho^\mathrm{real,final}_{K_A K_B \CfinalQKD \EfinalQKD } \in \worldreal(\coreQKDprotocol)$ obtained at the end of the core QKD protocol $\coreQKDprotocol$. Then, we have
    \begin{equation}
        \begin{aligned}
         & \tracedist{\rho^\mathrm{real,final}_{K_A K_B \Cfinal \Efinal } - \rho^\mathrm{ideal,final}_{K_A K_B \Cfinal \Efinal }}\\
         &= 
\tracedist{\rho^\mathrm{real,final}_{K_A K_B \Cfinal \Efinal \wedge \Onice} - \rho^\mathrm{ideal,final}_{K_A K_B \Cfinal \Efinal \wedge \Onice}} \\
    &=  \tracedist{ \authupdatemap \circ \authcommmap \circ \authmap \left[ \rho^\mathrm{real}_{K_A K_B \CfinalQKD \EfinalQKD \wedge \Onice} - \rho^\mathrm{ideal}_{K_A K_B \CfinalQKD \EfinalQKD\wedge \Onice} \right]} \\
    &\leq \tracedist{ \rho^\mathrm{real}_{K_A K_B \CfinalQKD \EfinalQKD \wedge \Onice} - \rho^\mathrm{ideal}_{K_A K_B \CfinalQKD \EfinalQKD\wedge \Onice} } 
\end{aligned}\end{equation}
where we use \cref{lemma:bothabort} for the first line, \cref{lemma:commutationidealauth} for the second line, and the fact that the one-norm cannot increase under the action of CPTP maps for the final line. This suffices to prove the required claim.  \end{proof}

Thus, \cref{corr:reduction} allows us to restrict our attention to bounding 
\begin{equation} \label{eq:targetofreduction}
  \tracedist{\rho^\mathrm{real}_{K_A K_B \CfinalQKD \EfinalQKD \wedge \Onice }  - \rho^\mathrm{ideal}_{K_A K_B \CfinalQKD \EfinalQKD \wedge \Onice} }  \leq \epssecure \qquad \forall \rho^\mathrm{real}_{K_A K_B \CfinalQKD \EfinalQKD } \in  \worldreal(\coreQKDprotocol)
\end{equation}
which is the distance between the real and ideal states after the core QKD protocol (partial on the event $\Onice$).

\subsection{Reducing to scenario where authentication satisfies honest behaviour}
The task now is to show that if the QKD protocol is analyzed under the assumption that authentication follows honest behaviour, then this analysis applies to output state (partial on $\Onice$) in the setting where authentication is not assumed to be honest. Under the communication model defined in \cref{subsec:classicalcommunicationsmodel}, the event $\Onice$ coincides with the event that authentication behaved honestly. Indeed, any attempt to tamper with the contents of a message or violate the temporal ordering of messages triggers an $\authabort$ (i.e., $\Onice^\complement$). Thus, it is natural to relate the two scenarios. 

While this may initially seem straightforward, making the connection formal is non-trivial. One might intuitively expect that conditioning on $\Onice$ should allow us to restrict attention to the honest-authentication setting. However, this naive reasoning fails upon realizing that the set of operations available to Eve in the real setting is strictly \emph{larger} than that in the honest authentication setting.\footnote{For example, consider an operation in which Eve - perhaps probabilistically, or via a unitary interaction between her quantum side information and the classical signals - preemptively sends messages to one party in order to induce that party to advance to the next stage of the protocol earlier than intended. Such an operation is simply not \emph{allowed} in the honest authentication setting, therefore the resulting output state cannot be identified in $\worldhonest(\coreQKDprotocol)$.
} Consequently, the set of possible output states in the {\realistic} authentication setting $\worldreal(\coreQKDprotocol)$ is strictly \emph{larger} than the set of possible output states in the honest authentication setting $\worldhonest(\coreQKDprotocol)$ (even if we consider partial states on $\Onice$).
To circumvent this problem, we introduce a \emph{virtual authentication setting}  (i.e, a new set of assumptions describing a virtual authentication scenario) which acts as a bridge that allows us to relate the two settings. This technical step is essential for our reduction argument that follows. Thus, the series of reductions that we build is (informally) described by:
 \begin{equation} \label{eq:reductionpath}
\begin{aligned}
   & \worldreal(\QKDprotocol) \xrightarrow{\cref{lemma:bothabort}}\worldreal(\QKDprotocol)_{\wedge \Onice} \xrightarrow{\cref{corr:reduction}} \worldreal(\coreQKDprotocol)_{\wedge \Onice}  \xrightarrow{\cref{lemma:reductionone}} \\
   &\worldvirtual(\coreQKDprotocol)_{\wedge \Onice} 
    \xrightarrow{\cref{lemma:reductiontwo}} 
    \worldvirtual(\coreQKDprotocol) \xrightarrow{\cref{lemma:reductionthree}}
    \worldhonest(\coreQKDprotocol),
    \end{aligned}
\end{equation}
where the set of output states in the virtual setting is denoted using $\worldvirtual$ and described shortly.

\subsubsection{Describing the virtual authentication setting correct messages}  \label{subsec:virtualsetting}
 In the virtual authentication setting, Eve still has the ability to implement any attack she wants on the classical channel; in particular, she controls the message timings $t^{(i)}_{E \rightarrow A}, t^{(i)}_{E \rightarrow B}$ and the contents of $C^{(i)}_{E \rightarrow A},C^{(i)}_{E \rightarrow B}$. Thus, the set of possible operations she can do is \emph{exactly the same} as in the {\realistic} authentication setting. However, the virtual setting is designed such that it is equivalent to the {\realistic} authentication setting whenever event $\Onice$ occurs, through the use of special registers $\bm{\Ccorr}$. The virtual authentication setting is described as follows:
\begin{enumerate}
\item Eve's attack results in the registers $\CAs,\CAr,\CBs,\CBr$ being sent and received in the exact same way as the {\realistic} authentication setting (see  \cref{subsec:classicalcommunicationsmodel}). As before,  each party labels their outgoing (incoming)
messages according to the order in which the messages are sent (received).
    \item \textbf{Timing:} If the $i$th message is received \textit{before} the $i$th message was sent, the receiving party gets the correct message in additional registers $\bm{\Ccorr}$, at some time after the message was sent. Stated formally:
 \begin{equation}
        \begin{aligned}
        &t^{(i)}_{A\rightarrow E} > t^{(i)}_{E \rightarrow B} \quad \implies \text{Bob receives correct message in $\Ccorr^{(i)}_{E \rightarrow B}$ at time after  $t^{(i)}_{A\rightarrow E}$} , \quad \forall i. \\
       &t^{(i)}_{B\rightarrow E} > t^{(i)}_{E \rightarrow A} \quad \implies \text{Alice receives correct message in $\Ccorr^{(i)}_{E \rightarrow A}$ at time after $t^{(i)}_{B\rightarrow E}$}, \quad \forall i.
        \end{aligned}
    \end{equation}

   \item \textbf{Modifying messages:} If $i$th message is received \textit{after} the $i$th message was sent, then the  receiving party gets the correct message in additional registers $\bm{\Ccorr}$, at the same time as the actual message is received. Stated formally:
     \begin{equation}
        \begin{aligned}
        &t^{(i)}_{A\rightarrow E} \leq t^{(i)}_{E \rightarrow B} \quad \implies \quad \text{Bob receives the correct message in $\Ccorr^{(i)}_{E \rightarrow B}$ at time $t^{(i)}_{E \rightarrow B}$}, \quad \forall i.  \\
        &t^{(i)}_{B\rightarrow E} \leq t^{(i)}_{E \rightarrow A} \quad \implies \quad \text{Alice receives the correct message in $\Ccorr^{(i)}_{E \rightarrow A}$ at time $t^{(i)}_{E \rightarrow A}$}, \quad \forall i.
        \end{aligned}
    \end{equation}
\item Alice and Bob use the registers $\bm{\Ccorr_{E \rightarrow A}},\bm{\Ccorr_{E \rightarrow B}}$ for the received message, and implement decisions based on the correct copies they receive. They entirely ignore the messages received in $\CAr, \CBr$.
\end{enumerate}
We use $\worldvirtual(\coreQKDprotocol)$ to denote the set of output states that can be obtained in the virtual setting for the given protocol $\coreQKDprotocol$. We will now prove a series of lemmas that reduce $\epssecure$-security statement we wish to prove (\cref{eq:targetofreduction}) to $\epssecure$-security of all states in $\worldhonest(\coreQKDprotocol)$.  We remark that, strictly speaking, the protocols in the virtual setting differ slightly from those in the {\realistic} authentication setting, as they make use of distinct classical communication registers (the virtual setting uses the corresponding magic registers). However, since these registers are equivalent by construction, and Alice and Bob  perform the same operations based on the received communication in both the {\realistic} and virtual authentication settings, we do not label these protocols differently across the two.
\subsubsection{Reducing security statements}
We first reduce the security analysis to states in $\worldvirtual(\coreQKDprotocol)$ partial on $\Onice$.
\begin{lemma} \label{lemma:reductionone}
        Consider the same setup as in \cref{theorem:reductionstatement,corr:reduction}. Then, the $\epssecure$-security for all output states in $\worldvirtual(\coreQKDprotocol)$ partial on $\Onice$, implies $\epssecure$-security for all output states in $\worldreal(\coreQKDprotocol)$ partial on $\Onice$. That is 
  \begin{equation}
      \begin{aligned}
    \tracedist{\rho^\mathrm{real,virt}_{K_A K_B \CfinalQKD \bm{\Ccorr} \EfinalQKD \wedge \Onice }  - \rho^\mathrm{ideal,virt}_{K_A K_B \CfinalQKD \bm{\Ccorr}  \EfinalQKD \wedge \Onice} }  &\leq \epssecure \qquad \forall \rho^\mathrm{real,virt}_{K_A K_B \CfinalQKD \bm{\Ccorr} \EfinalQKD } \in  \worldvirtual(\coreQKDprotocol) \\
    &\Downarrow \\
     \tracedist{\rho^\mathrm{real}_{K_A K_B \CfinalQKD \EfinalQKD \wedge \Onice}  - \rho^\mathrm{ideal}_{K_A K_B \CfinalQKD \EfinalQKD \wedge \Onice} }  &\leq \epssecure \qquad \forall \rho^\mathrm{real}_{K_A K_B \CfinalQKD \EfinalQKD  } \in  \worldreal(\coreQKDprotocol) 
      \end{aligned}
  \end{equation}
\end{lemma}
\begin{proof}
    Fix any attack by Eve, and consider the corresponding state $\rho^\mathrm{real}_{K_A K_B \CfinalQKD \EfinalQKD} \in \worldreal(\coreQKDprotocol)$:
    \begin{equation}
        \rho^\mathrm{real}_{K_A K_B \CfinalQKD \EfinalQKD} = \rho^\mathrm{real}_{K_A K_B \CfinalQKD \EfinalQKD \wedge \Onice }+ \rho^\mathrm{real}_{K_A K_B \CfinalQKD \EfinalQKD \wedge \Onice^\complement}.
    \end{equation}
    For the same attack, consider the corresponding state $\rho^\mathrm{real,virt}_{K_A K_B \CfinalQKD  \bm{\Ccorr} \EfinalQKD} \in \worldvirtual(\coreQKDprotocol)$ in the virtual setting:
    \begin{equation}
        \rho^\mathrm{real,virt}_{K_A K_B \CfinalQKD  \bm{\Ccorr}\EfinalQKD} = \rho^\mathrm{real,virt}_{K_A K_B \CfinalQKD  \bm{\Ccorr} \EfinalQKD \wedge \Onice }+ \rho^\mathrm{real,virt}_{K_A K_B \CfinalQKD  \bm{\Ccorr} \EfinalQKD \wedge \Onice^\complement}.
        \end{equation}
Recall that the virtual setting is the one where the correct messages are received by both parties in the registers $\bm{\Ccorr}$, and Alice and Bob use these values in the protocol. In the event $\Onice$, the actual messages (received in $\CfinalQKD$) are exactly the same as the  correct messages received in $\bm{\Ccorr}$, and are received at exactly the same time as the correct messages in $\bm{\Ccorr}$. Thus, these two registers are  equivalent, and lead to Alice and Bob performing the same operations at all times (conditioned on the event $\Onice$). Thus, we have 
\begin{equation}
\begin{aligned}
    \Tr_{\bm{\Ccorr}} \left[\rho^\mathrm{real,virt}_{K_A K_B \CfinalQKD  \bm{\Ccorr} \EfinalQKD \wedge \Onice}  \right] &= \rho^\mathrm{real}_{K_A K_B \CfinalQKD   \EfinalQKD \wedge \Onice} \\
     \Tr_{\bm{\Ccorr}} \left[\rho^\mathrm{ideal,virt}_{K_A K_B \CfinalQKD  \bm{\Ccorr} \EfinalQKD \wedge \Onice}  \right] &= \rho^\mathrm{ideal}_{K_A K_B \CfinalQKD   \EfinalQKD \wedge \Onice}
    \end{aligned}
\end{equation}
where the second equality follows from the first equality, and the fact that $\idealmap$ does not act on $\bm{\Ccorr}$.
With these two equalities, we obtain the desired claim by using the fact that the one-norm decreases under the action of CPTP maps, as follows:
\begin{equation}
\begin{aligned}
&\tracedist{\rho^\mathrm{real}_{K_A K_B \CfinalQKD \EfinalQKD \wedge \Onice}  - \rho^\mathrm{ideal}_{K_A K_B \CfinalQKD \EfinalQKD \wedge \Onice} }  \\
&= \tracedist{\Tr_{\bm{\Ccorr}} \left[\rho^\mathrm{real,virt}_{K_A K_B \CfinalQKD \bm{\Ccorr} \EfinalQKD \wedge \Onice} - \rho^\mathrm{ideal,virt}_{K_A K_B \CfinalQKD \bm{\Ccorr} \EfinalQKD \wedge \Onice}  \right] } \\
&\leq 
\tracedist{\rho^\mathrm{real,virt}_{K_A K_B \CfinalQKD \bm{\Ccorr} \EfinalQKD \wedge \Onice} - \rho^\mathrm{ideal,virt}_{K_A K_B \CfinalQKD \bm{\Ccorr} \EfinalQKD \wedge \Onice} } 
\end{aligned}
\end{equation}
The required claim follows from noting that the above inequality is true for all states $\rho^\mathrm{real}_{K_A K_B \CfinalQKD  \EfinalQKD} \in \worldreal(\coreQKDprotocol)$. In fact, the inequality above can actually be tightened to an equality, since $\bm{\Ccorr}$ is simply a copy of the messages in $\CfinalQKD$ (although the weaker inequality already suffices for our purposes).

\end{proof}

The following lemma reduces the security analysis to states in $\worldvirtual(\coreQKDprotocol)$ (without needing to consider partial states on event $\Onice$).

\begin{lemma} \label{lemma:reductiontwo}
 Consider the same setup as in \cref{theorem:reductionstatement,corr:reduction}. Then, the $\epssecure$-security for all output states in $\worldvirtual(\coreQKDprotocol)$, implies $\epssecure$-security for all output states in $\worldvirtual(\QKDprotocol)$, subnormalized conditioned on the event $\Onice$. That is 
\begin{equation}
      \begin{aligned}
\tracedist{\rho^\mathrm{real,virt}_{K_A K_B \CfinalQKD \bm{\Ccorr} \EfinalQKD }  - \rho^\mathrm{ideal,virt}_{K_A K_B \CfinalQKD \bm{\Ccorr}  \EfinalQKD } }  &\leq \epssecure \qquad \forall \rho^\mathrm{real,virt}_{K_A K_B \CfinalQKD \bm{\Ccorr} \EfinalQKD } \in  \worldvirtual(\coreQKDprotocol) \\
    &\Downarrow \\
    \tracedist{\rho^\mathrm{real,virt}_{K_A K_B \CfinalQKD \bm{\Ccorr} \EfinalQKD \wedge \Onice }  - \rho^\mathrm{ideal,virt}_{K_A K_B \CfinalQKD \bm{\Ccorr}  \EfinalQKD \wedge \Onice} }  &\leq \epssecure \qquad \forall \rho^\mathrm{real,virt}_{K_A K_B \CfinalQKD \bm{\Ccorr} \EfinalQKD  } \in  \worldvirtual(\coreQKDprotocol) \\
      \end{aligned}
  \end{equation}
\end{lemma}

\begin{proof}
For any state $\rho^\mathrm{real,virt}_{K_A K_B \CfinalQKD \bm{\Ccorr} \EfinalQKD  } \in  \worldvirtual(\coreQKDprotocol) $ we have
\begin{equation}
\begin{aligned}
&\tracedist{\rho^\mathrm{real,virt}_{K_A K_B \CfinalQKD \bm{\Ccorr} \EfinalQKD \wedge \Onice }  - \rho^\mathrm{ideal,virt}_{K_A K_B \CfinalQKD \bm{\Ccorr}  \EfinalQKD  \wedge \Onice } }  \\
&\leq \tracedist{\rho^\mathrm{real,virt}_{K_A K_B \CfinalQKD \bm{\Ccorr} \EfinalQKD \wedge \Onice }  - \rho^\mathrm{ideal,virt}_{K_A K_B \CfinalQKD \bm{\Ccorr}  \EfinalQKD  \wedge \Onice} }  \\
&+\tracedist{\rho^\mathrm{real,virt}_{K_A K_B \CfinalQKD \bm{\Ccorr} \EfinalQKD \wedge \Onice^\complement}  - \rho^\mathrm{ideal,virt}_{K_A K_B \CfinalQKD \bm{\Ccorr}  \EfinalQKD  \wedge \Onice^\complement} } \\
&= \tracedist{\rho^\mathrm{real,virt}_{K_A K_B \CfinalQKD \bm{\Ccorr} \EfinalQKD}  - \rho^\mathrm{ideal,virt}_{K_A K_B \CfinalQKD \bm{\Ccorr}  \EfinalQKD  } }
\end{aligned}
\end{equation}
where the first inequality follows from the fact that we add positive terms to the right hand side of the inequality, and the final equality follows from the fact that the states conditioned on $\Onice$ and $\Onice^\complement$ have support on orthogonal spaces.
\end{proof}

The following lemma reduces the security analysis to states in $\worldhonest(\coreQKDprotocol)$.

\begin{lemma} \label{lemma:reductionthree}
Consider the same setup as in \cref{theorem:reductionstatement,corr:reduction}. Then, the $\epssecure$-security for all output states in $\worldhonest(\coreQKDprotocol)$  implies $\epssecure$-security for all output states in $\worldvirtual(\coreQKDprotocol)$. That is 
\begin{equation}
      \begin{aligned}
\tracedist{\rho^\mathrm{real,hon}_{K_A K_B \CfinalQKD  \EfinalQKD }  - \rho^\mathrm{ideal,hon}_{K_A K_B \CfinalQKD   \EfinalQKD } }  &\leq \epssecure \qquad \forall \rho^\mathrm{real,hon}_{K_A K_B \CfinalQKD  \EfinalQKD } \in  \worldhonest(\coreQKDprotocol) \\
    &\Downarrow  \\
\tracedist{\rho^\mathrm{real,virt}_{K_A K_B \CfinalQKD \bm{\Ccorr} \EfinalQKD  }  - \rho^\mathrm{ideal,virt}_{K_A K_B \CfinalQKD \bm{\Ccorr}  \EfinalQKD } }  &\leq \epssecure \qquad \forall \rho^\mathrm{real,virt}_{K_A K_B \CfinalQKD \bm{\Ccorr} \EfinalQKD  } \in  \worldvirtual(\coreQKDprotocol) \\
      \end{aligned}
  \end{equation}
\end{lemma}
The idea behind the proof is based on the intuition that, in the virtual setting, the registers $\bm{\Ccorr}$ capture the communication that would occur in the honest model, and Alice and Bob use these registers for determining their actions. Consequently, the actual received messages are irrelevant — in fact, we can even assume that they are never delivered.

\begin{proof}

Consider any state $\rho^\mathrm{real,virt}_{K_A K_B \CfinalQKD  \bm{\Ccorr} \EfinalQKD} \in \worldvirtual(\coreQKDprotocol)$, and fix the corresponding attack strategy by Eve. From our construction of the virtual authentication setting, this implies that messages in $\bm{\Ccorr}$ are always correct and are received some time after being sent. Messages in $\CfinalQKD$ may include $\authabort$s and timing irregularities, but they are not used by Alice or Bob in the protocol at all. 

Next, consider the related attack strategy in the honest authentication setting, which we construct from the attack strategy in the virtual setting. Here Eve creates a copy of every message sent out from Alice’s or Bob’s laboratory. Each message is correctly delivered to the receiving party (according to the timing specified by $\bm{\Ccorr}$). Instead of attacking those messages, Eve performs her attack from the virtual setting on the \emph{copy} of the messages sent out, and does not forward the resulting (potentially tampered message) to the receiving party (see \cref{fig:connectingworlds}). Let the resulting state be $\rho^\mathrm{real,hon}_{K_A K_B \CfinalQKD \EfinalQKD} \in \worldhonest(\coreQKDprotocol)$. Then, we have the following equality with a slight abuse of notation:
\begin{equation}
    \rho^\mathrm{real,virt}_{K_A K_B \CfinalQKD  \bm{\Ccorr} \EfinalQKD} 
    = 
    \rho^\mathrm{real,hon}_{K_A K_B \CfinalQKD \EfinalQKD},
\end{equation}
where $\CfinalQKD$ and $\EfinalQKD$ on the left-hand side (which store the outcome of Eve’s attack) are identified with $\EfinalQKD$ on the right-hand side (where Eve implements the same attack on a copy of the classical messages and does not forward the resulting message to Bob), and $\bm{\Ccorr}$ on the left-hand side (which contains the correctly delivered, correctly timed messages) is identified with $\CfinalQKD$ on the right-hand side (the honest setting, where correct messages are received at the correct times).
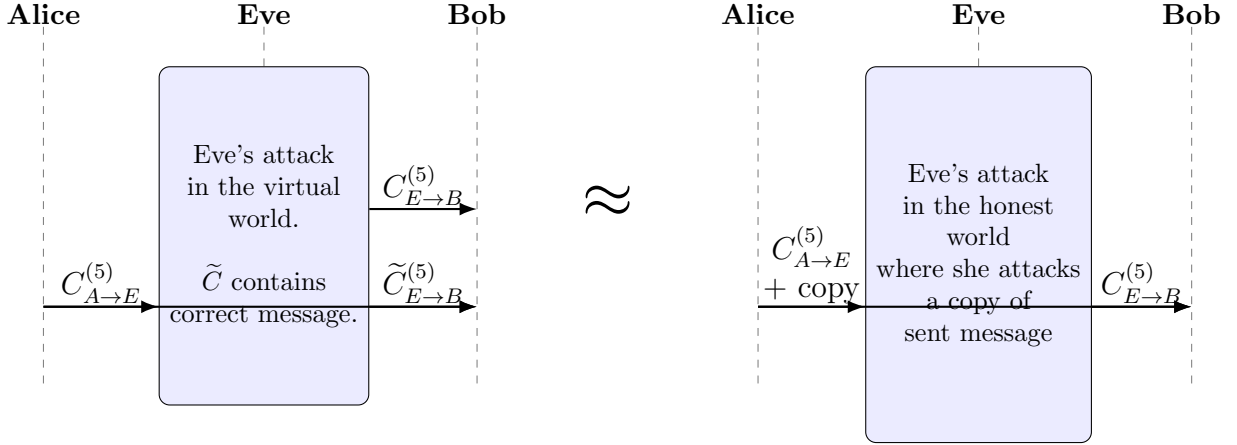
\begin{figure}[h!]
  \centering
  \begin{tikzpicture}[every node/.style={inner sep=1pt}]

    \begin{scope}[xshift=-5.5cm]
      \node[participant] (AliceL) {Alice};
      \node[participant] (EveL)   [right=2cm of AliceL] {Eve};
      \node[participant] (BobL)   [right=2cm of EveL]   {Bob};

      \draw[lifeline] (AliceL) -- ++(0,-5);
      \draw[lifeline] (EveL)   -- ++(0,-5);
      \draw[lifeline] (BobL)   -- ++(0,-5);

      \coordinate (A0L) at ($(AliceL)+(0,-\rowsep)$);
      \coordinate (B0L) at ($(BobL)+(0,-\rowsep)$);
       \coordinate (E0L) at ($(EveL)+(0,-\rowsep)$);

      \node[attackbox,align=center,minimum height=4.5cm,anchor=north] (EboxL)
        at ($(EveL)+(0,-0.7)$) {Eve's attack\\in the virtual \\world. \\\\ $\Ccorr$ contains \\ correct message.};

      \nextrow{A1L}{A0L}
      \nextrow{B1L}{B0L}
      \nextrow{E1L}{E0L}
      \draw[message] (EboxL.east |- A1L) -- node[midway,above] {$C_{E\to B}^{(5)}$} (B1L);

      \nextrow{A2L}{A1L}
      \nextrow{B2L}{B1L}
      \nextrow{E2L}{E1L}

       \draw[message]
        (A2L) -- (B2L);
        
      \draw[message] (A2L) -- node[midway,above] {$C_{A\to E}^{(5)}$} (EboxL.west |- B2L);
      \draw[message] (EboxL.east |- A2L) -- node[midway,above] {$\Ccorr_{E\to B}^{(5)}$} (B2L);
    \end{scope}

    \node[font=\Huge] at (2,-2.5) {$\approx$};

    \begin{scope}[xshift=4cm]
      \node[participant] (AliceR) {Alice};
      \node[participant] (EveR)   [right=2cm of AliceR] {Eve};
      \node[participant] (BobR)   [right=2cm of EveR]   {Bob};

      \draw[lifeline] (AliceR) -- ++(0,-5);
      \draw[lifeline] (EveR)   -- ++(0,-5);
      \draw[lifeline] (BobR)   -- ++(0,-5);

      \coordinate (A0R) at ($(AliceR)+(0,-\rowsep)$);
      \coordinate (B0R) at ($(BobR)+(0,-\rowsep)$);
    \coordinate (E0R) at ($(EveR)+(0,-\rowsep)$);

      \node[attackbox,align=center,minimum height=5.0cm,anchor=north] (EboxR)
        at ($(EveR)+(0,-0.7)$) {Eve's attack \\in the honest\\world \\ where she attacks \\ a copy of \\ sent message};

      \nextrow{A1R}{A0R}
      \nextrow{B1R}{B0R}
      \nextrow{E1R}{E0R}

      \nextrow{A2R}{A1R}
      \nextrow{B2R}{B1R}
      \nextrow{E2R}{E1R}

       \draw[message]
        (A2R) --  (B2R);

\draw[message] 
  (A2R) -- node[midway,above,align=center] {$C_{A\to E}^{(5)}$ \\ + copy} 
  (EboxR.west |- A2R);
      \draw[message] (EboxR.east |- A2R) -- node[midway,above] {$C_{E\to B}^{(5)}$} (B2R);

    \end{scope}

  \end{tikzpicture}
 \caption{For every operation that Eve performs in the virtual authentication setting, one can construct an equivalent operation in the honest authentication setting, where Eve creates a copy of each sent message and performs her original operation on this copy. She does not forward the result of her attack on the copy to the receiver; instead, she forwards the original, unmodified message to the receiving party.}
  \label{fig:connectingworlds}
\end{figure}

From these properties, we have  
\begin{equation}
\tracedist{
\rho^\mathrm{real,virt}_{K_A K_B \CfinalQKD  \bm{\Ccorr} \EfinalQKD} 
- 
\rho^\mathrm{ideal,virt}_{K_A K_B \CfinalQKD  \bm{\Ccorr} \EfinalQKD}
}
=
\tracedist{
\rho^\mathrm{real,hon}_{K_A K_B \CfinalQKD \EfinalQKD}
-
\rho^\mathrm{ideal,hon}_{K_A K_B \CfinalQKD \EfinalQKD}
}.
\end{equation}

This proves the required statement: if the right-hand side is upper bounded by $\epssecure$ for all states $\rho^\mathrm{real,hon}_{K_A K_B \CfinalQKD \EfinalQKD} \in \worldhonest(\coreQKDprotocol)$, then the same bound holds for the left-hand side for all states $\rho^\mathrm{real,virt}_{K_A K_B \CfinalQKD \bm{\Ccorr} \EfinalQKD} \in \worldvirtual(\coreQKDprotocol)$. This concludes the proof.
\end{proof}

Thus, combining all these reductions, we obtain the required statement which we restate below:

\reductionstatement*
\begin{proof}
    The proof follows from \cref{corr:reduction,lemma:reductionone,lemma:reductiontwo,lemma:reductionthree}.
\end{proof}

\section{Delayed Authentication}
\label{sec:delayedauthentication}

The protocols considered so far utilize an authenticated classical channel for every message sent and received, as described in \cref{subsec:classicalcommunicationsmodel}. However, the number of bits required to authenticate an $n$-bit message with information-theoretic security increases with $n$ (albeit only logarithmically)~\cite{wegman_new_1981,fung_practical_2010}. Consequently, it is more efficient (in terms of authentication key consumption) to batch multiple messages together before authenticating them.

In this section, we consider a modified version of the previously studied scenario, in which the authenticated classical channel is used only twice - once by each party - to authenticate the entire transcript of classical communication. For this section, we assume that Alice and Bob have synchronized clocks. We emphasize that this assumption is made only for this section and is necessary to ensure that Alice and Bob can compare transcripts along with their associated timestamps. We again let $\coreQKDprotocol$ denote a generic QKD protocol. However, we now analyze the security of a setting where all communication during $\coreQKDprotocol$ takes place over an insecure, unauthenticated classical channel. In this case, Eve is free to replace any message or arbitrarily modify its timing, and the delivery restrictions described in \cref{subsec:classicalcommunicationsmodel} no longer apply.

We then introduce a delayed Authentication Post-Processing Protocol, denoted by $\delayedAPPprotocol$, described in \cref{subsec:delAPPprotocol}. In this step, Alice and Bob exchange their entire communication transcripts (along with timestamps) over an authenticated channel and verify their consistency. The restrictions described in \cref{subsec:classicalcommunicationsmodel} apply to these messages. If the transcripts match\footnote{They don't need to match literally, since the time stamps will be different. But the messages need to match, and the timings must satisfy certain time-ordering.} - i.e., if they confirm that they received the correct messages at the appropriate times - they proceed; otherwise, they abort the protocol.

With these modifications, we are able to establish an analogue of \cref{theorem:reductionstatement}, stated in \cref{theorem:delayedreductionstatement}. The proof follows via analogous steps to that of \cref{theorem:reductionstatement}. We note that the idea of delayed authentication has been explored in the context of QKD in several prior works \cite{kon_quantumauthenticated_2024,kiktenko_lightweight_2020}. However, these studies typically focus on estimating authentication costs or perform the security analysis only for specific QKD protocols.

For ease of explanation and to maintain intuitive correspondence with earlier sections, we make a slight abuse of notation in this section. In particular, the real output states are denoted using the same symbols as before ($\rho^\mathrm{real}_{K_A K_B \CfinalQKD \EfinalQKD}$), but they now correspond to a different setting $\worlddelreal(\QKDprotocol)$ (described in greater detail in the following subsections). The honest and virtual settings, however, are defined exactly as before.

\subsection{Delayed Authentication Post-Processing Protocol (del-APP)} \label{subsec:delAPPprotocol}

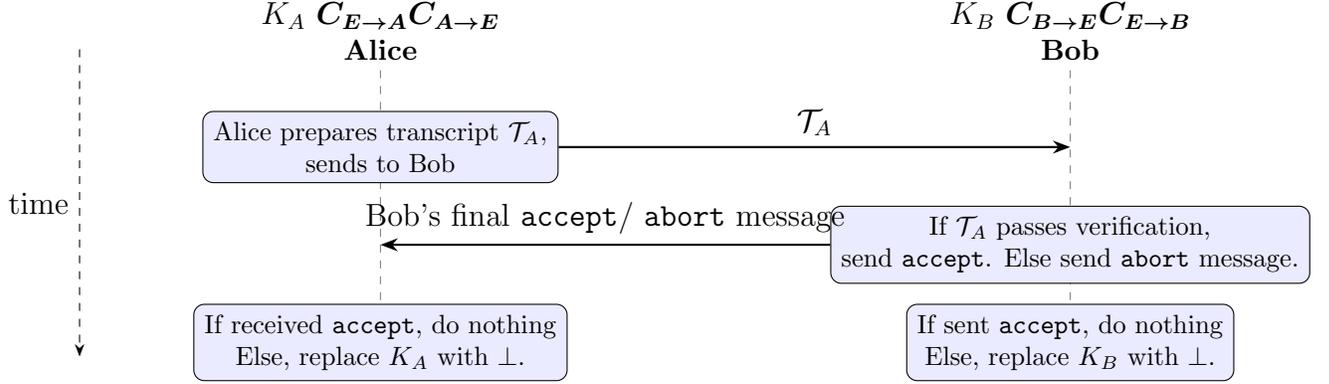
\begin{figure} 
\begin{tikzpicture}[
    >=Stealth,
    participant/.style={font=\small\bfseries},
    lifeline/.style  ={gray,dashed},
    message/.style   ={thick,-Stealth},
    localop/.style   ={draw,rectangle,rounded corners,
                       fill=blue!8,inner sep=4pt,
                       font=\footnotesize,align=center},
]

\def\rowsep{1.3} 

\node[participant] (Alice) {Alice};
\node[participant] (Bob)   [right=8cm of Alice] {Bob};

\node at ($(Alice)+(0,0.45)$) {$K_A\ \CAr \CAs$};
\node at ($(Bob)  +(0,0.45)$) {$K_B\ \CBs \CBr$};

\draw[lifeline] (Alice) -- ++(0,-4.4);
\draw[lifeline] (Bob)   -- ++(0,-4.4);

\coordinate (A0) at ($(Alice)+(0,0)$);
\coordinate (B0) at ($(Bob)  +(0,0)$);

\nextrow{A1}{A0}
\nextrow{B1}{B0}

\node[localop] (A1op) at (A1) {Alice prepares transcript $\transcript_A$, \\ sends to Bob};
\draw[message] (A1op) -- node[midway,above] {$\transcript_A$} (B1);

\nextrow{A2}{A1}
\nextrow{B2}{B1}

\node[localop] (B2op) at (B2) {If $\transcript_A$ passes verification, \\
send \accept. Else send \abort message.};
\draw[message] (B2op) -- node[midway,above] {Bob's final \accept / \abort message} (A2);

\nextrow{A3}{A2}
\nextrow{B3}{B2}

\node[localop] (A3op) at (A3) {If received \accept, do nothing \\
Else, replace $K_A$ with $\bot$.};
\node[localop] (B3op) at (B3) {If sent \accept, do nothing \\
Else, replace $K_B$ with $\bot$.};

\draw[->, dashed]
  ($(Alice)+(-4,0)$) -- ($(AliceEnd)+(-4,0)$)
  node[midway, left] {time};

\end{tikzpicture}
\caption{Schematic of \nameref{prot:delauthpp}  described in \cref{subsec:delAPPprotocol}. Alice and Bob first communicate and verify transcripts. If transcripts matches, they accept the protocol. Else, they abort the protocol and replace their key registers with $\bot$s.}
\label{fig:delappprotocol}
\end{figure}

As before, let $\rho^\mathrm{real}_{K_A K_B \CfinalQKD \EfinalQKD}$ denote the final state obtained after the execution of the core QKD protocol $\coreQKDprotocol$, where $\CfinalQKD = \CAs \CBr \CBs \CAr$ collects all classical communication that occurred during the core protocol, and $\EfinalQKD$ denotes all of Eve’s side information, which may include a copy of the classical communication. Recall our notation: Alice sends (receives) her $i$th message in the register $C^{(i)}_{A \rightarrow E}$ ($C^{(i)}_{E\rightarrow A}$) at time $t^{(i)}_{A \rightarrow E}$ ($t^{(i)}_{E \rightarrow A}$) with Bob's messages and timings defined analogously. During the authentication post-processing phase, Alice and Bob start with the state
$\rho^\mathrm{real}_{K_A K_B \CfinalQKD \EfinalQKD}$ and perform the following actions (see  \cref{fig:delappprotocol}).

\vspace{2em}
\begin{prot}[del-AuthPP Protocol]
\label{prot:delauthpp}
Starts with the state
$\rho^\mathrm{real}_{K_A K_B \CfinalQKD \EfinalQKD}$.
Classical communication is undertaken in the registers
$\CAsauth, \CBrauth, \CBsauth, \CArauth$.

\begin{enumerate}[label=\textbf{dAPP~\arabic*}, ref=dAPP~\arabic*]
\item \label{prot:delauthpp-alice}
Alice prepares a transcript $\transcript_A$ describing the messages she sent
and received during $\coreQKDprotocol$, and their timings. 
That is, she prepares
\[
\transcript_A =
\left(
\left\{ \left( C^{(i)}_{A \rightarrow E}, t^{(i)}_{A \rightarrow E} \right) \right\}_i,
\left\{ \left( C^{(j)}_{E \rightarrow A}, t^{(j)}_{E \rightarrow A} \right) \right\}_j
\right).
\]
She sends $\transcript_A$ to Bob by making one use of the  authenticated
classical channel.

\item \label{prot:delauthpp-bob}
If Bob receives an $\authabort$, he sends an $\abort$ message to Alice.
Otherwise, he checks whether the received transcript matches his own.
That is, he verifies that messages were received after they were sent, and that the message contents matched. Formally he checks whether:
\begin{equation}
  \begin{aligned}
    t^{(i)}_{A \rightarrow E} &\leq t^{(i)}_{E \rightarrow B} 
    &&\forall i, \\
    t^{(i)}_{B \rightarrow E} &\leq t^{(i)}_{E \rightarrow A} 
    &&\forall i, \\
    C^{(i)}_{A \rightarrow E} \text{ and }& C_{E\rightarrow B}^{(i)} \text{ contain identical messages }
    &&\forall i, \\
    C_{E \rightarrow A}^{(i)} \text{ and }&  C_{B \rightarrow E}^{(i)}  \text{ contain identical messages }
    &&\forall i.
  \end{aligned}
\end{equation}
If verification passes, he sends an $\accept$ message;
otherwise, he sends an $\abort$ message.

\item \label{prot:delauthpp-bob2}
If Bob sends an $\accept$ message, he does nothing.
If he sends an $\abort$ message, he replaces $K_B$ with $\bot$.
We denote by $l_B$ the final key length stored in $K_B$.

\item \label{prot:delauthpp-alice2}
If Alice receives an $\accept$ message, she does nothing.
If Alice receives either an $\authabort$ or an $\abort$ message,
she replaces her key register $K_A$ with $\bot$.
We denote by $l_A$ the final key length stored in $K_A$.

\end{enumerate}
\end{prot}

%

Mathematically, the above protocols can be described as follows:
\begin{enumerate}
    \item In \cref{prot:delauthpp-alice} and \cref{prot:delauthpp-bob}, Alice and Bob engage in two rounds of communication. This can be described by a map (influenced by Eve) $\delauthcommmap \in \CPTP(\CfinalQKD \EfinalQKD,\Cfinal \Efinal)$.
    \item In \cref{prot:delauthpp-alice2,prot:delauthpp-bob2}, Alice and Bob  use the result of the communication in the previous step $\Cauth$ to determine their final $\accept$ / $\abort$ status. This is described as a map 
    $\delauthupdatemap \in \CPTP( K_A K_B \Cauth, K_A K_B \Cauth )$
\end{enumerate}
The final output state is denoted by  $\rho^\mathrm{real,final}_{K_A K_B \Cfinal \Efinal} = \delauthupdatemap \circ \delauthcommmap \left[\rho^\mathrm{real}_{K_A K_B \CfinalQKD \EfinalQKD}\right]$, where $\Cfinal = \CfinalQKD \Cauth$. 

Notice that the structure of this protocol closely mirrors that of \nameref{prot:authpp}, except that the first two steps of \nameref{prot:authpp} are omitted here. This alignment ensures that the ensuing proofs proceed in essentially the same way.

\begin{remark} \label{remark:memory}
Notice that in this setting, there are only two uses of the authenticated classical channel. Consequently, the amount of secret key required for authentication is significantly lower than what was considered previously in this work. However, Alice and Bob must now store all messages sent and received (along with their corresponding timestamps) until the very end of the protocol, in order to compare and match their transcripts. This, in turn, reduces the benefits of on-the-fly announcements \cite[Remark 3.1]{tupkary2025qkdsecurityproofsdecoystate}, which allow Alice and Bob to perform public announcements on-the-fly and save  classical memory and storage requirements.
Nevertheless, one can still envision a use case where on-the-fly announcements are performed along with delayed authentication to enable Alice and Bob to begin their classical processing (such as sifting) earlier, while still retaining the necessary information in memory for transcript comparison at a later stage. Alternatively, one can accept the requirement for larger classical memory (and not implement on-the-fly announcements), in order to save authentication keys. Finally, note that Alice and Bob need only store sufficient information to recover the ordering of all messages; the precise timestamps are not necessary. This can simplify the storage requirements further.
\end{remark}

\subsection{Reduction Statement} \label{subsec:delayedreductionstatement}
We are interested in the security analysis of $\delayedQKDprotocol = \delayedAPPprotocol \circ \coreQKDprotocol$.
Let us denote the set of output states that are possible when we use insecure, unauthenticated communication during $\coreQKDprotocol$ and authenticated communication during $\delayedAPPprotocol$ as $\worlddelreal(\delayedQKDprotocol)$.  The analogous result  to \cref{theorem:reductionstatement} for the delayed authentication setting can now be stated and proved  in an analogous manner.

\begin{restatable}[Reduction of QKD security analysis to the honest authentication setting with delayed authentication]{theorem}{delayedreductionstatement}
\label{theorem:delayedreductionstatement}
Let $\coreQKDprotocol$ be an arbitrary QKD protocol. Let $\delayedAPPprotocol$ be the \nameref{prot:delauthpp} described in \cref{subsec:delAPPprotocol}, executed after the core QKD protocol $\coreQKDprotocol$. Let $\delayedQKDprotocol = \delayedAPPprotocol \circ \coreQKDprotocol$ denote the resulting QKD protocol. Let $\worlddelreal(\QKDprotocol)$ denote the set of output states in the delayed authentication setting, where communication during $\coreQKDprotocol$ is not authenticated while communication during $\delayedAPPprotocol$ is authenticated (see \cref{subsec:classicalcommunicationsmodel}). Let $\worldhonest(\coreQKDprotocol)$ denote the set of output states in the honest authentication setting (see \cref{subsec:classicalcommunicationsmodel}).
Then, the $\epssecure$-security for all output states in $\worldhonest(\coreQKDprotocol)$ implies $\epssecure$-security for all output states in $\worlddelreal(\delayedQKDprotocol)$. That is,
\begin{equation}
    \begin{aligned}
          \tracedist{ \rho^\mathrm{real,hon}_{K_A K_B \CfinalQKD \EfinalQKD} -  \rho^\mathrm{ideal,hon}_{K_A K_B \CfinalQKD \EfinalQKD}} &\leq \epssecure \qquad \forall \rho^\mathrm{real}_{K_A K_B \CfinalQKD \EfinalQKD} \in \worldhonest(\coreQKDprotocol) \\
          &\Downarrow \\
     \tracedist{ \rho^\mathrm{real}_{K_A K_B \Cfinal \Efinal} -  \rho^\mathrm{ideal}_{K_A K_B \Cfinal \Efinal}} &\leq \epssecure \qquad \forall \rho^\mathrm{real}_{K_A K_B \Cfinal \Efinal} \in \worlddelreal(\delayedQKDprotocol).
    \end{aligned}
\end{equation}
\end{restatable}

The proof is stated in \cref{appendix:delayedreductionproof}.

\section{Summary and Outlook} \label{sec:authenticationsummary}
In this chapter, we addressed the problem of analyzing QKD protocols in the {\realistic} authentication setting, where authentication may result in asymmetric aborts and message timing can be modified. This setting necessitates a modification of the standard QKD security definition and, crucially, renders nearly all existing analyses, which are undertaken in the honest authentication setting, invalid under {\realistic} authentication assumptions.

By introducing an authentication post-processing protocol, we demonstrated that the security of a complete QKD protocol (including the authentication post-processing) in the {\realistic} authentication setting can be reduced to that of the core QKD protocol in the honest authentication setting. The latter is the setting overwhelmingly considered in existing QKD security proofs. Hence, our results retroactively lift the security of such prior analyses to the {\realistic} authentication setting. Moreover, our construction is fully general—it applies to any QKD protocol and is equally suitable for device-dependent, measurement-device-independent\footnote{Such protocols also involve public announcements made by an untrusted third party. We emphasize that our analysis applies only to the communication between Alice and Bob themselves; it does not address how to handle announcements originating from an external third party. Typically in MDI protocols, Alice and Bob need to achieve consensus on the third party’s announcements --- while our result does not directly address those announcements, it does show that Alice and Bob can rely on the honest-authentication scenario for communication between themselves, in order to achieve such consensus.} (MDI), and device-independent (DI) protocols, whether of the prepare-and-measure or entanglement-based type. We also presented two variants of this construction: one in which all communication is authenticated, and another where only the final two messages are authenticated, thereby reducing the consumption of authentication keys.

In this chapter, our perspective was that Eve’s attack together with the protocol description jointly determine a single channel mapping input states to output states.  Looking ahead, an important next step is to formalize the new security definition and the {\realistic} authenticated channel within a composable security framework. Achieving this will likely require employing models such as causal boxes \cite{portmann_causal_2017} or related formalisms that explicitly capture the causal and temporal structure of classical communication.

\chapter{Summary and Outlook} \label{chap:conclusion}

As promised in \cref{chap:introduction}, this thesis set out to present \textit{rigorous} security proofs for \textit{practical} quantum key distribution protocols using a \textit{variety} of proof techniques. 

We encountered fully rigorous security proofs, most notably in \cref{chap:MEAT}, where we proved the security of a highly general QKD protocol with a strong emphasis on precision, explicit assumptions, and mathematical clarity. We also explored a wide range of proof techniques, including IID collective attack analyses (\cref{chap:variable}), postselection based uplifts to coherent attacks (\cref{chap:postselection}), entropic uncertainty relations (\cref{chap:EUR}), and MEAT-based methods (\cref{chap:MEAT}). In every case, the analysis was applied to one of the most practically relevant QKD protocols: decoy-state BB84.

Beyond serving as a unifying exposition of major proof techniques, this thesis makes several concrete contributions to the security analysis of QKD. In particular, \cref{chap:postselection} revamps the postselection technique introduced in Ref.~\cite{christandl_postselection_2009}, placing it on a more rigorous mathematical footing and ensuring its applicability to realistic optical implementations of QKD. In \cref{chap:variable}, we developed a security proof for variable-length QKD protocols, filling a gap in the existing literature. In \cref{chap:EUR}, we resolved a nearly two decades old open problem concerning basis-efficiency mismatch in phase error based security proofs, thereby allowing these methods to be applied to practical prepare-and-measure scenarios. This has now been extended to address additional scenarios, such as passive setups and correlated imperfections \cite{wang2025phase}, and combined with analyses of source imperfections \cite{curras_securityquantumkeydistribution_2025}. In \cref{chap:MEAT}, we presented a remarkably general and modular security analysis of practical QKD protocols using MEAT. Our analysis there can be straightforwardly applied to other QKD protocols beyond decoy-state BB84, and can be extended to include device imperfections and side-channel attacks as well. Therefore, it is meant to serve as the foundation for certification efforts for QKD. Finally, in \cref{chap:classicalauthentication}, we looked at some standard (unrealistic) assumptions commonly made about authentication in typical QKD security proofs, and showed that these assumptions can be relaxed to realistic ones with a minor protocol modification.

\paragraph*{Comparing proof techniques: }
At this point, a natural question arises: which proof technique is best?

 \begin{figure}[h!]
    \centering
    \includegraphics[width=0.85\linewidth]{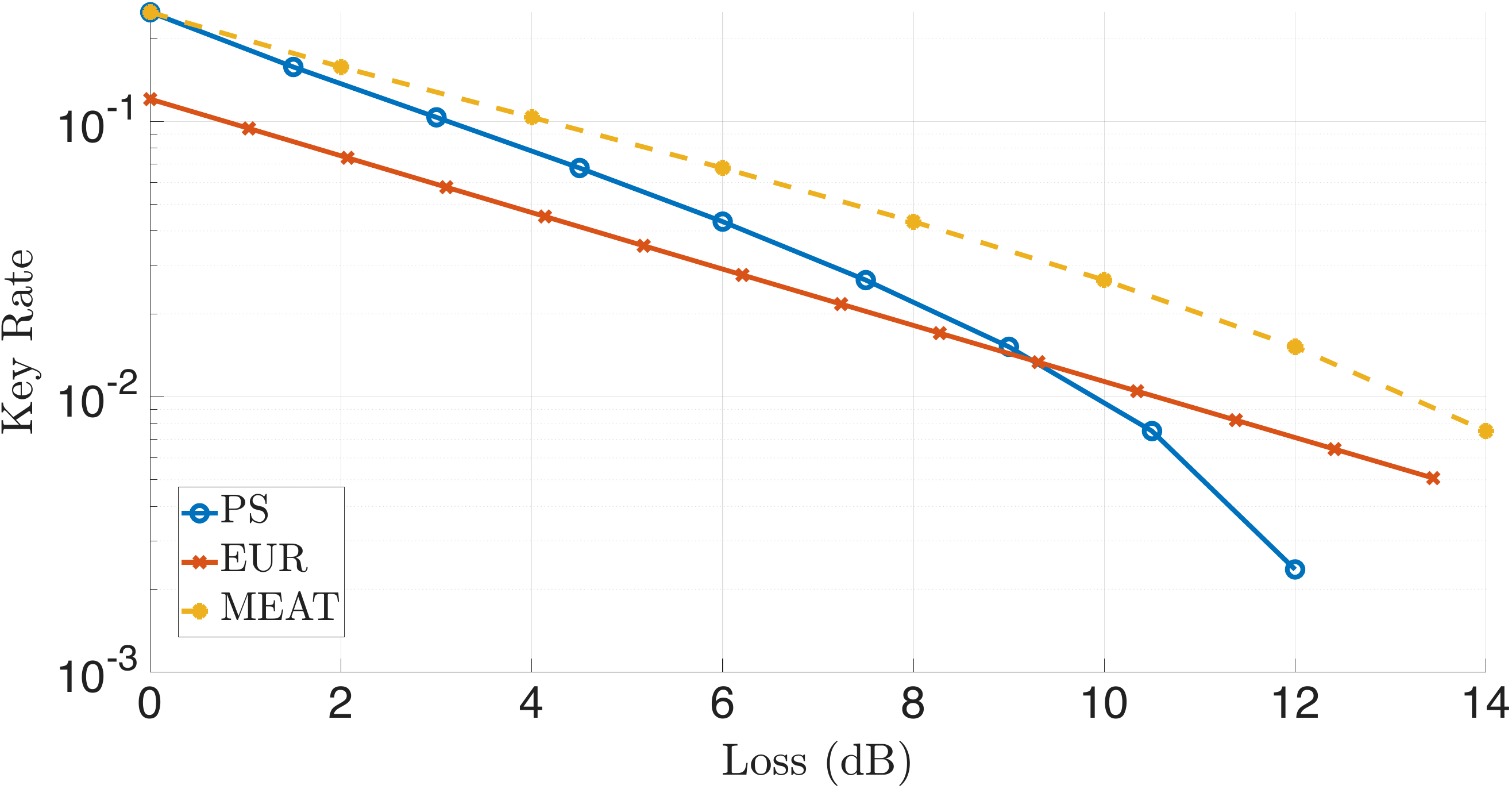}
   \caption{Key rate for the variable-length decoy-state BB84 protocol plotted as a function of loss, comparing the different proof techniques presented in this thesis. Note that the protocols are not strictly identical across techniques, due to differences inherent to the respective proof methods. For MEAT, the protocol is as described in \cref{sec:plotsMEAT}. The PS protocol is identical to the MEAT protocol, except that it does not allow on-the-fly announcements. For EUR, the protocol follows \cref{sec:eurresults}.
   Such comparison plots, over a wider range of parameters, were first presented in Refs.~\cite{kamin_phd_2026,kamin_renyi_2025}; the present figure is generated using code based on Ref.~\cite{kamin_renyi_2025} (MEAT),  Ref.~\cite{tupkary_phase_2024} (EUR) and Ref.~\cite{Kamin2025} (PS), which we duly acknowledge. In particular, Refs.~\cite{kamin_phd_2026,kamin_renyi_2025} should be cited for all results pertaining to the numerical comparison of key rates between these proof techniques.}
    \label{fig:comparison}
\end{figure}

In earlier chapters, we highlighted significant drawbacks of both postselection-based and phase error rate-based approaches. In particular, the postselection technique appears fundamentally difficult to extend to scenarios involving on-the-fly announcements and correlated imperfections. Phase error-based methods, on the other hand, suffer from a notable lack of modularity. (See also \cite[Section 7]{tupkary2025qkdsecurityproofsdecoystate} for a comparison between proof techniques).
In the opinion of this author, MEAT does not  suffer from comparably fundamental limitations, and is the ``correct" way to approach the security analysis of QKD protocols. Moreover, MEAT yields competitive key rates \cite{kamin_renyi_2025}, (see \cref{fig:comparison}). That said, MEAT is a relatively recent development and has not yet been explored to the same depth as more established techniques. Nevertheless, this author remain confident that it represents, overall, the most robust and conceptually apt approach to QKD security proofs currently available.

\paragraph{Future work: } While this thesis presents one highly rigorous security analysis for a practical protocol, it does not currently accommodate device imperfections into that analysis. The next major challenge in QKD lies in the systematic integration of imperfections and side-channel attacks into security analyses, with the aim of producing security proofs that apply directly to implemented systems. While many isolated results addressing individual imperfections exist—and while recent work has begun to combine several of these effects—a complete and unified treatment of all relevant imperfections and side channels for a given physical setup remains a formidable open problem. It is my hope that the methods developed in this thesis, and in particular those presented in \cref{chap:MEAT}, will serve as a foundation for future efforts to further close the gap between theoretical security proofs and real-world QKD systems. 

\newpage
Thank you for reading this thesis. Should you notice any shortcomings, I hope you can find solace in the following universal truth:

\epigraph{Papers and theses are never finished, only abandoned.}{Norbert L\"utkenhaus (original source unknown)}

\


\bibliographystyle{unsrturl}
\cleardoublepage 
\phantomsection  
\renewcommand*{\bibname}{References}

\addcontentsline{toc}{chapter}{\textbf{References}}

\bibliography{uw-ethesis.bib}



\appendix
\chapter*{APPENDICES}
\addcontentsline{toc}{chapter}{APPENDICES}
\appendix
\chapter{Variable-length input to privacy amplification}
\label{Appendix:variable}
We will now provide proof of the statements from  \cref{sec:variablelengthPA}.

\variableinputhashingfirstlemma*
\begin{proof}
As explained in \cref{subsec:problem}, the hashing procedure described above can be thought of as \textit{first} randomly sampling $f_i \in \hashfamily{i}{\lfixed}$ for every $i$, and then computing $f_{\mathrm{len}(\PAstring)}(\PAstring)$. However, as noted in that section, this process is not a valid universal$_2$ hashing procedure from $\{0,1\}^{\leq n}$ to $\lfixed$ bits.
	
Consider instead the following virtual hashing process, based on new hash families~\footnote{This specification of $\virthashfamily{i}{\lfixed}$ is not technically a set of\emph{functions}, since each element of $\virthashfamily{i}{\lfixed}$ is instead a tuple where the second term is an $\lfixed$-bit string. However, each such element uniquely specifies a function in a simple manner that we shall shortly specify.} $\virthashfamily{i}{\lfixed} \coloneqq \hashfamily{i}{\lfixed} \times \{0,1\}^{\lfixed}$ (for every $i$). This virtual process first randomly samples $(f_i,u_i) \in \virthashfamily{i{\lfixed}}$ for every $i$, i.e.~$f_i$ is sampled from the same universal$_2$ hash family $ \hashfamily{i}{\lfixed}$ as before, and $u_i$ is a random $\lfixed$-bit string. It then computes $  f_{\mathrm{len}(\PAstring)} (\PAstring) \oplus u_{\mathrm{len}(\PAstring)}$ as its hash output. 
Now, this virtual hashing procedure \textit{is} a valid universal$_2$ hashing procedure from $\{0,1\}^{\leq n}$ to $\lfixed$ bits, because each hash family $\virthashfamily{i}{\lfixed}$ is universal$_2$ \textit{and} satisfies the ``uniform output'' property (\cref{eq:uniformoutput}). 

Denote the output state of the virtual process (acting on $\rho_{\PAstring  \mathbf{C} \Eve}$) as $\rho^{(\suplabelsbrkt\text{virtual})}_{K_A   \mathbf{C}  \HPA  \Eve}$\footnote{The ``virtual" notation here has nothing to with the virtual states and worlds introduced in \cref{chap:MEAT}.}, where $\HPA$ stores the description of the hash function chosen in the virtual process (in particular, all the values $(f_i, u_i)$ from the virtual process). Let us analogously define $\rho^{(\suplabelsbrkt\text{ideal},\text{virtual})}_{K_A \mathbf{C} \HPA \Eve } \coloneqq \frac{\id_{K_A}}{|K_A|} \otimes \rho^{(\suplabelsbrkt\text{virtual})}_{  \mathbf{C} \HPA  \Eve}$.

Now, we construct a CPTP map $\mathcal{E}: K_A  \mathbf{C} \Eve \to K_A \mathbf{C} \Eve$ that will map the virtual output states to the actual output states. This map $\mathcal{E}$ does the following operations:
\begin{enumerate}
\item Look at $\mathbf{C}$ and determine the corresponding value $k_{\bar{C}}$ (as defined in the conditions of this lemma)\footnote{$\mathcal{E}$ cannot ``directly'' compute $\mathrm{len}(\PAstring)$ because the register $\PAstring$ is no longer present in the states it acts on.}, to be used in the subsequent steps.  
\item Look at $\HPA$ and determine $u_{k_{\bar{C}}}$, to be used in the subsequent steps.
\item Replace $K_A$ with $K_A \oplus u_{k_{\bar{C}}}$ (essentially removing the XOR that was applied during the hashing procedure).
\item Partial trace on the $\HPA$ register, on everything except the $f_{k_{\bar{C}}}$ information.
\end{enumerate}
It is straightforward to verify that this map $\mathcal{E}$ indeed satisfies
\begin{equation}
\begin{gathered}
\mathcal{E} \left( \rho^{(\suplabelsbrkt \text{virtual})}_{K_A \mathbf{C} \HPA \Eve} \right) = \rho^{\suplabelsbrkt}_{K_A \mathbf{C} \HPA \Eve}, \\
\mathcal{E} \left(  \rho^{(\suplabelsbrkt\text{ideal},\text{virtual})}_{K_A\mathbf{C} \HPA \Eve} \right) = 
\rho^{(\suplabelsbrkt\text{ideal})}_{\mathbf{C} \HPA \Eve},
\end{gathered}
\end{equation}
and analogously for the above states conditioned on the event $\Omega$ (since $\mathcal{E}$ does not disturb the register $\mathbf{C}$).

Therefore,  we have
\begin{equation}
\begin{aligned}
&	 \Pr(\Omega) \tracedist{ \rho^{\suplabelsbrkt}_{K_A \mathbf{C} \HPA \Eve| \Omega} - \rho^{(\suplabelsbrkt\text{ideal})}_{K_A \mathbf{C} \HPA \Eve| \Omega}}  \\
& =	\Pr(\Omega) \tracedist{ \mathcal{E} \left( \rho^{(\suplabelsbrkt \text{virtual})}_{K_A \mathbf{C} \HPA \Eve| \Omega} - \rho^{(\suplabelsbrkt\text{ideal},\text{virtual})}_{K_A\mathbf{C} \HPA \Eve | \Omega} \right)} \\
& \leq 	 \Pr(\Omega) \tracedist{  \rho^{(\suplabelsbrkt \text{virtual})}_{K_A \mathbf{C} \HPA \Eve| \Omega} - \rho^{(\suplabelsbrkt\text{ideal},\text{virtual})}_{K_A \mathbf{C} \HPA \Eve | \Omega} } \\
&\leq  \Pr(\Omega)  2^{- \left( \frac{\alpha-1}{\alpha}  \right)\left( \Halpha(\PAstring | \mathbf{C}  \Eve  )_{\rho | \Omega} - \lfixed +2 \right) } \\
&\leq  2^{- \left( \frac{\alpha-1}{\alpha}  \right)\left( \Halpha(\PAstring | \mathbf{C}  \Eve)_{\rho} - \lfixed + 2 \right)}
\end{aligned}
\end{equation}
where we used the fact that CPTP maps cannot increase trace norm in the third inequality, and
Leftover Hashing Lemma for {\Renyi} entropies (\cref{lemma:LHL}) for the fourth inequality, and \cref{lemma:conditioning} for the final inequality.
\end{proof}

\variableinputhashingsecondlemma*
\begin{proof} 
We intuitively expect \cref{eq:renyiinvariance} to be true, since \cref{eq:unitaryaction} essentially states that $\mathbf{C}$ can be used to isometrically convert $\rawkey$ to $\PAstring$. 
To formalize this, we first note that each isometry $V^{(\bar{c})}_{\rawkey_{\bar{c}} \rightarrow \PAstring}$ can always be extended to an isometry $V^{(\bar{c})}_{\rawkey \rightarrow \PAstring}$, i.e.~where the domain is the full Hilbert space of $\rawkey$ (padding the output space $\PAstring$ with extra dimensions if $\dim(\rawkey) > \dim(\PAstring)$). Furthermore, \cref{eq:unitaryaction} still holds with $V$ defined in terms of these new isometries instead, i.e.~we have
	\begin{equation}  \label{eq:unitaryactiontemp}
	\begin{gathered}
		V\rho_{\rawkey \mathbf{C} \Eve} V^\dagger =\rho_{\PAstring \mathbf{C} \Eve}, \text{ where} \\
		V \coloneqq \sum_{\bar{c}} V^{(\bar{c})}_{\rawkey \rightarrow \PAstring} \otimes \ket{\bar{c}}\bra{\bar{c}}_{\mathbf{C}}.
	\end{gathered}
\end{equation}
(It does not matter how we chose the extensions, since $\rho_{\rawkey \mathbf{C} \Eve}$ is only supported on a subspace that is unaffected by these choices of extensions.)

Furthermore, letting $\mathbf{C}_c$ be a copy of the register $\mathbf{C}$, using \cite[Lemma B.7]{dupuis_entropy_2020} we have 
\begin{equation}
	\begin{aligned}
		H_\alpha(\PAstring \mathbf{C}_c | \mathbf{C} E)_\rho &= H_\alpha( \PAstring | \mathbf{C} E)_\rho ,\\
		H_\alpha(\rawkey \mathbf{C}_c | \mathbf{C} E)_\rho &= H_\alpha(\rawkey | \mathbf{C} E)_\rho,
	\end{aligned}
\end{equation}
Thus, it is enough to show that $	H_\alpha(\PAstring \mathbf{C}_c | \mathbf{C} E)_\rho  = H_\alpha(\rawkey \mathbf{C}_c | \mathbf{C} E)_\rho$. This follows from \cref{eq:unitaryactiontemp}, and the fact that the {\Renyi} entropy is invariant under isometries on the first subsystem (\cref{lemma:dpinonconditioningregister} with $f=0$), since by defining the isometry $\widetilde{V}_{\rawkey\mathbf{C}_c \rightarrow \PAstring \mathbf{C}_c} \coloneqq \sum_{\bar{c}} V^{(\bar{c})}_{\rawkey \rightarrow \PAstring} \otimes \ket{\bar{c}}\bra{\bar{c}}_{\mathbf{C}_c}$ we have
\begin{equation}  
	\widetilde{V}\rho_{\rawkey \mathbf{C}_c \mathbf{C} E} \widetilde{V}^\dagger =\rho_{\PAstring  \mathbf{C}_c \mathbf{C} E}.
\end{equation}
which concludes the proof\footnote{An alternative proof would be to instead use \cite[Proposition 5.1]{tomamichel_quantum_2016} to split the conditional entropies into terms conditioned on each value of $\mathbf{C}$, and note that the equality holds for each term by invariance of {\Renyi} entropy under isometries on the first subsystem.}.
\end{proof}

\chapter{Proofs of statements for the postselection technique}
\label{Appendix:PS}
In this appendix, we will provide proofs of various statements utilized in \cref{chap:postselection}. 

\satisfyingpermutationinvariance*

\begin{proof}
The first part of the lemma follows from the fact that, in
\nameref{prot:abstractqkdprotocol}, the state preparation and measurement
procedures are identical in each round. After applying the source-replacement
scheme and obtaining the effective protocol map $\protMap{\lkey}$, Alice and
Bob perform the same measurements in every round. Since the measurements
are the same in each round, the permutation $\pi$ may equivalently be applied and announced
either to the post-measurement classical registers $X_1^n Y_1^n$ or to the
pre-measurement quantum systems $A_1^n B_1^n$.

For the second part, observe that the difference between the real and ideal
protocols can be written as
\begin{equation}
\label{eq:perm_difference}
 \Tr_{K_B} \circ  \protMap{\lkey,\mathrm{perm}}
-
 \Tr_{K_B} \circ  \idealmap \circ \protMap{\lkey,\mathrm{perm}}
=
\mathcal{F}
\circ
\sum_{\pi \in \permset{n}}  \frac{1}{n!} 
\mathcal{W}_{\pi}
\otimes
\ketbra{\pi}_{C_{\mathrm{perm}}},
\end{equation}
where $\mathcal{W}_{\pi}$ denotes the action of the permutation $\pi$ on the
systems $A_1^n B_1^n$.

To prove permutation invariance, consider any permutation $\pi' \in \permset{n}$
and define a map $G_{\pi'}$ that replaces the classical register
$\ketbra{\pi}_{C_{\mathrm{perm}}}$ by $\ketbra{\pi \circ \pi'}_{C_{\mathrm{perm}}}$.
Then
\begin{equation}
\begin{aligned}
G_{\pi'} \circ
\left(
\mathcal{F}
\circ
\sum_{\pi \in \permset{n}}
\mathcal{W}_{\pi}
\otimes
\ketbra{\pi}_{C_{\mathrm{perm}}}
\right)
\circ
\mathcal{W}_{\pi'}
&=
G_{\pi'} \circ
\left(
\mathcal{F}
\circ
\sum_{\pi \in \permset{n}}  \frac{1}{n!} 
\mathcal{W}_{\pi \circ \pi'}
\otimes
\ketbra{\pi}_{C_{\mathrm{perm}}}
\right)
\\
&=
\mathcal{F}
\circ
\sum_{\pi \in \permset{n}}  \frac{1}{n!} 
\mathcal{W}_{\pi \circ \pi'}
\otimes
\ketbra{\pi \circ \pi'}_{C_{\mathrm{perm}}}
\\
&=
\mathcal{F}
\circ
\sum_{\pi \in \permset{n}}  \frac{1}{n!} 
\mathcal{W}_{\pi}
\otimes
\ketbra{\pi}_{C_{\mathrm{perm}}},
\end{aligned}
\end{equation}
where the first two equalities follow from the definitions of
$\mathcal{W}_{\pi'}$ and $G_{\pi'}$, and the final equality follows by a simple
relabeling of the permutation variable in the summation. This establishes
permutation invariance.
\end{proof}

\permutationdoesnotmatter*

\begin{proof}
The claim follows by straightforward algebra. Let $\mathcal{W}_{\pi}$ denote the
permutation $\pi$ acting on the $A$ and $B$ systems. Then
\begin{equation}
\begin{aligned}
&\tracedist{
\left(
\left(
\Tr_{K_B} \circ \protMap{\lkey,\mathrm{perm}}
-
\Tr_{K_B} \circ \protMapId{\lkey,\mathrm{perm}}
\right)
\otimes \idmap_{E_1^n}
\right)
\left[
\rho_{ABE}^{\otimes n}
\right]
}
\\
&=
\sum_{\pi \in \permset{n}}
\frac{1}{n!}
\tracedist{
\left(
\left(
\Tr_{K_B} \circ \protMap{\lkey,\mathrm{perm}}
-
\Tr_{K_B} \circ \protMapId{\lkey,\mathrm{perm}}
\right)
\otimes \idmap_{E_1^n}
\right)
\circ \mathcal{W}_{\pi}
\left[
\rho_{ABE}^{\otimes n}
\otimes
\ketbra{\pi}_{C_{\mathrm{perm}}}
\right]
}
\\
&=
\tracedist{
\left(
\left(
\Tr_{K_B} \circ \protMap{\lkey}
-
\Tr_{K_B} \circ \protMapId{\lkey}
\right)
\otimes \idmap_{E_1^n}
\right)
\left[
\rho_{ABE}^{\otimes n}
\right]
},
\end{aligned}
\end{equation}
where the final equality follows from the fact that
$\rho_{ABE}^{\otimes n}$ is invariant under permutations of the $A$ and $B$
systems.
\end{proof}

\pslemmaone*

\begin{proof}
    Construct 
    \begin{equation}
    	\bar{\rho}_{A_1^nB_1^nR^{\prime\prime} \widetilde{R}} = \frac{1}{n!} \sum_{\perm \in \permset{n}} \left( \mathcal{W}_\perm \otimes \id_{R^{\prime\prime}} \right) \left( \rho_{A_1^nB_1^n R^{\prime\prime}} \right) \otimes \ketbra{\perm}{\perm}_{\widetilde{R}}
    \end{equation}
    as an extension of $\bar{\rho}_{A_1^nB_1^n}$. Therefore, there exists $\Phi \in \CPTP( R^{\prime} , R^{\prime \prime} \widetilde{R})$ such that \\
    $ \left( \idmap_{A_1^nB_1^n} \otimes \Phi \right) \left( \bar{\rho}_{A_1^nB_1^nR^\prime} \right) = \bar{\rho}_{A_1^nB_1^n R^{\prime \prime} \widetilde{R}}$. Since trace norm cannot increase under CPTNI maps, we have
    \begin{equation} \label{eq:dataprocessing2}
    	\tracedist{ \left( \left(\mathcal{F} - \mathcal{F}^\prime \right) \otimes \idmap_{ R^{\prime \prime} \widetilde{R}}\right) \left( \bar{\rho}_{A_1^nB_1^n R^{\prime \prime} \widetilde{R} } \right) } \leq	\tracedist{ \left( \left(\mathcal{F} - \mathcal{F}^\prime \right) \otimes \idmap_{R^\prime}\right) \left( \bar{\rho}_{A_1^nB_1^nR^\prime} \right) } .
    \end{equation}
    Next, making use of the permutation-invariance of $\mathcal{F}-\mathcal{F}^\prime$, we have
    \begin{equation} \label{eq:perminvariancetracedist}
    	\begin{aligned} 
    			\tracedist{ \left( \left(\mathcal{F} - \mathcal{F}^\prime \right) \otimes \idmap_{R^{\prime\prime} }\right) \left( \rho_{A_1^nB_1^nR^{\prime\prime}} \right) }  & = \frac{1}{n!} \sum_{\perm \in \permset{n}}	\tracedist{ \left( G_\perm \circ \left( \mathcal{F} - \mathcal{F}^\prime \right) \circ \mathcal{W}_\perm \otimes \idmap_{R^{\prime\prime} }\right) \left( \rho_{A_1^nB_1^nR^{\prime\prime}} \right) } \\
    			& \leq \frac{1}{n!} \sum_{\perm \in \permset{n}}	\tracedist{ \left(  \left( \mathcal{F} - \mathcal{F}^\prime \right) \circ \mathcal{W}_\perm \otimes \idmap_{R^{\prime \prime} }\right) \left( \rho_{A_1^nB_1^nR^{\prime \prime}} \right) } \\
    			& = \tracedist{ \left(  \left( \mathcal{F} - \mathcal{F}^\prime \right) \otimes \idmap_{ R^{\prime \prime} \widetilde{R} }\right) \left( \bar{\rho}_{A_1^nB_1^n R^{\prime \prime} \widetilde{R}} \right) } ,
    	\end{aligned}
    \end{equation}
    where the inequality again follows from the fact that CPTNI maps cannot increase trace norm, and the final inequality follows from the fact that the states \\
    $\left(  \left( \mathcal{F} - \mathcal{F}^\prime \right) \circ \mathcal{W}_\perm \otimes \idmap_{R^{\prime \prime} }\right) \left( \rho_{A_1^nB_1^nR^{\prime \prime}} \right)$ are orthogonal for different $\perm$.
    Putting Eqs.~\eqref{eq:dataprocessing2} and \eqref{eq:perminvariancetracedist} together, we get the desired result.
\end{proof}

\pslemmatwo*

\begin{proof}
Let us consider $\rho_{A_1^nB_1^n}  ,  \tau_{A_1^n B_1^n}$  to have trace $1$; the extension to positive operators follows naturally.  Since $\rho_{A_1^nB_1^n}  \leq \cost  \tau_{A_1^n B_1^n}$, there exists a $\omega_{A_1^nB_1^n} \in \dop{=}(A_1^nB_1^n)$ such that $\rho_{A_1^nB_1^n} + (\cost -1) \omega_{A_1^nB_1^n} = \cost \tau_{A_1^nB_1^n}$. One can then construct an extension of of $\tau_{A_1^nB_1^n}$ as,
    	\begin{equation}
    		\tau_{A_1^nB_1^n R^\prime M } = \frac{1}{\cost} \rho_{A_1^nB_1^n R^\prime} \otimes \ketbra{0}{0}_M + \left( 1- \frac{1}{\cost} \right) \omega_{A_1^nB_1^nR^\prime} \otimes \ketbra{1}{1}_M.
    	\end{equation}
        Since the map $(\mathcal{F}-\mathcal{F}^\prime) \otimes \idmap_{R^\prime M}$ acts identically on $M$, the two terms above are orthogonal before and after the action of the map. Therefore,
    	\begin{equation} \label{eq:triangle}
    		\begin{aligned}
    		\tracedist{ \left( \left(\mathcal{F} - \mathcal{F}^\prime \right) \otimes \idmap_{R^\prime M}\right) \left( \tau_{A_1^nB_1^nR^\prime M} \right) } &= 	\frac{1}{\cost} \tracedist{ \left( \left(\mathcal{F} - \mathcal{F}^\prime \right) \otimes \idmap_{R^\prime M}\right) \left( \rho_{A_1^nB_1^nR^\prime} \otimes \ketbra{0}{0}  \right) }\\
    		&+ \left( 1- \frac{1}{\cost} \right)	\tracedist{ \left( \left(\mathcal{F} - \mathcal{F}^\prime \right) \otimes \idmap_{R^\prime M}\right) \left( \omega_{A_1^nB_1^nR^\prime } \otimes \ketbra{1}{1} \right) } \\
    		&\geq 	\frac{1}{\cost} \tracedist{ \left( \left(\mathcal{F} - \mathcal{F}^\prime \right) \otimes \idmap_{R^\prime}\right) \left( \rho_{A_1^nB_1^nR^\prime}  \right) }
    		\end{aligned}
    	\end{equation}
    	Finally, for any purification $\tau_{A_1^nB_1^nR}$ of $\tau_{A_1^nB_1^n}$, and extension $\tau_{A_1^nB_1^nR^\prime M}$ of $\tau_{A_1^nB_1^n}$, there exists a CPTP map $\Phi \in \CPTP( R ,R^\prime M)$ such that $ \left( \idmap_{A_1^nB_1^n} \otimes \Phi \right) \left( \tau_{A_1^nB_1^nR} \right) = \tau_{A_1^nB_1^nR^\prime M}$ (\cref{lemma:pur_to_ext}). Therefore, 
    	\begin{equation} \label{eq:dataprocessing}
    			\tracedist{ \left( \left(\mathcal{F} - \mathcal{F}^\prime \right) \otimes \idmap_{R}\right) \left( \tau_{A_1^nB_1^nR} \right) } \geq 	\tracedist{ \left( \left(\mathcal{F} - \mathcal{F}^\prime \right) \otimes \idmap_{R^\prime M}\right) \left( \tau_{A_1^nB_1^nR^\prime M} \right) } 
    	\end{equation}
        follows from the fact that CPTP maps cannot increase trace norm. 
    	Putting Eq.~\eqref{eq:triangle} and \eqref{eq:dataprocessing} together, the desired result is obtained.
 \end{proof}

\chapter{Proofs of statements for EUR analysis
}
\label{Appendix:EUR}
In this chapter, we will provide the proof of various statements utilized in \cref{chap:EUR}. 

\section{Variable-length security}

We start with the proof of the following theorem, which is suitable for EUR based analysis. 
\eurvariabletheorem*
      
\begin{proof}
Again, the proof of correctness follows from \cref{lemma:correctnessissatisfied}, and thus we are only required to show the $(2\epsAT + \epsPA)$-secrecy of the protocol.
Recall that the \nameref{prot:qkdprotocol} is such that it produces an output key of length $\lkey(\cobs)$, and uses an error-correction protocol with number of possible transcripts given by $2^{\leak(\cobs)}$, when the event $\Omega(\cobs) \wedge \OmegaEV$ occurs. Otherwise the protocol produces a key of length $0$. Thus, we are required to obtain an upper bound on

\begin{equation}
\begin{aligned}
    \Delta =& \sum_{\cobs
    : \lkey(\cobs) > 0} \Pr(\Omega(\cobs) \wedge \OmegaEV ) \times \\
    & \tracedist{ \rho_{K_A \allpublic \Eve | \Omega(\cobs) \wedge \OmegaEV} - \rho^{\mathrm{ideal}}_{K_A \allpublic \Eve | \Omega(\cobs) \wedge \OmegaEV} }
    \end{aligned}
\end{equation}
where we have used our trick of noting that the states conditioned on different $\Omega(\cobs)$ have orthogonal support.\footnote{Regardless, the above expression is certainly larger than the expression we seek to bound for the secrecy requirement, via the triangle inequality}. Using the triangle inequality, and expanding in terms of the unobserved events $\cnotobs$, we obtain 
\begin{equation} \label{eq:Delta}
\begin{aligned}
    \Delta \leq &\sum_{ \cobs  
    : \lkey(\cobs) > 0} \sum_{\cnotobs} \Pr(\Omega(\cobs,\cnotobs) \wedge \OmegaEV )  \times \\
    &\tracedist{ \rho_{K_A \allpublic \Eve | \Omega(\cobs,\cnotobs) \wedge \OmegaEV} - \rho^{\mathrm{ideal}}_{K_A \allpublic \Eve | \Omega(\cobs,\cnotobs) \wedge \OmegaEV} } \\
    &=\sum_{\cobs
    : \lkey(\cobs) > 0} \sum_{ \cnotobs \in \EURset(\cobs)} \Pr(\Omega(\cobs,\cnotobs) \wedge \OmegaEV )  \times \\
    &\tracedist{ \rho_{K_A \allpublic \Eve | \Omega(\cobs,\cnotobs) \wedge \OmegaEV} - \rho^{\mathrm{ideal}}_{K_A \allpublic \Eve | \Omega(\cobs,\cnotobs) \wedge \OmegaEV} } \\
    &+ \sum_{\cobs
    : \lkey(\cobs) > 0}  \sum_{ \cnotobs \notin \EURset(\cobs) }\Pr(\Omega(\cobs,\cnotobs) \wedge \OmegaEV )  \times \\
    &\tracedist{ \rho_{K_A \allpublic \Eve | \Omega(\cobs,\cnotobs) \wedge \OmegaEV} - \rho^{\mathrm{ideal}}_{K_A \allpublic \Eve | \Omega(\cobs,\cnotobs) \wedge \OmegaEV} } 
    \end{aligned}
\end{equation}
where we have simply grouped the sum over $\cnotobs$ into two different terms. The first contribution can be bounded as follows:
\begin{align}
\Delta_1 
&\coloneq
\sum_{\cobs:\,\lkey(\cobs)>0}
\sum_{\cnotobs\in\EURset(\cobs)}
\Pr\!\left(\Omega(\cobs,\cnotobs)\wedge\OmegaEV\right)
\label{eq:Delta1-def}
\\
&\quad\times
\tracedist{
\rho_{K_A \allpublic \Eve \mid \Omega(\cobs,\cnotobs)\wedge\OmegaEV}
-
\rho^{\mathrm{ideal}}_{K_A \allpublic \Eve \mid \Omega(\cobs,\cnotobs)\wedge\OmegaEV}
}
\nonumber
\\
&\le
\sum_{\cobs:\,\lkey(\cobs)>0}
\sum_{\cnotobs\in\EURset(\cobs)}
\Pr\!\left(\Omega(\cobs,\cnotobs)\right)
\label{eq:Delta1-LHL}
\\
&\quad\times
\left(
2^{-\frac{1}{2}\!\left(
\Hmin[\sqrt{\serfbound(\cobs,\cnotobs)}]
(\PAstring \mid \CP_1^n \CEC \CEV \HEV \Eve)_
{(\rho \mid \Omega(\cobs,\cnotobs) )\wedge\OmegaEV}
-\lkey(\cobs)+2
\right)}
+2\sqrt{\serfbound(\cobs,\cnotobs)}
\right)
\nonumber
\\
&\le
\sum_{\cobs:\,\lkey(\cobs)>0}
\sum_{\cnotobs\in\EURset(\cobs)}
\Pr\!\left(\Omega(\cobs,\cnotobs)\right)
\label{eq:Delta1-ECsplit}
\\
&\quad\times
\left(
2^{-\frac{1}{2}\!\left(
\Hmin[\sqrt{\serfbound(\cobs,\cnotobs)}]
(\PAstring \mid \CP_1^n \HEV \Eve)_
{(\rho \mid \Omega(\cobs,\cnotobs) )\wedge\OmegaEV}
-\leak(\cobs)-\EVcost-\lkey(\cobs)+2
\right)}
+2\sqrt{\serfbound(\cobs,\cnotobs)}
\right)
\nonumber
\\
&\le
\sum_{\cobs:\,\lkey(\cobs)>0}
\sum_{\cnotobs\in\EURset(\cobs)}
\Pr\!\left(\Omega(\cobs,\cnotobs)\right)
\label{eq:Delta1-removeOmegaEV}
\\
&\quad\times
\left(
2^{-\frac{1}{2}\!\left(
\Hmin[\sqrt{\serfbound(\cobs,\cnotobs)}]
(\PAstring \mid \CP_1^n \HEV \Eve)_
{\rho \mid \Omega(\cobs,\cnotobs)}
-\leak(\cobs)-\EVcost-\lkey(\cobs)+2
\right)}
+2\sqrt{\serfbound(\cobs,\cnotobs)}
\right)
\nonumber
\\
&=
\sum_{\cobs:\,\lkey(\cobs)>0}
\sum_{\cnotobs\in\EURset(\cobs)}
\Pr\!\left(\Omega(\cobs,\cnotobs)\right)
\label{eq:Delta1-removeHEV}
\\
&\quad\times
\left(
2^{-\frac{1}{2}\!\left(
\Hmin[\sqrt{\serfbound(\cobs,\cnotobs)}]
(\PAstring \mid \CP_1^n  \Eve)_
{\rho \mid \Omega(\cobs,\cnotobs)}
-\leak(\cobs)-\EVcost-\lkey(\cobs)+2
\right)}
+2\sqrt{\serfbound(\cobs,\cnotobs)}
\right)
\nonumber \\
&=
\sum_{\cobs:\,\lkey(\cobs)>0}
\sum_{\cnotobs\in\EURset(\cobs)}
\Pr\!\left(\Omega(\cobs,\cnotobs)\right)
\label{eq:Delta1-EURbeta}
\\
&\quad\times
\left(
2^{-\frac{1}{2}\!\left(
\Hmin[\sqrt{\serfbound(\cobs,\cnotobs)}]
(\PAstring \mid \CP_1^n \HEV \Eve)_
{\rho \mid \Omega(\cobs,\cnotobs)}
-\EURbeta(\cobs)
+2\log\frac{1}{\epsPA}
\right)}
+2\sqrt{\serfbound(\cobs,\cnotobs)}
\right)
\nonumber
\\
&\le
\sum_{\cobs:\,\lkey(\cobs)>0}
\sum_{\cnotobs\in\EURset(\cobs)}
\Pr\!\left(\Omega(\cobs,\cnotobs)\right)
\label{eq:Delta1-epsPA}
\left(
\epsPA+2\sqrt{\serfbound(\cobs,\cnotobs)}
\right)
\\
&\le
\epsPA
+
2\sqrt{
\sum_{\cobs:\,\lkey(\cobs)>0}
\sum_{\cnotobs\in\EURset(\cobs)}
\Pr\!\left(\Omega(\cobs,\cnotobs)\right)
\serfbound(\cobs,\cnotobs)
}.
\label{eq:Delta1-final}
\end{align}
In \cref{eq:Delta1-def}, we explicitly rewrite the contribution $\Delta_1$ as a weighted sum of trace distances conditioned on the events $\Omega(\cobs,\cnotobs)$ and $\OmegaEV$.
The bound in \cref{eq:Delta1-LHL} follows from applying the Leftover Hashing Lemma (\cref{lemma:LHL}) to the resulting \emph{subnormalized} state, where the smoothing parameter depends on the conditioning event.\footnote{Here we point out a technical subtlety concerning the choice of smoothing parameter for subnormalized states. Namely, that the smoothing parameter for smooth min entropy of the state $\rho_{\wedge \Omega}$ must be smaller than $\sqrt{\Pr(\Omega)_\rho}$. In the above analysis, if this condition is not satisfied then the required bound follows trivially. Hence, the assumption that $\sqrt{\serfbound(\cobs,\cnotobs)} \leq \Pr(\OmegaEV | \Omega(\cobs, \cnotobs))_\rho$.}
In \cref{eq:Delta1-ECsplit}, we apply \cref{lemma:EC_cost} to remove the classical registers corresponding to error correction and error verification.
Since the state is already conditioned on $\OmegaEV$, the one bit verification response from Bob takes a deterministic value and therefore does not contribute to the entropy loss. The partial conditioning on $\OmegaEV$ is removed in \cref{eq:Delta1-removeOmegaEV} using \cref{lemma:gettingridofsubnormalizedconditioning}. In \cref{eq:Delta1-removeHEV} we remove the $\HEV$ register without penalty, since it is completely independent of the rest of the registers (by applying data processing \cref{lemma:DPIsmoothmin} in both directions). 
In \cref{eq:Delta1-EURbeta}, we substitute the definition of the key length $\lkey(\cobs)$, and in \cref{eq:Delta1-epsPA} we invoke the assumed lower bound involving $\EURbeta(\cobs)$ from the theorem statement. 
Finally, \cref{eq:Delta1-final} follows from the fact that the probabilities $\Pr(\Omega(\cobs,\cnotobs))$ sum to at most one, together with the concavity of the square root function and Jensen’s inequality, which allows the sum to be pulled inside the square root

To bound the remaining remaining term is a lot more straightforward:

\begin{equation}
\begin{aligned}
    \Delta_2 &\coloneq \sum_{\cobs
    : \lkey(\cobs) > 0} \sum_{ \cnotobs \notin \EURset(\cobs)} \Pr(\Omega(\cobs,\cnotobs) \wedge \OmegaEV )  \times \\
    &\tracedist{ \rho_{K_A \allpublic \Eve | \Omega(\cobs,\cnotobs) \wedge \OmegaEV} - \rho^{\mathrm{ideal}}_{K_A \allpublic \Eve | \Omega(\cobs,\cnotobs) \wedge \OmegaEV} } \\
       &\leq \sum_{\cobs
    : \lkey(\cobs) > 0} \sum_{ \cnotobs \notin \EURset(\cobs)} \Pr(\Omega(\cobs,\cnotobs)) 
    \end{aligned}
    \end{equation}
We can now simply combine the two bounds, and obtain
\begin{equation}
    \begin{aligned}
        \Delta &\leq \Delta_1 + \Delta_2 \\
        & \leq \epsPA + 2 \sqrt{\sum_{ \cobs
    : \lkey(\cobs) > 0}  \sum_{ \cnotobs \in \EURset(\cobs)}  \Pr(\Omega(\cobs,\cnotobs) )  \serfbound(\cobs,\cnotobs) } + \sum_{\cobs
    : \lkey(\cobs) > 0} \sum_{ \cnotobs \notin \EURset(\cobs)} \Pr(\Omega(\cobs,\cnotobs))  \\
    & \leq \epsPA + 2 \sqrt{\sum_{ \cobs
    : \lkey(\cobs) > 0}  \sum_{ \cnotobs \in \EURset(\cobs)}  \Pr(\Omega(\cobs,\cnotobs) )  \serfbound(\cobs,\cnotobs) + \sum_{\cobs
    : \lkey(\cobs) > 0} \sum_{ \cnotobs \notin \EURset(\cobs)} \Pr(\Omega(\cobs,\cnotobs))} \\
    &\leq \epsPA  + 2\epsAT.
    \end{aligned}
\end{equation}
Here we use the fact that  $2\sqrt{a} + b \leq 2\sqrt{a+b}$ for $ 0\leq a,b,a+b\leq 1$ in the third line, and the stated condition from the theorem for the final line. \end{proof}

\begin{remark}
			Note that the critical step here was using the concavity of the square root function to see that a bound on the \textit{average} failure probability of the phase error estimation procedure is enough to prove security. This is the same fundamental trick used by  Ref.~\cite{hayashi_concise_2012,kawakami_security_nodate,curras-lorenzo_tight_2021}. Our presentation here is in the EUR framework, where this manifests in the deliberate choice of our smoothing parameter in the first part of the proof. 
		\end{remark}

		\begin{remark}
			Notice that in the variable-length protocol for which we proved security, the number of bits on which privacy amplification is applied is variable. It depends on the number of key generation rounds obtained in each protocol run. Some subtle issues regarding two-universal hashing on a variable-length \textit{input} register were pointed out and addressed in \cref{sec:variablelengthPA}. In particular, it was noted that first looking at the number of bits in the raw key, and then choosing an appropriate two-universal hashing procedure for that many input bits, \emph{does not produce a valid two-universal hashing procedure on the input space of variable-length bit strings}. Due to this, the Leftover Hashing Lemma cannot be straightforwardly applied to such a scenario. However, this issue was addressed by showing that when the locations of the discard rounds are publicly announced,  the theoretical analyses of scenarios where the rounds are actually discarded, vs mapped to special symbols such as $0$ or $\bot$ (where leftover-hashing lemma can be applied), are equivalent \cref{lemma:variableinputhashing,lemma:renyiinvariance}. Due to this equivalence, the PA procedure described above can be applied in QKD protocols. 
			
			It is interesting to note that these issues are completely avoided by the above proof, in a very \textit{different} manner than \cref{sec:variablelengthPA}. This is because in this proof, we always apply the Leftover Hashing Lemma on a state conditioned on the specific length of the raw key register (\cref{eq:eurbound}). Therefore, the Leftover Hashing Lemma  can be applied in a  straightforward manner, and there are no issues is choosing the hashing family based on the specific length of the raw key register. In other words, the PA procedure described above is valid for this proof. In a similar sense, the variable-length security proof from \cref{chap:variable} critically relied on a technical \cref{lemma:renyiweightedaverage}, that necessitated the use of R\'enyi entropies instead of smooth min entropy in that analysis. However, the variable-length proof presented above takes a different approach, and does \textit{not} impose the same requirements on the behavior of smooth min entropy. Again, this is due to the use of \cref{eq:eurbound}.  Later, in \cref{chap:MEAT} we will see an analysis using {\Renyi} entropies that is much simpler than the one presented in \cref{chap:variable}.
		\end{remark}

\section{Technical Statements}

We first prove the two-step measurement lemma which was utilized extensively in \cref{chap:EUR} to reformulate the measurement process of the QKD protocol.

\twosteplemma*
		\begin{proof}
			Observe that $\{\tilde{F}_i | i \in \mathcal{P}_\mathcal{A}\}$ is a valid set of POVMs by construction. Moreover, $\{ G_{k} | k \in \mathcal{A}_i\}$ is a valid set of POVMs for each $i$, also by construction.  Thus we only need to show that $\rho_\text{final} = \rho^\prime_\text{final}$. Using the cyclicity of trace in \cref{eq:twostepsecondstate}, it suffices to prove
			\begin{equation}
				\sqrt{\tilde{F}_i} {G}_{k} \sqrt{\tilde{F}_i} = \Gamma_{k} \quad \forall i \in \mathcal{P}_\mathcal{A}, \forall k \in \mathcal{A}_i.
			\end{equation}
			Substituting the expression for $F_k$ into the above equation, we obtain 
			\begin{equation}
				\begin{aligned}
					\sqrt{\tilde{F}_i} G_{k} \sqrt{\tilde{F}_i} &= \sqrt{\tilde{F}_i} \left(  \sqrt{\tilde{F}}^+_i \Gamma_{k} \sqrt{\tilde{F}}^+_i + P_{k} \right)  \sqrt{\tilde{F}_i}  \\
					&= \sqrt{\tilde{F}_i} \left(  \sqrt{\tilde{F}}^+_i \Gamma_{k} \sqrt{\tilde{F}}^+_i \right)  \sqrt{\tilde{F}_i} \\
					&=\mathrm{\Pi}_{\tilde{F}_i} 			\Gamma_{k} \mathrm{\Pi}_{\tilde{F}_i} \\
					&=	\Gamma_{k},
				\end{aligned}
			\end{equation}
			where the second equality follows from the fact that $P_k$ and $\tilde{F}_i$ have orthogonal supports, and the final equality uses the fact that the support for $\tilde{F}_i$ is larger than the support for $\Gamma_k$ for $k\in \mathcal{A}_i$.
			This concludes the proof.
		\end{proof}

We will now turn our attention to technical statements needed to replace the smooth max entropy term (of two classical strings) with a suitable function of the error rate between those two strings.

		\begin{lemma}[{\cite[Lemma 7]{tomamichel_largely_2017}}] \label{lemma:smalleventstate}
			Let $\rho_{CQ} \in S{\leq} (CQ)$ be classical in $C$, and let $\Omega$ be any event on $C$ such that $\Pr(\Omega)_\rho \leq \varepsilon$. Then there exists a sub-normalized state $\tilde{\rho}_{CQ} \in  S{\leq}(CQ)$ with $\Pr(\Omega)_{\tilde{\rho}} =0$, and $P(\rho,\tilde{\rho}) \leq \sqrt{\varepsilon}$, where $P$ denotes the purified distance (see \cref{def:purifieddistance} ).
		\end{lemma}
		
We use the above lemma in the proof of the following statement. The following statement allows us to replace the smooth max entropy term in the EUR statement with our bound on the phase error rate. The proof is basically the same as the proof of \cite[Proposition 8]{tomamichel_largely_2017}.
		\begin{lemma} \label{lemma:smoothmaxerror}
			Let $\rho \in S{\leq}(XY)$ where $X,Y$ store $n$-bit strings, and let $\bm{e}_{XY}$ denote the error rate in these strings. Let $\Omega$ be any event such that $\bm{e}_{XY} > e_\text{max}$, and let $\Pr(\Omega)_{\rho} \leq \kappa$. For any $e_\text{max}<1/2$, we have
			\begin{equation}
				\Hmax[\sqrt{\kappa}](X|Y)_\rho \leq n h (e_\text{max})
			\end{equation}
			
		\end{lemma}
		\begin{proof}
			By \cref{lemma:smalleventstate}, there exists a state $\tilde{\rho}_{XY}$ such that $\Pr(\Omega)_{\tilde{\rho}}=0$ and $P(\rho,\tilde{\rho}) \leq \sqrt{\kappa}$. Therefore we have
			\begin{equation}
				\begin{aligned}
					\Hmax[\sqrt{\kappa}] (X | Y)_{\rho} &\leq H_\text{max}(X|Y)_{\tilde{\rho}} \\
					&= \log(   \sum_{y \in \{0,1\}^n} \Pr(Y=y)_{\tilde{\rho}} 2^{\Hmax[](X|Y)_{\tilde{\rho} | Y=y}  }         ) \\
					&\leq \max_{y \in \{0,1\}^n } \Hmax[](X|Y)_{\tilde{\rho } | Y=y}	\\
					& = \max_{y \in \{0,1\}^n } \log  \bigg \lvert \left\{ x \in \{0,1\}^n : \Pr(X=x \land Y=y)_{\tilde{\rho}} > 0 \right\}  \bigg  \rvert  \\
					&\leq \log ( \sum_{k=0}^{n e_\text{max}}  {n\choose k} ), \\
					&\leq \log(2^{ nh(e_\text{max})})
				\end{aligned}
			\end{equation}
			where we used the definition of the smooth max entropy in the first inequality, and \cite[Sec. 4.3.2]{tomamichel_framework_2013} for the second equality. The third inequality and the fourth equality follow from the definitions. The fifth inequality follows from the fact that the state $\tilde{\rho}$ is guaranteed to have $\leq n e_\text{max}$ errors, while the final inequality follows from the suitable bound on the sum of binomial coefficients.
		\end{proof}

		\section{Sampling}\label{appendix:sampling}
		In this section, we prove the technical statements needed to prove our sampling bounds.
		\subsection{Random Sampling} \label{appsubsec:randomsampling}
		We start with the usual Serfling \cite{serfling_probability_1974}  statement in the following lemma. The following lemma is obtained from \cite[Eq. 74, Lemma 6]{tomamichel_largely_2017}
		\begin{lemma}[Serfling] \label{lemma:serfling}
			Let $\X_1\dots \X_{m+n}$ be bit-valued random variables. Let $\bm{J}_m$ denote the choice of a uniformly random subset of $m$ positions, out of $m+n$ positions.  Then,
			\begin{equation}
				\begin{aligned}
					\Pr(\sum_{i \notin \bm{J}_m} \frac{\X_i}{n} \geq \sum_{i \in \bm{J}_m} \frac{\X_i}{m} + \gamma_\text{serf}) &\leq e^{-2 \gamma_\text{serf}^2 \fserf(m,n)}, \\
					\fserf(m,n) \coloneq \frac{nm^2}{(n+m)(m+1)}.
				\end{aligned}
			\end{equation}
		\end{lemma}
		Serfling basically states that if one chooses a random set of positions, then the fraction of $1$s in those positions gives us a good estimate of the fraction of $1$s in the remaining positions.
		However, observe that the sampling procedure in the protocol from \cref{subsec:samplingperfect,subsec:sampling} does not actually choose a random subset of fixed-length for testing. Instead, the protocol decides to map each conclusive round to test or key in an IID manner. Therefore, the application of the Serfling bound is not straightforward. In the following lemma, we show how the Serfling bound can still be rigorously used.

		\serflinglemma*
		
		\begin{proof}
			Since the sampling procedure randomly assigns each bit to test or key (or does nothing with them if $p_t+p_k <1$), \cref{lemma:serfling} cannot be directly applied. However, consider what happens if we condition on the event $\event{\nX,\nK}$. Then, for a given set of positions that form the $\nX+\nK$ positions selected for test or key, it is the case that each set of $\nX$ positions is equally likely. Therefore, the above sampling procedure is exactly equivalent to:
			\begin{enumerate}
				\item First determining the event $\event{\nX,\nK}$ by sampling from \textit{some} probability distribution. 
				\item Pick some $\nX+\nK$ positions at random.
				\item Then determining the exact positions of the $\nX$ test rounds, by choosing a random subset of fixed-size $\nX$ out of these $\nX+\nK$ positions.	\end{enumerate}
			The necessary claim follows by applying \cref{lemma:serfling} for step 3 of the above procedure.
		\end{proof}

	\subsection{Sampling with imperfect detectors} \label{subsec:pulkitlemmas}
		We now turn our attention to proving \cref{lemma:pulkitgeneric}, which is the main statement utilized in extending the EUR approach to imperfect detectors in \cref{sec:imperfectdetectors}. We start by proving \cref{lemma:orderingonPOVMs,lemma:smallPOVM} which we use later in the proof of \cref{lemma:pulkit}.
		
		Recall our notation: If  $\rho_{Q^n} \in S{=}(Q^{\otimes n})$ is an arbitrary state, and $\{P_1,P_2,\dots,P_m\}$ is a set of POVM elements, then we let $\num_{P_i}$ denote the classical random variable corresponding to the number of measurement outcomes corresponding to $P_i$ when the state $\rho_{Q^n}$ is measured. Moreover, let $ \bm{D}_P$ denote the classical random variable that describes the measurement outcomes when each subsystem of $\rho$ is measured using $\{P_1,P_2,\dots,P_m\}$.  We use $S \sim \bm{D}_P$ to denote the statement S is sampled from $\bm{D}_P$.
		\begin{lemma} \label{lemma:orderingonPOVMs}
			Let $\rho_{Q^n} \in S_\circ(Q^{\otimes n})$ be an arbitrary state. Let $\{P,I - P\}$ and $\{P^\prime,I-P^\prime\}$ be two sets of POVM elements such that $P \leq P^\prime $. Then, for any $e$, it is the case that
			\begin{equation}
				\Pr(\frac{\num_{P}}{n} \geq e ) \leq \Pr(\frac{\num_{P^\prime}}{n} \geq e) 
			\end{equation}
		\end{lemma}

	\begin{proof}
			We will describe a procedure to generate random strings $S,S^\prime$ such that  $S \sim \bm{D}_P, S^\prime \sim \bm{D}_P^\prime$. Consider the POVM $\{P,P^\prime-P,I-P^\prime \}$, and let $T$ be the classical string taking values in $\{0,1,2\}^n$ which stores the measurement outcomes when measured using this POVM. Then, $S,S^\prime$ can be obtained by first obtaining $T$, followed by the following remapping
			\[(S_i,S^\prime_i)=\begin{cases}
				(1,1) &\text{$T_i=0$}\\
				(0,1) &\text{$T_i=1$}\\
				(0,0) &\text{$T_i=2$}
			\end{cases} \] 
			where $i$ denotes the position in the string. The required claim follows from the observation that the above procedure maps more $S^\prime_i$ to $1$ than $S_i$. Thus, 
			\begin{equation}
				\Pr(w( \bm{S}^\prime) \geq w( \bm{S})) \geq 0 \implies \Pr(w( \bm{S}^\prime) > ne) \geq \Pr(w(\bm{S}) \geq ne)
			\end{equation}
			where $w$ denotes the hamming weight of the string (sum of each element of the string). The necessary statement follows after noting that $w(S) \sim \num_P$ and $w(S^\prime) \sim \num_{P^\prime}$ (which can be argued rigorously using the two-step measurement \cref{lemma:twostep}).  
		\end{proof}
		\begin{remark}
			Note a conceptual subtlety in the above proof: The procedure used to generate $S, S^\prime$ in the above proof has some joint probability distribution associated to it. This means that $(S,S^\prime)$ is a well-defined random variable. However, one cannot talk about the joint probability distribution of two different sets of measurement on the same quantum state. This subtle issue is avoided by noting that we are only interested in making statements on the marginal probability distribution of $S$ and $S^\prime$, and how they relate to one another. And it is indeed true that these distributions satisfy $S \sim \bm{D}_P$ and $S^\prime \sim \bm{D}_{P^\prime}$, which is enough to prove our claim. The fact that $S,S^\prime$ has some joint distribution associated with it is immaterial.
		\end{remark}
		
		\smallPOVM*
		
		\begin{proof}
			Since $\norm{P}_\infty \leq \delta$, we have $P \leq \delta I$. By \cref{lemma:orderingonPOVMs}, we have
			\begin{equation}
				\Pr(\frac{\num_{P}}{n} \geq \delta+c ) \leq \Pr(\frac{\num_{\delta I}}{n} \geq \delta+c) 
			\end{equation}
			Observe that measurement using $\{\delta I, (1-\delta) I\}$ is equivalent to Bernoulli sampling. Thus $\num_{\delta I}$ obeys the binomial distribution. Therefore, 
			\begin{equation}
				\begin{aligned}
					\Pr(\frac{\num_{\delta I}}{n} \geq \delta+c) &\leq \sum_{i = n (\delta+c)}^{n} {n \choose i} \delta^ i (1-\delta)^{n-i}.
				\end{aligned}
			\end{equation}
		\end{proof}
		
		\begin{lemma}
			\label{lemma:pulkit} Let $\rho_{Q^n} \in S_\circ(Q^{\otimes n})$ be an arbitrary state. Let $\{P,I - P\}$ and $\{P^\prime,I-P^\prime\}$ be two sets of POVM elements. Suppose there exists a $0 \leq \tilde{P} \leq I$ such that $P\leq \tilde{P}, P^\prime \leq \tilde{P} $, and $ \norm{\tilde{P}-{P}}_\infty  \leq \delta$. Then
			\begin{equation} \label{eq:pulkitequation}
				\Pr(\frac{\num_{P^\prime}}{n} \geq e +  (\delta+c)) \leq \Pr(\frac{\num_P}{n} \geq e) + \binfunction{n}{\delta}{c} ,
			\end{equation}
			where $\binfunction{n}{\delta}{c}$ was defined in \cref{lemma:smallPOVM}.
		\end{lemma}
		\begin{proof}
			We will describe a process to generate $S \sim \bm{D}_P$ and $\tilde{S} \sim  \bm{D}_{\tilde{P}}$, in a similar manner as in the proof of \cref{lemma:orderingonPOVMs}. In particular, let $T$ be the random variable taking values in $\{0,1,2\}^n$ that stores the measurement outcomes of $\{P,\tilde{P}-P,I-\tilde{P}\}$ measurements. We generate $S,\tilde{S}$ by first obtaining $T$, followed by the following remapping
			\[(S_i,\tilde{S}_i)=\begin{cases}
				(1,1) &\text{$T_i=0$}\\
				(0,1) &\text{$T_i=1$}\\
				(0,0) &\text{$T_i=2$}.
			\end{cases} \]
			Then,
			\begin{equation} \label{eq:pulkitwithordering} 
				\begin{aligned} 
					\Pr(\num_{\tilde{P}} \geq  ne +n (\delta+c)) &= 	\Pr(w(\bm{\tilde{S} })\geq  ne +n (\delta+c) ) \\
					&= 	\Pr(w( \bm{\tilde{S}} )\geq  ne +n (\delta+c)  \cap w(\bm{S}) \geq ne) \\
					&+ 	\Pr(w(\bm{\tilde{S} })\geq  ne +n (\delta+c) \cap w(\bm{S})  < ne) \\
					& \leq  \Pr( w(\bm{S})  \geq ne ) + \Pr(w(\bm{\tilde{S}}) - w(\bm{S}) \geq n(\delta+c)) \\
					& = \Pr(\num_P \geq ne) + \Pr(\num_{\tilde{P} - P} \geq n (\delta+c)) \\
					& \leq \Pr(\num_P \geq ne) +  \binfunction{n}{\delta}{c} . 
				\end{aligned}
			\end{equation}
			where we used \cref{lemma:smallPOVM} in the final inequality, the fact that $w(S) \sim \num_P$ and $w(\tilde{S}) - w(S) = w(\tilde{S}-S) \sim \num_{\tilde{P}-P}$ ($\tilde{S}_i - S_i = 1$ if and only if $T_i=1$) for the penultimate inequality, and basic properties of probabilities for the remaining steps. Next, we replace the $\tilde{P}$ with $P^\prime$ using \cref{lemma:orderingonPOVMs} and $\tilde{P}\geq P^\prime$, and obtain
			\begin{equation}\label{eq:pulkittemptwo} 
				\Pr(\num_{P^\prime} \geq ne + n(\delta+c)) \leq \Pr(\num_{\tilde{P}} \geq ne + n(\delta+c)).
			\end{equation}
			The proof follows after noting that \cref{eq:pulkitwithordering,eq:pulkittemptwo} $\implies$ \cref{eq:pulkitequation}.
		\end{proof}
		
		\cref{lemma:pulkit} above  requires an explicit construction of a $\tilde{P}$ satisfying the necessary requirements. However, this requirement can be removed, and we obtain a sightly worse result with greater generality below.

		\pulkitlemmageneric*

		\begin{proof}
			Let $G^\prime = (1-\delta) P^\prime $, and $G = (1-\delta) P$. 
			Using $0\leq G\leq P$ and \cref{lemma:orderingonPOVMs}, we obtain
			\begin{equation}
				\label{eq:pulkitgenerictempone}
				\Pr(\num_{G} \geq ne) \leq \Pr(\num_{P} \geq ne).
			\end{equation}
			Using $0 \leq G^\prime + \delta I \leq I$, $G^\prime+\delta I \geq G$,  $\norm{G^\prime+\delta I - G}_\infty \leq   \delta+\delta(1-\delta) \leq 2\delta $, and \cref{eq:pulkitwithordering}, we obtain
			\begin{equation} \label{eq:pulkitgenerictemptwo}
				\Pr(\num_{G^\prime+\delta I} \geq ne + n(2 \delta + c)) \leq \Pr(\num_{G} \geq ne) + \binfunction{n}{2\delta}{c},
			\end{equation}
			Finally, using $G^\prime + \delta I \geq P^\prime$,  and \cref{lemma:orderingonPOVMs}, we obtain
			\begin{equation} \label{eq:pulkitgenerictemp3}
				\Pr(\num_{P^\prime} \geq ne + n(2 \delta+c)) \leq \Pr(\num_{G^\prime + \delta I} \geq n e + n(2\delta+c)).
			\end{equation}
			The proof follows from the observation that \cref{eq:pulkitgenerictempone,eq:pulkitgenerictemptwo,eq:pulkitgenerictemp3} $\implies$ \cref{eq:pulkitequationgeneric}.
		\end{proof}
		
		\begin{remark}
		    The statements above are written for a measurement procedure where the same POVM is used to measure each round of the state. However the proofs do not actually use this fact. The same proofs are valid even if the measurement for each round is done using a different POVM element (as long it satisfies the required bounds on the $\infty$-norm). The proofs are identical, and the statements are stated formally in \cite[Appendix C3]{tupkary_phase_2024}.

		\end{remark}

	\section{Combining bounds} \label{appendix:combining}
		In this section, we will combine \cref{eq:boundone,eq:boundtwo,eq:boundthree,eq:boundfour} and obtain \cref{eq:finalbound}. This process is simply some cumbersome algebra and the use of the union bound for probabilities.
		
		Combining \cref{eq:boundone,eq:boundtwo}, we obtain
		
		\begin{equation}  \label{eq:bound12}
			\begin{aligned}
				&\Pr(\ephwierdrv \geq \eXrv + \gamma^{\epsATa}_\text{serf}   (\nXperf,\nKperf)   )_{ | \event{\nXperf,\nKperf}} \\
				&=   	\Pr(  \left(\ephwierdrv \geq \eXrv + \gamma^{\epsATa}_\text{serf}   (\nXperf,\nKperf)  \right) \bigcap \left(  \eXperfrv \geq \eXrv  \right)  )_{ | \event{\nXperf,\nKperf}}  \\
				&+	\Pr( \left( \ephwierdrv \geq \eXrv + \gamma^{\epsATa}_\text{serf}   (\nXperf,\nKperf)   \right)  \bigcap \left(   \eXperfrv < \eXrv  \right) )_{ | \event{\nXperf,\nKperf}}   \\
				&\leq  \Pr(\eXperfrv \geq \eXrv )_{ | \event{\nXperf,\nKperf}}  + \Pr(\ephwierdrv \geq \eXperfrv + \gamma^{\epsATa}_\text{serf}   (\nXperf,\nKperf)  )_{ | \event{\nXperf,\nKperf}} \\
				&\leq \epsATa^2.
			\end{aligned}
		\end{equation}
	We will now combine  \cref{eq:bound12,eq:boundthree}. To do so we will additionally need to condition on $\eX$. However, note that  \cref{eq:boundthree} remains true with this additional conditioning (because $\eX$ is observed on a different set of rounds). Thus we obtain

		\begin{equation} \label{eq:bound123}
			\begin{aligned}
				&	\Pr(\ephperfrv \geq  \eXrv + \gamma^{\epsATa}_\text{serf}   (\nXperf,\nKperf) + \deltaone + \gamma^{\epsATb}_{\text{bin}}(\nKperf,\deltaone) )_{ | \event{\nXperf,\nKperf}} \\
				&=\sum_{\eX} \Pr(\event{\eX} | \event{\nXperf,\nKperf})	\Pr(\ephperfrv \geq  \eX + \gamma^{\epsATa}_\text{serf}   (\nXperf,\nKperf) + \deltaone + \gamma^{\epsATb}_{\text{bin}}(\nKperf,\deltaone) )_{ | \event{\nXperf,\nKperf,\eX}} \\
				&\leq \sum_{\eX} \Pr(\event{\eX} | \event{\nXperf,\nKperf})	 \Pr(\ephwierdrv \geq  \eX + \gamma^{\epsATa}_\text{serf}   (\nXperf,\nKperf))_{| \event{\nXperf,\nKperf,\eX}}  +\frac{\epsATb^2}{2} \\
						&= \Pr(\ephwierdrv \geq  \eXrv + \gamma^{\epsATa}_\text{serf}   (\nXperf,\nKperf))_{| \event{\nXperf,\nKperf}}  +\frac{\epsATb^2}{2} \\
				&\leq \epsATa^2+\epsATb^2.
			\end{aligned}
		\end{equation}
		where the first equality follows from the definition of conditional probability. The second inequality is obtained by setting $e=\eX +  \gamma^{\epsATa}_\text{serf}   (\nXperf,\nKperf)$ in \cref{eq:boundthree}, the third equality follows from the definition of probability and the final inequality follows from \cref{eq:bound12}

		Combing \cref{eq:bound123,eq:boundfour}, we obtain
		
		\begin{equation}
			\begin{aligned}
				&\Pr (   \ephrv \geq \frac{\eXrv+  \gamma^{\epsATa}_\text{serf}   (\nXperf,\nKperf) + \deltaone + \gamma^{\epsATb}_{\text{bin}}(\nKperf,\deltaone)}{ (1-\deltatwo- \gamma^{\epsATc}_{\text{bin}}(\nKperf,\deltatwo) )}    )_{ | \event{\nXperf,\nKperf}} \\
				= 
&\Pr\Bigg(
    \ephrv \geq 
    \frac{
        \eXrv 
        + \gamma^{\epsATa}_{\text{serf}}(\nXperf,\nKperf)
        + \deltaone 
        + \gamma^{\epsATb}_{\text{bin}}(\nKperf,\deltaone)
    }{
        1 - \deltatwo 
        - \gamma^{\epsATc}_{\text{bin}}(\nKperf,\deltatwo)
    }
\nonumber \\
&\qquad\qquad
    \bigcap
    \left(
        \ephperfrv \leq 
        \ephrv \bigl(
            1 - \deltatwo
            - \gamma^{\epsATc}_{\text{bin}}(\nKperf,\deltatwo)
        \bigr)
    \right)
\Bigg)_{\mid \event{\nXperf,\nKperf}} \\
&\Pr\Bigg(
    \ephrv \geq 
    \frac{
        \eXrv 
        + \gamma^{\epsATa}_{\text{serf}}(\nXperf,\nKperf) 
        + \deltaone 
        + \gamma^{\epsATb}_{\text{bin}}(\nKperf,\deltaone)
    }{
        1 - \deltatwo 
        - \gamma^{\epsATc}_{\text{bin}}(\nKperf,\deltatwo)
    }
\nonumber \\
&\qquad\qquad
    \bigcap 
    \left(
        \ephperfrv >
        \ephrv \bigl(
            1 - \deltatwo 
            - \gamma^{\epsATc}_{\text{bin}}(\nKperf,\deltatwo)
        \bigr)
    \right)
\Bigg)_{\mid \event{\nXperf,\nKperf}}
= \\
				\leq &\Pr( \ephperfrv \leq \ephrv (1-\deltatwo- \gamma^{\epsATc}_{\text{bin}}(\nKperf,\deltatwo) ) )_{ | \event{\nXperf,\nKperf}} \\
				+ & \Pr(\ephperfrv \geq  \eXrv+  \gamma^{\epsATa}_\text{serf}   (\nXperf,\nKperf) + \deltaone + \gamma^{\epsATb}_{\text{bin}}(\nKperf,\deltaone)) 
				\leq \epsATa^2 + \epsATb^2 + \epsATc^2,
			\end{aligned}
		\end{equation}
		which is the required result.

	\section{Decoy Analysis}  \label{appendixdecoysection}
    In this section, we perform the decoy analysis from \cref{chap:EUR}. Let $O$ denote a specific outcome of a given round, and let $\nO$ denote the number of rounds that resulted in the outcome $O$. For instance, it could denote that both Alice and Bob measured in the $X$ basis and obtained a detection (in which case $\nO = \nX$). We will perform a general decoy analysis for any outcome $O$.  Let $\nOmu{k}$ denote the number of rounds that resulted in the outcome $O$ where Alice used intensity $\mu_k$. We have access to this information during the protocol. Let $\nOph{m}$ denote the number of rounds that resulted in the outcome $O$ where Alice prepared a state of $m$ photons. We wish to obtain bounds on $\nOph{m}$ using $\nOmu{k}$.
		
		In practice, Alice first chooses an intensity $\mu_k$ of the pulse, which then determines the photon number $m$ of the pulse, via the Poissonian distribution, independently for each round. Thus we have
		\begin{equation} \label{eq:poissonian}
			p_{m | \mu_k} = e^{-\mu_k}  \frac{\mu_k^m}{m!}.
		\end{equation}
		The probability of $m$-photons being emitted, can be obtained via
		\begin{equation} 
			\tau_{m} = \sum_{\mu_k} p_{\mu_k} p_{m | \mu_k}  = \sum_{\mu_k}  p_{\mu_k} e^{-\mu_k}  \frac{\mu_k^m}{m!}.
		\end{equation}
		Now, without loss of generality, we can view Alice as \textit{first} choosing the photon number $m$, and \textit{then} choosing a intensity setting $\mu_k$ with probability given by 
		\begin{equation}
			p_{\mu_k | m} = p_{\mu_k} p_{m | \mu_k} / \tau_m.
		\end{equation} 
		This is the fundamental idea used by \cite{lim_concise_2014,hayashi_security_2014,curty_finitekey_2014}.
		In this case, due to the fact that each signal is mapped to an intensity independently of other signals, one can apply the Hoeffdings inequality to these independent events, and obtain
		\begin{equation} \label{eq:decoy1}
			\Pr( \abs{ \nOmurv{k} - \sum_{m=0}^{\infty} p_{\mu_k | m} \nOphrv{m} } \geq \sqrt{ \frac{\nOrv}{2} \ln( \frac{2}{\epsdecoy^2}) } ) \leq  \epsdecoy^2.
		\end{equation}
		
		\begin{remark}
			The application of Hoeffdings inequality here is subtle, and is made rigorous in \cref{lemma:hoeffdings,lemma:decoy} in \cref{app:decoyanalysis} (see also Ref.~\cite{curty_finitekey_2014}). Note that in general, the photon numbers of every pulse in the protocol are chosen independently, since Alice chooses intensity independently for each pulse. However, here we are interested in photon numbers corresponding to rounds that led to a specific outcome $O$. Since we postselect pulses based on the outcome, we can no longer claim 
	that the photon numbers of these pulses (pulses that led to outcome $O$) are sampled independently, or that intensities of these pulses are chosen independently. This is because they now depend on Eve's attack.
		Rather, \cref{lemma:hoeffdings,lemma:decoy} rely on exploiting the fact that conditioned on  \textit{any fixed sequence} of photon numbers of the pulses, the intensities are chosen independently of one another. One can therefore apply Hoeffdings inequality. Then, since the resulting statements holds for any fixed sequence of photon numbers, the conditioning on this event can be removed. 
		\end{remark}	
		We can now combine \cref{eq:decoy1} for all intensities $\mu_k$ using the union bounds for probabilities ($\Pr(\Omega_1 \land \Omega_2) \geq 1 - \Pr(\Omega_1^c) - \Pr(\Omega_2^c)$). Reformulating the expressions, we obtain
		\begin{equation} 
        \begin{aligned}
        \label{eq:decoy1allintensities}
			&\Pr(  \nOmurv{k} - \sqrt{ \frac{\nOrv}{2} \ln( \frac{2}{\epsdecoy^2}) } \leq  \sum_{m=0}^{\infty} p_{\mu_k | m} \nOphrv{m}    \leq   \nOmurv{k} +  \sqrt{ \frac{\nOrv}{2} \ln( \frac{2}{\epsdecoy^2}) }\quad \forall k \in \{1,2,3\} ) \\
            &\geq  1- 3  \epsdecoy^2.
            \end{aligned}
		\end{equation}
		To obtain \cref{eq:decoyreqstepone}, we will apply decoy analysis (\cref{eq:decoy1allintensities}) for three separate events: conclusive $Z$ basis rounds selected for key generation (denoted by $K$), conclusive $X$ basis rounds (denoted by $X$), and conclusive $X$ basis rounds leading to an error (denoted by $X_{\neq}$).	Then, \cref{eq:decoy1allintensities} can be applied these events (again using the union bound for probabilities) to obtain:
		\begin{equation} \label{eq:decoy3}
			\begin{aligned}
				&\Pr \Bigg(  \nOmurv{k} - \sqrt{ \frac{\nOrv}{2} \ln( \frac{2}{\epsdecoy^2}) }   \leq  \sum_{m=0}^{\infty} p_{\mu_k | m} \nOphrv{m}    \leq  \nOmurv{k} + \sqrt{ \frac{\nOrv}{2} \ln( \frac{2}{\epsdecoy^2}) }\quad \\
                & \forall k \in \{1,2,3\} , \quad \forall O \in \{X_{\neq},X,K\} \Bigg) \geq  1-9 \epsdecoy^2.
			\end{aligned}
		\end{equation} 
		Let $\constraints$ denote the set of inequalities inside the probability in the above expressions. Therefore we have $\Pr(\constraints) \geq 1 - 9 \epsdecoy^2$.

\subsection{Bounds on zero and one photon statistics} 
		For any event $O \in \{X,X_{\neq} , K\}$, the relevant bounds on the zero-photon and single-photon components can be obtained by algebraic manipulation of the expressions in $\constraints$. In general, any method for bounding the relevant zero-photon and single-photon components using $\constraints$ suffices. In this work, we follow exactly the steps taken by Ref.~\cite[Appendix A]{lim_concise_2014} to obtain these bounds.  Thus, we only write the final expressions here. We define
		\begin{equation}\label{eq:decoydefexpressions} 
			\begin{aligned}
				\nOmurv{k}^{\pm} &\coloneq \frac{e^{\mu_k}}{p_{\mu_k}}  \left( \nOmurv{k} \pm \sqrt{ \frac{\nOrv}{2} \ln( \frac{2}{\epsdecoy^2}) }  \right)
			\end{aligned}
		\end{equation}

		The lower bound on the zero-photon component is given by \cite[Eq. 2]{lim_concise_2014}
		\begin{equation} \label{eq:decoyboundone}
			\constraints \implies \nOphrv{0} \geq \Bounddecoymin{0} (\nOmurv{\allk}) \coloneq \tau_0 \frac{ \mu_2 \nOmurv{3}^- - \mu_3 \nOmurv{2}^+} {\mu_2 - \mu_3}.
		\end{equation}
		
		The lower bound on the one-photon component is given by \cite[Eq. 3]{lim_concise_2014}
		
		\begin{equation} \label{eq:decoyboundtwo}
			\begin{aligned}
				\constraints \implies \nOphrv{1}  \geq \Bounddecoymin{1} (\nOmurv{\allk}) &\coloneq \left( \frac{\mu_1 \tau_1} {\mu_1 (\mu_2 - \mu_3) - \mu^2_2 + \mu_3^2}  \right) \times \\
				& \left(      \nOmurv{2}^- - \nOmurv{3}^+ - \frac{\mu_2^2 - \mu_3^2}{\mu_1^2} \left(  \nOmurv{1}^+ -   \Bounddecoymin{0} (\nOmurv{\allk}) / \tau_0   \right)            \right)  .
			\end{aligned}
		\end{equation}

		The upper bound on the one-photon component is given by \cite[Eq.  4]{lim_concise_2014}

		\begin{equation} \label{eq:decoyboundthree}
			\begin{aligned}
				\constraints \implies \nOphrv{1}  \leq \Bounddecoymax{1} (\nOmurv{\allk}) &\coloneq \tau_1 \frac{\nOmurv{2}^+ -\nOmurv{3}^-}{\mu_2 - \mu_3}.
			\end{aligned}
		\end{equation}

		Since $\Pr(\constraints) \geq 1 - 9 \epsdecoy^2$, and \cref{eq:decoyboundone,eq:decoyboundtwo,eq:decoyboundthree} follow from the expressions in $\constraints$, we obtain
		
		\begin{equation}
			\Pr \Big(   \eXphrv{1} \geq \frac{	\Bounddecoymax{1} (\nXneqmurv{\allk})  }{  	\Bounddecoymin{1} (\nXmurv{\allk})}
			\quad \lor \quad   \nXphrv{1} \leq    \Bounddecoymin{1}( \nXmurv{\allk})   \quad \lor \quad 
			\nKphrv{1} \leq   \Bounddecoymin{1}(\nKmurv{\allk}  ) \Big) \leq  9 \epsdecoy^2
		\end{equation}

    \subsection{Rigorous use of Hoeffdings}
    
    \label{app:decoyanalysis}

		In this section, we will rigorously justify the application of Hoeffdings concentration inequality \cite{hoeffding_probability_1963} in the decoy analysis of this work. To do so, we will first state the following general lemma.

		\begin{lemma} \label{lemma:hoeffdings}
			Let $\X_1 \dots \X_n$ be random variables. Let $X_i$ be a specific value taken by the random variable $\X_i$. For each $i$, a new random variable $\Y_i$ is generated from $X_i$ via the probability distribution $\Pr(\Y_i | X_i)$. Then
			\begin{equation}
				\Pr( \abs{ \sum_i (\Y_i )_{| \event{X_1 \dots X_{\nO}}}  - \mathrm{E}\left(\sum_i (\Y_i )_{| \event{X_1 \dots X_{\nO}}} \right) } \geq t ) \leq 2 \exp{\frac{-2t^2}{\sum_i (b_i-a_i)^2}}
			\end{equation}
			where $[a_i,b_i]$ denotes the range of $\Y_i$, and $\mathrm{E}$ denotes the expectation value. (Note that we do not require the $\X_i$s to be independent random variables, nor do we require the $\Y_i$s to be independent random variables).
		\end{lemma} 
		\begin{proof}
			Fix a specific sequence $X_1 \dots X_n$ of values taken by the random variables $\X_i$s.  The variables $\Y_i$, conditioned on this specific input $X_1 \dots X_n$, are then \textit{independent} random variables (since they are generated by $\Pr(\Y_i | X_i)$). Thus, Hoeffding's inequality applies.
			
		\end{proof}
		The above lemma is utilized to perform decoy analysis in the following lemma.
		\begin{lemma} \label{lemma:decoy}
			In the decoy-state QKD protocol of \cref{sec:decoy}, fix an outcome $O$, and intensity  $\mu_k$. 
			Then, we have
			\begin{equation}
				\Pr ( \abs{\nOmurv{k} - \sum_{m=0}^{\infty} p_{\mu_k | m} \nOphrv{m} }  \geq \sqrt{ \frac{\nOrv}{2} \ln( \frac{2}{\epsdecoy^2}) } ) \leq \epsdecoy^2.
			\end{equation}
		\end{lemma}
		
		\begin{proof}
			Consider all the rounds where $O$ is observed.  Condition on the event that $\nO$ such rounds are observed. Let $X_1 \dots X_{\nO}$ be the sequence of photon numbers of Alice's signals corresponding to these rounds. Condition further on the event that a specific sequence $X_1 \dots X_{\nO}$ is observed.

			Fix an intensity $\mu_k$ of interest. Define $\Y_i$ as
			\begin{equation}
				\Y_i \coloneq 
				\left\{
				\begin{array}{ll}
					1  & \mbox{if intensity $\mu_k$ is assigned to the $i$th round}   \\
					0 & \mbox{if intensity $\mu_k$ is not assigned to the $i$th round} .
				\end{array}
				\right.
			\end{equation}
			Since the intensity of each round is chosen from a probability distribution that only depends on the photon number of each round, each $\Y_i$ is generated independently via $\Pr(\Y_i | X_i)$. 
			By the construction of $\Y_i$s, $\sum_i \Y_i = \nOmurv{k}$. By the construction of $X_i$s, $ |\{  i | X_i = m\} | = \nOph{m}$. Then $ \mathrm{E}\left(\sum_i (\Y_i | X_i) \right)  = \sum_{m=0}^\infty p_{\mu_k | m} \nOph{m}$. Applying \cref{lemma:hoeffdings}, we directly obtain
			\begin{equation}
				\Pr( \abs{ \nOmurv{k}   - \sum_{m=0}^\infty p_{\mu_k | m} \nOph{m}  } \geq t )_{| \Omega(X_1 \dots X_{\nO},\nO) } \leq 2 \exp{\frac{-2t^2}{\nO}}.
			\end{equation}
			
			The above statement is valid for all $X_1 \dots X_{\nO}$ compatible with $\nO,\nOph{\vec{m}}$.  We now obtain a statement that only conditions on $\nO$ via
			
			\begin{equation}
				\begin{aligned}
					&\Pr( \abs{ \nOmurv{k}   - \sum_{m=0}^\infty p_{\mu_k | m} \nOphrv{m}  } \geq t )_{| \Omega(\nO) }  \hspace{-1em} \\ &= \sum_{X_1 \dots X_{\nO}} \Pr(\event{X_1 \dots X_{\nO}} | \event{\nO}) \Pr( \abs{ \nOmurv{k}   - \sum_{m=0}^\infty p_{\mu_k | m} \nOph{m}  } \geq t )_{| \Omega(X_1 \dots X_{\nO},\nO) }  \\
					&\leq \sum_{X_1 \dots X_{\nO}} \Pr(\event{X_1 \dots X_{\nO}} | \event{\nO})  2 \exp{\frac{-2t^2}{\nO}} \\
					&=   2 \exp{\frac{-2t^2}{\nO}} .
				\end{aligned}
			\end{equation}

			Setting $t=\sqrt{ \frac{\nO}{2} \ln( \frac{2}{\epsdecoy^2}) }$, we obtain
			\begin{equation}
				\Pr ( \abs{\nOmurv{k} - \sum_{m=0}^{\infty} p_{\mu_k | m} \nOphrv{m} }  \geq \sqrt{ \frac{\nO}{2} \ln( \frac{2}{\epsdecoy^2}) } )_{| \event{\nO}} \leq \epsdecoy^2,
			\end{equation}
			which directly implies the required statement.
			
		\end{proof}

\chapter{Proofs of statements related to delayed Authentication }
\label{appendix:delayedreductionproof}

In this chapter, we prove \cref{theorem:delayedreductionstatement}, using an approach that very closely mirrors the approach taken the prove \cref{theorem:reductionstatement}. To prove this result, we begin by defining the event $\Onicedel$ as the event in which Alice’s transcript $\transcript_A$ is such that it satisfies the verification conditions.
(Note that this refers to the actual transcript itself, not to whether Bob successfully receives it). Thus, $\Onicedel$ is already determined by the time $\coreQKDprotocol$ concludes.  
Intuitively, conditioned on the event $\Onicedel$, we may assume that authentication behaved honestly during the execution of $\coreQKDprotocol$. Analogous to \cref{lemma:bothabort}, we prove the following lemma, which states that if $\Onicedel$ does not occur, then both parties abort.

\begin{lemma} \label{lemma:delbothabort}
Let $\Onicedel$ be the event that Alice’s transcript $\transcript_A$ is such that it satisfies the verification conditions, and let $\rho^\mathrm{real,final}_{K_A K_B \Cfinal \Efinal}$ denote the final output state at the end of the full QKD protocol $\QKDprotocol$. Then the following equality holds
\begin{equation} \label{eq:delniceeventequality}
\rho^\mathrm{real,final}_{K_A K_B \Cfinal \Efinal | \Onicedel^\complement}= \idealmap \left[ \rho^\mathrm{real,final}_{K_A K_B \Cfinal \Efinal | \Onicedel^\complement} \right].
\end{equation}
Therefore, 
\begin{equation} \label{eq:delniceeventinequality}
\tracedist{\rho^\mathrm{real,final}_{K_A K_B \Cfinal \Efinal } - \rho^\mathrm{ideal,final}_{K_A K_B \Cfinal \Efinal }} =
    \tracedist{\rho^\mathrm{real,final}_{K_A K_B \Cfinal \Efinal \wedge \Onicedel} - \rho^\mathrm{ideal,final}_{K_A K_B \Cfinal \Efinal \wedge \Onicedel}} 
\end{equation}
\end{lemma}
\begin{proof}
    The proof follows in a similar manner as the proof of \cref{lemma:bothabort}, by noting that \nameref{prot:delauthpp} ensures that both parties abort when $\Onicedel^\complement$ occurs. 
\end{proof}

Analogous to \cref{lemma:commutationidealauth}, we now show that conditioned on the event $\Onicedel$, the final real and ideal output states (at the end of $\delayedQKDprotocol$) can be obtained by the action of $\delauthupdatemap \circ \delauthcommmap$ on the real and ideal states the end of the core QKD protocol $\coreQKDprotocol$.

\begin{lemma}[Commutation of $\idealmap$ and \nameref{prot:delauthpp}] \label{lemma:delcommutationidealauth}
Let $\Onicedel$ be the event in which Alice’s transcript $\transcript_A$ is such that it satisfies the verification conditions. Let $\rho^\mathrm{real}_{K_A K_B \CfinalQKD \EfinalQKD | \Onicedel}$ denote the real state at the end of the core QKD protocol  conditioned on $\Onicedel$. Let the following states denote its evolution through  \nameref{prot:delauthpp}: 
\begin{equation}
\begin{aligned}
    \rho^\mathrm{real,comm}_{K_A K_B \CfinalQKD \Cauth  \Efinal| \Onicedel} &\coloneq \delauthcommmap \left[ \rho^\mathrm{real}_{K_A K_B \CfinalQKD \EfinalQKD| \Onicedel} \right], \\
     \rho^\mathrm{real,final}_{K_A K_B \Cfinal\Efinal| \Onicedel} &\coloneq \delauthupdatemap \left[ \rho^\mathrm{real,comm}_{K_A K_B \CfinalQKD \Cauth  \Efinal| \Onicedel} \right].
   \end{aligned}
\end{equation}
Define the corresponding ideal states by the action of the map $\idealmap$ on the real states, i.e 
\begin{equation}
\begin{aligned}
   \rho^\mathrm{ideal}_{K_A K_B \CfinalQKD \EfinalQKD| \Onicedel} &\coloneq \idealmap \left[   \rho^\mathrm{real}_{K_A K_B \CfinalQKD \EfinalQKD| \Onicedel} \right], \\
   \rho^\mathrm{ideal,comm}_{K_A K_B \CfinalQKD \Cauth \Efinal| \Onicedel} &\coloneq \idealmap \left[   \rho^\mathrm{real,comm}_{K_A K_B \CfinalQKD \Cauth\Efinal| \Onicedel} \right], \\
     \rho^\mathrm{ideal,final}_{K_A K_B \Cfinal\Efinal| \Onicedel} &\coloneq \idealmap \left[  \rho^\mathrm{real,final}_{K_A K_B \Cfinal\Efinal| \Onicedel}  \right].
   \end{aligned}
\end{equation}
Then, the ideal states defined above are the same as those obtained by evolving  $ \rho^\mathrm{ideal}_{K_A K_B \CfinalQKD \EfinalQKD| \Onicedel}  $ through  the \nameref{prot:delauthpp}, i.e,
\begin{equation}
\begin{aligned}
    \rho^\mathrm{ideal,comm}_{K_A K_B \CfinalQKD \Cauth  \Efinal| \Onicedel} &= \delauthcommmap \left[ \rho^\mathrm{ideal}_{K_A K_B \CfinalQKD \EfinalQKD| \Onicedel} \right], \\
     \rho^\mathrm{ideal,final}_{K_A K_B \Cfinal\Efinal| \Onicedel} &= \delauthupdatemap \left[\rho^\mathrm{ideal,comm}_{K_A K_B \CfinalQKD \Cauth  \EfinalQKD| \Onicedel} \right].
     \end{aligned}
\end{equation} 
\end{lemma}
\begin{proof}
The fact that
\begin{equation}
    \delauthcommmap \circ \idealmap \left[ \rho^\mathrm{real}_{K_A K_B \CfinalQKD   \EfinalQKD| \Onicedel} \right] = \idealmap \circ \delauthcommmap \left[ \rho^\mathrm{real,comm}_{K_A K_B \CfinalQKD \EfinalQKD  | \Onicedel} \right]
\end{equation}
follows from the fact that $\delauthcommmap \in \CPTP(\CfinalQKD \EfinalQKD, \Cfinal \Efinal)$ and $\idealmap \in \CPTP(K_A K_B , K_A K_B)$: they act on disjoint registers. (This is analogous to the commutation corresponding to the second arrow in the proof of \cref{lemma:commutationidealauth}).

The fact that
\begin{equation}
    \delauthupdatemap \circ \idealmap \left[ \rho^\mathrm{real,comm}_{K_A K_B \CfinalQKD \Cauth  \Efinal| \Onicedel} \right] = \idealmap \circ \delauthupdatemap \left[ \rho^\mathrm{real,comm}_{K_A K_B \CfinalQKD \Cauth  \Efinal| \Onicedel} \right]
\end{equation}
follows from analogous arguments to the commutation corresponding to the third arrow in the proof of \cref{lemma:commutationidealauth}. In particular, we consider each possible combination of final accept / abort decisions, and find that the required commutation holds for all of them.
\end{proof}

Using these results, we obtain the following corollary which is analogous to \cref{corr:reduction}.

\begin{corollary} \label{corr:delayedreduction}
Consider the same setting as in \cref{theorem:delayedreductionstatement}. The $\epssecure$-security for all output states in $\worlddelreal(\coreQKDprotocol)$ partial on $\Onicedel$, implies $\epssecure$-security for all output states in $\worlddelreal(\delayedQKDprotocol)$. That is 
  \begin{equation}
      \begin{aligned}
    \tracedist{\rho^\mathrm{real}_{K_A K_B \CfinalQKD \EfinalQKD \wedge \Onicedel }  - \rho^\mathrm{ideal}_{K_A K_B \CfinalQKD \EfinalQKD \wedge \Onicedel} }  &\leq \epssecure \qquad \forall \rho^\mathrm{real}_{K_A K_B \CfinalQKD \EfinalQKD } \in  \worlddelreal(\coreQKDprotocol) \\
    &\Downarrow \\
     \tracedist{\rho^\mathrm{real,final}_{K_A K_B \Cfinal \Efinal }  - \rho^\mathrm{ideal,final}_{K_A K_B \Cfinal \Efinal } }  &\leq \epssecure \qquad \forall \rho^\mathrm{real,final}_{K_A K_B \Cfinal \Efinal } \in  \worlddelreal(\delayedQKDprotocol) 
      \end{aligned}
  \end{equation}
\end{corollary}
\begin{proof}
    The claim follows from \cref{lemma:delbothabort,lemma:delcommutationidealauth} using arguments identical to those used in the proof of \cref{corr:reduction}.
\end{proof}

We then construct the virtual authentication setting in exactly the same manner as before (see \cref{subsec:virtualsetting}). In the virtual setting, Eve can do any operation she wants on the classical messages, and the correct messages are delivered at correct timings in the special registers $\bm{\Ccorr}$. Moreover, Alice and Bob use these special registers for further steps in the QKD protocol. We then obtain the following lemma. (Note that \textit{technically}, this virtual setting differs slightly from the one considered in \cref{subsec:virtualsetting}, as the classical communication here is unauthenticated. However, this technical distinction does not affect the proof steps or underlying logic.)

\begin{lemma} \label{lemma:delayedreductionone}
        Consider the same setup as in \cref{theorem:delayedreductionstatement,corr:delayedreduction}. Then, the $\epssecure$-security for all output states in $\worldvirtual(\coreQKDprotocol)$ partial on $\Onicedel$, implies $\epssecure$-security for all output states in $\worlddelreal(\coreQKDprotocol)$ partial $\Onicedel$. That is 
  \begin{equation}
      \begin{aligned}
    \tracedist{\rho^\mathrm{real,virt}_{K_A K_B \CfinalQKD \bm{\Ccorr} \EfinalQKD \wedge \Onicedel }  - \rho^\mathrm{ideal,virt}_{K_A K_B \CfinalQKD \bm{\Ccorr}  \EfinalQKD \wedge \Onicedel} }  &\leq \epssecure \qquad \forall \rho^\mathrm{real,virt}_{K_A K_B \CfinalQKD \bm{\Ccorr} \EfinalQKD } \in  \worldvirtual(\coreQKDprotocol) \\
    &\Downarrow \\
     \tracedist{\rho^\mathrm{real}_{K_A K_B \CfinalQKD \EfinalQKD \wedge \Onicedel}  - \rho^\mathrm{ideal}_{K_A K_B \CfinalQKD \EfinalQKD \wedge \Onicedel} }  &\leq \epssecure \qquad \forall \rho^\mathrm{real}_{K_A K_B \CfinalQKD \EfinalQKD  } \in  \worlddelreal(\coreQKDprotocol) 
      \end{aligned}
  \end{equation}
\end{lemma}
\begin{proof}
  The claim follows by arguments identical to those used in the proof of \cref{lemma:reductionone}.
\end{proof}

The following lemma reduces the security analysis to states in $\worldvirtual(\coreQKDprotocol)$.

\begin{lemma} \label{lemma:delayedreductiontwo}
 Consider the same setup as in \cref{theorem:delayedreductionstatement,corr:delayedreduction}. Then, the $\epssecure$-security for all output states in $\worldvirtual(\coreQKDprotocol)$, implies $\epssecure$-security for all output states in $\worldvirtual(\QKDprotocol)$ partial on $\Onicedel$. That is 
\begin{equation}
      \begin{aligned}
\tracedist{\rho^\mathrm{real,virt}_{K_A K_B \CfinalQKD \bm{\Ccorr} \EfinalQKD }  - \rho^\mathrm{ideal,virt}_{K_A K_B \CfinalQKD \bm{\Ccorr}  \EfinalQKD } }  &\leq \epssecure \qquad \forall \rho^\mathrm{real,virt}_{K_A K_B \CfinalQKD \bm{\Ccorr} \EfinalQKD } \in  \worldvirtual(\coreQKDprotocol) \\
    &\Downarrow \\
    \tracedist{\rho^\mathrm{real,virt}_{K_A K_B \CfinalQKD \bm{\Ccorr} \EfinalQKD \wedge \Onicedel }  - \rho^\mathrm{ideal,virt}_{K_A K_B \CfinalQKD \bm{\Ccorr}  \EfinalQKD \wedge \Onice} }  &\leq \epssecure \qquad \forall \rho^\mathrm{real,virt}_{K_A K_B \CfinalQKD \bm{\Ccorr} \EfinalQKD  } \in  \worldvirtual(\coreQKDprotocol) \\
      \end{aligned}
  \end{equation}
\end{lemma}
\begin{proof}
    The claim follows by arguments identical to those used in the proof of \cref{lemma:reductiontwo}.
\end{proof}

The following lemma reduces the security analysis to states in $\worldhonest(\coreQKDprotocol)$.

\begin{lemma} \label{lemma:delayedreductionthree}
Consider the same setup as in \cref{theorem:delayedreductionstatement,corr:delayedreduction}. Then, the $\epssecure$-security for all output states in $\worldhonest(\coreQKDprotocol)$  implies $\epssecure$-security for all output states in $\worldvirtual(\coreQKDprotocol)$. That is 
\begin{equation}
      \begin{aligned}
\tracedist{\rho^\mathrm{real,hon}_{K_A K_B \CfinalQKD  \EfinalQKD }  - \rho^\mathrm{ideal,hon}_{K_A K_B \CfinalQKD   \EfinalQKD } }  &\leq \epssecure \qquad \forall \rho^\mathrm{real,virt}_{K_A K_B \CfinalQKD  \EfinalQKD } \in  \worldhonest(\coreQKDprotocol) \\
    &\Downarrow  \\
\tracedist{\rho^\mathrm{real,virt}_{K_A K_B \CfinalQKD \bm{\Ccorr} \EfinalQKD  }  - \rho^\mathrm{ideal,virt}_{K_A K_B \CfinalQKD \bm{\Ccorr}  \EfinalQKD } }  &\leq \epssecure \qquad \forall \rho^\mathrm{real,virt}_{K_A K_B \CfinalQKD \bm{\Ccorr} \EfinalQKD  } \in  \worldvirtual(\coreQKDprotocol) \\
      \end{aligned}
  \end{equation}
\end{lemma}
\begin{proof}
    The claim follows by arguments identical to those used in the proof of \cref{lemma:reductionthree}.
\end{proof}

Bringing it all together, we obtain the proof of the required reduction statement.

\delayedreductionstatement*
\begin{proof}
 The proof follows from \cref{corr:delayedreduction,lemma:delayedreductionone,lemma:delayedreductiontwo,lemma:delayedreductionthree}.
\end{proof}

\chapter{Miscellaneous}
This appendix collects various results and observations from across the thesis. \label{Appendix:misc}
\section{Proof of the Correctness and Secrecy Decomposition of the Security Definition \texorpdfstring{(\cref{subsec:correctnessandsecrecy})}{(see subsection)}}
\correctnessandsecrecy*
\begin{proof} The proof is just simple algebra, and can be found in many works \cite{renner_security_2005,portmann_security_2022}. For a version explicitly written for variable-length protocols, see \cite[Lemma 11]{tupkary_security_2024}. We include it here for the sake of completeness. 

Consider the same setting as in \cref{def:qkdcorrectnessandsecrecysymmetric,def:qkdsecuritysymmetric}, where $\mathcal{W}(\QKDprotocol)$ denotes the set of all possible output states for the given protocol, and let $\rho^\mathrm{real}_{K_AK_B \Esecdef } \in \mathcal{W}(\QKDprotocol) $ denote the output state, and let $\Omega_{\mathrm{len}=l}$ denote the event that a final key of length $l$ is produced. Fix the real output state $\rho^\mathrm{real}_{K_AK_B \Esecdef }$.
Let us write  
   \begin{equation}
   \begin{aligned}
\rho^\mathrm{real}_{K_AK_B  \Esecdef| \Omega_{\mathrm{len}=l} } &=\sum_{k_A , k_B \in\{0,1\}^l } \Pr(k_A ,k_B| \Omega_{\mathrm{len}=l} ) \ketbra{k_A,k_B}_{K_A K_B} \otimes \rho^{(k_A,k_B)}_{\Esecdef| \Omega_{\mathrm{len}=l}} \\
\rho^\mathrm{correct}_{K_AK_B  \Esecdef| \Omega_{\mathrm{len}=l} } &=\sum_{k_A , k_B \in\{0,1\}^l } \Pr(k_A ,k_B| \Omega_{\mathrm{len}=l} ) \ketbra{k_A,k_A}_{K_A K_B} \otimes \rho^{(k_A,k_B)}_{\Esecdef| \Omega_{\mathrm{len}=l}} \\
\rho^\mathrm{ideal}_{K_AK_B  \Esecdef| \Omega_{\mathrm{len}=l} } &=\sum_{k_A  \in\{0,1\}^l } \frac{1}{2^k} \ketbra{k_A,k_A}_{K_A K_B} \otimes \rho^\mathrm{real}_{\Esecdef|\Omega_{\mathrm{len}=l}},
   \end{aligned}
   \end{equation}
where the ideal output state written above can be easily verified to be the output of the $\idealmap$ acting on the real output state.  Here the states $\rho^{(k_A,k_B)}_{\Esecdef| \Omega_{\mathrm{len}=l}}$ are simply Eve's conditional states when the key registers hold keys $k_A, k_B$ respectively, each of length $l$. The states labeled ``correct" are obtained by simply replacing Bob's key with Alice's key in the real output state.
Then, using the triangle inequality, the trace distance from the security condition can be upper bounded as
\begin{equation} \label{eq:secrecycorrectsplit}
\begin{aligned}
 & \sum_l \Pr(\Omega_{\mathrm{len}=l} ) \tracedist{ \rho^\mathrm{real}_{K_AK_B  \Esecdef| \Omega_{\mathrm{len}=l} } - \rho^{\mathrm{ideal}}_{K_A K_B  \Esecdef| \Omega_{\mathrm{len}=l} } } \\
 &\leq  \sum_l \Pr(\Omega_{\mathrm{len}=l} ) \tracedist{ \rho^\mathrm{real}_{K_AK_B  \Esecdef| \Omega_{\mathrm{len}=l} } - \rho^{\mathrm{correct}}_{K_A K_B  \Esecdef| \Omega_{\mathrm{len}=l} } } \\
 &+ \sum_l \Pr(\Omega_{\mathrm{len}=l} ) \tracedist{ \rho^\mathrm{correct}_{K_AK_B  \Esecdef| \Omega_{\mathrm{len}=l} } - \rho^{\mathrm{ideal}}_{K_A K_B  \Esecdef| \Omega_{\mathrm{len}=l} } }
    \end{aligned}
\end{equation}
The first term in \cref{eq:secrecycorrectsplit} can be upper bounded by the correctness condition as follows:
\begin{equation}
    \begin{aligned}
        &\sum_l \Pr(\Omega_{\mathrm{len}=l} ) \tracedist{ \rho^\mathrm{real}_{K_AK_B  \Esecdef| \Omega_{\mathrm{len}=l} } - \rho^{\mathrm{correct}}_{K_A K_B  \Esecdef| \Omega_{\mathrm{len}=l} } } \\
        &\leq \sum_l \Pr(\Omega_{\mathrm{len}=l} ) \sum_{k_A , k_B \in\{0,1\}^l }  \Pr(k_A, k_B |  \Omega_{\mathrm{len}=l} ) \\
        &\times \tracedist{\ketbra{k_A,k_B}_{K_A K_B} \otimes \rho^{(k_A,k_B)}_{\Esecdef| \Omega_{\mathrm{len}=l}}  - \ketbra{k_A,k_A}_{K_A K_B} \otimes \rho^{(k_A,k_B)}_{\Esecdef| \Omega_{\mathrm{len}=l}} } \\
        & \leq \sum_l \Pr(\Omega_{\mathrm{len}=l} ) \sum_{k_A , k_B \in\{0,1\}^l, k_A \neq k_B }  \Pr(k_A, k_B |  \Omega_{\mathrm{len}=l} )  \\
        &= \Pr(K_A \neq K_B) \leq \epscorr,
    \end{aligned}
\end{equation}
where we apply the triangle inequality for the first inequality. In the second inequality, we replace the one-norm with $0$ when $k_A = k_B$ and with $2$ otherwise. The  equality in the final line then follows from basic properties of probability, and the final inequality follows form $\epscorr$-correctness of the protocol

The second term in \cref{eq:secrecycorrectsplit} is identical to the left-hand-side \cref{eq:tracedistsecrecy}, since $K_A = K_B$ and hence $K_B$ can be traced out without affecting the one-norm.
 Thus, upper bounding it by $\epssecret$ is identical to  the secrecy requirement. 

 Thus, we have shown that for all $\rho^\mathrm{real}_{K_AK_B \Esecdef } \in \mathcal{W}(\QKDprotocol) $, 
 \begin{equation}
\tracedist{\rho^\mathrm{real}_{K_AK_B \Esecdef }  - \rho^\mathrm{ideal}_{K_AK_B \Esecdef } } \leq \epscorr + \epssecret,
 \end{equation}
 which is identical to the required security statement. This concludes the proof. 
\end{proof}

\section{Explanation of Decoy-State Plots from \texorpdfstring{\cref{sec:variableapplicationtoDecoybb84}}{the referenced section}}
To perform finite-size security proof for decoy-state protocols against IID collective attacks using Ref.~\cite{Kamin2025}, we perform the following steps.
\begin{itemize}

\item 
We use the tagging source map from \cref{subsec:taggingsourcemap} and instead consider a protocol in which Alice emits the tagged states defined in \cref{eq:taggedStates}, which we recall here for convenience:
\begin{equation}
   ( \xi_{(a,\mu,\testgenflag)} )_{A'} 
   = 
   \sum_{N=0}^{\ndecoy} \poissonian{\mu}{N}\ketbra{N}_a 
   + 
   \left(1-\sum_{N=0}^{\ndecoy} \poissonian{\mu}{N}\right) 
   \ketbra{a,\mu}. 
\end{equation}
where $\poissonian{\mu}{N}   \coloneq e^{-\mu}\frac{\mu^N}{N!} $ is the poissonian distribution.
Source maps are required when we apply the postselection technique in \cref{chap:postselection}; we defer a rigorous justification for their use to that discussion.

\item 
Second, we employ the source-replacement scheme. 
The block-diagonal structure of the states allows us to write the following shared state:
\begin{equation}\label{eq:Decoy shared state}
	\sigma_{\Ameas \Ashield} 
	=
	\sum_{N = 0}^{\ndecoy} \Pr(N) \ketbra{N}_{\Ashield} \otimes \sigma_{A}^{(N)} 
	+ 
	\Pr(\mathrm{flag}) \ketbra{\mathrm{flag}}_{\Ashield} \otimes \sigma_{A}^{(\mathrm{flag})},
\end{equation}
where $\Pr(N)$ denotes the probability that Alice emits an $N$-photon pulse, and $\sigma_{A}^{(N)}$ is the source-replaced state corresponding to that emission event. We set $N=\mathrm{flag}$ when Alice emits a pulse containing more than $\ndecoy$ photons, which, after the application of the source map, manifests as the output of a classical flag.

\item In fact, we can treat the protocol as one where Alice first decides to prepare an $N$-photon pulse according to some known probability distribution, \emph{then} samples the encoding choice $a$, prepares and sends that state, and \emph{only afterward} assigns the pulse a particular intensity $\mu$ or a $\test,\gen$ label. We use $y$ to denote Bob's outcome. Alice assigns her labels using the distribution  
\begin{equation}
   \Pr(\test,\mu | a,y) =  \Pr(\test,\mu | a) = \Pr(\test,\mu,a) / \Pr(a).
\end{equation}
where the first equality follows from the fact that Alice decides the $\test,\gen$ label. Thus, we have
\begin{equation}
   \Fobs_{a,\mu,\test,y} \approx  \Pr(a,\mu,\test,y) = \Pr(\test,\mu | a,y)  \Tr[\Gamma_{a,y} \rho]
\end{equation}
where $\Gamma_{a,y}$ the POVM corresponding to Alice sending state $a$, and Bob measuring outcome $y$.\footnote{In this case, the $\test,\gen$ flags do not appear in the source-replacement scheme directly, but instead appear as classical postprocessing by Alice. Both perspectives are equivalent.}

\item 
We then turn to the decoy-state analysis. For the time being, we fix a true state $\rho_{AB}$ describing Eve’s attack, and let $y$ denote Bob’s measurement outcome (Recall that Alice's label for her state $x=(a,\mu,\testgenflag))$. 
Alice’s setting choice is specified by $(a,\mu)$. 
We define
\begin{equation}
	\widehat{Y}_N^{y|a} \coloneq \Pr(y|a,N,\test)_\rho = \Pr(y|a,N,\test,\mu)_\rho ,
	\qquad \forall N \in \{0,\dots,\ndecoy\}, N = \mathrm{flag}
\end{equation}
and obtain
\begin{equation}
    \begin{aligned}
        \Pr(y|a,\mu,\test) 
        &= 
        \sum_{N=0}^{\ndecoy} \Pr(N|a,\mu,\test)\Pr(y|a,N,\mu,\test) \\
        &+  
\Pr(\mathrm{flag}|a,\mu,\test)\Pr(y|a,\mathrm{flag},\mu,\test) \\
        &=
        \sum_{N=0}^{\ndecoy} \Pr(N|a,\mu,\test) \widehat{Y}_N^{y|a} 
        + 
\Pr(\mathrm{flag}|a,\mu,\test) \widehat{Y}_{\mathrm{flag}}^{y|a},
    \end{aligned}
\end{equation}
which holds for all intensities $\mu$. 
This system of linear equations can be solved via linear programming by imposing the bounds
\(
0 \le \widehat{Y}_N^{y|a}, \widehat{Y}_{\mathrm{flag}}^{y|a} \le 1
\),
provided that the quantities $\Pr(y|a,\mu,\test)$ are accessible. Note that the remaining terms $\Pr(N|a,\mu,\test),\Pr(\mathrm{flag}|a,\mu,\test)  $ are known exactly.

\item 
Finally, the probabilities $\Pr(y|a,\mu,\test)$ are related to the observed frequencies of the announcements $\Fobs_k$, including their worst-case values as obtained from the set constructions introduced earlier. In particular, we have 
\begin{equation}
F_{(a, \mu, \test, y)}  \approx  \Pr(y|a,\mu,\test) \Pr(a,\mu,\test).
\end{equation}

\end{itemize}

Putting everything together yields the following key-rate formula. 
Two remarks are in order. 
First, the decoy-state analysis in Ref.~\cite{Kamin2025} is carried out without the use of source maps, but it can be straightforwardly adapted to incorporate them. 
In any case, the analysis without source maps provides a valid lower bound on the one that includes source maps.\footnote{The differences arise from a modified treatment of terms corresponding to pulses containing more than $N_\mathrm{ph}$ photons. 
Since such events occur with very small probability, their impact on the key rate is negligible.}
Second, Ref.~\cite{Kamin2025} also derives an improved second-order correction term, which we do not include here. 
Subject to these modifications, we obtain the following theorem.

\begin{corollary}(Variable-length Decoy-state BB84 \cite{Kamin2025}) \label{cor:varibalelengthdecoystatebb84}
Consider the variable-length decoy-state BB84 variant of the \nameref{prot:abstractqkdprotocol} as specified in \cref{sec:statesandmeasurements}, with the confidence set $\variableset(\Fobs)$   (which depends on $\epsAT$)  as constructed in \cref{lemma:constructingconfidenceset}. Then this variable-length deocy-state BB84 protocol is $(\epsAT+\epsPA + \epsEV)$-secure against IID collective attacks, as long as the $\lkey(\Fobs)$ satisfies\footnote{See Ref.~\cite{Kamin2025} for an analysis that improves the second term in the key rate expression to  $n_{\mathrm{sift}}(\alpha-1)\log^2(1+2\dim(S))$, where $n_{\mathrm{sift}}$ denotes the number of key generation rounds that are not discarded and are actually used for key generation. This refinement effectively increases the key rate in the high-loss regime, where most rounds are discarded, while yielding only a negligible improvement in low-loss regimes, since this term is already small compared to the leading-order contribution. Both fixed-length and variable-length results can be obtained using this improved bound, but we do not pursue this here.}
  \begin{equation} \label{eq:lvariablevaluedecoybb84IID}
        \begin{aligned}
            \lkey(\Fobs) &\leq \Bigg \lfloor
                n \inf_{\nu^{(1)} \in \Sigma^{(1)} (\Fobs) } \Pr(N=1)
                H(S | \CP E)_{\nu^{(1)}} - n(\alpha-1)\log^2(1+2\dim(S))
                - \leakfixed \\
                &
                - \EVcost 
                - \frac{\alpha}{\alpha - 1}
                 \log\left( \frac{1}{ \epsPA} \right) 
                  + 2 \Bigg\rfloor,\\
    \Sigma^{(1)}(\Fobs) &\coloneq \big\{ \QKDGmap{}[\omega^{(1)}_{ABE}] \; \mid  \; \omega^{(1)}_A \in \dop{=}(AB), \omega^{(1)}_A = \sigma^{(1)}_A,  \\
    & \Pr(a, \test | N=1) \widetilde{Y}^{y|a}_{N=1,L} \leq \Pr(\test | a,N=1) \Tr[\Gamma_k \omega^{(1)}_{AB}]   \leq  \Pr(a, \test | N=1) \widetilde{Y}^{y|a}_{N=1,L}\big\}
        \end{aligned}
    \end{equation}
where $\alpha \in (1,1+1/\log(1+2\dim(S)))$, and $ \widetilde{Y}^{y|a}_{N,L}, \widetilde{Y}^{y|a}_{N,L}$ are given by
\begin{equation}\label{eq:LP decoy}
    \begin{aligned}
    &    \widehat{Y}_{N,L}^{y|a} := \min_{\{\mathbf{\widehat{Y}}_m\}_m}  \widehat{Y}^{y|a}_m, \qquad  \qquad \widehat{Y}_{N,L}^{y|a} := \min_{\{\mathbf{\widehat{Y}}_m\}_m}  Y^{y|a}_N \\
        \textrm{s.t.}\; & \frac{\Fobs_{(a,\mu,\test,y)}  - \kappalowervariable{(a,\mu,\test,y)}}{\Pr(a,\mu,\test)} \leq \sum_{m=1}^{\ndecoy} \poissonian{\mu}{m}  \widehat{Y}_m^{y|a}   + \left( 1- \sum_{m=0}^{\ndecoy} \poissonian{\mu}{m} \right) \widehat{Y}^{y|a}_\mathrm{flag}, \\
        &\frac{\Fobs_{(a,\mu,\test,y)} + \kappalowervariable{(a,\mu,\test,y)}}{\Pr(a,\mu,\test)} \geq \sum_{m=1}^{\ndecoy}  \poissonian{\mu}{m} \widehat{Y}_m^{y|a} + \left( 1- \sum_{m=0}^{\ndecoy} \poissonian{\mu}{m} \right) \widehat{Y}^{y|a}_\mathrm{flag}, \\
        &\forall  (a,\mu,\test,y) \in \mathcal{\CP}_\mathrm{test},  \\
        &0 \leq Y^{y|a}_m \leq 1 \; \forall m \in \{1,\dots,\ndecoy,\mathrm{flag} \}.
    \end{aligned}
\end{equation}

\end{corollary}

\cleardoublepage
\phantomsection		

\end{document}